# Hill's Lunar Equations, Series, Convergence, Motion of the Perigee

Thomas S. Ligon, ORCID 0000-0002-4067-876X
2025-11-28

## Abstract

We investigate Hill's lunar equations, series and the motion of the perigee, and we use computers to go farther than has previously been known, calculating the coefficients of Hill's series up to order 24 in $m$, and the coefficients that do not depend on $a_0$ up to order 30. Numerical calculations indicate that the radius of convergence of Hill's series is somewhere near the value of $m$ of the cusped orbit ($\approx 0.560958$), which we formulate as a conjecture. We calculate the motion of the perigee using a linearization of the equation for the anomalistic period, as in Hill's documentation, but with some discrepancies.

## Contents

## 1. Introduction

In 1878, G. W. Hill introduced a new method for calculating lunar orbits that takes three bodies (moon, earth, and sun) into account, but removes the explicit reference to the sun (G.W. Hill 1878). The influence of the sun is moved into the potential energy part of the equations. After this, Hill makes an important step and, instead of using Kepler orbits as a first approximation, uses symmetric, closed orbits (G. Hill 1895). In order to investigate these orbits, the equations are applied to Fourier series. These series result in Hill's series and an infinite set of infinite sums for finding the coefficients of the series, which are power series on the ratio of the period of the moon to the period of the earth, which Hill refers to as $m$. However, Hill was not able to prove the convergence of these series within the scope of the original papers. In 1896, A. Lyapunov proved the convergence of Hill's series for $m = 0.08084\ 89338\ 08312$, the value for the earth's moon (Lyapunov 2020). Independently, in 1929, A. Wintner proved convergence for $m = 1/12$ (Wintner 1929). Later, Merman and Petrovskaya extended this work to cover cases of $|m| = 2.1$ (Petrovskaya 1963). Proofs for positive values of $m$ refer to direct orbits, whereas negative $m$ refers to retrograde orbits. The earth's moon has a direct orbit, meaning that it revolves around the earth in the same direction as the earth's revolution around the sun. Examples of direct (prograde) orbits are the Earth's moon and the moons of Mars (Phobos and Deimos), whereas retrograde orbits include Saturn's Phoebe and Neptune's Triton[1].

　　Immediately after publication, Hill's equations and series generated enthusiasm among astronomers because the series converge much faster than those previously used, for example the series introduced by Delaunay (G.W. Hill 1878), page 142. In fact, the power series based on $m$ are a significant part of Hill's contribution, and the alternative methods and their justification are discussed in the paper. In addition, Hill's work significantly influenced the work of mathematicians, including Poincaré. In fact, Poincaré proved the existence of a continuum of periodic solutions and that the periodic solution can be used as a first approximation, or intermediate orbit (Poincaré 1892) and (Barrow-Green 1997), page 43 and provided a solid mathematical foundation for infinite matrices, proving the convergence of the determinant of an infinite matrix (Poincaré 1886) and (Barrow-Green 1997), page 27. Another of Hill's innovations is the examination of curves and surfaces of zero velocity, which can be used to show that the orbit of the moon cannot escape from the Earth(G.W. Hill 1878), page 23, (Barrow-Green 1997), page 24, and (Szebehely 1963).

　　Today, computers can be used to numerically solve differential equations, making it possible to calculate lunar orbits including three or more bodies and other factors such as tides and the distribution of the mass of the earth, so calculations using power series are less attractive. On the other hand, Hill's lunar equations are still fascinating for mathematicians, because they are an example of a "minimal" three-body system. For example, even though Hill's equations explicitly encompass only two spatial dimensions, they are nevertheless capable of generating chaotic solutions(Simó and Stuchi 2000; Waldvogel and Spirig 1995). Hill's equations also provide a basis for investigating other open questions of the three-body problem, such as how to find closed orbits, or the Birkhoff conjecture (Birkhoff 1915), page 328 and (Frauenfelder, van Koert, and Zhao 2016), page 3.

　　In section 2. Notation and Nondimensionalization we introduce some notation and state the scaling assumptions made to achieve nondimensionalization. This is all well-known, but we chose to state it explicitly in hopes of making the transition between astronomy and pure mathematics easier.

　　The basic model of this paper, Hill's Lunar Problem, is derived in section 3. Hill's Lunar Equations. We have chosen to base this on the exposition in (Frauenfelder and van Koert 2018), because it is a very modern formulation that preserves the symplectic differential geometry of the problem. The derivation begins with the restricted three-body problem and "moves the sun to infinity" while modifying the other parameters in such a way that the meaning is preserved.

---

[1] An extensive list, which links to more detailed descriptions, can be found at
https://en.wikipedia.org/wiki/Category:Moons_with_a_retrograde_orbit.

The result is a set of differential equations for a two-body problem (earth and moon), where the effect of the sun is now part of the Hamiltonian.

In section 4. Hill's Series we present the derivation of Hill's series, closely following Hill's original exposition(G.W. Hill 1878). Based on Hill's decision to investigate symmetric, periodic orbits, the derivation begins by applying a Fourier expansion (using only odd terms) of $q_1$ and $q_2$. Then, these variables are combined into a complex variable, using Euler's formula. Based on this, Hill's (differential) equations, including an equation for the energy, or Jacobi's integral, can be applied to the terms of the Fourier expansion. The resulting equations are then combined in a few clever ways to produce what is known as Hill's equation, an infinite set of infinite sums that determine the coefficients of the original Fourier expansion. The coefficients of this set of equations are quotients of second-order polynomials. In Hill's exposition, they are denoted by special symbols involving round and square brackets, but we chose to use a functional notation instead, because that fits much better with modern mathematical notation and is also much easier to use in software. Finally, by making a Taylor expansion of these functions, we arrive at set of equations that determine the coefficients of the Fourier expansion as power series in $m$. To be precise, we also need to know that the coefficients of the Fourier expansion can be written as power series in $m$, but that is proven in the book by Wintner (Wintner 2014). This gives us an infinite set of infinite sums that determine the coefficients of $m^n$ in Hill's series.

The next step in finding a formal solution to find these coefficients is presented in section 5. Wintner's Recursion Formula, which makes it possible to find the coefficient of $m^n$ in Hill's series in terms of the $m^{n-1}$. This is based on Wintner (Wintner 2014). The result of this step is a formulation of the equations in terms of the individual coefficients of the Fourier expansion, instead of the full (partial) sums. This is a big reduction in the computation required for a software implementation.

The equations still contain infinite sums, but we have observed that the coefficients with an index above a certain number are all zero. We formulate that limit and prove it in section 6. Transition to Finite Sums. As a result, we can now express Hill's equations as a recursive procedure for solving an infinite set of finite sums. This formulation is something that can be implemented directly in software. After this, section 7. Initial conditions briefly covers initial conditions.

A very important part of this paper is covered in section 8. The motion of the perigee, which is based on two of Hill's papers. The first is Hill's paper on the motion of the perigee (G.W. Hill 1886), which we sometimes refer to as Hill's numeric perigee, and the second is (G.W. Hill 1894), which we like to call Hills literal perigee. Hill's numeric perigee develops the equation for the motion of the perigee (to the first order), based on the use of a symmetric, closed orbit as a first approximation. In (Meyer and Schmidt 1982) "Hill proposed to construct a lunar theory by first finding a periodic solution of the system defined by the Hamiltonian (19) and then continue this solution into the full problem." In solving the equation, the variable $m$ is converted to its numerical value very early, thus reducing the number of calculations required. In contrast, Hill's literal perigee keeps $m$ as a (literal) parameter until the very last step, which make the results appropriate for moons other that the earth's moon, and provides a more general approach, interesting not only to astronomers, but also to mathematicians. The two papers use a different nomenclature: Hill's numerical perigee bases the equations of dynamics on the effective potential, whereas Hill's literal perigee uses expressions that refer directly to the spatial coordinates, as in the presentation in sections 3 and 4. We have chosen to present this material in one uniform style, and whenever the formulas diverge from those in Hill's numeric perigee we note how to convert between the two styles. The 1886 paper Hill's numeric perigee is important because it makes it possible to compare highly precise observations of the moon with the available theory, and it is also important because it makes use of two techniques, the symmetric, closed orbit as a first approximation and infinite matrices and determinants, which stimulated the development of mathematics (Poincaré 1886).

Now that we have established all of Hill's equations and their series expansions, both for the symmetric, closed orbits and for the motion of the perigee, we make a number of plausibility checks in section 9. Some Plausibility Checks. These checks consist mainly of comparing our results to results calculated (manually) by Hill, or by checking to see if the coefficients in our calculations actually solve the equations they were intended for. After calculations of the series coefficients and checking plausibility of them, we turn to an examination of their plausibility in section 9.11. Bourne's calculations of coefficients.

Now, after this check of our calculations of a number of series, we have devoted the next chapter to looking at the convergence of the series, see section 10. Hill's Series Convergence. This begins with an easy case and then goes into some techniques used for checking the convergence. Then we look at a number of selected orbits, using numerical evidence to see how far the series converge. Finally, we make a statement about the radius of convergence of Hill's series.

Finally, we address the calculation of the motion of the moon's perigee in section 11. Calculations of the motion of the moon's perigee. Here, we explain exactly how we made the calculations, and where this diverges from Hill's calculations. The details of the calculations are included in appendices.

## 2. Notation and Nondimensionalization

Here we summarize the notation used in this manuscript. We have followed the original as much as possible but have introduced some differences in order to avoid conflicts. The discussion of nondimensionalization has been made explicit with the goal of making it easier to apply the results to concrete examples in celestial mechanics.

When considering three bodies, we always refer to the largest as the *sun*, the next one as the *earth*, and the smallest as the *moon*, even though our intention is to encompass many different situations, not only the earth and its moon. In the restricted three-body problem, the sun is often referred to as the first primary, the earth as the second primary, and the moon as the secondary, or satellite.

$a'$: The distance to the sun[2]. We will need this symbol only when comparing results to Hill's paper.

$m$: ratio of the sidereal month to the sidereal year.[3] This is used throughout this document, not only for the earth's moon, but for all other celestial objects that fit this model.

$\mu$: In Hill (G.W. Hill 1878), this is the sum of the masses of the earth and the moon, assumed to be 1, i.e., $m_e + m_m = 1$[4]. As a result, our unit of mass is $m_e + m_m$. We will need this symbol only when comparing results to Hill's paper. In the restricted three-body problem, $\mu$ refers to the mass of the moon, $m_m$, and we will use this meaning of $\mu$ (only) in 3. Hill's Lunar Equations.

$n$: sidereal mean motion of the moon.[5] We will need this symbol only when comparing results to Hill's paper.

$n'$: mean angular velocity of the sun around the earth[6]. This is also set to 1, meaning that our unit of time is the period of the earth (a year) divided by $2\pi$, i.e., $\frac{T_e}{2\pi}$. We will need this symbol only when comparing results to Hill's paper.

$\nu$: $\frac{1}{m}$[7]. $\frac{2\pi}{\nu}$ is defined as the period of a closed orbit, so the period is $2\pi m$. We will need this symbol only when comparing results to Hill's paper.

$T_e$, $T_m$: the period of the earth/moon, i.e., a year/month.

We have one more symbol for nondimensionalization, the constant of gravity. Going all the way back to Newton's second law and law of gravity, we have $F = ma$ (force = mass times acceleration), giving $F$ units of $\frac{mass \cdot length}{time^2}$. The force of gravity is $F = \frac{Gm_1m_2}{r^2}$, where $G$ is the gravitational constant, $m_1$ and $m_2$ are the masses of the objects, and $r$ is the distance between them. That give $G$ units of $\frac{length^3}{mass \cdot time^2}$. Setting $G$ to 1 and solving for length tells us that our unit of length is $(mass \cdot time^2)^{1/3}$.[8]

In summary, our nondimensionalized units of measurement are:

- unit of mass: $m_e + m_m$
- unit of time: $\frac{T_e}{2\pi} = \frac{year}{2\pi}$
- unit of length: $(mass \cdot time^2)^{1/3} = \left( (m_e + m_m) \cdot \left(\frac{T_e}{2\pi}\right)^2 \right)^{1/3}$

$q_1, q_2$: spatial coordinates.[9]
$A_j, B_j$: coefficients of the Fourier expansion of $q_1, q_2$, see equations (4.2), (4.3).

---

[2] See (G.W. Hill 1878) pages 9.

[3] See (G.W. Hill 1878) page 6.

[4] See also (G.W. Hill 1878) pages 9 and 14.

[5] See (G.W. Hill 1878) page 18. Note that $m = \frac{n'}{n-n'}$ see (G.W. Hill 1878) page 18. Since we have $n' = 1$, we can also write $m = \frac{1}{n-1}$ and $n = \frac{m+1}{m}$.

[6] See also (G.W. Hill 1878) pages 9 and 14.

[7] See (G.W. Hill 1878) page 130.

[8] (G.W. Hill 1878) page 14 simply states that the unit of length is $\sqrt[3]{\frac{\mu}{n'^2}}$.

[9] (G.W. Hill 1878) uses $x, y$.

$a_j$: coefficients of the Hill series for $q_1, q_2$, see equations (4.4), (4.5).
$u_1, u_2$: spatial coordinates in the complex plane, see equation (4.12).[10]
$\zeta$: $e^{\frac{it}{m}}$, see equation (4.23).
$D$: $\zeta \frac{d}{d\zeta}$, see equation (4.24).
$E, F, G$: Hill symbols, see equations (4.60), (4.61), (4.62).[11]
$\bar{a}_i$: $b_i \coloneqq \frac{a_i}{a_0}$, see equation (5.1).[12]
$\bar{\bar{a}}_i$: $c_i \coloneqq \frac{b_i}{m} = \frac{a_i}{ma_0}$, see equation (5.1).[13]

---

[10] (G.W. Hill 1878) uses $u, s$.
[11] (G.W. Hill 1878) uses $[j, i], [j], (j)$.
[12] (G.W. Hill 1878) introduces this definition for $b_i$, but often writes in the text that he will assume that $a_0 = 1$, thus causing a possible confusion between $a_i$ and $\bar{a}_i$.
[13] (Wintner 2014) introduces $c_i$, but we prefer to reserve that symbol for the period of the perigee, i.e. the anomalistic period.

## 3. Hill's Lunar Equations
### 3.1. Hamiltonian and Newtonian Equations

In the original derivation of these equations, Hill began with the equations for a three-body system and then made some assumptions (approximations) about this system in the case of the earth's moon: The solar parallax, the eccentricity of the earth's orbit around the sun, and the lunar inclination all vanish. This puts us in the realm of the circular, planar restricted three-body problem. Then he expands the equations around the reciprocal of the distance to the sun and eliminates higher-order terms. This "moves the sun to infinity" and leads to the equations.

A more modern derivation of the equations can be found in Meyer and Schmidt, who use the technique of symplectic scaling, which preserves the symplectic structure and thus also the Hamiltonian structure of the problem (Meyer and Schmidt 1982). They also show that non-degenerate periodic solutions to Hill's equations can be extended to solutions of the full three-body problem.

The exposition of (Frauenfelder and van Koert 2018) uses mappings that are symplectic diffeomorphisms, thus preserving the full symplectic differential geometry of the problem. This is the most modern version of the derivation, and is also fairly concise, so we will reproduce it here. Readers who are not interested in symplectic differential geometry can just ignore that aspect of the description.

**Notation**: The configuration space for this is $\mathbb{R}^2$ and we choose the origin of the coordinates $q \in \mathbb{R}^2$ to be at the moon. Then, the phase space, consisting of configuration vectors $q$ and momentum vectors $p$ is the cotangent bundle $(q, p) \in T^*\mathbb{R}^2$. For simplicity, we can also think of the phase space as $\mathbb{R}^2 \times \mathbb{R}^2$. In addition, for the purpose of this derivation, we will use $\mu = m_m$ to denote the mass of the moon and use the nondimensionalization assumption $m_e + m_m = 1$, so we have $m_e = 1 - \mu$. In the rest of the paper, $\mu$ refers to $m_e + m_m$ to match (G.W. Hill 1878) page 9.

In the Hamiltonian for the restricted three-body problem, we change the sign of the angular momentum term compared with (Frauenfelder and van Koert 2018). This way, the result of the calculation is a set of differential equations that agrees with those of (G.W. Hill 1878), where (Frauenfelder and van Koert 2018), page 77 differs by a minus sign in front of $q_2$. This is a convention that decides which side of the plane is used to look at the orbit, and we choose to use the version of (G.W. Hill 1878), where positive values of the parameter $m$ lead to direct orbits, and negative values to retrograde orbits.

We begin our derivation with the Hamiltonian for the restricted three-body problem in a rotating frame:

$$H_r = \frac{1}{2}|p|^2 - \frac{\mu}{|q|} - (1-\mu)\left(\frac{1}{\sqrt{(q_1+1)^2 + q_2{}^2}} + q_1\right) - q_1 p_2 + q_2 p_1 \qquad (3.1.1)$$

Then we define the following mapping:

$$\phi: T^*\mathbb{R}^2 \to T^*\mathbb{R}^2: (q,p) \mapsto (\mu^{1/3}q, \mu^{1/3}p).$$

It is clear that this is a conformally symplectic diffeomorphism with a constant conformal factor of $\mu^{2/3}$. Next, we define:

$$H^\mu: T^*\big(\mathbb{R}^2 \backslash \{(0,0), (-\mu^{-1/3}, 0)\}\big) \to \mathbb{R}: H^\mu \coloneqq \mu^{-2/3}(H_r \circ \phi + 1 - \mu).$$

Now we combine the definition of $\phi$ with the definition of $H^\mu$ to give an explicit version of $H^\mu$.[14]

---

[14] This equation agrees with the corresponding equation in (Frauenfelder and van Koert 2018), page 78, except that the latter has a small typographical error.

$$H^\mu(q,p) = \frac{1}{2}|p|^2 - \frac{1}{|q|} - \mu^{-2/3}\left(\frac{1}{\sqrt{(\mu^{1/3}q_1+1)^2 + \mu^{2/3}q_2{}^2}} + \mu^{1/3}q_1\right)$$
$$+ \mu^{1/3}\left(\frac{1}{\sqrt{(\mu^{1/3}q_1+1)^2 + \mu^{2/3}q_2{}^2}} + \mu^{1/3}q_1\right) - q_1 p_2 + q_2 p_1 + \mu^{-2/3} - \mu^{-2/3}\mu$$

Completing the square gives us

$$H^\mu(q,p) = \frac{1}{2}|p|^2 - \frac{1}{|q|} - \mu^{-2/3}\left(\frac{1}{\sqrt{1 + 2\mu^{1/3}q_1 + \mu^{2/3}|q|^2}} + \mu^{1/3}q_1 - 1\right)$$
$$+ \mu^{1/3}\left(\frac{1}{\sqrt{1 + 2\mu^{1/3}q_1 + \mu^{2/3}|q|^2}} + \mu^{1/3}q_1 - 1\right) + \mu^{2/3}q_1 - q_1 p_2 + q_2 p_1$$

Now we would like to expand this around $\mu^{1/3}$, but doing this directly is quite tedious, so we use an intermediate step, using the Taylor expansion:

$$\frac{1}{\sqrt{1+r}} = 1 - \frac{1}{2}r + \frac{3}{8}r^2 + \mathcal{O}(r^3) \tag{3.1.2}$$

where $\mathcal{O}(r^3)$ indicates terms of the order $r^3$ and higher, i.e., $r^3, r^4, r^5$ etc.

Taking

$$r = 2\mu^{1/3}q_1 + \mu^{2/3}|q|^2$$

and using (3.1.2) gives

$$\frac{1}{\sqrt{1 + 2\mu^{1/3}q_1 + \mu^{2/3}|q|^2}}$$
$$= 1 - \mu^{1/3}q_1 - \frac{1}{2}\mu^{2/3}|q|^2 + \frac{3}{2}\mu^{2/3}q_1{}^2 + \frac{3}{2}\mu^{1/3}q_1\mu^{2/3}|q|^2 + \frac{3}{8}\mu^{4/3}|q|^4 + \mathcal{O}(r^3)$$

substituting this into $H^\mu(q,p)$ gives

$$H^\mu(q,p) = \frac{1}{2}|p|^2 - \frac{1}{|q|}$$
$$- \mu^{-2/3}\left(1 - \mu^{1/3}q_1 - \frac{1}{2}\mu^{2/3}|q|^2 + \frac{3}{2}\mu^{2/3}q_1{}^2 + \frac{3}{2}\mu^{1/3}q_1\mu^{2/3}|q|^2 + \frac{3}{8}\mu^{4/3}|q|^4\right.$$
$$\left. + \mu^{1/3}q_1 - 1 + \mathcal{O}(r^3)\right)$$
$$+ \mu^{1/3}\left(1 - \mu^{1/3}q_1 + \frac{1}{2}\mu^{2/3}|q|^2 + \frac{3}{2}\mu^{2/3}q_1{}^2 + \frac{3}{2}\mu^{1/3}q_1\mu^{2/3}|q|^2 + \frac{3}{8}\mu^{4/3}|q|^4\right.$$
$$\left. + \mu^{1/3}q_1 - 1 + \mathcal{O}(r^3)\right) + \mu^{2/3}q_1 - q_1 p_2 + q_2 p_1$$

Now we observe that $\mathcal{O}(r^3)$ implies $\mathcal{O}(\mu)$, so $\mu^{-2/3}\mathcal{O}(r^3)$ and $\mu^{1/3}\mathcal{O}(r^3)$ both imply $\mathcal{O}(\mu^{1/3})$, so we can write

$$H^\mu(q,p) = \frac{1}{2}|p|^2 - \frac{1}{|q|} + \frac{1}{2}|q|^2 - \frac{3}{2}q_1{}^2 - q_1 p_2 + q_2 p_1 + \mathcal{O}(\mu^{1/3})$$

or

$$H^\mu(q,p) = \frac{1}{2}|p|^2 - \frac{1}{|q|} - q_1{}^2 + \frac{1}{2}q_2{}^2 - q_1 p_2 + q_2 p_1 + \mathcal{O}(\mu^{1/3})$$

From this, we can deduce that, as $\mu \to 0$, $H^\mu(q,p)$ converges uniformly in the $C^\infty$-topology on each compact subset to the Hamiltonian (since the original Hamiltonian clearly has the property, and we have only modified it by removing higher-order terms)

$$H: T^*(\mathbb{R}^2 \setminus \{0\}) \to \mathbb{R}: H(q,p) = \frac{1}{2}|p|^2 - \frac{1}{|q|} - q_1{}^2 + \frac{1}{2}q_2{}^2 - q_1 p_2 + q_2 p_1.$$

We refer to this $H$ as *Hill's lunar Hamiltonian*. This immediately leads to the Hamiltonian equation[15]:

$$\begin{cases} \dot{q}_1 = p_1 + q_2 \\ \dot{q}_2 = p_2 - q_1 \\ \dot{p}_1 = p_2 + 2q_1 - \frac{q_1}{|q|^3} \\ \dot{p}_2 = -p_1 - q_2 - \frac{q_2}{|q|^3} \end{cases} \quad (3.1.3)$$

and the Newtonian equation[16],[17]:

$$\begin{cases} \ddot{q}_1 = 2\dot{q}_2 + 3q_1 - \frac{q_1}{|q|^3} \\ \ddot{q}_2 = -2\dot{q}_1 - \frac{q_2}{|q|^3} \end{cases} \quad (3.1.4)$$

To derive the Hamiltonian equations from the Hamiltonian, it suffices to calculate $\dot{q}_i = \frac{\partial H}{\partial p_i}$ and $\dot{p}_i = -\frac{\partial H}{\partial q_i}$. To go from the Hamiltonian equation to the Newtonian equation, just take the derivative and substitute values for $p_i$. To go from the Newtonian equation to the Hamiltonian equation, consider

$$p_1 = \dot{q}_1 - q_2, \qquad p_2 = \dot{q}_2 + q_1$$

as definitions and substitute them into the Newtonian equations.

These differential equations define Hill's lunar problem, but there is no analytical solution to them, and there is no symmetry, except for the Jacobian integral (Morales-Ruiz, Simó, and Simon 2005). Because of that, when these equations were first proposed, it was important to find power series solutions for them. Of course, today, we can use computers to calculate numerical solutions.

### 3.2. Lagrangian method

Here we present the effective potential and the Lagrange method (for conservative systems). This is just for completeness, since we don't use it explicitly.
We begin with the Hamiltonian

$$H(q,p) = \frac{1}{2}|p|^2 - \frac{1}{|q|} - q_1{}^2 + \frac{1}{2}q_2{}^2 - q_1 p_2 + q_2 p_1.$$

From (3.1.3), we have

$$= \frac{1}{2}[(\dot{q}_1 - q_2)^2 + (\dot{q}_2 + q_1)^2] - \frac{1}{|q|} - q_1{}^2 + \frac{1}{2}q_2{}^2 - q_1(\dot{q}_2 + q_1) + q_2(\dot{q}_1 - q_2)$$

giving us the Hamiltonian in terms of $q$ and $\dot{q}$

$$C = \frac{1}{2}(\dot{q}_1{}^2 + \dot{q}_2{}^2) - \frac{3}{2}q_1{}^2 - \frac{1}{|q|} \quad (3.2.1)$$

where $C$ is the total energy and

---

[15] This agrees with (Frauenfelder and van Koert 2018) (5.26) and (5.27) if we reverse the sign of $q_2$.
[16] This agrees with (G.W. Hill 1878) page 14, when we use $q_1 = x$ and $q_2 = y$.
[17] This agrees with (Hénon 1969) (2) and (5) when we set $q_1 = \xi$, $q_2 = \eta$ and $\Gamma = -2C$.

$$U = -\frac{3}{2}q_1{}^2 - \frac{1}{|q|} \tag{3.2.2}$$

is the *effective potential*.
Now we derive the Lagrange formulation.

$$\mathcal{L}(q,\dot{q}) = \sum_{i=1}^{2} \dot{q}_i h(\dot{q}_i) - H\big(q, h(\dot{q})\big) \tag{3.2.3}$$

where $h(\dot{q}_i)$ is the conjugate variable of $\dot{q}_i$, defined by $\dot{q}_i = \frac{\partial H}{\partial p_1} = p_1 + q_2$ or $p_1 = \dot{q}_1 - q_2$. This gives us

$$\mathcal{L}(q,\dot{q}) = \sum_{i=1}^{2} \dot{q}_i p_i - H(q, p_i)$$

$$\mathcal{L}(q,\dot{q}) = \dot{q}_1(\dot{q}_1 - q_2) + \dot{q}_2(\dot{q}_2 + q_1) - \frac{1}{2}(\dot{q}_1{}^2 + \dot{q}_2{}^2) + \frac{3}{2}q_1{}^2 + \frac{1}{|q|}$$

$$\mathcal{L}(q,\dot{q}) = \frac{1}{2}(\dot{q}_1{}^2 + \dot{q}_2{}^2) + q_1\dot{q}_2 - q_2\dot{q}_1 + \frac{3}{2}q_1{}^2 + \frac{1}{|q|} \tag{3.2.4}$$

Now we can apply the Euler-Lagrange equations

$$\frac{d}{dt}\left(\frac{\partial \mathcal{L}}{\partial \dot{q}_i}\right) - \frac{\partial \mathcal{L}}{\partial q_i} = 0 \tag{3.2.5}$$

which results in

$$\frac{d}{dt}(\dot{q}_1 - q_2) - \dot{q}_2 - 3q_1 + \frac{q_1}{|q|^3} = 0$$

$$\ddot{q}_1 = 2\dot{q}_2 + 3q_1 - \frac{q_1}{|q|^3}$$

$$\frac{d}{dt}(\dot{q}_2 + q_1) + \dot{q}_1 + \frac{q_2}{|q|^3} = 0$$

$$\ddot{q}_2 = -2\dot{q}_1 - \frac{q_2}{|q|^3}$$

which are exactly the Newtonian differential equations (3.1.4).

### 3.3. Jacobian matrix

We use the Jacobian matrix in the numeric solution of the Hamiltonian equations, so we want to calculate it here. First, we define the function $f$, based on (3.1.3):

$$f = \begin{pmatrix} p_1 + q_2 \\ p_2 - q_1 \\ p_2 + 2q_1 - \dfrac{q_1}{|q|^3} \\ -p_1 - q_2 - \dfrac{q_2}{|q|^3} \end{pmatrix} \tag{3.3.1}$$

Then, the Jacobian matrix is

$$Jac = \begin{pmatrix} \frac{\partial f_1}{\partial q_1} & \frac{\partial f_1}{\partial q_2} & \frac{\partial f_1}{\partial p_1} & \frac{\partial f_1}{\partial p_2} \\ \frac{\partial f_2}{\partial q_1} & \frac{\partial f_2}{\partial q_2} & \frac{\partial f_2}{\partial p_1} & \frac{\partial f_2}{\partial p_2} \\ \frac{\partial f_3}{\partial q_1} & \frac{\partial f_3}{\partial q_2} & \frac{\partial f_3}{\partial p_1} & \frac{\partial f_3}{\partial p_2} \\ \frac{\partial f_4}{\partial q_1} & \frac{\partial f_4}{\partial q_2} & \frac{\partial f_4}{\partial p_1} & \frac{\partial f_4}{\partial p_2} \end{pmatrix} \quad (3.3.2)$$

Most of the derivatives are simple, so we will just calculate two of them here

$$\frac{\partial f_3}{\partial q_1} = 2 + (-1)\frac{1}{|q|^3} + (-q_1)\left(-\frac{3}{2}\right)\frac{1}{|q|^3}(2q_1)$$

$$\frac{\partial f_3}{\partial q_1} = 2 + (-q_1{}^2 - q_2{}^2 + 3q_1{}^2)\frac{1}{|q|^5}$$

$$\frac{\partial f_3}{\partial q_1} = 2 + (2q_1{}^2 - q_2{}^2)\frac{1}{|q|^5}$$

$$\frac{\partial f_4}{\partial q_2} = -1 + (-1)\frac{1}{|q|^3} + (-q_2)\left(-\frac{3}{2}\right)\frac{1}{|q|^5}(2q_2)$$

$$\frac{\partial f_4}{\partial q_2} = -1 + (-q_1{}^2 - q_2{}^2 + 3q_2{}^2)\frac{1}{|q|^5}$$

$$\frac{\partial f_4}{\partial q_2} = -1 + (-q_1{}^2 + 2q_2{}^2)\frac{1}{|q|^5}$$

$$Jac = \begin{pmatrix} 0 & 1 & 1 & 0 \\ -1 & 0 & 0 & 1 \\ 2 + (2q_1{}^2 - q_2{}^2)\frac{1}{|q|^5} & 3q_1 q_2 \frac{1}{|q|^5} & 0 & 1 \\ 3q_1 q_2 \frac{1}{|q|^5} & -1 + (-q_1{}^2 + 2q_2{}^2)\frac{1}{|q|^5} & -1 & 0 \end{pmatrix} \quad (3.3.3)$$

We use this Jacobian matrix explicitly as part of our numeric solution of the equations of motion.

    Now that we have the dynamic equations (in the Hamiltonian, Newtonian, and Lagrangian form) for the lunar 3-body problem, the next step is to begin to solve the equations. This means choosing a closed, symmetric orbit as the intermediate orbit (first approximation) and finding a way of solving it using power series.

## 4. Hill's Series

In this section, we will reproduce Hill's derivation (G.W. Hill 1878) of Hill's series. In comparison, Lyapunov's exposition (Wintner 1926) (Lyapunov 2020) follows the same logic, but does not use the extra operator introduced by Hill.

We begin with the Newtonian version of the differential equations for Hill's lunar problem, see equation (3.1.4).

**Notation**: These equations agree with those in (G.W. Hill 1878) page 129, except that we have replaced $x$ by $q_1$, $y$ by $q_2$, and used our nondimensionalization assumptions to replace $\mu$ and $n'$ by 1.

We also define the Jacobian integral as the energy[18]

$$C = \frac{1}{2}(\dot{q}_1{}^2 + \dot{q}_2{}^2) - \frac{1}{|q|} - \frac{3}{2}q_1{}^2 \tag{4.1}$$

Here we look for periodic solutions that are symmetric with respect to both the $q_1$ and $q_2$ axes, and so we can do a Fourier expansion.

$$q_1 = \sum_{j=0}^{\infty} A_j \cos\left(\frac{2j+1}{m}t\right) \tag{4.2}$$

$$q_2 = \sum_{j=0}^{\infty} B_j \sin\left(\frac{2j+1}{m}t\right) \tag{4.3}$$

Due to the symmetry, only odd terms appear.[19] Now we define the series $a_j$ by

$$A_j = a_j + a_{-j-1} \tag{4.4}$$

$$B_j = a_j - a_{-j-1} \tag{4.5}$$

Solving for $a_j$ and $a_{-j-1}$

$$a_j = \frac{A_j}{2} + \frac{B_j}{2} \tag{4.6}$$

$$a_{-j-1} = \frac{A_j}{2} - \frac{B_j}{2} \tag{4.7}$$

Substituting this into equations (4.2) and (4.3) gives

$$q_1 = \sum_{j=0}^{\infty} (a_j + a_{-j-1}) \cos\left(\frac{2j+1}{m}t\right) \tag{4.8}$$

$$q_1 = \sum_{j=0}^{\infty} a_j \cos\left(\frac{2j+1}{m}t\right) + \sum_{j=0}^{\infty} a_{-j-1} \cos\left(\frac{2j+1}{m}t\right)$$

Now set $k = -j - 1$, i.e., $j = -k - 1$

$$q_1 = \sum_{j=0}^{\infty} a_j \cos\left(\frac{2j+1}{m}t\right) + \sum_{k=-1}^{-\infty} a_k \cos\left(\frac{2(-k-1)+1}{m}t\right)$$

$$q_1 = \sum_{j=0}^{\infty} a_j \cos\left(\frac{2j+1}{m}t\right) + \sum_{k=-1}^{-\infty} a_k \cos\left(\frac{-(2k+1)}{m}t\right)$$

---

[18] This definition agrees with (G.W. Hill 1878) page 14 except for a factor of $-1$.
[19] Here, (4.2) and (4.3) correspond to (G.W. Hill 1878) page 130 and (Wintner 2014) §503 (2) if we assume that $t_0 = 0$ and $m = 1/\nu$.

Then we observe that $\cos(-x) = \cos(x)$ and replace $k$ by $j$.

$$q_1 = \sum_{j=-\infty}^{\infty} a_j \cos\left(\frac{2j+1}{m}t\right) \tag{4.9}$$

Now we do the same for $q_2$.

$$q_2 = \sum_{j=0}^{\infty} (a_j - a_{-j-1}) \sin\left(\frac{2j+1}{m}t\right) \tag{4.10}$$

Now set $k = -j - 1$, i.e., $j = -k - 1$

$$q_2 = \sum_{j=0}^{\infty} a_j \sin\left(\frac{2j+1}{m}t\right) + \sum_{k=-1}^{-\infty} -a_k \sin\left(\frac{2(-k-1)+1}{m}t\right)$$

$$q_2 = \sum_{j=0}^{\infty} a_j \sin\left(\frac{2j+1}{m}t\right) + \sum_{k=-1}^{-\infty} -a_k \sin\left(\frac{-(2k+1)}{m}t\right)$$

Observe that $\sin(-x) = -\sin(x)$ and replace $k$ by $j$.

$$q_2 = \sum_{j=-\infty}^{\infty} a_j \sin\left(\frac{2j+1}{m}t\right) \tag{4.11}$$

This gives us the coordinates in terms of $a_i$ instead of $A_i$ and $B_i$.[20]

Now define the complex variables

$$u_1 = q_1 + iq_2, \quad u_2 = q_1 - iq_2 \tag{4.12}$$

where $i = \sqrt{-1}$, so that

$$q_1 = \frac{1}{2}(u_1 + u_2), \quad q_2 = -\frac{i}{2}(u_1 - u_2).\text{[21]}$$

$$u_1 = \sum_{j=-\infty}^{\infty} a_j \cos\left(\frac{2j+1}{m}t\right) + i \sum_{j=-\infty}^{\infty} a_j \sin\left(\frac{2j+1}{m}t\right) \tag{4.13}$$

$$u_1 = \sum_{j=-\infty}^{\infty} a_j \left[\cos\left(\frac{2j+1}{m}t\right) + i \sin\left(\frac{2j+1}{m}t\right)\right]$$

Applying Euler's formula and defining $\zeta = e^{it/m}$:

$$u_1 = \sum_{j=-\infty}^{\infty} a_j e^{i\frac{2j+1}{m}t}$$

$$u_1 = \sum_{j=-\infty}^{\infty} a_j \zeta^{2j+1} \tag{4.14}$$

Now we do the same for $u_2$.

---

[20] Here, (4.9) and (4.11) correspond to (G.W. Hill 1878) page 130 and Wintner (Wintner 2014) §503 (3) if we assume that $t_0 = 0$ and $m = 1/\nu$.

[21] We are using $(q_1, q_2)$ instead of $(x, y)$ and $(u_1, u_2)$ instead of $(u, s)$.

$$u_2 = \sum_{j=-\infty}^{\infty} a_j \cos\left(\frac{2j+1}{m}t\right) - i \sum_{j=-\infty}^{\infty} a_j \sin\left(\frac{2j+1}{m}t\right) \tag{4.15}$$

Now if we change $j$ to $-j-1$, that will change $2j+1$ to $2(-j-1)+1 = -(2j+1)$ and the index still goes from $-\infty$ to $\infty$, so we have

$$u_2 = \sum_{j=-\infty}^{\infty} a_{-j-1} \cos\left(-\frac{2j+1}{m}t\right) - i \sum_{j=-\infty}^{\infty} a_{-j-1} \sin\left(-\frac{2j+1}{m}t\right)$$

Since $\cos(-x) = \cos(x)$ and $\sin(-x) = -\sin(x)$, this gives

$$u_2 = \sum_{j=-\infty}^{\infty} a_{-j-1} \cos\left(\frac{2j+1}{m}t\right) + i \sum_{j=-\infty}^{\infty} a_{-j-1} \sin\left(\frac{2j+1}{m}t\right)$$

Applying Euler's formula and defining $\zeta = e^{it/m}$:

$$u_2 = \sum_{j=-\infty}^{\infty} a_{-j-1} e^{i\frac{2j+1}{m}t}$$

$$u_2 = \sum_{j=-\infty}^{\infty} a_{-j-1} \zeta^{2j+1} \tag{4.16}$$

First, we note

$$|q|^2 = q_1^2 + q_2^2 = \frac{1}{4}(u_1+u_2)^2 - \frac{1}{4}(u_1-u_2)^2 = \frac{1}{4}(u_1^2 + 2u_1u_2 + u_2^2 - u_1^2 + 2u_1u_2 - u_2^2)$$
$$= u_1 u_2$$

Now we substitute $q_1 = \frac{1}{2}(u_1+u_2)$ and $q_2 = -\frac{i}{2}(u_1-u_2)$ into equations (3.1.4) and (4.1)

$$\frac{\ddot{u}_1}{2} + \frac{\ddot{u}_2}{2} + i\dot{u}_1 - i\dot{u}_2 - \frac{3}{2}u_1 - \frac{3}{2}u_2 + \frac{u_1+u_2}{2(u_1u_2)^{3/2}} = 0 \tag{4.17}$$

$$-\frac{i\ddot{u}_1}{2} + \frac{i\ddot{u}_2}{2} + \dot{u}_1 + \dot{u}_2 - \frac{i(u_1-u_2)}{2(u_1u_2)^{3/2}} = 0 \tag{4.18}$$

$$\frac{1}{8}(\dot{u}_1 + \dot{u}_2)^2 - \frac{1}{8}(\dot{u}_1 - \dot{u}_2)^2 - \frac{1}{(u_1u_2)^{1/2}} - \frac{3}{8}(u_1+u_2)^2 = C \tag{4.19}$$

Now we would like to separate $\ddot{u}_1$ from $\ddot{u}_2$, so we define $(4.20) = (4.17) + i(4.18)$ and $(4.21) = -(4.17) + i(4.18)$, and $(4.22) = (4.19)$ simplified.

$$\ddot{u}_1 + 2i\dot{u}_1 - \frac{3}{2}u_1 - \frac{3}{2}u_2 + \frac{u_1}{(u_1u_2)^{3/2}} = 0 \tag{4.20}$$

$$-\ddot{u}_2 + 2i\dot{u}_2 + \frac{3}{2}u_1 + \frac{3}{2}u_2 - \frac{u_2}{(u_1u_2)^{3/2}} = 0 \tag{4.21}$$

$$\frac{1}{2}\dot{u}_1\dot{u}_2 - \frac{1}{(u_1u_2)^{1/2}} - \frac{3}{8}(u_1+u_2)^2 = C \tag{4.22}$$

Now we define

$$\zeta := e^{\frac{it}{m}}, \quad \text{so } \frac{d\zeta}{dt} = \frac{i}{m}\zeta, \quad \frac{d}{dt} = \frac{d\zeta}{dt}\frac{d}{d\zeta} = \frac{i}{m}\zeta\frac{d}{d\zeta}. \tag{4.23}$$

Then we define

$$D := \zeta \frac{d}{d\zeta}, \qquad \text{so } \frac{d}{dt} = \frac{i}{m}\zeta\frac{d}{d\zeta} = \frac{i}{m}D. \tag{4.24}$$

Now we can replace $\frac{d}{dt}$ by $\frac{i}{m}D$ in equations (4.20) through (4.22).

$$-\frac{D^2}{m^2}u_1 - 2\frac{D}{m}u_1 - \frac{3}{2}u_1 - \frac{3}{2}u_2 + \frac{u_1}{(u_1u_2)^{3/2}} = 0 \tag{4.25}$$

$$\frac{D^2}{m^2}u_2 - 2\frac{D}{m}u_2 + \frac{3}{2}u_1 + \frac{3}{2}u_2 - \frac{u_2}{(u_1u_2)^{\frac{3}{2}}} = 0 \tag{4.26}$$

$$-\frac{1}{2}\frac{D}{m}u_1\frac{D}{m}u_2 - \frac{1}{(u_1u_2)^{\frac{1}{2}}} - \frac{3}{8}(u_1+u_2)^2 = C \tag{4.27}$$

Multiplying all three equations by $m^2$, the first by $-1$, and the last one by $-2$.

$$\left[D^2 + 2Dm + \frac{3}{2}m^2 - m^2\frac{1}{(u_1u_2)^{3/2}}\right]u_1 + \frac{3}{2}m^2u_2 = 0 \tag{4.28}$$

$$\left[D^2 - 2Dm + \frac{3}{2}m^2 - m^2\frac{1}{(u_1u_2)^{3/2}}\right]u_2 + \frac{3}{2}m^2u_1 = 0 \tag{4.29}$$

$$Du_1Du_2 + 2m^2\frac{1}{(u_1u_2)^{1/2}} + \frac{3}{4}m^2(u_1+u_2)^2 = -2m^2C \tag{4.30}$$

Now we have two equations for the two Newtonian equations, and one for the Jacobi integral.[22] At this point we can see that equations (4.28) − (4.30) are all linear in $u_1$ and $u_2$ with constant coefficients, except for the term $m^2\frac{1}{(u_1u_2)^{3/2}}$ in (4.28) and (4.30), so we would like to eliminate it, and then see what we can do with the result. To do that, we define equations (4.31) $:= u_1 *$ (4.29) $+ u_2 *$ (4.28) and (4.32) $:= u_1 *$ (4.29) $- u_2 *$ (4.28)[23].

$$u_1D^2u_2 + u_2D^2u_1 - 2m(u_1Du_2 - u_2Du_1) - 2m^2\frac{1}{(u_1u_2)^{1/2}} + \frac{3}{2}m^2(u_1+u_2)^2 = 0 \tag{4.31}$$

$$u_1D^2u_2 - u_2D^2u_1 - 2m(u_1Du_2 + u_2Du_1) + \frac{3}{2}m^2(u_1{}^2 - u_2{}^2) = 0 \tag{4.32}$$

Then we define (4.33) = (4.31) + (4.30) and (4.34) = (4.32) and observe that, since $D$ is a derivation, it follows the Leibnitz rule and we have

$$D^2(u_1u_2) = DD(u_1u_2) = D(u_1Du_2 + u_2Du_1) = Du_1Du_2 + u_1D^2u_2 + Du_2Du_1 + u_2D^2u_1$$
$$= 2Du_1Du_2 + u_1D^2u_2 + u_2D^2u_1$$

and also

$$D(u_1u_2) = u_1Du_2 + u_2Du_1$$

$$D(u_1Du_2 - u_2Du_1) = Du_1Du_2 + u_1D^2u_2 - Du_2Du_1 - u_2D^2u_1 = u_1D^2u_2 - u_2D^2u_1$$

The combination leads to

$$D^2(u_1u_2) - Du_1Du_2 - 2m(u_1Du_2 - u_2Du_1) + \frac{9}{4}m^2(u_1+u_2)^2 = -2m^2C \tag{4.33}$$

$$D(u_1Du_2 - u_2Du_1 - 2m(u_1u_2)) + \frac{3}{2}m^2(u_1{}^2 - u_2{}^2) = 0 \tag{4.34}$$

---

[22] Here, equations (4.28) − (4.30) agree with those on (G.W. Hill 1878) page 131, except that we have $-2m^2C$ instead of $C$ in the last equation, but this is explained in (G.W. Hill 1878) page 146.

[23] Here, equations (4.31) − (4.32) correspond to (G.W. Hill 1878) page 132 top.

This gives us two equations for the Newtonian plus Jacobi.[24] Now we want to substitute the Fourier series expansion into these equations and see what we can learn about the coefficients of the series. As a preliminary step, we begin with the series expansions for $u_1$ and $u_2$ from equations (4.14) and (4.16) and calculate several intermediate steps, including some combinations of them with $D$.

**Notation**. In the remainder of this chapter, we no longer need $i = \sqrt{-1}$, so we can use $i$ and $j$ as summation indices.

Based on (4.14) and (4.16), we have

$$u_1 u_2 = \sum_{i=-\infty}^{\infty} a_i \zeta^{2i+1} \sum_{j=-\infty}^{\infty} a_{-j-1} \zeta^{2j+1} = \sum_{i=-\infty}^{\infty} \sum_{j=-\infty}^{\infty} a_i \zeta^{2i+1} a_{-j-1} \zeta^{2j+1}$$

The second equality sign is valid because $a_i \zeta^{2i+1}$ does not depend on $j$. For fixed $i$, set $-j-1 = i - J$, so $j = -i + J - 1$ and $2j + 1 = -2i + 2J - 1$, then replace $J$ by $j$.

$$u_1 u_2 = \sum_{j=-\infty}^{\infty} \sum_{i=-\infty}^{\infty} a_i a_{i-j} \zeta^{2i+1} \zeta^{-2i+2j-1}$$

$$u_1 u_2 = \sum_{j=-\infty}^{\infty} \sum_{i=-\infty}^{\infty} a_i a_{i-j} \zeta^{2j} \qquad (4.35)$$

Based on (4.14), we have

$$u_1^2 = \sum_{j=-\infty}^{\infty} \sum_{i=-\infty}^{\infty} a_j a_i \zeta^{2j+1} \zeta^{2i+1}$$

With fixed $i$, set $j = -i + J - 1$, so $2j + 1 = -2i + 2J - 1$, then replace $J$ by $j$.

$$u_1^2 = \sum_{j=-\infty}^{\infty} \sum_{i=-\infty}^{\infty} a_i a_{-i+j-1} \zeta^{-2i+2j-1} \zeta^{2i+1}$$

$$u_1^2 = \sum_{j=-\infty}^{\infty} \sum_{i=-\infty}^{\infty} a_i a_{-i+j-1} \zeta^{2j} \qquad (4.36)$$

Based on (4.15), we have

$$u_2^2 = \sum_{j=-\infty}^{\infty} a_{-j-1} \sum_{i=-\infty}^{\infty} a_{-i-1} \zeta^{2i+1} \zeta^{2j+1}$$

Set $-i - 1 = I$, so $i = -I - 1$, and $2i + 1 = -2I - 1$, then replace $I$ by $i$.

$$u_2^2 = \sum_{j=-\infty}^{\infty} \sum_{i=-\infty}^{\infty} a_i \, a_{-j-1} \zeta^{-2i-1} \zeta^{2j+1}$$

For fixed $i$, set $-j - 1 = -i - J - 1$, so $j = i + J$ and $2j + 1 = 2i + 2J + 1$, then replace $J$ by $j$.

$$u_2^2 = \sum_{j=-\infty}^{\infty} \sum_{i=-\infty}^{\infty} a_i a_{-i-j-1} \zeta^{-2i-1} \zeta^{2i+2j+1}$$

$$u_2^2 = \sum_{j=-\infty}^{\infty} \sum_{i=-\infty}^{\infty} a_i \, a_{-i-j-1} \zeta^{2j} \qquad (4.37)$$

Based on (4.14), we have

$$Du_1 = \zeta \frac{d}{d\zeta} \sum_{i=-\infty}^{\infty} a_i \zeta^{2i+1}$$

---

[24] Here, equations (4.33) − (4.34) correspond to (G.W. Hill 1878) page 132 bottom and (Wintner 2014) §505 (10).

$$Du_1 = \sum_{i=-\infty}^{\infty} (2i+1)a_i \zeta^{2i+1} \tag{4.38}$$

$$D^2 u_1 = \zeta \frac{d}{d\zeta} \sum_{i=-\infty}^{\infty} (2i+1)a_i \zeta^{2i+1}$$

$$D^2 u_1 = \sum_{i=-\infty}^{\infty} (2i+1)^2 a_i \zeta^{2i+1} \tag{4.39}$$

Based on (4.16), we have

$$Du_2 = \zeta \frac{d}{d\zeta} \sum_{i=-\infty}^{\infty} a_{-i-1} \zeta^{2i+1}$$

$$Du_2 = \sum_{i=-\infty}^{\infty} (2i+1)a_{-i-1} \zeta^{2i+1} \tag{4.40}$$

$$D^2 u_2 = \zeta \frac{d}{d\zeta} \sum_{i=-\infty}^{\infty} (2i+1)a_{-i-1} \zeta^{2i+1}$$

$$D^2 u_2 = \sum_{i=-\infty}^{\infty} (2i+1)^2 a_{-i-1} \zeta^{2i+1} \tag{4.41}$$

Based on (4.38) and (4.39), we have

$$Du_1 Du_2 = \sum_{i=-\infty}^{\infty} (2i+1)a_i \zeta^{2i+1} \sum_{i=-\infty}^{\infty} (2i+1)a_{-i-1} \zeta^{2i+1}$$

$$Du_1 Du_2 = \sum_{j=-\infty}^{\infty} \sum_{i=-\infty}^{\infty} (2j+1)(2i+1)a_i a_{-j-1} \zeta^{2j+1} \zeta^{2i+1}$$

For fixed $i$, set $-j - 1 = i - J$, so $j = -i + J - 1$ and $2j + 1 = -2i + 2J - 1$ and replace $J$ by $j$

$$Du_1 Du_2 = \sum_{j=-\infty}^{\infty} \sum_{i=-\infty}^{\infty} (-2i+2j-1)(2i+1)a_i a_{i-j} \zeta^{-2i+2j-1} \zeta^{2i+1}$$

$$Du_1 Du_2 = -\sum_{j=-\infty}^{\infty} \sum_{i=-\infty}^{\infty} (2i-2j+1)(2i+1)a_i a_{i-j} \zeta^{2j} \tag{4.42}$$

Based on (4.14) and (4.16), we have

$$D(u_1 u_2) = \zeta \frac{d}{d\zeta} \sum_{j=-\infty}^{\infty} \sum_{i=-\infty}^{\infty} a_i a_{i-j} \zeta^{2j}$$

$$D(u_1 u_2) = \sum_{j=-\infty}^{\infty} \sum_{i=-\infty}^{\infty} 2j a_i a_{i-j} \zeta^{2j} \tag{4.43}$$

$$D^2(u_1 u_2) = \zeta \frac{d}{d\zeta} \sum_{j=-\infty}^{\infty} \sum_{i=-\infty}^{\infty} 2j a_i a_{i-j} \zeta^{2j}$$

$$D^2(u_1 u_2) = \sum_{j=-\infty}^{\infty} \sum_{i=-\infty}^{\infty} 4j^2 a_i a_{i-j} \zeta^{2j} \tag{4.44}$$

Based on (4.14) and (4.39), we have

$$u_1 D u_2 = \sum_{i=-\infty}^{\infty} a_i \zeta^{2i+1} \sum_{i=-\infty}^{\infty} (2i+1) a_{-i-1} \zeta^{2i+1}$$

$$u_1 D u_2 = \sum_{i=-\infty}^{\infty} a_i \zeta^{2i+1} \sum_{j=-\infty}^{\infty} (2j+1) a_{-j-1} \zeta^{2j+1}$$

$$u_1 D u_2 = \sum_{j=-\infty}^{\infty} \sum_{i=-\infty}^{\infty} a_i \zeta^{2i+1} (2j+1) a_{-j-1} \zeta^{2j+1}$$

For fixed $i$, set $-j - 1 = i - J$, so $j = -i + J - 1$ and $2j + 1 = -2i + 2J - 1$ and replace $J$ by $j$

$$u_1 D u_2 = \sum_{j=-\infty}^{\infty} \sum_{i=-\infty}^{\infty} a_i \zeta^{2i+1} (-2i + 2j - 1) a_{i-j} \zeta^{-2i+2j-1}$$

$$u_1 D u_2 = \sum_{j=-\infty}^{\infty} \sum_{i=-\infty}^{\infty} (-2i + 2j - 1) a_i a_{i-j} \zeta^{2j} \tag{4.45}$$

Based on (4.16) and (4.33), we have

$$u_2 D u_1 = \sum_{i=-\infty}^{\infty} a_{-i-1} \zeta^{2i+1} \sum_{i=-\infty}^{\infty} (2i+1) a_i \zeta^{2i+1}$$

$$u_2 D u_1 = \sum_{j=-\infty}^{\infty} a_{-j-1} \zeta^{2j+1} \sum_{i=-\infty}^{\infty} (2i+1) a_i \zeta^{2i+1}$$

For fixed $i$, set $-j - 1 = i - J$, so $j = -i + J - 1$ and $2j + 1 = -2i + 2J - 1$ and replace $J$ by $j$

$$u_2 D u_1 = \sum_{j=-\infty}^{\infty} a_{i-j} \zeta^{-2i+2j-1} \sum_{i=-\infty}^{\infty} (2i+1) a_i \zeta^{2i+1}$$

$$u_2 D u_1 = \sum_{j=-\infty}^{\infty} \sum_{i=-\infty}^{\infty} (2i+1) a_i a_{i-j} \zeta^{2j} \tag{4.46}$$

Based on (4.44) and (4.45), we have

$$u_1 D u_2 - u_2 D u_1 = \sum_{j=-\infty}^{\infty} \sum_{i=-\infty}^{\infty} [(-2i + 2j - 1) - (2i + 1)] a_i a_{i-j} \zeta^{2j}$$

$$u_1 D u_2 - u_2 D u_1 = \sum_{j=-\infty}^{\infty} \sum_{i=-\infty}^{\infty} (-4i + 2j - 2) a_i a_{i-j} \zeta^{2j} \tag{4.47}$$

Based on (4.44) and (4.45), we have

$$D(u_1 D u_2 - u_2 D u_1) = \zeta \frac{d}{d\zeta} \sum_{j=-\infty}^{\infty} \sum_{i=-\infty}^{\infty} (-4i + 2j - 2) a_i a_{i-j} \zeta^{2j}$$

$$D(u_1 Du_2 - u_2 Du_1) = \sum_{j=-\infty}^{\infty} \sum_{i=-\infty}^{\infty} 2j(-4i + 2j - 2) a_i a_{i-j} \zeta^{2j} \tag{4.48}$$

Based on (4.39) and (4.40), we have

$$u_1 Du_2 + u_2 Du_1 = \sum_{j=-\infty}^{\infty} \sum_{i=-\infty}^{\infty} [(-2i + 2j - 1) + (2i + 1)] a_i a_{i-j} \zeta^{2j}$$

$$u_1 Du_2 + u_2 Du_1 = \sum_{j=-\infty}^{\infty} \sum_{i=-\infty}^{\infty} 2j a_i a_{i-j} \zeta^{2j} \tag{4.49}$$

$$D(u_1 Du_2 + u_2 Du_1) = \zeta \frac{d}{d\zeta} \sum_{j=-\infty}^{\infty} \sum_{i=-\infty}^{\infty} 2j a_i a_{i-j} \zeta^{2j}$$

$$D(u_1 Du_2 + u_2 Du_1) = \sum_{j=-\infty}^{\infty} \sum_{i=-\infty}^{\infty} 4j^2 a_i a_{i-j} \zeta^{2j} \tag{4.50}$$

$$D(u_1 u_2) = u_1 Du_2 + u_2 Du_1 = \sum_{j=-\infty}^{\infty} \sum_{i=-\infty}^{\infty} 2j a_i a_{i-j} \zeta^{2j}$$

$$D(u_1 u_2) = \sum_{j=-\infty}^{\infty} \sum_{i=-\infty}^{\infty} 2j a_i a_{i-j} \zeta^{2j} \tag{4.51}$$

$$D^2(u_1 u_2) = D\bigl(D(u_1 u_2)\bigr) = D(u_1 Du_2 + u_2 Du_1) = \sum_{j=-\infty}^{\infty} \sum_{i=-\infty}^{\infty} 4j^2 a_i a_{i-j} \zeta^{2j}$$

$$D^2(u_1 u_2) = \sum_{j=-\infty}^{\infty} \sum_{i=-\infty}^{\infty} 4j^2 a_i a_{i-j} \zeta^{2j} \tag{4.52}$$

Now we are ready to put the Fourier series into the differential equations in order to determine the values of the coefficients. We do that by putting the expressions we just calculated into equations (4.33) and (4.34). We begin with (4.33).

$$D^2(u_1 u_2) - Du_1 Du_2 - 2m(u_1 Du_2 - u_2 Du_1) + \frac{9}{4} m^2 (u_1 + u_2)^2 = -2m^2 C$$

Substituting the expressions (4.52), (4.42), (4.45), (4.46), (4.36), (4.35), and (4.37) above gives

$$\sum_{j=-\infty}^{\infty} \sum_{i=-\infty}^{\infty} 4j^2 a_i a_{i-j} \zeta^{2j} + \sum_{j=-\infty}^{\infty} \sum_{i=-\infty}^{\infty} (2i - 2j + 1)(2i + 1) a_i a_{i-j} \zeta^{2j}$$

$$- 2m \left( \sum_{j=-\infty}^{\infty} \sum_{i=-\infty}^{\infty} (-4i + 2j - 2) a_i a_{i-j} \zeta^{2j} \right) + \frac{9}{4} m^2 \sum_{j=-\infty}^{\infty} \sum_{i=-\infty}^{\infty} a_i a_{-i+j-1} \zeta^{2j}$$

$$+ 2 \frac{9}{4} m^2 \sum_{j=-\infty}^{\infty} \sum_{i=-\infty}^{\infty} a_i a_{i-j} \zeta^{2j} + \frac{9}{4} m^2 \sum_{j=-\infty}^{\infty} \sum_{i=-\infty}^{\infty} a_i a_{-i-j-1} \zeta^{2j} = -2m^2 C$$

Collecting similar terms

$$\sum_{j=-\infty}^{\infty} \sum_{i=-\infty}^{\infty} \left[(2i - 2j + 1)(2i + 1) + 4j^2 + 4m(2i - j + 1) + \frac{9}{2}m^2\right] a_i a_{i-j} \zeta^{2j}$$

$$+ \frac{9}{4} m^2 \left( \sum_{j=-\infty}^{\infty} \sum_{i=-\infty}^{\infty} [a_i a_{-i+j-1} + a_i a_{-i-j-1}] \zeta^{2j} \right) = -2m^2 C$$

For each $j$, the coefficient of $\zeta^{2j}$ must be zero (because the $\zeta^{2j}$ form a basis), so that gives

$$\sum_{i=-\infty}^{\infty} \left[(2i - 2j + 1)(2i + 1) + 4j^2 + 4m(2i - j + 1) + \frac{9}{2}m^2\right] a_i a_{i-j}$$

$$+ \frac{9}{4} m^2 \left( \sum_{i=-\infty}^{\infty} [a_i a_{-i+j-1} + a_i a_{-i-j-1}] \right) = -2\delta_{j,0} m^2 C \qquad (4.53)$$

Where $\delta_{j,0}$ is the Kronecker delta symbol, i.e., $\delta_{j,0} = 1 \; if \; j = 0$ and $\delta_{j,0} = 0 \; if \; j \neq 0$. [25]
Now we continue with (4.34).

$$D(u_1 D u_2 - u_2 D u_1 - 2m(u_1 u_2)) + \frac{3}{2} m^2 (u_1^2 - u_2^2) = 0$$

Substituting the expressions (4.48), (4.51), (4.36), and (4.37) above gives

$$\sum_{j=-\infty}^{\infty} \sum_{i=-\infty}^{\infty} 2j(-4i + 2j - 2) a_i a_{i-j} \zeta^{2j} - 2m \sum_{j=-\infty}^{\infty} \sum_{i=-\infty}^{\infty} 2j a_i a_{i-j} \zeta^{2j}$$

$$+ \frac{3}{2} m^2 \sum_{j=-\infty}^{\infty} \sum_{i=-\infty}^{\infty} a_i a_{-i+j-1} \zeta^{2j} - \frac{3}{2} m^2 \sum_{j=-\infty}^{\infty} \sum_{i=-\infty}^{\infty} a_i a_{-i-j-1} \zeta^{2j} = 0$$

Collecting similar terms

$$\sum_{j=-\infty}^{\infty} \sum_{i=-\infty}^{\infty} [4j(-2i + j - 1) - 4mj] a_i a_{i-j} \zeta^{2j} + \frac{3}{2} m^2 \sum_{j=-\infty}^{\infty} \sum_{i=-\infty}^{\infty} [a_i a_{-i+j-1} - a_i a_{-i-j-1}] \zeta^{2j} = 0$$

For each $j$, the coefficient of $\zeta^{2j}$ must be zero (because the $\zeta^{2j}$ form a basis), so that gives

$$4j \sum_{i=-\infty}^{\infty} [(-2i + j - 1) - m] a_i a_{i-j} + \frac{3}{2} m^2 \sum_{i=-\infty}^{\infty} [a_i a_{-i+j-1} - a_i a_{-i-j-1}] = 0 \qquad (4.54)$$

With this, we have reduced the equation from a double sum to a single sum.[26] Now we define $(4.55) = (4.53) * 2 + (4.54) * 3$ and $(4.56) = (4.53) * 2 - (4.54) * 3$.
Starting with $(4.55) = (4.53) * 2 + (4.54) * 3$

$$2 \sum_{i=-\infty}^{\infty} \left[(2i - 2j + 1)(2i + 1) + 4j^2 + 4m(2i - j + 1) + \frac{9}{2}m^2\right] a_i a_{i-j}$$

$$+ \frac{9}{2} m^2 \left( \sum_{i=-\infty}^{\infty} [a_i a_{-i+j-1} + a_i a_{-i-j-1}] \right) + 12j \sum_{i=-\infty}^{\infty} [(-2i + j - 1) - m] a_i a_{i-j}$$

$$+ \frac{9}{2} m^2 \sum_{i=-\infty}^{\infty} [a_i a_{-i+j-1} - a_i a_{-i-j-1}] = -4\delta_{j,0} m^2 C$$

Collecting similar terms

---

[25] Equation (4.53) corresponds to (G.W. Hill 1878) page 133, first equation.
[26] Equation (4.54) corresponds to (G.W. Hill 1878) page 133, second equation.

$$\sum_{i=-\infty}^{\infty} [2(2i-2j+1)(2i+1) + 8j^2 + 8m(2i-j+1) + 9m^2 + 12j(-2i+j-1) - 12jm]a_i a_{i-j}$$

$$+ 9m^2 \left( \sum_{i=-\infty}^{\infty} a_i a_{-i+j-1} \right) = -4\delta_{j,0} m^2 C$$

Expanding

$$\sum_{i=-\infty}^{\infty} [8i^2 - 32ij + 8i - 16j + 20j^2 + 2 + 4m(4i - 5j + 2) + 9m^2] a_i a_{i-j}$$

$$+ 9m^2 \left( \sum_{i=-\infty}^{\infty} a_i a_{-i+j-1} \right) = -4\delta_{j,0} m^2 C \quad (4.55)$$

This gives us a combined equation that includes the Jacobi integral.[27] Now we continue with $(4.56) = (4.53) * 2 - (4.54) * 3$.

$$2 \sum_{i=-\infty}^{\infty} \left[ (2i-2j+1)(2i+1) + 4j^2 + 4m(2i-j+1) + \frac{9}{2}m^2 \right] a_i a_{i-j}$$

$$+ \frac{9}{2} m^2 \left( \sum_{i=-\infty}^{\infty} [a_i a_{-i+j-1} + a_i a_{-i-j-1}] \right) - 12j \sum_{i=-\infty}^{\infty} [(-2i+j-1) - m] a_i a_{i-j}$$

$$- \frac{9}{2} m^2 \sum_{i=-\infty}^{\infty} [a_i a_{-i+j-1} - a_i a_{-i-j-1}] = -4\delta_{j,0} m^2 C$$

Collecting similar terms

$$\sum_{i=-\infty}^{\infty} [2(2i-2j+1)(2i+1) + 8j^2 + 8m(2i-j+1) + 9m^2 - 12j(-2i+j-1) + 12mj]a_i a_{i-j}$$

$$+ 9m^2 \left( \sum_{i=-\infty}^{\infty} a_i a_{-i-j-1} \right) = -4\delta_{j,0} m^2 C$$

Expanding

$$\sum_{i=-\infty}^{\infty} [8i^2 + 16ij + 8i + 8j - 4j^2 + 2 + 4m(4i + j + 2) + 9m^2] a_i a_{i-j}$$

$$+ 9m^2 \left( \sum_{i=-\infty}^{\infty} a_i a_{-i-j-1} \right) = -4\delta_{j,0} m^2 C \quad (4.56)$$

This gives us a combined equation that includes the Jacobi integral.[28] Now we give the following explanation for why (4.55) and (4.56) are not distinct, meaning that one formula will suffice[29]: If we take equation (4.55) and, in the first sum, replace $i$ by $i-1$ and $j$ by $-j$, then the result is the same as (4.56).

Now we are ready to make the final manipulation of equations (4.55) and (4.56). We begin by defining some shorthand: $P_1 - P_4$ are the coefficients in (4.55) and (4.56). (None of these polynomials can be factorized.)

$$P_1 = 8i^2 - 32ij + 16im + 8i + 20j^2 - 20jm - 16j + 9m^2 + 8m + 2$$

---

[27] Equation (4.55) corresponds to (G.W. Hill 1878) page 133, third equation.
[28] Equation (4.56) corresponds to (G.W. Hill 1878) page 133, fourth equation.
[29] This is Hill (G.W. Hill 1878) page 133, last paragraph.

$$P_2 = 9m^2$$
$$P_3 = 8i^2 + 16ij + 16im + 8i - 4j^2 + 4jm + 8j + 9m^2 + 8m + 2$$
$$P_4 = 9m^2$$

With these polynomials, equations (4.55) and (4.56) can be written as

$$\sum_{i=-\infty}^{\infty} [P_1 a_i a_{i-j} + P_2 a_i a_{-i+j-1}] = 0 \qquad (4.57)$$

$$\sum_{i=-\infty}^{\infty} [P_3 a_i a_{i-j} + P_4 a_i a_{-i-j-1}] = 0 \qquad (4.58)$$

Now we consider the special case of $i = 0$ and then $i = j$.

$$P_5 := P_3|_{i=0} = -4j^2 + 4jm + 8j + 9m^2 + 8m + 2$$
$$P_6 := P_1|_{i=0} = 20j^2 - 20jm - 16j + 9m^2 + 8m + 2$$
$$P_7 := -(P_1 P_5 - P_3 P_6)|_{i=j} = 48j^2(8j^2 + m^2 - 4m - 2)$$

Now define the new equation (4.59) by multiplying (4.57) by $P_5$, multiplying (4.58) by $P_6$, adding the two, and dividing the result by $P_7$. Our choice of $P_7$ will produce a simple value of $E_{j,j} = -1$ later on.

$$(4.59) := [(4.57) * P_5 - (4.57) * P_6]/P_7$$

$$[P_1 P_5 a_i a_{i-j} + P_2 P_5 a_j a_{-i+j-1} - P_3 P_6 a_i a_{i-j} - P_4 P_6 a_i a_{-i-j-1}]/P_7$$

$$\sum_{i=-\infty}^{\infty} \left[ \left(\frac{P_1 P_5 - P_3 P_6}{P_7}\right) a_i a_{i-j} + \left(\frac{P_2 P_5}{P_7}\right) a_i a_{-i+j-1} - \left(\frac{P_4 P_6}{P_7}\right) a_i a_{-i-j-1} \right] = 0 \qquad (4.59)$$

Then we make some more definitions

$$E_{j,i} := \frac{P_1 P_5 - P_3 P_6}{P_7}$$

$$F_j := \frac{P_2 P_5}{P_7}$$

$$G_j := \frac{P_4 P_6}{P_7}$$

And apply them

$$E_{j,i} = -\frac{i(4ij - 4im - 4i + 4j^2 + 4jm + 4j + m^2 - 4m - 2)}{j(8j^2 + m^2 - 4m - 2)} \qquad (4.60)$$

$$F_j = -\frac{3m^2(4j^2 - 4jm - 8j - 9m^2 - 8m - 2)}{16j^2(8j^2 + m^2 - 4m - 2)} \qquad (4.61)$$

$$G_j = -\frac{3m^2(20j^2 - 20jm - 16j + 9m^2 + 8m + 2)}{16j^2(8j^2 + m^2 - 4m - 2)} \qquad (4.62)$$

**Notation.** Here we have defined the letters $E_{j,i}, F_j, G_j$ to match Hill's symbols $[j, i], [j], (j)$.

$$E_{j,i} = [j, i]$$
$$F_j = [j]$$
$$G_j = (j)$$

Altogether, this gives us

$$\sum_{i=-\infty}^{\infty} \left[ E_{j,i} a_i a_{i-j} + F_j a_i a_{-i+j-1} + G_j a_i a_{-i-j-1} \right] = 0 \tag{4.63}$$

This is *Hill's equation* for calculating the coefficients of the infinite sum.[30] We also look at two important special cases:

$$E_{j,0} = 0 \tag{4.64}$$

$$E_{j,j} = -1 \tag{4.65}$$

**Notation**. In fact, our symbols are intended as functions:

$$E: \mathbb{Z}^2 \times \mathbb{R} \to \mathbb{R}, \qquad E_{j,i} \coloneqq E(j, i, m)$$

$$F: \mathbb{Z} \times \mathbb{R} \to \mathbb{R}, \qquad F_j \coloneqq F(j, m)$$

$$G: \mathbb{Z} \times \mathbb{R} \to \mathbb{R}, \qquad G_j \coloneqq G(j, m)$$

This functional notation replaces Hill's unorthodox use of square and round brackets by more modern mathematical notation, and it makes it much easier to convert these expressions into software.

Our next goal is to get an explicit formula for $C$. For this, we start with equation (4.53) and set $j = 0$. On the right-hand side, we can ignore the $= 0$, because this is only for $j = 0$.

$$\sum_{i=-\infty}^{\infty} \left[ (2i+1)(2i+1) + 4m(2i+1) + \frac{9}{2}m^2 \right] a_i a_i + \frac{9}{2} m^2 \left( \sum_{i=-\infty}^{\infty} a_i a_{-i-1} \right) = -2m^2 C$$

Rearranging some

$$\sum_{i=-\infty}^{\infty} \left\{ \left[ (2i+1)^2 + 8im + 4m + \frac{9}{2}m^2 \right] a_i^2 + \frac{9}{2} m^2 a_i a_{-i-1} \right\} = -2m^2 C \tag{4.66}$$

We can use this for calculating the Jacobi integral (energy expression) $C$ based on the coefficients $a_i$. (Wintner 1926)[31]

At this point, we look at the determination of $a_0$.[32] Similar to how we substituted the Fourier series into equations (4.33) and (4.34) to get (4.53) and (4.54), we now want to substitute the Fourier series into (4.28). We start by separating out the term involving $(u_1 u_2)^{3/2}$ and deal with it a bit later.

$$m^2 \frac{u_1}{(u_1 u_2)^{3/2}} = \left[ D^2 + 2Dm + \frac{3}{2} m^2 \right] u_1 + \frac{3}{2} m^2 u_2 = 0$$

Applying equations (4.14) and (4.16)

$$m^2 \frac{u_1}{(u_1 u_2)^{3/2}} = \sum_{i=-\infty}^{\infty} \left[ (2i+1)^2 + 2m(2i+1) + \frac{3}{2} m^2 \right] a_i \zeta^{2i+1} + \frac{3}{2} m^2 \sum_{i=-\infty}^{\infty} a_{-i-1} \zeta^{2i+1} = 0$$

Rearranging

$$m^2 \frac{u_1}{(u_1 u_2)^{3/2}} = \sum_{i=-\infty}^{\infty} \left\{ \left[ (2i+1+m)^2 + \frac{1}{2} m^2 \right] a_i + \frac{3}{2} m^2 a_{-i-1} \right\} \zeta^{2i+1} = 0$$

---

[30] Equation (4.63) corresponds to (G.W. Hill 1878) page 135 and (Wintner 2014) §506 (14).
[31] This reproduces (Wintner 2014) §506 (12) and (Wintner 1926, 1929), except that we have a different version of $C$, which is intentional.
[32] This is explained in (G.W. Hill 1878) page 144 and (Wintner 2014) §506.

For the left-hand side, we use a simple method.[33] By looking at time $t = 0$, we have $\zeta = 1$ and $u_1 = u_2 = \sum_{i=-\infty}^{\infty} a_i$, which gives us

$$m^2 \frac{1}{u_1{}^2} = \sum_{i=-\infty}^{\infty} \left[(2i + 1 + m)^2 + \frac{1}{2}m^2\right] a_i + \frac{3}{2}m^2 \sum_{i=-\infty}^{\infty} a_{-i-1} = 0$$

Now we can replace $-i - 1$ in the second sum by $i$, giving

$$m^2 \frac{1}{u_1{}^2} = \sum_{i=-\infty}^{\infty} [(2i + 1 + m)^2 + 2m^2] a_i = 0$$

Substituting $\sum_{i=-\infty}^{\infty} a_i$ for $u_1$ gives

$$m^2 = \sum_{i=-\infty}^{\infty} [(2i + 1 + m)^2 + 2m^2] a_i \left(\sum_{i=-\infty}^{\infty} a_i\right)^2 = 0$$

Dividing by $a_0{}^3$ and inverting

$$a_0{}^3 = \frac{m^2}{\sum_{i=-\infty}^{\infty}[(2i + 1 + m)^2 + 2m^2]\frac{a_i}{a_0}\left(\sum_{i=-\infty}^{\infty}\frac{a_i}{a_0}\right)^2}$$

or

$$a_0 = \left[\frac{m^2}{\sum_{i=-\infty}^{\infty}[(2i + 1 + m)^2 + 2m^2]\frac{a_i}{a_0}\left(\sum_{i=-\infty}^{\infty}\frac{a_i}{a_0}\right)^2}\right]^{1/3} \quad (4.67)$$

But $\kappa = \frac{\mu}{n^2}(1 + m)^2$ and our dimensionalization sets $\kappa = m^2$, so the last equation is the same as

$$a_0 = \left[\frac{\mu}{n^2}\right]^{1/3} \left[\frac{(1 + m)^2}{\sum_{i=-\infty}^{\infty}[(2i + 1 + m)^2 + 2m^2]\frac{a_i}{a_0}\left(\sum_{i=-\infty}^{\infty}\frac{a_i}{a_0}\right)^2}\right]^{1/3} \quad (4.68)$$

Equation (4.68) is just the same as (4.67) but includes the parameters before nondimensionalization.[34]

With this, we have derived Hill's equations, equation (4.63), which are infinitely many equations, each of which contains an infinite sum. Hill solved these equations for the first coefficients on the series, but didn't provide a general procedure for solving them, leaving it open where to start and what to include. First, this can be done by calculating everything until enough powers of $m$ have been determined and then eliminating higher-order terms. To make the calculations more efficient, and better suited for solving on a computer, we will derive a recursive formula, based on input from Wintner.

---

[33] This is a suggestion from (G.W. Hill 1878) on page 145.
[34] This reproduces the equation for $a_0$ on (G.W. Hill 1878) page 144 and 145, (Wintner 2014) §506 (13) and (Wintner 1929) (5).

## 5. Wintner's Recursion Formula

Up to now, we have derived Hill's Lunar Equations in Hamiltonian and Newtonian format, defined Hill's series based on a Fourier series for symmetric, closed orbits, and derived Hill's equation, our equation (4.63), for calculating the coefficients of the series. However, Hill's equation is an infinite number of equations, each of which is an infinite sum. This makes it very difficult to define a method for solving these equations, so it was very welcome to see that (Wintner 2014) describes a recursion formula for calculating the coefficients.

The coefficients (and the other parts of the equation) are power series in $m$, and the recursion formula is a procedure for calculating the $k^{th}$ partial sum of a coefficient based on the $(k-1)^{th}$ partial sums of the coefficients. For this purpose, we begin by introducing Wintner's version of Hill's symbols, and our version of denoting coefficients in the resulting series in powers of $m$.

**Notation.** $A_i$, $B_i$, and $a_i$ denote the coefficients defined in equations (4.2) through (4.5). For convenience, we also define the variants[35]

$$\bar{a}_i = b_i \coloneqq \frac{a_i}{a_0}, \qquad \bar{\bar{a}}_i = c_i \coloneqq \frac{b_i}{m} = \frac{a_i}{ma_0}. \tag{5.1}$$

Then we define Wintner's version of the functions $F$ and $G$, which we call $\tilde{F}$ and $\tilde{G}$. Finally, all of these are power series in $m$, and we will denote the coefficient of $m^k$ by adding an additional index, so we have, for example

$$a_i = \sum_{n=0}^{\infty} a_{i,n} m^n$$

and the partial sum up to power $m^k$ is

$$\sum_{n=0}^{k} a_{i,n} m^n$$

The symbols defined as functions are:

$$a\colon \mathbb{Z}^2 \times \mathbb{R} \to \mathbb{R}, \qquad a_{i,k} \coloneqq a(i,k,m), \qquad a_i \coloneqq a_{i,\infty} = a(i,\infty)$$

$$\bar{a}\colon \mathbb{Z}^2 \times \mathbb{R} \to \mathbb{R}, \qquad \bar{a}_{i,k} \coloneqq \bar{a}(i,k,m), \qquad \bar{a}_i \coloneqq \frac{a_j}{a_0}$$

$$\bar{\bar{a}}\colon \mathbb{Z}^2 \times \mathbb{R} \to \mathbb{R}, \qquad \bar{\bar{a}}_{i,k} \coloneqq \bar{\bar{a}}(i,k,m), \qquad \bar{\bar{a}}_j \coloneqq \frac{a_j}{ma_0}$$

$$E\colon \mathbb{Z}^3 \times \mathbb{R} \to \mathbb{R}, \qquad E_{j,i,k} \coloneqq E(j,i,k,m)$$

$$F\colon \mathbb{Z}^2 \times \mathbb{R} \to \mathbb{R}, \qquad F_j \coloneqq F(j,k,m)$$

$$G\colon \mathbb{Z}^2 \times \mathbb{R} \to \mathbb{R}, \qquad G_j \coloneqq G(j,k,m)$$

$$\tilde{F} \coloneqq F/m^2$$

$$\tilde{G} \coloneqq G/m^2$$

$$E_{j,i} \coloneqq E(j,i)$$

$$\tilde{F}_j \coloneqq \tilde{F}(j)$$

$$\tilde{G}_j \coloneqq \tilde{G}(j)$$

---

[35] The $b_i$ were defined by Hill but not used much. Instead, he often writes in the text that he temporarily assumes $a_0$ to be equal to 1. The $c_i$ were defined by Wintner in the context of his recursion formula. We avoid both in the text but use them in the software.

Our first step here is to write Hill's equation, equation (4.58), in such a way that $a_0$ never occurs in a sum.

$$a_0 a_j = \sum_{\substack{i=-\infty \\ i \notin \{0,j\}}}^{\infty} E_{j,i} a_i a_{i-j} + m^2 \tilde{F}_j \sum_{\substack{i=-\infty \\ i \notin \{0,j-1\}}}^{\infty} a_i a_{-i+j-1} + m^2 \tilde{G}_j \sum_{\substack{i=-\infty \\ i \notin \{0,-j-1\}}}^{\infty} a_i a_{-i-j-1} \\ + (2 - \delta_{1,j}) m^2 \tilde{F}_j a_0 a_{j-1} + (2 - \delta_{-1,j}) m^2 \tilde{G}_j a_0 a_{-j-1} \tag{5.2}$$

Where $\delta_{1,j}$ is the Kronecker delta symbol, i.e., $\delta_{1,j} = 1 \; if \; 1 = j$ and $\delta_{1,j} = 0 \; if \; 1 \neq j$, and $j = \pm 2, \pm 3, \ldots$ (Note that we have temporarily left $j = \pm 1$ out).

Each sum omits the two terms that would lead to $a_0$. In the first sum, in one of those terms $E_{j,i}$ is zero and in the other it is $-1$, leading to the left-hand side of the equation. In the other two sums, the terms that have been omitted appear as the last two terms of the equation.

For $j = 1$, the term $(2 - \delta_{1,j}) m^2 \tilde{F}_j a_0 a_{j-1}$ changes to $m^2 \tilde{F}_j a_0 a_{j-1}$ and for $j \neq 1$, it is $2 m^2 \tilde{F}_j a_0 a_{j-1}$. This is because, for $j = 1$, the choices $i = 0, i = j - 1$ both result in 0, so this occurs only once.

For $j = -1$, the term $(2 - \delta_{-1,j}) m^2 \tilde{G}_j a_0 a_{-j-1}$ changes to $m^2 \tilde{G}_j a_0 a_{-j-1}$ and for $j \neq -1$, it is $2 m^2 \tilde{G}_j a_0 a_{-j-1}$. This is because, for $j = -1$, the choices $i = 0, i = -j - 1$ both result in 0, so this occurs only once.[36]

Note that $a_0 a_j$ appears on the left-hand side of the equation because $E_{j,0} = 0$ and $E_{j,j} = -1$.

Next, we divide both sides of equation (5.1) by $a_0^2$ and apply the definition of $\bar{\bar{a}}_i$.

$$\bar{\bar{a}}_j = m \left\{ \sum_{\substack{i=-\infty \\ i \notin \{0,j\}}}^{\infty} E_{j,i} \bar{\bar{a}}_i \bar{\bar{a}}_{i-j} + m^2 \tilde{F}_j \sum_{\substack{i=-\infty \\ i \notin \{0,j-1\}}}^{\infty} \bar{\bar{a}}_i \bar{\bar{a}}_{-i+j-1} + m^2 \tilde{G}_j \sum_{\substack{i=-\infty \\ i \notin \{0,-j-1\}}}^{\infty} \bar{\bar{a}}_i \bar{\bar{a}}_{-i-j-1} \right. \\ \left. + (2 - \delta_{1,j}) m \tilde{F}_j \bar{\bar{a}}_{j-1} + (2 - \delta_{-1,j}) m \tilde{G}_j \bar{\bar{a}}_{-j-1} \right\} \tag{5.3}$$

Now we introduce a new symbol $W_j$ that will only be used here.[37]

$$W_j := \bar{\bar{a}}_j / m$$

With this, we can rewrite equation (5.2) as

$$\bar{\bar{a}}_j = m W_j \tag{5.4}$$

where

$$W_j = \sum_{\substack{i=-\infty \\ i \notin \{0,j\}}}^{\infty} E_{j,i} \bar{\bar{a}}_i \bar{\bar{a}}_{i-j} + m^2 \tilde{F}_j \sum_{\substack{i=-\infty \\ i \notin \{0,j-1\}}}^{\infty} \bar{\bar{a}}_i \bar{\bar{a}}_{-i+j-1} + m^2 \tilde{G}_j \sum_{\substack{i=-\infty \\ i \notin \{0,-j-1\}}}^{\infty} \bar{\bar{a}}_i \bar{\bar{a}}_{-i-j-1} \\ + (2 - \delta_{1,j}) m \tilde{F}_j \bar{\bar{a}}_{j-1} + (2 - \delta_{-1,j}) m \tilde{G}_j \bar{\bar{a}}_{-j-1} \tag{5.5}$$

**Claim (5.1)**[38]. When computing the $k^{th}$ partial sum of $\bar{\bar{a}}_j$, $\sum_{n=0}^{k} \bar{\bar{a}}_{j,n} m^n$, the following hold:
(i) The $k^{th}$ partial sum of $\bar{\bar{a}}_j$ only depends on the $(k-1)^{th}$ partial sums of all $\bar{\bar{a}}_j$.
(ii) In the $k^{th}$ partial sum, all terms of order up to $k - 1$ are the same as in the $(k-1)^{th}$ partial sum.

---

[36] These two special cases are missing in (Wintner 2014), §509, (25) and (27). This appears to be a typographical error, since our calculations of $\bar{\bar{a}}_j$ based on this formula agree with those in (Wintner 2014) §514 (23 bis).

[37] $W_j$ corresponds to $G_j$ (Wintner 2014) §513 (31) except for the special case of $j = \pm 1$ (see above).

[38] The recursive formula was taken from (Wintner 2014)
§507-§514, which also treats questions of existence and convergence, but we are currently only looking at how to calculate the $\bar{\bar{a}}_j$, so our formula is much simpler.

**Proof.** (i) follows directly from equation (5.3), because the prefix $m$ means that $W_j$ can only contain sums of order 1 less than the left-hand side.
(ii) is a corollary of this, because, if a lower order term were to change during computation of the $k^{th}$ partial sum, it would depend on a higher-order term. ∎

Now, with the recursive formula, we have a clear procedure for successively calculating the coefficients of powers of $m$ in each partial sum of $\bar{\bar{a}}_j$. In addition, we only need to calculate the highest-order term. Taking account of that, we can reformulate equation (5.2) so that it only includes the relevant terms (coefficients of powers of $m$), and no partial sums.

We only need the terms with a power of exactly $m^k$, because the lower-order terms are part of the $(k-1)^{th}$ partial sum and the higher-order terms are not included in $\bar{\bar{a}}_{j,k}$. Let's take a closer look at the first sum. We start with the partial sum of the Taylor series for $E_{j,i}$. This sum begins with the term $m^0$, we have one explicit factor of $m$, and the series for the $\bar{\bar{a}}_i$ begin with a term for $m^1$, so we always have a factor of $m^3$ and as a result we only need the partial sum of $E_{j,i}$ up to $k-3$, which we have formulated as a sum from $k_1 = 0$ to $k-3$. Then, for $\bar{\bar{a}}_{i,k-2}$, we need the terms from $k_2 = 1$ to $k - k_1 - 2$, where the 2 comes from the explicit $m$ and the minimum of 1 in $\bar{\bar{a}}_{i-j,k-2}$. Finally, for $\bar{\bar{a}}_{i-j,k-2}$, we need the term for $k - k_1 - k_2 - 1$, so that the three factors together have a power of $m^{(k-1)}$ or, including the explicit $m$, a power of $m^k$. The other two sums and the two terms after the sums are analogous.

$$\bar{\bar{a}}_{j,k} = \sum_{\substack{i=-\infty \\ i \notin \{0,j\}}}^{+\infty} \sum_{k_1=0}^{k-3} \sum_{k_2=1}^{k-k_1-2} E_{j,i,k_1} \bar{\bar{a}}_{i,k_2} \bar{\bar{a}}_{i-j,k-k_1-k_2-1}$$

$$+ \sum_{\substack{i=-\infty \\ i \notin \{0,j-1\}}}^{+\infty} \sum_{k_1=0}^{k-5} \sum_{k_2=1}^{k-k_1-4} \tilde{F}_{j,k_1} \bar{\bar{a}}_{i,k_2} \bar{\bar{a}}_{-i+j-1,k-k_1-k_2-3}$$

$$+ \sum_{\substack{i=-\infty \\ i \notin \{0,-j-1\}}}^{+\infty} \sum_{k_1=0}^{k-5} \sum_{k_2=1}^{k-k_1-4} \tilde{G}_{j,k_1} \bar{\bar{a}}_{i,k_2} \bar{\bar{a}}_{-i-j-1,k-k_1-k_2-3} \quad (5.6)$$

$$+ \delta_{i,1} \tilde{F}_{j,k-1} + 2(1 - \delta_{i,1}) \sum_{k_1=0}^{k-3} \tilde{F}_{j,k_1} \bar{\bar{a}}_{j-1,k-k_1-2}$$

$$+ \delta_{i,-1} \tilde{G}_{j,k-1} + 2(1 - \delta_{i,-1}) \sum_{k_1=0}^{k-3} \tilde{G}_{j,k_1} \bar{\bar{a}}_{-j-1,k-k_1-2}$$

With this, we have put Hill's equations into a form that provides a recursive procedure for calculating the individual coefficients, and contains only the required terms (i.e., no higher- or lower-order terms that need to be eliminated later). Our next step will be to move to finite sums instead of infinite sums, based on an observation that the sums automatically truncate after a certain point.

## 6. Transition to Finite Sums

Now that we have a version of Hill's equations that are much more conducive to calculation, we still need to eliminate the infinite sum. Fortunately, that is possible and shown here.

**Claim (6.1)**: for all $k \in \{1,2,3,\ldots\}$, define
$j' = j'(k) := (k + k\bmod 2)/2 + 1$, then
i) for all $k$ and all $J \geq j'$, $\bar{\bar{a}}_{J,k} = \bar{\bar{a}}_{-J,k} = 0$, and
ii) for all $k$ and all $K \leq j'$, $\bar{\bar{a}}_{j',K} = \bar{\bar{a}}_{-j',K} = 0$, and
i) and ii) are equivalent.

As a table:

| $k$ | $j'$ | i) | ii) |
|---|---|---|---|
| 1 | 2 | $\bar{\bar{a}}_{\pm 2,1} = \bar{\bar{a}}_{\pm 3,1} = \cdots = 0$ | $\bar{\bar{a}}_{\pm 2,1} = 0$ |
| 2 | 2 | $\bar{\bar{a}}_{\pm 2,2} = \bar{\bar{a}}_{\pm 3,2} = \cdots = 0$ | $\bar{\bar{a}}_{\pm 2,2} = \bar{\bar{a}}_{\pm 2,1} = 0$ |
| 3 | 3 | $\bar{\bar{a}}_{\pm 3,3} = \bar{\bar{a}}_{\pm 4,3} = \cdots = 0$ | $\bar{\bar{a}}_{\pm 3,3} = \bar{\bar{a}}_{\pm 3,2} = \cdots = 0$ |
| 4 | 3 | $\bar{\bar{a}}_{\pm 3,4} = \bar{\bar{a}}_{\pm 4,4} = \cdots = 0$ | $\bar{\bar{a}}_{\pm 3,4} = \bar{\bar{a}}_{\pm 3,3} = \cdots = 0$ |
| 5 | 4 | $\bar{\bar{a}}_{\pm 4,5} = \bar{\bar{a}}_{\pm 5,5} = \cdots = 0$ | $\bar{\bar{a}}_{\pm 4,5} = \bar{\bar{a}}_{\pm 4,4} = \cdots = 0$ |
| 6 | 4 | $\bar{\bar{a}}_{\pm 4,6} = \bar{\bar{a}}_{\pm 5,6} = \cdots = 0$ | $\bar{\bar{a}}_{\pm 4,6} = \bar{\bar{a}}_{\pm 4,5} = \cdots = 0$ |
| 7 | 5 | $\bar{\bar{a}}_{\pm 5,7} = \bar{\bar{a}}_{\pm 6,7} = \cdots = 0$ | $\bar{\bar{a}}_{\pm 5,7} = \bar{\bar{a}}_{\pm 5,6} = \cdots = 0$ |
| 8 | 5 | $\bar{\bar{a}}_{\pm 5,8} = \bar{\bar{a}}_{\pm 6,8} = \cdots = 0$ | $\bar{\bar{a}}_{\pm 5,8} = \bar{\bar{a}}_{\pm 5,7} = \cdots = 0$ |
| 9 | 6 | $\bar{\bar{a}}_{\pm 6,9} = \bar{\bar{a}}_{\pm 7,9} = \cdots = 0$ | $\bar{\bar{a}}_{\pm 6,9} = \bar{\bar{a}}_{\pm 6,8} = \cdots = 0$ |
| 10 | 6 | $\bar{\bar{a}}_{\pm 6,10} = \bar{\bar{a}}_{\pm 7,10} = \cdots = 0$ | $\bar{\bar{a}}_{\pm 6,10} = \bar{\bar{a}}_{\pm 6,9} = \cdots = 0$ |

**Proof:** First, we show that i) and ii) are equivalent, then that i) is true.
i) $\Rightarrow$ ii):
for $n \in \mathbb{N}, k - n \leq k \Rightarrow j'(k) \geq j'(k-n)$ & i) $\Rightarrow \bar{\bar{a}}_{j'(k),k-n} = 0 \Rightarrow$ ii).
ii) $\Rightarrow$ i):
for $n \in \mathbb{N}, k + 2n \geq k$, ii) $\Rightarrow \bar{\bar{a}}_{j'(k+2n),k} = 0$, but $j'(k+2n) = j'(k) + n$, so $\bar{\bar{a}}_{j'(k)+n,k} = 0 \Rightarrow$ i).

### 6.1. Preliminary calculations

$$(a + b)\bmod 2 = a\bmod 2 + b\bmod 2$$

**For the first sum:**
$$\max(k_2) = k - 2 \Rightarrow$$
$$j'(k_2) \leq j'(k-2) = \big((k-2) + (k-2)\bmod 2\big)/2 + 1 = (k + k\bmod 2)/2 = j'(k) - 1 \Rightarrow$$
$$j'(k) \geq j'(k_2) + 1$$

$$\max(k_3) = k - 2 \Rightarrow$$
$$j'(k) \geq j'(k_3) + 1$$

**For the second and third sum:**
$$\max(k_2) = k - 4 \Rightarrow$$
$$j'(k) \geq j'(k_2) + 2$$

$$\max(k_3) = k - 4 \Rightarrow$$
$$j'(k) \geq j'(k_3) + 2$$

**For the first sum:**
$$k_3 = k - k_1 - k_2 - 1 \Rightarrow k_3 \leq k - k_2 - 1 < k - k_2 \Rightarrow$$
$$j'(k_3) < j'(k - k_2) = \big((k - k_2) + (k - k_2)\bmod 2\big)/2 + 1$$
$$= (k + k\bmod 2)/2 - (k_2 + k_2\bmod 2)/2 + 1 - 1 + 1$$
$$= j'(k) - j'(k_2) + 1 \Rightarrow$$
$$-j'(k_2) > -j'(k) + j'(k_3) - 1$$

**For the second and third sum:**
$$k_3 = k - k_1 - k_2 - 3 \Rightarrow k_3 \leq k - k_2 - 3 < k - k_2 \Rightarrow$$
$$j'(k_3) < j'(k - k_2 - 2) = \big((k - k_2 - 2) + (k - k_2 - 2)\bmod 2\big)/2 + 1$$
$$= (k + k\bmod 2)/2 - (k_2 + k_2\bmod 2)/2 - 1 + 1 - 1 + 1$$
$$= j'(k) - j'(k_2) + 1 \Rightarrow$$
$$-j'(k_2) > -j'(k) + j'(k_3)$$

**For the fourth and fifth sum:**
$$k_2 := k - k_1 - 2$$
$$\max(k_2) = k - 2 \Rightarrow$$
$$j'(k) \geq j'(k_2) + 1$$

### 6.2. Proof of i):

The proof is done via induction on $k$. We start with the conditions of the claim and then look at various cases for each sum in the formula. Notice that the conditions of the claim are never fulfilled for $j \in \{1, -1\}$, so we can always assume $j \notin \{1, -1\}$ and only look at the last of the three formulas.

### 6.3. First sum:

Recall that, in $\sum_{\substack{i=-\infty \\ i \notin \{0,j\}}}^{+\infty}$ , $i \notin \{0, j\}$.

From the conditions of the claim, we know that $j \geq j'(k)$.

*Case $j > 0, i < 0$:*
$|i - j| = |i| + |j| \geq j + \min(|i|) = j + 1 \geq j'(k) + 1 \geq j'(k_3) + 2 > j'(k_3)$
Since $k_3 < k$, the induction hypothesis implies
$\bar{\bar{a}}_{i-j,k_3} = 0$

*Case $j > 0, i \geq j'(k_2)$:*
Since $k_2 < k$, the induction hypothesis implies
$\bar{\bar{a}}_{i,k_2} = 0$

*Case $j > 0, 0 < i < j'(k_2)$:*
$|i - j| = j - i > j'(k) - j'(k_2) > j'(k) - j'(k) + j'(k_3) - 1 = j'(k_3) - 1 \Rightarrow$
$|i - j| \geq j'(k_3)$
Since $k_3 < k$, the induction hypothesis implies
$\bar{\bar{a}}_{i-j,k_3} = 0$

*Case $j < 0, i > 0$:*
$|i - j| = -j + i \geq |j| + \min(|i|) = |j| + 1 \geq j'(k) + 1 \geq j'(k_3) + 2 > j'(k_3)$
Since $k_3 < k$, the induction hypothesis implies
$\bar{\bar{a}}_{i-j,k_3} = 0$

*Case $j < 0, i \leq -j'(k_2)$:*
$-i \geq -j'(k_2), |i| \geq j'(k_2)$
Since $k_2 < k$, the induction hypothesis implies
$\bar{\bar{a}}_{i,k_2} = 0$

*Case $j < 0, i < 0, i > -j'(k_2)$:*
$|i - j| = |-j + i| = |j| + i > j'(k) - j'(k_2) \geq j'(k) - j'(k) + j'(k_3) - 1 = j'(k_3) - 1 \Rightarrow$
$|i - j| \geq j'(k_3)$
Since $k_3 < k$, the induction hypothesis implies
$\bar{\bar{a}}_{i-j,k_3} = 0$

### 6.4. Second sum:

Recall that, in $\sum_{\substack{i=-\infty \\ i \notin \{0,j-1\}}}^{+\infty}$ , $i \notin \{0, j - 1\}$.

*Case $j > 0, i < 0$:*

$|-i+j-1| = |i|+|j|-1 \geq j + \min(|i|) - 1 = j \geq j'(k) \geq j'(k_3) + 2 > j'(k_3)$
Since $k_3 < k$, the induction hypothesis implies
$\bar{\bar{a}}_{-i+j-1,k_3} = 0$

*Case $j > 0, i \geq j'(k_2)$:*
Since $k_2 < k$, the induction hypothesis implies
$\bar{\bar{a}}_{i,k_2} = 0$

*Case $j > 0, 0 < i < j'(k_2)$:*
$|-i+j-1| = j-i-1 > j'(k) - j'(k_2) - 1 > j'(k) - j'(k) + j'(k_3) - 1 = j'(k_3) - 1 \Rightarrow$
$|-i+j-1| \geq j'(k_3)$
Since $k_3 < k$, the induction hypothesis implies
$\bar{\bar{a}}_{-i+j-1,k_3} = 0$

*Case $j < 0, i > 0$:*
$|-i+j-1| = |-j+i+1| = |j| + i - 1 \geq j + \min(|i|) - 1 = j \geq j'(k) \geq j'(k_3) + 2 > j'(k_3)$
Since $k_3 < k$, the induction hypothesis implies
$\bar{\bar{a}}_{-i+j-1,k_3} = 0$

*Case $j < 0, i \leq -j'(k_2)$:*
$|i| \geq j'(k_2)$
Since $k_2 < k$, the induction hypothesis implies
$\bar{\bar{a}}_{i,k_2} = 0$

*Case $j < 0, i < 0, i > -j'(k_2)$:*
$|-i+j-1| = |-j+i+1| = |j| + i + 1 > j'(k) - j'(k_2) + 1 > j'(k) - j'(k) + j'(k_3) + 1$
$\qquad = j'(k_3) + 1 \Rightarrow$
$|-i+j-1| \geq j'(k_3)$
Since $k_3 < k$, the induction hypothesis implies
$\bar{\bar{a}}_{-i+j-1,k_3} = 0$

### 6.5. Third sum:

Recall that, in $\sum_{\substack{i=-\infty \\ i \notin \{0,-j-1\}}}^{+\infty}$, $i \notin \{0, -j-1\}$.

*Case $j > 0, i > 0$:*
$|-i-j-1| = |j+i+1| = j+i+1 \geq j+2 \geq j'(k) + 2 \geq j'(k_3) + 4 > j'(k_3)$
Since $k_3 < k$, the induction hypothesis implies
$\bar{\bar{a}}_{-i-j-1,k_3} = 0$

*Case $j > 0, i < 0, i \leq -j'(k_2)$:*
$|i| \geq j'(k_2)$
Since $k_2 < k$, the induction hypothesis implies
$\bar{\bar{a}}_{i,k_2} = 0$

*Case $j > 0, i < 0, i > -j'(k_2)$:*
$|i| < j'(k_2)$
$|-i-j-1| = |j+i+1| = j+i+1 = j - |i| + 1 > j'(k) - j'(k_2) + 1$
$\qquad > j'(k) - j'(k) + j'(k_3) + 1 = j'(k_3) + 1 > j'(k_3)$
Since $k_3 < k$, the induction hypothesis implies
$\bar{\bar{a}}_{-i-j-1,k_3} = 0$

*Case $j < 0, i < 0$:*
$|-i-j-1| = |j| + |i| - 1 \geq |j| + \min(|i|) - 1 = |j| \geq j'(k) \geq j'(k_3) + 2 > j'(k_3)$
Since $k_3 < k$, the induction hypothesis implies
$\bar{\bar{a}}_{-i-j-1,k_3} = 0$

*Case $j < 0, i \geq j'(k_2)$:*
$|i| \geq j'(k_2)$
Since $k_2 < k$, the induction hypothesis implies
$\bar{\bar{a}}_{i,k_2} = 0$

*Case $j < 0, i > 0, i < j'(k_2)$:*
$|i| < j'(k_2)$

$$|-i-j-1| = |j+i+1| = |j| - |i| + 1 > j'(k) - j'(k_2) + 1 > j'(k) - j'(k) + j'(k_3) + 1$$
$$= j'(k_3) + 1 > j'(k_3)$$

Since $k_3 < k$, the induction hypothesis implies
$$\bar{\bar{a}}_{-i-j-1,k_3} = 0$$

### 6.6. Fourth sum:

*Case $j > 0$:*
$$|j-1| = j - 1 \geq j'(k) - 1 \geq j'(k_2)$$

Since $k_2 < k$, the induction hypothesis implies
$$\bar{\bar{a}}_{j-1,k_2} = 0$$

*Case $j < 0$:*
$$|j-1| = |j| + 1 \geq j'(k) + 1 \geq j'(k_2) + 2 \geq j'(k_2)$$

Since $k_2 < k$, the induction hypothesis implies
$$\bar{\bar{a}}_{j-1,k_2} = 0$$

### 6.7. Fifth sum:

*Case $j > 0$:*
$$|-j-1| = j + 1 \geq j'(k) + 1 \geq j'(k_2) + 2 > j'(k_2)$$

Since $k_2 < k$, the induction hypothesis implies
$$\bar{\bar{a}}_{-j-1,k_2} = 0$$

*Case $j < 0$:*
$$|-j-1| = |j| - 1 \geq j'(k) - 1 \geq j'(k_2)$$

Since $k_2 < k$, the induction hypothesis implies
$$\bar{\bar{a}}_{-j-1,k_2} = 0. \blacksquare$$

Now we can use Claim (6.1) to change the infinite sums over $i$ to finite sums. In the first sum, $\max(k_2) = k - 2$, so if $|i| \geq j'(k) - 1 = j'(\max(k_2))$, then $|i| \geq j'(k_2)$ and claim (6.1) tells us that $c_{i,k_2} = 0$. In the second and third sum, $\max(k_2) = k - 4$, so if $|i| \geq j'(k) - 2 = j'(\max(k_2))$, then $|i| \geq j'(k_2)$ and claim (6.1) tells us that $\bar{\bar{a}}_{i,k_2} = 0$.

$$\bar{\bar{a}}_{j,k} = \sum_{\substack{i=-(j'(k)-2) \\ i \notin \{0,j\}}}^{+(j'(k)-2)} \sum_{k_1=0}^{k-3} \sum_{k_2=1}^{k-k_1-2} E_{j,i,k_1} \bar{\bar{a}}_{i,k_2} \bar{\bar{a}}_{i-j,k-k_1-k_2-1}$$
$$+ \sum_{\substack{i=-(j'(k)-3) \\ i \notin \{0,j-1\}}}^{+(j'(k)-3)} \sum_{k_1=0}^{k-5} \sum_{k_2=1}^{k-k_1-4} \tilde{F}_{j,k_1} \bar{\bar{a}}_{i,k_2} \bar{\bar{a}}_{-i+j-1,k-k_1-k_2-3}$$
$$+ \sum_{\substack{i=-(j'(k)-3) \\ i \notin \{0,-j-1\}}}^{+(j'(k)-3)} \sum_{k_1=0}^{k-5} \sum_{k_2=1}^{k-k_1-4} \tilde{G}_{j,k_1} \bar{\bar{a}}_{i,k_2} \bar{\bar{a}}_{-i-j-1,k-k_1-k_2-3}$$
$$+ \delta_{i,1} \tilde{F}_{j,k-1} + 2(1 - \delta_{i,1}) \sum_{k_1=0}^{k-3} \tilde{F}_{j,k_1} \bar{\bar{a}}_{j-1,k-k_1-2}$$
$$+ \delta_{i,-1} \tilde{G}_{j,k-1} + 2(1 - \delta_{i,-1}) \sum_{k_1=0}^{k-3} \tilde{G}_{j,k_1} \bar{\bar{a}}_{-j-1,k-k_1-2} \quad (6.1)$$

When doing the calculations, we will see a special case for $\bar{\bar{a}}_0$. This is because, by definition, the full sum is $\bar{\bar{a}}_0 = \frac{1}{m}\frac{a_0}{a_0} = \frac{1}{m}$ and we don't have a Taylor expansion for this. So, for this case, we have chosen to extend the power series by one term and write

$$\bar{\bar{a}}_0 = \sum_{n=-1}^{\infty} \bar{\bar{a}}_{0,n} m^n$$

where $\bar{\bar{a}}_{0,-1} = 1$ and $\bar{\bar{a}}_{0,n} = 0$ for all $n \neq -1$.

Now we want to apply the same transition to finite sums to the calculation of $a_0$ and $C$. Equation (4.67) and $\frac{a_i}{a_0} = m\bar{\bar{a}}_i$ lead to

$$a_0 = \left[ m^{-2} \sum_{i=-(j'(k)-1)}^{(j'(k)-1)} [(2i+1+m)^2 + 2m^2] m\bar{\bar{a}}_i \left( \sum_{i=-(j'(k)-1)}^{(j'(k)-1)} m\bar{\bar{a}}_i \right)^2 \right]^{-1/3} \quad (6.2)$$

This gives us a finite sum for $a_0$.[39]

Equation (4.66) and $\frac{a_i}{a_0} = m\bar{\bar{a}}_i$ lead to

$$C = -\frac{a_0^2}{2m^2} \sum_{i=-(j'(k)-1)}^{(j'(k)-1)} \left\{ \left[ (2i+1)^2 + 8im + 4m + \frac{9}{2}m^2 \right] m^2 \bar{\bar{a}}_i^2 + \frac{9}{2} m^4 \bar{\bar{a}}_i \bar{\bar{a}}_{-i-1} \right\} \quad (6.3)$$

This gives us a finite sum for $C$.[40]

Now we are ready to calculate some partial sums for $a_0$ and $C$.

$$\sum_{n=0}^{5} a_{0,n} = m^{\frac{2}{3}} \left( 1 - \frac{2m}{3} + \frac{7m^2}{18} - \frac{4m^3}{81} + \frac{19565 m^4}{62208} - \frac{47161 m^5}{93312} \right)$$

This is the partial sum of $a_0$ up to a power of $m^5$.[41]

$$\sum_{n=0}^{5} C_n = m^{-\frac{2}{3}} \left( -\frac{1}{2} - \frac{4m}{3} - \frac{7m^2}{36} + \frac{70m^3}{81} + \frac{39533 m^4}{15552} + \frac{1271 m^5}{729} \right)$$

This is the partial sum of $C$ up to a power of $m^5$.[42]

Now that we have an efficient method for calculating the coefficients, using a recursive formula with finite sums, we will make a brief discursion into the initial conditions for our solutions.

---

[39] This agrees with (Wintner 2014) §506, equation (13), and (G.W. Hill 1878) page 146. However, (Wintner 1929) (5) differs by a factor of $m^2$.

[40] This differs from (Wintner 2014) §506, equation (12), and (G.W. Hill 1878) page 146 by a factor of $-\frac{1}{2m^2}$.

[41] This agrees with (Wintner 2014) §514, equation (34bis).

[42] This differs from (Wintner 2014) §514, equation (35) by a factor of $-\frac{1}{2}$. The means that (Wintner 2014) §506, equation (13) and §514 and equation (35) do not agree with each other, making it look like there is some confusion about which of Hill's versions of $C$ are being used.

## 7. Initial conditions

Based on the Fourier series, equations (4.2) and (4.3), we can formulate the initial conditions. This will be useful when we calculate specific orbits.

$$q_1(0) = \sum_{i=0}^{\infty} A_i \tag{7.1}$$

$$\dot{q}_1(0) = 0 \tag{7.2}$$

$$\ddot{q}_1(0) = -\sum_{i=0}^{\infty} \left(\frac{2i+1}{m}\right)^2 A_i \tag{7.3}$$

$$q_2(0) = 0 \tag{7.4}$$

$$\dot{q}_2(0) = \sum_{i=0}^{\infty} \frac{2i+1}{m} B_i \tag{7.5}$$

$$\ddot{q}_2(0) = 0 \tag{7.6}$$

The value of 0 for equations (7.2), (7.4), and (7.6) is simply a result of our assumptions about the symmetry of the orbits.

Now we are ready to calculate the partial sums.

$$\sum_{n=0}^{5} q_1(0)_n = m^{\frac{2}{3}}\left(1 - \frac{2m}{3} - \frac{11m^2}{18} - \frac{89m^3}{162} + \frac{1477m^4}{7776} - \frac{38051m^5}{116640}\right)$$

$$\sum_{n=0}^{5} \dot{q}_2(0)_n = m^{-\frac{1}{3}}\left(1 - \frac{2m}{3} + \frac{77m^2}{36} + \frac{158m^3}{81} + \frac{36029m^4}{15552} + \frac{12901m^5}{7290}\right)$$

Wintner Notes:

$$C = m^{\frac{4}{3}}\left(1 + \frac{8m}{3} + \frac{7m^2}{18} - \frac{140m^3}{81} - \frac{39533m^4}{7776} - \frac{2542m^5}{729} - \frac{424859m^6}{209952} - \frac{421861m^7}{157464} \right.$$
$$\left. - \frac{1890215575m^8}{483729408} + \frac{10199959357m^9}{816293376} + \frac{9665055967729m^{10}}{195910410240}\right)$$

With this, we have fully covered the definition and calculations of the basic calculations, i.e. the closed, symmetric orbits. This puts us in a position to move closer to the real motion of the moon, where the orbits are characterized by three different periods.

## 8. The motion of the perigee

In this chapter, we will look at Hill's calculations of the anomalistic period, which determines, to a large part, the motion of the perigee.

### 8.1. Summary of Hill's Numeric Perigee

The first paper, Hill's numeric perigee (G.W. Hill 1886), calculates the moon's orbit including the motion of the perigee. Specifically, it includes the synodic and anomalistic periods but does not cover the draconic period. The paper makes use of two methods that were not previously known: 1) The orbit is calculated as the continuation of a symmetric, periodic, closed orbit of the same energy, and 2) calculations involve the use of determinants of infinite matrices. Both of these methods provided significant stimulus to the further development of pure mathematics (Poincaré 1892) and (Poincaré 1886).

The introduction provides the motivation for the paper, noting that experiments have measured the orbit more accurately than theory has calculated it, and stating a goal of theoretical calculations that will determine whether current theory is sufficient to calculate orbits up to the accuracy of existing measurements.

Section I (page 2) is difficult to reproduce, since it does not state what the equations of motion or the potential function are. Between equations (3) and (4) is a discussion of formulating $C$ as a power series in $c$, but $c$ is not defined. The conclusion is that $\delta C = 0$, but that could also be an assumption, stating that the goal is to calculate an orbit that is a continuation of a symmetric, periodic, closed orbit of the same energy. Equation (8) is also known as Hill's differential equation. We can reproduce equations (3) to (9). This section is not used in the following.

Section II (page 6) explicitly states the potential and the equations of motion. The coordinates are mapped to the complex plane and the "operator $\delta$" (a derivation) is applied in equation (12). Before this step, the equations agree with our version of Hill's paper 2 (G.W. Hill 1878) (after implicit substitution of the potential and conversion back to Cartesian coordinates).

Between equations (15) and (16), "Hill's differential equation" is derived, which has variables $\theta$ and $w$. We understand this equation as a linearization of the anomalistic period. Between equations (17) and (18), a Fourier expansion is applied to the coordinates, yielding coefficients $a_i$ as in paper 2. This is used to derive a newer version of the function $\theta$. Then the functions $R_i$ are defined and used to derive the equations defining the $a_i$. We cannot reproduce all of this, but it is treated more thoroughly in paper 2. Between equations (18) and (19), the variables $U_i$ and $h_i$ are defined, and equations for calculating them are derived. This is an infinite set of infinite sums, analogous to the case for $R_i$ and $a_i$. The section concludes with a discussion of some properties of the $U_i$ and a reference to paper 2. The $R_i$, $a_i$, $U_i$, and $h_i$ are used in the last section for the calculations of the moon's orbit.

Section III (page 17) begins by applying Fourier expansions to the elements of Hill's differential equation ($\theta$ and $w$). This leads to equations for the coefficients $\theta_i$ and $w_i$[43]. By treating this infinite set of infinite sums as a matrix, the solutions can be found by defining determinants of this and other matrices. The solution involves a constant $c$, which is the ratio of the synodic to the anomalistic month.

Section IV (page 33) is devoted to calculating the numerical values of the moon's orbit. First, the $a_i$ are used to calculate the $U_i$, and $h_i$ and, from that calculate the $\theta_i$, as in Hill's paper 3. This is then applied to the matrix in equation (24) of section III, leading to a calculation of $c$. Finally, the $w_i$ can be calculated via equation (21) of section III, leading to the motion of the perigee in the orbit of the moon.

---

[43] We use $w_i$ where Hill uses $b_i$, because this is just a coefficient in the series for $w$.

## 8.2. Summary of Hill's Literal Perigee

This is an outline of how Hill's literal perigee (G.W. Hill 1894) solves the equations for $c$ (and $w$), where $c$ is the anomalistic period and $w$ is.

The paper begins with an introduction and a review of some topics from Hill's Numeric Perigee paper (page 31-32). Then, approximations are calculated for the $R_i$, $U_i$, and $\theta_i$. A comparison with our calculations shows the name of the coefficient, the maximum power of $m$ calculated, the number of coefficients calculated, and the number of mismatches with our calculations:

| coefficients | max. power of $m$ | number of coefficients | mismatches | see section |
|---|---|---|---|---|
| $R_i$ | 11 | 23 | 1 ($R_{2,8}$) | 9.5. Hill's literal perigee page 33, $R_i$ |
| $U_i$ | 10 | 30 | 2 ($U_{1,10}, U_{-1,10}$) | 9.6. Hill's literal perigee page 34, $U_i$ |
| $\theta_i$ | 11 | 26 | 3 ($\theta_{0,10}, \theta_{1,10}, \theta_{2,8}$) | 9.9. Hill's literal perigee page 34, $\theta_i$ |

Next, the matrix (21) from Hill's Numeric Perigee is cited and is followed by a discussion of the order (power of $m$) with increasing index of $s_i = [i]$, $\theta_i$, and $w_i = b_i$.

Order of $s_i = [i]$:
We know:

$$\theta_0 = 1 + 2m - \frac{1}{2}m^2 + \mathcal{O}(3)$$

$$c^2 = 1 + 2m - \frac{1}{2}m^2 + \mathcal{O}(3)$$

which implies:

$$c = 1 + m - \frac{3}{4}m^2 + \mathcal{O}(3)$$

The equation for $c^2$ results from our first approximation, using only a single term in the middle of the matrix, which then yields $s_0 = 0$ or $c^2 = \theta_0 + \mathcal{O}(3)$.

$$s_0 = c^2 - \theta_0 = \mathcal{O}(3)$$
$$s_{-1} = (c-2)^2 - \theta_0 = c^2 - 4c + 4 - \theta_0 = (1 - 4 + 4 - 1)m^0 + (2 - 4 - 2)m^1 + \mathcal{O}(2) = \mathcal{O}(1)$$
$$s_1 = (c+2)^2 - \theta_0 = c^2 + 4c + 4 - \theta_0 = (1 + 4 + 4 - 1)m^0 + \mathcal{O}(1) = \mathcal{O}(0)$$

This is used in Hill numeric perigee page 22 and Hill literal perigee pages 35 and 37.

Order of $\theta_i = \mathcal{O}(2i)$ is clear from the expansion.

Order of $w_i = b_i$. Claim: $w_{-1}, w_1, w_{-2}, w_2, w_{-3}$ have order 1,2,3,4,5. We don't see why this is the case, nor where it is used, so we don't need the statement.

Page 35: Claim: Three approximations are necessary. This first is on page 32, stating that we will need $m^{11}$. We don't need this statement.

Page 36 big equation: This is equation 0 after elimination of all $w_i$. We need to understand which terms to eliminate based on their order, but this is clear from the order of the $\theta_i$.

Page 36-37: This is a calculation of products of $\theta_i$ as power series: it is straightforward and in our software.

Page 37: Approximation of $c^2$ and $c$ up to $m^6$. Here, we have a discrepancy in the big equation on page 36. Our calculations give a third term of

$$\left(\frac{1}{s_{-1}s_1s_2} + \frac{1}{s_{-1}s_1s_{-2}}\right)\theta_1^4$$

see our appendix Hill's perigee results 2 page 7.

Then, an approximation is calculated by neglecting all terms involving $w_i$ with $|i| > 3$ and then eliminating all $w_i$ results in an equation involving products of the $\theta_i, i = 1,2,3,4$ of order up to five. After that, these $\theta_i$ are replaced by their respective power series in $m$, terms of 7th order are neglected, resulting in an expression for $c$, which is solved, producing a power series in $m$.

Page 38: first, calculation of reciprocals of $s_i$, which is straightforward (Taylor) in software. We don't need this step.

Page 38: Important substitution of approximation of $c$ up to $m^6$ in main equation. We don't understand why this substitution is valid, and do not use it in our calculations. This is a key question and might be the basis of a recursive procedure. This results in a new equation for $c^2$, which can be solved by taking a cube root, then a square root and applying Taylor, resulting in a power series up to $m^{11}$ for $c$.

It seems noteworthy that this paper does not make use of a determinant. Instead, it does the following:
1. neglect higher order terms
2. eliminate the remaining $w_i$ (in the equation for $j = 0$)
3. replace all $\theta_i$ by power series in $m$
4. replace all $\frac{1}{s_j} = \frac{1}{[j]}$ by power series in $m$
5. solve resulting expression for $c$

However, the calculations involve making multiple approximations and solving one equation for $c^2$ (page 37 and 39) and another for $(c - \theta_0)^3$ (page 39).

## 8.3. Details of Hill's Numeric Perigee

### 8.3.1. Hill's Introduction

Section I in Hill's numeric perigee will be left out, because it is not used in the remainder of Hill's paper.

### 8.3.2. Hill's Section II (without $\Omega$)

This section derives the key equation for the motion of the perigee, $w$, involving the parameter $\Theta$. As done in the other cases, these variables are expanded in Fourier series, the coefficients of which are assumed to be power series in $m$. With this, two intermediate variables, $R_i$ and $U_i$, are defined and used as an aid in finding the parameter $\Theta$ and solutions for $w$.[44]

For use when we convert from Hill's notation to ours, we will need some partial derivatives of $\Omega$, as defined in (3.2.2).

$$\Omega := \frac{m^2}{\sqrt{u_1 u_2}} + \frac{3}{8} m^2 (u_1 + u_2)^2. \tag{8.3.2.1}$$

With the above observations,

$$\Omega = m^2 \left( \frac{1}{r} + \frac{3}{2} q_1^{\ 2} \right) = -m^2 U$$

where $U$ is our version of the effective potential (3.2.2).
The derivatives of $\Omega$ are:

---

[44] Here we will repeat parts of section II but replace the effective potential $\Omega$ and its derivatives by their coordinates and replace $u$ and $s$ by $u_1$ and $u_2$.

$$\left.\begin{aligned}\frac{d\Omega}{du_1} &= \left(-\frac{1}{2}m^2\frac{u_2}{r^3} + \frac{3}{4}m^2(u_1+u_2)\right) \\ \frac{d\Omega}{du_2} &= \left(-\frac{1}{2}m^2\frac{u_1}{r^3} + \frac{3}{4}m^2(u_1+u_2)\right) \\ \frac{d^2\Omega}{du_1{}^2} &= \left(\frac{3}{4}m^2\frac{u_2{}^2}{r^5} + \frac{3}{4}m^2\right) \\ \frac{d^2\Omega}{du_2{}^2} &= \left(\frac{3}{4}m^2\frac{u_1{}^2}{r^5} + \frac{3}{4}m^2\right) \\ \frac{d^2\Omega}{du_1 du_2} &= \frac{3}{4}\frac{m^2 u_1 u_2}{r^5} - \frac{1}{2}\frac{m^2}{r^3} + \frac{3}{4}m^2 \\ \frac{d^2\Omega}{du_1 du_2} &= \left(\frac{1}{4}\frac{m^2}{r^3} + \frac{3}{4}m^2\right) \\ d\Omega &= \frac{d\Omega}{du_1}Du_1 + \frac{d\Omega}{du_2}Du_2 \end{aligned}\right\} \quad (8.3.2.2)$$

where $r := \sqrt{u_1 u_2}$.

Here we start with equations (4.28) and (4.29).

$$\left.\begin{aligned}D^2 u_1 + 2mDu_1 - m^2\frac{u_1}{r^3} + \frac{3}{2}m^2(u_1+u_2) &= 0 \\ D^2 u_2 - 2mDu_2 - m^2\frac{u_2}{r^3} + \frac{3}{2}m^2(u_1+u_2) &= 0\end{aligned}\right\} \quad (8.3.2.3)$$

rearranging

$$\left.\begin{aligned}\left[D^2 + 2mD + \frac{3}{2}m^2 - m^2\frac{1}{r^3}\right]u_1 + \frac{3}{2}m^2 u_2 &= 0 \\ \left[D^2 - 2mD + \frac{3}{2}m^2 - m^2\frac{1}{r^3}\right]u_2 + \frac{3}{2}m^2 u_1 &= 0\end{aligned}\right\} \quad (8.3.2.4)$$

where $r := \sqrt{u_1 u_2}$. We will solve these equations for the second derivative for usage later.

Now we multiply the first equation in (8.3.2.3) by $Du_2$, the second by $Du_1$ and add, giving

$$D^2 u_1 Du_2 + Du_1 D^2 u_2 + 2mDu_1 Du_2 - 2mDu_1 Du_2 - m^2\frac{u_1 Du_2}{r^3} - m^2\frac{u_2 Du_1}{r^3}$$
$$+ \frac{3}{2}m^2(u_1+u_2)(Du_1+Du_2) = 0$$

rearranging

$$D^2 u_1 Du_2 + Du_1 D^2 u_2 - \frac{m^2}{r^3}(u_1 Du_2 + u_2 Du_1) + \frac{3}{2}m^2(u_1+u_2)(Du_1+Du_2) = 0$$

in other words

$$D(Du_1 Du_2) + 2m^2 D\left[\frac{1}{r} + \frac{3}{8}(u_1+u_2)^2\right] = 0$$

integrating

$$Du_1 Du_2 + 2m^2\left[\frac{1}{r} + \frac{3}{8}(u_1+u_2)^2\right] = \text{constant}$$

defining the constant to be $2C$

$$2C = Du_1 Du_2 + 2m^2\left[\frac{1}{r} + \frac{3}{8}(u_1+u_2)^2\right] \quad (8.3.2.5)$$

Based on (8.3.2.1), we can also write

$$C = \frac{1}{2}Du_1 Du_2 + \Omega$$

Since $C$ is a constant, $DC = 0$, and we have

$$D\Omega = -\frac{1}{2}D(Du_1 Du_2) \tag{8.3.2.6}$$

(8.3.2.5) matches our (4.30) if we change $2C$ to $-2m^2 C$. In summary,

$$\left. \begin{aligned} D^2 u_1 &= -2mDu_1 - \frac{3}{2}m^2 u_1 + m^2 \frac{1}{r^3}u_1 - \frac{3}{2}m^2 u_2 \\ D^2 u_2 &= 2mDu_2 - \frac{3}{2}m^2 u_2 + m^2 \frac{1}{r^3}u_2 - \frac{3}{2}m^2 u_1 \\ 2C &= Du_1 Du_2 + \frac{2m^2}{r} + \frac{3}{4}m^2(u_1 + u_2)^2 \end{aligned} \right\} \tag{8.3.2.7}$$

Now we'll try applying the operator $\delta$ to (8.3.2.7). For that, we only need to know that $\delta$ is a derivation, i.e. follows the Leibnitz rules. The operator $\delta$ is sometimes called a variation and will help us find a linear approximation.

$$\left. \begin{aligned} D^2 \delta u_1 + 2mD\delta u_1 - m^2 \frac{\delta u_1}{r^3} + \frac{3}{2}m^2 \frac{u_1}{r^5}(u_1 \delta u_2 + u_2 \delta u_1) + \frac{3}{2}m^2(\delta u_1 + \delta u_2) &= 0 \\ D^2 \delta u_2 - 2mD\delta u_2 - m^2 \frac{\delta u_2}{r^3} + \frac{3}{2}m^2 \frac{u_2}{r^5}(u_1 \delta u_2 + u_2 \delta u_1) + \frac{3}{2}m^2(\delta u_1 + \delta u_2) &= 0 \\ Du_1 D\delta u_2 + Du_2 D\delta u_1 - m^2 \frac{u_1 \delta u_2 + u_2 \delta u_1}{r^3} + \frac{3}{2}m^2(u_1 + u_2)(\delta u_1 + \delta u_2) &= 0 \end{aligned} \right\} \tag{8.3.2.8}$$

where we have applied the law of conservation (the Jacobi integral), i.e. $\delta C = 0$. If we change $\delta$ to $D$ in these equations, they still hold, so

$$\delta u_1 = Du_1, \qquad \delta u_2 = Du_2$$

is a particular solution. Then we define $v_1$ and $v_2$[45] such that

$$\delta u_1 = (Du_1)v_1, \qquad \delta u_2 = (Du_2)v_2$$

Substituting this into (8.3.2.8) gives us

$$D^2(Du_1)v_1 + 2mD((Du_1)v_1) - m^2 \frac{(Du_1)v_1}{r^3} + \frac{3}{2}m^2 \frac{u_1}{r^5}(u_1(Du_2)v_2 + u_2(Du_1)v_1)$$
$$+ \frac{3}{2}m^2((Du_1)v_1 + (Du_2)v_2) = 0 \tag{8.3.2.9a}$$

$$D^2(Du_2)v_2 - 2mD((Du_2)v_2) - m^2 \frac{(Du_2)v_2}{r^3} + \frac{3}{2}m^2 \frac{u_2}{r^5}(u_1(Du_2)v_2 + u_2(Du_1)v_1)$$
$$+ \frac{3}{2}m^2((Du_1)v_1 + (Du_2)v_2) = 0 \tag{8.3.2.9b}$$

$$Du_1 D((Du_2)v_2) + Du_2 D((Du_1)v_1) - m^2 \frac{u_1(Du_2)v_2 + u_2(Du_1)v_1}{r^3}$$
$$+ \frac{3}{2}m^2(u_1 + u_2)((Du_1)v_1 + (Du_2)v_2) = 0 \tag{8.3.2.9c}$$

taking the derivatives

$$D^3 u_1 v_1 + 2D^2 u_1 Dv_1 + Du_1 D^2 v_1 + 2m(D^2 u_1)v_1 + 2mDu_1 Dv_1 - m^2 \frac{(Du_1)v_1}{r^3}$$
$$+ \frac{3}{2}m^2 \frac{u_1}{r^5}(u_1(Du_2)v_2 + u_2(Du_1)v_1) + \frac{3}{2}m^2((Du_1)v_1 + (Du_2)v_2) = 0 \tag{8.3.2.10a}$$

---

[45] We use $v_1$ for Hill's $v$ and $v_2$ for Hill's $w$.

$$D^3 u_2 v_2 + 2D^2 u_2 D v_2 + D u_2 D^2 v_2 - 2m(D^2 u_2) v_2 - 2m D u_2 D v_2 - m^2 \frac{(Du_2)v_2}{r^3}$$
$$+ \frac{3}{2} m^2 \frac{u_2}{r^5}(u_1(Du_2)v_2 + u_2(Du_1)v_1) + \frac{3}{2} m^2 \big((Du_1)v_1 + (Du_2)v_2\big) = 0 \quad (8.3.2.10b)$$

$$Du_1(D^2 u_2)v_2 + Du_1 Du_2 Dv_2 + Du_2(D^2 u_1)v_1 + Du_1 Du_2 Dv_1$$
$$- m^2 \frac{u_1(Du_2)v_2 + u_2(Du_1)v_1}{r^3} + \frac{3}{2} m^2 (u_1 + u_2)\big((Du_1)v_1 + (Du_2)v_2\big) = 0 \quad (8.3.2.10c)$$

rearranging

$$Du_1 D^2 v_1 + 2[D^2 u_1 + m Du_1] Dv_1$$
$$+ \left[D^3 u_1 + 2m(D^2 u_1) - m^2 \frac{(Du_1)}{r^3} + \frac{3}{2} m^2 \frac{u_1 u_2 (Du_1)}{r^5} + \frac{3}{2} m^2 Du_1\right] v_1$$
$$+ \left[\frac{3}{2} m^2 \frac{u_1 u_1 (Du_2)}{r^5} + \frac{3}{2} m^2 Du_2\right] v_2 = 0 \quad (8.3.2.11a)$$

$$Du_2 D^2 v_2 + 2[D^2 u_2 - m Du_2] Dv_2$$
$$+ \left[D^3 u_2 - 2m(D^2 u_2) - m^2 \frac{(Du_2)}{r^3} + \frac{3}{2} m^2 \frac{u_1 u_2 (Du_2)}{r^5} + \frac{3}{2} m^2 Du_2\right] v_2$$
$$+ \left[\frac{3}{2} m^2 \frac{u_2 u_2 (Du_1)}{r^5} + \frac{3}{2} m^2 Du_1\right] v_1 = 0 \quad (8.3.2.11b)$$

$$Du_1 Du_2 D(v_1 + v_2)$$
$$+ \left[Du_2(D^2 u_1) + \left(-m^2 \frac{u_2}{r^3} + \frac{3}{2} m^2 (u_1 + u_2)\right)(Du_1)\right] v_1$$
$$+ \left[Du_1(D^2 u_2) + \left(-m^2 \frac{u_1}{r^3} + \frac{3}{2} m^2 (u_1 + u_2)\right)(Du_2)\right] v_2 = 0 \quad (8.3.2.11c)$$

(8.3.2.11)[46]. From (8.3.2.3), we can deduce

$$\left. \begin{array}{l} D^2 u_1 = -2m Du_1 + m^2 \dfrac{u_1}{r^3} - \dfrac{3}{2} m^2 (u_1 + u_2) \\[4pt] D^2 u_2 = 2m Du_2 + m^2 \dfrac{u_2}{r^3} - \dfrac{3}{2} m^2 (u_1 + u_2) \end{array} \right\} \quad (8.3.2.12)$$

taking $D$

$$\left. \begin{array}{l} D^3 u_1 = -2m D^2 u_1 + m^2 \dfrac{Du_1}{r^3} - \dfrac{3}{2} m^2 \dfrac{u_1}{r^5}(u_1 Du_2 + u_2 Du_1) - \dfrac{3}{2} m^2 (Du_1 + Du_2) \\[4pt] D^3 u_2 = 2m D^2 u_2 + m^2 \dfrac{Du_2}{r^3} - \dfrac{3}{2} m^2 \dfrac{u_2}{r^5}(u_1 Du_2 + u_2 Du_1) - \dfrac{3}{2} m^2 (Du_1 + Du_2) \end{array} \right\} \quad (8.3.2.13)$$

substituting $D^2 u_1$ and $D^2 u_2$ from (8.3.2.3)

$$D^3 u_1 = -2m\left(-2m Du_1 + m^2 \frac{u_1}{r^3} - \frac{3}{2} m^2 (u_1 + u_2)\right) + m^2 \frac{Du_1}{r^3}$$
$$- \frac{3}{2} m^2 \frac{u_1}{r^5}(u_1 Du_2 + u_2 Du_1) - \frac{3}{2} m^2 (Du_1 + Du_2) \quad (8.3.2.14a)$$

---

[46] This agrees with the last equation before (13) in (G.W. Hill 1878), page 7 (when we replace $\Omega$ and its derivatives by coordinates and their derivatives).

$$D^3 u_2 = 2m\left(2mDu_2 + m^2\frac{u_2}{r^3} - \frac{3}{2}m^2(u_1+u_2)\right) + m^2\frac{Du_2}{r^3}$$

$$-\frac{3}{2}m^2\frac{u_2}{r^5}(u_1 Du_2 + u_2 Du_1) - \frac{3}{2}m^2(Du_1 + Du_2) \quad (8.3.2.14b)$$

Now we can use these equations for $D^2u_1$, $D^2u_2$, $D^3u_1$, and $D^3u_2$ (8.3.2.12) and (8.3.2.14) to eliminate these second and third derivatives of $u_1$ and $u_2$ from (8.3.2.11).

$$Du_1 D^2 v_1 + 2\left[-2mDu_1 + m^2\frac{u_1}{r^3} - \frac{3}{2}m^2(u_1+u_2) + mDu_1\right]Dv_1$$

$$+ \left[-2m\left(-2mDu_1 + m^2\frac{u_1}{r^3} - \frac{3}{2}m^2(u_1+u_2)\right) + m^2\frac{Du_1}{r^3}\right.$$

$$- m^2\frac{3}{2}\frac{u_1}{r^5}(u_1 Du_2 + u_2 Du_1) - \frac{3}{2}m^2(Du_1 + Du_2)$$

$$+ 2m\left(-2mDu_1 + m^2\frac{u_1}{r^3} - \frac{3}{2}m^2(u_1+u_2)\right) - m^2\frac{(Du_1)}{r^3} + \frac{3}{2}m^2\frac{u_1 u_2(Du_1)}{r^5}$$

$$\left. + \frac{3}{2}m^2 Du_1\right]v_1 + \left[\frac{3}{2}m^2\frac{u_1 u_1(Du_2)}{r^5} + \frac{3}{2}m^2 Du_2\right]v_2 = 0$$

$$Du_2 D^2 v_2 + 2\left[2mDu_2 + m^2\frac{u_2}{r^3} - \frac{3}{2}m^2(u_1+u_2) - mDu_2\right]Dv_2$$

$$+ \left[2m\left(2mDu_2 + m^2\frac{u_2}{r^3} - \frac{3}{2}m^2(u_1+u_2)\right) + m^2\frac{Du_2}{r^3}\right.$$

$$- m^2\frac{3}{2}\frac{u_2}{r^5}(u_1 Du_2 + u_2 Du_1) - \frac{3}{2}m^2(Du_1 + Du_2)$$

$$- 2m\left(2mDu_2 + m^2\frac{u_2}{r^3} - \frac{3}{2}m^2(u_1+u_2)\right) - m^2\frac{(Du_2)}{r^3} + \frac{3}{2}m^2\frac{u_1 u_2(Du_2)}{r^5}$$

$$\left. + \frac{3}{2}m^2 Du_2\right]v_2 + \left[\frac{3}{2}m^2\frac{u_2 u_2(Du_1)}{r^5} + \frac{3}{2}m^2 Du_1\right]v_1 = 0$$

$$Du_1 Du_2 D(v_1+v_2)$$

$$+ \left[Du_2\left(-2mDu_1 + m^2\frac{u_1}{r^3} - \frac{3}{2}m^2(u_1+u_2)\right)\right.$$

$$+ \left(-m^2\frac{u_2}{r^3} + \frac{3}{2}m^2(u_1+u_2)\right)(Du_1)\right]v_1$$

$$+ \left[Du_1\left(2mDu_2 + m^2\frac{u_2}{r^3} - \frac{3}{2}m^2(u_1+u_2)\right)\right.$$

$$+ \left. \left(-m^2\frac{u_1}{r^3} + \frac{3}{2}m^2(u_1+u_2)\right)(Du_2)\right]v_2 = 0$$

simplifying

$$Du_1 D^2 v_1 + 2\left[-mDu_1 + m^2\frac{u_1}{r^3} - \frac{3}{2}m^2(u_1+u_2)\right]Dv_1$$

$$- \left[\frac{3}{2}m^2\frac{u_1^2}{r^5} + \frac{3}{2}m^2\right]Du_2(v_1-v_2) = 0 \quad (8.3.2.15a)$$

$$Du_2 D^2 v_2 + 2\left[mDu_2 + m^2\frac{u_2}{r^3} - \frac{3}{2}m^2(u_1+u_2)\right]Dv_2$$

$$- \left[\frac{3}{2}m^2\frac{u_2^2}{r^5} + \frac{3}{2}m^2\right]Du_1(v_2-v_1) = 0 \quad (8.3.2.15b)$$

$$Du_1 Du_2 D(v_1 + v_2) - 2 \begin{bmatrix} \left(-\frac{1}{2}m^2 \frac{u_2}{r^3} + \frac{3}{4}m^2(u_1 + u_2)\right) Du_1 \\ + \left(-\frac{1}{2}m^2 \frac{u_1}{r^3} + \frac{3}{4}m^2(u_1 + u_2)\right) Du_2 \\ + mDu_1 Du_2 \end{bmatrix} (v_1 - v_2) = 0 \qquad (8.3.2.15c)$$

Now we take (8.3.2.15a) times $Du_2$ minus (8.3.2.15b) times $Du_1$ and subtract giving

$$\left\{ Du_1 D^2 v_1 + 2\left[-mDu_1 + m^2 \frac{u_1}{r^3} - \frac{3}{2}m^2(u_1 + u_2)\right]Dv_1 - \left[\frac{3}{2}m^2 \frac{u_1^2}{r^5} + \frac{3}{2}m^2\right]Du_2(v_1 - v_2)\right\} Du_2$$

$$- \left\{ Du_2 D^2 v_2 + 2\left[mDu_2 + m^2 \frac{u_2}{r^3} - \frac{3}{2}m^2(u_1 + u_2)\right]Dv_2 \right.$$

$$\left. - \left[\frac{3}{2}m^2 \frac{u_2^2}{r^5} + \frac{3}{2}m^2\right]Du_1(v_2 - v_1)\right\} Du_1 = 0$$

combining

$$Du_1 Du_2 D^2(v_1 - v_2) - 2\frac{3}{2}m^2(u_1 + u_2)(Du_2 Dv_1 - Du_1 Dv_2)$$

$$-\frac{3}{2}m^2 \left(\left(\frac{u_1^2}{r^5} + 1\right)Du_2^2 - \left(\frac{u_2^2}{r^5} + 1\right)Du_1^2\right)(v_1 - v_2) + 2mDu_1 Du_2 D(v_2 - v_1)$$

$$+ 2m^2 \left(\frac{u_1}{r^3} - \frac{u_2}{r^3}\right)(Du_1 Dv_2 - Du_2 Dv_1) = 0 \qquad (8.3.2.16)$$

then we expand the second term of this equation, giving

$$-2 \begin{bmatrix} \left(-\frac{1}{2}m^2 \frac{u_1}{r^3} + \frac{3}{4}m^2(u_1 + u_2)\right)Du_2 Dv_1 + \left(-\frac{1}{2}m^2 \frac{u_1}{r^3} + \frac{3}{4}m^2(u_1 + u_2)\right)Du_2 Dv_1 \\ -\left(-\frac{1}{2}m^2 \frac{u_2}{r^3} + \frac{3}{4}m^2(u_1 + u_2)\right)Du_1 Dv_2 - \left(-\frac{1}{2}m^2 \frac{u_2}{r^3} + \frac{3}{4}m^2(u_1 + u_2)\right)Du_1 Dv_2 \\ +\left(-\frac{1}{2}m^2 \frac{u_1}{r^3} + \frac{3}{4}m^2(u_1 + u_2)\right)Du_2 Dv_2 - \left(-\frac{1}{2}m^2 \frac{u_1}{r^3} + \frac{3}{4}m^2(u_1 + u_2)\right)Du_2 Dv_2 \\ +\left(-\frac{1}{2}m^2 \frac{u_2}{r^3} + \frac{3}{4}m^2(u_1 + u_2)\right)Du_1 Dv_1 - \left(-\frac{1}{2}m^2 \frac{u_2}{r^3} + \frac{3}{4}m^2(u_1 + u_2)\right)Du_1 Dv_1 \end{bmatrix} \qquad (8.3.2.17)$$

and rearranging

$$-2 \begin{bmatrix} \left(-\frac{1}{2}m^2 \frac{u_1}{r^3} + \frac{3}{4}m^2(u_1 + u_2)\right)Du_2 \\ + \left(-\frac{1}{2}m^2 \frac{u_2}{r^3} + \frac{3}{4}m^2(u_1 + u_2)\right)Du_1 \end{bmatrix} D(v_1 - v_2)$$

$$-2 \begin{bmatrix} \left(-\frac{1}{2}m^2 \frac{u_1}{r^3} + \frac{3}{4}m^2(u_1 + u_2)\right)Du_2 \\ - \left(-\frac{1}{2}m^2 \frac{u_2}{r^3} + \frac{3}{4}m^2(u_1 + u_2)\right)Du_1 \end{bmatrix} D(v_1 + v_2) \qquad (8.3.2.18)$$

Combining this second term with the rest of the equation (8.3.2.16) then gives us

$$DuDu_2 D^2(v_1 - v_2)$$
$$-2\left[\left(-\frac{1}{2}m^2\frac{u_1}{r^3} + \frac{3}{4}m^2(u_1+u_2)\right)Du_2 + \left(-\frac{1}{2}m^2\frac{u_2}{r^3} + \frac{3}{4}m^2(u_1+u_2)\right)Du_1\right]D(v_1-v_2)$$
$$-2\left[\begin{array}{l}\left(-\frac{1}{2}m^2\frac{u_1}{r^3} + \frac{3}{4}m^2(u_1+u_2)\right)Du_2 - \left(-\frac{1}{2}m^2\frac{u_2}{r^3} + \frac{3}{4}m^2(u_1+u_2)\right)Du_1 \\ +mDu_2Du_1\end{array}\right]D(v_1+v_2)$$
$$-2\left[\left(\frac{3}{4}m^2\frac{u_2^2}{r^5} + \frac{3}{4}m^2\right)(Du_1)^2 + \left(\frac{3}{4}m^2\frac{u^2}{r^5} + \frac{3}{4}m^2\right)(Du_2)^2\right](v_1-v_2) = 0 \qquad (8.3.2.19)$$

Now we define[47]
$$\Delta := \left(-\frac{1}{2}m^2\frac{u_1}{r^3} + \frac{3}{4}m^2(u_1+u_2)\right)Du_2 - \left(-\frac{1}{2}m^2\frac{u_2}{r^3} + \frac{3}{4}m^2(u_1+u_2)\right)Du_1$$
$$+mDu_2Du_1 \qquad (8.3.2.20)$$

and[48]
$$\rho_1 := v_1 + v_2, \qquad \rho_2 := v_1 - v_2 \qquad (8.3.2.21)$$

With these definitions, the last equation of (8.3.2.15c), along with the one we just derived, are

$$Du_1 Du_2 D\rho_1 - 2\Delta\rho_2 = 0$$
$$Du_1 Du_2 D^2\rho_2 - 2\left[\begin{array}{l}\left(-\frac{1}{2}m^2\frac{u_1}{r^3} + \frac{3}{4}m^2(u_1+u_2)\right)Du_2 \\ +\left(-\frac{1}{2}m^2\frac{u_2}{r^3} + \frac{3}{4}m^2(u_1+u_2)\right)Du_1\end{array}\right]D\rho_2$$
$$-2\Delta D\rho_1 - 2\left[\begin{array}{l}\left(\frac{3}{4}m^2\frac{u_2^2}{r^5} + \frac{3}{4}m^2\right)(Du_1)^2 \\ +\left(\frac{3}{4}m^2\frac{u^2}{r^5} + \frac{3}{4}m^2\right)(Du_2)^2\end{array}\right]\rho_2 = 0 \qquad (8.3.2.22)$$

But the first and second terms of the second equation of (8.3.2.22) can be reformulated due to (8.2.2)

$$Du_1 Du_2 D^2\rho_2 - 2\left[\frac{d\Omega}{du_2}Du_2 + \frac{d\Omega}{du_1}Du_1\right]D\rho_2$$
$$= Du_1 Du_2 D^2\rho_2 - 2[d\Omega]D\rho_2$$
$$= Du_1 Du_2 D^2\rho_2 + D(Du_1 Du_2)D\rho_2$$
$$= D(Du_1 Du_2 D\rho_2)$$

and our two equations become

$$\left.\begin{array}{l}Du_1 Du_2 D\rho_1 - 2\Delta\rho_2 = 0 \\ D[Du_1 Du_2 D\rho_2] - 2\Delta D\rho_1 - 2\left[\begin{array}{l}\left(\frac{3}{4}m^2\frac{u_2^2}{r^5} + \frac{3}{4}m^2\right)(Du_1)^2 \\ +\left(\frac{3}{4}m^2\frac{u_1^2}{r^5} + \frac{3}{4}m^2\right)(Du_2)^2\end{array}\right]\rho_2 = 0\end{array}\right\} \qquad (8.3.2.23)$$

We can eliminate $D\rho_1$ from these two equations by solving the first for $D\rho_1$ and substituting into the second, giving

---
[47] This is the second equation on (G.W. Hill 1886) page 8, after inserting the derivatives of $\Omega$.
[48] We are using $\rho_1$ and $\rho_2$ where Hill uses $\rho$ and $\sigma$.

$$D[Du_1 Du_2 D\rho_2] - 2\begin{bmatrix}\left(\frac{3}{4}m^2\frac{u_2^2}{r^5} + \frac{3}{4}m^2\right)(Du_1)^2 \\ +\left(\frac{3}{4}m^2\frac{u_1^2}{r^5} + \frac{3}{4}m^2\right)(Du_2)^2 \\ +\frac{2\Delta^2}{Du_1 Du_2}\end{bmatrix}\rho_2 = 0 \qquad (8.3.2.24)$$

In order to eliminate the term involving $D\rho_2$, we define

$$\rho_2 = \frac{w}{\sqrt{Du_1 Du_2}} \qquad (8.3.2.25)$$

As a result,

$$D\rho_2 = -\frac{1}{2}(Du_1 Du_2)^{-\frac{3}{2}}D(Du_1 Du_2)w + (Du_1 Du_2)^{-\frac{1}{2}}Dw \qquad (8.3.2.26)$$

Substituting this into (8.2.2) yields

$$D\begin{bmatrix}-(Du_1 Du_2)\frac{1}{2}(Du_1 Du_2)^{-\frac{3}{2}}D(Du_1 Du_2)w \\ +(Du_1 Du_2)(Du_1 Du_2)^{-\frac{1}{2}}Dw\end{bmatrix}$$
$$-2\begin{bmatrix}\left(\frac{3}{4}m^2\frac{u_2^2}{r^5} + \frac{3}{4}m^2\right)Du_1^2 \\ +\left(\frac{3}{4}m^2\frac{u_1^2}{r^5} + \frac{3}{4}m^2\right)Du_2^2 \\ +\frac{2\Delta^2}{Du_1 Du_2}\end{bmatrix}(Du_1 Du_2)^{-1/2}w = 0 \qquad (8.3.2.27)$$

expanding

$$D\begin{bmatrix}-\frac{1}{2}(Du_1 Du_2)^{-\frac{1}{2}}D(Du_1 Du_2)w \\ +(Du_1 Du_2)^{\frac{1}{2}}Dw\end{bmatrix}$$
$$+\begin{bmatrix}-2\left(\frac{3}{4}m^2\frac{u_2^2}{r^5} + \frac{3}{4}m^2\right)Du_1^2 \\ -2\left(\frac{3}{4}m^2\frac{u_1^2}{r^5} + \frac{3}{4}m^2\right)Du_2^2 \\ -\frac{4\Delta^2}{Du_1 Du_2}\end{bmatrix}(Du_1 Du_2)^{-1/2}w = 0 \qquad (8.3.2.28)$$

taking $D$

$$\frac{1}{4}(Du_1 Du_2)^{-3/2}[D(Du_1 Du_2)]^2 w - \frac{1}{2}(Du_1 Du_2)^{-1/2}D^2(Du_1 Du_2)w$$
$$-\frac{1}{2}(Du_1 Du_2)^{-1/2}D(Du_1 Du_2)Dw + \frac{1}{2}(Du_1 Du_2)^{-1/2}D(Du_1 Du_2)Dw + (Du_1 Du_2)^{\frac{1}{2}}D^2 w$$
$$-2(Du_1 Du_2)^{-1/2}\left(\frac{3}{4}m^2\frac{u_2^2}{r^5} + \frac{3}{4}m^2\right)Du_1^2 w - 2(Du_1 Du_2)^{-1/2}\left(\frac{3}{4}m^2\frac{u_1^2}{r^5} + \frac{3}{4}m^2\right)Du_2^2 w$$
$$-(Du_1 Du_2)^{-1/2}\frac{4\Delta^2}{Du_1 Du_2}w = 0 \qquad (8.3.2.29)$$

eliminating redundant terms for $Dw$, multiplying by $(Du_1 Du_2)^{1/2}$, and rearranging

$$D^2 w = \left[ \frac{2}{Du_1 Du_2} \left[ \left( \frac{3}{4} m^2 \frac{u_2{}^2}{r^5} + \frac{3}{4} m^2 \right) Du_1{}^2 + \left( \frac{3}{4} m^2 \frac{u_1{}^2}{r^5} + \frac{3}{4} m^2 \right) Du_2{}^2 \right] + \left( \frac{2\Delta}{Du_1 Du_2} \right)^2 + \frac{D^2(Du_1 Du_2)}{2 Du_1 Du_2} - \left( \frac{D(Du_1 Du_2)}{2 Du_1 Du_2} \right)^2 \right] w \quad (8.3.2.30)$$

New we define[49]

$$\Theta = \frac{2}{Du_1 Du_2} \left[ \left( \frac{3}{4} m^2 \frac{u_2{}^2}{r^5} + \frac{3}{4} m^2 \right) Du_1{}^2 + \left( \frac{3}{4} m^2 \frac{u_1{}^2}{r^5} + \frac{3}{4} m^2 \right) Du_2{}^2 \right] + \left( \frac{2\Delta}{Du_1 Du_2} \right)^2 + \frac{D^2(Du_1 Du_2)}{2 Du_1 Du_2} - \left( \frac{D(Du_1 Du_2)}{2 Du_1 Du_2} \right)^2 \quad (8.3.2.31)$$

so we can write.

$$\frac{-m^2 d^2}{dt^2} w = \Theta w, \qquad D^2 w = \Theta w \quad (8.3.2.32)$$

This second-order, linear, ordinary differential equation is sometimes called *Hill's equation*, see https://en.wikipedia.org/wiki/Hill_differential_equation and (Magnus and Winkler 1966) (Magnus and Winkler 2018), (Teschl 2000).

Since

$$Du_1 Du_2 = 2\Omega + 2C$$

and

$$DC = 0$$

we have

$$D\Omega = -\frac{D(Du_1 Du_2)}{2}, \qquad D^2 \Omega = -\frac{D^2(Du_1 Du_2)}{2}$$

and we can write

$$\Theta = \frac{2}{Du_1 Du_2} \left[ \left( \frac{3}{4} m^2 \frac{u_2{}^2}{r^5} + \frac{3}{4} m^2 \right) Du_1{}^2 + \left( \frac{3}{4} m^2 \frac{u_1{}^2}{r^5} + \frac{3}{4} m^2 \right) Du_2{}^2 \right] + \left( \frac{2\Delta}{Du_1 Du_2} \right)^2 - \frac{D^2 \Omega}{Du_1 Du_2} - \left( \frac{D\Omega}{Du_1 Du_2} \right)^2 \quad (8.3.2.33)$$

Now we begin with

$$D\Omega = \left( -\frac{1}{2} m^2 \frac{u_2}{r^3} + \frac{3}{4} m^2 (u_1 + u_2) \right) Du_1 + \left( -\frac{1}{2} m^2 \frac{u_1}{r^3} + \frac{3}{4} m^2 (u_1 + u_2) \right) Du_2 \quad (8.3.2.34)$$

and apply $D$ giving

$$D^2 \Omega = D \left( -\frac{1}{2} m^2 \frac{u_2}{r^3} + \frac{3}{4} m^2 (u_1 + u_2) \right) Du_1 + \left( -\frac{1}{2} m^2 \frac{u_2}{r^3} + \frac{3}{4} m^2 (u_1 + u_2) \right) D^2 u_1$$
$$+ D \left( -\frac{1}{2} m^2 \frac{u_1}{r^3} + \frac{3}{4} m^2 (u_1 + u_2) \right) Du_2 + \left( -\frac{1}{2} m^2 \frac{u_1}{r^3} + \frac{3}{4} m^2 (u_1 + u_2) \right) D^2 u_2 \quad (8.3.2.35)$$

applying $D$

$$D^2 \Omega = \left( \frac{3}{4} m^2 \frac{u_2{}^2}{r^5} + \frac{3}{4} m^2 \right) Du_1{}^2 + \left( \frac{1}{4} \frac{m^2}{r^3} + \frac{3}{4} m^2 \right) Du_1 Du_2$$

---

[49] This is the first equation on (G.W. Hill 1886) page 9, after inserting the derivatives of $\Omega$.

$$+\left(-\frac{1}{2}m^2\frac{u_2}{r^3}+\frac{3}{4}m^2(u_1+u_2)\right)D^2u_1+\left(\frac{1}{4}\frac{m^2}{r^3}+\frac{3}{4}m^2\right)Du_1Du_2$$

$$+\left(\frac{3}{4}m^2\frac{u_1^2}{r^5}+\frac{3}{4}m^2\right)Du_2{}^2+\left(-\frac{1}{2}m^2\frac{u_1}{r^3}+\frac{3}{4}m^2(u_1+u_2)\right)D^2u_2 \quad (8.3.2.36)$$

rearranging

$$D^2\Omega=\left(\frac{3}{4}m^2\frac{u_2^2}{r^5}+\frac{3}{4}m^2\right)Du_1{}^2+2\left(\frac{1}{4}\frac{m^2}{r^3}+\frac{3}{4}m^2\right)Du_1Du_2+\left(\frac{3}{4}m^2\frac{u_1^2}{r^5}+\frac{3}{4}m^2\right)Du_2{}^2$$

$$+\left(-\frac{1}{2}m^2\frac{u_2}{r^3}+\frac{3}{4}m^2(u_1+u_2)\right)D^2u_1+\left(-\frac{1}{2}m^2\frac{u_1}{r^3}+\frac{3}{4}m^2(u_1+u_2)\right)D^2u_2 \quad (8.3.2.37)$$

Now we use (8.2.3) to eliminate $Du_1{}^2$ and $Du_2{}^2$

$$D^2\Omega=\left(\frac{3}{4}m^2\frac{u_2^2}{r^5}+\frac{3}{4}m^2\right)Du_1{}^2+2\left(\frac{1}{4}\frac{m^2}{r^3}+\frac{3}{4}m^2\right)Du_1Du_2+\left(\frac{3}{4}m^2\frac{u_1^2}{r^5}+\frac{3}{4}m^2\right)Du_2{}^2$$

$$-2m\left(-\frac{1}{2}m^2\frac{u_2}{r^3}+\frac{3}{4}m^2(u_1+u_2)\right)Du_1$$

$$-2\left(-\frac{1}{2}m^2\frac{u_2}{r^3}+\frac{3}{4}m^2(u_1+u_2)\right)\left(-\frac{1}{2}m^2\frac{u_1}{r^3}+\frac{3}{4}m^2(u_1+u_2)\right)$$

$$+2m\left(-\frac{1}{2}m^2\frac{u_1}{r^3}+\frac{3}{4}m^2(u_1+u_2)\right)Du_2$$

$$-2\left(-\frac{1}{2}m^2\frac{u_2}{r^3}+\frac{3}{4}m^2(u_1+u_2)\right)\left(-\frac{1}{2}m^2\frac{u_1}{r^3}+\frac{3}{4}m^2(u_1+u_2)\right) \quad (8.3.2.38)$$

rearranging and applying the definition of $\Delta$

$$D^2\Omega=\left(\frac{3}{4}m^2\frac{u_2^2}{r^5}+\frac{3}{4}m^2\right)Du_1{}^2+2\left(\frac{1}{4}\frac{m^2}{r^3}+\frac{3}{4}m^2\right)Du_1Du_2+\left(\frac{3}{4}m^2\frac{u_1^2}{r^5}+\frac{3}{4}m^2\right)Du_2{}^2$$

$$+2m\Delta-2m^2Du_1Du_2-4\left(-\frac{1}{2}m^2\frac{u_2}{r^3}+\frac{3}{4}m^2(u_1+u_2)\right)\left(-\frac{1}{2}m^2\frac{u_1}{r^3}+\frac{3}{4}m^2(u_1+u_2)\right) \quad (8.3.2.39)$$

Now we substitute the definition of $\Delta$ into this last equation

$$D^2\Omega=\left(\frac{3}{4}m^2\frac{u_2^2}{r^5}+\frac{3}{4}m^2\right)Du_1{}^2+2\left(\frac{1}{4}\frac{m^2}{r^3}+\frac{3}{4}m^2\right)Du_1Du_2+\left(\frac{3}{4}m^2\frac{u_1^2}{r^5}+\frac{3}{4}m^2\right)Du_2{}^2$$

$$+2m\left(-\frac{1}{2}m^2\frac{u_1}{r^3}+\frac{3}{4}m^2(u_1+u_2)\right)Du_2-2m\left(-\frac{1}{2}m^2\frac{u_2}{r^3}+\frac{3}{4}m^2(u_1+u_2)\right)Du_1$$

$$-4\left(-\frac{1}{2}m^2\frac{u_2}{r^3}+\frac{3}{4}m^2(u_1+u_2)\right)\left(-\frac{1}{2}m^2\frac{u_1}{r^3}+\frac{3}{4}m^2(u_1+u_2)\right) \quad (8.3.2.40)$$

Then we take the square of $D\Omega$

$$(D\Omega)^2=\left[\left(-\frac{1}{2}m^2\frac{u_2}{r^3}+\frac{3}{4}m^2(u_1+u_2)\right)\right]^2 Du_1{}^2$$

$$+2\left(-\frac{1}{2}m^2\frac{u_2}{r^3}+\frac{3}{4}m^2(u_1+u_2)\right)\left(-\frac{1}{2}m^2\frac{u_1}{r^3}+\frac{3}{4}m^2(u_1+u_2)\right)Du_1Du_2$$

$$+\left[\left(-\frac{1}{2}m^2\frac{u_1}{r^3}+\frac{3}{4}m^2(u_1+u_2)\right)\right]^2 Du_2{}^2 \quad (8.3.2.41)$$

Then we combine these to get

$$D^2\Omega + \frac{(D\Omega)^2}{Du_1 Du_2}$$

$$= \left(\frac{3}{4}m^2\frac{u_2{}^2}{r^5} + \frac{3}{4}m^2\right)Du_1{}^2 + 2\left(\frac{1}{4}\frac{m^2}{r^3} + \frac{3}{4}m^2\right)Du_1 Du_2$$

$$+ \left(\frac{3}{4}m^2\frac{u_1{}^2}{r^5} + \frac{3}{4}m^2\right)Du_2{}^2 + 2m\left(-\frac{1}{2}m^2\frac{u_1}{r^3} + \frac{3}{4}m^2(u_1+u_2)\right)Du_2$$

$$-2m\left(-\frac{1}{2}m^2\frac{u_2}{r^3} + \frac{3}{4}m^2(u_1+u_2)\right)Du_1$$

$$-2\left(-\frac{1}{2}m^2\frac{u_2}{r^3} + \frac{3}{4}m^2(u_1+u_2)\right)\left(-\frac{1}{2}m^2\frac{u_1}{r^3} + \frac{3}{4}m^2(u_1+u_2)\right)$$

$$+\left[\left(-\frac{1}{2}m^2\frac{u_2}{r^3} + \frac{3}{4}m^2(u_1+u_2)\right)\right]^2\frac{Du_1}{Du_2}$$

$$+\left[\left(-\frac{1}{2}m^2\frac{u_1}{r^3} + \frac{3}{4}m^2(u_1+u_2)\right)\right]^2\frac{Du_2}{Du_1} \tag{8.3.2.42}$$

We will also need the square of $\Delta$

$$\frac{\Delta^2}{Du_1 Du_2}$$

$$= \left[\left(-\frac{1}{2}m^2\frac{u_1}{r^3} + \frac{3}{4}m^2(u_1+u_2)\right)\right]^2\frac{Du_2}{Du_1} + \left[\left(-\frac{1}{2}m^2\frac{u_2}{r^3} + \frac{3}{4}m^2(u_1+u_2)\right)\right]^2\frac{Du_1}{Du_2}$$

$$+m^2 Du_1 Du_2 - 2\left(-\frac{1}{2}m^2\frac{u_1}{r^3} + \frac{3}{4}m^2(u_1+u_2)\right)\left(-\frac{1}{2}m^2\frac{u_2}{r^3} + \frac{3}{4}m^2(u_1+u_2)\right)$$

$$+2m\left(-\frac{1}{2}m^2\frac{u_1}{r^3} + \frac{3}{4}m^2(u_1+u_2)\right)Du_2 - 2m\left(-\frac{1}{2}m^2\frac{u_2}{r^3} + \frac{3}{4}m^2(u_1+u_2)\right)Du_1 \tag{8.3.2.43}$$

Comparing these last two equations yields

$$D^2\Omega + \frac{(D\Omega)^2}{Du_1 Du_2}$$

$$= \left(\frac{3}{4}m^2\frac{u_2{}^2}{r^5} + \frac{3}{4}m^2\right)Du_1{}^2 + 2\left(\frac{1}{4}\frac{m^2}{r^3} + \frac{3}{4}m^2\right)Du_1 Du_2$$

$$+ \left(\frac{3}{4}m^2\frac{u_1{}^2}{r^5} + \frac{3}{4}m^2\right)Du_2{}^2 + \frac{\Delta^2}{Du_1 Du_2} - m^2 Du_1 Du_2 \tag{8.3.2.44}$$

Substitution of this last equation into (8.3.2.33), our last equation for $\Theta$ gives us

$$\Theta = \frac{2}{Du_1 Du_2}\left[\left(\frac{3}{4}m^2\frac{u_2{}^2}{r^5} + \frac{3}{4}m^2\right)Du_1{}^2 + \left(\frac{3}{4}m^2\frac{u_1{}^2}{r^5} + \frac{3}{4}m^2\right)Du_2{}^2\right]$$

$$+ \left(\frac{2\Delta}{Du_1 Du_2}\right)^2 - \left(\frac{3}{4}m^2\frac{u_2{}^2}{r^5} + \frac{3}{4}m^2\right)\frac{Du_1}{Du_2} - 2\left(\frac{1}{4}\frac{m^2}{r^3} + \frac{3}{4}m^2\right)$$

$$- \left(\frac{3}{4}m^2\frac{u_1{}^2}{r^5} + \frac{3}{4}m^2\right)\frac{Du_2}{Du_1} - \frac{\Delta^2}{Du_1 Du_2} + m^2 \tag{8.3.2.45}$$

rearranging

$$\Theta = \frac{1}{Du_1 Du_2} \begin{bmatrix} \left(\frac{3}{4}m^2\frac{u_2^2}{r^5} + \frac{3}{4}m^2\right)Du_1^2 \\ -2\left(\frac{1}{4}\frac{m^2}{r^3} + \frac{3}{4}m^2\right)Du_1 Du_2 \\ +\left(\frac{3}{4}m^2\frac{u_1^2}{r^5} + \frac{3}{4}m^2\right)Du_2^2 \end{bmatrix} + 3\left(\frac{\Delta}{Du_1 Du_2}\right)^2 + m^2 \quad (8.3.2.46)$$

rearranging

$$\Theta = \frac{1}{Du_1 Du_2}\frac{3}{4}\begin{bmatrix} m^2\frac{u_2^2}{r^5}Du_1^2 + m^2 Du_1^2 \\ -\frac{2}{3}\frac{m^2}{r^3}Du_1 Du_2 - 2m^2 Du_1 Du_2 \\ +m^2\frac{u_1^2}{r^5}Du_2^2 + m^2 Du_2^2 \end{bmatrix} + 3\left(\frac{\Delta}{Du_1 Du_2}\right)^2 + m^2 \quad (8.3.2.47)$$

completing the squares and rearranging

$$\Theta = \frac{1}{Du_1 Du_2}\frac{3}{4}\begin{bmatrix} \frac{m^2}{r^5}(u_1 Du_2 - u_2 Du_1)^2 + 2\frac{m^2}{r^5}u_1 u_2 Du_1 Du_2 \\ +m^2(Du_1 - Du_2)^2 + 2m^2 Du_1 Du_2 \\ -\frac{2}{3}\frac{m^2}{r^3}Du_1 Du_2 - 2m^2 Du_1 Du_2 \end{bmatrix} + 3\left(\frac{\Delta}{Du_1 Du_2}\right)^2 + m^2$$

rearranging

$$\Theta = \frac{1}{Du_1 Du_2}\frac{3}{4}\begin{bmatrix} \frac{m^2}{r^5}(u_1 Du_2 - u_2 Du_1)^2 \\ +m^2(Du_1 - Du_2)^2 + \frac{4}{3}\frac{m^2}{r^3}Du_1 Du_2 \end{bmatrix} + 3\left(\frac{\Delta}{Du_1 Du_2}\right)^2 + m^2$$

rearranging[50]

$$\Theta = m^2\frac{1}{r^3} + \frac{3}{4(Du_1 Du_2)}\begin{bmatrix} \frac{m^2}{r^5}(u_1 Du_2 - u_2 Du_1)^2 \\ +m^2(Du_2 - Du_1)^2 \end{bmatrix} + 3\left(\frac{\Delta}{Du_1 Du_2}\right)^2 + m^2 \quad (8.3.2.48)$$

Then starting with the definition of $\Delta$ in equation (8.3.2.20)

$$\Delta := \left(-\frac{1}{2}m^2\frac{u_1}{r^3} + \frac{3}{4}m^2(u_1 + u_2)\right)Du_2 - \left(-\frac{1}{2}m^2\frac{u_2}{r^3} + \frac{3}{4}m^2(u_1 + u_2)\right)Du_1 + mDu_2 Du_1$$

rearranging

$$\Delta = \left(-\frac{1}{2}m^2\frac{1}{r^3} + \frac{3}{4}m^2\right)(u_1 Du_2 - u_2 Du_1) - \frac{3}{4}m^2(u_1 Du_1 - u_2 Du_2) + m(Du_1 Du_2) \quad (8.3.2.49)$$

Now we look at the closed, periodic orbit, which will serve as a first approximation of the true lunar orbit. Since this was treated in more depth in (G.W. Hill 1878), we could skip the rest of section II, but will do enough here to make the connection to section III. Since we are looking at a symmetric, closed, periodic orbit, we can take a Fourier expansion, which is

$$u_1 = \sum_{j=-\infty}^{\infty} a_j \varsigma^{2j+1}, \qquad u_2 = \sum_{j=-\infty}^{\infty} a_j \varsigma^{-2j-1} \quad (8.3.2.50)$$

These expressions match our (4.14) and (4.16), after replacing $j \to -j - 1$ we have

$$u_2 = \sum_{j=-\infty}^{\infty} a_{-j-1} \varsigma^{2j+1} \tag{8.3.2.51}$$

---

[50] This is equation (16) on (G.W. Hill 1886) page 9, after inserting the derivatives of $\Omega$.

The derivatives are

$$Du_1 = \sum_{j=-\infty}^{\infty} (2j+1) a_j \varsigma^{2j+1}, \qquad Du_2 = -\sum_{j=-\infty}^{\infty} (2j+1) a_j \varsigma^{-2j-1} \tag{8.3.2.52}$$

$$D^2 u_1 = \sum_{j=-\infty}^{\infty} (2j+1)^2 a_j \varsigma^{2j+1}, \qquad D^2 u_2 = \sum_{j=-\infty}^{\infty} (2j+1)^2 a_j \varsigma^{-2j-1} \tag{8.3.2.53}$$

Here, we have the inconvenience that, for large $j$, the coefficient $a_j$ is multiplied by a large number. We will avoid that by using the inverse of $D$, $D^{-1}$, and the following equation

$$u_1 Du_2 - u_2 Du_1 = 2mr^2 - \frac{3}{2} m^2 D^{-1} (u_1^2 - u_2^2)$$

To demonstrate the validity of this equation, let's take a derivative

$$D(u_1 Du_2 - u_2 Du_1 - 2mr^2)$$
$$= Du_1 Du_2 + u_1 D^2 u_2 - Du_2 Du_1 - u_2 D^2 u_1 - 2mu_1 Du_2 - 2mu_2 Du_1$$

since $r^2 = u_1 u_2$. Then we eliminate the second-order derivatives via the equations (8.2.3)

$$= Du_1 Du_2 + u_1 \left( 2mDu_2 - \frac{3}{2} m^2 u_2 + m^2 (u_1 u_2)^{-\frac{3}{2}} u_2 - \frac{3}{2} m^2 u_1 \right) - Du_2 Du_1$$
$$- u_2 \left( -2mDu_1 - \frac{3}{2} m^2 u_1 + m^2 (u_1 u_2)^{-\frac{3}{2}} u_1 - \frac{3}{2} m^2 u_2 \right) - 2mu_1 Du_2 - 2mu_2 Du_1$$

combining similar terms

$$= -\frac{3}{2} m^2 (u_1^2 - u_2^2)$$

which demonstrates the validity of the equation, up to a constant of integration, but that can be found by using the Jacobi integral, e.g., from equation (4.30).

$$-2 \sum_{j=-\infty}^{\infty} (2j+1) a_j^2$$

Now, wherever $Du_1$ and $Du_2$ appear in $\Theta$, they are multiplied by $m^2$, so that should suffice for convergence.

Next, we claim to have a new expression for $\Delta$[51]

$$\Delta = \frac{1}{2} [Du_1 D^2 u_2 - Du_2 D^2 u_1] - mDu_1 Du_2 \tag{8.3.2.54}$$

In order to demonstrate this, we substitute for $D^2 u_1$ and $D^2 u_2$ the values via the equations (8.3.2.7).

$$\Delta = \frac{1}{2} \left[ \begin{array}{l} Du_1 \left( 2mDu_2 - \frac{3}{2} m^2 u_2 + m^2 (u_1 u_2)^{-\frac{3}{2}} u_2 - \frac{3}{2} m^2 u_1 \right) \\ -Du_2 \left( -2mDu_1 - \frac{3}{2} m^2 u_1 + m^2 (u_1 u_2)^{-\frac{3}{2}} u_1 - \frac{3}{2} m^2 u_2 \right) \end{array} \right] - mDu_1 Du_2 \tag{8.3.2.55}$$

rearranging

---

[51] This is the last equation before (18) on (G.W. Hill 1886), page 11.

$$\Delta = \left[-\frac{1}{2}\frac{m^2}{r^3} + \frac{3}{4}m^2\right](u_1 Du_2 - u_2 Du_1) - \frac{3}{4}m^2(u_1 Du_1 - u_2 Du_2) + m Du_1 Du_2 \quad (8.3.2.56)$$

which is $\Delta$ in equation (8.3.2.49). Now we can substitute this version of $\Delta$ from (8.3.2.54) into $\Theta$ of equation (8.3.2.48).

With this, our next step will be to calculate a new version of $\Theta$. We begin with a preliminary calculation. Let's start with the first equation of (8.3.2.13).

$$D^3 u_1 = -2m D^2 u_1 + \frac{m^2}{r^3} Du_1 - \frac{3}{2}\frac{m^2}{r^5} u_1(u_1 Du_2 + u_2 Du_1) - \frac{3}{2}m^2(Du_1 + Du_2)$$

Divide by $Du_1$

$$\frac{D^3 u_1}{Du_1} = -2m\frac{D^2 u_1}{Du_1} + \frac{m^2}{r^3} - \frac{3}{2}\frac{m^2}{r^3} - \frac{3}{2}\frac{m^2}{r^3}\frac{u_1 Du_2}{u_2 Du_1} - \frac{3}{2}m^2 - \frac{3}{2}m^2\frac{Du_2}{Du_1}$$

Combining similar terms

$$\frac{D^3 u_1}{Du_1} = -2m\frac{D^2 u_1}{Du_1} - \frac{1}{2}\frac{m^2}{r^3} - \frac{3}{2}\frac{m^2}{r^3}\frac{u_1 Du_2}{u_2 Du_1} - \frac{3}{2}m^2 - \frac{3}{2}m^2\frac{Du_2}{Du_1} \quad (8.3.2.57)$$

Let's start with the second equation of (8.3.2.13).

$$D^3 u_2 = 2m D^2 u_2 + \frac{m^2}{r^3} Du_2 - \frac{3}{2}\frac{m^2}{r^5} u_2(u_1 Du_2 + u_2 Du_1) - \frac{3}{2}m^2(Du_1 + Du_2)$$

Divide by $Du_2$

$$\frac{D^3 u_2}{Du_2} = 2m\frac{D^2 u_2}{Du_2} + \frac{m^2}{r^3} - \frac{3}{2}\frac{m^2}{r^3} - \frac{3}{2}\frac{m^2}{r^3}\frac{u_2 Du_1}{u_1 Du_2} - \frac{3}{2}m^2 - \frac{3}{2}m^2\frac{Du_1}{Du_2}$$

Combining similar terms

$$\frac{D^3 u_2}{Du_2} = 2m\frac{D^2 u_2}{Du_2} - \frac{1}{2}\frac{m^2}{r^3} - \frac{3}{2}\frac{m^2}{r^3}\frac{u_2 Du_1}{u_1 Du_2} - \frac{3}{2}m^2 - \frac{3}{2}m^2\frac{Du_1}{Du_2} \quad (8.3.2.58)$$

Now we can combine the results of (8.3.2.57) and (8.3.2.58):

$$\frac{1}{2}\left(\frac{D^3 u_1}{Du_1} + \frac{D^3 u_2}{Du_2}\right)$$
$$= -m\left(\frac{D^2 u_1}{Du_1} - \frac{D^2 u_2}{Du_2}\right) - \frac{1}{2}\frac{m^2}{r^3} - \frac{3}{4}\frac{m^2}{r^3}\left(\frac{u_1 Du_2}{u_2 Du_1} + \frac{u_2 Du_1}{u_1 Du_2}\right) - \frac{3}{2}m^2 - \frac{3}{4}m^2\left(\frac{Du_2}{Du_1} + \frac{Du_1}{Du_2}\right) \quad (8.3.2.59)$$

rearranging

$$\frac{1}{2}\left(\frac{D^3 u_1}{Du_1} + \frac{D^3 u_2}{Du_2}\right)$$
$$= -m\left(\frac{D^2 u_1}{Du_1} - \frac{D^2 u_2}{Du_2}\right) - \frac{3}{2}m^2 - \frac{3}{4}m^2\left(\frac{Du_1}{Du_2} + \frac{Du_2}{Du_1}\right) - \frac{1}{2}m^2(u_1 u_2)^{-\frac{3}{2}}$$
$$- \frac{3}{4}\frac{m^2}{r^3}\left(\frac{u_1 Du_2}{u_2 Du_1} + \frac{u_2 Du_1}{u_1 Du_2}\right)$$

rearranging

$$-\frac{1}{2}\left(\frac{D^3u_1}{Du_1}+\frac{D^3u_1}{Du_1}\right)+\frac{1}{2}\left(\left(\frac{D^2u_1}{Du_1}\right)^2+\left(\frac{D^2u_2}{Du_2}\right)^2\right)$$

$$=m\left(\frac{D^2u_1}{Du_1}-\frac{D^2u_2}{Du_2}\right)+\frac{3}{2}m^2+\frac{3}{4}m^2\left(\frac{Du_1}{Du_2}+\frac{Du_2}{Du_1}\right)+\frac{1}{2}\frac{m^2}{r^3}$$

$$+\frac{3}{4}\frac{m^2}{r^3}\left(\frac{u_1Du_2}{u_2Du_1}+\frac{u_2Du_1}{u_1Du_2}\right)+\frac{1}{2}\left(\left(\frac{D^2u_1}{Du_1}\right)^2+\left(\frac{D^2u_2}{Du_2}\right)^2\right) \qquad (8.3.2.60)$$

Now we can take $\Theta$ from equation (8.3.2.48) and substitute $\Delta$ from (8.3.2.54).

$$\Theta=\frac{m^2}{r^3}+\frac{3}{4Du_1Du_2}\left[\frac{m^2}{r^5}(u_1Du_2-u_2Du_1)^2+m^2(Du_1-Du_2)^2\right]$$

$$+3\left(\frac{\frac{1}{2}[Du_1D^2u_2-Du_2D^2u_1]-mDu_1Du_2}{Du_1Du_2}\right)^2+m^2 \qquad (8.3.2.61)$$

expanding

$$\Theta=\frac{m^2}{r^3}+\frac{3}{4Du_1Du_2}\left[\begin{array}{c}\frac{m^2}{r^5}(u_1{}^2(Du_2)^2-2u_1u_2Du_1Du_2+u_2{}^2(Du_1)^2)\\ +m^2((Du_1)^2-2Du_1Du_2+(Du_2)^2)\end{array}\right]$$

$$+\frac{3}{(Du_1)^2(Du_2)^2}\left[\begin{array}{c}\frac{1}{4}(Du_1)^2(D^2u_2)^2-\frac{1}{2}Du_1Du_2D^2u_1D^2u_2+\frac{1}{4}(Du_2)^2(D^2u_1)^2\\ +m^2(Du_1)^2(Du_2)^2-m(Du_1)^2Du_2D^2u_2+m(Du_2)^2Du_1D^2u_1\end{array}\right]+m^2 \quad (8.3.2.62)$$

rearranging

$$\Theta=\frac{m^2}{r^3}-\frac{3}{2}\frac{m^2}{r^3}+\frac{3}{2}m^2+m^2+\frac{3}{4}\left[\frac{m^2}{r^5}\left(\frac{u_1{}^2u_2Du_2}{u_2Du_1}+\frac{u_2{}^2u_1Du_1}{u_1Du_2}\right)+m^2\left(\frac{Du_1}{Du_2}+\frac{Du_2}{Du_1}\right)\right]$$

$$+3\left(\frac{1}{4}\frac{(D^2u_2)^2}{(Du_2)^2}-\frac{1}{2}\frac{D^2u_1D^2u_2}{Du_1Du_2}+\frac{1}{4}\frac{(D^2u_1)^2}{(Du_1)^2}-m\frac{D^2u_2}{Du_2}+m\frac{D^2u_1}{Du_1}\right) \qquad (8.3.2.63)$$

combining

$$\Theta=-\frac{1}{2}\frac{m^2}{r^3}+\frac{5}{2}m^2+\frac{3}{4}\left[\frac{m^2}{r^3}\left(\frac{u_1Du_2}{u_2Du_1}+\frac{u_2Du_1}{u_1Du_2}\right)+m^2\left(\frac{Du_1}{Du_2}+\frac{Du_2}{Du_1}\right)\right]$$

$$+3\left(\frac{1}{4}\frac{(D^2u_2)^2}{(Du_2)^2}-\frac{1}{2}\frac{D^2u_1D^2u_2}{Du_1Du_2}+\frac{1}{4}\frac{(D^2u_1)^2}{(Du_1)^2}-m\frac{D^2u_2}{Du_2}+m\frac{D^2u_1}{Du_1}\right) \qquad (8.3.2.64)$$

expanding

$$\Theta=-\frac{1}{2}\frac{m^2}{r^3}+\frac{5}{2}m^2+\frac{3}{4}\frac{m^2}{r^3}\frac{u_1Du_2}{u_2Du_1}+\frac{3}{4}\frac{m^2}{r^3}\frac{u_2Du_1}{u_1Du_2}+\frac{3}{4}m^2\frac{Du_1}{Du_2}+\frac{3}{4}m^2\frac{Du_2}{Du_1}$$

$$+\frac{3}{4}\frac{(D^2u_2)^2}{(Du_2)^2}-\frac{3}{2}\frac{D^2u_1D^2u_2}{Du_1Du_2}+\frac{3}{4}\frac{(D^2u_1)^2}{(Du_1)^2}-3m\frac{D^2u_2}{Du_2}+3m\frac{D^2u_1}{Du_1} \qquad (8.3.2.65)$$

rearranging

$$\Theta=-\frac{m^2}{r^3}-m^2+\frac{1}{2}\frac{(D^2u_1)^2}{(Du_1)^2}+\frac{1}{2}\frac{(D^2u_2)^2}{(Du_2)^2}+2m^2-\frac{D^2u_1D^2u_2}{Du_1Du_2}+2m\frac{D^2u_1}{Du_1}-2m\frac{D^2u_2}{Du_2}$$

$$-\frac{1}{4}\left(\frac{(D^2u_1)^2}{(Du_1)^2}+\frac{(D^2u_2)^2}{(Du_2)^2}\right)-\frac{1}{2}\frac{D^2u_1D^2u_2}{Du_1Du_2}+m\left(\frac{D^2u_1}{Du_1}-\frac{D^2u_2}{Du_2}\right)+\frac{3}{2}m^2$$

$$+\frac{3}{4}m^2\left(\frac{Du_1}{Du_2}+\frac{Du_2}{Du_1}\right)+\frac{1}{2}\frac{m^2}{r^3}+\frac{3}{4}\frac{m^2}{r^3}\left(\frac{u_1Du_2}{u_2Du_1}+\frac{u_2Du_1}{u_1Du_2}\right)$$

$$+\frac{1}{2}\left(\frac{(D^2u_1)^2}{(Du_1)^2}+\frac{(D^2u_2)^2}{(Du_2)^2}\right) \tag{8.3.2.66}$$

rearranging

$$\Theta=-\left[\frac{m^2}{r^3}+m^2\right]+2\left[\frac{1}{2}\left(\frac{D^2u_1}{Du_1}-\frac{D^2u_2}{Du_2}\right)+m\right]^2-\left[\frac{1}{2}\left(\frac{D^2u_1}{Du_1}+\frac{D^2u_2}{Du_2}\right)\right]^2$$
$$+m\left(\frac{D^2u_1}{Du_1}-\frac{D^2u_2}{Du_2}\right)+\frac{3}{2}m^2+\frac{3}{4}m^2\left(\frac{Du_1}{Du_2}+\frac{Du_2}{Du_1}\right)-\frac{1}{2}\frac{m^2}{r^3}$$
$$+\frac{3}{4}\frac{m^2}{r^3}\left(\frac{u_1Du_2}{u_2Du_1}+\frac{u_2Du_1}{u_1Du_2}\right)+\frac{1}{2}\left(\frac{D^2u_1}{Du_1}\right)^2+\frac{1}{2}\left(\frac{D^2u_2}{Du_2}\right)^2 \tag{8.3.2.67}$$

rearranging and using (8.3.2.20)

$$\Theta=-\left[\frac{m^2}{r^3}+m^2\right]+2\left[\frac{1}{2}\left(\frac{D^2u_1}{Du_1}-\frac{D^2u_2}{Du_2}\right)+m\right]^2-\left[\frac{1}{2}\left(\frac{D^2u_1}{Du_1}+\frac{D^2u_2}{Du_2}\right)\right]^2$$
$$-\frac{1}{2}\frac{D^3u_1}{Du_1}+\frac{1}{2}\left(\frac{D^2u_1}{Du_1}\right)^2-\frac{1}{2}\frac{D^3u_2}{Du_2}+\frac{1}{2}\left(\frac{D^2u_2}{Du_2}\right)^2 \tag{8.3.2.68}$$

rearranging[52]

$$\Theta=-\left[\frac{m^2}{r^3}+m^2\right]+2\left[\frac{1}{2}\left(\frac{D^2u_1}{Du_1}-\frac{D^2u_2}{Du_2}\right)+m\right]^2-\left[\frac{1}{2}\left(\frac{D^2u_1}{Du_1}+\frac{D^2u_2}{Du_2}\right)\right]^2$$
$$-D\left[\frac{1}{2}\left(\frac{D^2u_1}{Du_1}+\frac{D^2u_2}{Du_2}\right)\right] \tag{8.3.2.69}$$

From (8.3.2.7), we can write[53]

$$\left.\begin{array}{l}\dfrac{m^2}{r^3}+m^2=\dfrac{D^2u_1+2mDu_1+\frac{3}{2}m^2u_2}{u_1}+\dfrac{5}{2}m^2\\[2mm]\dfrac{m^2}{r^3}+m^2=\dfrac{D^2u_2-2mDu_2+\frac{3}{2}m^2u_1}{u_2}+\dfrac{5}{2}m^2\end{array}\right\} \tag{8.3.2.70}$$

rearranging[54]

$$\left.\begin{array}{l}\dfrac{m^2}{r^3}+m^2=\left(\dfrac{Du_1}{u_1}\right)^2+\dfrac{2mDu_1}{u_1}+m^2+\dfrac{D^2u_1}{u_1}-\left(\dfrac{Du_1}{u_1}\right)^2+\dfrac{3}{2}m^2+\dfrac{3}{2}m^2\dfrac{u_2}{u_1}\\[2mm]\dfrac{m^2}{r^3}+m^2=\left(\dfrac{Du_2}{u_2}\right)^2-\dfrac{2mDu_2}{u_2}+m^2+\dfrac{D^2u_2}{u_2}-\left(\dfrac{Du_2}{u_2}\right)^2+\dfrac{3}{2}m^2+\dfrac{3}{2}m^2\dfrac{u_1}{u_2}\end{array}\right\} \tag{8.3.2.71}$$

rearranging

$$\left.\begin{array}{l}\dfrac{m^2}{r^3}+m^2=\left[\dfrac{Du_1}{u_1}+m\right]^2+D\left[\dfrac{Du_1}{u_1}+m\right]+\dfrac{3}{2}m^2\left[1+\dfrac{u_2}{u_1}\right]\\[2mm]\dfrac{m^2}{r^3}+m^2=\left[\dfrac{Du_2}{u_2}-m\right]^2+D\left[\dfrac{Du_2}{u_2}-m\right]+\dfrac{3}{2}m^2\left[1+\dfrac{u_1}{u_2}\right]\end{array}\right\} \tag{8.3.2.72}$$

Now we take one half of the sum of these two equations (8.3.2.72), and substitute that into the first term of (8.3.2.69):

---

[52] This is the equation (18) on (G.W. Hill 1886), page 11.
[53] This is the first pair of equations on (G.W. Hill 1886), page 12.
[54] This is the second pair of equations on (G.W. Hill 1886), page 12.

$$\Theta = +\frac{1}{2}\left[\frac{Du_1}{u_1} + m\right]^2 + \frac{1}{2}\left[\frac{Du_2}{u_2} - m\right]^2 + \frac{1}{2}D\left[\frac{Du_1}{u_1} + \frac{Du_2}{u_2} + 2m\right] + \frac{3}{4}\left[2 + \frac{u_2}{u_1} + \frac{u_1}{u_2}\right]$$
$$+ 2\left[\frac{1}{2}\left(\frac{D^2u_1}{Du_1} - \frac{D^2u_2}{Du_2}\right) + m\right]^2 - \left[\frac{1}{2}\left(\frac{D^2u_1}{Du_1} + \frac{D^2u_2}{Du_2}\right)\right]^2 - D\left[\frac{1}{2}\left(\frac{D^2u_1}{Du_1} + \frac{D^2u_2}{Du_2}\right)\right]$$
(8.3.2.73)

The purpose of this last equation is to show that it is symmetrical in $u_1$ and $u_2$. It is not used in the following.

Next, we take the first of these two equations (8.3.2.70) and substitute $u_1, u_2, Du_1, D^2u_1$ in terms of $\zeta$:

$$m^2(u_1u_2)^{-\frac{3}{2}} + m^2 = \frac{D^2u_1 + 2mDu_1 + \frac{3}{2}m^2u_2}{u_1} + \frac{5}{2}m^2 \qquad (8.3.2.74)$$

$$m^2 r^{-3} + m^2 = \frac{\sum_{i=-\infty}^{\infty}\left\{[(2i+1)^2 + 2m(2i+1)]a_i + \frac{3}{2}m^2 a_{-i-1}\right\}\varsigma^{2i+1}}{\sum_{i=-\infty}^{\infty} a_i \varsigma^{2i+1}} + \frac{5}{2}m^2 \qquad (8.3.2.75)$$

but

$$(2i+1)^2 + 2m(2i+1) = 4i(i+1+m) + 1 + 2m$$

so, we can write[55]

$$\frac{m^2}{r^3} + m^2 = 1 + 2m + \frac{5}{2}m^2 + \frac{\sum_{i=-\infty}^{\infty}\left\{4i(i+1+m)a_i + \frac{3}{2}m^2 a_{-i-1}\right\}\varsigma^{2i}}{\sum_{i=-\infty}^{\infty} a_i \varsigma^{2i}} \qquad (8.3.2.76)$$

On the right-hand side, we have written $\varsigma^{2i}$ instead of $\varsigma^{2i+1}$, which is legitimate, since it is just a matter of dividing numerator and denominator by $\varsigma$.

The next few pages have the goal of making it easier to solve for $\Theta$. For two parts of $\Theta$, functions $R_i$ and $U_i$ are defined and put into Fourier expansions, the coefficients of which are treated as power series in $m$. These then lead to recursive solutions which can be inserted into $\Theta$ to provide a full solution for it.

Next, we define a Fourier expansion of the last term (the quotient) of equation (8.3.2.76) to be

$$\sum_{i=-\infty}^{\infty} R_i \varsigma^{2i} \qquad (8.3.2.77)$$

in other words

$$\sum_{j=-\infty}^{\infty} R_j \varsigma^{2j} = \frac{\sum_{i=-\infty}^{\infty}\left\{4i(i+1+m)a_i + \frac{3}{2}m^2 a_{-i-1}\right\}\varsigma^{2i}}{\sum_{i=-\infty}^{\infty} a_i \varsigma^{2i}} \qquad (8.3.2.78)$$

Now let's multiply by the denominator of the right-hand side

$$\sum_{j=-\infty}^{\infty}\sum_{i=-\infty}^{\infty} a_i R_j \varsigma^{2i+2j} = \sum_{i=-\infty}^{\infty}\left\{4i(i+1+m)a_i + \frac{3}{2}m^2 a_{-i-1}\right\}\varsigma^{2i} \qquad (8.3.2.79)$$

Changing $i$ to $i - j$ gives us

$$\sum_{j=-\infty}^{\infty}\sum_{i=-\infty}^{\infty} a_{i-j} R_j \varsigma^{2i} = \sum_{i=-\infty}^{\infty}\left\{4i(i+1+m)a_i + \frac{3}{2}m^2 a_{-i-1}\right\}\varsigma^{2i} \qquad (8.3.2.80)$$

---

[55] This is the fifth equation on (G.W. Hill 1886), page 12.

Due to independence, this must hold for all $2i$, giving[56]

$$\sum_{j=-\infty}^{\infty} a_{i-j} R_j = 4i(i+1+m)a_i + \frac{3}{2}m^2 a_{-i-1} \tag{8.3.2.81}$$

We will continue with this equation at (8.3.3.5).

This is also a convenient place to look at what happens when we replace $j$ by $-j$. The result is an equation defining $R_{-j}$, and it is exactly the same as the equation defining $R_{-j}$, so we can conclude:

$$R_{-j} = R_j \tag{8.3.2.82}$$

Our next step is to find a series expansion of $\Theta$. First, we recall the Fourier expansion of the coordinates, as in $(8.3.2.50) - (8.3.2.53)$.

Now we assume a Fourier expansion of $\frac{D^2 u_1}{D u_1}$ as in Hill[57]

$$\sum_{j=-\infty}^{\infty} U_j \varsigma^{2j} = \frac{D^2 u_1}{D u_1} = \frac{\sum_{i=-\infty}^{\infty}(2i+1)^2 a_i \varsigma^{2i}}{\sum_{i=-\infty}^{\infty}(2i+1) a_i \varsigma^{2i}} \tag{8.3.2.83}$$

rearranging

$$\sum_{j=-\infty}^{\infty}\sum_{i=-\infty}^{\infty}(2i+1)a_i U_j \varsigma^{2i+2j} = \sum_{i=-\infty}^{\infty}(2i+1)^2 a_i \varsigma^{2i}$$

on the left-hand side, changing $i \to i - j$

$$\sum_{j=-\infty}^{\infty}\sum_{i=-\infty}^{\infty}(2i-2j+1)a_{i-j} U_j \varsigma^{2i} = \sum_{i=-\infty}^{\infty}(2i+1)^2 a_i \varsigma^{2i}$$

Since the $\varsigma^{2i}$ form a basis, we have[58]

$$\sum_{j=-\infty}^{\infty}(2i-2j+1)a_{i-j} U_j = (2i+1)^2 a_i \tag{8.3.2.84}$$

The remark after equation (8.3.2.78) also applies here.

$$\sum_{i=-\infty}^{\infty} h_{j-i} U_i = (2j+1) h_j \tag{8.3.2.85}$$

Let's move one little piece to the left-hand side:[59]

$$\sum_{i=-\infty}^{\infty} h_{j-i} U_i - h_j = 2j h_j \tag{8.3.2.86}$$

Our next step is to prove $U_0 = 1$, for which Hill provides two proofs. The first proof consists of an analysis of the form of (8.3.2.86)[60] and the second uses an integral of (8.3.2.83). We were not able to reproduce the result of the first proof but can reproduce the second one including some simplifications.

---

[56] This is the last equation on (G.W. Hill 1886), page 12.

[57] This is last equations on (G.W. Hill 1886), page 13.

[58] This is equal to the equation on (G.W. Hill 1886), page 14, when we note that we need $h_{i-j}$ on the left-hand side.

[59] This is now the format of the rows in (19) on (G.W. Hill 1886), page 14, except that we have switched the role of $i$ and $j$. Now, (19) has one row for each $j$ and one column for each $i$. Also, the column for $i = 0$ has the form $h_j(U_0 - 1)$.

[60] Or, equivalently, (19) on (G.W. Hill 1886), page 14.

For the proof via an integral, we recall that $D = -i\frac{d}{d\tau}$, and use the definition of $U_i$ from our (8.3.2.83)[61].

$$\sum_{j=-\infty}^{\infty} U_j \varsigma^{2j} = \frac{D^2 u_1}{D u_1} \tag{8.3.2.87}$$

Taking the integral of the left-hand side gives us

$$\frac{1}{2\pi}\int_0^{2\pi} \sum_{j=-\infty}^{\infty} U_j \varsigma^{2j}\, d\tau = U_0 \tag{8.3.2.88}$$

since all constant terms in the sum yield 0 when integrated from 0 to $2\pi$. Then taking the integral of the right-hand side gives us

$$\frac{1}{2\pi}\int_0^{2\pi} \frac{D^2 u_1}{D u_1}\, d\tau = \frac{1}{2\pi i}\int_0^{2\pi} \frac{\frac{d}{d\tau}\left(\frac{du_1}{d\tau}\right)}{\frac{du_1}{d\tau}}\, d\tau = \frac{1}{2\pi i}\ln\left(\frac{du_1}{d\tau}\right)\Big|_{\tau=0}^{2\pi} \tag{8.3.2.89}$$

$$\frac{1}{2\pi}\int_0^{2\pi} \frac{D^2 u_1}{D u_1}\, d\tau = \frac{1}{2\pi i}\ln(\rho e^{i\varphi})\Big|_{\tau=0}^{2\pi} \tag{8.3.2.90}$$

since an arbitrary complex number, such as $\frac{du_1}{d\tau}$, can always be written as $\rho e^{i\varphi}$ with real $\rho$ and $\varphi$. Now we can apply the logarithm

$$= \frac{1}{2\pi i}\ln(\rho)\Big|_{\tau=0}^{2\pi} + \frac{1}{2\pi i} i\varphi\Big|_{\tau=0}^{2\pi} \tag{8.3.2.91}$$

but the first term is 0 because we are looking at a periodic orbit, so $\rho$, as a function of $\tau$, must have the same value at $\tau = 0$ and $\tau = 2\pi$. The second term is 1 because, when $\tau$ traverses one orbit, then so must $\varphi$. As a result, the integral of the right-hand side is 1, and

$$U_0 = 1 \blacksquare \tag{8.3.2.92}$$

Next, we can see that

$$\frac{D^2 u_2}{D u_2} = -\frac{\sum_{i=-\infty}^{\infty}(2i+1)^2 a_i \varsigma^{-2i}}{\sum_{i=-\infty}^{\infty}(2i+1) a_i \varsigma^{-2i}} = -\sum_{j=-\infty}^{\infty} U_j \varsigma^{-2j} = -\sum_{j=-\infty}^{\infty} U_{-j}\varsigma^{2j} \tag{8.3.2.93}$$

As a result:

$$\frac{D^2 u_1}{D u_1} - \frac{D^2 u_2}{D u_2} = \sum_{j=-\infty}^{\infty}(U_j + U_{-j})\varsigma^{2j} \tag{8.3.2.94}$$

$$\frac{D^2 u_1}{D u_1} + \frac{D^2 u_2}{D u_2} = \sum_{j=-\infty}^{\infty}(U_j - U_{-j})\varsigma^{2j} \tag{8.3.2.95}$$

$$D\left(\frac{D^2 u_1}{D u_1} + \frac{D^2 u_2}{D u_2}\right) = \sum_{j=-\infty}^{\infty} 2j(U_j - U_{-j})\varsigma^{2j} \tag{8.3.2.96}$$

With this, we have a Fourier expansion of all parts of $\Theta$.

---

[61] From (G.W. Hill 1886), page 13.

### 8.3.3. Hill's Section III

The preceding section has determined that the lunar inequalities which have the simple power of the eccentricity as a factor depends on the following equation. (8.3.2.32). This is a linear approximation of the anomalistic orbit.

$$D^2 w = \Theta w \tag{8.3.3.1}$$

Now we make a Fourier expansion of the form

$$\Theta = \sum_{i=-\infty}^{\infty} \Theta_i \varsigma^{2i} \tag{8.3.3.2}$$

This is legitimate because we already know that $\Theta$ is periodic. As a first approximation, we can take

$$D^2 w = \Theta_0 w$$

and, since $\Theta_0$ is constant, this can be integrated as

$$w = K\varsigma^c + K'\varsigma^{-c}$$

where $c = \sqrt{\Theta_0}$.

Adding additional terms to $\Theta$ will change $c$ and add new terms to $w$ of the form $A\varsigma^{\pm c+2i}$ and we can assume

$$w = Kf(\varsigma, c) + K'f(\varsigma, -c)$$

Let us take, as an ansatz

$$w = \sum_{i=-\infty}^{\infty} w_i \varsigma^{c+2i} \tag{8.3.3.3}$$

[62]We can put this into the equation (8.3.3.1), yielding

$$D^2 w - \Theta w = \sum_{j=-\infty}^{\infty} (c+2j)^2 w_j \varsigma^{c+2j} - \sum_{i=-\infty}^{\infty} \Theta_i \varsigma^{2i} \sum_{j=-\infty}^{\infty} w_j \varsigma^{c+2j} = 0$$

In the sum over $i$, we can replace $i$ by $j - i$, giving

$$D^2 w - \Theta w = \sum_{j=-\infty}^{\infty} (c+2j)^2 w_j \varsigma^{c+2j} - \sum_{i=-\infty}^{\infty} \Theta_{j-i} \varsigma^{2j-2i} \sum_{i=-\infty}^{\infty} w_i \varsigma^{c+2i} = 0$$

$$D^2 w - \Theta w = \sum_{j=-\infty}^{\infty} (c+2j)^2 w_j \varsigma^{c+2j} - \sum_{j=-\infty}^{\infty} \sum_{i=-\infty}^{\infty} \Theta_{j-i} w_i \varsigma^{c+2j} = 0$$

which leads to[63]

$$(c+2j)^2 w_j - \sum_{i=-\infty}^{\infty} \Theta_{j-i} w_i = 0 \tag{8.3.3.4}$$

In summary, with equation (8.3.2.86), we should be able to find the $U_i$ as a function of the $a_i$. Then, the discussion following (8.3.2.86) should make it possible to find the $\Theta_i$ as a function of the $U_i$ (and hence of the $a_i$). Finally, (8.3.3.5) should give us $c$ and $w_i$ as a function of $\Theta_i$ (and hence of the $a_i$).

---

[62] Here, we deviate from Hill's notation by writing $w_i$ instead of $b_i$, since $b_i$ is used earlier, and $w_i$ better fits our conventions.
[63] This is equation (20) on (G.W. Hill 1886), page 18.

In order to solve (8.3.4) for $w_j$, we will need $\Theta_{j-i}$, but we can find that in terms of $R_j$ and $U_j$, so we will now focus on finding them. In (8.3.2.81), we will separate the term involving $a_0$ from the sum. Taking (8.3.2.81) and rearranging gives

$$a_0 R_j = - \sum_{\substack{i=-\infty \\ i \neq j}}^{\infty} a_{j-i} R_i + 4j(j+1+m)a_j + \frac{3}{2} m^2 a_{-j-1} \qquad (8.3.3.5)$$

Then we divide by $a_0$ and apply the definition of $\bar{\bar{a}}_i$.

$$R_j = m \left\{ - \sum_{\substack{i=-\infty \\ i \neq j}}^{\infty} \bar{\bar{a}}_{j-i} R_i + 4j(j+1+m)\bar{\bar{a}}_j + \frac{3}{2} m^2 \bar{\bar{a}}_{-j-1} \right\} \qquad (8.3.3.6)$$

Now we consider this as power series over $m$ and add an index to $R_{j,k}$ to denote the coefficients in that power series. Based on equation (8.3.3.6) we see that this gives us a recursive procedure by virtue of the analog of claim (5.1).

$$R_j = \sum_{k=0}^{\infty} R_{j,k} m^k$$

Then we can write a formula for the coefficient of $m^k$ in terms of the coefficients of $m^{k-1}$.

$$R_{j,k} = - \sum_{\substack{i=-\infty \\ i \neq j}}^{\infty} \sum_{k_1=1}^{k-1} \bar{\bar{a}}_{j-i,k_1} R_{i,k-k_1-1} + 4j(j+1)\bar{\bar{a}}_{j,k-1} + 4j\bar{\bar{a}}_{j,k-2} + \frac{3}{2} \bar{\bar{a}}_{-j-1,k-3} \qquad (8.3.3.7)$$

The sum over $k_1$ starts at 1 because $c$ starts at $m^1$, and ends at $k-1$ because $R$ might include a term in $m^0$ and we need to calculate the coefficient of $m^{k-1}$. Finally, we can apply claim (6.1) and transition to finite sums. However, claim (6.1) applies to $\bar{\bar{a}}_{i,k}$, and we have $\bar{\bar{a}}_{j-i,k_1}$, so we need

$$j - i \in [-j'(k), j'(k)]$$
$$-i \in [-j'(k) - j, j'(k) - j]$$
$$i \in [-j'(k) + j, j'(k) + j]$$

$$R_{j,k} = - \sum_{\substack{i=-j'(k)+j \\ i \neq j}}^{j'(k)+j} \sum_{k_1=1}^{k-1} \bar{\bar{a}}_{j-i,k_1} R_{i,k-k_1-1} + 4j(j+1)\bar{\bar{a}}_{j,k-1} + 4j\bar{\bar{a}}_{j,k-2} + \frac{3}{2} \bar{\bar{a}}_{-j-1,k-3} \qquad (8.3.3.8)$$

[64]Now we start with (8.3.2.83) and, similar to (8.3.3.5), we will separate the term involving $a_0$ from the sum. Note: if (8.3.2.84) uses $\frac{a_i}{a_0} = b_i = \bar{a}_i$ instead of $a_i$, then (8.3.3.10) is not changed.

$$a_0 U_j = - \sum_{\substack{i=-\infty \\ i \neq j}}^{\infty} (2j - 2i + 1) a_{j-i} U_i + (2j+1)^2 a_j \qquad (8.3.3.9)$$

Then we divide by $a_0$ and apply the definition of $\bar{\bar{a}}_j$.

---

[64] Our calculations of $R_j$ agree with Hill Literal Expression page 33 (G.W. Hill 1894) except for the coefficient $R_{2,8}$. Hill calculates to $m^{11}$, and 23 coefficients, and this is 1 of 23, see 9.5. Hill's literal perigee page 33, $R_i$.

$$U_j = m \left\{ -\sum_{\substack{i=-\infty \\ i \neq j}}^{\infty} (2j - 2i + 1)\bar{\bar{a}}_{j-i} U_i + (2j + 1)^2 \bar{\bar{a}}_j \right\} \quad (8.3.3.10)$$

Now we consider this as power series over $m$ and add an index to $U_{j,k}$ to denote the coefficients in that power series. Based on equation (8.3.3.10) we see that this gives us a recursive procedure by virtue of the analog of claim (5.1).

$$U_j = \sum_{k=0}^{\infty} U_{j,k} m^k$$

Then we can write a formula for the coefficient of $m^k$ in terms of the coefficients of $m^{k-1}$.

$$U_{j,k} = -\sum_{\substack{i=-\infty \\ i \neq j}}^{\infty} \sum_{k_1=1}^{k-1} (2j - 2i + 1)\bar{\bar{a}}_{j-i,k_1} U_{i,k-k_1-1} + (2j + 1)^2 \bar{\bar{a}}_{j,k-1} \quad (8.3.3.11)$$

Finally, we can apply claim (6.1), transition to finite sums, and use the same logic as that leading up to equation (8.3.3.8).

The term in $\bar{\bar{a}}_{j-i,k}$ applies only when

$$j - i \in [-j'(k), j'(k)]$$
$$-i \in [-j'(k) - j, j'(k) - j]$$
$$i \in [-j'(k) + j, j'(k) + j]$$

$$U_{j,k} = -\sum_{\substack{i=-j'(k)+j \\ i \neq j}}^{j'(k)+j} \sum_{k_1=1}^{k-1} (2j - 2i + 1)\bar{\bar{a}}_{j-i,k_1} U_{i,k-k_1-1} + (2j + 1)^2 \bar{\bar{a}}_{j,k-1} \quad (8.3.3.12)$$

With these pieces, we should be able to assemble the full $\Theta$. Starting with (8.3.2.69), we have

$$\Theta = -\left[\frac{m^2}{r^3} + m^2\right] + 2\left[\frac{1}{2}\left(\frac{D^2 u_1}{Du_1} - \frac{D^2 u_2}{Du_2}\right) + m\right]^2$$
$$- \left[\frac{1}{2}\left(\frac{D^2 u_1}{Du_1} + \frac{D^2 u_2}{Du_2}\right)\right]^2 - D\left[\frac{1}{2}\left(\frac{D^2 u_1}{Du_1} + \frac{D^2 u_2}{Du_2}\right)\right] \quad (8.3.3.13)$$

Substituting (8.3.2.76), (8.3.2.78), (8.3.2.93), (8.3.2.94), and (8.3.2.95) gives us

$$\Theta = -1 - 2m - \frac{5}{2}m^2 - \sum_{j=-\infty}^{\infty} R_j \varsigma^{2j} + \left[\frac{1}{2}\left(\sum_{j=-\infty}^{\infty} (U_j + U_{-j})\varsigma^{2j}\right)\right]^2$$
$$+ 2m\left[\left(\sum_{j=-\infty}^{\infty} (U_j + U_{-j})\varsigma^{2j}\right)\right] + 2m^2 - \frac{1}{4}\left[\left(\sum_{j=-\infty}^{\infty} (U_j - U_{-j})\varsigma^{2j}\right)\right]^2$$
$$- \frac{1}{2}\left[\left(\sum_{j=-\infty}^{\infty} 2j(U_j - U_{-j})\varsigma^{2j}\right)\right] \quad (8.3.3.14)$$

Completing the squares

$$\Theta = -1 - 2m - \frac{1}{2}m^2 - \sum_{j=-\infty}^{\infty} R_j \varsigma^{2j} + 2m\left[\left(\sum_{j=-\infty}^{\infty} (U_j + U_{-j})\varsigma^{2j}\right)\right]$$

$$-\left[\left(\sum_{j=-\infty}^{\infty} j(U_j - U_{-j})\varsigma^{2j}\right)\right]$$

$$+\frac{1}{2}\left[\left(\sum_{j_1=-\infty}^{\infty}\sum_{j_2=-\infty}^{\infty} (U_{j_1} + U_{-j_1})(U_{j_2} + U_{-j_2})\varsigma^{2j_1+2j_2}\right)\right]$$

$$-\frac{1}{4}\left[\left(\sum_{j_1=-\infty}^{\infty}\sum_{j_2=-\infty}^{\infty} (U_{j_1} - U_{-j_1})(U_{j_2} - U_{-j_2})\varsigma^{2j_1+2j_2}\right)\right]$$

In the sums, we can now replace $j_1$ by $j - j_2$, giving

$$\Theta = -1 - 2m - \frac{1}{2}m^2 - \sum_{j=-\infty}^{\infty} R_j \varsigma^{2j} + 2m\left[\left(\sum_{j=-\infty}^{\infty} (U_j + U_{-j})\varsigma^{2j}\right)\right]$$

$$-\left[\left(\sum_{j=-\infty}^{\infty} j(U_j - U_{-j})\varsigma^{2j}\right)\right]$$

$$+\frac{1}{2}\left[\left(\sum_{j=-\infty}^{\infty}\sum_{j_2=-\infty}^{\infty} (U_{j-j_2} + U_{-(j-j_2)})(U_{j_2} + U_{-j_2})\varsigma^{2j}\right)\right]$$

$$-\frac{1}{4}\left[\left(\sum_{j=-\infty}^{\infty}\sum_{j_2=-\infty}^{\infty} (U_{j-j_2} - U_{-(j-j_2)})(U_{j_2} - U_{-j_2})\varsigma^{2j}\right)\right]$$

Since the $\varsigma^{2j}$ form a basis, we can write

$$\Theta_j = \delta_{j,0}\left(-1 - 2m - \frac{1}{2}m^2\right) - R_j + 2m(U_j + U_{-j}) - j(U_j - U_{-j})$$

$$+\frac{1}{2}\left[\left(\sum_{j_2=-\infty}^{\infty} (U_{j-j_2} + U_{-(j-j_2)})(U_{j_2} + U_{-j_2})\right)\right]$$

$$-\frac{1}{4}\left[\left(\sum_{j_2=-\infty}^{\infty} (U_{j-j_2} - U_{-(j-j_2)})(U_{j_2} - U_{-j_2})\right)\right] \tag{8.3.3.15}$$

This is a good opportunity to look at $\Theta_{-j}$, by substituting $-j$ for $j$ in the equation. The term beginning with $2m$ is unchanged, as well as $-j(U_j - U_{-j})$. The first sum is clearly unchanged, and the second as well, because it involves two minus signs. As a result, we can say

$$\Theta_{-j} = \Theta_j \tag{8.3.3.16}$$

Finally, we would like to look at the formula for the power series in $m$. At the same time, we will change the infinite sum to a finite sum, assuming that the $U_j$ are zero whenever the $\bar{\bar{a}}_j$ are. This should be clear from (8.3.2.83).

$$\Theta_{j,k} = \delta_{j,0}\left(-\delta_{k,0} - 2\delta_{k,1} - \frac{1}{2}\delta_{k,2}\right) - R_{j,k} + 2(U_{j,k-1} + U_{-j,k-1}) - j(U_{j,k} - U_{-j,k})$$

$$+ \frac{1}{2}\left[\left(\sum_{j_2=-j'(k)+j}^{j'(k)+j} \sum_{k_1=0}^{k} (U_{j-j_2,k_1} + U_{-(j-j_2),k_1})(U_{j_2,k-k_1} + U_{-j_2,k-k_1})\right)\right]$$

$$- \frac{1}{4}\left[\left(\sum_{j_2=-j'(k)+j}^{j'(k)+j} \sum_{k_1=0}^{k} (U_{j-j_2,k_1} - U_{-(j-j_2),k_1})(U_{j_2,k-k_1} - U_{-j_2,k-k_1})\right)\right] \quad (8.3.3.17)$$

Starting here[65], Hill discusses the properties of three different determinants. At that time, taking a determinant of the equation and solving it was a welcome reduction in the number of (manual) calculations required to find *c*. However, today, with the help of computers, we can solve the complete equation, so we will skip over this part. We present these calculation steps in section 11. Calculations of the motion of the moon's perigee.

Now that we have described Hill's method of calculating the motion of the perigee, we will digress once more, this time to make a number of plausibility checks. We check the coefficients of Hill's series for the intermediate orbits, and also series calculated as part of our examination of the motion of the perigee. We compare our calculations to those made by Hill and other authors. As a result, we see some errors in Hill's extensive manual calculations and show that our software calculations fulfill the defining equations and also match other more modern papers.

---

[65] This is (G.W. Hill 1886), pages 18-32.

## 9. Some Plausibility Checks

The plausibility checks we make here compare our calculations with those published by Hill and a few more modern authors. We also check the results against the defining equations.

### 9.1. Hill's Calculations of $\bar{a}_i$ up to Order $m^9$

Calculation of denominators for Hill's $\bar{a}_i$ up to $m^9$ on (G.W. Hill 1878) page 142. We are using the notation of $\bar{a}_i = b_i = \frac{a_i}{a_0}$ where Hill assumes $a_0 = 1$ and uses $a_i$. We are also using $\bar{a}_i$ for the partial sum. In order to arrive at these values, we took the formulas on (G.W. Hill 1878) page 142 and calculated the products of the prime factors in the denominator.

$$\bar{a}_1 = \frac{3m^2}{16} + \frac{m^3}{2} + \frac{7m^4}{12} + \frac{11m^5}{36} - \frac{30749m^6}{110592} - \frac{1010521m^7}{829440} - \frac{18445871m^8}{6220800} - \frac{2114557853m^9}{373248000}$$

$$\bar{a}_{-1} = -\frac{19m^2}{16} - \frac{5m^3}{3} - \frac{43m^4}{36} - \frac{14m^5}{27} - \frac{7381m^6}{82944} + \frac{3574153m^7}{2488320} + \frac{55218889m^8}{9331200} + \frac{13620153029m^9}{1119744000}$$

$$\bar{a}_2 = \frac{25m^4}{256} + \frac{803m^5}{1920} + \frac{6109m^6}{7200} + \frac{897599m^7}{864000} + \frac{237203647m^8}{368640000} - \frac{44461407673m^9}{58060800000}$$

$$\bar{a}_{-2} = \frac{23m^5}{640} + \frac{299m^6}{2400} + \frac{56339m^7}{288000} + \frac{79400351m^8}{368640000} + \frac{8085846833m^9}{29030400000}$$

$$\bar{a}_3 = \frac{833m^6}{12288} + \frac{27943m^7}{71680} + \frac{12275527m^8}{11289600} + \frac{27409853579m^9}{14224896000}$$

$$\bar{a}_{-3} = \frac{m^6}{192} + \frac{71m^7}{1920} + \frac{46951m^8}{403200} + \frac{14086643m^9}{63504000}$$

$$\bar{a}_4 = \frac{3537m^8}{65536} + \frac{37269889m^9}{96337920}$$

$$\bar{a}_{-4} = \frac{23m^8}{6144} + \frac{1576553m^9}{57802752}$$

Of the total of 39 coefficients on (G.W. Hill 1878) page 142, 8 disagree with our calculations. They are $\bar{a}_{2,9}, \bar{a}_{-2,8}, \bar{a}_{-2,9}, \bar{a}_{-3,7}, \bar{a}_{-3,8}, \bar{a}_{-3,9}, \bar{a}_{4,9}, \bar{a}_{-4,9}$. This means that all coefficients up to $m^6$ agree with our calculations. They also agree with (Wintner 2014) §415 (page 399), covering a total of 18 coefficients.

### 9.2. Brown's Calculations of $\bar{a}_i$ up to Order $m^9$

Next, we investigate (G.M. Brown et al. 2024) (2024) table 2, page 15, which gives us the coefficients of Hill's series for all values of $j$ between $-4$ and $4$ and up to $m^9$. However, Brown displays these coefficients as power series in $M = \frac{m}{1-m/3}$, so we need to convert that to a power series in $m$. The results of that conversion are:

$$\bar{a}_1 = \frac{3m^2}{16} + \frac{m^3}{2} + \frac{7m^4}{12} + \frac{11m^5}{36} - \frac{30749m^6}{110592} - \frac{1010521m^7}{829440} - \frac{18445871m^8}{6220800} - \frac{2114557853m^9}{373248000} + O(m^{10})$$

$$\bar{a}_{-1} = -\frac{19m^2}{16} - \frac{5m^3}{3} - \frac{43m^4}{36} - \frac{14m^5}{27} - \frac{7381m^6}{82944} + \frac{3574153m^7}{2488320} + \frac{55218889m^8}{9331200} + \frac{13620153029m^9}{1119744000} + O(m^{10})$$

$$\bar{a}_2 = \frac{25m^4}{256} + \frac{803m^5}{1920} + \frac{6109m^6}{7200} + \frac{897599m^7}{864000} + \frac{237203647m^8}{368640000} - \frac{11098919887m^9}{14515200000} + O(m^{10})$$

$$\bar{a}_{-2} = \frac{23m^5}{640} + \frac{299m^6}{2400} + \frac{56339m^7}{288000} + \frac{26850677m^8}{110592000} + \frac{93086381m^9}{268800000} + O(m^{10})$$

$$\bar{a}_{-2} = \frac{23m^5}{640} + \frac{299m^6}{2400} + \frac{56339m^7}{288000} + \frac{238200053m^8}{1105920000} + \frac{146886277m^9}{537600000} + O(m^{10})$$

$$\bar{a}_3 = \frac{833m^6}{12288} + \frac{27943m^7}{71680} + \frac{12275527m^8}{11289600} + \frac{27409853579m^9}{14224896000} + O(m^{10})$$

$$\bar{a}_{-3} = \frac{m^6}{192} + \frac{7477m^7}{215040} + \frac{65239m^8}{627200} + \frac{2674679587m^9}{14224896000} + O(m^{10})$$

$$\bar{a}_4 = \frac{3537m^8}{65536} + \frac{18638507m^9}{48168960} + O(m^{10})$$

$$\bar{a}_{-4} = \frac{23m^8}{6144} + \frac{795829m^9}{28901376} + O(m^{10})$$

These values are in 100% agreement with ours.

## 9.3. Checking Against Hill's Equation for the Series of $\bar{a}_i$ up to Order $m^{30}$

Now we would like to do a plausibility check on the $\bar{a}_i$ on (G.W. Hill 1878) page 142. For this, it would be nice to simply insert the values into the original (Newtonian) ODEs, which is our (3.4) or their equivalent in complex coordinates, our (4.17) − (4.18) or (4.20) − (4.21). However, these equations use $a_i$ and not $\bar{a}_i = \frac{a_i}{a_0}$, and the term $\frac{q_i}{|q|^3}$ or $\frac{u_i}{(u_i u_2)^{3/2}}$ prevents us from using $\frac{a_i}{a_0}$. This can be done by following Hill's steps that remove this term, leading to (4.29) and following. But, once we take this step, the easiest method is to use Hill's equation, our (4.58), which can be converted to $\frac{a_i}{a_0}$ simply by dividing the whole equation by $a_0^2$. Now we calculate this equation for values of $j$ between $-4$ and 4, remembering that we need to use the partial sums up to $m^9$. The results of this are:

```
equation j=-4, Hill version
```
$$0$$
```
equation j=-3, Hill version
```
$$-\frac{32048563m^9}{948326400} - \frac{70163m^8}{5644800} - \frac{95m^7}{43008}$$
```
equation j=-2, Hill version
```
$$-\frac{703m^9}{232243200} - \frac{m^8}{1105920}$$
```
equation j=-1, Hill version
```
$$0$$
```
equation j=1, Hill version
```
$$0$$
```
equation j=2, Hill version
```
$$0$$
```
equation j=3, Hill version
```
$$0$$
```
equation j=4, Hill version
```
$$0$$
```
equation j=-4, our version
```
$$0$$
```
equation j=-3, our version
```
$$0$$
```
equation j=-2, our version
```
$$0$$
```
equation j=-1, our version
```

equation j=1, our version
$$0$$
equation j=2, our version
$$0$$
equation j=3, our version
$$0$$
equation j=4, our version
$$0$$

These results demonstrate that the values taken from (G.W. Hill 1878) page 142 do not all solve Hill's equation, with discrepancies for $j = -3$ and $j = -2$, for $m^7, m^8, m^9$ but our values solve the equation for all values of $j$ between $-4$ and $4$ and up to $m^9$. In fact, they solve the equation for all values up to $m^{30}$.

### 9.4. Hill's numeric perigee page 13, $R_i$

Equations on Hill's numeric perigee (G.W. Hill 1886) page 13, indK=9

$$R_0 = -\frac{9m^4}{32} + 4m^5 + \frac{34m^6}{3} + 15m^7 + \frac{2704801m^8}{221184} + \frac{122957m^9}{12960}$$

$$R_1 = \frac{3m^2}{2} + \frac{19m^3}{4} + \frac{20m^4}{3} + \frac{43m^5}{9} + \frac{18709m^6}{13824} + \frac{759413m^7}{414720} + \frac{6675059m^8}{1555200} - \frac{41991161m^9}{11664000}$$

$$R_2 = \frac{33m^4}{16} + \frac{2937m^5}{320} + \frac{23051m^6}{1200} + \frac{97051m^7}{4000} + \frac{167206573m^8}{8640000} + \frac{1059187213m^9}{172800000}$$

$$R_3 = \frac{1393m^6}{512} + \frac{562041m^7}{35840} + \frac{40986497m^8}{940800} + \frac{11343987331m^9}{148176000}$$

$$R_4 = \frac{7193m^8}{2048} + \frac{300204311m^9}{12042240}$$

These all agree with our calculated values for $R_i$, thus confirming the first 5 equations on page 13.

Next, we calculate the results of the next equation on (G.W. Hill 1878), page 13. Here, we need to combine the coefficients of $(\zeta^{2i} + \zeta^{-2i})$ because that is what corresponds to $R_i$, according to the discussion on page 12, and when observing that $R_{-i} = R_i$.

$$R_0 = -\frac{9m^4}{32} + 4m^5 + \frac{34m^6}{3} + 15m^7$$

$$R_1 = \frac{3m^2}{2} + \frac{19m^3}{4} + \frac{20m^4}{3} + \frac{43m^5}{9}$$

$$R_2 = \frac{33m^4}{16} + \frac{2937m^5}{320} + \frac{23051m^6}{1200} + \frac{97051m^7}{4000}$$

These results agree with our calculated values up to the order of the errors mentioned in the text on page 13, thus confirming the second set of equations on page 13.

### 9.5. Hill's literal perigee page 33, $R_i$

Here we compare the values of $R_i$ given in (G.W. Hill 1894) page 33. We show a power series of $R_i$, maximum power $m^{11}$, first Hill's version (G.W. Hill 1894) page 33, then our version:

$$R_0 = -\frac{9m^4}{32} + 4m^5 + \frac{34m^6}{3} + 15m^7 + \frac{2704801m^8}{221184} + \frac{122957m^9}{12960} + \frac{1260881m^{10}}{207360}$$
$$- \frac{291394307m^{11}}{11664000}$$

$$R_0 = -\frac{9m^4}{32} + 4m^5 + \frac{34m^6}{3} + 15m^7 + \frac{2704801m^8}{221184} + \frac{122957m^9}{12960} + \frac{1260881m^{10}}{207360}$$
$$- \frac{291394307m^{11}}{11664000}$$

$$R_1 = \frac{3m^2}{2} + \frac{19m^3}{4} + \frac{20m^4}{3} + \frac{43m^5}{9} + \frac{18709m^6}{13824} + \frac{759413m^7}{414720} + \frac{6675059m^8}{1555200} - \frac{41991161m^9}{11664000}$$
$$- \frac{4528083484913m^{10}}{179159040000}$$

$$R_1 = \frac{3m^2}{2} + \frac{19m^3}{4} + \frac{20m^4}{3} + \frac{43m^5}{9} + \frac{18709m^6}{13824} + \frac{759413m^7}{414720} + \frac{6675059m^8}{1555200} - \frac{41991161m^9}{11664000}$$
$$- \frac{4528082998913m^{10}}{179159040000} - \frac{1391582016661969m^{11}}{37623398400000}$$

$$R_2 = \frac{33m^4}{16} + \frac{2937m^5}{320} + \frac{23051m^6}{1200} + \frac{97051m^7}{4000} + \frac{334413271m^8}{17280000}$$

$$R_2 = \frac{33m^4}{16} + \frac{2937m^5}{320} + \frac{23051m^6}{1200} + \frac{97051m^7}{4000} + \frac{167206573m^8}{8640000} + \frac{1059187213m^9}{172800000}$$
$$- \frac{12134100557m^{10}}{486000000} - \frac{7476551840719m^{11}}{68040000000}$$

$$R_3 = \frac{1393m^6}{512}$$

$$R_3 = \frac{1393m^6}{512} + \frac{562041m^7}{35840} + \frac{40986497m^8}{940800} + \frac{11343987331m^9}{148176000} + \frac{4562756952133m^{10}}{49172480000}$$
$$+ \frac{4728398067354011m^{11}}{69701990400000}$$

No Hill's version

$$R_4 = \frac{7177m^8}{2048} + \frac{99759441m^9}{4014080} + \frac{466011937m^{10}}{5419008} + \frac{114711969628733m^{11}}{597445632000}$$

No Hill's version

$$R_5 = \frac{582527m^{10}}{131072} + \frac{34020726071m^{11}}{908328960}$$

All coefficients agree except $R_{2,8}$ (1 coefficient out of 23).

Now we check the plausibility of these values against the equation defining $R_i$ i.e., our (8.3.2.81) or the last equation on page 12 of Hill's numeric perigee (G.W. Hill 1886). However, since this equation includes a sum over $R_i$, and Hill's approximation for $R_3$ has a maximum order of $m^6$, we can only use it as a plausibility check up to $m^6$. Since the first (and only) difference between Hill's approximations and our calculations is at $R_{2,8}$, we would expect the plausibility to fail for $m^8$, but the test cannot be used that far. In fact, for Hill's version, all tests up to $m^6$ succeed.

For our version, all tests up to $m^{30}$ succeed.

### 9.6. Hill's literal perigee page 34, $U_i$

Here we compare the values of $U_i$ given in (G.W. Hill 1894) page 34. In order to document Hill's approximation, we take the $A_i$ and $B_i$ from page 34, expand the prime factors, and convert the result to $U_i$. We show a power series of $U_i$, maximum power $m^{10}$, first Hill's version (G.W. Hill 1894) page 34, then our version:

$$U_1 = \frac{9m^2}{8} + 3m^3 + \frac{7m^4}{2} + \frac{11m^5}{6} - \frac{38273m^6}{18432} - \frac{611123m^7}{69120} - \frac{20324711m^8}{1036800} - \frac{1014978499m^9}{31104000}$$
$$+ \frac{114261285013m^{10}}{5308416000}$$

$$U_1 = \frac{9m^2}{8} + 3m^3 + \frac{7m^4}{2} + \frac{11m^5}{6} - \frac{38273m^6}{18432} - \frac{611123m^7}{69120} - \frac{20324711m^8}{1036800} - \frac{1014978499m^9}{31104000}$$
$$- \frac{9546264328121m^{10}}{238878720000}$$

$$U_{-1} = -\frac{19m^2}{8} - \frac{10m^3}{3} - \frac{43m^4}{18} - \frac{28m^5}{27} - \frac{292693m^6}{165888} - \frac{738053m^7}{124416} - \frac{22461197m^8}{1866240}$$
$$- \frac{950625733m^9}{55987200} - \frac{72083221902403m^{10}}{2149908480000}$$

$$U_{-1} = -\frac{19m^2}{8} - \frac{10m^3}{3} - \frac{43m^4}{18} - \frac{28m^5}{27} - \frac{292693m^6}{165888} - \frac{738053m^7}{124416} - \frac{22461197m^8}{1866240}$$
$$- \frac{950625733m^9}{55987200} - \frac{7675497823457m^{10}}{429981696000}$$

$$U_2 = \frac{169m^4}{128} + \frac{479m^5}{96} + \frac{6143m^6}{720} + \frac{354899m^7}{43200} + \frac{12122443m^8}{4608000}$$

$$U_2 = \frac{169m^4}{128} + \frac{479m^5}{96} + \frac{6143m^6}{720} + \frac{354899m^7}{43200} + \frac{12122443m^8}{4608000} - \frac{122035633m^9}{11340000}$$
$$- \frac{2453358364771m^{10}}{57153600000}$$

$$U_{-2} = \frac{361m^4}{128} + \frac{4007m^5}{480} + \frac{15269m^6}{1200} + \frac{2759051m^7}{216000} + \frac{2701214713m^8}{207360000}$$

$$U_{-2} = \frac{361m^4}{128} + \frac{4007m^5}{480} + \frac{15269m^6}{1200} + \frac{2759051m^7}{216000} + \frac{2701212463m^8}{207360000} + \frac{5775890957m^9}{226800000}$$
$$+ \frac{16223813466421m^{10}}{285768000000}$$

$$U_3 = \frac{3185m^6}{2048}$$

$$U_3 = \frac{3185m^6}{2048} + \frac{19887m^7}{2560} + \frac{4924057m^8}{268800} + \frac{4535613997m^9}{169344000} + \frac{18807789518657m^{10}}{758661120000}$$

$$U_{-3} = -\frac{6539m^6}{2048}$$

$$U_{-3} = -\frac{6539m^6}{2048} - \frac{247773m^7}{17920} - \frac{57423407m^8}{1881600} - \frac{53420639383m^9}{1185408000} - \frac{33104764593493m^{10}}{590069760000}$$

All coefficients agree except $U_{1,10}$ and $U_{-1,10}$, (2 coefficients out of 25).

Now we check the plausibility of these values against the equation defining $U_i$ i.e., our (18) or the fourth equation on page 14 of Hill's numeric perigee (G.W. Hill 1886). However, since this equation includes a sum over $U_i$, and Hill's approximation for $U_3$ has a maximum order of $m^6$, we can only use it as a plausibility check up to $m^6$. Since the only differences between Hill's approximations and our calculations are at $U_{1,10}$ and $U_{-1,10}$, we would expect the plausibility to fail for $m^{10}$, but the test cannot be used that far.

For our version, all tests up to $m^{30}$ succeed.

## 9.7. Hill's numeric perigee page 16, $U_i$

Now, we calculate the results of the equation for $U_i$ on (G.W. Hill 1886) page 16. Here, we need to combine the coefficients of $\left(\zeta^{2i} + \zeta^{-2i}\right)$ because that is what corresponds to $U_i$, according to the discussion on page 14. We show a power series of $U_i$, first Hill's version (G.W. Hill 1886) page 16, then our version:

$$U_1 = \frac{9m^2}{8} + 3m^3 + \frac{7m^4}{2} + \frac{11m^5}{6} - \frac{38273m^6}{18432} - \frac{611123m^7}{69120} - \frac{20324711m^8}{1036800} - \frac{1014978499m^9}{31104000}$$

$$U_1 = \frac{9m^2}{8} + 3m^3 + \frac{7m^4}{2} + \frac{11m^5}{6} - \frac{38273m^6}{18432} - \frac{611123m^7}{69120} - \frac{20324711m^8}{1036800} - \frac{1014978499m^9}{31104000}$$

$$U_{-1} = -\frac{19m^2}{8} - \frac{10m^3}{3} - \frac{43m^4}{18} - \frac{28m^5}{27} - \frac{292693m^6}{165888} - \frac{738053m^7}{124416} - \frac{22461197m^8}{1866240} - \frac{950625733m^9}{55987200}$$

$$U_{-1} = -\frac{19m^2}{8} - \frac{10m^3}{3} - \frac{43m^4}{18} - \frac{28m^5}{27} - \frac{292693m^6}{165888} - \frac{738053m^7}{124416} - \frac{22461197m^8}{1866240} - \frac{950625733m^9}{55987200}$$

$$U_2 = \frac{169m^4}{128} + \frac{479m^5}{96} + \frac{6143m^6}{720} + \frac{354899m^7}{43200} + \frac{12122443m^8}{4608000} - \frac{122035633m^9}{11340000}$$

$$U_2 = \frac{169m^4}{128} + \frac{479m^5}{96} + \frac{6143m^6}{720} + \frac{354899m^7}{43200} + \frac{12122443m^8}{4608000} - \frac{122035633m^9}{11340000}$$

$$U_{-2} = \frac{361m^4}{128} + \frac{4007m^5}{480} + \frac{15269m^6}{1200} + \frac{2759051m^7}{216000} + \frac{2701212463m^8}{207360000} + \frac{5775890957m^9}{226800000}$$

$$U_{-2} = \frac{361m^4}{128} + \frac{4007m^5}{480} + \frac{15269m^6}{1200} + \frac{2759051m^7}{216000} + \frac{2701212463m^8}{207360000} + \frac{5775890957m^9}{226800000}$$

$$U_3 = \frac{3185m^6}{2048} + \frac{19887m^7}{2560} + \frac{4924057m^8}{268800} + \frac{4535613997m^9}{169344000}$$

$$U_3 = \frac{3185m^6}{2048} + \frac{19887m^7}{2560} + \frac{4924057m^8}{268800} + \frac{4535613997m^9}{169344000}$$

$$U_{-3} = -\frac{6539m^6}{2048} - \frac{247773m^7}{17920} - \frac{57423407m^8}{1881600} - \frac{53420639383m^9}{1185408000}$$

$$U_{-3} = -\frac{6539m^6}{2048} - \frac{247773m^7}{17920} - \frac{57423407m^8}{1881600} - \frac{53420639383m^9}{1185408000}$$

$$U_4 = \frac{60049m^8}{32768} + \frac{13808173m^9}{1204224}$$

$$U_4 = \frac{60049m^8}{32768} + \frac{13808173m^9}{1204224}$$

$$U_{-4} = \frac{387251m^8}{98304} + \frac{19858463m^9}{860160}$$

$$U_{-4} = \frac{387251m^8}{98304} + \frac{19858463m^9}{860160}$$

All of these values agree with our calculated values, thus confirming agreement of them with the approximation on page 16, up to $m^9$

### 9.8. Hill's numeric perigee page 16, $\theta_i$

Now, we calculate the results of the equation for $\theta_i$ on (G.W. Hill 1886) page 16. Here, we need to combine the coefficients of $(\zeta^{2i} + \zeta^{-2i})$ because that is what corresponds to $\theta_i$, according to the discussion on page 14, and when observing that $\theta_{-i} = \theta_i$. We show a power series of $\theta_i$, first Hill's version (Hill numeric perigee page 16), then our version:

Hill's version:

$$\theta_0 = 1 + 2m - \frac{m^2}{2} + \frac{255m^4}{32} + 19m^5 + \frac{80m^6}{3} + \frac{533m^7}{18} + \frac{4970221m^8}{221184} - \frac{164342761m^9}{6635520}$$

Our version:

$$\theta_0 = 1 + 2m - \frac{m^2}{2} + \frac{255m^4}{32} + 19m^5 + \frac{80m^6}{3} + \frac{533m^7}{18} + \frac{11230225m^8}{221184} + \frac{1576037m^9}{10368}$$

Hill's version:

$$\theta_1 = -\frac{15m^2}{2} - \frac{57m^3}{4} - 11m^4 - \frac{23m^5}{6} - \frac{49465m^6}{18432} - \frac{75757m^7}{27648} + \frac{302455m^8}{20736} + \frac{63139921m^9}{1244160}$$

Our version:

$$\theta_1 = -\frac{15m^2}{2} - \frac{57m^3}{4} - 11m^4 - \frac{23m^5}{6} - \frac{68803m^6}{4608} - \frac{1792417m^7}{27648} - \frac{7172183m^8}{51840} \\ - \frac{596404499m^9}{3110400}$$

Hill's version:

$$\theta_2 = \frac{197m^4}{32} + \frac{5783m^5}{240} + \frac{6199m^6}{150} + \frac{2327167m^7}{54000} + \frac{6870944459m^8}{207360000} + \frac{238224622589m^9}{14515200000}$$

Our version:

$$\theta_2 = \frac{111m^4}{16} + \frac{1397m^5}{64} + \frac{8807m^6}{240} + \frac{319003m^7}{7200} + \frac{126191191m^8}{1728000} + \frac{149693929741m^9}{725760000}$$

Here we have major differences. Hill's version disagrees with our version of $\theta_1$ in 4 coefficients out of 8. Hill's version disagrees with our version of $\theta_2$ in all 6 coefficients.

### 9.9. Hill's literal perigee page 34, $\theta_i$

Now, we calculate the results of the equation for $\theta_i$ on (G.W. Hill 1894) page 34. Here, we need to expand the prime factors. We show a power series of $\theta_i$, first Hill's version (Hill literal perigee page 34), then we convert the prime factors to explicit integers and calculate the numeric version (based on the earth's moon for $m$). Finally, we show our version:

Hill's version:

$$\theta_0 = 1 + 2m - \frac{m^2}{2} + 255m^4 \cdot 2^{-5} + 19m^5 + \frac{80m^6}{3} + \frac{533m^7 \cdot 3^{-2}}{2} + 11230225m^8 \cdot 2^{-13} \cdot 3^{-3} \\ + 1576037m^9 \cdot 2^{-7} \cdot 3^{-4} + 49359583m^{10} \cdot 2^{-9} \cdot 3^{-5} \\ + \frac{720508007m^{11} \cdot 2^{-8} \cdot 3^{-6}}{5}$$

$$\theta_0 = 1 + 2m - \frac{m^2}{2} + \frac{255m^4}{32} + 19m^5 + \frac{80m^6}{3} + \frac{533m^7}{18} + \frac{11230225m^8}{221184} + \frac{1576037m^9}{10368} \\ + \frac{49359583m^{10}}{124416} + \frac{720508007m^{11}}{933120}$$

$$\theta_0 = 1.0 + 2.0m - 0.5m^2 + 7.96875m^4 + 19.0m^5 + 26.6666666666667m^6 \\ + 29.6111111111111m^7 + 50.7732250072338m^8 + 152.009741512346m^9 \\ + 396.7301874357m^{10} + 772.149355924211m^{11}$$

$$\theta_0 = 1.1588439394757$$

Our version:

$$\theta_0 = 1 + 2m - \frac{m^2}{2} + \frac{255m^4}{32} + 19m^5 + \frac{80m^6}{3} + \frac{533m^7}{18} + \frac{11230225m^8}{221184} + \frac{1576037m^9}{10368}$$
$$+ \frac{49539583m^{10}}{124416} + \frac{720508007m^{11}}{933120}$$

Hill's version:

$$\theta_1 = -\frac{15m^2}{2} - 57m^3 \cdot 2^{-2} - 11m^4 - \frac{23m^5}{6} - 68803m^6 \cdot 2^{-9} \cdot 3^{-2} - 1792417m^7 \cdot 2^{-10} \cdot 3^{-3}$$
$$- \frac{7172183m^8 \cdot 2^{-7} \cdot 3^{-4}}{5} - 596404499m^9 \cdot 2^{-9} \cdot 3^{-5} \cdot 5^{-2}$$
$$- 2641291011773m^{10} \cdot 2^{-17} \cdot 3^{-6} \cdot 5^{-3}$$

$$\theta_1 = -\frac{15m^2}{2} - \frac{57m^3}{4} - 11m^4 - \frac{23m^5}{6} - \frac{68803m^6}{4608} - \frac{1792417m^7}{27648} - \frac{7172183m^8}{51840}$$
$$- \frac{596404499m^9}{3110400} - \frac{2641291011773m^{10}}{11943936000}$$

$$\theta_1 = 8.3724494170096$$
$$\cdot 10^{-11}m^2(-89579520000.0 - 170201088000.0m - 131383296000.0m^2$$
$$- 45785088000.0m^3 - 178337376000.0m^4 - 774324144000.0m^5$$
$$- 1652470963200.0m^6 - 2290193276160.0m^7 - 2641291011773.0m^8)$$

$$\theta_1 = -0.0570440180881171$$

Our version:

$$\theta_1 = -\frac{15m^2}{2} - \frac{57m^3}{4} - 11m^4 - \frac{23m^5}{6} - \frac{68803m^6}{4608} - \frac{1792417m^7}{27648} - \frac{7172183m^8}{51840}$$
$$- \frac{596404499m^9}{3110400} - \frac{2813929549973m^{10}}{11943936000}$$

Hill's version:

$$\theta_2 = 111m^4 \cdot 2^{-4} + 1397m^5 \cdot 2^{-6} + \frac{8807m^6 \cdot 2^{-4}}{15} + 319003m^7 \cdot 2^{-5} \cdot 3^{-2} \cdot 5^{-2}$$
$$+ 252382507m^8 \cdot 2^{-10} \cdot 3^{-3} \cdot 5^{-3}$$

$$\theta_2 = \frac{111m^4}{16} + \frac{1397m^5}{64} + \frac{8807m^6}{240} + \frac{319003m^7}{7200} + \frac{252382507m^8}{3456000}$$

$$\theta_2 = 2.89351851851852$$
$$\cdot 10^{-7}m^4(23976000.0 + 75438000.0m + 126820800.0m^2 + 153121440.0m^3$$
$$+ 252382507.0m^4)$$

$$\theta_2 = 0.000383200169831117$$

Our version:

$$\theta_2 = \frac{111m^4}{16} + \frac{1397m^5}{64} + \frac{8807m^6}{240} + \frac{319003m^7}{7200} + \frac{126191191m^8}{1728000} + \frac{149693929741m^9}{725760000}$$

Hill's version:

$$\theta_3 = -11669m^6 \cdot 2^{-9}$$

$$\theta_3 = -\frac{11669m^6}{512}$$

$$\theta_3 = -22.791015625 m^6$$

$$\theta_3 = -6.36516199613764 \cdot 10^{-6}$$

Our version:

$$\theta_3 = -\frac{11669 m^6}{512}$$

Here we have discrepancies in coefficients $\theta_{1,10}, \theta_{1,10}$ and $\theta_{2,8}$ (3 out of 26). Our values agree 100% with those in (Bourne 1972), (see 9.11. Bourne's calculations of coefficients).

## 9.10. Hill's literal perigee page 39, $c$

Here, we document the series for c, the anomalistic period of the earth's moon, as in (G.W. Hill 1894). Our version will be documented later, after a discussion of the methods used to solve the relevant equations.

Hill's version:

$$c = 1 + m - 3m^2 \cdot 2^{-2} - 201 m^3 \cdot 2^{-5} - 2367 m^4 \cdot 2^{-7} - 111749 m^5 \cdot 2^{-11} - \frac{4095991 m^6 \cdot 2^{-13}}{3}$$
$$- 332532037 m^7 \cdot 2^{-16} \cdot 3^{-2} - 15106211789 m^8 \cdot 2^{-18} \cdot 3^{-3}$$
$$- 5975332916861 m^9 \cdot 2^{-23} \cdot 3^{-4} - \frac{1547804933375567 m^{10} \cdot 2^{-25} \cdot 3^{-5}}{5}$$
$$- 818293211836767367 m^{11} \cdot 2^{-28} \cdot 3^{-6} \cdot 5^{-2}$$

$$c = 1 + m - \frac{3m^2}{4} - \frac{201 m^3}{32} - \frac{2367 m^4}{128} - \frac{111749 m^5}{2048} - \frac{4095991 m^6}{24576} - \frac{332532037 m^7}{589824}$$
$$- \frac{15106211789 m^8}{7077888} - \frac{5975332916861 m^9}{679477248} - \frac{1547804933375567 m^{10}}{40768634880}$$
$$- \frac{818293211836767367 m^{11}}{4892236185600}$$

$$c = 1.0 + m - 0.75 m^2 - 6.28125 m^3 - 18.4921875 m^4 - 54.56494140625 m^5$$
$$- 166.666300455729 m^6 - 563.781801011827 m^7 - 2134.28240020187 m^8$$
$$- 8794.01471417774 m^9 - 37965.5815783736 m^{10} - 167263.635849259 m^{11}$$

$$c = 1.07158336881014$$

Our version (to order 7):

$$c = 1 + m - \frac{3m^2}{4} - \frac{201 m^3}{32} - \frac{2367 m^4}{128} - \frac{111749 m^5}{2048} - \frac{4399741 m^6}{24576} - \frac{42332413 m^7}{65536} + O(m^8)$$

$$c = 1.07158387041564$$

Our version (to order 11):

$$c = 1.07157608626447$$

## 9.11. Bourne's calculations of coefficients

In (Bourne 1972), the coefficients $\bar{a}_{2,9}, \bar{a}_{-2,8}, \bar{a}_{-2,9}, \bar{a}_{-3,7}, \bar{a}_{-3,8}, \bar{a}_{-3,9}, \bar{a}_{4,9}, \bar{a}_{-4,9}$ are reported as being erroneous in (G.W. Hill 1878). We identified exactly the same coefficients (see 9.1. Hill's Calculations of $\bar{a}_i$ up to Order $m^9$). Our values are in complete agreement with those reported as correct in (Bourne 1972).

After this, the coefficients $\theta_{0,10}, \theta_{1,10}, \theta_{2,8}$, are reported as being erroneous in (G.W. Hill 1894) page 35. We identified exactly the same coefficients (see 9.9. Hill's literal perigee page 34, $\theta_i$). Our values are in complete agreement with those reported as correct in (Bourne 1972).

Now, after this check of our calculations of a number of series, we have devoted the next chapter to looking at the convergence of the series.

## 10. Hill's Series Convergence

Here, we will look at the convergence of the Hill series and other series defined for the purpose of calculating the intermediate orbit. As a basis for examining the radius of convergence of Hill's series, we take a closer look at seven orbits, calculated numerically, giving us useful input to the search for the radius of convergence.

According to (Petrovskaya 1963), the following statements about the radius of convergence of the Hill series have been proven:

Lyapunov (1896): $m = \sigma = 0.0808 \ldots$

Wintner (1929): $m \leq \frac{1}{12} = 0,0833 \ldots$

Merman (1952): $\frac{1}{7} = 0.14287 \ldots, 0.179$

Petrovskaya (1959): $|m| \leq 0.21$

Petrovskaya (1963): 0.21 and residual

(Wintner 1929) proves convergence for $0 < m < \frac{1}{12}$. This involves multiple estimates of majorants, mainly by subdividing the problem and applying absolute values to the polynomials. In this paper, he cites "Cauchy Majorantenmethode", Lindelöf, Hölder, Riemann, Lyapunov, Moulton (1906), Brown, Lindstedt-Poincaré, Weierstraß Doppenreihensatz, Schwarzes Lemma.

(Wintner 2014) – pg. 441 cites Lyapunov on topic of convergence and Wintner (1925) for a proof of existence, referring to §503-§515. The existence proof is §507-§515. Wintner (1925) proves existence (and convergence), similar to Wintner (1929), but without finding the value $\frac{1}{12}$. Note - §519 – for curve of zero velocity $m \approx 0.56096$. Proves existence in §507-§515, proves the recursive solution, power series, regular analytic, converges for $m$ in a "sufficiently small circle" (§507). Uses majorant §508. Cites Riemann's theory of trigonometric series (§511). §514: $|c_j(m)| \leq j^{-4}$ in this circle – always? §516 cites Hölder and Schwartz Lemma! §518 different convergence for $m < 0$ & $m > 0$? Euler transformation $m = \frac{m}{1-km}$. Potential surface: §471: $P_h, Z_h, N_h$ are points above, on, or below $U = -h(plane)$. §472: topology of $Z_h$. Lunar systems §489-§502, §497 Hill's region for lunar systems, $h = \frac{1}{2}C$.

We formulate a conjecture stating that the radius of convergence is equal to the value of m at the cusped orbit:

$$\hat{m} \approx 0.560958$$

## 10.1. Convergence of E, F, and G

We recall the definition of these functions $((4.37) - (4.39))$

$$E: \mathbb{Z}^2 \times \mathbb{R} \to \mathbb{R}, E_{j,i} := E(j, i, m)$$

$$F: \mathbb{Z} \times \mathbb{R} \to \mathbb{R}, F_j := F(j, m)$$

$$G: \mathbb{Z} \times \mathbb{R} \to \mathbb{R}, G_j := G(j, m)$$

$$E_{j,i} = -\frac{i(4ij - 4im - 4i + 4j^2 + 4jm + 4j + m^2 - 4m - 2)}{j(8j^2 + m^2 - 4m - 2)}$$

$$F_j = -\frac{3m^2(4j^2 - 4jm - 8j - 9m^2 - 8m - 2)}{16j^2(8j^2 + m^2 - 4m - 2)}$$

$$G_j = -\frac{3m^2(20j^2 - 20jm - 16j + 9m^2 + 8m + 2)}{16j^2(8j^2 + m^2 - 4m - 2)}$$

All three functions are the quotient of two second-order polynomials. These can be extended to include complex values of $m$:

$$E: \mathbb{Z}^2 \times \mathbb{C} \to \mathbb{C}, E_{j,i} := E(j, i, m)$$

$$F: \mathbb{Z} \times \mathbb{C} \to \mathbb{C}, F_j := F(j, m)$$

$$G: \mathbb{Z} \times \mathbb{C} \to \mathbb{C}, G_j := G(j, m)$$

The result is a meromorphic function of $m$. Now we know that the radius of convergence is determined by the nearest singularity, i.e., the zero of the denominator, see for example, (Ahlfors 1979). Theorem 3, page 179 and the subsequent discussion. For example, for $j = 1$, we want to know the solution of

$$8j^2 + m^2 - 4m - 2 = 0$$

$$m^2 - 4m + 6 = 0$$

solving the quadratic equation

$$m = \frac{-b \pm \sqrt{b^2 - 4ac}}{2a} = \frac{4 \pm \sqrt{16 - 24}}{2} = 2 \pm \sqrt{-2} = 2 \pm \sqrt{2}\mathrm{i}$$

$$|m| = \sqrt{6}$$

This gives us[66],[67]

| $j$ | radius of convergence of $E, F, G$ |
|---|---|
| $\pm 1$ | $\sqrt{6} \approx 2.449$ |
| $\pm 2$ | $\sqrt{30} \approx 5.477$ |
| $\pm 3$ | $\sqrt{70} \approx 8.366$ |

---

[66] These values agree with our numerical calculations (see Fig. 10.1).
[67] The values of the denominators, and their meaning for the approximations, are mentioned in (G.W. Hill 1878) page 138-139, but the radius of convergence is not discussed.

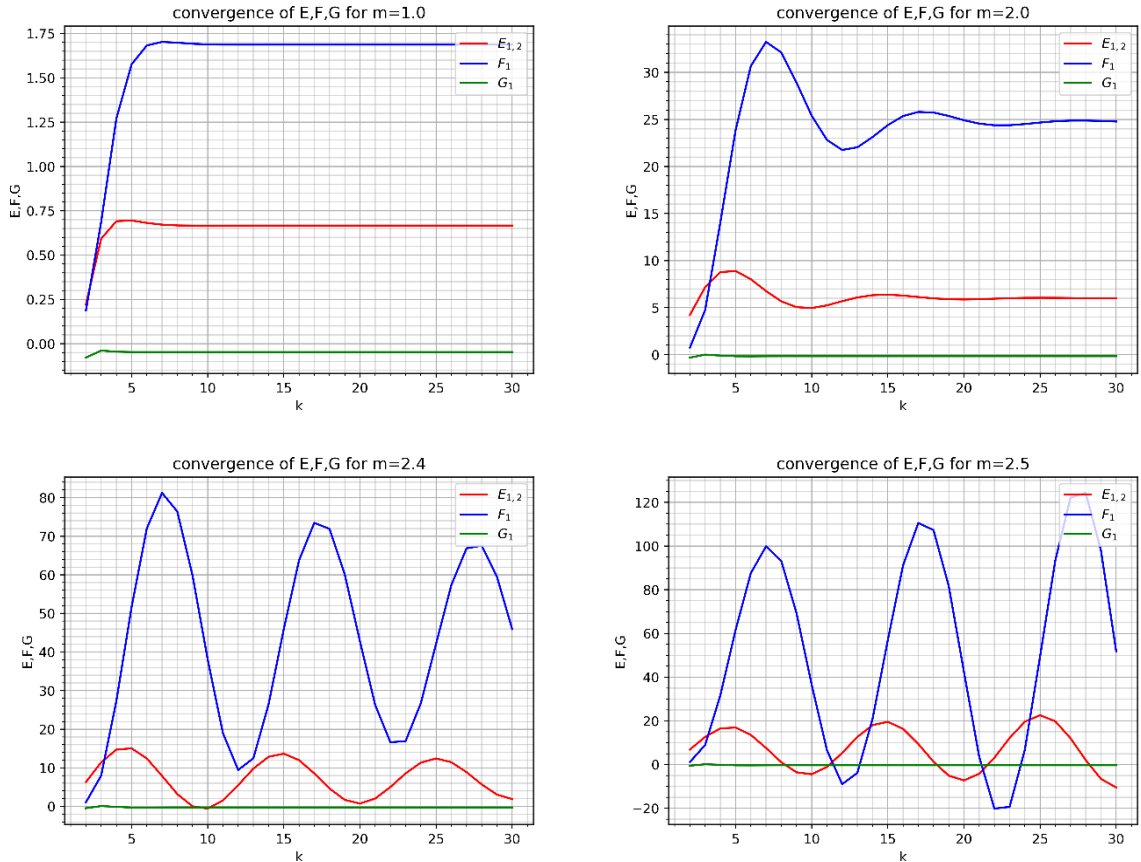

Fig. 10.1. Convergence of $E_{j,i}$, $F_j$ and $F_j$ for $j = 1$ and a few values of $m$.

This convergence behavior is illustrative of things we see elsewhere in the power series we are investigating in this paper. The series converges (diverges) for values of $|m|$ less than (greater than) the radius of convergence (here $\sqrt{6} \approx 2.449$), and the behavior on the boundary is unknown. The rate of convergence is good (useful in practice) when it is far from the radius of convergence, but painfully slow when it is close.

### 10.2. Hill's Regions

Hill's regions identify the points of zero velocity, and are useful in analyzing their respective orbits.

We follow (Frauenfelder and van Koert 2018) section 5.5 and 5.8.2.
The Hamiltonian is (4.1)

$$H = \frac{1}{2}((p_1 - q_2)^2 + (p_2 - q_1)^2) - \frac{1}{|q|} - \frac{3}{2}q_1^2$$

where

$$\dot{q}_1 = p_1 - q_2$$
$$\dot{q}_2 = p_2 - q_1$$

The energy (Jacobian integral) is (4.1)

$$C = \frac{1}{2}(\dot{q}_1^2 + \dot{q}_2^2) - \frac{1}{|q|} - \frac{3}{2}q_1^2$$

This can be split into kinetic energy and potential energy.
The effective potential of Hill's lunar Problem is

$$U(q) = -\frac{1}{|q|} - \frac{3}{2}q_1^2$$

The Lagrange points are the critical points of $U$ and are located at

$$\text{crit}(U) = \left\{ \left(3^{-\frac{1}{3}}, 0\right), \left(-3^{-\frac{1}{3}}, 0\right) \right\}$$

where

$$3^{-\frac{1}{3}} \approx 0.6933612744$$

They have an energy of

$$C = -\frac{3^{\frac{4}{3}}}{2}$$

where

$$-\frac{3^{\frac{4}{3}}}{2} \approx -2.1633743555$$

The footprint projection is

$$\pi: T^*(\mathbb{R}^2 \setminus \{0\}) \to \mathbb{R}^2 \setminus \{0\} : (q, p) \mapsto q$$

The three-dimensional hypersurface of constant energy is

$$\Sigma_C = H^{-1}(C)$$

The boundary of Hill's region for energy $C$ is

$$\partial \mathfrak{K}_C = \pi(\Sigma_C) = \{q \in \mathbb{R}^2 \setminus \{0\} : U(q) = C\}$$

Next, we'll look at some of the details for calculating the points on the boundary of a Hill's region.

$$U(q) = C$$

$$-\frac{1}{|q|} - \frac{3}{2} q_1^2 = C$$

$$\frac{1}{|q|} + \frac{3}{2} q_1^2 + C = 0$$

Multiply by $|q|$.

$$\left(\frac{3}{2} q_1^2 + C\right) |q| + 1 = 0$$

$$\left(\frac{3}{2} q_1^2 + C\right) |q| = -1$$

Square both sides.

$$\left(C^2 + 3C q_1^2 + \frac{9}{4} q_1^4\right)(q_1^2 + q_2^2) = 1$$

$$\left(C^2 + 3C q_1^2 + \frac{9}{4} q_1^4\right) q_2^2 + \left(C^2 + 3C q_1^2 + \frac{9}{4} q_1^4\right) q_1^2 - 1 = 0$$

$$a q_2^2 + b q_2 + c = 0$$

where

$$a = C^2 + 3C q_1^2 + \frac{9}{4} q_1^4$$

$$b = 0$$

$$c = \left(C^2 + 3Cq_1{}^2 + \frac{9}{4}q_1{}^4\right)q_1^2 - 1$$

$$q_2 = \frac{-b \pm \sqrt{b^2 - 4ac}}{2a} = \pm\frac{\sqrt{-4ac}}{2a}$$

Now we can put this into software and plot some Hill's regions.

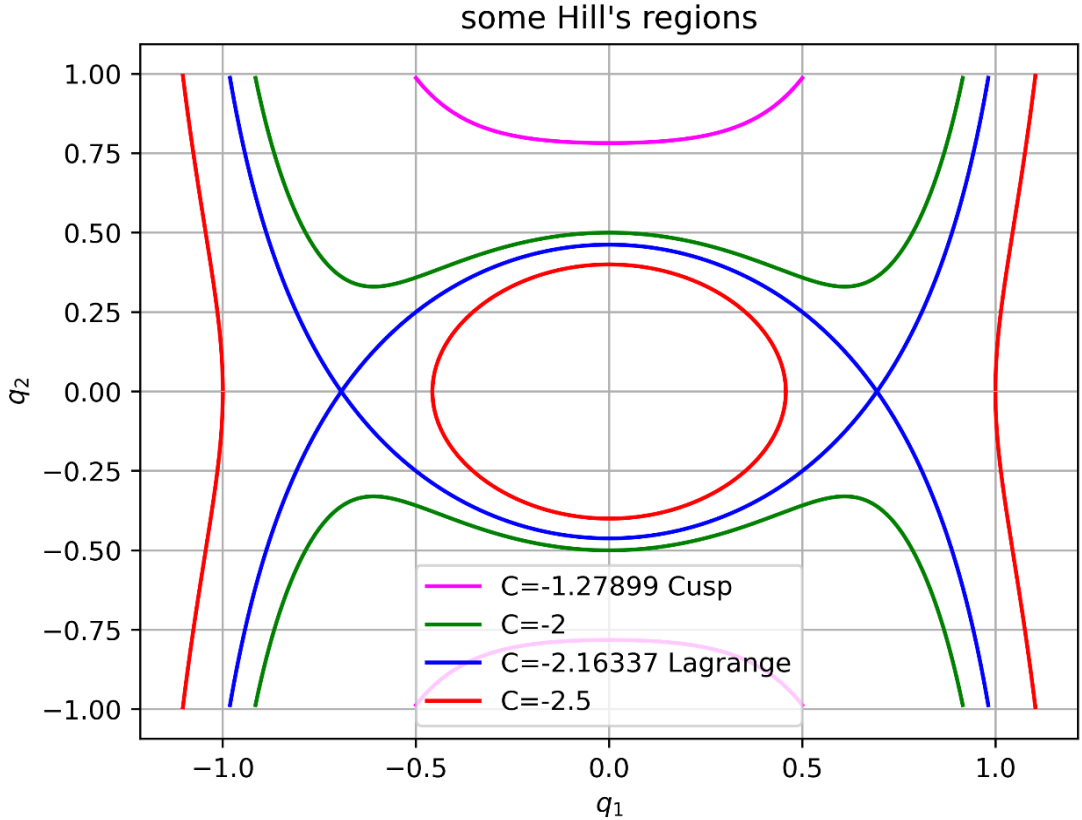

Fig. 10.2. Some Hill's Regions

One of Hill's innovations is the examination of curves and surfaces of zero velocity, which can be used to show that the orbit of the moon cannot escape from the Earth(G.W. Hill 1878), page 23, (Barrow-Green 1997), page 24, and (Szebehely 1963). Since the Earth's moon has an energy of $C = -3.25444$, the corresponding Hill's region is inside of the red oval in the figure, so we can conclude that the moon cannot leave this region.

### 10.3. Equations

Here, we look at some equations that will be used to simulate orbits. We have taken the idea for making these calculations and plots from (Wintner 2014) §238-§240 (see below). First, we note a few formulas that will be useful for the cusped orbit and some orbits close to it.

#### 10.3.1. Derivatives

From (3.4), we have

$$\ddot{q}_1 = 2\dot{q}_2 + 3q_1 - \frac{q_1}{|q|^3} \tag{10.3.1.1}$$

$$\ddot{q}_2 = -2\dot{q}_1 - \frac{q_2}{|q|^3} \tag{10.3.1.2}$$

We take the derivative of $\ddot{q}_1$

$$\dddot{q}_1 = 2\ddot{q}_2 + 3\dot{q}_1 - \frac{\dot{q}_1}{|q|^3} + \frac{3}{2}\frac{q_1}{|q|^5}(2q_1\dot{q}_1 + 2q_2\dot{q}_2) \tag{10.3.1.3}$$

and replace $\ddot{q}_2$ by the right-hand side of the ODE

$$\dddot{q}_1 = -4\dot{q}_1 - 2\frac{q_2}{|q|^3} + 3\dot{q}_1 - \frac{\dot{q}_1}{|q|^3} + 3\frac{q_1}{|q|^5}(q_1\dot{q}_1 + q_2\dot{q}_2)$$

$$\dddot{q}_1 = -\dot{q}_1 - 2\frac{q_2}{|q|^3} - \frac{\dot{q}_1}{|q|^3} + 3\frac{q_1}{|q|^5}(q_1\dot{q}_1 + q_2\dot{q}_2) \tag{10.3.1.4}$$

Then we take the derivative of $\ddot{q}_2$

$$\dddot{q}_2 = -2\ddot{q}_1 - \frac{\dot{q}_2}{|q|^3} + \frac{3}{2}\frac{q_2}{|q|^5}(2q_1\dot{q}_1 + 2q_2\dot{q}_2) \tag{10.3.1.5}$$

and replace $\ddot{q}_1$ by the right-hand side of the ODE

$$\dddot{q}_2 = -4\dot{q}_2 - 6q_1 + 2\frac{q_1}{|q|^3} - \frac{\dot{q}_2}{|q|^3} + 3\frac{q_2}{|q|^5}(q_1\dot{q}_1 + q_2\dot{q}_2) \tag{10.3.1.6}$$

This will provide the basis for a Taylor expansion. It can be done indefinitely: calculate the next derivative and replace the second-order terms by the ODEs. Using software, we will calculate this to order 12 and plot the resulting orbits.

### 10.3.2. Taylor Expansion of ODEs

Now we look at a Taylor expansion of $q_1$ and $q_2$ (based on (10.3.1.1) and (10.3.1.2)).

$$q_1 = q_1(0) + \dot{q}_1(0)t + \frac{1}{2}\ddot{q}_1(0)t^2 + \frac{1}{6}\dddot{q}_1(0)t^3 + \mathcal{O}(t^4) \tag{10.3.2.1}$$

substituting $\ddot{q}_1$ and $\dddot{q}_1$

$$q_1 = q_1(0) + \dot{q}_1(0)t + \frac{1}{2}\left(2\dot{q}_2(0) + 3q_1(0) - \frac{q_1(0)}{|q(0)|^3}\right)t^2$$

$$+\frac{1}{6}\left(-\dot{q}_1(0) - 2\frac{q_2(0)}{|q(0)|^3} - \frac{\dot{q}_1(0)}{|q(0)|^3} + 3\frac{q_1(0)}{|q(0)|^5}(q_1(0)\dot{q}_1(0) + q_2(0)\dot{q}_2(0))\right)t^3$$

$$+\mathcal{O}(t^4) \tag{10.3.2.2}$$

$$q_2 = q_2(0) + \dot{q}_2(0)t + \frac{1}{2}\ddot{q}_2(0)t^2 + \frac{1}{6}\dddot{q}_2(0)t^3 + \mathcal{O}(t^4) \tag{10.3.2.3}$$

$$q_2 = q_2(0) + \dot{q}_2(0)t + \frac{1}{2}\left(-2\dot{q}_1(0) - \frac{q_2(0)}{|q(0)|^3}\right)t^2$$

$$+\frac{1}{6}\begin{pmatrix}-4\dot{q}_2(0) - 6q_1(0) + 2\frac{q_1(0)}{|q(0)|^3} - \frac{\dot{q}_2(0)}{|q(0)|^3} + \\ 3\frac{q_2(0)}{|q(0)|^5}(q_1(0)\dot{q}_1(0) + q_2(0)\dot{q}_2(0))\end{pmatrix}t^3 + \mathcal{O}(t^4) \tag{10.3.2.4}$$

### 10.3.3. Taylor Expansion of Hill Series

We also look at a Taylor expansion of the orbits based on Hill Series. We begin with the values of in terms of the coefficients of the Hill Series (our (4.9) and (4.11))

$$q_1 = \sum_{j=-\infty}^{\infty} a_j \cos\left(\frac{2j+1}{m}t\right) \tag{10.3.3.1}$$

$$q_2 = \sum_{j=-\infty}^{\infty} a_j \sin\left(\frac{2j+1}{m}t\right) \tag{10.3.3.2}$$

First, we calculate the Taylor expansion around the point at the right, where the angle is $\left(\frac{2j+1}{m}t\right) = 0$, and $q_2 = \dot{q}_1 = 0$.

$$q_1 = \sum_{j=-\infty}^{\infty} a_j - 0t - \frac{1}{2!}\left[\sum_{j=-\infty}^{\infty} a_j \left(\frac{2j+1}{m}\right)^2\right] t^2 + 0t^3 + \frac{1}{4!}\left[\sum_{j=-\infty}^{\infty} a_j \left(\frac{2j+1}{m}\right)^4\right] t^4 + \cdots$$

$$q_1 = \sum_{n=0}^{\infty} (-1)^n \frac{1}{(2n)!} \left[\sum_{j=-\infty}^{\infty} a_j \left(\frac{2j+1}{m}\right)^{2n}\right] t^{2n} \tag{10.3.3.3}$$

$$q_2 = 0 + \left[\sum_{j=-\infty}^{\infty} a_j \left(\frac{2j+1}{m}\right)^1\right] t - 0t^2 - \frac{1}{3!}\left[\sum_{j=-\infty}^{\infty} a_j \left(\frac{2j+1}{m}\right)^3\right] t^3 + \cdots$$

$$q_2 = \sum_{n=0}^{\infty} (-1)^n \frac{1}{(2n+1)!} \left[\sum_{j=-\infty}^{\infty} a_j \left(\frac{2j+1}{m}\right)^{(2n+1)}\right] t^{(2n+1)} \tag{10.3.3.4}$$

Next, we calculate the Taylor expansion around the point at the top, where the angle is $\left(\frac{2j+1}{m}t\right)\Big|_{j=0} = \frac{\pi}{2}$, and $q_1 = \dot{q}_2 = 0$.

$$q_1 = 0 - \left[\sum_{j=-\infty}^{\infty} (-1)^j a_j \left(\frac{2j+1}{m}\right)^1\right] t - 0t^2 + \frac{1}{3!}\left[\sum_{j=-\infty}^{\infty} (-1)^j a_j \left(\frac{2j+1}{m}\right)^3\right] t^3 + \cdots$$

$$q_1 = -\sum_{n=0}^{\infty} (-1)^n \frac{1}{(2n+1)!} \left[\sum_{j=-\infty}^{\infty} (-1)^j a_j \left(\frac{2j+1}{m}\right)^{(2n+1)}\right] t^{(2n+1)} \tag{10.3.3.5}$$

$$q_2 = \sum_{j=-\infty}^{\infty} (-1)^j a_j + 0t - \frac{1}{2!}\left[\sum_{j=-\infty}^{\infty} (-1)^j a_j \left(\frac{2j+1}{m}\right)^2\right] t^2 - 0t^3 + \frac{1}{4!}\left[\sum_{j=-\infty}^{\infty} (-1)^j a_j \left(\frac{2j+1}{m}\right)^4\right] t^4$$
$$+ \cdots$$

$$q_2 = \sum_{n=0}^{\infty} (-1)^n \frac{1}{(2n)!} \left[\sum_{j=-\infty}^{\infty} (-1)^j a_j \left(\frac{2j+1}{m}\right)^{2n}\right] t^{2n} \tag{10.3.3.6}$$

The $a_j$ are power series in $m$. Note that the factor of $(-1)^j$ in the formulas for the point at the top come from the fact that increasing $j$ by 1 increases $\left(\frac{2j+1}{m}t\right)$ by $\pi$, which changes the sign.

## 10.4. Selected Orbits

We have chosen six orbits in the family $g$ and investigate their behavior. In section 10.4.4, we pay special attention to the cusped orbit. Part of this is based on (Wintner 2014), §238 and §240, which derive the differential equations near the point of zero velocity and show the shape of the orbit there. However, Wintner's discussion is more general than ours, and applies to the general case of two degrees of freedom. Here are some of Wintner's observations: The cusp is on the curve of zero velocity (§168). The curve of zero velocity has a definite normal (§166). The curve of zero velocity reflects the path in the transverse direction (§170). The path has a tangent and no cusp unless the point is an equilibrium point (§166).

   Our discussion is restricted to the case of Hill's lunar problem, which makes it much simpler and more explicit, because all of the equations can be deduced directly from the defining ODE. In the plots, the orbit is based on our numerical solution of the ODE, and the approximation is an analytical calculation based on the approximation of the ODE near the point of zero velocity. We also choose two more pre- and post-cusped orbits that are not included in (Hénon 1969) Fig. 4, but are included in Table 4 and are closer to the cusped orbit.

We will begin by using manual calculations to reproduce Wintner's results and plot partial orbits up to order 3 of the Taylor expansion. Then, we use software to do the calculations and plot orbits up to order 12 of the Taylor expansion. Finally, we plot the convergence behavior of the expansion ($q_1$ and $q_2$ vs. power of $t$ and $\bar{\bar{a}}_{\pm 1}$ vs. $k$ (power of $m$).

For the moment, we know of the existence of the cusped orbit due to numerical solutions of the differential equations. In order to look at the orbit near the cusp, we only need to assume that the orbit touches the point of zero velocity. Then, we will look at the path near the point of zero velocity and set the time to $t = 0$ at that point.

Here, we introduce some shorthand notation:

$$T = 2\pi m \quad (10.4.1)$$

$$Q := q_2(0) \neq 0 \quad (10.4.2)$$

Next, we look at orbits other than the cusped orbit, but begin with the point closest to the Hill region (points of zero velocity). Here, we can no longer assume that the orbit passes through a point of zero velocity. However, we assume that it crosses the $q_2$ axis (perpendicularly due to symmetry), and $\dot{q}_1$ is small at that point. We define this point as the origin of our coordinates, i.e., $t = 0$. At this point, the orbit crosses the $q_2$ axis perpendicularly, so $\dot{q}_2(0) = 0$.

At this point, based on the Newtonian equations, we have

$$q_1(0) = 0 \quad (10.4.3)$$

$$\dot{q}_1(0) \neq 0 \quad (10.4.4)$$

$$\ddot{q}_1(0) = 0 \quad (10.4.5)$$

$$\dddot{q}_1(0) = -\dot{q}_1(0) - \frac{2}{Q^2} - \dot{q}_1(0)\frac{1}{Q^3}$$

$$\dddot{q}_1(0) = -\dot{q}_1(0)\left(1 + \frac{1}{Q^3}\right) - \frac{2}{Q} \quad (10.4.6)$$

$$q_2(0) = Q$$

$$\dot{q}_2(0) = 0 \quad (10.4.7)$$

$$\ddot{q}_2(0) = -2\dot{q}_1(0) - \frac{1}{Q^2} \quad (10.4.8)$$

$$\dddot{q}_2(0) = 0 \quad (10.4.9)$$

Now we define

$$\alpha := \frac{1}{2}\frac{1}{Q} \quad (10.4.10)$$

$$\beta := \alpha \quad (10.4.11)$$

$$\gamma := \frac{1}{6}\left(1 + \frac{1}{Q^3}\right) \quad (10.4.12)$$

Now we can insert the initial conditions into the Taylor expansion

$$q_1 = q_1(0) + \dot{q}_1(0)t + \frac{1}{2}\ddot{q}_1(0)t^2 + \frac{1}{6}\left[-\dot{q}_1(0)\left(1 + \frac{1}{Q^3}\right) - \frac{2}{Q}\right]t^3 + \mathcal{O}(t^4)$$

$$q_1 = 0 + \dot{q}_1(0)t + 0 - \left(\dot{q}_1(0)\gamma + \frac{2}{3}\alpha\right)t^3 + \mathcal{O}(t^4) \quad (10.4.13)$$

our version:

$$q_2 = Q + 0 + \frac{1}{2}\left(-2\dot{q}_1(0) - \frac{1}{Q^2}\right)t^2 + \mathcal{O}(t^4)$$

$$q_2 = Q - (\dot{q}_1(0) + \alpha)t^2 + \mathcal{O}(t^4) \tag{10.4.14}$$

Wintner's version[68]:

$$q_2 = Q + (\dot{q}_1(0) + \alpha)t^2 + \dot{q}_1(0)\beta t^3 + \mathcal{O}(t^4) \tag{10.4.15}$$

### 10.4.1. Round Orbit $g01$ $C = -4.0$ $m = 0.054165202$

When $C$ approaches $-\infty$, we expect the orbit in family g to approach a circular shape. In our numerical solution, the orbit looks very round, and we estimate an elliptical shape with semi-axes of

$$a = 0.13858$$
$$b = 0.13772$$

For the first plot, we simply take the average of these values. For the round orbit, based on the numeric solution of the Newtonian equations, we have the following values:

$$name = g01 \tag{10.4.1.1}$$
$$C = -4.0 \tag{10.4.1.2}$$
$$a = b = 0.13815 \tag{10.4.1.3}$$
$$m = 0.054165202 \tag{10.4.1.4}$$
$$T = 0.34033 \tag{10.4.1.5}$$

At the top:

$$q_1(0) = 0 \tag{10.4.1.6}$$
$$\dot{q}_1(0) = p_1(0) + q_2(0) = -2.674687 + 0.1385826 = -2.5361044 \tag{10.4.1.7}$$
$$q_2(0) = 0.1385826 \tag{10.4.1.8}$$
$$\dot{q}_2(0) \tag{10.4.1.9}$$

This results in the following equations (up to order 12):

$$q_1 = -2.5361t + 141.88t^3 - 2087.2t^5 - 6233.9t^7 + 9.0323 \cdot 10^5 t^9$$
$$-1.8868 \cdot 10^7 t^{11} \tag{10.4.1.10}$$

$$q_2 = 0.13858 - 23.499t^2 + 637.31t^4 - 4234.6t^6 - 1.2765 \cdot 10^5 t^8$$
$$+4.6899 \cdot 10^6 t^{10} - 4.912 \cdot 10^7 t^{12} \tag{10.4.1.11}$$

On the right:

$$q_1(0) = 0.13772$$
$$\dot{q}_1(0) = 0$$
$$q_2(0) = 0$$
$$\dot{q}_2(0) = 2.565$$

This results in the following equations (up to order 12):

$$q_1 = 0.13772 - 23.59t^2 + 701.06t^4 - 11215.0t^6 + 2.585 \cdot 10^5 t^8$$
$$-8.6712 \cdot 10^6 t^{10} + 2.8538 \cdot 10^8 t^{12} \tag{10.4.1.12}$$

$$q_2 = 2.565t - 147.93t^3 + 2867.8t^5 - 49639.0t^7 + 1.486 \cdot 10^6 t^9$$
$$-4.9962 \cdot 10^7 t^{11} \tag{10.4.1.13}$$

---

[68] (10.4.10) − (10.4.12) and (10.4.15) use Wintner's notation.

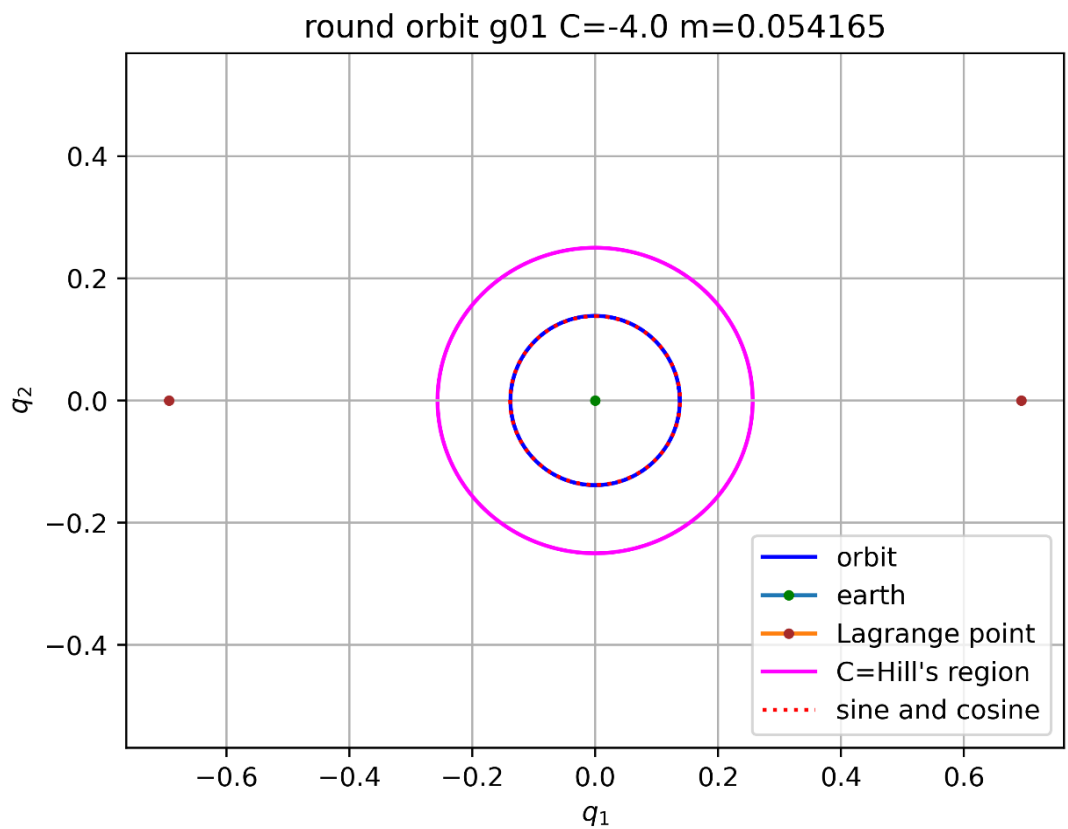

Fig. 10.3. The Round Orbit $g01\ C = -4.0\ m = 0.054165$ (plot assuming circular shape)

Now we plot the approximation of this orbit using the Taylor expansion.

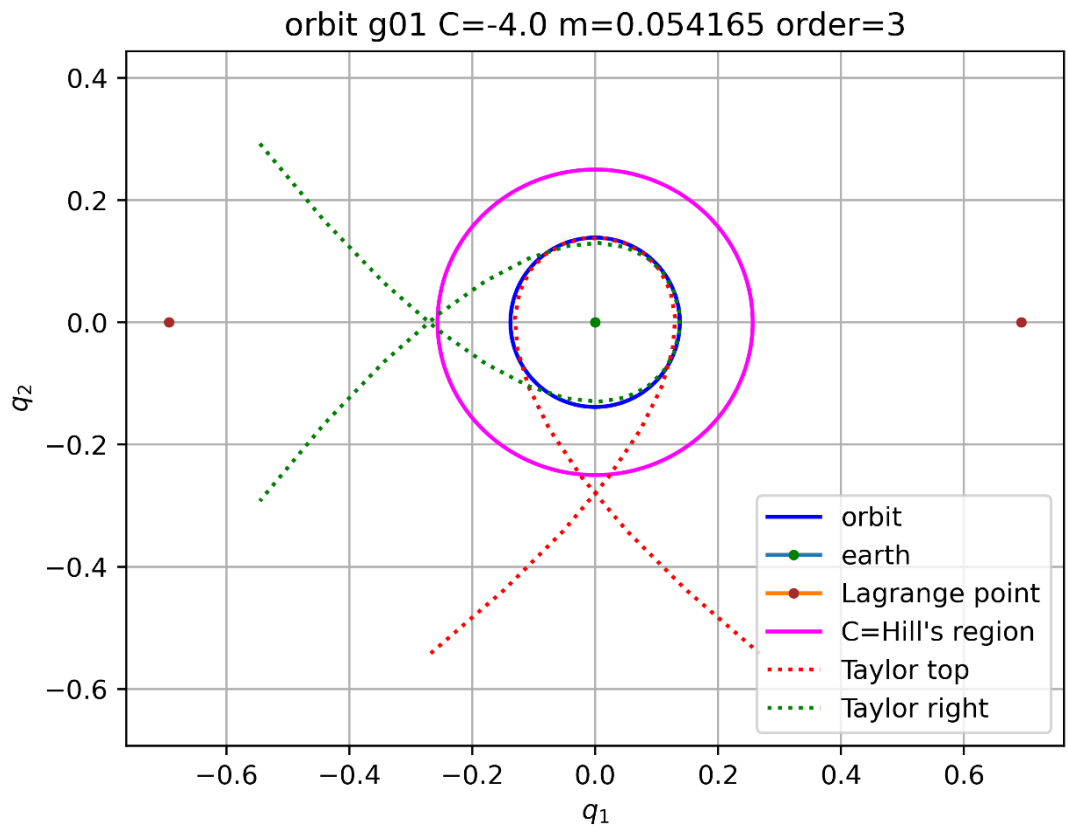

Fig. 10.4. The Round Orbit $g01$ $C = -4.0$ $m = 0.054165$ (plot based on Taylor expansion to order 3)

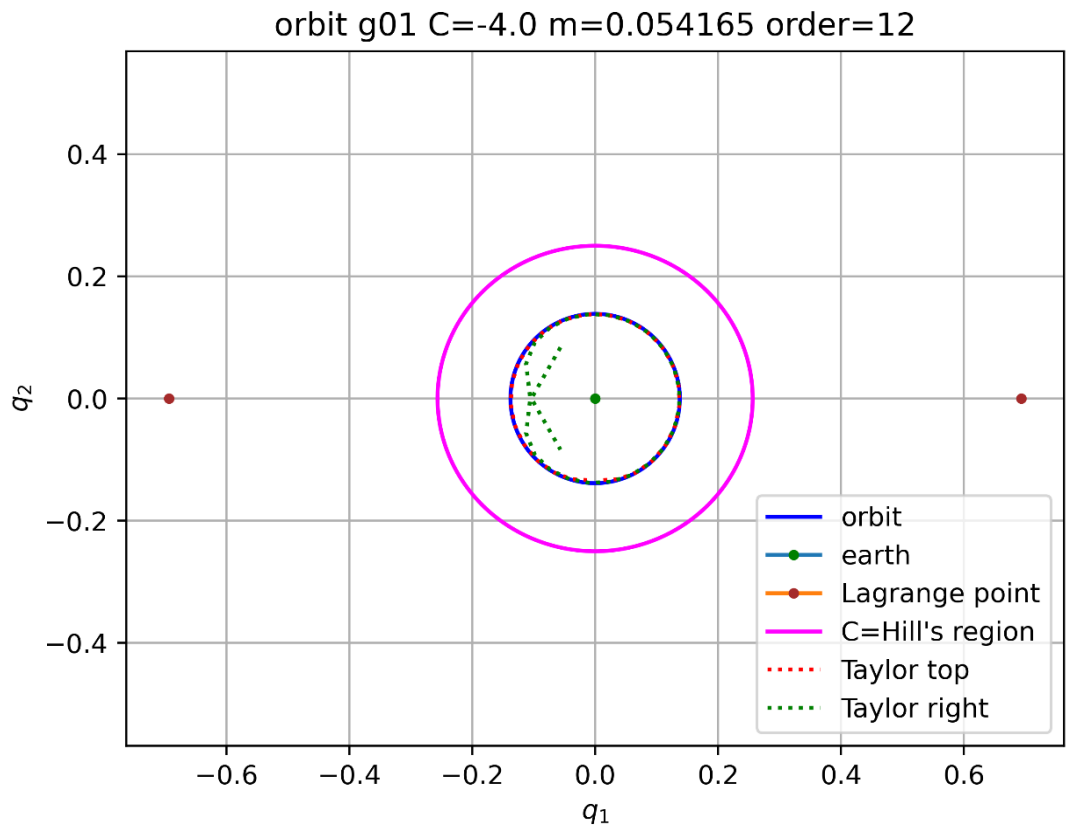

Fig. 10.5. The Round Orbit $g01\ C = -4.0\ m = 0.054165$ (plot based on Taylor expansion to order 12)

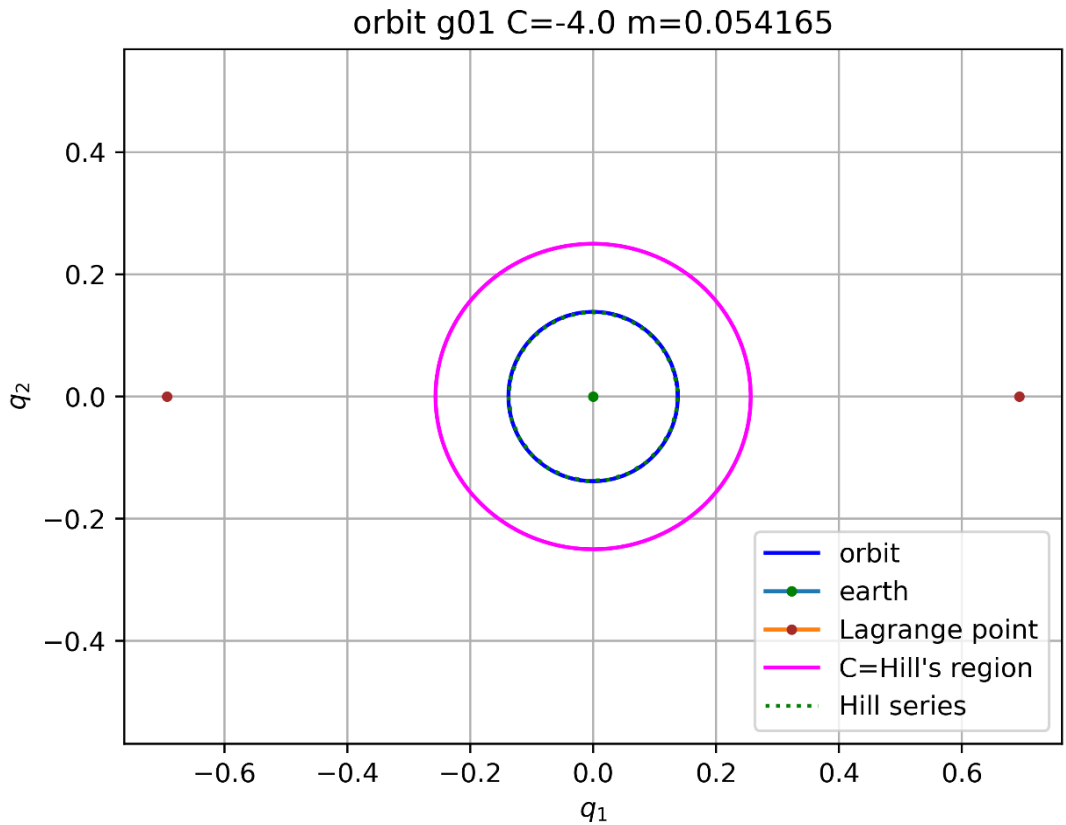

Fig. 10.6. The Round Orbit $g01$ $C = -4.0$ $m = 0.054165$ (plot based on Hill series)

### 10.4.2. Orbit of earth's moon $g04$ $C = -3.25444$ $m = 0.080849$

For the orbit of the earth's moon (our g04), based on the numeric solution of the Newtonian equations, we have the following values:

$$name = g04 \qquad (10.4.2.1)$$
$$C = -3.25444 \qquad (10.4.2.2)$$
$$m = 0.080849 \qquad (10.4.2.3)$$
$$T = 0.507989 \qquad (10.4.2.4)$$

At the top:

$$q_1(0) = 0 \qquad (10.4.2.5)$$
$$\dot{q}_1(0) = p_1(0) + q_2(0) = -2.34 + 0.178644 = -2.161356 \qquad (10.4.2.6)$$
$$q_2(0) = 0.178644 \qquad (10.4.2.7)$$
$$\dot{q}_2(0) = 0 \qquad (10.4.2.8)$$

This results in the following equations (up to order 12):

$$q_1 = -2.1614t + 53.1t^3 - 259.7t^5 - 3164.3t^7 + 59741.0t^9 - 78867.0t^{11} \qquad (10.4.2.9)$$

$$q_2 = 0.17864 - 13.506t^2 + 151.96t^4 + 76.067t^6 - 16717.0t^8 + 1.4409 \cdot 10^5 t^{10}$$
$$+ 1.5221 \cdot 10^6 t^{12} \qquad (10.4.2.10)$$

At the right:

$$q_1(0) = 0.176097 \qquad (10.4.2.11)$$
$$\dot{q}_1(0) = 0 \qquad (10.4.2.12)$$

$$q_2(0) = 0 \quad (10.4.2.13)$$
$$\dot{q}_2(0) = 2.223 \quad (10.4.2.14)$$

This results in the following equations (up to order 12):

$$q_1 = 0.1761 - 13.637t^2 + 193.38t^4 - 1934.7t^6 + 32355.0t^8 - 6.44 \cdot 10^5 t^{10} \\ + 1.2888 \cdot 10^7 t^{12} \quad (10.4.2.15)$$

$$q_2 = 2.223t - 58.756t^3 + 597.48t^5 - 7499.5t^7 + 1.4452 \cdot 10^5 t^9 - 2.8662 \cdot 10^6 t^{11} \quad (10.4.2.16)$$

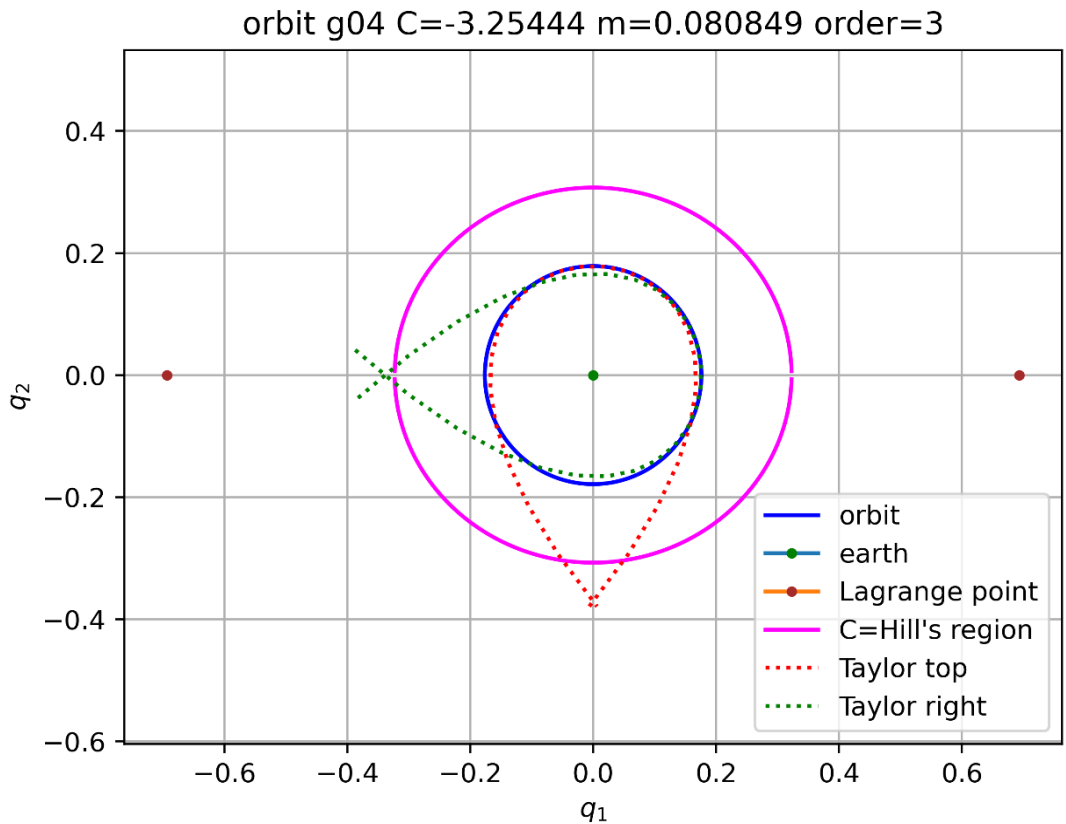

Fig. 10.7. Orbit of earth's moon $g04\ C = -3.25444\ m = 0.080849$ (plot based on Taylor expansion to order 3)

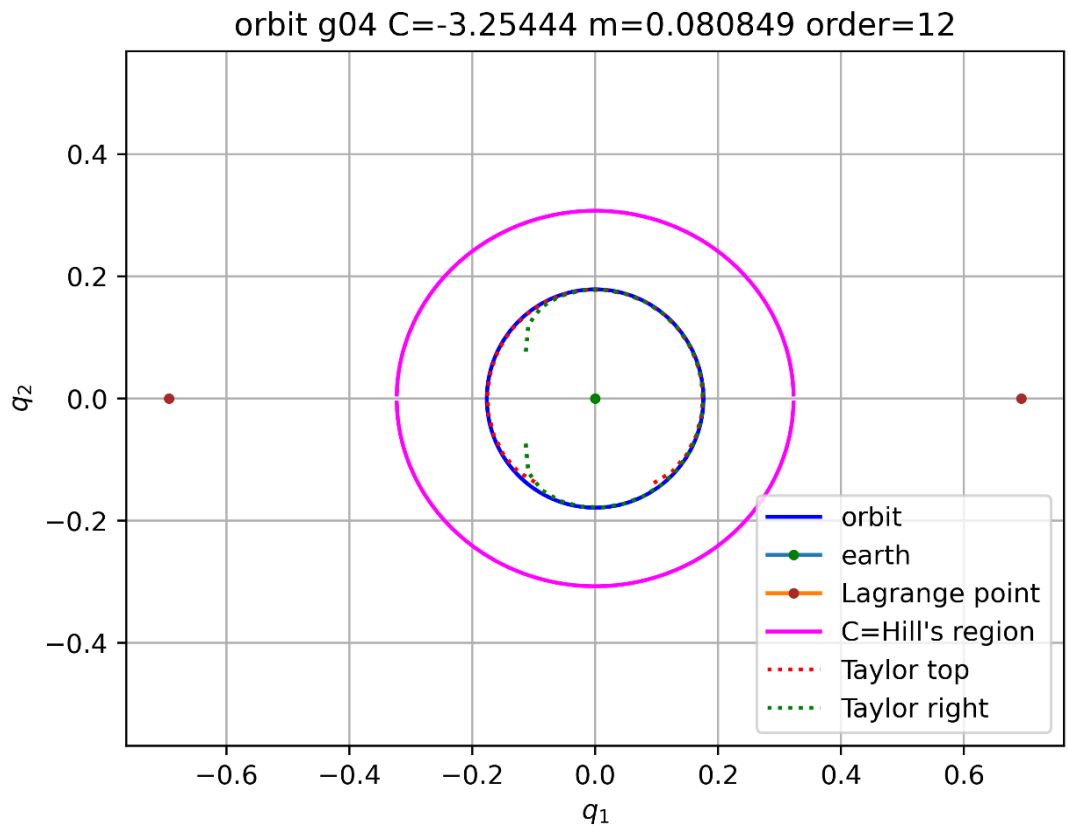

Fig. 10.8. Orbit of earth's moon $g04\ C = -3.25444\ m = 0.080849$ (plot based on Taylor expansion to order 12)

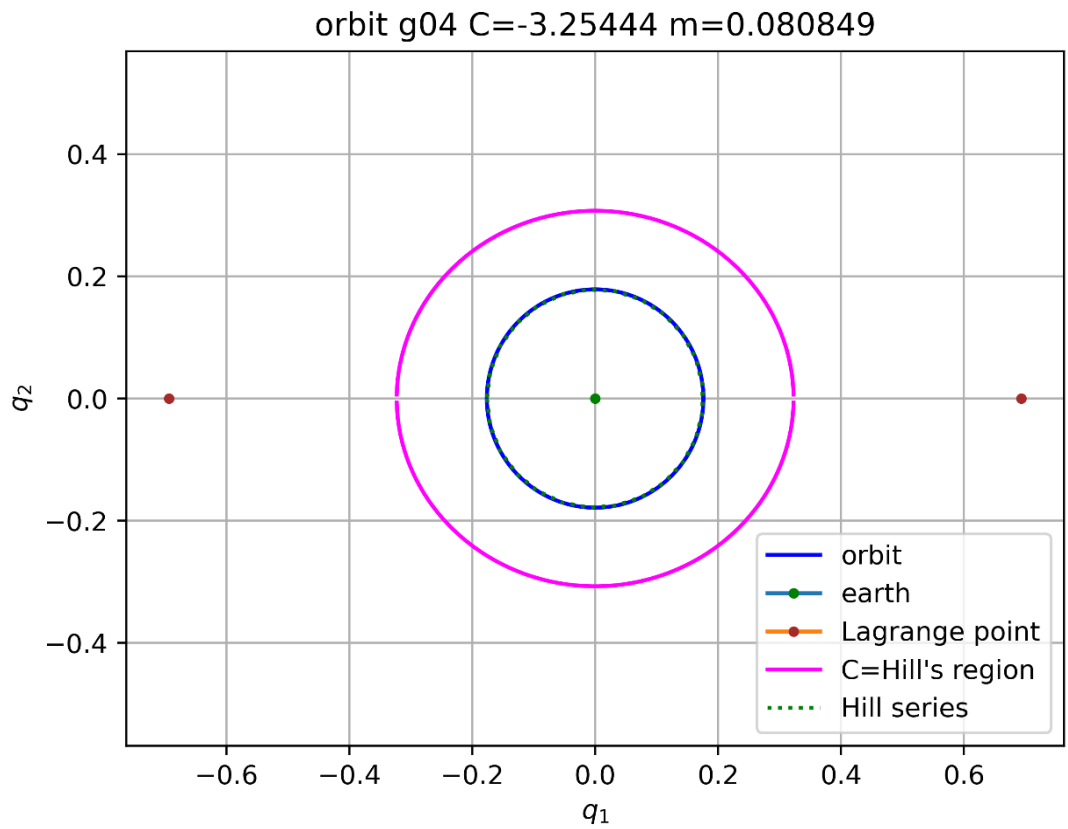

Fig. 10.9. Orbit of earth's moon $g04\ C = -3.25444\ m = 0.080849$ (plot based on Hill series)

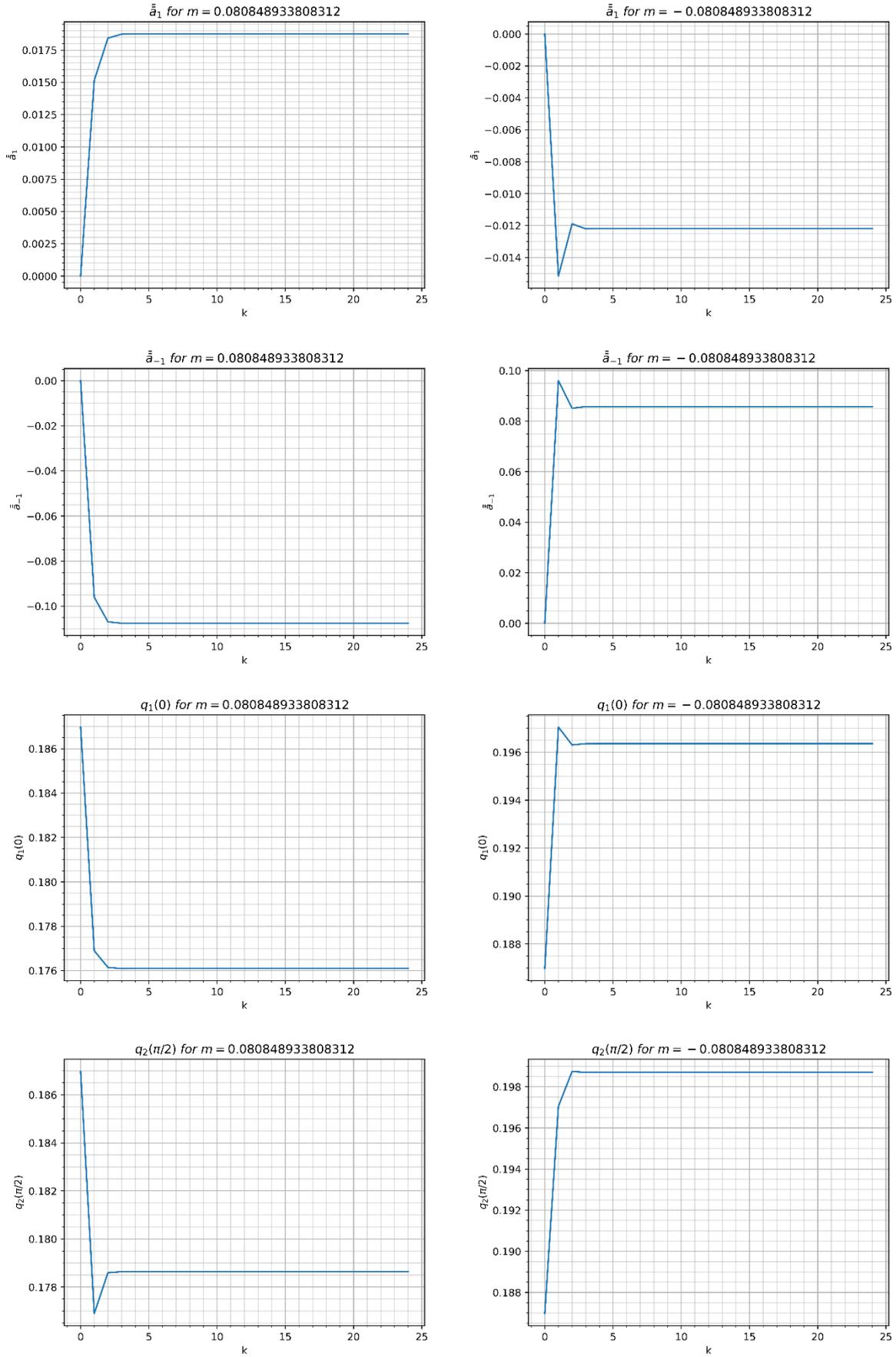

Fig. 10.10. Convergence of Orbit of earth's moon $g04\ C = -3.25444\ m = 0.080849$

### 10.4.3. Pre-cusped Orbit $g20$ $C = -1.75$ $m = 0.380571$

For the pre-cusped orbit (our g20), based on the numeric solution of the Newtonian equations, we have the following values:

$$name = g20 \tag{10.4.3.1}$$
$$C = -1.75 \tag{10.4.3.2}$$
$$m = 0.380571 \tag{10.4.3.3}$$
$$T = 2.3912 \tag{10.4.3.4}$$

At the top:

$$q_1(0) = 0 \tag{10.4.3.5}$$
$$\dot{q}_1(0) = p_1(0) + q_2(0) = -1.126086 + 0.5165991 = -0.6094869 \tag{10.4.3.6}$$
$$q_2(0) = 0.5165991 \tag{10.4.3.7}$$
$$\dot{q}_2(0) = 0 \tag{10.4.3.8}$$

This results in the following equations (up to order 12):

$$q_1 = -0.60949t - 0.41064t^3 + 0.98007t^5 + 1.735t^7 + 1.6685t^9 - 0.78824t^{11} \tag{10.4.3.9}$$

$$q_2 = 0.5166 - 1.2641t^2 - 0.67082t^4 - 0.44441t^6 + 0.97223t^8 + 3.4192t^{10} \\ + 5.5802t^{12} \tag{10.4.3.10}$$

At the right:

$$q_1(0) = 0.33173 \tag{10.4.3.11}$$
$$\dot{q}_1(0) = 0 \tag{10.4.3.12}$$
$$q_2(0) = 0 \tag{10.4.3.13}$$
$$\dot{q}_2(0) = 1.6909 \tag{10.4.3.14}$$

This results in the following equations (up to order 12):

$$q_1 = 0.33173 - 2.3551t^2 + 15.096t^4 - 136.31t^6 + 1437.6t^8 - 16711.0t^{10} \\ + 2.0677 \cdot 10^5 t^{12} \tag{10.4.3.15}$$

$$q_2 = 1.6909t - 6.1498t^3 + 43.317t^5 - 425.98t^7 + 4763.9t^9 - 57494.0t^{11} \tag{10.4.3.16}$$

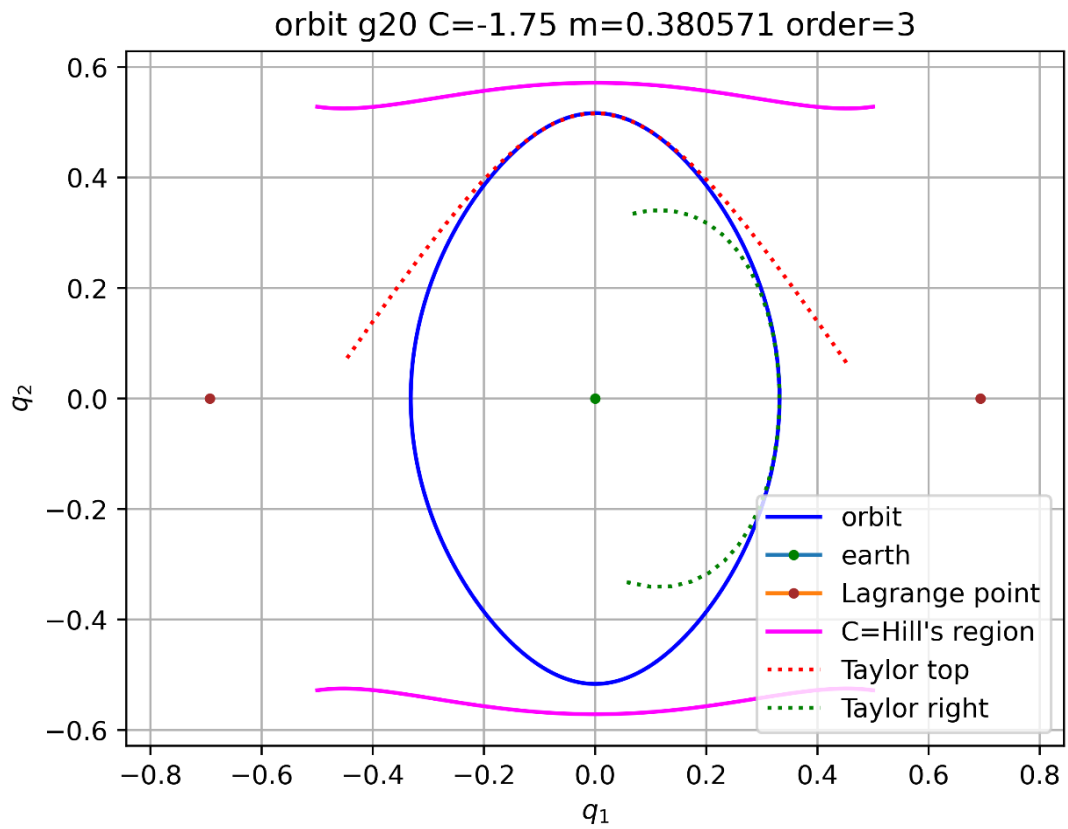

Fig. 10.11. The Pre-cusped Orbit $g20\ C = -1.75\ m = 0.380571$ (plot based on Taylor expansion to order 3)

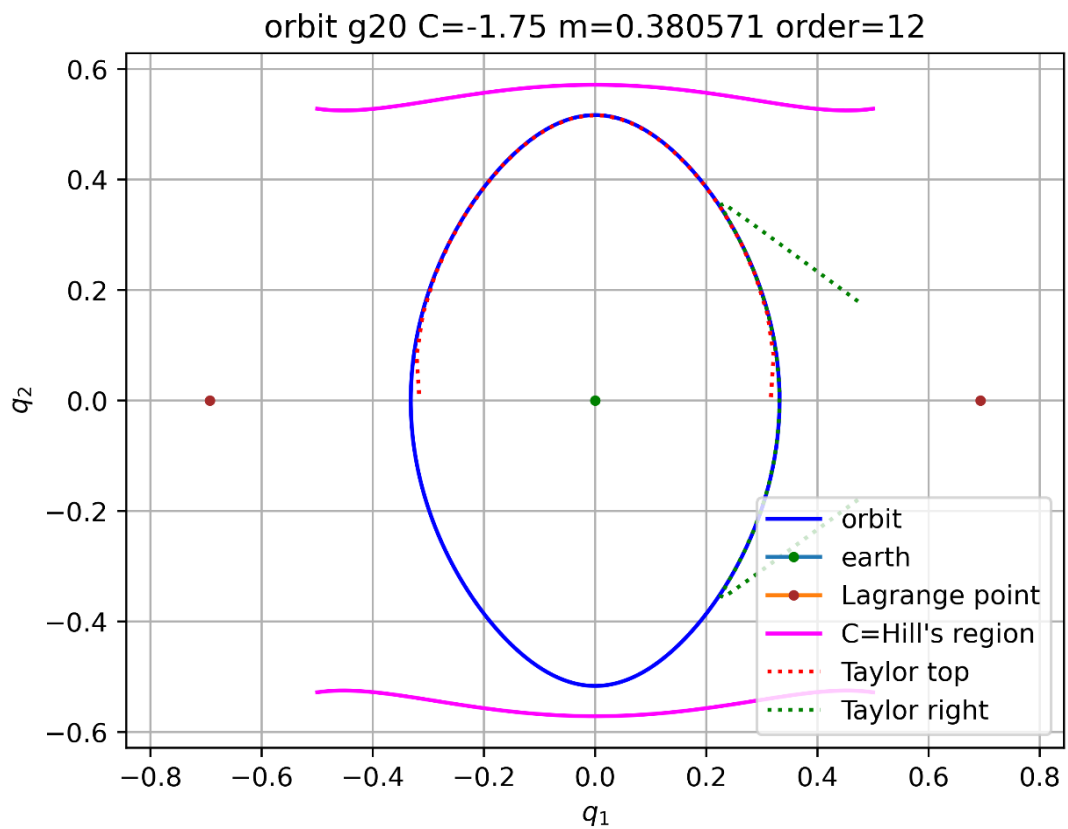

Fig. 10.12. The Pre-cusped Orbit $g20\ C = -1.75\ m = 0.380571$ (plot based on Taylor expansion to order 12)

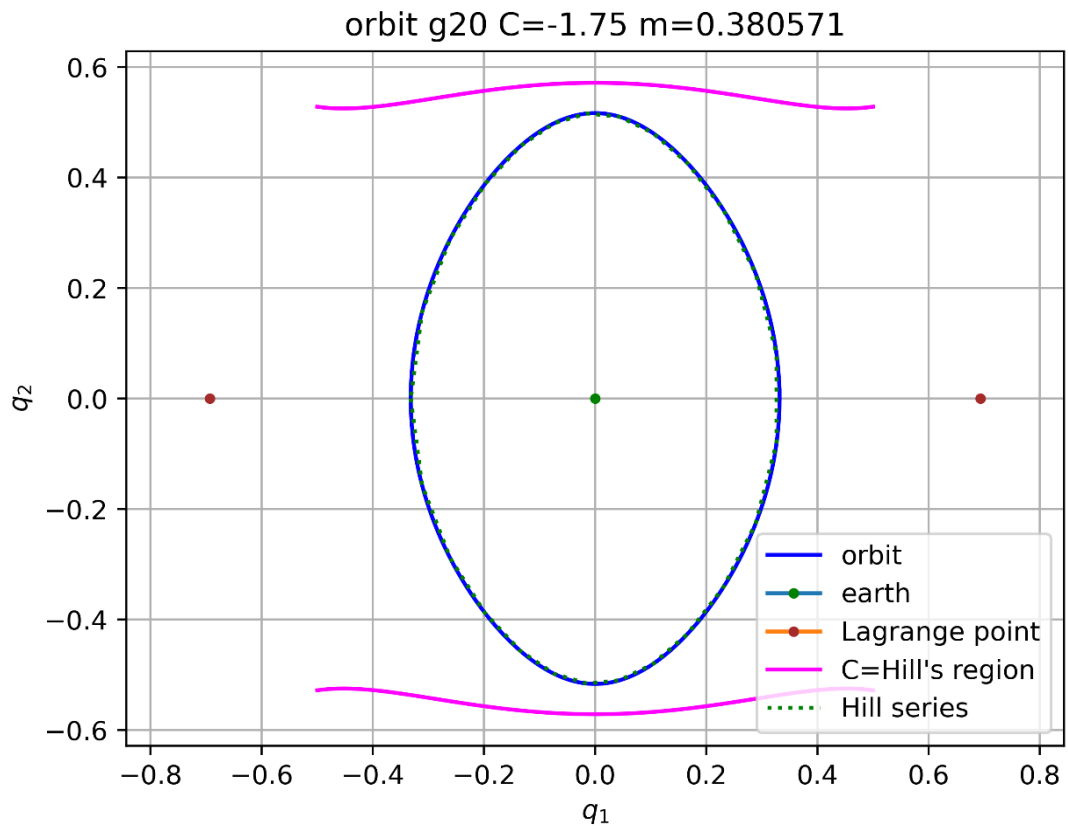

Fig. 10.13. The Pre-cusped Orbit $g20\ C = -1.75\ m = 0.380571$ (plot based on Hill series)

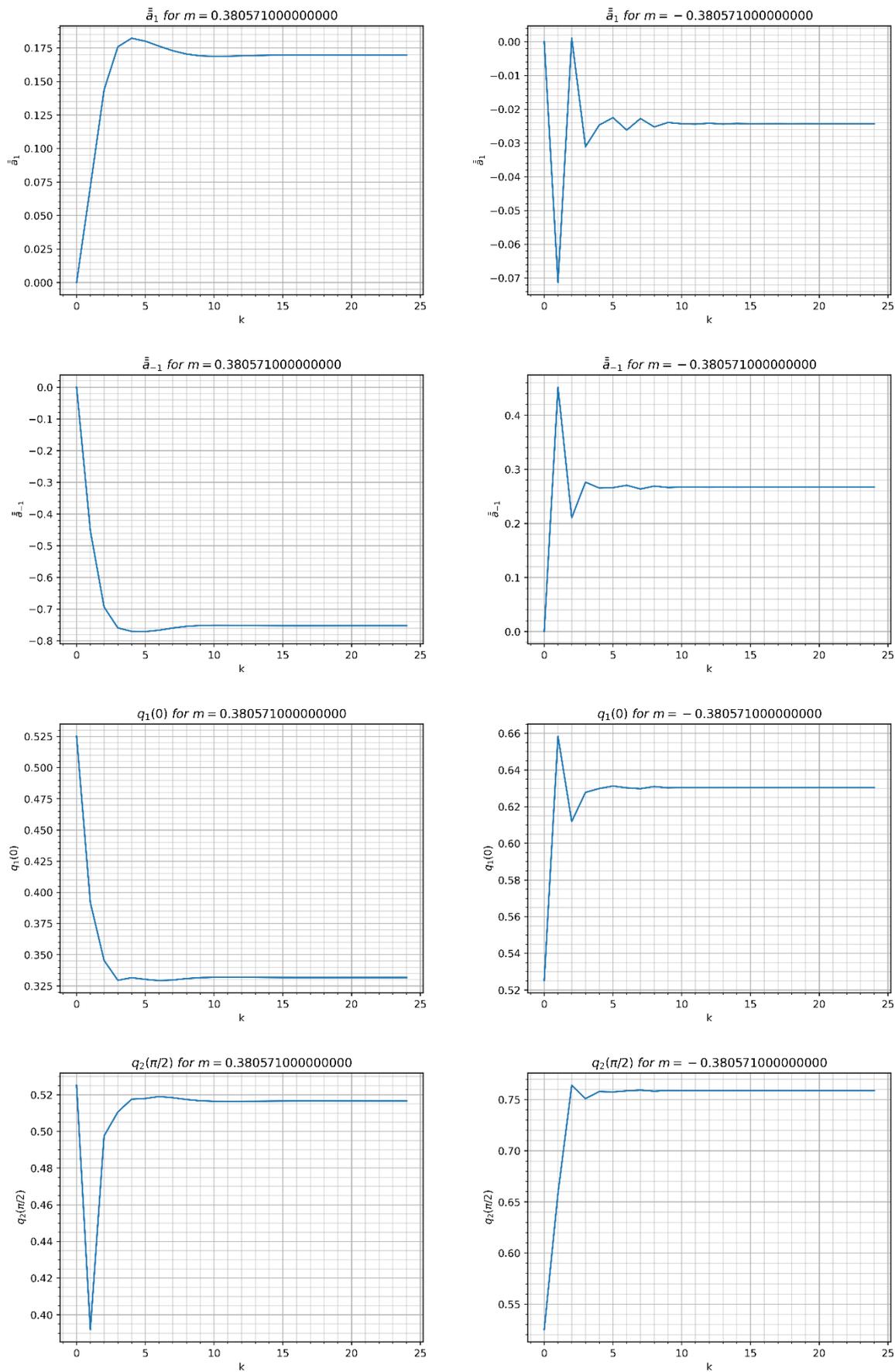

Fig. 10.14. Convergence of the Pre-cusped Orbit $g20\ C = -1.75\ m = 0.380571$

### 10.4.4. Pre-cusped Orbit $g22$ $C = -1.445$ $m = 0.500001169$

For the pre-cusped orbit 2 (g22), based on the numeric solution of the Newtonian equations, we have the following values:

$$name = g22 \tag{10.4.4.1}$$

$$C = -1.445 \tag{10.4.4.2}$$

$$m = 0.500001169 \tag{10.4.4.3}$$

$$T = 3.1416 \tag{10.4.4.4}$$

At the top:

$$q_1(0) = 0 \tag{10.4.4.5}$$

$$\dot{q}_1(0) = p_1(0) + q_2(0) = -0.8658682 + 0.6842303 = -0.1816379 \tag{10.4.4.6}$$

$$q_2(0) = 0.6842303 \tag{10.4.4.7}$$

$$\dot{q}_2(0) = 0 \tag{10.4.4.8}$$

This results in the following equations (up to order 12):

$$q_1 = -0.18164t - 0.58721t^3 + 0.051261t^5 + 0.19996t^7 + 0.32111t^9 \\ + 0.40748t^{11} \tag{10.4.4.9}$$

$$q_2 = 0.68423 - 0.88635t^2 - 0.14873t^4 - 0.31947t^6 - 0.24291t^8 - 0.11934t^{10} \\ + 0.062491t^{12} \tag{10.4.4.10}$$

At the right:

$$q_1(0) = 0.298855 \tag{10.4.4.11}$$

$$\dot{q}_1(0) = 0 \tag{10.4.4.12}$$

$$q_2(0) = 0 \tag{10.4.4.13}$$

$$\dot{q}_2(0) = 2.0175 \tag{10.4.4.14}$$

This results in the following equations (up to order 12):

$$q_1 = 0.29885 - 3.1324t^2 + 38.185t^4 - 625.35t^6 + 12192.0t^8 - 2.6267 \cdot 10^5 t^{10} \\ + 6.028 \cdot 10^6 t^{12} \tag{10.4.4.15}$$

$$q_2 = 2.0175t - 10.509t^3 + 143.92t^5 - 2634.7t^7 + 54736.0t^9 - 1.2266 \cdot 10^6 t^{11} \tag{10.4.4.16}$$

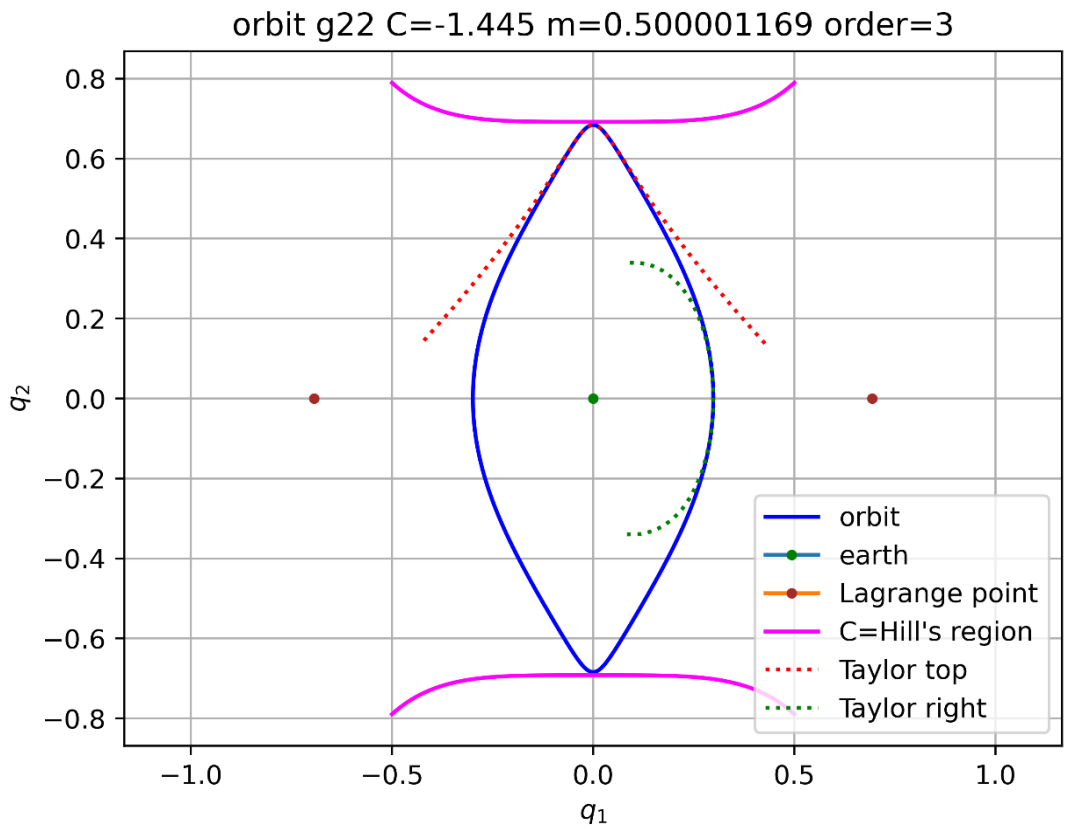

Fig. 10.15. The Pre-cusped Orbit 2 $g22\ C = -1.445\ m = 0.500001169$ (plot based on Taylor expansion to order 3)

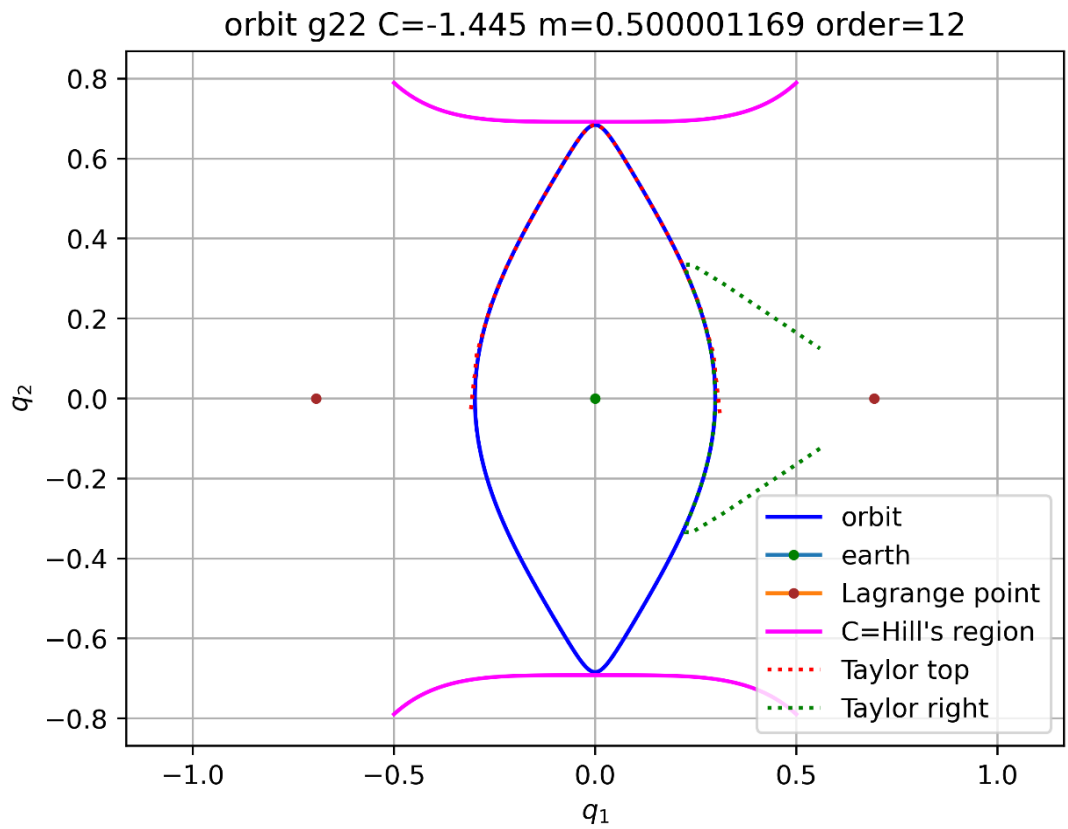

Fig. 10.16. The Pre-cusped Orbit 2 $g22\ C = -1.445\ m = 0.500001169$ (plot based on Taylor expansion to order 12)

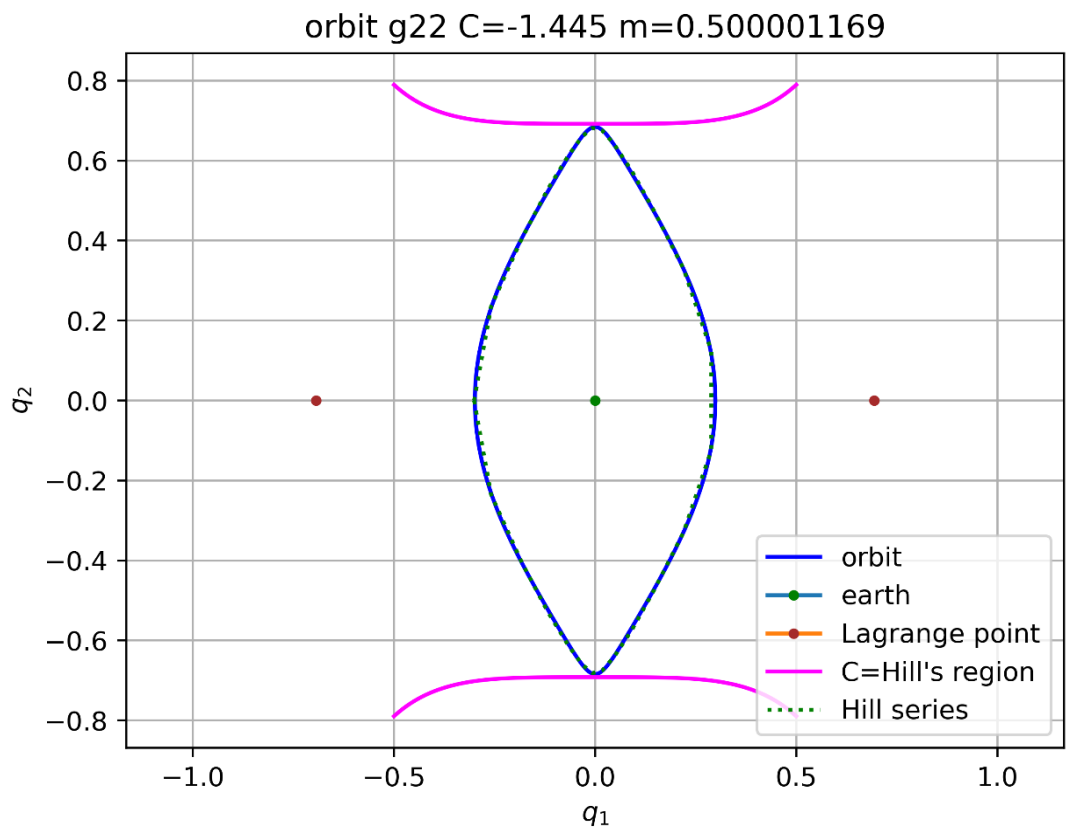

Fig. 10.17. The Pre-cusped Orbit 2 $g22\ C = -1.445\ m = 0.500001169$ (plot based on Hill series)

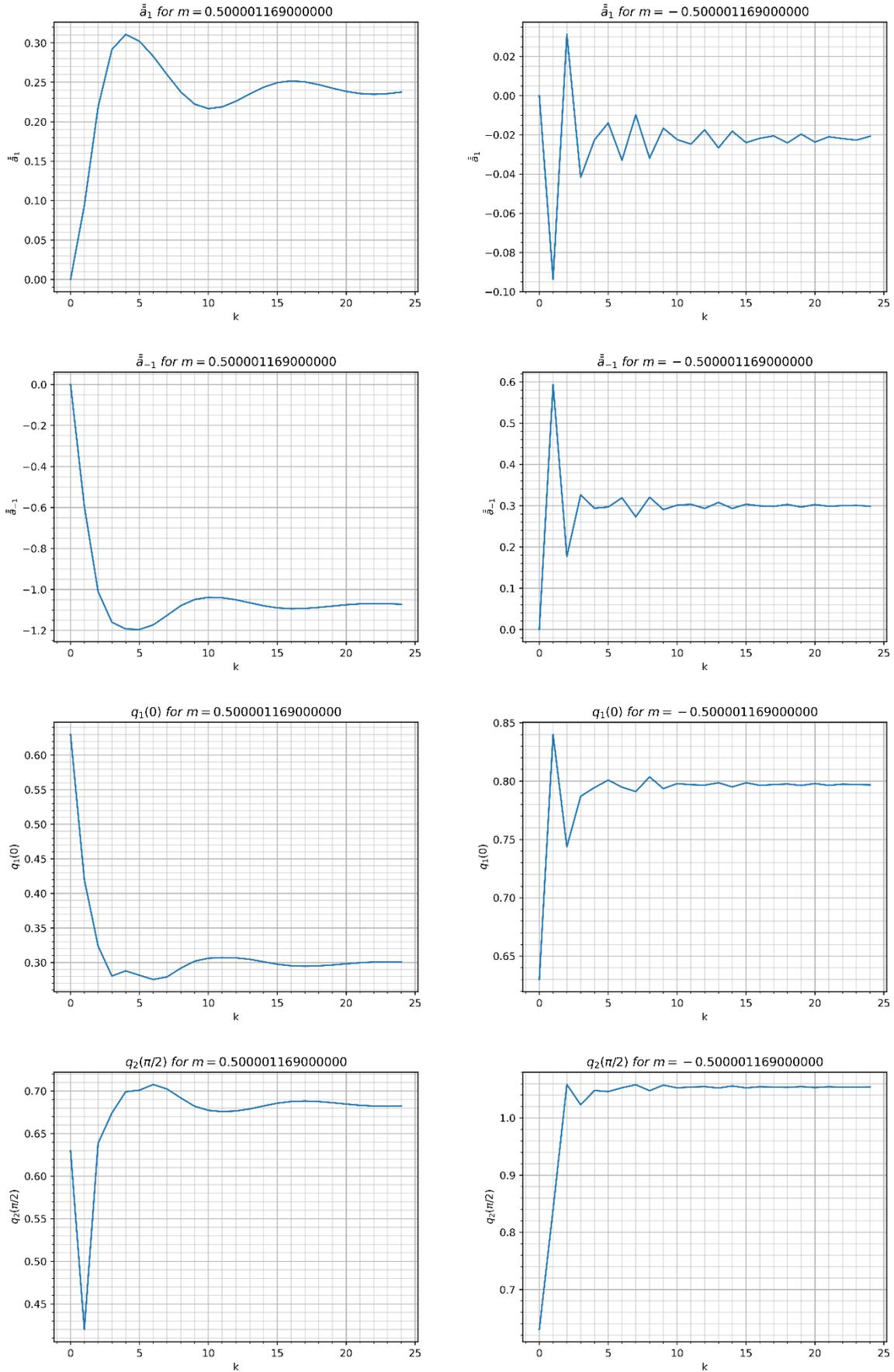

Fig. 10.18. Convergence of the Pre-cusped Orbit 2 $g22\ C = -1.445\ m = 0.500001169$

### 10.4.5. Cusped Orbit at $g23$ $C = -1.27899$ $m = 0.560958$

We need to pay special attention to the cusped orbit, known in (G.W. Hill 1878) as the "orbit of maximum lunation", because he initially thought that this was the largest possible value of $m = \frac{T}{2\pi}$. The cusps are points of zero velocity. For values of $m > \frac{1}{3}$, he no longer uses the series, but mechanical quadrature, see page 252 and following, and the plot of the orbit is on page 261 (Plate III). In (Hénon 1969) it is part of Fig. 4. and Table 4. This is the only (known) cusped orbit. In (Hénon 1969), Fig. 2 and Fig. 6, there are some orbits which appear to have a cusp, but they are in family a and family g', which are not symmetric, and thus not included in the Hill series. Here, we follow the discussion in Wintner §238, but restrict ourselves to the case of Hill's equations, in contrast to the more general case presented by Wintner.

Then we can look at the energy $C$ (equation (4.1)) at this point.

We introduce some new notation for the cusped orbit, where the circumflex reminds us of the cusp.

$$\hat{q} = Q = q_2(0) \tag{10.4.5.1}$$

$$\hat{m} = 0.560958 \tag{10.4.5.2}$$

$$C = -\frac{1}{\hat{q}} \tag{10.4.5.3}$$

We note some values at this point.

$$q_1(0) = 0 \tag{10.4.5.4}$$

$$\dot{q}_1(0) = 0 \tag{10.4.5.5}$$

$$\ddot{q}_1(0) = 0 \tag{10.4.5.6}$$

$$\dddot{q}_1(0) = -\frac{2}{\hat{q}^2} \tag{10.4.5.7}$$

$$q_2(0) = \hat{q} \tag{10.4.5.8}$$

$$\dot{q}_2(0) = 0 \tag{10.4.5.8}$$

$$\ddot{q}_2(0) = -\frac{1}{\hat{q}^2} \tag{10.4.5.10}$$

$$\dddot{q}_2(0) = 0 \tag{10.4.5.11}$$

Now we look at a Taylor expansion of $q_1$ and $q_2$ around the point of zero velocity.

$$q_1 = q_1(0) + \dot{q}_1(0)t + \frac{1}{2}\ddot{q}_1(0)t^2 + \frac{1}{6}\dddot{q}_1(0)t^3 + \mathcal{O}(t^4)$$

$$q_1 = 0 + 0 + 0 - \frac{1}{3}\frac{1}{\hat{q}^2}t^3 + \mathcal{O}(t^4) \tag{10.4.5.12}$$

$$q_2 = q_2(0) + \dot{q}_2(0)t + \frac{1}{2}\ddot{q}_2(0)t^2 + \frac{1}{6}\dddot{q}_2(0)t^3 + \mathcal{O}(t^4)$$

$$q_2 = \hat{q} + 0 - \frac{1}{2}\frac{1}{\hat{q}^2}t^2 + 0 + \mathcal{O}(t^4) \tag{10.4.5.13}$$

Then we define

$$\alpha := \frac{1}{2}\frac{1}{\hat{q}^2} \tag{10.4.5.14}$$

which results in

$$q_1 = -\frac{2}{3}\alpha t^3 + \mathcal{O}(t^4) = -\frac{1}{3\hat{q}^2}t^3 = -\frac{C^2}{3}t^3 \tag{10.4.5.15}$$

$$q_2 = \hat{q} - \alpha t^2 + \mathcal{O}(t^4) = \hat{q} - \frac{1}{2\hat{q}^2}t^2 = \hat{q} - \frac{C^2}{2}t^2 \tag{10.4.5.16}$$

Now we recall that, in (Hénon 1969), the Jacobi integral is $\Gamma = -2C$, and note some numerical values. For the cusped orbit, (Hénon 1969) shows it in Fig. 4, but does not include it in Table 4, so we use our own simulation values for the point of zero velocity.

$$name = g23 \tag{10.4.5.17}$$

$$C = -1.27899 \tag{10.4.5.18}$$

$$\hat{q} = -\frac{1}{C} = 0.7818669 \tag{10.4.5.19}$$

$$\hat{q}^2 = \frac{1}{C^2} \tag{10.4.5.20}$$

$$\hat{m} = 0.560958 \tag{10.4.5.21}$$

$$\hat{T} = 2\pi\hat{m} = 3.524603 \tag{10.4.5.22}$$

At the top:

$$q_1(0) = 0 \tag{10.4.5.23}$$

$$\dot{q}_1(0) = p_1(0) + q_2(0) = 0 \tag{10.4.5.24}$$

$$q_2(0) = 0.7818669 \tag{10.4.5.25}$$

$$\dot{q}_2(0) = 0 \tag{10.4.5.26}$$

In the simulation, $q_2(0) = 0.7818014$.

This results in the following equations (up to order 12):

$$q_1 = -0.54527t^3 - 0.029777t^5 + 0.035789t^7 + 0.077497t^9 + 0.10365t^{11} \tag{10.4.5.27}$$

$$q_2 = 0.78187 - 0.81791t^2 - 0.012567t^4 - 0.17084t^6 - 0.13711t^8 - 0.10255t^{10}$$
$$-0.065243t^{12} \tag{10.4.5.28}$$

At the right:

$$q_1(0) = 0.271795 \tag{10.4.5.29}$$

$$\dot{q}_1(0) = 0 \tag{10.4.5.30}$$

$$q_2(0) = 0 \tag{10.4.5.31}$$

$$\dot{q}_2(0) = 2.241 \tag{10.4.5.32}$$

This results in the following equations (up to order 12):

$$q_1 = 0.27179 - 4.1197t^2 + 71.879t^4 - 1728.0t^6 + 49807.0t^8 - 1.5883 \cdot 10^6 t^{10}$$
$$+5.3974 \cdot 10^7 t^{12} \tag{10.4.5.33}$$

$$q_2 = 2.241t - 15.856t^3 + 326.05t^5 - 8835.8t^7 + 2.7178 \cdot 10^5 t^9$$
$$-9.0199 \cdot 10^6 t^{11} \tag{10.4.5.34}$$

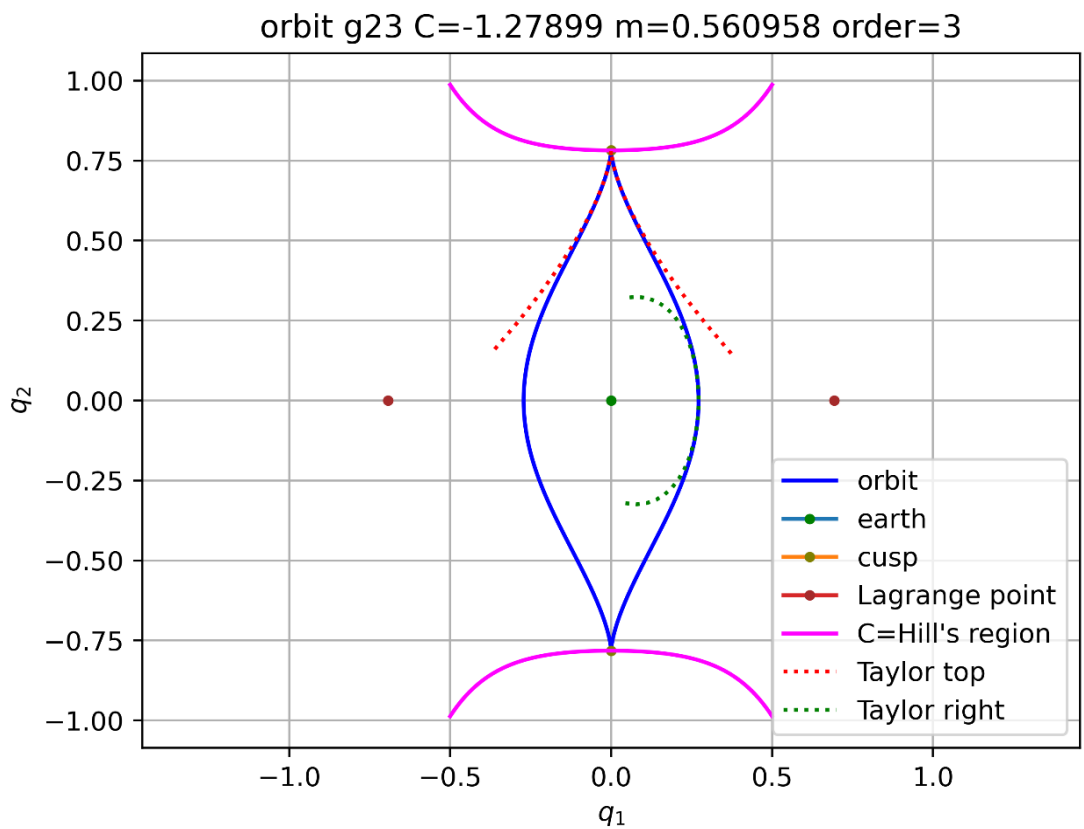

Fig. 10.19. The Cusped Orbit $g23\ C = -1.27899\ m = 0.560958$ (plot based on Taylor expansion to order 3)

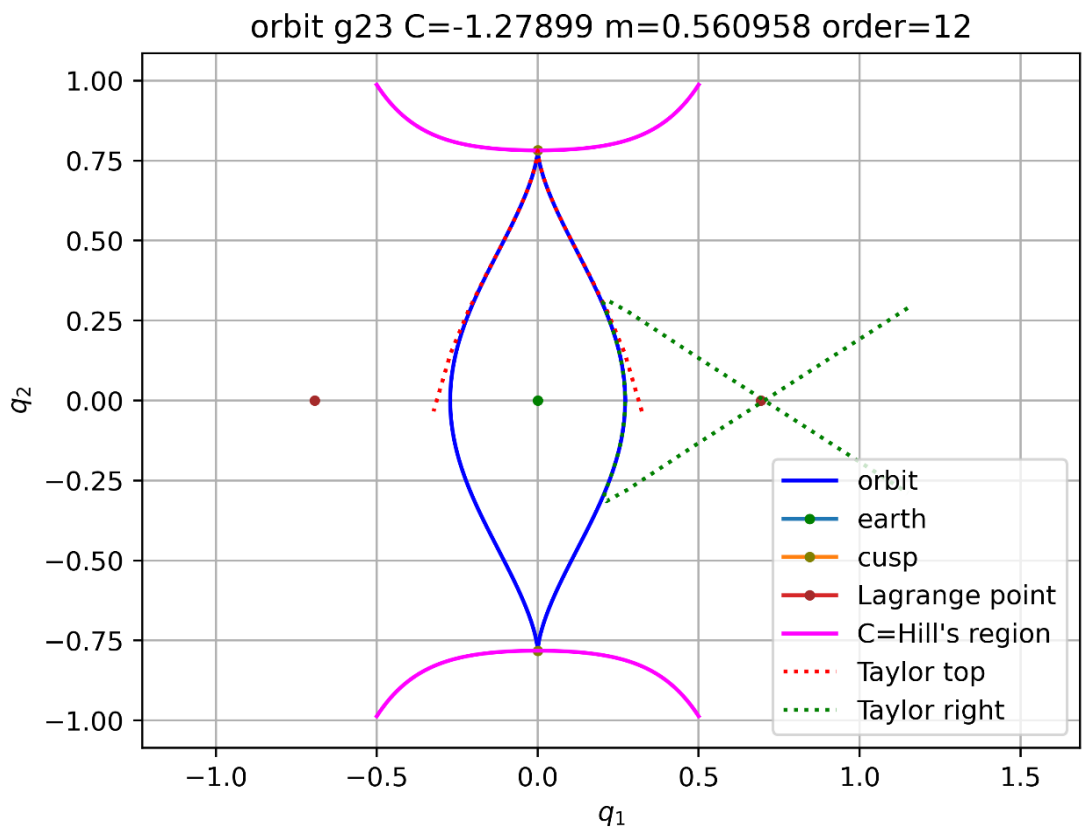

Fig. 10.20. The Cusped Orbit $g23\ C = -1.27899\ m = 0.560958$ (plot based on Taylor expansion to order 12)

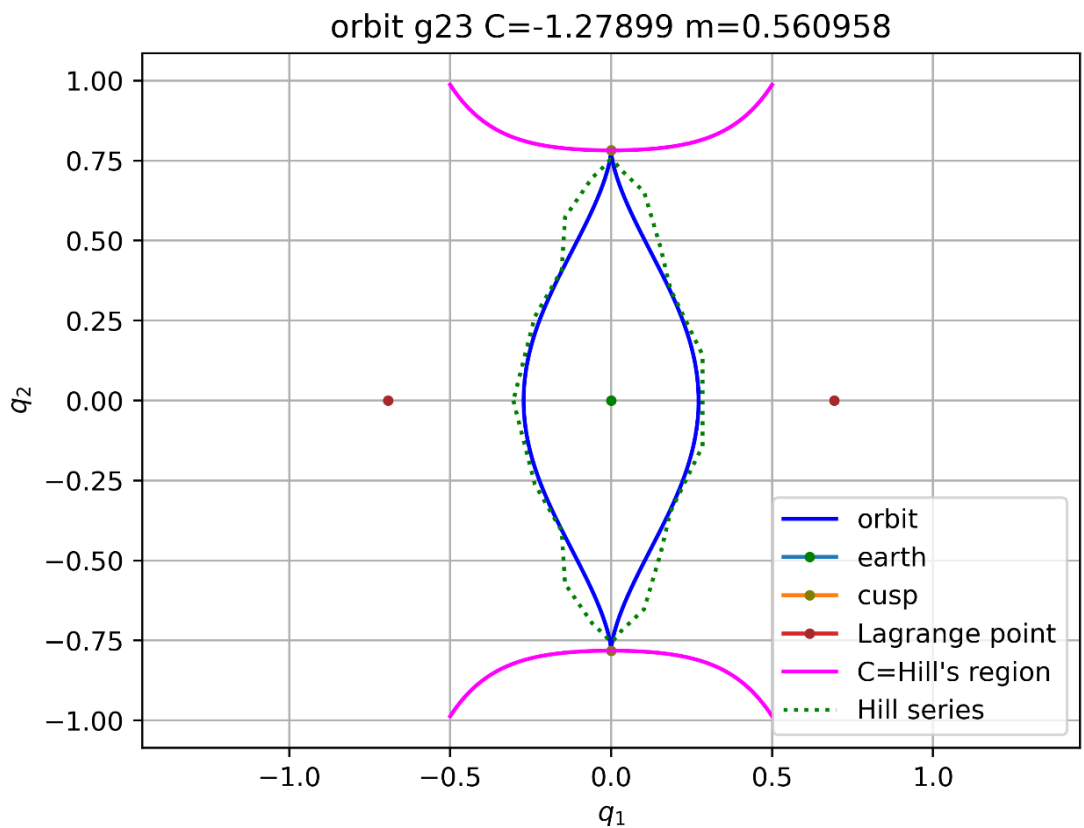

Fig. 10.21. The Cusped Orbit $g23\ C = -1.27899\ m = 0.560958$ (plot based on Hill series)

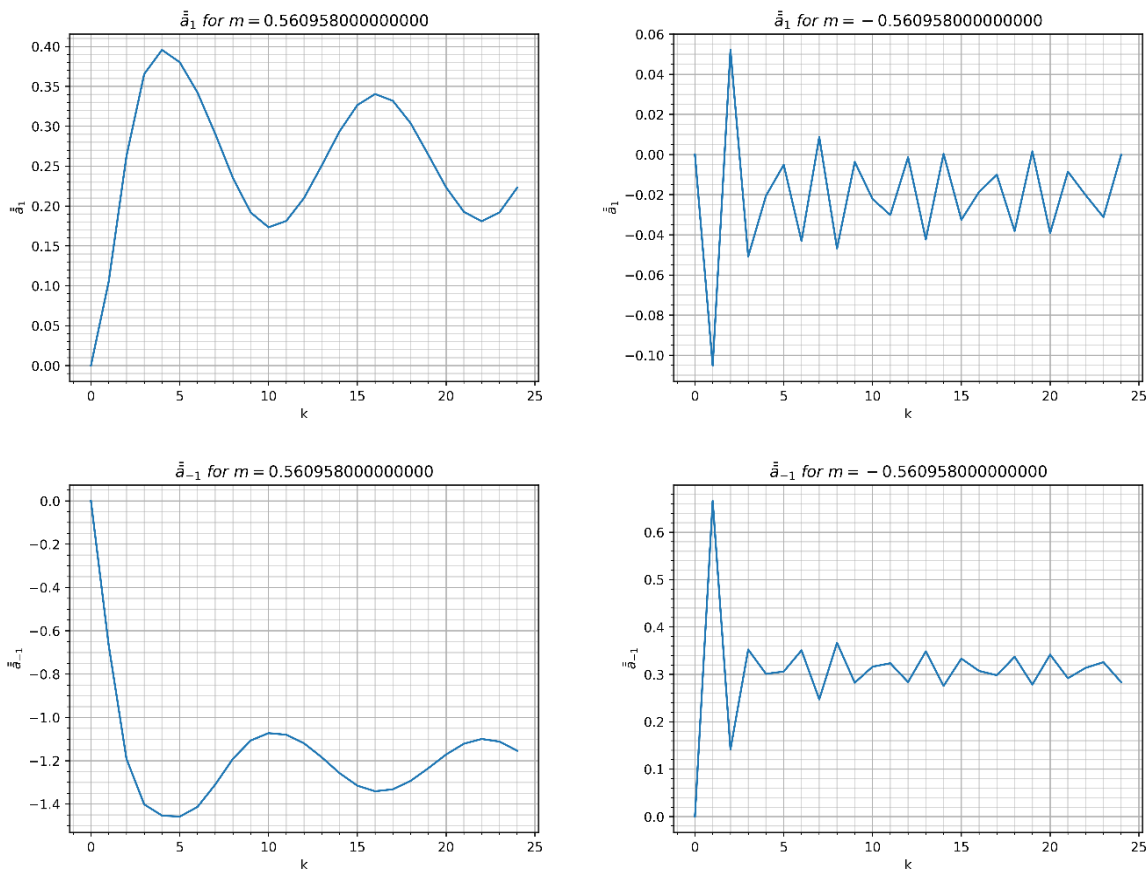

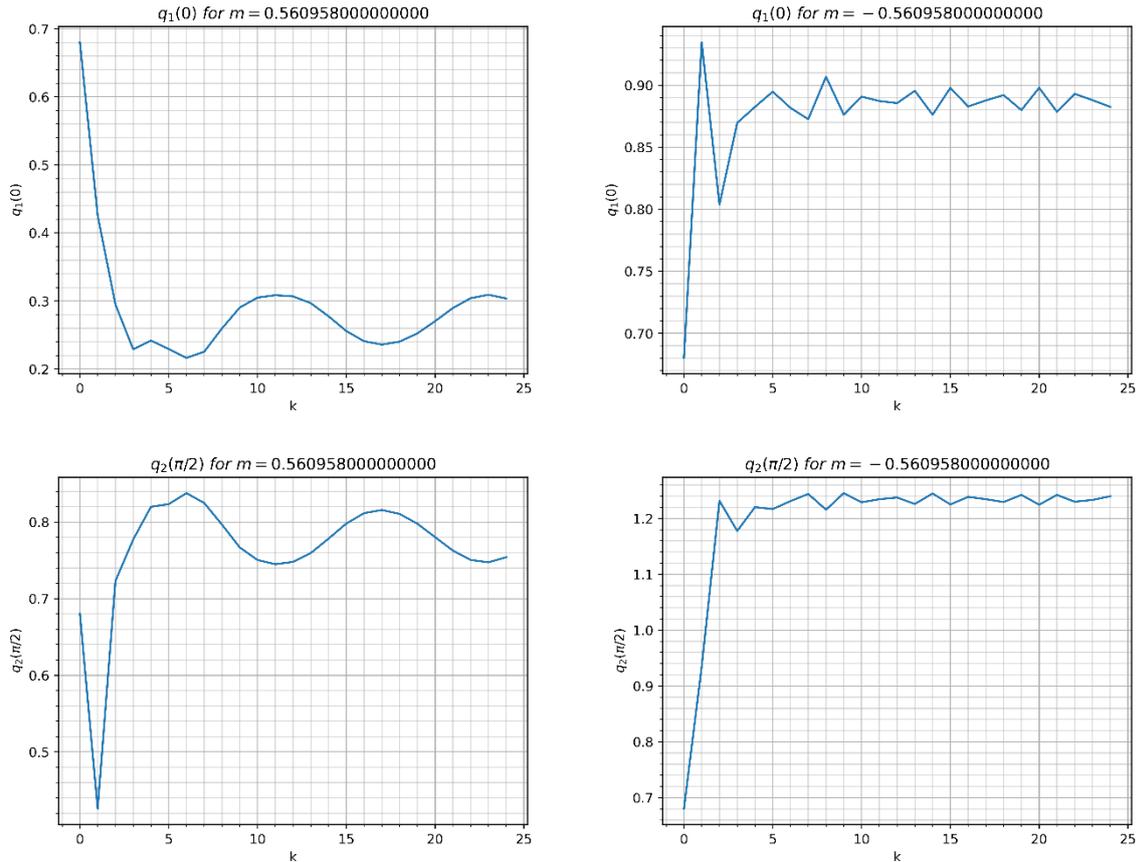

Fig. 10.22. Convergence of the Cusped Orbit $g23\ C = -1.27899\ m = 0.560958$

### 10.4.6. Post-cusped Orbit $g24\ C = -1.25\ m = 0.57168$

For the post-cusped orbit 2 (g24), based on the numeric solution of the Newtonian equations, we have the following values:

$$name = g24 \tag{10.4.6.1}$$

$$C = -1.25 \tag{10.4.6.2}$$

$$m = 0.57168 \tag{10.4.6.3}$$

$$T = 3.59196 \tag{10.4.6.4}$$

At the top:

$$q_1(0) = 0 \tag{10.4.6.5}$$

$$\dot{q}_1(0) = p_1(0) + q_2(0) = -0.7709934 + 0.7997354 = 0.0288014 \tag{10.4.6.6}$$

$$q_2(0) = 0.7997354 \tag{10.4.6.7}$$

$$\dot{q}_2(0) = 0 \tag{10.4.6.8}$$

This results in the following equations (up to order 12):

$$q_1 = 0.028801t - 0.53536t^3 - 0.035t^5 + 0.023152t^7 + 0.058387t^9 + 0.07916t^{11} \tag{10.4.6.9}$$

$$q_2 = 0.79974 - 0.81057t^2 + 0.0038161t^4 - 0.15181t^6 - 0.12085t^8 - 0.091283t^{10}$$
$$-0.061324t^{12} \tag{10.4.6.10}$$

At the right:

$$q_1(0) = 0.26679 \tag{10.4.6.11}$$

$$\dot{q}_1(0) = 0 \tag{10.4.6.12}$$

$$q_2(0) = 0 \qquad (10.4.6.13)$$
$$\dot{q}_2(0) = 2.2826 \qquad (10.4.6.14)$$

This results in the following equations (up to order 12):

$$q_1 = 0.26679 - 4.342t^2 + 80.792t^4 - 2083.9t^6 + 64513.0t^8 - 2.21 \cdot 10^6 t^{10}$$
$$+ 8.0682 \cdot 10^7 t^{12} \qquad (10.4.6.15)$$

$$q_2 = 2.2826t - 17.139t^3 + 379.3t^5 - 11041.0t^7 + 3.6483 \cdot 10^5 t^9$$
$$- 1.3008 \cdot 10^7 t^{11} \qquad (10.4.6.16)$$

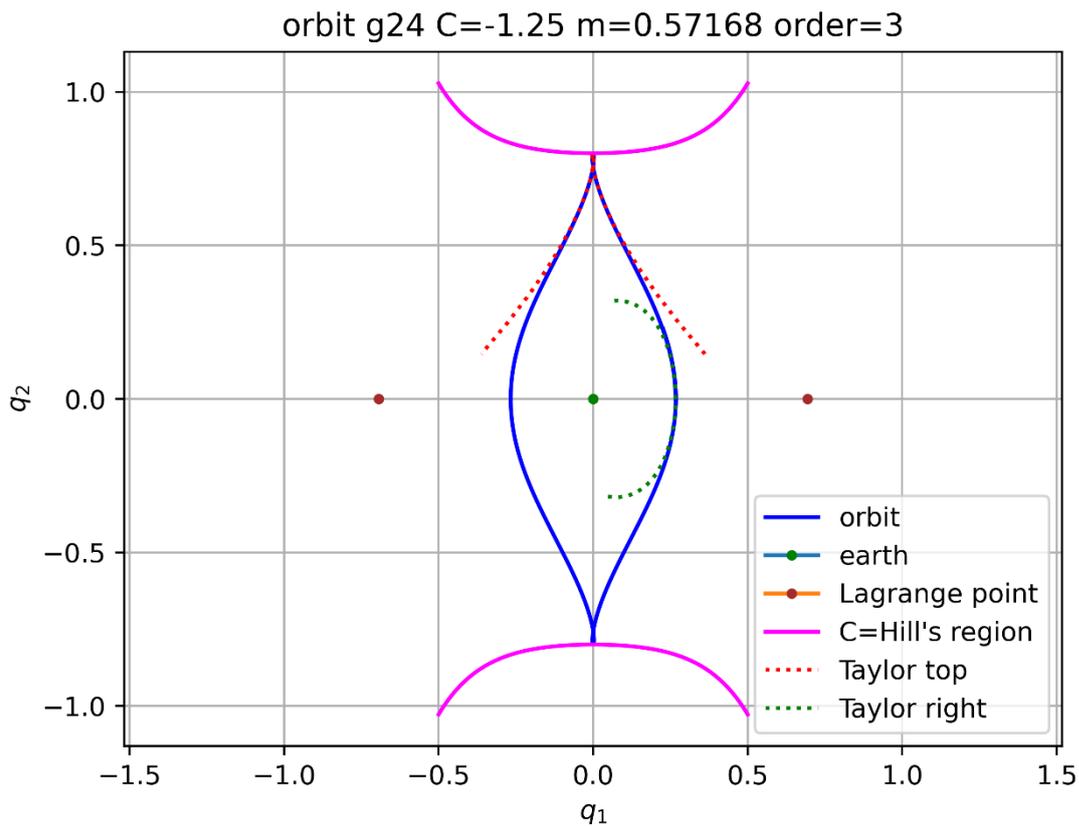

Fig. 10.23. The Post-cusped Orbit 2 $g24\ C = -1.25\ m = 0.57168$ (plot based on Taylor expansion to order 3)

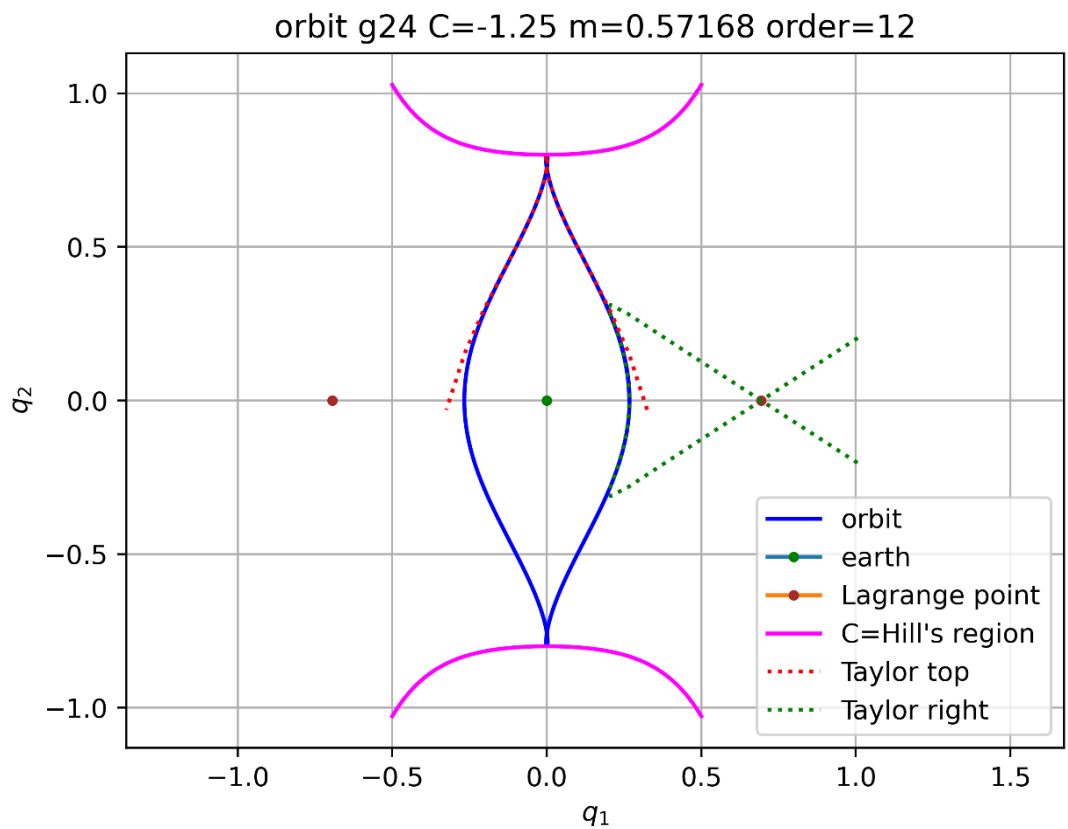

Fig. 10.24. The Post-cusped Orbit 2 $g24\ C = -1.25\ m = 0.57168$ (plot based on Taylor expansion to order 12)

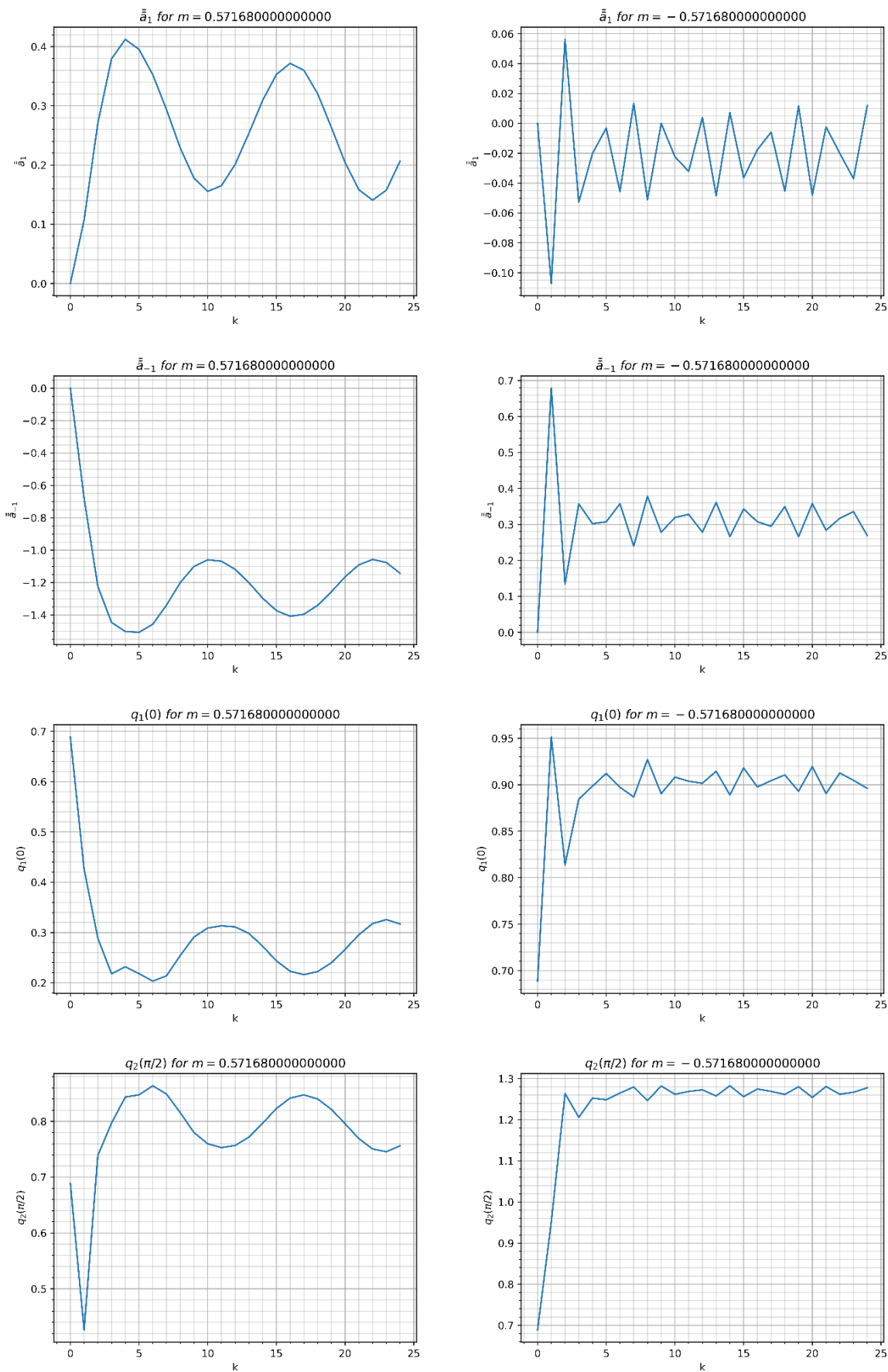

Fig. 10.25. Convergence of the Post-cusped Orbit 2 $g24\ C = -1.25\ m = 0.57168$

### 10.4.7. Post-cusped Orbit $g25$ $C = -1.0$ $m = 0.669562$

For the post-cusped orbit (g25), based on the numeric solution of the Newtonian equations, we have the following values:

$$name = g25 \tag{10.4.7.1}$$
$$C = -1.0 \tag{10.4.7.2}$$
$$m = 0.669562 \tag{10.4.7.3}$$
$$T = 4.20098 \tag{10.4.7.4}$$

At the top:

$$q_1(0) = 0 \tag{10.4.7.5}$$
$$\dot{q}_1(0) = p_1(0) + q_2(0) = -0.7227615 + 0.9702782 = 0.2475167 \tag{10.4.7.6}$$
$$q_2(0) = 0.9702782 \tag{10.4.7.7}$$
$$\dot{q}_2(0) = 0 \tag{10.4.7.8}$$

This results in the following equations (up to order 12):

$$\begin{aligned}q_1 = {} & 0.24752t - 0.44048t^3 - 0.038529t^5 - 0.012679t^7 - 0.00028395t^9 \\ & +0.0041101t^{11}\end{aligned} \tag{10.4.7.9}$$

$$\begin{aligned}q_2 = {} & 0.97028 - 0.77862t^2 + 0.086817t^4 - 0.050711t^6 - 0.034095t^8 \\ & -0.023459t^{10} - 0.01589t^{12}\end{aligned} \tag{10.4.7.10}$$

At the right:

$$q_1(0) = 0.221684 \tag{10.4.7.11}$$
$$\dot{q}_1(0) = 0 \tag{10.4.7.12}$$
$$q_2(0) = 0 \tag{10.4.7.13}$$
$$\dot{q}_2(0) = 2.6776 \tag{10.4.7.14}$$

This results in the following equations (up to order 12):

$$\begin{aligned}q_1 = {} & 0.22168 - 7.1641t^2 + 241.59t^4 - 11993.0t^6 + 7.1954 \cdot 10^5 t^8 \\ & -4.7841 \cdot 10^7 t^{10} + 3.3919 \cdot 10^9 t^{12}\end{aligned} \tag{10.4.7.15}$$

$$\begin{aligned}q_2 = {} & 2.6776t - 36.187t^3 + 1567.3t^5 - 88396.0t^7 + 5.6677 \cdot 10^6 t^9 \\ & -3.9239 \cdot 10^8 t^{11}\end{aligned} \tag{10.4.7.16}$$

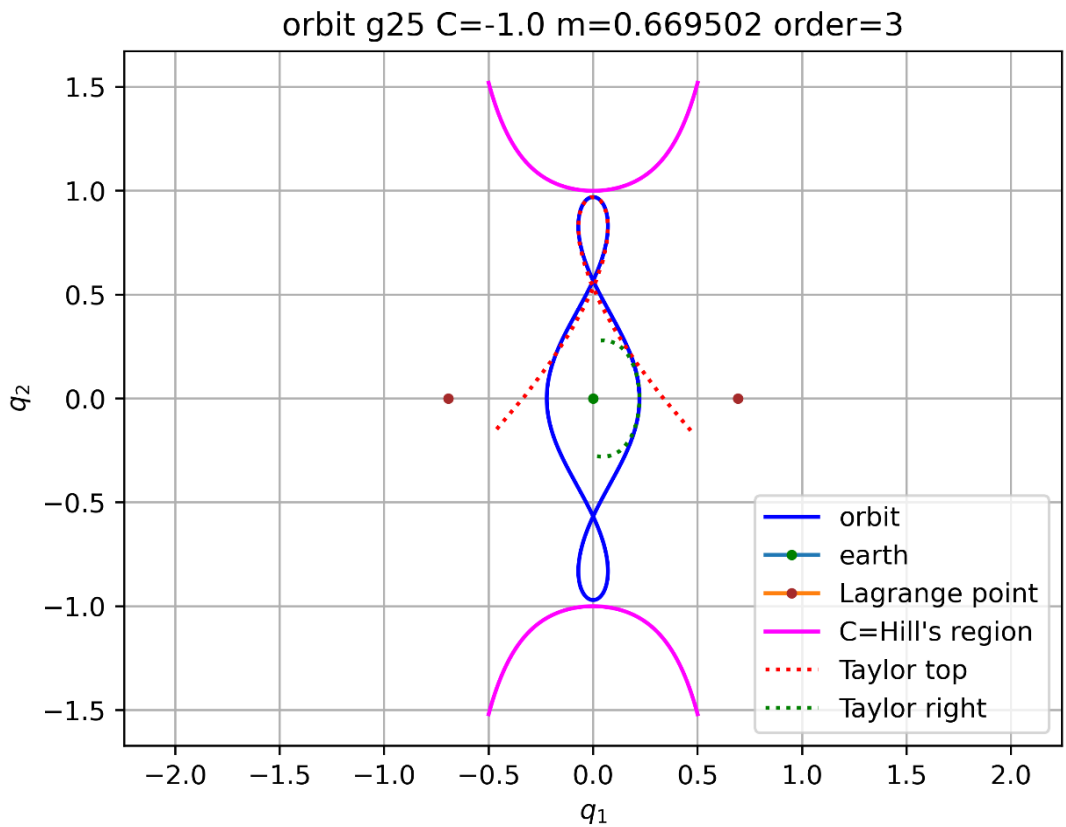

Fig. 10.26. The Post-cusped Orbit $g25$ $C = -1.0$ $m = 0.669562$ (plot based on Taylor expansion to order 3)

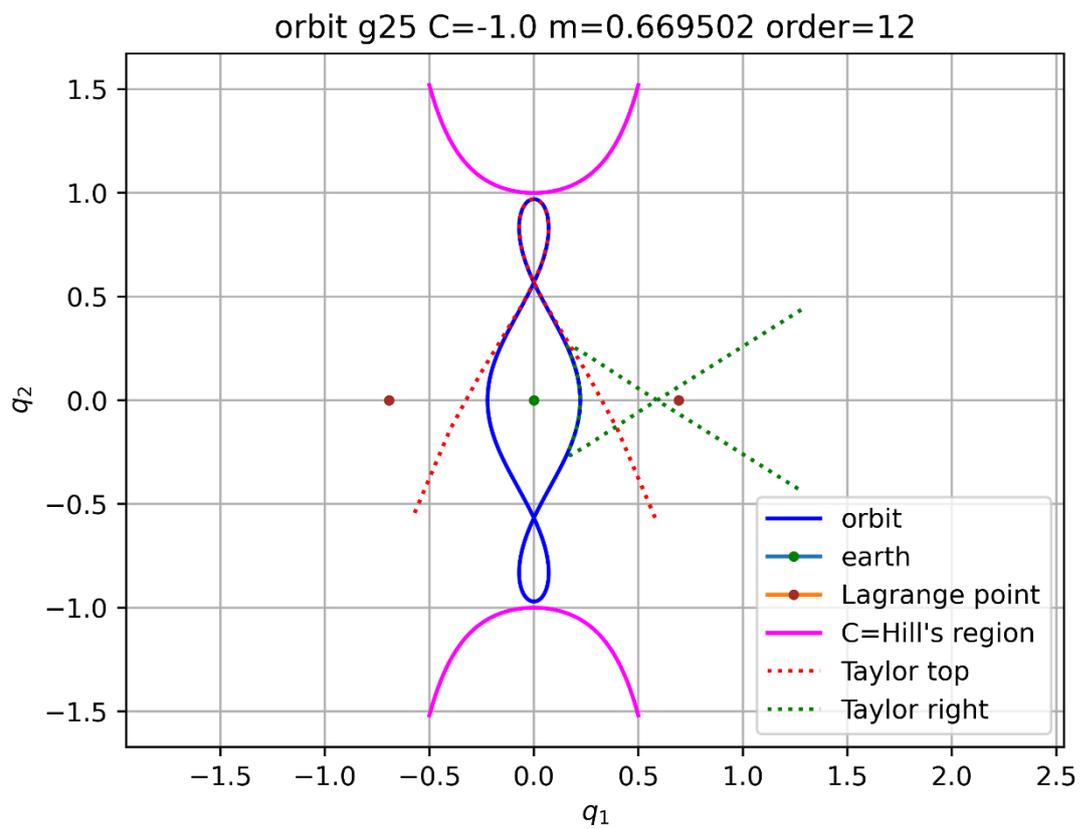

Fig. 10.27. The Post-cusped Orbit $g25$ $C = -1.0$ $m = 0.669562$ (plot based on Taylor expansion to order 12)

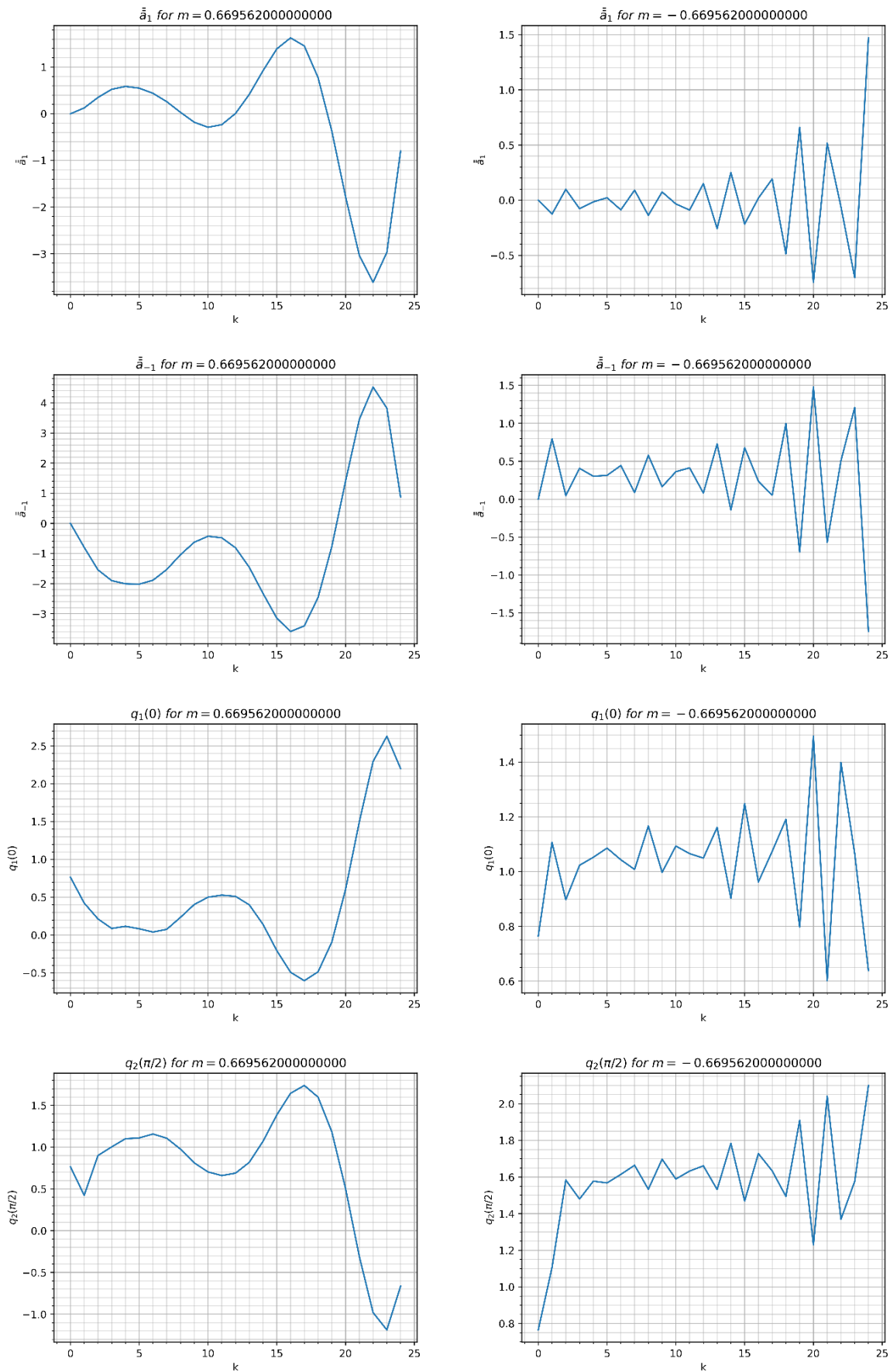

Fig. 10.28. Convergence of the Post-cusped Orbit $g25\ C = -1.0\ m = 0.669562$

## 10.5 Radius of Convergence

**Conjecture** (10.5.1): The radius of convergence of Hill's series is equal to the value of $m$ of the cusped orbit.

$$\hat{m} \approx 0.560958$$

**Numerical Evidence:**
Numerical calculation of all series and their coefficients shows a radius of convergence at (or near) the value of $m$ of the cusped orbit. In this manuscript, we have included plots of 6 selected orbits, see Figures 10.10, 10.14, 10.18, 10.22, 10.25, and 10.28. The $\bar{\bar{a}}_i$ are coefficients of Fourier series, summed over powers of $m$. The $q_i$ are summed over Fourier series and powers of $m$.

**Singularity**:
First, we observe that the radius of convergence is determined by the nearest singularity, see for example, (Ahlfors 1979) Theorem 3, page 179 and the subsequent discussion. Then, we just need to demonstrate that the cusp is a singularity.

We begin with our Taylor expansion of the Newtonian equations, i.e., our equations (10.4.5.12) and (10.4.5.13).

$$q_1 = -\frac{C^2}{3}t^3 + \mathcal{O}(t^4) \tag{10.5.1}$$

$$q_2 = -\frac{C^2}{2}t^2 + \mathcal{O}(t^4) \tag{10.5.2}$$

where we have chosen the coordinate system with the origin at the top cusp, i.e. at $(0,\hat{q})$. Then we have

$$\dot{q}_1 = -C^2 t^2 \tag{10.5.3}$$

$$\dot{q}_2 = -C^2 t \tag{10.5.4}$$

This gives us

$$\frac{dq_2}{dq_1} = \frac{\dot{q}_2}{\dot{q}_1} = \frac{1}{t} \tag{10.5.5}$$

and

$$\lim_{t \to 0^+} \frac{dq_2}{dq_1} = +\infty, \qquad \lim_{t \to 0^-} \frac{dq_2}{dq_1} = -\infty$$

This shows that the derivative of $q_2$ along the tangent goes to infinity and changes direction when $q_1$ crosses zero, so it is clearly a singularity.

As a plausibility test, we calculate this result directly. From (10.5.1) we have

$$t = \left(-\frac{3}{C^2}q_1\right)^{\frac{1}{3}}$$

$$t^2 = \left(-\frac{3}{C^2}\right)^{\frac{2}{3}}(q_1)^{\frac{2}{3}} \tag{10.5.6}$$

and from (10.5.2)

$$q_2 = -\frac{C^2}{2}\left(-\frac{3}{C^2}\right)^{\frac{2}{3}}(q_1)^{\frac{2}{3}}$$

taking the derivative

$$\frac{dq_2}{dq_1} = \frac{2}{3}\left(-\frac{C^2}{2}\right)\left(-\frac{3}{C^2}\right)^{\frac{2}{3}}(q_1)^{-\frac{1}{3}} \tag{10.5.7}$$

and from (10.5.1)

$$\frac{dq_2}{dq_1} = \frac{2}{3}\left(-\frac{C^2}{2}\right)\left(-\frac{3}{C^2}\right)^{\frac{2}{3}}\left(-\frac{C^2}{3}\right)^{-\frac{1}{3}}(t^3)^{-\frac{1}{3}}$$

$$\frac{dq_2}{dq_1} = \frac{1}{t} \tag{10.5.8}$$

This shows that the derivative of $q_2$ along the tangent goes to infinity and changes direction when $q_1$ crosses zero, so it is clearly a singularity.

Next, we will make one more observation about these equations.

$$q_1{}^2 = \frac{C^4}{9}t^6 \tag{10.5.9}$$

$$q_2{}^3 = -\frac{C^6}{8}t^6 \tag{10.5.10}$$

Then we define a rescaled version of $q_2$

$$Q_2 := \left(\frac{8}{9C^2}\right)^{1/3} q_2 \tag{10.5.11}$$

which implies

$$Q_2{}^3 = \frac{8}{9C^2}\left(-\frac{C^6}{8}\right)t^6 = -\frac{C^4}{9}t^6$$

and thus

$$q_1{}^2 + Q_2{}^3 = \frac{C^4}{9}t^6 - \frac{C^4}{9}t^6 = 0 \tag{10.5.12}$$

which shows that our equations define a **semi-cubic parabola**, a well-known cusp.

Now we need to ask if we have really proven that the Hill series have a singularity at the cusp. We are thinking of a function like this:

$$f: \mathbb{R} \times \mathbb{C} \to \mathbb{C}: (t, m) \mapsto u_1 = q_1 + iq_2$$

We have defined the domain of $m$ to be complex, even though we only need it to be real, because that is sometimes necessary in complex analysis. For example, it was necessary in section 10.1 where we needed to find a complex zero of the denominator.

First, the function $f_1$ is defined by the Newtonian differential equations of Hill's lunar problem. Then, we applied a Taylor expansion around the cusp, defining $f_2$. This is a good approximation of $f_1$ when we are close to the cusp, but when we take the limit, all higher powers of $t$ disappear, because they ultimately approach zero faster than the lowest power. As a result, the Taylor series identifies the singularity at the cusp of the function $f_1$ as defined by the Newtonian differential equation.

After this, we need to look at the series, $f_3$, which was derived in section 4 of this paper. Assuming that this derivation is correct, and Hill's series $f_3$ is a correct power-series representation of the function defined by the Newtonian equations, $f_1$, the singularity at the cusp applies to the power series as well. This has been confirmed by section 4 and multiple numeric calculations. As a result, we have three different descriptions of the same function, where the Taylor expansions are limited in range of applicability and the Hill series are limited in range of convergence. The singularity is valid for all three descriptions. ∎

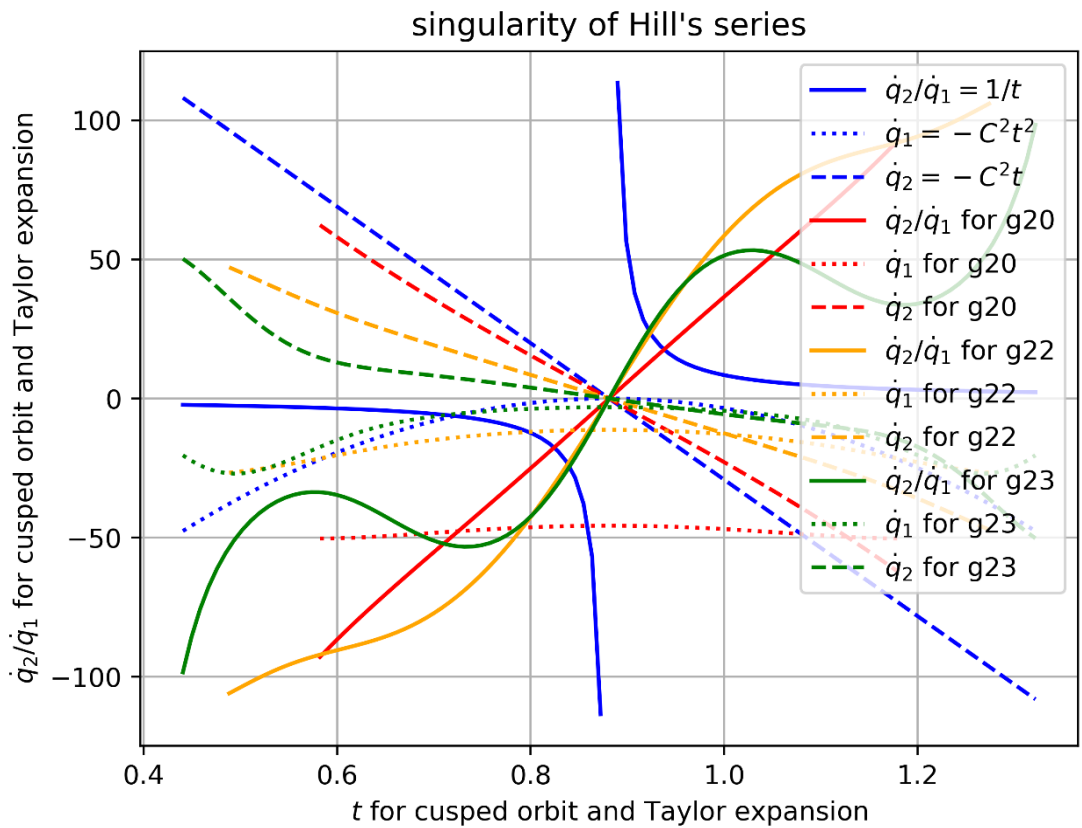

Fig. 10.29. Singularity at the cusp.

The plot shows the behavior of $\frac{dq_2}{dq_1} = \frac{\dot{q}_2}{\dot{q}_1}$ in the context of the singularity at the cusp and two related orbits. The first curve (blue) shows $\frac{1}{t}$ based on the Taylor expansion of the Newton equations at the cusp as discussed above. The second curve (red) shows $\frac{\dot{q}_2}{\dot{q}_1}$ based on Hill's series of the $m$ value of orbit $g20$, and the next curve (orange) shows the same thing but for the $m$ value of orbit $g22$. The last curve (yellow) shows the result for the $m$ value of cusped orbit $g23$. However, the series does not converge fast enough at $\hat{m}$, the value of $m$ at the cusp, if it converges at all, so this curve is not useful (as in Fig. 10.21.). In order to make a good comparison of the results, we have made two arbitrary changes to the curves of the Hill series results: We have shifted the origin of $g20$ and $g22$ to match the origin of the first curve, since these origins are different due to the different periods, and we have applied an arbitrary scaling factor to all three Hill series results in order to keep them nearly the same size as the first curve. All curves cover the range of $\frac{T}{4} - \frac{T}{8}$ to $\frac{T}{4} + \frac{T}{8}$ where $T$ is the period of the respective orbit.

The plot illustrates the following: The first curve shows that $\dot{q}_1$ touches zero at the cusp, which is what causes the singularity (infinite value) of $\frac{\dot{q}_2}{\dot{q}_1}$ and demonstrates the importance of the "point of zero velocity". The other three curves show $\dot{q}_1$ moving closer to zero but not reaching it until the cusped orbit. The curve of the Hill series for the cusped orbit $g23$ does not look plausible, which we attribute to insufficient convergence of the series. Either the series does not converge at the cusp, or convergence is too slow to be useful.

With this basis, we are now ready to dive into the calculations we have made for the motion of the moon's perigee.

# 11. Calculations of the motion of the moon's perigee

This section describes the calculations used to determine the motion of the moon's perigee. We follow Hill's method to a large extent, but our results differ, and we intentionally modify one of the methods used in solving the equations.

## 11.1 Overall method

We begin by showing the equations that need to be solved[69].

$$
\begin{array}{cccccccccc}
\cdots & \cdots & \cdots & \cdots & \cdots & \cdots & \cdots & \cdots & \cdots & \cdots \\
\cdots & +s(-3)w_{-3} & -\theta_1 w_{-2} & -\theta_2 w_{-1} & -\theta_3 w_0 & -\theta_4 w_1 & -\theta_4 w_2 & -\theta_6 w_3 & = 0 & \cdots \\
\cdots & -\theta_1 w_{-3} & +s(-2)w_{-2} & -\theta_1 w_{-1} & -\theta_2 w_0 & -\theta_3 w_1 & -\theta_3 w_2 & -\theta_5 w_3 & = 0 & \cdots \\
\cdots & -\theta_2 w_{-3} & -\theta_1 w_{-2} & +s(-1)w_{-1} & -\theta_1 w_0 & -\theta_2 w_1 & -\theta_2 w_2 & -\theta_4 w_3 & = 0 & \cdots \\
\cdots & -\theta_3 w_{-3} & -\theta_2 w_{-2} & -\theta_1 w_{-1} & +s(0)w_0 & -\theta_1 w_1 & -\theta_2 w_2 & -\theta_3 w_3 & = 0 & \cdots \quad (11.1) \\
\cdots & -\theta_4 w_{-3} & -\theta_3 w_{-2} & -\theta_2 w_{-1} & -\theta_1 w_0 & +s(1)w_1 & -\theta_1 w_2 & -\theta_2 w_3 & = 0 & \cdots \\
\cdots & -\theta_5 w_{-3} & -\theta_4 w_{-2} & -\theta_3 w_{-1} & -\theta_2 w_0 & -\theta_1 w_1 & +s(2)w_2 & -\theta_1 w_3 & = 0 & \cdots \\
\cdots & -\theta_6 w_{-3} & -\theta_5 w_{-2} & -\theta_4 w_{-1} & -\theta_3 w_0 & -\theta_2 w_1 & -\theta_1 w_2 & +s(3)w_3 & = 0 & \cdots \\
\cdots & \cdots & \cdots & \cdots & \cdots & \cdots & \cdots & \cdots & \cdots & \cdots \\
\end{array}
$$

where we are using different symbols than Hill:

$$s_i = s(i) = [i] = (c + 2i)^2 - \theta_0 \qquad (11.2)$$

$$w_n = b_n \qquad (11.3)$$

We also recall equation (8.3.3.16):

$$\theta_{-j} = \theta_j$$

This means that each row in (11.1) is an example of equation (8.3.4)[70]. Each row in (11.1) is one equation and we will refer to it by using a number for the equation, where equation 0 is the equation with $s(0)$ in it, or $j = 0$ in (8.3.4).

Starting here, (G.W. Hill 1886) makes numerous calculations involving determinants of infinite matrices and replaces $m$ by its numeric value very early in the calculation, where (G.W. Hill 1894) skips the determinants and keeps $m$ as a symbolic variable (literal value), so that is what we will follow here.

Since we will calculate the results using successive approximations, it will be useful to know the order (maximum exponent of m) for each of the symbols in (11.1).

| symbol | order |
|---|---|
| $s(0)$ | 3 |
| $s(-1)$ | 1 |
| $s(i), i \notin \{-1,0\}$ | 0 |
| $\theta_j$ | $2j$ |

Table 11.1

The first step in the calculation is to eliminate all $w_i$ in (11.1). After choosing an approximation, in other words a maximum order, we take a piece of (11.1), specifically an $n \times n$ matrix in the center of (11.1), giving us a list of $n$ equations. Then we loop through all equations, except for equation 0, solve equation $j$ for $w_j$ and substitute this value in all other equations, thus eliminating $w_j$. This leaves us with equation 0, which always contains a factor of $w_0$, so we can divide by $w_0$, completing the process of eliminating all $w_j$. The details of this step are documented in Appendix 2, and the details of solving the resulting equation are in Appendix 3. In summary, the steps involved in this are:

---

[69] This is (G.W. Hill 1886) (21) and (G.W. Hill 1894) page 35.
[70] Our equation (8.3.4) is (G.W. Hill 1886) equation (20).

- Choose a subset of (11.1) and a maximum order and list the equations from this subset.
- Eliminate all $w_j$ from these equations, leaving only equation 0. See appendix 2 for details.
- Eliminate higher-order terms (1), primarily by neglecting terms based on the order of $\Theta_i$ in Table 11.1.
- Replace square brackets $s_i$ by definition $(c+2i)^2 - \theta_0$.
- Eliminate higher-order terms (2). For example, for order 7, neglect $s(0)^3$, which has order 9 by replacing $c^6$ by $\theta_0^3 - 3c^2\theta_0^2 + 3c^4\theta_0$. This means that we will need to solve a quadratic equation in $c^2$. For more details, see the beginning of Appendix 3.
- Replace $\Theta_i$ by the corresponding partial sum of series in $m$.
- Eliminate higher-order terms (3). These terms are higher order in $m$ caused by calculating products of the partial sums of $\Theta_i$.
- Solve for $c$.

## 11.2. 1x1 matrix, order 3

The very first approximation uses the $1 \times 1$ matrix and assumes that we want an approximation of order 3, i.e. a maximum power of $m^3$. The matrix is $s(0)w_0 = 0$. After dividing by $w_0$, we have $s_0 = 0$, which can be solved as $c=\sqrt{\theta_0}$, which is the approximation we made between (8.3.2) and (8.3.3).

$$s_0 = 0 \tag{11.4}$$
$$c=\sqrt{\theta_0}$$

$$c = 1 + m - \frac{3m^2}{4} + \frac{3m^3}{4} + O(m^4)$$

## 11.3. 3x3 matrix, order 3

The second simple approximation uses the $3 \times 3$ matrix and assumes that we want an approximation of order 3, i.e. a maximum power of $m^3$. Elimination of the $w_i$ gives us the equation 0:

$$s_0 - \left(\frac{1}{s_1} + \frac{1}{s_{-1}}\right)\theta_1^2 = 0 \tag{11.5}$$

Since the order of $\theta_1^2$ is 4, we can neglect this part, yielding the same equation as before.

## 11.4. 5x5 matrix, order 7

Our next step is to calculate the approximation of order 7. For this, we use the $5 \times 5$ matrix. The resulting equation 0 is

$$s_0 - \left(\frac{1}{s_1} + \frac{1}{s_{-1}}\right)\theta_1^2 + \left(\frac{1}{s_{-1}s_1s_2} + \frac{1}{s_{-1}s_{-2}s_1}\right)\theta_1^4 - 2\left(\frac{1}{s_1s_2} + \frac{1}{s_{-1}s_1} + \frac{1}{s_{-1}s_{-2}}\right)\theta_1^2\theta_2$$
$$- \left(\frac{1}{s_2} + \frac{1}{s_{-2}}\right)\theta_2^2 = 0 \tag{11.6}$$

This is the first occurrence of a significant difference between our results and Hill's. The coefficient of $\theta_1^4$ our calculation (marked in yellow) is

$$\frac{1}{s_{-1}s_1s_2} + \frac{1}{s_{-1}s_{-2}s_1}$$

In Hill's calculation, it is

$$\frac{1}{s_{-1}^2 s_{-2}} + \frac{1}{s_1^2 s_2}$$

We consider this to be an error in Hill's paper, since this power of $s_i$ cannot occur in our elimination calculation, because $w_i$ is eliminated only once for each $i$.

Now, when we look at the equation 0, we see that the term involving $\theta_1^4$ and the two terms after it all have order 8, so they can be neglected. That leads us to

$$s_0 - \left(\frac{1}{s_1} + \frac{1}{s_{-1}}\right)\theta_1^2 = 0$$

and after multiplying by $s_1$ and $s_{-1}$, we have

$$-s_1\theta_1^2 - s_{-1}\theta_1^2 + s_{-1}s_0 s_1 = 0$$

In this equation, we can replace the $s_i$ by their definitions, eliminate higher-order terms, and replace the $\theta_i$ by their partial sums in $m$, up to $m^7$. The result is a quadratic equation in $c^2$, which can be solved directly, yielding two solutions, one of which (with a plus sign before the square root), yields a numeric value that fits our situation. After converting the square root to a power series in $m$, we have

$$c = 1 + m - \frac{3m^2}{4} - \frac{201m^3}{32} - \frac{2367m^4}{128} - \frac{111749m^5}{2048} - \frac{4399741m^6}{24576} - \frac{42332413m^7}{65536} + O(m^8)$$

## 11.5. 7x7 matrix, order 11

Our next step is to calculate the approximation of order 11. For this, we use the 7 × 7 matrix

$$\begin{aligned}
s_0 - \theta_1^2\left(\frac{1}{s_1} + \frac{1}{s_{-1}}\right) + \theta_1^4\Bigg(&\frac{1}{s_{-1}s_1 s_2} + \frac{1}{s_{-1}s_{-2}s_1} + \frac{1}{s_1 s_2 s_3} + \frac{1}{s_{-2}s_{-3}s_1} + \frac{1}{s_{-1}s_2 s_3} + \frac{1}{s_{-1}s_{-2}s_{-3}}\Bigg) \\
&- 2\theta_1^2\theta_2\left(\frac{1}{s_1 s_2} + \frac{1}{s_{-1}s_1} + \frac{1}{s_{-1}s_{-2}}\right) - \theta_2^2\left(\frac{1}{s_2} + \frac{1}{s_{-2}}\right) \\
-\theta_1^6\Bigg(&\frac{1}{s_{-2}s_{-3}s_1 s_2 s_3} + \frac{1}{s_{-1}s_{-2}s_1 s_2 s_3} + \frac{1}{s_{-1}s_{-2}s_{-3}s_2 s_3} + \frac{1}{s_{-1}s_{-2}s_{-3}s_1 s_2}\Bigg) \\
+2\theta_1^4\theta_2\Bigg(&\frac{1}{s_{-2}s_{-3}s_1 s_2} + \frac{2}{s_{-1}s_1 s_2 s_3} + \frac{1}{s_{-1}s_{-2}s_2 s_3} + \frac{2}{s_{-1}s_{-2}s_1 s_2} + \frac{2}{s_{-1}s_{-2}s_{-3}s_1}\Bigg) \\
+\theta_1^2\theta_2^2\Bigg(&-\frac{2}{s_1 s_2 s_3} + \frac{1}{s_{-2}s_2 s_3} + \frac{1}{s_{-2}s_1 s_2} + \frac{1}{s_{-2}s_{-3}s_2} + \frac{1}{s_{-1}s_1 s_3} - \frac{2}{s_{-1}s_1 s_2} \\
&+ \frac{1}{s_{-1}s_{-3}s_1} + \frac{1}{s_{-1}s_{-2}s_2} - \frac{2}{s_{-1}s_{-2}s_1} - \frac{2}{s_{-1}s_{-2}s_{-3}}\Bigg) \\
-\theta_3^2\left(\frac{1}{s_3} + \frac{1}{s_{-3}}\right) &- 2\theta_1^3\theta_3\left(\frac{1}{s_{-1}s_{-2}s_{-3}} + \frac{1}{s_{-1}s_{-2}s_1} + \frac{1}{s_{-1}s_1 s_2} + \frac{1}{s_1 s_2 s_3}\right) \\
-2\theta_1\theta_2\theta_3\Bigg(&\frac{1}{s_{-1}s_2} + \frac{1}{s_{-1}s_{-3}} + \frac{1}{s_2 s_3} + \frac{1}{s_1 s_3} + \frac{1}{s_{-2}s_1} + \frac{1}{s_{-2}s_{-3}}\Bigg) = 0 \quad (11.7)
\end{aligned}$$

Since $\theta_1^6$ and the following terms have order 12, we can neglect them. The resulting equation 0 is

$$s_0 - \theta_1^2\left(\frac{1}{s_1} + \frac{1}{s_{-1}}\right) + \theta_1^4\left(\frac{1}{s_{-1}s_1 s_2} + \frac{1}{s_{-1}s_{-2}s_1} + \frac{1}{s_1 s_2 s_3} + \frac{1}{s_{-2}s_{-3}s_1} + \frac{1}{s_{-1}s_2 s_3} + \frac{1}{s_{-1}s_{-2}s_{-3}}\right)$$
$$- 2\theta_1^2\theta_2\left(\frac{1}{s_1 s_2} + \frac{1}{s_{-1}s_1} + \frac{1}{s_{-1}s_{-2}}\right) - \theta_2^2\left(\frac{1}{s_2} + \frac{1}{s_{-2}}\right) = 0 \quad (11.8)$$

For the moment, we just want to compare it with Hill's results. Everything marked in yellow differs from Hill's results, (G.W. Hill 1894) page 36, by the same type of difference as mentioned above. The parts marked in green do not appear in Hill's version, but this might be caused by terms being neglected in his calculations earlier than ours.

In Hill's calculations to solve this equation, approximations of order 7 are substituted into parts of this equation, thus reducing it to a manageable form. We don't understand that remark, so we decided not to use it. Instead, we neglect all terms of order 12 and higher ($\theta_1^6$ and the following). This is in fact a big reduction in the size of the equation. The result is then

$$s_{-1}s_{-2}s_{-3}s_0s_1s_2s_3 - (s_{-2}s_{-3}s_1s_2s_3 + s_{-1}s_{-2}s_{-3}s_2s_3)\theta_1^2$$
$$+(s_1s_2s_3 + s_{-3}s_2s_3 + s_{-2}s_{-3}s_3 + s_{-2}s_{-3}s_1 + s_{-1}s_2s_3 + s_{-1}s_{-2}s_{-3})\theta_1^4$$
$$-2(s_{-3}s_1s_2s_3 + s_{-2}s_{-3}s_2s_3 + s_{-1}s_{-2}s_{-3}s_3)\theta_1^2\theta_2$$
$$-(s_{-1}s_{-3}s_1s_2s_3 + s_{-1}s_{-2}s_{-3}s_1s_3)\theta_2^2 = 0 \qquad (11.9)$$

Here, we can replace the $s_i$ by their definitions, eliminate higher-order terms, and replace the $\theta_i$ by their partial sums in $m$, up to $m^{11}$. The result is a cubic equation in $c^2$. We tried using the Cardano formula, but that led to complex expressions that were difficult to manage. In contrast, the trigonometric formula worked well. The reason for this is explained in (Fuchs and Tabachnikov 2007) Theorem 4.3. Since both the Cardano and trigonometric formulas come in multiple variants, we will choose one of them and use it to define the notation we use in the calculations in software, and document that in Appendix 4.

Using the notation of (Fuchs and Tabachnikov 2007) Lecture 4, we will need some symbols that overlap symbols used for other purposes in this document, for example $a$ and $c$. For the document, those symbols can be thought of as local to this description. Even though it isn't necessary, in the software (function cSolve), we coerce this uniqueness by giving the symbols new names, e.g., aCeq.

First, we replace $c^2$ by $X$, thus creating a cubic equation in $X$, giving us this:

$$aX^3 + bX^2 + cX + d = 0 \qquad (11.10)$$

where

$$X = c^2 \qquad (11.11)$$

For the lhs of this equation, we define a name $eqX$. Then we use upper-case symbols to convert to the notation of (Fuchs and Tabachnikov 2007). Then we check plausibility, which should be equal to 0:

$$aX^3 + bX^2 + cX + d - eqX$$

Then we convert to

$$X^3 + AX^2 + BX + C = 0 \qquad (11.12)$$

by defining

$$A := \frac{b}{a}, \qquad B := \frac{c}{a}, \qquad C := \frac{d}{a}$$

and check plausibility, which should be equal to 0:

$$a(X^3 + AX^2 + BX + C) - eqX$$

The factor of $a$ is necessary in order to convert the $A, B, C$ to $a, b, c, d$.

Then we introduce the reduced form

$$Y^3 + pY + q = 0 \qquad (11.13)$$

by defining

$$Y = X + \frac{A}{3}$$

$$p = B - \frac{A^2}{3} \qquad (11.14)$$

$$q = \frac{2}{27}A^3 - \frac{AB}{3} + C \qquad (11.15)$$

where numeric calculations confirm that $p$ and $q$ are both negative. This is described in (Fuchs and Tabachnikov 2007) page 65. We will now show this conversion, using our notation.

Assume $X = Y - \frac{A}{3}$:

$$X^3 + AX^2 + BX + C$$

$$= \left(Y - \frac{A}{3}\right)^3 + A\left(Y - \frac{A}{3}\right)^2 + B\left(Y - \frac{A}{3}\right) + C$$

$$= Y^3 - 3\frac{A}{3}Y^2 + 3\frac{A^2}{9}Y - \frac{A^3}{27} + A\left(Y^2 - 2\frac{A}{3}Y + \frac{A^2}{9}\right) + B\left(Y - \frac{A}{3}\right) + C$$

$$= Y^3 - AY^2 + 3\frac{A^2}{9}Y - \frac{A^3}{27} + AY^2 - 2\frac{A^2}{3}Y + \frac{A^3}{9} + BY - \frac{AB}{3} + C$$

$$= Y^3 + \left(B - \frac{A^2}{3}\right)Y + \frac{2A^3}{27} - \frac{AB}{3} + C$$

which is

$$Y^3 + pY + q$$

We check plausibility, which should be equal to 0:

$$a\left(\left(X + \frac{A}{3}\right)^3 + p\left(X + \frac{A}{3}\right) + q\right) - eqX$$

where the factor of a is explained by

$$a(Y^3 + pY + q) = a(X^3 + AX^2 + BX + C) = aX^3 + bX^2 + cX + d$$

Next, we calculate the discriminant

$$D = \frac{1}{27}p^3 + \frac{1}{4}q^2 \tag{11.16}$$

where numeric calculations confirm that the discriminant $D$ is negative, so we expect to have three real solutions.

Now we can calculate α for the trigonometric solution

$$\alpha_1 = \frac{9}{4}\frac{q}{p^2} \tag{11.17}$$

$$\alpha_2 = \sqrt{-\frac{4p}{3}} \tag{11.18}$$

$$\alpha = \arcsin(\alpha_1 \alpha_2) = \arcsin\left(\frac{9}{4}\frac{q}{p^2}\sqrt{-\frac{4p}{3}}\right) \tag{11.19}$$

where numeric calculations confirm that $\alpha_1\alpha_2$ is between $-1$ and 1.

Then, the three solutions are

$$X = \alpha_2 \sin\left(\frac{1}{3}(\alpha + 2k\pi)\right) - \frac{A}{3}, k = 0,1,2 \tag{11.20}$$

The term $-\frac{A}{3}$ appears here because the solutions are solutions for $Y$, but we need solutions for $X$.

Since $X = c^2$, we can immediately calculate $c$ as a power series of $m$. However, this time, the coefficients are no longer rational. They include irrational numbers such as $\sqrt{3}$, so we can lo longer compare this series with all previous series with rational coefficients. As a workaround, we could augment the field of coefficients to include $\sqrt{3}$, or calculate a rational equivalent of $\sqrt{3}$ (up to a power of 11), but the result is not easy to compare with the previous series for $c$, so we simply resort to documenting only a series with real (floating-point) coefficients.

Here we have a small difficulty. We would like to use the well-known series expansion to calculate arcsin in (11.19). The series expansion is:

$$\arcsin(x) = \sum_{n=0}^{\infty} \frac{(2n)!}{2^{2n}(n!)^2} \frac{x^{(2n+1)}}{2n+1}, \qquad |x| \leq 1 \tag{11.21}$$

This series converges for $x \in [-1,1]$ but does not converge well when $x$ is close to $\pm 1$, so we need a way to move the calculation away from 1. We begin with the following identity, which can be found in https://en.wikipedia.org/wiki/List_of_trigonometric_identities and in (Abramowitz and Stegun 1965) page 80, 4.4.32.

$$\arcsin(x) \pm \arcsin(y) = \arcsin\left(x\sqrt{1-y^2} \pm y\sqrt{1-x^2}\right) \tag{11.22}$$

This identity is valid when the argument of the right-hand side is within the domain of arcsine, i.e. $[-1,1]$, and the left-hand side is within the principal range of arcsine, i.e. $\left[-\frac{\pi}{2},\frac{\pi}{2}\right]$. In our case, we want to apply the identity to a power series of which the constant term is equal to $-1$ and the numeric value for the Earth's moon is equal to $-0.97$. Then the identity for a sum (two + signs) fulfills the conditions, but the difference (two − signs) does not. So, we choose the two + signs and a value of $y = 1$. This gives us

$$\arcsin(x) = -\frac{\pi}{2} + \arcsin\left(\sqrt{1-x^2}\right) \tag{11.23}$$

## 11.6. Summary of numeric values

| method | $c$ | $\dfrac{1}{n}\dfrac{d\omega}{dt}$ |
|---|---|---|
| 1x1 matrix, order 3 | 1.0763428761 | 0.0041689986 |
| Observation (according to Hill) |  | 0.008452 |
| Current observation (Wikipedia) | 1.0717137091 | 0.0084518977 |
| Bourne's calculation, order 7 | 1.0715891774 | 0.0085671143 |
| 3x3 matrix, order 7 | 1.0715838704 | 0.0085720244 |
| 5x5 matrix, order 7 | 1.0715838704 | 0.0085720244 |
| Bourne's calculation, order 11 | 1.0715833688 | 0.0085724885 |
| Hill's calculation (literal perigee) |  | 0.008572573 |
| Hill's calculation (numeric perigee) |  | 0.008572573 |
| 7x7 matrix, order 11 | 1.0715759535 | 0.0085793491 |
| Hill's estimate for order 7 |  | 0.008591 |

Here, we can see that our calculation for order 7 (after correcting what we consider to be an error in Hill's calculation) is closer to the observed value. On the other hand, our calculation for order 11 (which differs significantly from Hill's calculation) is farther from the observed value. The calculation in (Bourne 1972), and is out of the scope of this manuscript, which covers Hill's papers, but we show it here for comparison and discuss it more in our sections 9.11. Bourne's calculations of coefficients and 13. Outlook.

We calculate the values for Bourne's calculations by taking the series as quoted in (G.W. Hill 1894) page 39 and correct the coefficients of $m^{10}$ and $m^{11}$ as quoted in (Bourne 1972) page 180. We also note that the values of $c$ decrease and the values of $\dfrac{1}{n}\dfrac{d\omega}{dt}$ increase when we go from our calculations of order 7 to order 11, and the same is true of Bourne's calculations. However, our calculations of order 7 are the closest to the observed values, and our calculations of order 11 are the farthest.

The calculations in (Bourne 1972) identify a power series in $m$ for $c$ (their $c_0$), where errors in $c_{10}, c_{11}$ are identified and corrected. Here is the series and the resulting numeric values based on it:

$$c = 1 + m - \frac{3m^2}{4} - \frac{201m^3}{32} - \frac{2367m^4}{128} - \frac{111749m^5}{2048} - \frac{4095991m^6}{24576} - \frac{332532037m^7}{589824} \\ - \frac{15106211789m^8}{7077888} - \frac{5975332916861m^9}{679477248} - \frac{1547775442175567m^{10}}{40768634880} \\ - \frac{818429336556024967m^{11}}{4892236185600}$$

$$c = 1.07158336879192$$

$$\frac{1}{n}\frac{d\omega}{dt} = 0.00857248846399006$$

For comparison with Hill's observation, we now show a current value for the period:
Note: Wikipedia https://en.wikipedia.org/wiki/Lunar_month

| Month type | Length in days |
|---|---|
| draconitic | $27.212220815 + 4.14 \times 10^{-6} \times T$ |
| tropical | $27.321582252 + 1.82 \times 10^{-7} \times T$ |
| sidereal | $27.321661554 + 2.17 \times 10^{-7} \times T$ |
| anomalistic | $27.554549886 - 1.007 \times 10^{-6} \times T$ |
| synodic | $29.530588861 + 2.52 \times 10^{-7} \times T$ |

| synodic/anomalistic | $c = 1.07171370910341,\qquad \dfrac{1}{n}\dfrac{d\omega}{dt} = 0.0084518977$ |
|---|---|

# 12. Software Documentation

First, we will briefly describe the software used for numerical solution of the ODEs and then cover the Python software used for calculating the coefficients of all series and for calculating the motion of the moon's perigee..

## 12.1. HillsLunarProblem (numeric calculation of orbits)

This is the software for numerically calculating orbits based on the Hamiltonian ODEs. We use the software CVODE of Sundials version 4.4.0, which is pure ANSI C language (Hindmarsh et al. 2005). Based on an existing sample, we created a command-line program in C with the following properties:

```
Usage: HillsLunarProblem directory fileNameOrbit fileNameZeros
fileNameStats tInc tMax q1(0) q2(0) p1(0) p2(0) c d/a.
where
directory = name of a directory for output
fileNameOrbit = name of a tsv file for the orbit data
fileNameZeros = name of a tsv file for the zeros data
fileNameStats = name of a tsv file for the statistics data
tInc = time increment
tMax = maximum time
```
q1(0) = initial value of $q_1$
q2(0) = initial value of $q_2$
p1(0) = initial value of $p_1$
p2(0) = initial value of $p_2$
c = C (energy) for calculating an initial value of $p_2$ from the others
d = "r" if retrograde orbit (will set $\dot{q}_2$ to negative)

## 12.2. Python

Python, including Sympy, is used for calculating the coefficients of all series and for calculating the motion of the perigee.

### 12.2.1. Data Files and their Functions

The data of all calculations is saved in files, so that it can be reused for other calculations. Each of these files is a Python pickle file and contains one data structure, which is a dictionary containing a name for the variable and its value.

For example, the coefficients of the Hill series are implemented as Python functions, where each function returns the value retrieved from the file, or, if it hasn't been calculated, calculated the value, saves it, and returns it. For these coefficients we use the older names in the software where the function
def a(jIn, kIn, mIn=None):
calculates and returns the coefficient $a(j, k, m) = a_{j,k,m}$.

#### 12.2.1.1. HillData.pckl

This contains all of the basic Hill data, such as the coefficients of the Hill series.

| function | Data |
|---|---|
| def E(jIn, iIn, kIn, mIn=None): | $E_{j,i,k}$ |
| def PSE(jIn, iIn, kIn, mIn=None): | partial sum of $E_{j,i,k}$ up to $E_{j,i,k}m^k$ |
| def F(jIn, kIn, mIn=None): | $F_{j,k}$ |
| def PSF(jIn, kIn, mIn=None): | partial sum of $F_{j,k}$ up to $F_{j,k}m^k$ |
| def G(jIn, kIn, mIn=None): | $G_{j,k}$ |
| def PSG(jIn, kIn, mIn=None): | partial sum of $G_{j,k}$ up to $G_{j,k}m^k$ |
| def a(jIn, kIn, mIn=None): | $a_{j,k}$ |

| function | Data |
|---|---|
| def PSa(jIn, kIn, mIn=None): | partial sum of $a_{j,k}$ up to $a_{j,k}m^k$ |
| def b(jIn, kIn, mIn=None): | $\bar{a}_{j,k}$ |
| def PSb(jIn, kIn, mIn=None): | partial sum of $\bar{a}_{j,k}$ up to $\bar{a}_{j,k}m^k$ |
| def c(jIn, kIn, mIn=None): | $\bar{\bar{a}}_{j,k}$ |
| def PSc(jIn, kIn, mIn=None): | partial sum of $\bar{\bar{a}}_{j,k}$ up to $\bar{\bar{a}}_{j,k}m^k$ |
| def a0(kIn, mIn=None): | $a_{0,k}$ |
| def PSa0(kIn, mIn=None): | partial sum of $a_{0,k}$ up to $a_{0,k}m^k$ |
| def C(kIn, mIn=None): | energy function $C_{k,m}$ |
| def PSC(kIn, mIn=None): | partial sum of $C_{k,m}$ up to $C_k m^k$ |
| def A(jIn, kIn, mIn=None): | $A_{j,k,m}$ |
| def PSA(jIn, kIn, mIn=None): | partial sum of $A_{j,k}$ up to $A_{j,k}m^k$ |
| def B(jIn, kIn, mIn=None): | $B_{j,k,m}$ |
| def PSB(jIn, kIn, mIn=None): | partial sum of $B_{j,k}$ up to $B_{j,k}m^k$ |
| def PSq(qIn, kIn, mIn=None): | partial sum of $q_{j,k}$ up to $q_{j,k}m^k$ |
| def PSqd(qIn, kIn, mIn=None): | partial sum of $\dot{q}_{j,k}$ up to $q_{j,k}m^k$ |
| def PSqdd(qIn, kIn, mIn=None): | partial sum of $\ddot{q}_{j,k}$ up to $q_{j,k}m^k$ |
| def PSqs(kIn, mIn=None): | partial sum of $q^2{}_k$ up to $q^2{}_{j,k}m^k$ |
| def PSqam3(kIn, mIn=None): | partial sum of $\|q\|^{-3}{}_k$ up to $\|q\|^{-3}{}_k m^k$ |

### 12.2.1.2. CuspData.pckl

This contains data relevant to our investigation of the behavior of the cusped orbit.

| function | Data |
|---|---|
| def qd(qi, nDer): | nth derivative of $q_i$ with respect to $t$ |
| def TaylorForOrbit(qi, maxO): | Taylor expansion for calculating orbits |
| def TaylorForOrbitQdot(qi, maxO): | Taylor expansion for calculating orbits |
| def TaylorForOrbitIC(qi=1, maxO=3, oName='g01', point0='r'): | Taylor expansion for calculating an orbit |
| def TaylorHill(qi, maxO, point0='r'): | Taylor expansion at point at right/top |

### 12.2.1.3. PerigeeData.pckl

This contains data created as part of calculations of the movement of the moon's perigee.

| function | Data |
|---|---|
| def R(jIn, kIn, mIn=None): | $R_{j,k}$ |
| def PSR(jIn, kIn, mIn=None): | partial sum of $R_j$ up to $R_{j,k}m^k$ |
| def U(jIn, kIn, mIn=None): | $U_{j,k}$ |
| def PSU(jIn, kIn, mIn=None): | partial sum of $U_j$ up to $U_{j,k}m^k$ |
| def thet(jIn, kIn, mIn=None): | $\theta_{j,k}$ |
| def PSthet(jIn, kIn, mIn=None): | partial sum of $\theta_j$ up to $\theta_{j,k}m^k$ |
| def c1(maxOmega, maxLines, mIn=None, printM=False): | $c$ (period of perigee) based on maxLines x maxLines matrix |
| def cElim(maxLines=1, maxTheta=1, maxK=5, mIn=None, printM=False): | eliminate all $w_i$ |
| def cSolve(maxLines=1, maxTheta=1, maxK=5, solNum=1, mIn=None, printM=False, variant="normal"): | solve for $c$ |

### 12.2.2. Modules

Each module contains an entry point called `main()`, which consists mainly of a list of subroutines, typically called `sub1()` at the end of the file. This is the best place to see what logic is included in the module. Normally, most of the calls to these subroutines are commented out, and only the call to the currently used subroutine is uncommented.

#### 12.2.2.1. HillSeries2Lx

This is not a module, but the main unique entry point to the whole project. The name is historically motivated and stands for version 2 and the variant created for Linux. However, it is now used for all purposes, including Windows and Linux.

I do almost all of my work in VS (Visual Studio) on Windows and use Linux for calculations that require high performance (see below). The Visual Studio project HillSeries2Lx contains a Startup File called `runmain.py`, which does nothing other than to call `HillSeries2.main()`. In Linux, the code is called by starting the file run.sh containing the lines
```
python3.9 -m pip install -r HillSeries2Lx/requirements.txt
python3.9 -c "import HillSeries2Lx;HillSeries2Lx.main()"
```
I have two Linux machines. One runs on WSL (Windows Subsystem for Linux). As a result, this Linux machine shares the same files as the VS version making it easy to move between Windows and Linux. The high-performance Linux machine is a VM (virtual machine) running on Azure with one CPU and over 200 GB or RAM. This is needed for creating Taylor expansions to a high order in $m$, because each derivative makes the expression longer and more complex. I do not use the high-availability and scale-out features of Azure. Instead, I limit the cost by running the VM as a "spot price" offering. The VM is a basic Ubuntu machine with Python software included. For all tasks on this machine, I use standard Linux commands, including SSH, SCP, and
```
nohup ./HillSeries2Lx/run.sh >./HillSeries2Lx/pickle/output.txt 2>&1 &
```
This redirects Standard Output and Standard Error to the pickle subdirectory, and I retrieve (download) the file using SCP. When moving between Windows and Linux, I need to make one small change in the Python code:
```
# from . import config # for use with run.bat or run.sh
import config # for use in Visual Studio
```

#### 12.2.2.2. config

This is a small module containing some names and constants used globally.

#### 12.2.2.3. HillSeries2

This module contains the logic for calculating the coefficients for the Hill series. For example, there is a Python function called
```
def a(jIn, kIn, mIn=None):
```
which calculates $a_{j,k}$. It checks the file HillData.pckl to see if the coefficient has already been calculated and updates the file if necessary.

#### 12.2.2.4. HillPlots

This module plots orbits in numerous ways, such as spatial coordinates, velocity, etc.

#### 12.2.2.5. HillSeriesPlots

This module plots parameters (for investigating convergence), orbits, recreating diagrams from the Hénon paper,

#### 12.2.2.6. HillCusp

This one contains investigations around the cusped orbit.

#### 12.2.2.7. HillPerigee

This module contains all calculations relevant to the motion of the moon's perigee.

## 12.3. Files

All files can be found here:
arXiv Password: HillSeries#2015

## 13. Outlook

In this manuscript, we have followed Hill's original documents as closely as possible. One method that is painfully missing in the calculations of the motion of the perigee is a recursive procedure, which was developed after Hill's original paper. In section 5. Wintner's Recursion Formula and section 6. Transition to Finite Sums, we have documented how to find a recursive procedure for finding the coefficients of the series and provided a description of the fact that we only need finite sums, not infinite sums. We don't have a similar procedure for the perigee. Another important topic is the fact that Hill calculated the motion of the perigee to the first order (linear approximation) and didn't include the node. In order to show the gap between Hill's work and what is available today, we will provide a quick review of some more recent papers, with the question of whether it would be worthwhile to make an update of these results based on today's technology.

Shortly after Hill's work, Ernest W. Brown, (Ernest W Brown 1899; Ernest W. Brown 1904; Ernest W Brown 1905, 1908) provided a calculation using a recursive procedure and including calculations of the perigee and node up to the 6th order. For some time, the accuracy of this work was superior to the accuracy of the observations, and the work necessary to improve this accuracy was of the order of a scientist's lifetime (Henrard 1981). This data was used as a basis for the lunar ephemerides for some time (Dieter S. Schmidt 1980b) and (Gutzwiller 1998).

A recursive procedure, also called solution by successive approximations, is presented in (Bourne 1972), and as described there, is much better suited to implementation by a computer. The calculations were done on the Cambridge Algebra system (CAMAL), and the results were reported, making this paper very useful for us for comparing our results with others, see section 9.11. Bourne's calculations of coefficients.

The method for finding the recursive procedure by using Lie transformations is described in (Dieter S Schmidt 1979) based on work by (Deprit 1969) and (Henrard 1970). (Dieter S Schmidt 1980a) compares the results of Brown with calculations up to order 6 carried out on the software POLYPAX and programming language PL/I. (Dieter S. Schmidt 1980b) reports calculations up to order 9 and compares them with the results of Brown.

(Henrard 1981) is a review of the results achieved up to 1982, including the work by Brown, higher order calculations the motion of the perigee and node, and work that includes the shape of the earth, effects of other planets, relativity, the shape of the moon, and tides. (Gutzwiller 1998) is a very informative 51-page review of all the work on the moon-earth-sun system from antiquity to the end of the 20th century.

With this, we have covered the important work based on manual calculations, and the work using early computers in the 1970s. It is not clear whether it would be worth revisiting some of this with more modern computers.

In addition, to our knowledge, little is known about the radius of convergence of series such as the series for the perigee and node. This might be an interesting topic.

Now we will take a very brief look at some more modern developments, including Floquet, monodromy matrices, symplectic splitting, Conley–Zehnder indices, GIT quotients (geometric invariant theory quotient), and bifurcation graphs of orbit families. An overview can be found in (Aydin 2023a, 2023b, 2023c, 2023d)

## Acknowledgments

I would like to thank Urs Frauenfelder for many useful discussions without which this project would not have been possible. My thanks also go to the people who helped me understand how to use Sympy, especially Aaron Meurer.

## Symbols

The following symbols are used in this paper.

$A_i$: coefficient of $q_1$ in a Fourier series, see equation (4.2).

$B_i$: coefficient of $q_2$ in a Fourier series, see equation (4.3).
$a_i$: coefficient of $q_1$ and $q_1$ in a series, see equation (4.4), (4.5), (4.9) and (4.11).
$\bar{a}_i = b_i := \frac{a_i}{a_0}$
$\bar{\bar{a}}_i = c_i := \frac{b_i}{m} = \frac{a_i}{ma_0}$.
$C$: the Jacobi integral, which we have chosen to be the energy constant, see (4.1).[71]
$C^\infty$: infinitely differentiable
$i = \sqrt{-1}$.
$c$: ratio of the anomalistic to the synodic month[72], see (8.3.2.32) and (8.3.3).
$w$: evection or anomalistic orbit of the moon caused by the influence of the sun
$i, j, k$: indices of summation. Note that a sum over $i$ cannot contain an instance of $i$ meaning $\sqrt{-1}$. This is not a contradiction, since the index of summation is a "bound variable", meaning that it is determined by the summation and cannot have another meaning. Hill(G.W. Hill 1878) uses summation over $i, j$ and this other meaning does not cause any problems. Lyapunov (Wintner 1926) avoids the double use of $i$ by summing over $s, j$. We also use the programming language Python and the package SymPy for symbolic mathematics in Python. Python uses $j$ for $\sqrt{-1}$ based on the observation that $i$ is often used to denote electrical current, and SymPy uses $I$.
$m$: ratio of the sidereal month to the sidereal year[73].
$m_e$: mass of the earth
$m_m$: mass of the moon
$m_s$: mass of the sun
$\mathcal{O}(r^3)$ terms of the order $r^3$ and higher, i.e., $r^4, r^5$ etc.
$p = (p_1, p_2)$: momentum coordinates
$q = (q_1, q_2)$: spatial coordinates, $(x, y)$ in Hill
$r = |q| = \sqrt{q_1^2 + q_2^2}$: Euclidean distance between two bodies.
$\mathbb{R}^2$: configuration space $\mathbb{R}^2$.
$T^*\mathbb{R}^2$: cotangent bundle of $\mathbb{R}^2$, i.e., phase space, with $(q, p) \in T^*\mathbb{R}^2$.
$\mu$: the sum of the masses of the moon and the earth, i.e., $m_e + m_m$[74]. However, $\mu$ also denotes the mass of the moon[75], used only in section 3, otherwise $m_m$.

---

[71] Note that (G.W. Hill 1878) defines two variants of $C$. The first is $C$ of chapter I and agrees with our $C$ by a factor of $-1$. The other is $C$ of chapter II which matches $C$ of chapter I after being multiplied by $\frac{1}{2}v^2 = \frac{1}{2}\frac{n^2}{(1+m)^2} = \frac{1}{2m^2}$, see (G.W. Hill 1878) page 146.
[72] See (G.W. Hill 1878) page 8 and 18.
[73] See (G.W. Hill 1878) page 6.
[74] See (G.W. Hill 1878) pages 9,
[75] See (Frauenfelder and van Koert 2018).

## A1. Appendix 1 – Series Data

This section contains data, such as Hill's series, to the maximum power of $m$ which we have calculated. We recall: $a_i, \bar{a}_i, \bar{\bar{a}}_i$, are defined in (4.6), (4.7), (5.1). Our notation avoids a conflict with the other meanings of $b$ and $c$ in Hill's perigee papers.

$$\bar{\bar{a}}_i = c_i = \frac{a_i}{ma_0}, \qquad \bar{a}_i = b_i = \frac{a_i}{a_0}$$

The $\bar{\bar{a}}_i = c_i$ form the basis for our recursive procedure for calculating all of Hill's power series in $m$ (following Wintner), so we calculated them first. Since calculating the $a_i$ requires a complicated calculation of $a_0$, the papers of Hill and following often assume $a_0 = 1$, which means using $\bar{a}_i = b_i$ instead of $a_i$, so we find it useful to have all three of them at hand. Calculating $a_0$ involves taking the 2/3 power of a polynomial expression (partial sum of powers of $m$), and we solved this by taking a Taylor expansion of it, which requires taking successive derivatives of an expression that grows exponentially in size and complexity. Using a server with large memory (RAM), we were able to take these calculations, and thus the $a_i$, to the power of $m^{24}$. We calculated $\bar{\bar{a}}_i$ and $\bar{a}_i$ to the power of $m^{30}$.

The $A_i$ (our (4.2)) and $B_i$ (our (4.3)) are the coefficients of the Fourier expansion of the coordinates, so the indices begin with 0. We show three expressions for each series:
- the series as power series of $m$ with rational coefficients
- the series as power series of $m$ with floating-point coefficients
- the numeric value assuming the earth's moon for $m$

$a_i$

$$\begin{aligned}a_0 = m^{\frac{2}{3}}\Bigg(&1 - \frac{2m}{3} + \frac{7m^2}{18} - \frac{4m^3}{81} + \frac{19565m^4}{62208} - \frac{47161m^5}{93312} - \frac{2284055m^6}{3359232} - \frac{1152145m^7}{1259712} - \frac{65557603m^8}{1934917632} + \frac{75549738823m^9}{261213880320} + \frac{13098033424219m^{10}}{7836416409600} + \frac{179546710339933m^{11}}{29386561536000} + \frac{162516609782650657m^{12}}{10579162152960000} \\ &+ \frac{61750709189572144619m^{13}}{28436787867156480000} + \frac{509038377602479058963 11m^{14}}{37323284075642880000000} - \frac{21780360499154271421871921m^{15}}{117568344838275072000000000} - \frac{10719604824291913728780111299689m^{16}}{12640948437011335741440000000000} \\ &- \frac{284074179721873362408256860454 4413m^{17}}{132729958588619025285120000000000000} - \frac{21128851782254679097731460336846 72009m^{18}}{50171924346497991557775360000000000000} - \frac{5879522384355233385676800000000000 0 m^{19}}{35145239445699761551047460655671099 57}\\ &- \frac{14454496957100913697186964252929196966744069m^{20}}{354013098188889828431662940160000000000000000} + \frac{5154950239361107440987901132274990223200121583m^{21}}{76665961576531453469732005478400000000000000} \\ &+ \frac{919890682067625128757311871738812896371373181634617m^{22}}{283357393986860252024129492248166400000000000000000} + \frac{1283789589407029163121970163167252718074972372 07307341m^{23}}{1636388950274117955439347817733160960000000000000000} \\ &+ \frac{20563291706399885263971969165347900238425473669682871182 11897m^{24}}{1451542454451153591192919088242023097958400000000000000000}\Bigg)\end{aligned}$$

$$a_0 = m^{\frac{2}{3}}(1.0 - 0.666666666666667m + 0.388888888888889m^2 - 0.0493827160493827m^3 + 0.314509387860082m^4 - 0.505411951303155m^5 - 0.679933687223746m^6 - 0.914609847330183m^7$$
$$- 0.0338813404332035m^8 + 0.289225590655626m^9 + 1.67143152425811m^{10} + 6.10982370700224m^{11} + 15.361954702356m^{12} + 21.7150788893749m^{13}$$
$$+ 13.6386277416214m^{14} - 18.5257013944654m^{15} - 84.8006372125217m^{16} - 214.024160590849m^{17} - 421.128989120177m^{18} - 597.75670791249m^{19}$$
$$- 408.304015615503m^{20} + 672.390997693964m^{21} + 3246.39731162365m^{22} + 7845.25946103447m^{23} + 14166.5106958068m^{24})$$

$$a_0 = 0.17736945990121$$

$$a_1 = m^{\frac{2}{3}}\left(\frac{3m^2}{16} + \frac{3m^3}{8} + \frac{31m^4}{96} + \frac{11m^5}{108} - \frac{12199m^6}{55296} - \frac{2190833m^7}{2488320} - \frac{553809001m^8}{223948800} - \frac{1360134391m^9}{279936000} - \frac{80713776199259m^{10}}{12899450880000} - \frac{22928331146472407m^{11}}{6094990540800000} + \frac{4996935771615979081m^{12}}{853298675712000000}\right.$$
$$+ \frac{925392139641411684649m^{13}}{335986353356160000000} + \frac{737460772963452448460492657m^{14}}{108375758204829696000000000} + \frac{29682869502280921175804 88821m^{15}}{23707197107306496000000000} + \frac{39429018940230634076255381628779m^{16}}{2389685468416494796800000000000}$$
$$+ \frac{23948651256480158319431208 7715479m^{17}}{235234663297248706560000000000} - \frac{34530736806323615561935379 6092295 01409m^{18}}{16861620665146787286220800000000000} - \frac{10947175959812294060273185525 85386927 77861m^{19}}{11685103120946723589351014400000000000000}$$
$$- \frac{13645411560733701981576677646145223582628700883m^{20}}{6073332347112059585565189734400000000000000000} - \frac{23286524739344880432352096099963753325100 38982781m^{21}}{584558238409535735110649511936000000000000000000}$$
$$- \frac{21059738088969739554071355374068083479039578459905 00951m^{22}}{414821231839035662778040344541675520000000000000000000} - \frac{5051043103697271712819310551682041284559414776 17674201213451m^{23}}{1868562238818936142983683136998797737984000000000000000000000}$$
$$\left. + \frac{1542572294099332155895146 0221133377 0596116697754973438 440868747m^{24}}{1870430801057755079126666820135796535721984000000000000000000}\right)$$

$$a_1 = m^{\frac{2}{3}}(0.1875m^2 + 0.375m^3 + 0.322916666666667m^4 + 0.101851851851852m^5 - 0.220612702546296m^6 - 0.880446646733539m^7 - 2.47292685203046m^8 - 4.8587333926326m^9$$
$$- 6.2571482267049m^{10} - 3.76183211327231m^{11} + 5.85602194618021m^{12} + 27.5425513516206m^{13} + 68.0466540838086m^{14} + 125.206153084764m^{15}$$
$$+ 164.996688733091m^{16} + 101.807492657738m^{17} - 204.788955297157m^{18} - 936.848896111869m^{19} - 2246.7750455353m^{20} - 3983.61073529693m^{21}$$
$$- 5076.82260997229m^{22} - 2703.1709186684m^{23} + 8247.14976478673m^{24})$$

$$a_1 = 0.000268840217018271$$

$$a_{-1} = m^{\frac{2}{3}}\left(-\frac{19m^2}{16} - \frac{7m^3}{8} - \frac{157m^4}{288} - \frac{101m^5}{324} - \frac{165553m^6}{331776} + \frac{5333827m^7}{3732480} + \frac{8364744161m^8}{1343692800} + \frac{4804304819m^9}{419904000} + \frac{58244382031039m^{10}}{4837294080000} + \frac{62546595003093763m^{11}}{9142485811200000} - \frac{1104608617647891613m^{12}}{159993501696000000}\right.$$
$$- \frac{11355345504207119748 33m^{13}}{25198976517120000000} - \frac{2132817544250670295375247 63m^{14}}{16256363730724454400000000} - \frac{287581088773891318781219253923m^{15}}{11379454611507118080000000000} - \frac{116103025526838203427576798330211m^{16}}{3584528202624742195200000000000}$$
$$- \frac{11139762747845315476406011905665051m^{17}}{56456319191339689574400000000000} + \frac{16752197058510535739683424261765 48887m^{18}}{5058486199544036185866240000000000} + \frac{27993672236968345840605928153819749 24125299m^{19}}{175276546814200858384026521600000000000000}$$
$$+ \frac{72868537475950014179168650881781067544466036319m^{20}}{1821999704133617875669556692032000000000000000} + \frac{51457449728499403362583952779281818272746805692757m^{21}}{70146988609144288213277941432320000000000000000000}$$
$$+ \frac{2963363415311740097485289348662058707639517071050794 5367m^{22}}{31111592387927674708353032584062566400000000000000000000} + \frac{306015191494949020759554265529757102306408428050971278 9360843m^{23}}{5605686716456808428951049410996393213952000000000000000000000}$$
$$\left. - \frac{37683409809114938011100374594264970944821300 8333290596785 3549073m^{24}}{2805646201586632618690000230203694803582976000000000000000000000}\right)$$

$$a_{-1} = m^{\frac{2}{3}}(-1.1875m^2 - 0.875m^3 - 0.545138888888889m^4 - 0.311728395061728m^5 - 0.498990282600309m^6 + 1.42903029621056m^7 + 6.22519087770657m^8 + 11.4414361830323m^9$$
$$+ 12.0406948735767m^{10} + 6.84131168423265m^{11} - 6.90408426553932m^{12} - 45.0627232121605m^{13} - 131.198931068087m^{14} - 252.719571009215m^{15}$$
$$- 323.900438115741m^{16} - 197.316490118508m^{17} + 331.170164307755m^{18} + 1597.11454531579m^{19} + 3999.37153176432m^{20} + 7335.66055347264m^{21}$$
$$+ 9524.94934480314m^{22} + 5459.01344426847m^{23} - 13431.2764694283m^{24})$$

$$a_{-1} = -0.0015423599420059$$

$$a_2 = m^{\frac{2}{3}}\Bigg(\frac{25m^4}{256} + \frac{113m^5}{320} + \frac{23333m^6}{38400} + \frac{12779m^7}{20250} + \frac{482545279m^8}{1658880000} - \frac{43506090533m^9}{58060800000} - \frac{1166252840494783m^{10}}{329204736000000} - \frac{26425869901517621m^{11}}{2880541440000000} - \frac{5104081814391464666651m^{12}}{309715815628800000000}$$
$$- \frac{88247516167129484539 52069m^{13}}{43902216865 3824000000000} - \frac{6107237633379151520 54952673m^{14}}{6146310361153 5360000000000} + \frac{248387314840303568145 07143611m^{15}}{8067032349014 01600000000000} + \frac{50029711594119879077 354098957212847m^{16}}{39031529317469415014 4000000000000}$$
$$+ \frac{9182632853324632923264011 0091354300257m^{17}}{30054277574451449561088 0000000000} + \frac{619255324580834247578537 4986393872 9157171m^{18}}{1157089686616380808101888 00000 00000} + \frac{11825967333919770624453 77515735503 83252420439m^{19}}{18041920938565917750328688640 0000000 00000}$$
$$+ \frac{100014774224275936894240953 66377814 30360679903 14779m^{20}}{355642345541011370694479 11047168000 00000000000} - \frac{6725084881056703119831 717784516465 2081978175867 02663407m^{21}}{53399698182928573097760384 7322752 00000000000 000000}$$
$$- \frac{21078014731709113625264 979077623172 9733050 26876860396 727166951m^{22}}{432970092837443305353 3950972536566 0549 12000000000 00000000} - \frac{6054800690120348825698 4380880786576 901891 61949594259 743497938891m^{23}}{541753828662850935823 4356154386378 27620 86400000000 00000000}$$
$$- \frac{7919404847709758533250073 99929335 46101499597 308385163543 1998761305615939m^{24}}{416483007353482588231110 95120953 0725464436 37555200000 000000000000}\Bigg)$$

$a_2 = m^{\frac{2}{3}}(0.09765625m^4 + 0.353125m^5 + 0.607630208333333m^6 + 0.631061728395062m^7 + 0.290886187668789m^8 - 0.749319515628445m^9 - 3.54263688507441m^{10} - 9.17392457354046m^{11}$
$- 16.4798875512021m^{12} - 20.1009248434363m^{13} - 9.93642897042535m^{14} + 30.7904200818858m^{15} + 128.177687292739m^{16} + 305.534971871379m^{17}$
$+ 535.183514072873m^{18} + 655.471630442682m^{19} + 281.222906884531m^{20} - 1259.38630926566m^{21} - 4868.23803315735m^{22} - 11176.2951543226m^{23}$
$- 19014.9530902429m^{24})$

$$a_2 = 1.04269414226024 \cdot 10^{-6}$$

$$a_{-2} = m^{\frac{2}{3}}\Bigg(\frac{23m^5}{640} + \frac{161m^6}{1600} + \frac{3037m^7}{24000} + \frac{436771039m^8}{3317760000} + \frac{3060682759m^9}{14515200000} + \frac{1925065980737m^{10}}{8128512000000} - \frac{401054268149309m^{11}}{1440270720000000} - \frac{8196836707655786 4643m^{12}}{51619302604800000000} - \frac{33686681367765728 428709m^{13}}{10840053547008000000000}$$
$$- \frac{12725425296485089065 0497917m^{14}}{3073155180576768000 0000000} - \frac{2561677276803099433 498540183m^{15}}{537802156600934400 0000000000} - \frac{1001325754612829076 28295831961499m^{16}}{2891224393886623334 4000000000000}$$
$$+ \frac{1049789661276617824148972 781574243553m^{17}}{15027138787225724780544 00000000000} + \frac{84535879055126076911576 092539442354 00661m^{18}}{23141793732327616162 03776000000000 0000} + \frac{31655112054160314641 3522904089 651409166867m^{19}}{371232941122755509266 022400000000000000}$$
$$+ \frac{48289587331788717396386873 27889796289 5207976 91207m^{20}}{3556423455410113706944791 10471680000 00000 000000} + \frac{276868077698918455880526 28732481675 59470367 2479437773m^{21}}{17799899394327619103258 679479107584 00000 00000000 0}$$
$$+ \frac{4066763192652811388268942117 89578362 183393537 81684360 40807m^{22}}{5345309788116584016708 514475760074 75200 000000000 0000000} - \frac{3637975121761837897312 4676135874 322313204 374258179 8687827179507m^{23}}{13543845716571273955858 90385965945 690521 60000000000 00000000}$$
$$- \frac{158855469860902846532342139329 49882998358 728771 93597667650 110187439829m^{24}}{138827669117827529410 3703170698 43575154812 12518400000 000000000000000}\Bigg)$$

$a_{-2} = m^{\frac{2}{3}}(0.0359375m^5 + 0.100625m^6 + 0.126541666666667m^7 + 0.13164636351032m^8 + 0.210860529582782m^9 + 0.236828829278594m^{10} - 0.278457558416038m^{11} - 1.5879402266263m^{12}$
$- 3.1076120815901m^{13} - 4.14083394711516m^{14} - 4.76323355226696m^{15} - 3.46332770548731m^{16} + 6.9859583793092m^{17} + 36.5295275002969m^{18}$
$+ 85.2702132478403m^{19} + 135.781320580173m^{20} + 155.544743015317m^{21} + 76.0809635709764m^{22} - 268.607247741362m^{23} - 1144.263754267m^{24})$

$$a_{-2} = 2.90514300107151 \cdot 10^{-8}$$

$$a_3 = m^{\frac{2}{3}}\left(\frac{833m^6}{12288} + \frac{55583m^7}{161280} + \frac{231339263m^8}{270950400} + \frac{3601367267m^9}{2667168000} + \frac{802747258116191m^{10}}{573547806720000} + \frac{27808003808704897m^{11}}{120445039411200000} - \frac{154725263453910715709m^{12}}{37940187414528000000} - \frac{294939686566541193965 57m^{13}}{1991859839262720000000}\right.$$
$$- \frac{36184609448413491387383 0359m^{14}}{10708238495876382720000 0} - \frac{6178622109819889943123 9874083161m^{15}}{1113121391646349983744 0000000000} - \frac{1551483292976540127992 5559240485517m^{16}}{2571310414703068462448 640000000000}$$
$$- \frac{45795620219270842995218 90063213922613m^{17}}{74246588224551101853204 4800000000000} + \frac{12052285722678403532191 21313259180084 70847651m^{18}}{65859693618705809387866 50193920000000000}$$
$$+ \frac{60481517512974949325923 88860033456807872735260 003m^{19}}{98888329968486772795881 55266170880000000000000} + \frac{13223760053942767284163 592215245933228949498016 7128411m^{20}}{98987218298455259568677 43421437050880000000000 00}$$
$$+ \frac{12034411119809258485872 70378560271756524150610 274114624603m^{21}}{55735990603173964590888 43780232899461120000000 00000000} + \frac{17671474556128824687789 11782338426255569376983 0206282110886615 4489m^{22}}{77126482843237516339094 62059435754175267340288 000000000000000000}$$
$$- \frac{48615763467593821485492 33882301650764413330051 48643555087844267988273 m^{23}}{69483248393472678469890 34369345670936498346865 459200000000000000000}$$
$$\left. - \frac{78508208600377083735587 10886751000928087675337 56394926088049192903736 18481m^{24}}{97808528871374275052381 57911474242104205251079 83155200000000000000000 00}\right)$$

$$a_3 = m^{\frac{2}{3}}(0.0677897135416667m^6 + 0.344636656746032m^7 + 0.853806685651691m^8 + 1.35025887645623m^9 + 1.39961699567284m^{10} + 0.23087711992661m^{11} - 4.07813650901112m^{12}$$
$$- 14.807251030058m^{13} - 33.7913742417725m^{14} - 55.5071724987826m^{15} - 60.3382339255878m^{16} - 6.16804371949951m^{17} + 182.999419834156m^{18}$$
$$+ 611.614308101359m^{19} + 1335.90581503886m^{20} + 2159.18134576475m^{21} + 2291.23303756075m^{22} - 69.9676031153443m^{23} - 8026.72420353254m^{24})$$

$$a_3 = 5.32669426010788 \cdot 10^{-9}$$

$$a_{-3} = m^{\frac{2}{3}}\left(\frac{m^6}{192} + \frac{20191m^7}{645120} + \frac{2806421m^8}{33868800} + \frac{1407714799m^9}{10668672000} + \frac{5455071491971m^{10}}{35846737920000} + \frac{17070360834997451m^{11}}{120445039411200000} + \frac{2728450467623099m^{12}}{9485046853632000} - \frac{1217778435512281857211m^{13}}{1991859839262720000000}\right.$$
$$- \frac{21728339399329952271302 0969m^{14}}{10708238495876382720000 0} - \frac{22316973785120377330601 41576459m^{15}}{55656069582317499187200 00000} - \frac{14293436370653028639496 645257246667m^{16}}{25713104147030684624486 40000000000}$$
$$- \frac{18911915690926079050453 81059605904307m^{17}}{37123294112275509266022 400000000000} + \frac{51967953291637043570477 23206337139610 59161m^{18}}{65859693618705809387866 50193920000000000}$$
$$+ \frac{39346371096791761806281 327645051121966243045308 3m^{19}}{19777665993697354559176 310532341760000000000 00} + \frac{18582229580293741392042 458982219438420926721827 569243m^{20}}{29696165489536577870603 23026431115264000000000 00}$$
$$+ \frac{86656073264609147435100 46081577393268643036097 83152040663m^{21}}{66883188723808757509066 12536279479353344000000 00000000} + \frac{11067419075393782920614 73868517876140273309826 54480258007 4113559m^{22}}{57130728032027789880810 83006989447537235066880 000000000000000000}$$
$$+ \frac{42311527383455914006510 55823307277755391005910 98226195880665265030257 m^{23}}{23161082797824226156630 11456448556978832782288 486400000000000000000}$$
$$\left. - \frac{32852693710662068342022 53692484412523162172771 33939384010334329113286 218941m^{24}}{69552731641866151148360 23403715016607434845212 32465920000000000000000 00}\right)$$

$$a_{-3} = m^{\frac{2}{3}}(0.00520833333333333m^6 + 0.0312980530753968m^7 + 0.082861542186319m^8 + 0.131948456096504m^9 + 0.152177626431315m^{10} + 0.14172738801404m^{11}$$
$$+ 0.000287658090648052m^{12} - 0.611377573616344m^{13} - 2.02912359560326m^{14} - 4.00980053974395m^{15} - 5.55881401518907m^{16} - 5.09435278931039m^{17}$$
$$+ 0.789070680961638m^{18} + 19.8943450199485m^{19} + 62.574508438947m^{20} + 129.563310180159m^{21} + 193.72095292028m^{22} + 182.683718860634m^{23}$$
$$- 47.2342249328521m^{24})$$

$$a_{-3} = 4.3639850375903 \cdot 10^{-10}$$

$$a_4 = m^{\frac{2}{3}}\left(\frac{3537m^8}{65536} + \frac{16905377m^9}{48168960} + \frac{4868581429m^{10}}{4335206400} + \frac{223400417119m^{11}}{95591301120} + \frac{55029952624187m^{12}}{16052649984000} + \frac{3936363491156966458273m^{13}}{1335494596991385600000} - \frac{177507193271387691069397m^{14}}{6610698255107358720000000}\right.$$
$$- \frac{16711910670107908393542906133m^{15}}{80171243088814492876800000000} - \frac{123762608352151564699063901895317791m^{16}}{2031859984829145074696192000000000} - \frac{39802098367323375056534482295404464432169m^{17}}{32033796556037179396139148902400000000000}$$
$$- \frac{15226612613337366469305765058392837434530061333m^{18}}{8245499233523969976566216927477760000000000} - \frac{12876283744394775469662217459736886187947609626399m^{19}}{812477997130815810362805215777390592000000000000}$$
$$+ \frac{14804995230069835139936772995523880188250975138690785199m^{20}}{10707550026827365866874166402104337250969600000000000000} + \frac{6902150576072588516162562638930550680323907090310491067505877m^{21}}{6752502273418141736626855558159058247325895884800000000000000000}$$
$$+ \frac{24861914844646947123439022877948553281336013455772910917810409879m^{22}}{8690470425889148415038763103350707964308428003737600000000000000000}$$
$$+ \frac{973472695335861940339386233641425608009236779471398346208522965485699243m^{23}}{17126473014620232030496724245814280110369498499907584000000000000000000000}$$
$$+ \left.\frac{1870270507645948370536608549957655018258027938980614104800254228981331226892677m^{24}}{22570773268291825941836726692286768998609463763904620606259200000000000000000000}\right)$$

$$a_4 = m^{\frac{2}{3}}(0.0539703369140625m^8 + 0.350959975054475m^9 + 1.12303336445527m^{10} + 2.33703709962641m^{11} + 3.42809147891697m^{12} + 2.94749488318773m^{13} - 2.68515043980789m^{14}$$
$$- 20.8452682361359m^{15} - 60.9109925267414m^{16} - 124.250331357687m^{17} - 184.66574530054m^{18} - 158.481630147106m^{19} + 138.266878912323m^{20}$$
$$+ 1022.16190333523m^{21} + 2860.8249756633m^{22} + 5684.02317572827m^{23} + 8286.24914802262m^{24})$$

$$a_4 = 3.10872388533166 \cdot 10^{-11}$$

$$a_{-4} = m^{\frac{2}{3}}\left(\frac{23m^8}{6144} + \frac{723701m^9}{28901376} + \frac{129255937m^{10}}{1625702400} + \frac{1147545898319m^{11}}{7169347584000} + \frac{239958697016107m^{12}}{1032386052096000} + \frac{987943158398958659459m^{13}}{4006483790974156800000} + \frac{74499419911036521214577m^{14}}{12395059228326297600000}\right.$$
$$- \frac{17660864558849998624174424223m^{15}}{24051372926644347863040000000} - \frac{131146293998680050264763050662023m^{16}}{4762171839475580876881920000000000} - \frac{14965205284665062620713393384357737880017m^{17}}{240253747170278845471043616768000000000000}$$
$$- \frac{6278763548045194542124800797482359127585663m^{18}}{618412442514297748242466295608320000000000} - \frac{22395378173576595480271110862279279075012418151111m^{19}}{19499471931139579448707325178657374208000000000000}$$
$$- \frac{31136095986042207845827948235686295524488945844364754639m^{20}}{9177900022994885028749285487518003421511680000000000000} + \frac{2842219633434960642833873503392610714479295268067692951565481m^{21}}{10128753410127212604940283337238587370988843827200000000000000000}$$
$$+ \frac{54093627620807098738567079346441798903718767656732528460178861588m^{22}}{5214282255533489049232578620104247785850568022425600000000000000000}$$
$$+ \frac{60885256426667787477911071477328199941456901262820436575619453492959267m^{23}}{256897095219304530457450863687214201655542477498613760000000000000000000}$$
$$+ \left.\frac{13484439086881997641812918535262606260296839721461499282357672800626162683339329m^{24}}{338561599024377389127550900384301534979141956458569309093888000000000000000000000}\right)$$

$$a_{-4} = m^{\frac{2}{3}}(0.00374348958333333m^8 + 0.0250403648601368m^9 + 0.0795077481585806m^{10} + 0.160062806953314m^{11} + 0.232431169066001m^{12} + 0.246586086439338m^{13}$$
$$+ 0.0601041257961751m^{14} - 0.734297564330917m^{15} - 2.75391771694492m^{16} - 6.22892357180172m^{17} - 10.1530356060066m^{18} - 11.4851203420603m^{19}$$
$$- 3.39250764423581m^{20} + 28.0609026436874m^{21} + 103.741272470246m^{22} + 237.002510186782m^{23} + 398.286135395736m^{24})$$

$$a_{-4} = 2.17873990984449 \cdot 10^{-12}$$

$$a_5 = m^{\frac{2}{3}}\Bigg(\frac{732413 m^{10}}{15728640} + \frac{28066785623 m^{11}}{76299632640} + \frac{36184504862629 m^{12}}{25178878771200} + \frac{1686851966082485563 m^{13}}{458035278446592000} + \frac{26431182399563667538391 m^{14}}{3869482032316809216000} + \frac{40522615311068690946306549583 m^{15}}{45753964763249051172864000000}$$
$$+ \frac{282316521473281382044515231502667 m^{16}}{78513803533735371812634624000000} + \frac{21965287475746571027233769300447292 2423 m^{17}}{9283717485464913286162832424960000000} - \frac{40118462329858736871490606253634177790092568481 m^{18}}{4078299956494794546958187632955228160000 0000}$$
$$- \frac{9470380074133315722355132334736541798577199552457567 m^{19}}{38578474523464684277223628004558333018112000000000} - \frac{7466838916960266401920662618031411211817324859223820 4103 m^{20}}{16550165570566349554928936413955524864770048000 0000}$$
$$- \frac{45779409727487963334754788432695386878422652154475299101898 93 m^{21}}{78277731853246927648686263780495794841024515 27680000000000} - \frac{57614502159940445767732065090984604262255528397976234570174 57098941 m^{22}}{2149193405762747645522330058357292543155169091439820800 0000000000}$$
$$+ \frac{8920222044511385442433635426105405418169100706471698512857 1706658635043362 89 m^{23}}{69122637711284238932483097239794829220987158490869403746304000 0000000} $$
$$+ \frac{1370301936002282601090456746136868426241869088083136337662300237 87639725011661 m^{24}}{25719969225938569108908123886215494362686701371117885791928320000000000000}\Bigg)$$

$a_5 = m^{\frac{2}{3}}(0.0465655644734701 m^{10} + 0.367849551195428 m^{11} + 1.4370975447889 m^{12} + 3.68279921975306 m^{13} + 6.83067712391942 m^{14} + 8.8566347246081 m^{15} + 3.5957565264556 m^{16}$
$- 23.6600128236739 m^{17} - 98.370553313444 m^{18} - 245.483529121223 m^{19} - 451.16400105626 m^{20} - 584.833114650205 m^{21} - 268.074999697354 m^{22}$
$+ 1290.49213685536 m^{23} + 5327.77439959117 m^{24})$

$a_5 = 1.96269419519761 \cdot 10^{-13}$

$$a_{-5} = m^{\frac{2}{3}}\Bigg(\frac{447 m^{10}}{163840} + \frac{542476609 m^{11}}{25433210880} + \frac{21227372363 m^{12}}{262279987200} + \frac{30323638290005213 m^{13}}{152678426148864000} + \frac{142972142525815954987 m^{14}}{403071045033000960000} + \frac{5755887120975301271182682269 m^{15}}{12201057270199746979430400000}$$
$$+ \frac{4326394059409525164236346395893 m^{16}}{13085633922289228635439104000000} - \frac{817777561046258527493024652227119387 m^{17}}{12378289980619884381504432332800000000} - \frac{769766394838819594294416811089348416585473 m^{18}}{21241145606743721598740560588308480000000 0}$$
$$- \frac{6249602555218508706485454513771698381607326452 0351 m^{19}}{64297457539107807128706046674263888363520000000} - \frac{3753047186902190317479630398013949019771109792 8593 m^{20}}{2010467148999799508661624592006262449 7664000000000}$$
$$- \frac{17167485850975692139347499896493991510330481361 77061736739 m^{21}}{65231443211039106373905219817079829034187096064000 0000000} - \frac{9200013975583538873144806233292914884624104643057844 22072525409 m^{22}}{447748626200572426150485428824435946490660 227383296000000000000}$$
$$+ \frac{117674773391524057045546978374624910515226414898826740019926751072 89059903 m^{23}}{46081758474189492621655398159863219480658105660579602497536000 0000000000}$$
$$+ \frac{2659810531699447860049713497312309999206166964988227734151424913301504990465577 m^{24}}{1680371322761319848448664093899412298362197822913035205072650240000000000000000}\Bigg)$$

$a_{-5} = m^{\frac{2}{3}}(0.002728271484375 m^{10} + 0.0213294582252919 m^{11} + 0.0809340147893678 m^{12} + 0.198611153225009 m^{13} + 0.354707052981467 m^{14} + 0.471753143478284 m^{15}$
$+ 0.330621663811046 m^{16} - 0.660654712667594 m^{17} - 3.62394010704597 m^{18} - 9.71982842621312 m^{19} - 18.6675379837433 m^{20} - 26.3178078023428 m^{21}$
$- 20.5472746028311 m^{22} + 25.5360857067627 m^{23} + 158.287069986926 m^{24})$

$a_{-5} = 1.13867743970411 \cdot 10^{-14}$

$$a_6 = m^{\frac{2}{3}} \left( \frac{31979701 m^{12}}{754974720} + \frac{68656059199267 m^{13}}{174573559480320} + \frac{41102482881908729 m^{14}}{22694562732441600} + \frac{64904079515872522296863 m^{15}}{11807294479166708121600} + \frac{108106847852022080683292548183 m^{16}}{884130210600003104145408000} \right.$$
$$+ \frac{103315841143666044002432050830552157 m^{17}}{5110957818181232944366170931200000} + \frac{726694923418767594135012357927911148323 m^{18}}{35878923883632255269450199370240000000} - \frac{73811040342354593488459081721072249210839027 m^{19}}{5185199655335106261381445734723747840000000}$$
$$- \frac{2602306477388945917550638043098092039666570088799473 m^{20}}{186368520091652328907647517573483436441600000000} - \frac{160833873382714472235241832274554357297060628108815283199969 m^{21}}{366300523789678571571579760666562902621805346816000000000}$$
$$- \frac{21132595967290889968524515916063016556663368645432555410749646461 m^{22}}{21857152254530120365676164318973808399443125044510720000000000}$$
$$- \frac{68196893351008377216638410670679737415967058517467065244170104985388579 m^{23}}{42959434970221661229343153788945377865098882600921695518720000000000}$$
$$\left. - \frac{1348951936487748672499438342312450940533950304381405019675141259378667155484427 m^{24}}{8202846350954004881775699157076386231073441039350392229126471680000000000000} \right)$$

$a_6 = m^{\frac{2}{3}}(0.0423586381806268 m^{12} + 0.393278680939119 m^{13} + 1.81111587680662 m^{14} + 5.49694763947762 m^{15} + 12.2274803593304 m^{16} + 20.2145751968721 m^{17} + 20.2540891632004 m^{18}$
$- 14.2349466266761 m^{19} - 139.632298207095 m^{20} - 439.076285555796 m^{21} - 966.850380195844 m^{22} - 1587.47184171022 m^{23} - 1644.49250756826 m^{24})$

$$a_6 = 1.3051957903789 \cdot 10^{-15}$$

$$a_{-6} = m^{\frac{2}{3}} \left( \frac{24931 m^{12}}{11796480} + \frac{1671401227159 m^{13}}{87286779740160} + \frac{1447865626142297 m^{14}}{17020922049331200} + \frac{21855506520764154278773 m^{15}}{88554708593750310912000} + \frac{10746270771101924681941427 m^{16}}{20465977097222294077440000} + \frac{156399496097265084198547120106610631 m^{17}}{183994481454524385997182153523200000} \right.$$
$$+ \frac{5496389874027692072075974696810673863 m^{18}}{5979820647272042544908419989504000000} - \frac{29710614764591938344385892778330695172737107 m^{19}}{18666718759206382540973204645005492224000000}$$
$$- \frac{845078890783806017945699382614099603618862808601 m^{20}}{19413387509574637842612132830805711912960000} - \frac{96270132196084174750658693133017253316739082119586129533357 m^{21}}{6593409428214214288284356919981322471924962426880000000}$$
$$- \frac{239948479785995698549556088787667051641086933337054271156127749 m^{22}}{7285717418176706788558721439657936133147708348170240000000}$$
$$- \frac{107714218572987125546673818678183299712143841192542744350960980360813803 m^{23}}{1933174573659974755320441920502542003929449717041476298342400000000000}$$
$$\left. - \frac{1100897307397092657226275300691617186149469891680900393718843112400610633221 m^{24}}{17089263231154176837032706577242471314736335498646650477346816000000000000} \right)$$

$a_{-6} = m^{\frac{2}{3}}(0.0021134270562066 m^{12} + 0.0191483891619615 m^{13} + 0.0850638773825527 m^{14} + 0.246802308627399 m^{15} + 0.525079780948276 m^{16} + 0.85002275536139 m^{17}$
$+ 0.919156308899517 m^{18} - 0.159163563494193 m^{19} - 4.35307279766097 m^{20} - 14.6009637721161 m^{21} - 32.9340909087913 m^{22} - 55.7188264529352 m^{23}$
$- 64.4204078611258 m^{24})$

$$a_{-6} = 6.37845509639894 \cdot 10^{-17}$$

$$a_7 = m^{\frac{2}{3}} \left( \frac{75164925 m^{14}}{1879048192} + \frac{5162357562053249 m^{15}}{12103766790635520} + \frac{5744081754386214269 m^{16}}{2541791026033459200} + \frac{97185844492388716697364493 m^{17}}{12279586258333376446464000} + \frac{6751740323681162062383142001699 m^{18}}{33007527862400158880952320000} \right.$$
$$+ \frac{17527286469934908672828572861110615515941 m^{19}}{433736324282132152590690729905356800000} + \frac{43969258612388750938463236440761450069123021 m^{20}}{774219338843605892374382952881061888000000}$$
$$+ \frac{168163137213915994843694751930797229559863091970709 m^{21}}{6358531522302679714716893906516564175224832000000} - \frac{102829128428853900723314835431379439348564192197483923147 m^{22}}{638592211962952202733413591103694322090272358400000000}$$
$$- \frac{516473537145432336358722824461671899701793898639016814961503510628601 m^{23}}{725439854756843785164363701576265988502335073255090552832000000000}$$
$$\left. - \frac{11559073404555712862684211674442490602814303133872357068965391151706944261 m^{24}}{61508231685195892434623487347397652500141735023615990248243200000000000} \right)$$

$$a_7 = m^{\frac{2}{3}}(0.0400015951267311 m^{14} + 0.426508346645218 m^{15} + 2.25985602103176 m^{16} + 7.91442337289132 m^{17} + 20.4551529936669 m^{18} + 40.4100037020048 m^{19} + 56.7917338232114 m^{20}$$
$$+ 26.4468512303635 m^{21} - 161.024714211234 m^{22} - 711.945358059417 m^{23} - 1879.27259292967 m^{24})$$

$$a_7 = 9.00551419659353 \cdot 10^{-18}$$

$$a_{-7} = m^{\frac{2}{3}}\left(\frac{228281 m^{14}}{132120576} + \frac{541054656941891 m^{15}}{30259416976588800} + \frac{2608901554069868621 m^{16}}{28595149042876416000} + \frac{93868502380252063580020913 m^{17}}{30698965645833441116160000} + \frac{7013523260022612677518923549111 m^{18}}{9283367211300032593526784000000}\right.$$
$$+ \frac{34728899261963964962097441859279780815 7 m^{19}}{240964624601184529217050405502976000000} + \frac{70444606445988349287813235084758670986919601 m^{20}}{3483987024796226515684723287964778496000000 0}$$
$$+ \frac{21443245783878134956334187240544109699382001039201 m^{21}}{1766258756195188809643581640699045604229120000000 0} - \frac{34959085096947661310402820919649987574495896053913 5675609 m^{22}}{81719847124634039693541520486550882779989540864000000 00000}$$
$$- \frac{51873342381240340894223222107147436649706873325345174580290551 9517 m^{23}}{25188883845723742540429295193620346822997745599135088640000000000 0}$$
$$- \frac{5454997918275821097439477800893262688734064681235103985703013246 73316077 m^{24}}{98852515208350541412877475226033700895135027165256986132480000000 0000}\right)$$

$$a_{-7} = m^{\frac{2}{3}}(0.00172782322717091 m^{14} + 0.0178805380606142 m^{15} + 0.0912358089184289 m^{16} + 0.305770896202792 m^{17} + 0.755493464858905 m^{18} + 1.44124471878157 m^{19}$$
$$+ 2.02195375426545 m^{20} + 1.21404894433873 m^{21} - 4.2779185628713 m^{22} - 20.5937439304389 m^{23} - 55.1831979872072 m^{24})$$

$$a_{-7} = 3.78382268899528 \cdot 10^{-19}$$

$$a_8 = m^{\frac{2}{3}}\left(\frac{52553071771 m^{16}}{1352914698240} + \frac{138523389109435488089 m^{17}}{296300211034757529600} + \frac{5182626134459290672992163 m^{18}}{1851135568439647666176000} + \frac{241144222592085415988868921 9411297 m^{19}}{217186586233890762533319475 20000}\right.$$
$$+ \frac{4534095563301346730811857481605 612184821 m^{20}}{1389438154236142119861159343423 48800000} + \frac{5268343835739717165566521598156186 314793678326553 m^{21}}{7079613318174766945515578008416831 09150720000000}$$
$$+ \frac{151650696402441103510559500311379630931532020923 18988621 m^{22}}{117651491986089638269008805796173844969548677120000 00000} + \frac{3538384100692540590871305863157952422133182739448964 175094848811 m^{23}}{2622683073909619187506247936135011006672180766897602 5600000000000}$$
$$- \frac{3278117267665029599020259428347456420391648166920240 7014717515210566060 3 m^{24}}{3187906944826947293234794451344037314759726463822465 7190748160000000000}\right)$$

$$a_8 = m^{\frac{2}{3}}(0.0388443350045395 m^{16} + 0.467510261385491 m^{17} + 2.79970101748291 m^{18} + 11.1030902093172 m^{19} + 32.6325828139793 m^{20} + 74.415700391642 m^{21} + 128.898234814008 m^{22}$$
$$+ 134.914665667853 m^{23} - 102.829766501951 m^{24})$$

$$a_8 = 6.38699799198811 \cdot 10^{-20}$$

$$a_{-8} = m^{\frac{2}{3}}\left(\frac{1386117 m^{16}}{939524096} + \frac{16181140607161939 m^{17}}{9406355905865318 40} + \frac{8164782244638521219513 m^{18}}{82272691930651007385600} + \frac{2602006780036780123445 8636619 m^{19}}{68948122619333575407403 008000} + \frac{4098847586606007130736073719 0820119 m^{20}}{3859550428433728110725442620 6208000}\right.$$
$$+ \frac{25654682582506763431773062075559 3567131955149 m^{21}}{11012731828271859693024232457537 29280901120000 0} + \frac{34319760218163788589463923253878592 2410997606794441 m^{22}}{87149253323029361680747263552721366 6441101312000 0}$$
$$+ \frac{12289705423785079437870785737625578 69377492988016194689 16829 m^{23}}{29140923043440213194513865957055677 85191311963219584000 0000} - \frac{112383904719065431006877372959771599669 0123737935329701 66331859653 m^{24}}{55345606681023390507548514780278425603 4674733302511409561 600000000}\right)$$

$$a_{-8} = m^{\frac{2}{3}}(0.00147533948932375 m^{16} + 0.0172023478264013 m^{17} + 0.0992404898033572 m^{18} + 0.377386168236428 m^{19} + 1.06200130367759 m^{20} + 2.32954756209043 m^{21}$$
$$+ 3.93804409212244 m^{22} + 4.21733567103035 m^{23} - 2.03058402389144 m^{24})$$

$$a_{-8} = 2.34846033759916 \cdot 10^{-21}$$

$$a_9 = m^{\frac{2}{3}}\left(\frac{7509100901081 m^{18}}{194819716546560} + \frac{237364216876940337342 5249 m^{19}}{459383847188288073891 8400} + \frac{110652688091166586302422902069 m^{20}}{32076477129922214759497728000} + \frac{8840188360619904721006673817327111269 m^{21}}{5788481717629797619125177318113 28000}\right.$$

$$+ \frac{19408962156256506525820749150147941167213 5871 m^{22}}{38628900202284432469850911316115707658 24000} + \frac{378843377342943509126353653894567149096728242293196563987 m^{23}}{292778805816812273519140643612138377377282313420 8000000}$$

$$\left. + \frac{3208597779474038449916036079301788454807025560427015125201 2157 m^{24}}{12265968069695350199084397264130537320221242520764416 0000000}\right)$$

$$a_9 = m^{\frac{2}{3}}(0.0385438447103294 m^{18} + 0.516701269166854 m^{19} + 3.44965214362475 m^{20} + 15.2720329645261 m^{21} + 50.2446666993349 m^{22} + 129.395765614257 m^{23} + 261.585368659266 m^{24})$$

$$a_9 = 4.62719722382528 \cdot 10^{-22}$$

$$a_{-9} = m^{\frac{2}{3}}\left(\frac{1134559157 m^{18}}{869730877440} + \frac{3112464432265340299699 m^{19}}{1837535388753152295 56736} + \frac{6997029916415054212461803767 m^{20}}{6415295425984442951899 5456000} + \frac{142124378654355850187077863603 97943 m^{21}}{3062688739486665406944538263 5520000}\right.$$

$$+ \frac{53055731560850909080706165754020242302 66673 m^{22}}{36214593939641655440485229358858475 92960000} + \frac{637359192195094722937323266245862887038954546605441 74111 m^{23}}{1756672834900873641114843861672830264263693880 5248000000}$$

$$\left. + \frac{1742915929128103552963977445214994784266114673141275693 971681 m^{24}}{245319361393907003981687945282610746404424850415288 3320000000}\right)$$

$$a_{-9} = m^{\frac{2}{3}}(0.00130449451253186 m^{18} + 0.016938255727294 m^{19} + 0.109067929873882 m^{20} + 0.464051004667804 m^{21} + 1.46503731753221 m^{22} + 3.62821795574166 m^{23} + 7.10468150261291 m^{24})$$

$$a_{-9} = 1.51078316391356 \cdot 10^{-23}$$

$$a_{10} = m^{\frac{2}{3}}\left(\frac{1069535470353 m^{20}}{27487790694400} + \frac{4247394494999632645185343 m^{21}}{73890311928698187546624 00} + \frac{7387820451500375330470209203 m^{22}}{17456586193154946807889 92000} + \frac{5391564621810862283733541874644352633047 m^{23}}{26069606550510495943119096993546240000}\right.$$

$$\left. + \frac{14237272409825115083865834109251074517268 07117 m^{24}}{18920277650098497536253507583403611914 240000}\right)$$

$$a_{10} = m^{\frac{2}{3}}(0.0389094737457708 m^{20} + 0.574824274540651 m^{21} + 4.23211065998533 m^{22} + 20.681419228037 m^{23} + 75.248749902732 m^{24})$$

$$a_{10} = 3.38631129342355 \cdot 10^{-24}$$

$$a_{-10} = m^{\frac{2}{3}}\left(\frac{144484346557 m^{20}}{121762322841600} + \frac{31383562652037913015417 m^{21}}{1847257798217454688665 600} + \frac{62007935094222450445785621601 m^{22}}{513223634078755436151963 648000} + \frac{115312248744311021104453103390187668627 m^{23}}{2027636065039705240082037421060915 20000}\right.$$

$$\left. + \frac{3101689598744795126505302729197281722 11517 m^{24}}{15584548225773708634958678583693803 5200000}\right)$$

$$a_{-10} = m^{\frac{2}{3}}(0.00118660964397796 m^{20} + 0.0169892706271545 m^{21} + 0.120820498076882 m^{22} + 0.568702888711209 m^{23} + 1.99023388667451 m^{24})$$

$$a_{-10} = 9.94530064685134 \cdot 10^{-26}$$

$$a_{11} = m^{\frac{2}{3}}\left(\frac{109266428162197927 m^{22}}{2743061608975564800} + \frac{23127992831767576470432693 63403 m^{23}}{3597453282685371438169954 713600} + \frac{6180342775739879597977593429535187 m^{24}}{1194579330681711153187310587 0848000}\right)$$

$$a_{11} = m^{\frac{2}{3}}(0.039833749196397 m^{22} + 0.642899045919044 m^{23} + 5.17365621269618 m^{24})$$

$$a_{11} = 2.18629312994725 \cdot 10^{-26}$$

$$a_{-11} = m^{\frac{2}{3}}\left(\frac{83522332107m^{22}}{75591424409600} + \frac{82316151696584800977001711m^{23}}{4758536088208163278002585600} + \frac{11290159508863008837633232775 0507m^{24}}{838301284688920107499867078656000}\right)$$

$$a_{-11} = m^{\frac{2}{3}}(0.00110491808772415m^{22} + 0.0172986292781445m^{23} + 0.134679019525213m^{24})$$

$$a_{-11} = 5.8887881852064 \cdot 10^{-28}$$

$$a_{12} = m^{\frac{2}{3}}\left(\frac{2172954185 08894375m^{24}}{52666782892330 84416}\right)$$

$$a_{12} = m^{\frac{2}{3}} \cdot 0.0412585327175046m^{24}$$

$$a_{12} = 4.69331574678584 \cdot 10^{-29}$$

$$a_{-12} = m^{\frac{2}{3}}\left(\frac{863391067766779m^{24}}{822918482692669440}\right)$$

$$a_{-12} = m^{\frac{2}{3}} \cdot 0.00104918176699796m^{24}$$

$$a_{-12} = 1.1934843495302 \cdot 10^{-30}$$

$\bar{a}_i = b_i$

$$\bar{a}_1 = \frac{3m^2}{16} + \frac{m^3}{2} + \frac{7m^4}{12} + \frac{11m^5}{36} - \frac{30749m^6}{110592} - \frac{1010521m^7}{829440} - \frac{18445871m^8}{6220800} - \frac{2114557853m^9}{373248000} - \frac{5617623210853m^{10}}{716636160000} - \frac{225152471718641m^{11}}{37623398400000} + \frac{9094202047857023m^{12}}{1975228416000000} + \frac{24566758289071529423m^{13}}{829595934720000000}$$
$$+ \frac{335234246577973178630521 9m^{14}}{44599077450547200000000} + \frac{3268578330537669103632619 43m^{15}}{23414515661537280000000000} + \frac{4714998789833747210329626586 7m^{16}}{24585241444614144000000000000} + \frac{746324868804940966209393923671m^{17}}{516290070336897024000000000000}$$
$$- \frac{10998738615910394188334371558 3573303m^{18}}{69389385453278960025600000000000000} - \frac{37157567447200822007307916426699398 7951m^{19}}{40072370099268599414784000000000000000} - \frac{33862763042233240453196080974967914595087m^{20}}{144636210827047601012736000000000000000000}$$
$$- \frac{457882779033816036621801385066761200779 0282787m^{21}}{106915087043353586668614451200000000000000000} - \frac{36286188185855738902445641356279371814918 22419155771m^{22}}{6322530587395757701235184186163200000000000000000}$$
$$- \frac{18567593795969045239687545049433551289206967192 227655441m^{23}}{474663983848736509420231452776202240000000000000000000} + \frac{4646322125988144965913594727493942888217369651 43984004472353m^{24}}{712707971748877868894477526343467663360000000000000000000000}$$
$$+ \frac{70530920481089930381357300276664775474766773028389528 37540865527m^{25}}{214026203916188024029011609143339307008000000000000000000000} + \frac{66912368350781013590121793236672465149401437955507854 45539372289883929m^{26}}{8226824836612001742836759530064804453618496307200000000000000000000000}$$
$$+ \frac{27067289029035336183751327910589244049300308885406188066086 729395085170697277m^{27}}{18528866238259380925304091651588455830662258307891200000000000000000000000000} + \frac{79049944033470955248419571992769050217656080740605320678763 01910411849877 84627701m^{28}}{41731638985119690689016140422290099644609071273947955200000000000000000000000000}$$
$$+ \frac{828576843913305445562701887920518870432390826717644977994857867 2391620942784 46 74258929m^{29}}{7519206712338865868346928181288230153965662462139994256793600000000000000000000000000} - \frac{12308579172155132925259232703537672645823301461450517429276609588 262761 9544814957 8888415798963m^{30}}{4335394129373493926907998463054442589252905799770799526050267136000000000000000000000000000000}$$

$\bar{a}_1 = 0.1875m^2 + 0.5m^3 + 0.583333333333333m^4 + 0.305555555555556m^5 - 0.278040002893519m^6 - 1.21831717785494m^7 - 2.96519274048354m^8 - 5.66528917234654m^9$
$\quad - 7.8388776961143m^{10} - 5.9843735891397m^{11} + 4.60412678057433m^{12} + 29.6129202915672m^{13} + 75.1661840874827m^{14} + 139.596239263959m^{15} + 191.78167521583m^{16}$
$\quad + 144.555340434477m^{17} - 158.507508663786m^{18} - 927.261535944913m^{19} - 2341.23687620146m^{20} - 4282.67695136559m^{21} - 5739.18744785416m^{22}$
$\quad - 3911.73428525516m^{23} + 6519.25095574106m^{24} + 32954.3388568951m^{25} + 81334.3783047374m^{26} + 146081.733663474m^{27} + 189424.489322497m^{28}$
$\quad + 110194.715428375m^{29} - 283909.116561309m^{30}$

$$\bar{a}_1 = 0.00151570747956276$$

$$\bar{a}_{-1} = -\frac{19m^2}{16} - \frac{5m^3}{3} - \frac{43m^4}{36} - \frac{14m^5}{27} - \frac{7381m^6}{82944} + \frac{3574153m^7}{2488320} + \frac{55218889m^8}{9331200} + \frac{13620153029m^9}{1119744000} + \frac{32912081196529m^{10}}{2149908480000} + \frac{1236405122017013m^{11}}{112870195200000} - \frac{24873006197479589m^{12}}{5925685248000000}$$
$$- \frac{10187973535191919666539m^{13}}{2488787804160000000} - \frac{1991520225951219608001449m^{14}}{16724654043955200000000} - \frac{33546705573622080072975096 73m^{15}}{1404870939692236800000000} + \frac{6206133453956157340011081 4889m^{16}}{18438931083460608000000000}$$
$$- \frac{83960826762009002728406260157981m^{17}}{30977404220213821440000000000} + \frac{70434632469766055486336634659715 1883m^{18}}{41633631271967376015360000000000} + \frac{77582585049252722614152469875526587 1859m^{19}}{60108555148902899122176000000000000}$$
$$+ \frac{47981836843266782832784059149557816126186 37m^{20}}{138850762393965696972226560000000000000} + \frac{265807871288483041374228856284318663325765 64827m^{21}}{4009315764125759500073041920000000000000}$$
$$+ \frac{441899963363166314234918737208808739296854214 6342557m^{22}}{47418979405468182759263881396224000000000000000} + \frac{2079989585891454813488966550855477979732413248015253 54451m^{23}}{2847983903092419056521388716657213440000000000000000}$$
$$- \frac{76187453272941315866755842161861865632931689604689504128 7077m^{24}}{1069061957623316803341716289515201495040000000000000000000} - \frac{7098368335947774573139458224077684169719231587462610146792076 537m^{25}}{15853792882680594372519377863773580689408000000000000000000000000}$$
$$- \frac{42826085096059769097697049994645591355881501560588900825436886500014 32183m^{26}}{37020711764754007842765417885291620041283233382400000000000000000}$$
$$- \frac{897876594231192769982843266167054489423454385518684123256477184734754970 31327m^{27}}{41689949036083607081934206216074025618990081192755200000000000000000000}$$
$$- \frac{34454749673787507911245449080178895039829194996883597501345895618696065966593 0747m^{28}}{1173702346456491300628578949376909052504630129579786240000000000000000000000}$$
$$- \frac{21932580479516052210501961253982430897067900644811020329264611310913924905622267 53369m^{29}}{105738844392226530127362867754936573654014212837384294236160000000000000000000000000}$$
$$+ \frac{31036685679013756776810315362112675172095807560288682513129220735942282182214132569 33567277163m^{30}}{975463679109036133554299654187249582581903804948429893361310105600000000000000000000000000000}$$

$\bar{a}_{-1} = -1.1875m^2 - 1.66666666666667m^3 - 1.19444444444444m^4 - 0.518518518518518m^5 - 0.0889877507716049m^6 + 1.43637192965535m^7 + 5.91766214420439m^8 + 12.1636311773048m^9$
$+ 15.3085963903584m^{10} + 10.9542215270043m^{11} - 4.19749027437368m^{12} - 40.9354848097648m^{13} - 119.07691607355m^{14} - 238.788522317723m^{15}$
$- 336.577723831451m^{16} - 271.038935880953m^{17} + 169.177250020925m^{18} + 1290.70786774133m^{19} + 3455.6408633268m^{20} + 6629.75647033985m^{21}$
$+ 9319.05260095514m^{22} + 7303.37549883251m^{23} - 7126.57042275803m^{24} - 44773.9439292307m^{25} - 115681.420087749m^{26} - 215370.038820163m^{27}$
$- 293556.111375336m^{28} - 207422.169265928m^{29} + 318173.668007423m^{30}$

$$\bar{a}_{-1} = -0.00869574696153979$$

$$\bar{a}_2 = \frac{25m^4}{256} + \frac{803m^5}{1920} + \frac{6109m^6}{7200} + \frac{897599m^7}{864000} + \frac{237203647m^8}{368640000} - \frac{11098919887m^9}{14515200000} - \frac{19388340038959m^{10}}{4572288000000} - \frac{10750132421861267m^{11}}{960180480000000} - \frac{543219484755139552363m^{12}}{25809651302400000000}$$
$$- \frac{15151012074663624045 0569m^{13}}{54200267735040000000 00} - \frac{35778967428986740840 52887m^{14}}{18970093707264000000 0000} + \frac{11150004513564865721 27651069m^{15}}{39837196785254400000 00000000} + \frac{57992078843389352089 848692402831m^{16}}{40155894359536435200 0000000000000}$$
$$+ \frac{133939212706430257173 9920040118589507m^{17}}{3710404638821166612480 000000000000} + \frac{14089030006243676114975 29544447798485819m^{18}}{21427586789192237187072 00000000000000} + \frac{8552114777526462609378176 8080988565153 01959m^{19}}{98999545096606813580427 264000000000000000}$$
$$+ \frac{39216232202745405683563858 15419407066608 04114609m^{20}}{7317743735411756598651833 5488000000000 0000000} - \frac{7508345012164292186000688 897014281692944 5684128240157m^{21}}{6592555331232451519725436 844113920000000 0000000000}$$
$$- \frac{196291056845792618510314789 462608307592083 7154707705369629m^{22}}{3712020686192072233825403 7830388940800 00000000000000000} - \frac{34302407197512598375377457 89500142758299610 43016948182999 1299689m^{23}}{26753275489523503003626450 1451179174133760 0000000000 000000000}$$
$$- \frac{1542529656864750326781892411 88715083505346805760 0000000000000000000}{3514494489934164419939924521 6942896309966810 790940180460 7922495733 52067m^{24}}$$
$$- \frac{20131890121952251172085856506 89441402381823081 99044120506654912 75875074005817m^{25}}{694832483934726784698903436 93456709369498346 8645470936853 6885898138 27935855601 971m^{26}}$$
$$- \frac{23658137614710796055839742351 3256063401528614956 43647553688589813 827935855601971m^{26}}{1564936461941988400838105265835 8787366728401727 7304832000000 000000 000000000}$$
$$+ \frac{13845867376892492913146966637350 137720347243655991843 293522835407 34133757457 89270556491m^{27}}{28197025171270747006300980679830 8630773712342330 24784629760 00000000000000000000}$$
$$+ \frac{333471566318869256053804671623 97906696644145794323751824 917026795882424 11283247036197 294820223m^{28}}{162577279851506022259049942 36454159709698396749140 4982226885017 6000000000 000000000000}$$
$$+ \frac{300537592743334654946040244163412 16932309933678862776409 64264502663875186 75777494441 8386342746 29083m^{29}}{62247995352744254567600689557391598 05048596393302786 810748530275123 2000000000 00000000000000}$$
$$+ \frac{79788388752142309272238697609160176 6702182293393792293167 3931281248 2168832531 044271224 8509649189121205697m^{30}}{9533467232258840819791748077831954162 2607683423501708 42656965722 5943449 6000000000 00000000000000000000}$$

$$\bar{a}_2 = 0.09765625m^4 + 0.418229166666667m^5 + 0.848472222222222m^6 + 1.03888773148148m^7 + 0.643456073676215m^8 - 0.764641195918761m^9 - 4.24040218791095m^{10}$$
$$- 11.1959497675492m^{11} - 21.0471454414662m^{12} - 27.9537587318386m^{13} - 18.860722556835m^{14} + 27.9889284722262m^{15} + 144.417350848063m^{16}$$
$$+ 360.982765343308m^{17} + 657.518279816278m^{18} + 863.889673118192m^{19} + 535.906060948427m^{20} - 1138.91270302932m^{21} - 5287.98391603672m^{22}$$
$$- 12821.7597919721m^{23} - 22783.9670653562m^{24} - 28973.73192449m^{25} - 15117.6346069364m^{26} + 49104.0004851281m^{27} + 205115.725040703m^{28}$$
$$+ 482806.861554756m^{29} + 836929.385797421m^{30}$$

$$\bar{a}_2 = 5.87865657842669 \cdot 10^{-6}$$

$$\bar{a}_{-2} = \frac{23m^5}{640} + \frac{299m^6}{2400} + \frac{56339m^7}{288000} + \frac{238200053m^8}{1105920000} + \frac{146886277m^9}{537600000} + \frac{164530763567m^{10}}{508032000000} - \frac{42341873843687m^{11}}{320060160000000} - \frac{56433070995456755597m^{12}}{34412868403200000000} - \frac{13955946209954151767143m^{13}}{361335118233600000000}$$

$$- \frac{6792588382013329642243 1m^{14}}{11856308567040000000000} - \frac{1769569850149311339827968 57m^{15}}{2655813119016960000000000} - \frac{30262906523808366713614880791m^{16}}{55772075499356160000000000} + \frac{56592035770884835710817098072 18077m^{17}}{12368015462737222041600000000000}$$

$$+ \frac{25468402335543205625307683804961273509m^{18}}{71425289297307457290240000000000000} + \frac{31017606563829941289423181359263401588 6049m^{19}}{329984836553560452680908800000000000000}$$

$$+ \frac{8044555622167387197750004701559333775902 7842273m^{20}}{48784958236078377324345556992000000000000000} + \frac{22978402614060127632618528520994197271254 1510245199m^{21}}{109875922187207525328757280735232000000000000000}$$

$$+ \frac{741598357108127780747357845309561114594506179 21054782699m^{22}}{4949360914922762978433871107185254400000000000000000} - \frac{81926314644191282273461798118409844146532062 42323113195623977m^{23}}{4458879248253917167271075024186319568896000000000000000000}$$

$$- \frac{561952056450220052488610146989291374155437562572229627516043845403737m^{24}}{514176552288250108927974706290502783511560192000000000000000000000} - \frac{6595623896547884733067238663040693681636469402102244744571156770294 2721237m^{25}}{2316108279824226156630114564485569788327822884864000000000000000000000}$$

$$- \frac{27324789631216319636870190599734067247408098917826199997610067586797516 14868431m^{26}}{5216454873139961336127017552786262455576133909243494400000000000000000000000} - \frac{65421824252220935178684066602708254581335252225140341706788707628 9394271953123 61849m^{27}}{9399008390423582335433660226610287692457078077674928209920000000000000000000000000}$$

$$- \frac{131463125141274116254927546347649442845584450914474936743331394868124082777081 7845933786461m^{28}}{27096213308584337043174990394090266182830661248567497037814169600000000000000000000000} + \frac{597112347861586077178348060674196262316157463758013118620317536730940053534 472103266450 2601753777m^{29}}{829973271369923394234675860765221307339812852440371574766470703349760000000000000000000000000}$$

$$+ \frac{12287078363841925217960784147470541288239928915093294863415171418976837248 0660690847834545835 8944333667m^{30}}{31778224107529469399305828025943984720753589447450056947552321907531448320000000 000000000000000}$$

$\bar{a}_{-2} = 0.0359375m^5 + 0.124583333333333m^6 + 0.195621527777778m^7 + 0.215386332646123m^8 + 0.273225961681548m^9 + 0.32385905527014m^{10} - 0.13229348458642m^{11}$
$- 1.63988280006932m^{12} - 3.86232765809709m^{13} - 5.72909210620278m^{14} - 6.66300590760058m^{15} - 5.4261754207368m^{16} + 4.57567634366138m^{17}$
$+ 35.6574017215822m^{18} + 93.997066313062m^{19} + 164.89827834306m^{20} + 209.130464224085m^{21} + 149.837195115867m^{22} - 183.737459758017m^{23}$
$- 1092.91653606014m^{24} - 2847.71828421057m^{25} - 5238.19151046707m^{26} - 6960.50280355932m^{27} - 4851.71575984107m^{28} + 7194.356354103m^{29} + 38665.088150507m^{30}$

$$\bar{a}_{-2} = 1.6379048584179 \cdot 10^{-7}$$

$$\bar{a}_3 = \frac{833m^6}{12288} + \frac{27943m^7}{71680} + \frac{12275527m^8}{11289600} + \frac{27409853579m^9}{14224896000} + \frac{287958210882233m^{10}}{127455068160000} + \frac{12754813704802943m^{11}}{13382782156800000} - \frac{9116329553696361023m^{12}}{2107788189696000000} - \frac{521693700525977598 2269m^{13}}{295090346557440000000}$$

$$- \frac{8425194073315008400481 9941m^{14}}{19830071288659968000000} - \frac{56518636752596153101138794007m^{15}}{76345774461340876800000 0000} - \frac{21211049395299963803832 13666960879m^{16}}{23808429765769152430080000000000}$$

$$- \frac{5625698346576135969810139957599345619m^{17}}{164992418267802263404554000000000000} + \frac{13749726171186454080896579435804482836130169m^{18}}{7317743735411756598651833548800000 0000000}$$

$$+ \frac{23523089861947432500065463480601525970615116184 71m^{19}}{3296277665612257598627184220569 6000000000000} + \frac{121782771705212736936676253800446443804479827818438767m^{20}}{7424041372384144467650807566077788160000000000 0000}$$

$$+ \frac{3730760314382127625183537277399016391304104098447960016 1687m^{21}}{13376637744761751501813225072558958706688000000000000000} + \frac{50439900943296840669863467785107306255714750804990399423898679 0963m^{22}}{154252965686475032678189241188715083505346805760000000 00000000000}$$

$$+ \frac{318138411197282377088693911671700698094435612901886728686810274 1435271m^{23}}{34741624196736339234945171846728354682491734327296000000000 000000000}$$

$$- \frac{12764540050290279556168111874045748513891409990588057318881295552424 62958341m^{24}}{15649364619419884008381052658358787366728401727730483200000000000000000}$$

$$- \frac{40957807449318721860759704652048921621128450064363919024787954120306556 8210846713m^{25}}{14098512585637350315049033991543153868561711651239231488000000000000000000 00}$$

$$- \frac{105338017714966334679854005000211950319296712801558912841596963640730049685462313 13140037m^{26}}{16257727985150602225904994236454159709683967491404982226885017600000000 00000000000}$$

$$- \frac{1335786455135840008912636661476569349549759380181310529245741289062863567727799668539428 8572211m^{27}}{124495990705488509135201379114783196100971927866055736214970655024640000000 00000000000000}$$

$$- \frac{22929318481535989633829870486213971363292106995725596338927048905206127944960024928316217427313486071m^{28}}{190669344645176816395834968155663908324521536684700341685313931445188669920000000000000000 00000000}$$

$$- \frac{5829988267721257688937657220910594084125089274677857898748542384209848927146207467052173523997355339568 97m^{29}}{36501977675553456201839141097430493189531808634839889287663605678281229284147200000000000000000 00000}$$

$$+ \frac{25820725056931279942115376820001466257650499697843710498948259586388043904295731319307173668862827894991 44850247299m^{30}}{71556958552883692514307455699133085740241473124500909620141365765706305389503950356480000000 00000000000000000000}$$

$\bar{a}_3 = 0.0677897135416667 m^6 + 0.389829799107143 m^7 + 1.08733055201247 m^8 + 1.92689307387555 m^9 + 2.25929196099716 m^{10} + 0.953076389898645 m^{11} - 4.32506909292968 m^{12}$
$- 17.6791178231386 m^{13} - 42.4869580682397 m^{14} - 74.0298165175015 m^{15} - 89.0905011543281 m^{16} - 34.0967082326101 m^{17} + 187.89570485571 m^{18}$
$+ 713.625860688829 m^{19} + 1640.38379632714 m^{20} + 2789.01199656325 m^{21} + 3269.94691601702 m^{22} + 915.726937220076 m^{23} - 8156.58677570224 m^{24}$
$- 29051.1550069825 m^{25} - 64792.5822176256 m^{26} - 107295.540006249 m^{27} - 120256.974314387 m^{28} - 15971.7052033205 m^{29} + 360841.567041291 m^{30}$

$$\bar{a}_3 = 3.00316315056419 \cdot 10^{-8}$$

$$\bar{a}_{-3} = \frac{m^6}{192} + \frac{7477 m^7}{215040} + \frac{65239 m^8}{627200} + \frac{2674679587 m^9}{14224896000} + \frac{944592921809 m^{10}}{3982970880000} + \frac{2991640633593949 m^{11}}{13382782156800000} + \frac{115453688848205717 m^{12}}{2107788189696000000} - \frac{185388670761024497527 m^{13}}{295090346557440000000}$$
$$- \frac{92609199806916558845338 67 m^{14}}{3966014257731993600000} - \frac{69360977335172540692880213023 m^{15}}{1374223940304135782400000 0000} - \frac{30287776432922689712803161636119 m^{16}}{396807162762819207168000 0000000}$$
$$- \frac{86399818815651747955295611 3375435073 m^{17}}{109994 94551785348422696960000000 0000} - \frac{7720259818748220684443409671 0749352571779 m^{18}}{48784958236078377324345556 99200000000000 0} + \frac{2731681946601336809805342710566 104252026490287 m^{19}}{1373449 02734009406660946600919040 000000000000}$$
$$+ \frac{1725270296930632245885695373 945490118146708133961529 m^{20}}{2474680457461381489216935855 35926272000000000000000} + \frac{857715042983264144663527950 15386507942076041476392888559 m^{21}}{55735 99060317396459088437802328994611200 0000000000000}$$
$$+ \frac{63627130974201329101547698 31441509388871094965480617 265963739883 m^{22}}{25708827614412 50544636487353145213917557800960000000000000000}$$
$$+ \frac{62062342354714649458931666 8237035416755201351494767983 6369963956175 37 m^{23}}{23161082797824226156630114 564485569788327822884864000 00000000000000000}$$
$$+ \frac{29065538724731741792056482 0688030233861287917492756904 7756489908778 99221 m^{24}}{652056859142495167015877194 09828280694701673865543680000 00000000000000000 00}$$
$$- \frac{67643601688893407203814872 99721563855861216828737788 97396849239863786951 7231291 m^{25}}{9399008390423582335433660 22661028769245707807767492820 9920000000000000000 0000}$$
$$- \frac{65653609206922924602186435 32517588982351762720775662344 96767879765046083 4079540 4370607 m^{26}}{270962133085843370431 7499039409026618283066124856749 70378141696000000000000000000}$$
$$- \frac{43774230739168041214106763 9102595714360990347320434055 72844772543717177482 5460206278654540087 m^{27}}{82997327136992339423467586076522130733981 285244037157476647070334976000000000000 00000}$$
$$- \frac{13551235284646250686643416 138873817551552077381481520 490089575039543229 527426514450893423408841033 m^{28}}{158891120537 647346996529140129719923603767947237250284737761609537657241 60000000000000000 00000}$$
$$- \frac{90038824970823534678803887 02221170081964016374001684 8073220508958120755743584 84519029077656998326765169 01 m^{29}}{9733860713480921653823770 5931479818387514896 29063714337696151420832780910592 00000000000000000000}$$
$$- \frac{40858245756950648443015922 281741835192999735 5070042121071267270291165 707055316189851403032561359 4712455359171349 m^{30}}{47 704639035255795009538303799 4220571601609820830006064 1342757177137536926335966 904320000000000000000000000}$$

$\bar{a}_{-3} = 0.00520833333333333 m^6 + 0.0347702752976191 m^7 + 0.104016262755102 m^8 + 0.188028059185811 m^9 + 0.237157877942859 m^{10} + 0.223543998440851 m^{11} + 0.0547748058427338 m^{12}$
$- 0.628243766438961 m^{13} - 2.33506976497548 m^{14} - 5.04728343764859 m^{15} - 7.6328703902521 m^{16} - 7.85488991415774 m^{17} - 1.58250823571246 m^{18}$
$+ 19.8892124296136 m^{19} + 69.7168918002641 m^{20} + 153.888902610519 m^{21} + 247.491375058004 m^{22} + 267.959589352812 m^{23} + 44.5751598456539 m^{24}$
$- 719.68870415962 m^{25} - 2422.9809700429 m^{26} - 5274.17354861511 m^{27} - 8528.62969232787 m^{28} - 9250.06301416723 m^{29} - 856.483700185943 m^{30}$

$$\bar{a}_{-3} = 2.46039258394366 \cdot 10^{-9}$$

$$\bar{a}_4 = \frac{3537 m^8}{65536} + \frac{18638507 m^9}{48168960} + \frac{1473975631 m^{10}}{1083801600} + \frac{5918813731091 m^{11}}{1911826022400} + \frac{54678083521820137 m^{12}}{11012117889024000} + \frac{3356400964488250212299 m^{13}}{667747298495692800000} - \frac{54030384478822544361277 m^{14}}{4131686409442099200000}$$

$$- \frac{37520906824758658390338346181 m^{15}}{16034248617762898575360000000} - \frac{675511955970505990561625523913529 m^{16}}{9070803503763011194060800000000} - \frac{514421297557886924378884056855791481 98149 m^{17}}{3203379655603717939613914890240000000000}$$

$$- \frac{103068740419049624707077711593472000000000 0}{26273783110295513452463872610639377132617783 1 m^{18}} - \frac{12999647954093052965804883452438249472000000 000 0}{33312819370889329761770583177716713680654279875 4749 m^{19}}$$

$$+ \frac{103068740419049624707077711593472000000000 0}{362665808331792193954872239757788288764544599 88 7 5 3834447 m^{20}} + \frac{12999647954093052965804883452438249472000000 000 0}{761874684013281496074524361110319441764056152 9 609 752455157929 m^{21}}$$

$$+ \frac{5353775013413682933437083201052186662548480 0000000 000 0}{747868486223350726683548490556524345062089982 322842099194908367261 m^{22}}$$

$$+ \frac{2172617606472287103759690775837676991077107 0000934400000000000 0}{493061435764800527942990279777085140101675789802 0093580425820304510989197 m^{23}}$$

$$+ \frac{6850589205848092081219868969832571204414779 3999963033600000000000 0}{629495787084753299419935889352616261124991877017 9799985171740097339000602071903 m^{24}}$$

$$+ \frac{5642693317072956485459181673071692249652365940 97615515156480000000000000 0}{12974873941606717509430313396381876478274296580 7334936850526473637886963397856172118151 m^{25}}$$

$$- \frac{12098735734255443064645420710863284363554123213 41649525898645340160000000000000 0}{11974857374570974520080995795192611792056719112 8490378685064867821706356219124162228087 07 013 m^{26}}$$

$$- \frac{2647082391297748388113771597329777985902006617 8633949977136461397360640000000000000000 0}{7243208189988404862644464654642930752403891844 984290288303302511042687603838621052127920601667 m^{27}}$$

$$- \frac{1385673958164535375163668115824865308222412777 438612004852862942739169280000000000000000 0}{1522291376408000086854173708145950862828516364 7678553403588088377203245729182072528545829850329 7952045159977 m^{28}}$$

$$- \frac{9934327162693834862461121157730540849675048069 55561518831232231807067248299868160000000000000000 0}{6313004526373094539428004259260452891748637717 5253499443921714723951234538694525046603440821846 26873204658718754887 m^{29}}$$

$$- \frac{2023557770570905024618375757930974345258783715 6005546006804344142972705344708967120502784000000 00000000000 0}{1411812156758528802951702065516976329784443872 8346028987994294794953904893486988816741032017512 05830 7 07250 2 967729918707 m^{30}}$$

$$\phantom{0} \frac{0}{3004267817086056391567440789098799879637220176281664814490918808777820208750307 9891433136128000000000000000 00}$$

$$\bar{a}_4 = 0.0539703369140625 m^8 + 0.38694019966385 m^9 + 1.36000503320903 m^{10} + 3.09589557927496 m^{11} + 4.96526499923497 m^{12} + 5.02645382774229 m^{13} - 1.30770777654731 m^{14}$$
$$- 23.4004771406579 m^{15} - 74.4710163427386 m^{16} - 160.587052695425 m^{17} - 254.915146954095 m^{18} - 256.25939632004 m^{19} + 67.7402033935208 m^{20}$$
$$+ 1128.28497224277 m^{21} + 3442.24627470306 m^{22} + 7197.35808043916 m^{23} + 11155.9454273388 m^{24} + 10724.1568264613 m^{25} - 4523.79473111157 m^{26}$$
$$- 52272.095808041 m^{27} - 153235.478505744 m^{28} - 311975.50266094 m^{29} - 469935.519306629 m^{30}$$

$$\bar{a}_4 = 1.75268272625015 \cdot 10^{-10}$$

$$\bar{a}_{-4} = \frac{23 m^{8}}{6144} + \frac{795829 m^{9}}{28901376} + \frac{44780807 m^{10}}{464486400} + \frac{6131572502021 m^{11}}{28677390336000} + \frac{87149512084429 m^{12}}{258096513024000} + \frac{774328898942856258667 m^{13}}{2003241895487078400000} + \frac{45411088063785748715129 m^{14}}{247901184566525952000000}$$

$$- \frac{3570160210680097723751446038 1 m^{15}}{481027458532886957260800000 00} - \frac{30555378107411066054852708130 1541 m^{16}}{9524343678951161753763840000 0000} - \frac{9354880158767503627401208420 57092321811 m^{17}}{12012673708513942273552180838 4000000000}$$

$$- \frac{5249477225153914739764341610899046568991163 m^{18}}{386507776571436092651541418475 20000000000} - \frac{32879018668241357076657954860516681781668165 29499 m^{19}}{19499471931139579448707325178 6573742080000000000}$$

$$- \frac{58216421583802665999776523415748339973974996165308708073 m^{20}}{64245300160964195201244998412626023950581760000000 00000} + \frac{13603837877412059330874202221419958600307827032935844940122699 m^{21}}{5064376705063606302470141668619293685494421913600000000000000}$$

$$+ \frac{15344282807459241271597317276614819207415349600649205765433271383 m^{22}}{130357056388337226225581446550260619464626420056064000000000000}$$

$$+ \frac{58843778862431793927492184995696398613118527184391872961914008188 0376211 m^{23}}{20551767617544276243659606909497713613244339819988910080 00000000 00}$$

$$+ \frac{7044954417038218014397181364569728105973512340268749757561445578 6012222700537 m^{24}}{1381884077650519955622656736262455244812824312075793098342400000000000}$$

$$+ \frac{47006833693380382963365480077650681010561507869395734104800640092376463422529665193717 m^{25}}{725924144055326583878725242651797061813247392804989715539187204096000000000000}$$

$$+ \frac{170511081673422294307679866679399801004191582719986830406807311979364626450389856817451633 m^{26}}{4963279483683278227713321744993333723566262408493865620713086512005120000000000000000}$$

$$- \frac{91690559510532204907179938748296536244594479268128593438796937169858471600252073400206028936985737 m^{27}}{8513580798962905236580557690362797245371850410458283215781598992018945605632000000000000000000}$$

$$- \frac{684832633104050118557328877247742975173315178047486226517963632997887307011813231338264193844704501993209 m^{28}}{149014907440407522936916817365958112745192572104333422782468483477106008724498022400000000000000000000}$$

$$- \frac{1070565841575732816489277511845553480594655307228516749740827487219685947153630301573475002857728543855722 3812441 m^{29}}{9713077298740344118168203638068676857242161834882660832660851886268985654603042178413363200000000000000000}$$

$$- \frac{491937747093504819178080210463733694242030725220503454161429351519431274088587315603618584089020280457610852846 3068710809 m^{30}}{25235849663522873689166502628429918988952649480765984441723717993733689753502587022880383434752000000000000000000000}$$

$\bar{a}_{-4} = 0.00374348958333333 m^{8} + 0.027536024582359 m^{9} + 0.0964092963755236 m^{10} + 0.21381208088254 m^{11} + 0.337662493240756 m^{12} + 0.386537891748007 m^{13} + 0.183182214894174 m^{14}$
$- 0.74219468085438 m^{15} - 3.20813476890156 m^{16} - 7.78750874764647 m^{17} - 13.5818152786473 m^{18} - 16.8614918313431 m^{19} - 9.06158449535508 m^{20}$
$+ 26.8618206536854 m^{21} + 117.709644821591 m^{22} + 286.319794761591 m^{23} + 509.807915944444 m^{24} + 647.544706679404 m^{25} + 343.545194732586 m^{26}$
$- 1076.99171095788 m^{27} - 4595.73236575623 m^{28} - 11021.9017994902 m^{29} - 19493.607453391 m^{30}$

$$\bar{a}_{-4} = 1.22836248757708 \cdot 10^{-11}$$

$$\bar{a}_5 = \frac{732413 m^{10}}{15728640} + \frac{6087081853 m^{11}}{15259926528} + \frac{1178453745829 m^{12}}{699413299200} + \frac{8525414240854708717 m^{13}}{1832141113786368000} + \frac{488694654491079992327 m^{14}}{52646014045126656000} + \frac{48377521509303991657992953 4689 m^{15}}{36603171810599240938291200000}$$

$$+ \frac{57119720930897778781049600163901 m^{16}}{65428169611446143177195520000000} - \frac{3389592064306727231903768879553517773323 m^{17}}{148539479767438612578605318799360000000} - \frac{78426754349466004849682725701809078644607 4121 m^{18}}{67971665941579909115969793882587136000000000}$$

$$- \frac{1484593433867161081296714601391017602020354690406959 m^{19}}{482230931543308553465295350056979162726400000000000} - \frac{412577929735438895847535550554498733591967875837460877 27 m^{20}}{68959023210693123145537235058148020269875200000000000}$$

$$- \frac{6590091547360945906490523176180031447780528721208817739 30441 m^{21}}{78277731853246927648686263780495794841024515276800000000} - \frac{412706584287397392431791780796711469426764018413395218502 7142430247 m^{22}}{7163978019209158818407766861190975143850563638132736000000 0000}$$

$$+ \frac{83487701233352032482372373693413265347999847565062708829089 9168917012984319 m^{23}}{69122637711284238932483097239794829220987158490869403746304 0000000000000}$$

$$+ \frac{1036124296109169767893285105039184741472549919603049078489 62503260859884 69316323 m^{24}}{168037132276131984848664093899412298362197822913035205072650 240000000000}$$

$$+ \frac{13001154796661517526746641181082603231020592623486308122517 129611474960711914035283132389 m^{25}}{8106648139638229190607934937148765789528569019936902544018485 469511680000000000000}$$

$$+ \frac{38378576214365891796971631483086625326141834313069085956850526 08392072349101265088247 3562857139 m^{26}}{12612647441574742503915449492613535765980128823978630454085720 43288505221120000000000000}$$

$$+ \frac{48266600093934113745888979845701096615809001571595011755041305 5296524784507060998731609605 34763749888133 m^{27}}{11561013789812686784132444574311311446355420698626113151785231 9362400174179195289600000000000000000}$$

$$+ \frac{203630145749968590751851501957875520664346680781969055276784096 34597836963811720722894991925264016 59632919767 m^{28}}{7628452370537974141032764034898073777081578952126307719522449182 557570023502326446161920000000000000000000}$$

$$- \frac{194847462058064240364505208843313707281429097964614530058665885 9937063533066965003142007763887680959 81222339020882103 m^{29}}{30591738580993796003110344490230261825521067619548901260255963152 25721728879364533467240988672000000000000000000000}$$

$$- \frac{442665192762522293009813390946225766431695891381453053610106205 72156311213737141432276391859794111347805590544614122720 3601513 m^{30}}{14469138568371429822337676304990674569820084144940484611740178640 85592159765398367879003548238951219200000000000000000000000}$$

$$\bar{a}_5 = 0.0465655644734701 m^{10} + 0.398893260844408 m^{11} + 1.68491755472327 m^{12} + 4.6532519666216 m^{13} + 9.28265251139782 m^{14} + 13.216756667873 m^{15} + 8.73014196639014 m^{16}$$
$$- 22.819469070537 m^{17} - 115.381539150257 m^{18} - 307.859437617562 m^{19} - 598.294335572118 m^{20} - 841.885858383822 m^{21} - 576.085776897675 m^{22}$$
$$+ 1207.81995591761 m^{23} + 6166.04367186253 m^{24} + 16037.6453655255 m^{25} + 30428.644257398 m^{26} + 41749.4528480927 m^{27} + 26693.5068686295 m^{28}$$
$$- 63692.8370521315 m^{29} - 305937.489416377 m^{30}$$

$$\bar{a}_5 = 1.10655701173394 \cdot 10^{-12}$$

$$\bar{a}_{-5} = \frac{447 m^{10}}{163840} + \frac{117747149 m^{11}}{5086642176} + \frac{199973223959 m^{12}}{2098239897600} + \frac{154681921246663037 m^{13}}{610713704595456000} + \frac{49052001826917693323 m^{14}}{100767761258250240000} + \frac{42495178607599701879125759 57 m^{15}}{61005286350998734897152000 00}$$

$$+ \frac{7874842776183337930361428219703 m^{16}}{130856339222892286354391040000 0} - \frac{128627713271033818570629269294 79863519 m^{17}}{247565799612397687631008864665 60000000} - \frac{2181982254730747698498842288493 00117716357 m^{18}}{5310286401685930399685140147077 1200000000}$$

$$- \frac{767779155762953565472398702822 5718017721946377513 1 m^{19}}{642974575391078071287060466742 6388836352000000} - \frac{839181795546239271666385902261 746605103237991704674 9 m^{20}}{344795116053465617276861752907 401013493760000000}$$

$$- \frac{191992430705721425132836026323 62906940641871740629097059636 7 m^{21}}{521851545688312850991241758536 63832273496768512000000 0} - \frac{154573790871983262051730042651 47689207219804845593469049021380 9 m^{22}}{447748626200572426150485428824 435946490660227383296000000000 0}$$

$$+ \frac{735694568290353282118738755429 6302382422613256649921326026649 922660210443 m^{23}}{460817584741894926216553981598 6321948065810566057960249753600 00000000000} + \frac{418516495282612037402130253693 34694826529752188026719080307529 6215038778665401 m^{24}}{240053046108759978349809156271 344614051742546130433600724664320 0000000000000}$$

$$+ \frac{277815239503789105659624039080 91294925957346211410972471139926 71711273001654779363394 03 m^{25}}{540443209309215279373862329143 25105265232793466246016960123236 84341200000000000000} + \frac{133170362251165619821973698631 29793808328218663369186162351107 535073410772446365736492178711 51 m^{26}}{126126474415747425039154494926 13535765980128823978630454085720 43288505221120000000000000 0}$$

$$+ \frac{124273846840778115858504307462 42949289154401438444420994224339 386299751939455743419446398795 63079762771 m^{27}}{770734252654179118942162971620 75409642369471324174087678390154 6241600116119463526400000000000 0000} + \frac{508694668549691275906593491900 9302439144659569719413068914040 384244329553274184410712644918 00085075180972 29 m^{28}}{333744791211036368670183426526 79072774731907915552596272910901 7368936885282267820195840000000 000000000}$$

$$- \frac{223915883059911554796382514579 9713994089519178325751770198847 7105325136650670913210124258334 386458432734579727093691 m^{29}}{326311878197267157366510341229 122792805558054608521613442730 27362407698441379888356983903879 16800000000000000000} - \frac{217983835003545392879163365094 87020500789915428500964265558583 183146954170921287267739416741887 02963975178366872285890052087 m^{30}}{289382771367428596446753526099 81349139640168298809692223480357 28171184319530796735758007096477 90243840000000000000000000}$$

$$\bar{a}_{-5} = 0.002728271484375 m^{10} + 0.0231483058815419 m^{11} + 0.0953052242442499 m^{12} + 0.253280579889928 m^{13} + 0.486782689368338 m^{14} + 0.69658190542866 m^{15} + 0.601792990920359 m^{16}$$
$$- 0.519569801129317 m^{17} - 4.10897283061419 m^{18} - 11.9410500064636 m^{19} - 24.3385638738604 m^{20} - 36.7906222166089 m^{21} - 34.5224489427558 m^{22}$$
$$+ 15.9649846848275 m^{23} + 174.343338718975 m^{24} + 514.050754488871 m^{25} + 1055.84781361762 m^{26} + 1612.40851062238 m^{27} + 1524.20257018492 m^{28}$$
$$- 686.202060117917 m^{29} - 7532.71640787392 m^{30}$$

$$\bar{a}_{-5} = 6.41980553102523 \cdot 10^{-14}$$

$$\bar{a}_6 = \frac{31979701 m^{12}}{754974720} + \frac{8176206447469 m^{13}}{19397062164480} + \frac{8832390156215257 m^{14}}{4255230512332800} + \frac{105775817514605724187357 m^{15}}{15743059305555610828800} + \frac{281277749662514619497467285991 m^{16}}{17682604212000062082908160000}$$
$$+ \frac{19215635301306405461114018891812 1191 m^{17}}{681461042424164392582156124160 0000} + \frac{58818677443649014103575136266 6159368379 m^{18}}{17939461941816127634725259968 512000000} - \frac{9114125783817372235891430969 3317311 55722029 m^{19}}{276543981617872333940343772 51859988480000000}$$
$$- \frac{71284295121066819949183745560547125788 4530365057317 m^{20}}{46592130022979130822269118793937085911 0400000000} - \frac{32511192499947857235069083371900122664 848244230979816020419 m^{21}}{61050087298279761928596626777760483770 300891136000000000}$$
$$- \frac{13570652890218502899581916265362654180 05352821140897546872704305 3 m^{22}}{10928576127265060182838082159486904199 72156252222553600000000 00}$$
$$- \frac{38766901052383846567521400001234644264 5356574387281402716384941850 7437 m^{23}}{17899764570925692178892980745393907443 79120108371737313280000000000}$$
$$- \frac{26098373881265364649509683438658433582 650506533571743847336374691732 1087343969 m^{24}}{10253557938692506102219623946345482788 418012991879902864080896000000 000000}$$
$$- \frac{42503370192520688549586658113678991088 6172076572087984109118334172208984 6002687337333 m^{25}}{16337360366893730597604644571950595900 398446371341863098858806530434662 40000 00000 0}$$
$$+ \frac{62263241468261474773694238393526595694 0167301736460697474462825771294816311 9517118547082 3297983 m^{26}}{69458083382844109464083676339590809088 09523976718028919150948022906133676 032000 00000 00 0}$$
$$+ \frac{54347346480226487486737399331653508504 2643448632774984291181650068018889071 5707620409873597078 92699 4663 m^{27}}{17292194838065305623277063393562049758 42846724957790513454514315059761491072 484638720 000 00000 000 0}$$
$$+ \frac{61570424981378066837392923864043604017 164515513310305749176499555663578228391 96419329935122818321428 8092627 009 m^{28}}{85547694780042671756849040354253538935 08264512667909114845613881073614815739 3804048 9902080000 000 0000 00000}$$
$$+ \frac{57628793122412589063295160936804551360 50500301900332391426656845518146913621 959314623262615252201037149 75507 35853982083 m^{29}}{46855256219008946072184855652932388945 66848796307678885584252938178397231039 40045196 6463 3769885 6960 0000 0000 00000 0}$$
$$+ \frac{28142732908777017329557581806676620332 17489099466063671726851555695984531253 453300705006096 283649194371 33046966 7744 34496369569 m^{30}}{19920453612441379661406801149774638994 76477557189214293607538156350372859779 8623029297 69466474691297280 000 00000 00000 0000 0}$$

$\bar{a}_6 = 0.0423586381806268 m^{12} + 0.421517773059536 m^{13} + 2.07565492177606 m^{14} + 6.7188857935178 m^{15} + 15.9070319218947 m^{16} + 28.1977018568083 m^{17} + 32.7873141538014 m^{18}$
$- 3.29572378704348 m^{19} - 152.996428980408 m^{20} - 532.533104188762 m^{21} - 1241.75855410495 m^{22} - 2165.77714744653 m^{23} - 2545.2993036477 m^{24}$
$- 260.160572075341 m^{25} + 8964.14620672362 m^{26} + 31428.8307465699 m^{27} + 71972.0445298799 m^{28} + 122993.230157672 m^{29} + 141275.562576549 m^{30}$

$\bar{a}_6 = 7.35862752944951 \cdot 10^{-15}$

$$\bar{a}_{-6} = \frac{24931 m^{12}}{11796480} + \frac{1794384055127 m^{13}}{87286779740160} + \frac{1667146256627447 m^{14}}{17020922049331200} + \frac{107756938499258356265887 m^{15}}{354218834375001243648000} + \frac{38136212223171656963007 1559 m^{16}}{552581381625001940090880000} + \frac{2191845447937355403321609 46768672021 m^{17}}{18399448145452438599718215 3523200000}$$

$$+ \frac{25850426212932965954476036 376666202179 m^{18}}{17939461941816127634725259968512000} + \frac{255392649404594356308608576206506171955377377 m^{19}}{74666875036825530163892818580021968896000000}$$

$$- \frac{7280020316090489190979549811552141967360000 0000 m^{20}}{33520370088526205489901376437279182470781457 4661} - \frac{659340942821421428828435691998132247192496 2426880000 00000 m^{21}}{11513269811371083015325800220693730530924285 0217410558366427}$$

$$- \frac{312245032207573148081088061699625834277758929207 2960000000000 m^{22}}{13076065093769632830240919492646120232556408541040 9152327984731}$$

$$- \frac{580550008721583312918852126186314267562017765645 11899849927 6454395652027 m^{23}}{77326982946398990212817676820101680157177988681659 05193369600000000}$$

$$- \frac{4908613218901104278552657128460731863776801300640 09437595627920300531544803 m^{24}}{51267789693462530511098119731727413944209006495939 9514320404480000000000000}$$

$$- \frac{4777911186849386461524217764364456457936577257173 934471990579914826376853901300724673 m^{25}}{11027718247653268153383135086066652232768951300655 757591729694408043397120000000000000}$$

$$+ \frac{1312990113011840003459488876391382901550431910997 011339539835936220702729574449358 2344883889 m^{26}}{63143712166221917694621523945082553716450217970163 899265008618390055760691200000000000000}$$

$$+ \frac{3983942439673083227796867996749368773158785748059 031934846040207749704383269311678973946054488 46102123 m^{27}}{46688926062776325182848071162617534347756861573860 343863271886506613560258957085245440000000000000000}$$

$$+ \frac{1217823410213069367584816446881597494340009700619 651096568282404136472187709433290455282938081017 868250869739 m^{28}}{58814040161279336832833715243549308017869318524591 875164563595432381101858208240283680768000000000000 0000}$$

$$+ \frac{2360721275801333076308719961092792623043240410850 816775493000135289744039684038792866165928871679024 4483508390997 3212493 m^{29}}{63254595895662077197449555131458725076652458750153 664955387414665408362619031906101547255589345689600 00000000000000000}$$

$$+ \frac{1909432993953115047068322615169763465691531099777 099599944390058988231005772379714218324711989703956 764946960795046896336 24611 m^{30}}{39840907224882759322813602299549277989529551143784 2858721507631270074571955972460580595389329493825945 6000000000000000000000}$$

$$\bar{a}_{-6} = 0.0021134270562066 m^{12} + 0.0205573405327659 m^{13} + 0.0979468827714274 m^{14} + 0.304210075924927 m^{15} + 0.690146528480976 m^{16} + 1.19125608040537 m^{17} + 1.44098113403706 m^{18}$$
$$+ 0.342042772352044 m^{19} - 4.60443359126878 m^{20} - 17.4617850396246 m^{21} - 41.8775760860687 m^{22} - 75.0772869444558 m^{23} - 95.7445844310903 m^{24}$$
$$- 43.3263806668813 m^{25} + 207.936794966294 m^{26} + 853.294940713867 m^{27} + 2070.63382633392 m^{28} + 3732.09447056673 m^{29} + 4792.644362176 m^{30}$$

$$\bar{a}_{-6} = 3.59614056448578 \cdot 10^{-16}$$

$$\bar{a}_7 = \frac{75164925 m^{14}}{1879048192} + \frac{5485137547831499 m^{15}}{12103766790635520} + \frac{269685852617699521 m^{16}}{105907959418060800} + \frac{46356789399262346956 5998527 m^{17}}{491183450333350 5785856000} + \frac{56699292473033854072566 12457831 m^{18}}{2200501857493334105920634880000}$$

$$+ \frac{23387444829895429067114565114328122363731 m^{19}}{43373632428213215259069072990535680000 0} + \frac{1640552941449381442916221777716322628164199 m^{20}}{19851777919066817753189306484129792000000}$$

$$+ \frac{307787913405950059864452157267500654861 44761119849127 m^{21}}{50868252178421437717735151252132513401798 6560000000} - \frac{20940824006331267147455396007312981368602598 2666895972521 m^{22}}{1383616459253063105922396114058004364528923 443200000000}$$

$$- \frac{15007172258892635705113848565012429235532 9494290615522307358742651429 m^{23}}{1813599636892109462910909253940664971255837 683137726382080000000 00}$$

$$- \frac{960257979723832976336091797568342930824742 87648542772264274767015633981 59 m^{24}}{41005487790130594956415658231598435000094 4900157439934988 2880000000000}$$

$$- \frac{186541522053273395281334564044804017224402 339838679356911672973489630820754699242159 m^{25}}{3839190367978107794796957559817152814025507 12156648644294206971445248000000000 0}$$

$$- \frac{827664266033208992089779942061559506153677 68851997731244593162377402334925617813303486 0237 m^{26}}{1111092406015808220676597893471562561014939 6503831370847897851192026333184000000000 00}$$

$$- \frac{372823751982277748289408923830935237728572 65182112744618565855387366759104472224841145 81304142906469 m^{27}}{594442422639579433675661875716346350576929 5012326245643968763422959060534669659340800 000000000}$$

$$+ \frac{350420634100611919703515863406374236503076 4751799559567256388046867463217857679755407 354371014049417751 m^{28}}{4276472826893744076294093213466425866277560 07276704082140619031598775049191668108492800 0000000000}$$

$$+ \frac{941197331918428659529666700107681882606878 579718476065229342416686621747399524091533728662638439609874789364083849 m^{29}}{1761716647736457025934013247003639652667912 05933310495584545153543143005881964215349300467807027200000000000000}$$

$$+ \frac{386544995238453178644092416648888409247808 8903129297244622722687333273935988728701476 5813821005087116473561876004492777521 m^{30}}{254927445794056277480755452894414672299657 546633737619530660218983069655231437498179051748935480639488000000000000000 0}$$

$\bar{a}_7 = 0.0400015951267311 m^{14} + 0.453176076729705 m^{15} + 2.54641722963561 m^{16} + 9.43777510577829 m^{17} + 25.766527885426 m^{18} + 53.9208812372436 m^{19} + 82.6401014628366 m^{20}$
$+ 60.5068781066779 m^{21} - 151.348474255908 m^{22} - 827.479888814369 m^{23} - 2341.77918974775 m^{24} - 4858.87659047016 m^{25} - 7449.10379687568 m^{26}$
$- 6271.8227667329 m^{27} + 8194.15083733869 m^{28} + 53425.0120828251 m^{29} + 151629.415198678 m^{30}$

$\bar{a}_7 = 5.07726291861888 \cdot 10^{-17}$

$$\bar{a}_{-7} = \frac{228281 m^{14}}{132120576} + \frac{1727729817811273 m^{15}}{90778250929766400} + \frac{2952510841426631951 m^{16}}{28595149042876416000} + \frac{4510163047019424028531 5187 m^{17}}{12279586258333376446 4640000} + \frac{891756511137501599808620144291 m^{18}}{9283367211300032593526784 00000}$$
$$+ \frac{420462580163999175810338836750 3811553553 m^{19}}{2168681621410660762953453649526 784000000} + \frac{10233778631940745199195435554926 2663200407131 m^{20}}{34839870247962265156847232879647 784960000000}$$
$$+ \frac{61473034322083526019248652651318540 2322570259 6935389 m^{21}}{254341260892107188588675756260662567 0089932800000000} - \frac{694990407742236900470907114818261692 0437386659696972391 m^{22}}{185726925283259181121685273833070188 1363398656000000000}$$
$$- \frac{8583128799595628025409502153893492499 223857480134922236023 9756604613 m^{23}}{362719927378421892582181850788132994 25116753662754527641600000000 00}$$
$$- \frac{1080954343289442970597313445627667050201 6542864326626470666302196724144067 m^{24}}{158164024333360866260460396036165392143 22160434644111778119680000000000 0}$$
$$- \frac{3140326330187745480064225787615160051485534 96174851467121118456866851003 36088196107 m^{25}}{219382306741606159702683289132408732230028 9783752277967395468408258560000 00000000}$$
$$- \frac{21088312883384036693334318772584920831795093 9297515418872600003531065007909 58067404092 3159 m^{26}}{937484217575838186195879472616630911351260533 00107719152913811943272126824000 00000000000}$$
$$- \frac{22495263001381610141577970521811283842297615615 427546655711414084996359520796061 25160466182434491057 m^{27}}{10402742396192640089324082825036061135096266271 570929876945335990178355935671903846 400000000000}$$
$$+ \frac{80391614101530701638260465188315184877226952769845 099984900493684943677870796526683830500 4745230053487843 m^{28}}{63505621506037209953296728421997642411422176708059 055619788192619241809480496271411180800000 0000000000}$$
$$+ \frac{554006987840907844992567753720799812911539221990144 33452276609244430491253605129312185046350171921176 64458150345829 m^{29}}{440429161934114256483503311750909913166978014833276 23896136288385785751470491053837325116951756800000000 00000000}$$
$$+ \frac{406304956642367555935218937428966100282486271221697 10615974804882409240863157699478529729195507801187880 76391011684335531 m^{30}}{10754751619436749206219370668983118987641802748610805 82394972798834825108007626945442874565821558947840000 00000000000000}$$

$\bar{a}_{-7} = 0.00172782322717091 m^{14} + 0.0190324202120615 m^{15} + 0.103252157804792 m^{16} + 0.367289495927329 m^{17} + 0.960595967863929 m^{18} + 1.93879348638783 m^{19} + 2.93737564437092 m^{20}$
$+ 2.41695091494261 m^{21} - 3.74200136400406 m^{22} - 23.6632402901893 m^{23} - 68.3438821088116 m^{24} - 143.144010874427 m^{25} - 224.945790958642 m^{26}$
$- 216.243584091968 m^{27} + 126.5897603945 m^{28} + 1257.87989471002 m^{29} + 3777.91111333584 m^{30}$

$\bar{a}_{-7} = 2.1332999366787 \cdot 10^{-18}$

$$\bar{a}_8 = \frac{52553071771 m^{16}}{1352914698240} + \frac{146196445549002066149 m^{17}}{29630021103475729600} + \frac{384238055673929255372 3557 m^{18}}{123409037895765110784000} + \frac{4513594747561991206132016052775417 m^{19}}{347498537974225220053311160320000}$$
$$+ \frac{37137944559411368495649663029502 52333619 m^{20}}{9262921028240947465741062289489 9200000} + \frac{2381511648134019743482645549120553933202380 4914083 m^{21}}{24778646613611684309304523029458908820275 2000000}$$
$$+ \frac{2173105620579789500357420765028196348358326 6788739971 m^{22}}{1225536374855100398635508393710144218432798720 0000000} + \frac{2261696056396147312550722959252465377887387125 8943448805203196099 m^{23}}{10490732295638476750024991744540044026688723067590410 2400000000}$$
$$- \frac{30414057214627552980008077823220980654832840693 7646754002008 23226961 m^{24}}{1135294496020992625795866969851865140583948170876946481 1152000000}$$
$$- \frac{15863533662102433887575642003174612031412907609216048219 037086837248438746186 45181 m^{25}}{1421294272474691060959009591613984810827195696915928876471 79129225 2160000000}$$
$$- \frac{41336029887935037694321877128848241186833683668 9997908433 23510666107646356548557515797 m^{26}}{10224920204753152464453747639306608000437284679585639239963200355670425 6000000000}$$
$$- \frac{5946433714888313708753001966737911701268788981341 7215805331611078488387751873 76961143556108290103 m^{27}}{60174463756488929423284687177246508745 0821109097764009290648278638509728928366592000000 0000}$$
$$- \frac{55629489431891276846916418064390475005802990695412 3520227886961530854383824910 9234353915918 3157623779449 m^{28}}{30476199611966409151434455806536975805069266219 808995120137888979237419135197625 76506880000000000}$$
$$- \frac{4382439700701818074563181086913108726080437666 707962379544990 26261633619673046934356973698 84213259863737092 19354349 m^{29}}{18750733698575936113390005998320838182500743729476254279999538 901351299691613626170954087818854400000 00000000}$$
$$- \frac{15261851783286208156892139844803186079789854244454 2634526501135850800237215894250417 60799999464194704 09676382322836097 m^{30}}{232165334411201048977290590935875844762996708610465155076866095749689800901499604804002986401044 88960000000000000000}$$

$\bar{a}_8 = 0.0388443350045395 m^{16} + 0.49340648472185 m^{17} + 3.1135325436846 m^{18} + 12.9888165109252 m^{19} + 40.0931244541377 m^{20} + 96.1114497200094 m^{21} + 177.318736935631 m^{22}$
$+ 215.589912377847 m^{23} - 26.7895751465575 m^{24} - 1116.13294792792 m^{25} - 4042.67505860042 m^{26} - 9881.98871028091 m^{27} - 18253.4207480543 m^{28}$
$- 23372.0971731077 m^{29} - 6573.69965330616 m^{30}$

$$\bar{a}_8 = 3.60094998345335 \cdot 10^{-19}$$

$$\bar{a}_{-8} = \frac{1386117 m^{16}}{939524096} + \frac{2443758832628437 m^{17}}{1343765129409 33120} + \frac{9115047895180925941003 m^{18}}{8227269193065100738 5600} + \frac{34305610416530447 68362605378873 m^{19}}{7722189732760560445629136896000} + \frac{169243201638435484360 8415454556469 m^{20}}{12865168094779093702418142068 73600}$$
$$+ \frac{8353960619642714835783845331280 33227286459031 m^{21}}{2753182957067964923256058114384232202252800000} + \frac{3236369205869851054927602148941948660 47216362669237 m^{22}}{59419945447520019327782225149582749984620544000 0}$$
$$+ \frac{1109579111319056697553261047175572 937222540003560 54993272523 m^{23}}{1665195602482297896829 36376897461016296646397898260 480000000} + \frac{38318024179909444107306802587854346 54121986915938 97375989515531799 m^{24}}{110691213362046781015097029560556 8512069349466 6050228191232000000 00}$$
$$- \frac{8947862914954125746263338998241946114386903 1056254199790391865502304733060209 m^{25}}{315843171661042457990891020358663291294932 3770924289215953620500480000 00}$$
$$- \frac{81871921932955583416117644676770410040855173 04445425455458236341257721849937 9788189 m^{26}}{7810702934186435910346612780025881111445148019 127918863807780494704640000000 0}$$
$$- \frac{85719030843207008597377709821107145596167992913398313345888045688996351 06297668790184493374067 m^{27}}{33430257642493849679602603987359171525045617 17209800051614712659102831827 3 79814400000000}$$
$$- \frac{12074950495141415595777463010111669067023 8105905985 39939 7008101082480020578623915 29595303851 35772503 m^{28}}{2539683300997200762619537983878081317 0891055183174162601149074 8269784927 93313548042240000 00000000}$$
$$- \frac{9111763054346090349236408824269339435 62177460363107041654003960 071311486955429473727 7393572732186563 4032019113 m^{29}}{14493320733198791198755560191938812 1217397052981458970280192880310223 22681477381733020605821472000000 0000}$$
$$- \frac{81205531533985657009193962397393254877 08172028012954276801513218736328986068957842870202005747769953692 3114241986439 m^{30}}{282636059283201277015832023748022 76753756121048230540618053263830397019240182560584835146170561 9865600000000000000}$$

$$\bar{a}_{-8} = 0.00147533948932375 m^{16} + 0.0181859074859505 m^{17} + 0.110790684992587 m^{18} + 0.444247183813583 m^{19} + 1.31551488788644 m^{20} + 3.03429185415973 m^{21} + 5.44660413518593 m^{22} + 6.66335600253215 m^{23} + 0.346170423252834 m^{24} - 28.3300818817664 m^{25} - 104.820171273718 m^{26} - 256.4115172545 m^{27} - 475.451033221357 m^{28} - 628.687049854244 m^{29} - 287.314830740043 m^{30}$$

$$\bar{a}_{-8} = 1.3240481910621 \cdot 10^{-20}$$

$$\bar{a}_{9} = \frac{7509100901081 m^{18}}{194819716546560} + \frac{2768538851732367209767 01 m^{19}}{510426496875875637657600} + \frac{15917735756830334179211827 51 m^{20}}{4193003546395060750875520 00} + \frac{20277803868640570348377234757 85042683 m^{21}}{115255102644362192594136863933988864} + \frac{12750115121663423740480368627922337 6818281853 m^{22}}{21070309201246054074464133445154022 35904000} + \frac{1362940269606156496144287846017521958706 74953361557230157 m^{23}}{8365108737623207814832589817489667925065208 95488000000} + \frac{85039733456730743241208403899157916259579 1297438936395267 62509 m^{24}}{24531936139390700398168794528261074640442485 0415288320000000} + \frac{13500675664775403729076016131195343944719241556 338164478899 75269998859339 m^{25}}{247912173861315549708587926777423491405180995638 9829066568 4999200000000} + \frac{43592837788875752219425958581673799773878765378 30349293980 68970021984620870461 m^{26}}{11078699225514469285377377271829500983914728333114 86813268130 92224960000000000} + \frac{298762139093314686817960752137655010894488676111 243562020196 97512203598332271 65318477659 m^{27}}{25750320332558619583395016460299346055800209576303 0328676536897 358432546324480000000000} + \frac{47201401042203857308740946929225300303490080384123 493601003765 57026067823661678824539242110 73141 m^{28}}{74437867252945492353797060807692636107534734833745 87377733424 1872337567545022218240000000000 0} + \frac{32026520273265847542568775583725969691825921609502 30940529935 58073695230630869468174634713575590 4783615911737 m^{29}}{17301660489359571181619083266880344238846837892648 4193417391610 30258557317041884863725043712000 000000000} + \frac{42676867947076262185710228550013463170073813190952 9103521207485 8805979748094120223624960867 17282915326527 86113220079 m^{30}}{10669837134489307134105871583818813762169107386 10408137690222 668680586828832086914960903997596 94561280000000000 0}$$

$$\bar{a}_{9} = 0.0385438447103294 m^{18} + 0.542397165640406 m^{19} + 3.79626098110879 m^{20} + 17.593844787255 m^{21} + 60.5122354868405 m^{22} + 162.931566385521 m^{23} + 346.649090285961 m^{24} + 544.574937749036 m^{25} + 393.483358483824 m^{26} - 1160.22688352953 m^{27} - 6341.0469407741 m^{28} - 18510.6627730685 m^{29} - 39997.6751370713 m^{30}$$

$$\bar{a}_{9} = 2.60898664776723 \cdot 10^{-21}$$

$$\bar{a}_{-9} = \frac{1134559157 m^{18}}{869730877440} + \frac{545378014613241004 9877 m^{19}}{30625589812552538 2594560} + \frac{25753684335415054352685 14539 m^{20}}{2138431808661480983966 5152000} + \frac{34845266328959114060971008 0239835619229 m^{21}}{6483099523745373333420198596 28687360000} + \frac{1892735044497049225181508787 7329598701671 m^{22}}{106513511587181339530838909878 995517440000} + \frac{10111551011666581116194410812 26917392132465333 1608758717 m^{23}}{21958410436260920513935548270910 3783032961 7350656000000} + \frac{38769011458551140147410600579152 50990994716 51343340689006161 m^{24}}{40886560232317833996479908804351244 007374750692147 20000000} + \frac{271628415059311051445667901037651 523006492890 5593001290137887 1394002633 m^{25}}{18593413039598666228144094508306761 8553885746 7292371799926374400000000} + \frac{287354799552097319887523585275434128 453305303615 104420561824144457099791759 m^{26}}{259657013097995373876032279808503929310 5014450737 97181849600000000} + \frac{60684704156763683761985284723877904705798 55849460184 864565686187835420642838080 2226289692 17 m^{27}}{231752882993027576250555148142694114502201 886186727958 08883207622589291692032000000 0} + \frac{110688884584976649114230392343044860503960248 531512854364 08182643851687943199411179543 26105919 m^{28}}{744378672529454923537970608076926361075347348 3374587377733 424187233756754502221824000000 00000} + \frac{448942796552629884768631730405005479474425588998 73355322637566 9678396984018562677989602194 99033876810250033 m^{29}}{103809962936157427089714499601282065433081027355 89051605043496 6181551344230225130918235026 227200000000000} + \frac{206534861796330768171988892413457082369139247511 7947578321529 5095940803485796407674810693746 90617152342790755 03407 m^{30}}{222288273635193898627205657996225286711856403877 168362018796 39308455589340018107283521666 1660303360000000000}$$

$$\bar{a}_{-9} = 0.00130449451253186 m^{18} + 0.0178079187356485 m^{19} + 0.120432572276107 m^{20} + 0.53747850393678 m^{21} + 1.776990558562 m^{22} + 4.60486474693492 m^{23} + 9.48209172849593 m^{24} + 14.6088517735189 m^{25} + 11.0667066575109 m^{26} - 26.1850913667337 m^{27} - 148.699699050817 m^{28} - 432.466002158894 m^{29} - 929.130711300063 m^{30}$$

$$\bar{a}_{-9} = 8.51824849016809 \cdot 10^{-23}$$

$$\bar{a}_{10} = \frac{1069535470353 m^{20}}{27487790694400} + \frac{88781267436067983408087 m^{21}}{147780623857396375093 2480} + \frac{3224223426189956650221144 1397 m^{22}}{6982634477261978723155968000} + \frac{7360399672834286446343486043442154 2674019 m^{23}}{312835278606125951326942916392255488000 0}$$
$$+ \frac{67474135310403080123359611253469464 5056700449 m^{24}}{756811106003939901450140303336144476569600 0} + \frac{27425014376363608635128077121623200339629356 83167687388259 m^{25}}{10315492115248936338855285362237699846923446 386688000000}$$
$$+ \frac{89871169450869585889905961705402921410451254600945888327 40552379 m^{26}}{141250549308709447734762365229388935888915097545238118400 00000}$$
$$+ \frac{518281468616380207116973441115709284068492386873735013354 93122333 8905907688259 m^{27}}{4366521978517478062688258803574653011703611908940997750901 6355471360 0000000}$$
$$+ \frac{2187679730048843877925087045290495615625130869653331750916611171928 0118627977837263 m^{28}}{15185016832492381710804688815311213313500480774533213778535527778720 1536000000000}$$
$$- \frac{37721457210739553437240621452844951314127572441291676816515610291 7578722434382354240824179 72301 m^{29}}{97795651751062283281357179445761397637444673223430088138303176855710094408841953 2800000000}$$
$$- \frac{1192748552037700285476653929060865261557444985842195054991681038 960283612093005671132648586 4628086471 m^{30}}{13603766341179767853569909089623193456959103824072018980390525113 336696972647551069061120 0000000000}$$

$$\bar{a}_{10} = 0.0389094737457708 m^{20} + 0.600763923704498 m^{21} + 4.61748848044275 m^{22} + 23.5280359223851 m^{23} + 89.1558471791927 m^{24} + 265.862394832549 m^{25} + 636.253592575078 m^{26}$$
$$+ 1186.94345560662 m^{27} + 1440.68311163655 m^{28} - 385.717120703476 m^{29} - 8767.78182813401 m^{30}$$

$$\bar{a}_{10} = 1.9227442098798 \cdot 10^{-23}$$

$$\bar{a}_{-10} = \frac{144484346557 m^{20}}{121762322841600} + \frac{35472468885361335067067 m^{21}}{19950384220748510637588 48} + \frac{17398623509246548342028 70899 m^{22}}{1315958036099372913210163 2000} + \frac{16012976029511572923074035 50300688982483 m^{23}}{2463577819023241866699675466 5890119680000}$$
$$+ \frac{68739600974595502743488646068445416199166 4423 m^{24}}{2897167515171332435238818348708678074368000 0} + \frac{11963466455887918407592010500480640961559234768562571 5673947 m^{25}}{17546652088038440712392840401166327439616782303756288000 0}$$
$$+ \frac{98240163288059881298465514752423776645989039906378 4574024699 58539 m^{26}}{62291492245140866451030203066160520727011558017450 01021440000 0}$$
$$+ \frac{7872681285423242605926598898440049517148333140818441 039598809632259566415078889 m^{27}}{27509088464660111794936030462520313973732755026328285 830680303946956800 0000}$$
$$+ \frac{10883187185053770119569722828375833755318748610810240381919280974082913 6664423591 m^{28}}{3170735048830085385636774511151631189040583343545997763415799124249804800 0000000}$$
$$- \frac{24759837153375727203555754961160590021228491441237147423255837901 710057837562820205634841869138019 m^{29}}{39431206786028312619043214752530995527417692243687011537363840908222 31006564507556249600000}$$
$$- \frac{48606082421539375158397880571538166652471690511697294619934059169 55379814498624032796576039 3058759 m^{30}}{25796615739853274954524982406793207406772294403232543732164695454856 73961531462857523200000000 0}$$

$$\bar{a}_{-10} = 0.00118660964397796 m^{20} + 0.0177803437231398 m^{21} + 0.132212601252983 m^{22} + 0.649988642772421 m^{23} + 2.37264847871699 m^{24} + 6.81809065106125 m^{25} + 15.7710402732764 m^{26}$$
$$+ 28.6184738383352 m^{27} + 34.3238618725629 m^{28} - 6.27924914591985 m^{29} - 188.420383943804 m^{30}$$

$$\bar{a}_{-10} = 5.64141163423046 \cdot 10^{-25}$$

$$\bar{a}_{11} = \frac{109266428162197927 m^{22}}{2743061608975564800} + \frac{240833265104892047633948656979 m^{23}}{35974532826853714381699954713600} + \frac{535598584357377460563115487012138771 m^{24}}{955663464545368922549848469667840000} + \frac{50167355081987075134265336925656708892372227826443 m^{25}}{161142112718123893666517498737090222312390656000}$$

$$+ \frac{5881414354726359428321675631242753223621068665869021310 1m^{26}}{4566122905980758650934439844214188539443901628416000 00} + \frac{486061396972321269268093026382672439221009360722817963 7204989228393687 m^{27}}{1154896131271053595957096848852953229494089545117169893 769216000000}$$

$$+ \frac{445232739327679978773969997056017711272491208011935568 254670944941505461 m^{28}}{3994767672814523276953128617054410189342584489141114820 67681280000000}$$

$$+ \frac{6132266542004255454850004988978217494313535960711999930 83337781455248914681132715180 2381 m^{29}}{258658539721308731585504908342248145339504489417243856 86081847307716335239168000000}$$

$$+ \frac{272414367386394059616453055318218704864294170955783107 07224544849435284682370219731190711326 71 m^{30}}{732934838154300421820686708278594344634019921212702192 800815225311450075337064448000000000}$$

$$\bar{a}_{11} = 0.039833749196397 m^{22} + 0.669454878716642 m^{23} + 5.60446856270868 m^{24} + 31.1323677192577 m^{25} + 128.805432438597 m^{26} + 420.870227037104 m^{27} + 1114.5397574873 m^{28} + 2370.79608839185 m^{29} + 3716.76107077126 m^{30}$$

$$\bar{a}_{11} = 1.43683921368383 \cdot 10^{-25}$$

$$\bar{a}_{-11} = \frac{83522332107 m^{22}}{75591424409600} + \frac{858213467598843538730424 79 m^{23}}{4758536088208163278002585600} + \frac{2329793219370826601676629356807853 m^{24}}{1592772440908948204249747494464000} + \frac{9765953998269811543853706425023643887000598 63 m^{25}}{12433804993682399202663387248232270240015360000}$$

$$+ \frac{44759929906120808507827630383121264301084307098153 93 m^{26}}{14269134081189870784170124513169339185762192588800 000} + \frac{42317128784039456779541967526293460544099526288655112446 1575998683 m^{27}}{4277393078781679983687813647723316010923736868561914775 4700800000}$$

$$+ \frac{5109216485223106136284284111478314393153615557044721595 980331325105261 m^{28}}{20170819018550650076923095186704714171039479866274662605 3898240000000}$$

$$+ \frac{5690513396793767010949864001310734050019124421193078501 5527430812946501317199954 87 m^{29}}{10886302176822758063363001192855561672538067736415987241 61693910257421516800000000}$$

$$+ \frac{91577817490919831056724568663865675890376802478255553550 1541194269793712315606410295972231 m^{30}}{11452106846160944090948229816853036634906561268948471762 5127378954914074271416320000000000}$$

$$\bar{a}_{11} = 0.00110491808772415 m^{22} + 0.0180352413366272 m^{23} + 0.146272823382183 m^{24} + 0.785435673410664 m^{25} + 3.13683574991186 m^{26} + 9.89320551201073 m^{27} + 25.3297423397844 m^{28} + 52.2722344499037 m^{29} + 79.9659125793252 m^{30}$$

$$\bar{a}_{11} = 3.8261875340291 \cdot 10^{-27}$$

$$\bar{a}_{12} = \frac{217295418508894375 m^{24}}{52666782892330844 16} + \frac{110279988206983628880575573392111 m^{25}}{1470958675586907432496159260672 00} + \frac{142750150124025982131566272434300615529 m^{26}}{2102459621999811629609666632692480000}$$

$$+ \frac{70617022409403236827231605780240792755575636132761837 7 m^{27}}{17295920098411964586872878197781017194863263744000000} + \frac{55171387815912731336144083452898603478829938145680572383273 83 m^{28}}{30136411179473007096167302971813644360329750747545600000 00}$$

$$+ \frac{40282952342405060524095009530947511783045347563478409124 9565159729070740151 m^{29}}{61979425711546542963641975551084899828494722546214509656 14592000000 0}$$

$$+ \frac{15125274492901967267039917335911893888248274079558556480 4277241424840438368960 41 m^{30}}{80351898333183553913571501325894422094205128010152709535 8993203200000000000 0}$$

$$\bar{a}_{12} = 0.0412585327175046 m^{24} + 0.749715067032609 m^{25} + 6.78967380064333 m^{26} + 40.8287168347216 m^{27} + 183.072189609266 m^{28} + 649.940716293221 m^{29} + 1882.37923517179 m^{30}$$

$$\bar{a}_{12} = 1.08551044904176 \cdot 10^{-27}$$

$$\bar{a}_{-12} = \frac{863391067766779 m^{24}}{822918482692669440} + \frac{20447021345073480742468981718069 m^{25}}{110321900669018057437211944550 4000} + \frac{152159912659897957553725432412688 0761 m^{26}}{9344264986665829464931851703418880000}$$
$$+ \frac{6150134885248141408822823580128955343975041074017841 m^{27}}{6485970036904486720077329324167881448073723904000000} + \frac{13830854910759656043927871666404633890638303873425207 0559 m^{28}}{336343874770904097055438649238991566521537396736000 00000}$$
$$+ \frac{65614804392587131773082671681150218885346833988263240225475319614268066931 m^{29}}{4648456928365990722272731481663313674871371041909660882242109440000000000}$$
$$+ \frac{4768471470097445275659591411673807746159797457371639828270370994797741359 35417 m^{30}}{12052784749977533087035725198884163314130769201522906430384898048000000000 00}$$

$$\bar{a}_{-12} = 0.00104918176699796 m^{24} + 0.0185339639918075 m^{25} + 0.162837754362733 m^{26} + 0.948221291534577 m^{27} + 4.11211737397637 m^{28} + 14.1153947221044 m^{29} + 39.5632342982507 m^{30}$$

$$\bar{a}_{-12} = 2.64567883136721 \cdot 10^{-29}$$

$$\bar{a}_{13} = \frac{4863942539190675027 m^{26}}{112702580674946662400} + \frac{33070411109374172695939835561 04449 m^{27}}{39225564682317531533230913617 92000} + \frac{1437020154711521990049002232079375 7689 m^{28}}{174994638405746626234017871036416 0000}$$
$$+ \frac{51416742415589565718752180556519558171923274425979 2131 m^{29}}{9676039215894805363285526264492876752371056640000000} + \frac{482584362170159215668089737196745290678153898000113916 0463658287 m^{30}}{18805120575991156428008397054411714080845764466468454 400000000}$$

$$\bar{a}_{13} = 0.0431573306490568 m^{26} + 0.843083111159951 m^{27} + 8.21179533157818 m^{28} + 53.138212101422 m^{29} + 256.623912736983 m^{30}$$

$$\bar{a}_{13} = 8.11435477645341 \cdot 10^{-30}$$

$$\bar{a}_{-13} = \frac{3152346664059167 m^{26}}{3112128080001368064} + \frac{305910248233582446031271173812 6863 m^{27}}{1588653369633860027095852001525 76000} + \frac{124717012321139440379550135157113 2368399 m^{28}}{6844877238928082383703123336953049 0880000}$$
$$+ \frac{17801885762102555344578906480375989114113507415170 63343 m^{29}}{15566328088570768128185590378002915475376937369600 00} + \frac{2620532775376254923611395476818458369590110935922307 06122693 m^{30}}{48971668166643636531271867329197172085535844964761 600000000}$$

$$\bar{a}_{-13} = 0.00101292317765334 m^{26} + 0.0192561650131261 m^{27} + 0.182204892750816 m^{28} + 1.14361496563683 m^{29} + 5.35112009347722 m^{30}$$

$$\bar{a}_{-13} = 1.82683767765794 \cdot 10^{-31}$$

$$\bar{a}_{14} = \frac{139641327379143943040461 m^{28}}{3067313435649348363878400} + \frac{4207848391002200221992344255 5323608 6783 m^{29}}{4422760869060666315434851972247 71584000} + \frac{892500403391488471792814362254809 4961 72355291 m^{30}}{89997698594698910751560977452950 43461120000}$$

$$\bar{a}_{14} = 0.0455256139643851 m^{28} + 0.951407619715124 m^{29} + 9.91692473616274 m^{30}$$

$$\bar{a}_{14} = 4.86793502701863 \cdot 10^{-32}$$

$$\bar{a}_{-14} = \frac{195638652129059907 m^{28}}{197229516181156659200} + \frac{82696469811186515043527567821016911 m^{29}}{40951489528339502920693073817108 48000} + \frac{348054863760834241631216003714 37275833595377 m^{30}}{16999565290109794253072629129666 8415426560000}$$

$$\bar{a}_{-14} = 0.000991933945370349 m^{28} + 0.0201937635880029 m^{29} + 0.204743390681484 m^{30}$$

$$\bar{a}_{-14} = 1.03013165462349 \cdot 10^{-33}$$

$$\bar{a}_{15} = \frac{356117996563085075664453 53 m^{30}}{7361552245558436073308160 00}$$

$$\bar{a}_{15} = 0.0483753948466436 m^{30}$$

$$\bar{a}_{15} = 8.21965205843189 \cdot 10^{-35}$$

$$\bar{a}_{-15} = \frac{2175362866726952084 89 m^{30}}{2212004881477895454 72000}$$

$$\bar{a}_{-15} = 0.000983434930429963 m^{30}$$

$$\bar{a}_{-15} = 1.67099265563996 \cdot 10^{-36}$$

$\bar{\bar{a}}_i = c_i$

$$\bar{a}_1 = \frac{3m}{16} + \frac{m^2}{2} + \frac{7m^3}{12} + \frac{11m^4}{36} - \frac{30749m^5}{110592} - \frac{1010521m^6}{829440} - \frac{18445871m^7}{6220800} - \frac{2114557853m^8}{373248000} - \frac{5617623210853m^9}{716636160000} - \frac{225152471718641m^{10}}{37623398400000} + \frac{9094202047857023m^{11}}{1975228416000000} + \frac{24566758289071529423m^{12}}{829595934720000000}$$

$$+ \frac{33523424657797317786305219m^{13}}{44599077450547200000000} + \frac{3268578330537669103632619 43 m^{14}}{234145156615372800000000} + \frac{471499878983374721032962658 67 m^{15}}{2458524144461414400000000 0} + \frac{74632486880490409662093939236 71 m^{16}}{516290070336897024000000000 0}$$

$$- \frac{10998738615910394188334371558 3573303 m^{17}}{69389385453278960025600000000 00} - \frac{37157567447200822007307916426 6993987951 m^{18}}{40072370099268599414784000000 0000} - \frac{33862763042233240453196080974967914 595087 m^{19}}{14463621082704760101273600000000000000}$$

$$- \frac{45788277903381603662180138506676120 07790282787 m^{20}}{10691508704353586668614451 20000000000000} - \frac{362861881858557389024456413562793 7181491822419155771 m^{21}}{63225305873957577012351841861632000 0000000000000}$$

$$- \frac{18565793795969045239687545049433551289206967192227655 441 m^{22}}{47466398384873650942023145277620224000000000000000000} + \frac{4646322125988144965913594727493942888217369651439840044 72353 m^{23}}{712707971748877868894477526343467663360000000 00000000000}$$

$$+ \frac{705309204810899930381357300276664775474766773028389528375408 65527 m^{24}}{21402620391618802402901160116094333930700800000000000000000000 000}$$

$$+ \frac{691236835078101359012179323626672465149401437955507854455393722 89883929 m^{25}}{822682483661200174283675953006480445361849630720000000000000000 00000000}$$

$$+ \frac{2706728902903533618375132791058924404930030888540618806608672939 5085170697277 m^{26}}{18528866238259380925304091651588455830662258307891200000000000000000 000000000000}$$

$$+ \frac{790499440334709552484195719927690502176560807406053206787630191041184 987784627701 m^{27}}{4173163898511969068901614042229009964460907127394795520000000000000000000 000000000000}$$

$$+ \frac{82857684391330544556270188792051887043239082671764497799485786723916209427 844674258929 m^{28}}{75192067123388658683469281812882301539656624621399425679360000000000000000000 0000000000000}$$

$$- \frac{123085791721551329252592327035376726458233014614505174292766095882627619544 81495788884 15798963 m^{29}}{4335394129373493926907998463054442589252905799770799526050267136000000000000 00000000000000000}$$

$$- \frac{209158843064327702498181477304440493189998498457685237890502226048934407161 840139049062747136 7923373 m^{30}}{165994654273984678846935172153044261467962570488074314953294140669952000000 00000000000000000000000}$$

$\bar{a}_1 = 0.1875m + 0.5m^2 + 0.583333333333333m^3 + 0.305555555555556m^4 - 0.278040002893519m^5 - 1.21831717785494m^6 - 2.96519274048354m^7 - 5.66528917234654m^8$
$- 7.8388776961143m^9 - 5.9843735891397m^{10} + 4.60412678057433m^{11} + 29.6129202915672m^{12} + 75.1661840874827m^{13} + 139.596239263959m^{14} + 191.78167521583m^{15}$
$+ 144.555340434477m^{16} - 158.507508663786m^{17} - 927.261535944913m^{18} - 2341.23687620146m^{19} - 4282.67695136559m^{20} - 5739.18744785416m^{21}$
$- 3911.73428525516m^{22} + 6519.25095574106m^{23} + 32954.3388568951m^{24} + 81334.3783047374m^{25} + 146081.733663474m^{26} + 189424.489322497m^{27}$
$+ 110194.715428375m^{28} - 283909.116561309m^{29} - 1260033.60758292m^{30}$

$$\bar{a}_1 = 0.0187474022002122$$

$$\bar{a}_{-1} = -\frac{19m}{16} - \frac{5m^2}{3} - \frac{43m^3}{36} - \frac{14m^4}{27} - \frac{7381m^5}{82944} + \frac{3574153m^6}{2488320} + \frac{55218889m^7}{9331200} + \frac{13620153029m^8}{1119744000} + \frac{32912081196529m^9}{2149908480000} + \frac{1236405122017013m^{10}}{112870195200000} - \frac{24873006197479589m^{11}}{5925685248000000}$$

$$- \frac{10187973535191966539m^{12}}{24887878041600000000} - \frac{1991520225951219608001449m^{13}}{167246540439552000000} - \frac{3354670557362208007297509673m^{14}}{140487093969223680000000} - \frac{6206133453956157340011081488 9m^{15}}{18438931083460608000000000}$$

$$- \frac{839608267620090027284062601579 81m^{16}}{30977404220213821440000000000} + \frac{70434632469766055486336634659715 1883m^{17}}{41633631271967376015360000000000} + \frac{77582585049252722614152469875526 5871859m^{18}}{60108555148902899122176000000000000}$$

$$+ \frac{479818368432667828327840591495578161261863 7m^{19}}{1388507623939656969722265600000000000000} + \frac{265807871288483041374228856284318633257656482 7m^{20}}{400931576412575950007304192000000000000000}$$

$$+ \frac{44189996336316631423491873720880873929685421463425 57m^{21}}{47418979405468182759263881396224000000000000000} + \frac{20799895858914548134889665508554779797324132480152535445 1m^{22}}{2847983903092419056521388716657213440000000000000000}$$

$$- \frac{761874532729413158667558421618618656329316896046895041287077m^{23}}{10690619576233168033417162895152014950400000000000000000} - \frac{1585379288268059437251937786377358068940800000000000000000m^{24}}{7098368335947774573139458224077684169719231587462610146792076537}$$

$$- \frac{42826085096059769097697049999464559135588150156058890082543688650001432183m^{25}}{37020711764754007842765417885291620041283233382400000000000000000000}$$

$$- \frac{8978765942311927699828432661670544894234543855186841232564771847347549703132 7m^{26}}{41689949036083607081934206216074025618990081192755200000000000000000000}$$

$$- \frac{34454749673787507911245449080178895039829194996883597501345895618696065966593074 7m^{27}}{1173702346456491300628578949376909052504630129579786240000000000000000000000}$$

$$- \frac{2193258047951605221050196125398243089706790064481102032926461131091392490562226753369 m^{28}}{10573884439226530127362867754936573654014212837384294236160000000000000000000000000}$$

$$+ \frac{310366856790137567768103153621126751720958075602886825131292207359422821822141325693356 7277163m^{29}}{975463679109036133554299654187249582581903804948429893361310105600000000000000000000000000}$$

$$+ \frac{1999631662802541307040409895164659690509435567826747224296389792317085129793415092573418425 37669933961m^{30}}{11951615107726896876979332395019186825693305075141350676637178128236544000000000000000000000000000000}$$

$\bar{a}_1 = -1.1875m - 1.66666666666667m^2 - 1.19444444444444m^3 - 0.518518518518518m^4 - 0.0889877507716049m^5 + 1.43637192965535m^6 + 5.91766214420439m^7 + 12.1636311773048m^8$
$\quad + 15.3085963903584m^9 + 10.9542215270043m^{10} - 4.19749027437368m^{11} - 40.9354848097648m^{12} - 119.07691607355m^{13} - 238.788522317723m^{14}$
$\quad - 336.577723831451m^{15} - 271.038935880953m^{16} + 169.177250020925m^{17} + 1290.70786774133m^{18} + 3455.6408633268m^{19} + 6629.75647033985m^{20}$
$\quad + 9319.05260095514m^{21} + 7303.37549883251m^{22} - 7126.57042275803m^{23} - 44773.9439292307m^{24} - 115681.420087749m^{25} - 215370.038820163m^{26}$
$\quad - 293556.111375336m^{27} - 207422.169265928m^{28} + 318173.668007423m^{29} + 1673105.80601759m^{30}$

$$\bar{\bar{a}}_1 = -0.107555493337202$$

$$\bar{\bar{a}}_2 = \frac{25m^3}{256} + \frac{803m^4}{1920} + \frac{6109m^5}{7200} + \frac{897599m^6}{864000} + \frac{237203647m^7}{368640000} - \frac{11098919887m^8}{14515200000} - \frac{19388340038959m^9}{4572288000000} - \frac{10750132421861267m^{10}}{960180480000000} - \frac{543219484755139552363m^{11}}{25809651302400000000}$$

$$- \frac{15151012074663624045 0569m^{12}}{542002677350400 0000000} - \frac{357789674289867408405 2887m^{13}}{189700937072640 0000000} + \frac{111500045135648657212765 1069m^{14}}{398371967852544 00000000000} + \frac{579920788433893520898486924 02831m^{15}}{40155894359536435 2000000000000}$$

$$+ \frac{133939212706430257173992004011858 9507m^{16}}{37104046388211 6612480000000000000} + \frac{140890300062436761149752954444779848 5819m^{17}}{21427586789192237187072 00000000000000} + \frac{85521147775264626093781768080988565153 01959m^{18}}{98995450966081358042726 4000000000000000}$$

$$+ \frac{39216232202745405683563858154194070666080411 4609m^{19}}{73177437354117565986518335488 00000000000000000} - \frac{7508345012164292186000688897014281692944568412 8240157m^{20}}{6592555331232451519725436844113920 000000000000000}$$

$$- \frac{1962910568457926185103147894626083075920837154707705369 629m^{21}}{37120206861920722338254037830388940 80000000000000000000} - \frac{34302407197512598375377457895001427582996104301694818299 91299689m^{22}}{2675327548952350300362645014511791741337600 00000000000000000}$$

$$- \frac{1542529656864750326781892411887150835053468057600000000 0000000000m^{23}}{35144944899341644199392452169428963099668107909401804607922 49573352067}$$

$$- \frac{6948324839347267846989034369345670936498346854520000 00000000000m^{24}}{20131890121952251172085856506894140238182308199044120506654912758750 74005817}$$

$$- \frac{23658137614710796055839742351325606340152861495643647553688589813827 935855601971m^{25}}{1564936461941988400838105265835878736672840172773048320000000000000000 0000000}$$

$$+ \frac{13845867376892492913146966637350137720347243655991843293522835407341 3375745789270556491m^{26}}{2819702517127074700630098067983086307737123423302478462976000000000 000000000000}$$

$$+ \frac{33347156631886925605380467162397906696644157943237518249170267958824 241128324703619729 4820223m^{27}}{16257727985150602225904994236454159709698396749140498222688501760000 00000000000000000000}$$

$$+ \frac{30053759274333465494604024416341216932309933678862776409642645026638 75186757774944418386342 74629083m^{28}}{62247995352744254567600689557391598050485963932786810748530275123200 000000000000000000000000}$$

$$+ \frac{79788388752142309722386976091601766702182293393792293167393128124821 688325310442712248509649189 121205697m^{29}}{95334672322588408197917484077831954162260768342350170842656965722594 344960000000000000000000000000}$$

$$+ \frac{74520256934559238651803445930433895719193187022019892959663778944762 4973425374442150566250116101 69349925985599m^{30}}{73003955351106912403678282194860986379063617269679778575327211356562 4585682944000000000000000000000 00000000}$$

$\bar{\bar{a}}_2 = 0.09765625m^3 + 0.418229166666667m^4 + 0.848472222222222m^5 + 1.03888773148148m^6 + 0.643456073676215m^7 - 0.764641195918761m^8 - 4.24040218791095m^9$
$\quad - 11.1959497675492m^{10} - 21.0471454414662m^{11} - 27.9537587318386m^{12} - 18.860722556835m^{13} + 27.9889284722262m^{14} + 144.417350848063m^{15}$
$\quad + 360.982765343308m^{16} + 657.518279816278m^{17} + 863.889673118192m^{18} + 535.906060948427m^{19} - 1138.91270302932m^{20} - 5287.98391603672m^{21}$
$\quad - 12821.7597919721m^{22} - 22783.9670653562m^{23} - 28973.73192449m^{24} - 15117.6346069364m^{25} + 49104.0004851281m^{26} + 205115.725040703m^{27}$
$\quad + 482806.861554756m^{28} + 836929.385797421m^{29} + 1020770.12918218m^{30}$

$\bar{\bar{a}}_2 = 7.27116153734895 \cdot 10^{-5}$

$$\bar{\bar{a}}_{-2} = \frac{23m^4}{640} + \frac{299m^5}{2400} + \frac{56339m^6}{288000} + \frac{238200053m^7}{1105920000} + \frac{146886277m^8}{537600000} + \frac{164530763567m^9}{508032000000} - \frac{42341873843687m^{10}}{320060160000000} - \frac{56433070995456755597m^{11}}{34412868403200000000} - \frac{13955946209954151767143m^{12}}{36133511823360000000000}$$

$$- \frac{6792588382013329642243 1m^{13}}{118563085670400000000000} - \frac{17695698501493113398279 6857m^{14}}{265581311901696000000000000} - \frac{30262906523808366713614 880791m^{15}}{55772075499356160000000000 00} + \frac{56592035770884835710817 09807218077m^{16}}{12368015462737222041600000 00000000}$$

$$+ \frac{25468402335543205625307 68380496 1273509m^{17}}{71425289297307457290240 0000000000000} + \frac{31017606563829941289423181359 2634015886049m^{18}}{329984836553356045268090 8800000000000000000}$$

$$+ \frac{80445556221673871977500 04701559333775 90278422 73m^{19}}{48784958236078377324345 5569920000000000000000} + \frac{22978402614060127632618 528520999419727125415102 45199m^{20}}{10987592218720752532875 7280735232000000000000000}$$

$$+ \frac{74159835710812778074735 7845309561114594506179 21054782699m^{21}}{49493609149227629784338 7171071852544000000000 00000000} - \frac{81926314644191282273461 798118409844146532062423 23113195623977m^{22}}{44588792482539171672710 75024186319568896000000 0000000000000}$$

$$- \frac{56195205645022005248861 01469892913741554375625 7222962751604384540373 7m^{23}}{51417655228825010892729 7470629050278351156019 200000000000000000000}$$

$$- \frac{65956238965478847330672 38663040693681636469402 10224474457115677029427 21237m^{24}}{23161082797824226156630 1145644855697883278228 848640000000000000000 00000}$$

$$- \frac{27324789631216319636870 19059973406724740809891 78261999976100675867975 1614868431m^{25}}{52164548731399613361270 17552786262455576133909 24349440000000000000000 0000000}$$

$$- \frac{65421824252220935178684 06660270825458133525222 51403417068788707628939 427195312361849m^{26}}{93990083904235823354336 60226610287692457080776 749282099200000000000 00000000000000}$$

$$- \frac{13146312514127411625492 75463476494428455844509 14474936743331394868124 0827770817845933786461m^{27}}{27096213308584337043174 99039409026618283066612 48567497037814169600000 0000000000000000000}$$

$$+ \frac{59711234786158607717834 80606741962623161574637 58013118620317536730940 05353347210326645026017 53777m^{28}}{82997327136992339423467 58607652213073398128524 40371547476647070334976 00000000000000000000000000}$$

$$+ \frac{12287078363841925217976 07841474705412882399289 15093294863415171418976 83724806609084783454583 58944333667m^{29}}{31778224107529469399305 82802594398472075358944 74500569475523219075314 4832000000000000000000000 00000}$$

$$+ \frac{95868633152177874333569 56832843058920164335160 31020949934617565264481 37279569805681573958134 94868593740213731m^{30}}{97338607134809216538237 70959314798183875148969 29063714337696151420832 78091059200000000000000 0000000000000}$$

$\bar{\bar{a}}_{-2} = 0.0359375m^4 + 0.124583333333333m^5 + 0.195621527777778m^6 + 0.215386332646123m^7 + 0.273225961681548m^8 + 0.32385905527014m^9 - 0.13229348458642m^{10}$
$\qquad - 1.63988280006932m^{11} - 3.86232765809709m^{12} - 5.72909210620278m^{13} - 6.66300590760058m^{14} - 5.4261754207368m^{15} + 4.57567634366138m^{16}$
$\qquad + 35.6574017215822m^{17} + 93.997066313062m^{18} + 164.89827834306m^{19} + 209.130464224085m^{20} + 149.837195115867m^{21} - 183.737459758017m^{22}$
$\qquad - 1092.91653606014m^{23} - 2847.71828421057m^{24} - 5238.19151046707m^{25} - 6960.50280355932m^{26} - 4851.71575984107m^{27} + 7194.356354103m^{28} + 38665.088150507m^{29}$
$\qquad + 98489.8345827001m^{30}$

$$\bar{\bar{a}}_{-2} = 2.02588306520068 \cdot 10^{-6}$$

$$\bar{a}_3 = \frac{833 m^5}{12288} + \frac{27943 m^6}{71680} + \frac{12275527 m^7}{11289600} + \frac{27409853579 m^8}{14224896000} + \frac{287958210882233 m^9}{127455068160000} + \frac{12754813704802943 m^{10}}{13382782156800000} - \frac{9116329553696361023 m^{11}}{2107788189696000000} - \frac{5216937005259775982269 m^{12}}{295090346557440000000}$$
$$- \frac{8425194073315008400481 9941 m^{13}}{1983007128865996800000} - \frac{7634577446134087680000000 m^{14}}{565186367525961531011387 94007} - \frac{2121104939529996380383213666960879 m^{15}}{2380842976576915243008000000000}$$
$$- \frac{56256983465761359698101399 57599345619 m^{16}}{1649924182767802263404544 00000000000} + \frac{13749726171186454080896579 43580448283 6130169 m^{17}}{73177437354117565986518335 488000000000000}$$
$$+ \frac{23523089861947432500065463 48060152597061511618471 m^{18}}{32962776656162257598627184 2205696000000000000} + \frac{12178277170521273693667625 38004464438044798278184 38767 m^{19}}{74240413723841444676508075 66077788160000000000000}$$
$$+ \frac{37307603143821276251835372 77399016391304104098447 9600161687 m^{20}}{13377637744761751501813225 07255895870668800000000 0000} + \frac{50439900943296840669863467 7851073062557147508049 9039942 38986790963 m^{21}}{15425296568647503267818924 11887150835053468057600 00000000000000}$$
$$+ \frac{31813841119728237708869391 16717006980944356129018 8672868681 02741435271 m^{22}}{34741624196736339234945171 84672835468249173432729 600000000000 00000000} - \frac{12764540050290279556168111 87404574851389140999058 8057318881 29555242462958341 m^{23}}{15649364619419884008381052 65835878736672840172773 0483200000 000000000000000}$$
$$- \frac{40957807449318721860759704 65204892162112845006436 3919024787 9541203065568210846713 m^{24}}{14098512585637350315049033 99154315386856171165123 9231488000 0000000000000000000} - \frac{10533801771496633467985400 50002119503192967128015 5891284159 6963640730049685462313 13140037 m^{25}}{16257727985150602225904994 23645415970969839674914 0498222688 5017600000000000000000000}$$
$$- \frac{13357864551358400891263666 14765693495497593801813 1052924574 128906286356772779966685 394288572211 m^{26}}{12449599070548850913520137 91147831961009719278660 5573621497 0605502464000000000000 00000000} - \frac{22929318481535989633829870 48621397136329210699572 5596338927 04890520612794496002492 8316217427313486071 m^{27}}{19066934464517681639583496 81556639083245215366847 0034168513 9314451886899200000000 00000000000}$$
$$- \frac{58299882677212576889376572 20910594084125089274677 8578987485 4238420984892714620746705 21735239973553395 6897 m^{28}}{36501977675553456201839141 09743049318953180863483 9889287663 605678281229284147200000 00000000000000000} + \frac{25820725056931279942115376 82000146625765049969784 3710498948 25958638804390429573131 930717366886282789499144850 247299 m^{29}}{71556958552883692514307455 69913308574024147312450 0909620141 36576570630538950395035 64800000000000000000000000}$$
$$+ \frac{25119602009023168587980914 27667445501478627369549 5236453050 54595852994161560225011 2309229772831428186229766078 533306465361 m^{30}}{20822409459174612703323086 54910972601271288443352 9706841001 67012311891379970552716 9973647360000000000000000000000000}$$

$\bar{a}_3 = 0.0677897135416667 m^5 + 0.389829799107143 m^6 + 1.08733055201247 m^7 + 1.92689307387555 m^8 + 2.25929196099716 m^9 + 0.953076389898645 m^{10} - 4.32506909292968 m^{11}$
$- 17.6791178231386 m^{12} - 42.4869580682397 m^{13} - 74.0298165175015 m^{14} - 89.0905011543281 m^{15} - 34.0967082326101 m^{16} + 187.89570485571 m^{17}$
$+ 713.625860688829 m^{18} + 1640.38379632714 m^{19} + 2789.01199656325 m^{20} + 3269.94691601702 m^{21} + 915.726937220076 m^{22} - 8156.58677570224 m^{23}$
$- 29051.1550069825 m^{24} - 64792.5822176256 m^{25} - 107295.540006249 m^{26} - 120256.974314387 m^{27} - 15971.7052033205 m^{28} + 360841.567041291 m^{29}$
$+ 1206373.45347924 m^{30}$

$$\bar{a}_3 = 3.71453649306558 \cdot 10^{-7}$$

$$\bar{a}_{-3} = \frac{m^5}{192} + \frac{7477 m^6}{215040} + \frac{65239 m^7}{627200} + \frac{2674679587 m^8}{14224896000} + \frac{944592921809 m^9}{3982970880000} + \frac{2991640633593949 m^{10}}{13382782156800000} + \frac{115453688848205717 m^{11}}{2107788189696000000} - \frac{185388670761024497527 m^{12}}{295090346557440000000}$$
$$- \frac{9260919980691655884533867 m^{13}}{3966014257731993600000000} - \frac{693609773351725406928802130 23 m^{14}}{1374223940304135782400000000 0} + \frac{302877764329226897128031616 36119 m^{15}}{39680716276281920716800000000 00}$$
$$- \frac{8639981881565174795529561133 75435073 m^{16}}{109994945517853484226969600 0000000000} - \frac{772025981874822068444340967107 49352571779 m^{17}}{487849582360783773243455569920 00000000000} + \frac{2731681946601336809805342710566 104252026490287 m^{18}}{137344902734009406660946009190 40000000000000}$$
$$+ \frac{17252702969306322458856953739 4549011814670813396152 9 m^{19}}{24746804574613814892169358553 59262720000000000000 00} + \frac{85771504298326414466352795015 386507942076041476392888559 m^{20}}{55735990603173964590884378023 2899461120000000000000 0}$$
$$+ \frac{636271309742013291015476983144 150938887109496548061726596373 9883 m^{21}}{257088276144125054463648735314 52513917557800960000000000000000}$$
$$+ \frac{620623423547146494589316668237 035416755201351494767983636996 395617537 m^{22}}{231610827978242261566301145644 855697883278228848640000000000000000}$$
$$+ \frac{290655387247317417920564820688 030233861287917492756904775648 990877799221 m^{23}}{652056859142495167015877194098 282806947016738655436800000000000000}$$
$$- \frac{676436016888934072038148729972 156385586121682873778897396849 2398637869517231291 m^{24}}{939900839042358233543366022661 028769245707807767492820992000 0000000000000000}$$
$$- \frac{656536092069229246021864353251 758898235176272077566234496767 8797650460834079540437 0607 m^{25}}{270962133085843370431749903940 902661828306612485674970378141 696000000000000000000}$$
$$- \frac{437742307391680412141067639102 595714360990347320434055728447 7254371717748254602062786 8654540087 m^{26}}{829973271369923394234675860765 2213073398128524403715747664707 033497600000000000000000}$$
$$- \frac{135512352846462506886434161388 738175515520773814815204900895 75039543229527462651445089342 3408841033 m^{27}}{158891120537647346996529140129 7199236037679472372502847377616 095376572416000000000000000000}$$
$$- \frac{900388249708253546788038870222 117008196401637400168480732205 0895812075574358484519029077656 998326676516901 m^{28}}{973386071348092165382377095931 4798183875148969290637143376961 5142083278091059200000000000000 000000}$$
$$- \frac{408582457569506484430159222817 418351929997355070042120712672 70291165707055316189851403032561 3594712455359171349 m^{29}}{477046390352557950095383037994 220571601609820830006413427577 17771375369263359669043200000000 00000000000}$$
$$+ \frac{386530087803285423537906774902 070856318497699778318557461925 955722506049031947960812979933279 2676834811286232250 13391357 m^{30}}{138816063061164084688205769940 64840084752562890198045606677800 8207927586647035144646490982400 00000000000000000000}$$

$\bar{a}_{-3} = 0.00520833333333333 m^5 + 0.0347702752976191 m^6 + 0.104016262755102 m^7 + 0.188028059185811 m^8 + 0.237157877942859 m^9 + 0.223543998440851 m^{10} + 0.0547748058427338 m^{11}$
$\quad - 0.628243766438961 m^{12} - 2.33506976497548 m^{13} - 5.04728343764859 m^{14} - 7.6328703902521 m^{15} - 7.85488991415774 m^{16} - 1.58250823571246 m^{17}$
$\quad + 19.8892124296136 m^{18} + 69.7168918002641 m^{19} + 153.888902610519 m^{20} + 247.491375058004 m^{21} + 267.959589352812 m^{22} + 44.5751598456539 m^{23}$
$\quad - 719.68870415962 m^{24} - 2422.9809700429 m^{25} - 5274.17354861511 m^{26} - 8528.62969232787 m^{27} - 9250.06301416723 m^{28} - 856.483700185943 m^{29}$
$\quad + 27844.7666127064 m^{30}$

$$\bar{a}_{-3} = 3.04319731633951 \cdot 10^{-8}$$

$$\bar{a}_4 = \frac{3537 m^7}{65536} + \frac{18638507 m^8}{48168960} + \frac{1473975631 m^9}{1083801600} + \frac{5918813731091 m^{10}}{1911826022400} + \frac{54678083521820137 m^{11}}{11012117889024000} + \frac{335640096448825021 2299 m^{12}}{6677472984956928000 00} - \frac{54030384478822544361 277 m^{13}}{413168640944209920000 00}$$

$$- \frac{3752090682475865839033833346181 m^{14}}{16034248617762898575360000000} - \frac{6755119559705059905616255239 13529 m^{15}}{90708035037630111940608000000} - \frac{5144212975578869243788840568557 9148198149 m^{16}}{32033796556037179396139148902400 0000000}$$

$$- \frac{26273783110295513452463872610639 7713261 77831 m^{17}}{10306874041904962470707711593472 0000000000} - \frac{333128193708893297617705 83177716713680 6542798 754749 m^{18}}{129996479540930529658048 83452438249472 00000 00000}$$

$$+ \frac{3626658083317921939548722397577882 8876454459988753834447 m^{19}}{535377501341368293343708320105216866 2548480000000000} + \frac{7618746840132814960745243611103 194417640561529609752455157929 m^{20}}{675250227341814173662685555815905 8247325895884800000000}$$

$$+ \frac{747868486223350726683548490556524345 06208998232284209919 490836 7261 m^{21}}{21726176064722871037596907758376 76991077107000934400000000000000} + \frac{4930614357648005279429902797770851 40101675789802009358042582030451098 9197 m^{22}}{6850589205848092081219868969832571 20441477993999630336000000000000}$$

$$+ \frac{6294957870847532994199358893526162 611249918770179799985171740097339006 02071903 m^{23}}{56426933170729564854591816730716922 4965236594097615515156480000000000} + \frac{129748739416067175094303133963818764782 7429658073349368505264736378869633978561 72118151 m^{24}}{12098735734255443064645420710863284363554 1232134164952589864534016000000000000}$$

$$- \frac{119748573745709745200809957951926117920 5671911284903786850648678217063562191241 62228 08707013 m^{25}}{264708239129774838811377159732977798590200 66178633949977136461397360640000000000000} - \frac{724320818998840486264446465456429307524 038918444984290288303302511042687603838621 0521279206 01667 m^{26}}{13856739581645353575163668115824865308222 41277743861200485286294273916928000000000000 00000}$$

$$- \frac{152229137640800086854173708145950862828516 3647678553403588088377203245729182072528545 82985032979 52045159977 m^{27}}{9934327162698348624611211577305408496795048 09555615188312322318070672482998681600000000 0000000000} - \frac{6313004526373094539428004259260452891748637717 5253499443921714723951234538694525046603440821 8462687320465 8718754887 m^{28}}{20235577705709050246183757579309743452587837156 00554600680434414297270534470896712050278400000 000000000000000}$$

$$- \frac{141181215675852880295170206551697632978444387 28346028987994294794953904893486988881671410320 17512058307072502967 729918707 m^{29}}{30042678170860563915674407890987998796372201762 81664814490918808777820208750307989143313612800 00000000000000000} - \frac{72358015931724107156777779731254216401844793694 72274844115015158983688789139483560131285607062 18345527131868523310971319099 6729 m^{30}}{17134594303594332788585000371596878064256788724946 371408014068019113064711321560605582397838384787 60738816000000000000000000000}$$

$\bar{a}_4 = 0.0539703369140625 m^7 + 0.38694019966385 m^8 + 1.36000503320903 m^9 + 3.09589557927496 m^{10} + 4.96526499923497 m^{11} + 5.02645382774229 m^{12} - 1.30770777654731 m^{13}$
$- 23.4004771406579 m^{14} - 74.4710163427386 m^{15} - 160.587052695425 m^{16} - 254.915146954095 m^{17} - 256.25939632004 m^{18} + 67.7402033935208 m^{19}$
$+ 1128.28497224277 m^{20} + 3442.24627470306 m^{21} + 7197.35808043916 m^{22} + 11155.9454273388 m^{23} + 10724.1568264613 m^{24} - 4523.79473111157 m^{25}$
$- 52272.095808041 m^{26} - 153235.478505744 m^{27} - 311975.50266094 m^{28} - 469935.519306629 m^{29} - 422291.970557748 m^{30}$

$$\bar{a}_4 = 2.16784890497834 \cdot 10^{-9}$$

$$\bar{a}_{-4} = \frac{23 m^{7}}{6144} + \frac{795829 m^{8}}{28901376} + \frac{44780807 m^{9}}{464486400} + \frac{6131572502021 m^{10}}{28677390336000} + \frac{87149512084429 m^{11}}{258096513024000} + \frac{774328898942856258667 m^{12}}{2003241895487078400000} + \frac{45411088063785748715129 m^{13}}{247901184566525952000000}$$
$$- \frac{3570160210680097723751446038 1m^{14}}{48102745853288695726080000000} - \frac{305553781074110660548527081301541 m^{15}}{95243436789511617537638400000000} - \frac{93548801587675036274012084205709 2321811 m^{16}}{120126737085139422735521808384000000000}$$
$$- \frac{5249477225153914739764341610899046568991163 m^{17}}{38650777657143609265154141847552000000000} - \frac{32879018668241357076657954860516681781668165 29499 m^{18}}{194994719311395794487073251786573742080000000000}$$
$$- \frac{5821642158380266599977652341574833997397499616530870 8073 m^{19}}{64245300160964195201244998412626023950581760000000000} + \frac{136038378774120593308748202214199586003078270 29358449 40122699 m^{20}}{506437670506360630247014166861929368549442191360000000000 00}$$
$$+ \frac{15344282807459241271597317276614819207415349600649205765 433271383 m^{21}}{130357056388337226225581446550260619464626400560640000000 00000}$$
$$+ \frac{58843778862431793927492184995696398613118527184391872961 9140081880376211 m^{22}}{205517676175442762436596069094977136132443398199889100800 00000000000000}$$
$$+ \frac{70449544170382180143971813645697281059735123402687497575 61445578601222270 0537 m^{23}}{138188407765051995562265673626245524481282431207579 30983420 49897155391872040960000000000000}$$
$$+ \frac{47006833693380382963365480077650681010561507869395734104 800640092376463422529665193717 m^{24}}{72592414405532658387872524265179706181324739280498971553 91872040960000000000000}$$
$$+ \frac{17051108167342229430767986667939980100419158271998683040 68073119793646264503898568174516 33 m^{25}}{49632794836832782277133217449933337235662624084938656207 130865120051200000000000 0000}$$
$$- \frac{91690559510532204907179938748296536244594479268128593438 7969371698584716002520734002060289 36985737 m^{26}}{85135807989629052365805576903627972453718504104582832157 8159899201894560563200000000000 00000}$$
$$- \frac{68483263310405011855732887724774297517331517804748622651 7963632997887307011813231338264193844704501993209 m^{27}}{14901490744040752293691681736595811274519257210433342278 246848347710600872449802240000000000000000000}$$
$$- \frac{10705658415757328164892775118455534805946553072285167497 4082748721968594715363030157347500285772854385572238124 41 m^{28}}{97130772987403441181682036380686768572421618348826620832 66085188626898565460304217841336320000000000000000}$$
$$- \frac{49193774709350481917808021046373369424203072522050345416 1429351519431274088587315603618584089020280457610852846306 8710809 m^{29}}{25235849663522873689166502628429918988952649480765984441 72371799373368975350258702288038343475200000000000000000}$$
$$- \frac{12606016912978839211811612663201576632634970838882241949 796008529195688752155233411999446385518275241455841699881 4376 3671465194317 m^{30}}{51403782910782998365755001114790634192770366174839112442 20405733919413396468181674719215154362822164480000000000 00000000000 000}$$

$\bar{a}_{-4} = 0.00374348958333333 m^{7} + 0.027536024582359 m^{8} + 0.0964092963755236 m^{9} + 0.21381208088254 m^{10} + 0.337662493240756 m^{11} + 0.386537891748007 m^{12} + 0.183182214894174 m^{13}$
$- 0.74219468085438 m^{14} - 3.20813476890156 m^{15} - 7.78750874764647 m^{16} - 13.5818152786473 m^{17} - 16.8614918313431 m^{18} - 9.06158449535508 m^{19}$
$+ 26.8618206536854 m^{20} + 117.709644821591 m^{21} + 286.319794761591 m^{22} + 509.807915944444 m^{23} + 647.544706679404 m^{24} + 343.545194732586 m^{25}$
$- 1076.99171095788 m^{26} - 4595.73236575623 m^{27} - 11021.9017994902 m^{28} - 19493.607453391 m^{29} - 24523.5198640108 m^{30}$

$$\bar{a}_{-4} = 1.51933047192614 \cdot 10^{-10}$$

$$\bar{a}_5 = \frac{732413 m^9}{15728640} + \frac{6087081853 m^{10}}{15259926528} + \frac{1178453745829 m^{11}}{699413299200} + \frac{8525414240854708717 m^{12}}{1832141113786368000} + \frac{488694654491079992327 m^{13}}{52646014045126656000} + \frac{483775215093039916579929534689 m^{14}}{3660317181059924093829120000}$$

$$+ \frac{5711972093089777878104960016390 1 m^{15}}{65428169611446143177195520000} + \frac{33895920643067272319037688795535 17773323 m^{16}}{14853947967438612578605318799360000000} + \frac{78426754349466004849682725701809078 6446074121 m^{17}}{67971665941579909115969793882587136000000}$$

$$- \frac{6542816961144614317719552000 0}{1484593433867161081296714601391 01760202 0354690406959 m^{18}} - \frac{148539479674386125786053187993600000 00}{4125779297354388958473535055449 87335919 67 87583746087727 m^{19}}$$

$$- \frac{48223093154330855346529535005679 16272640 00000000}{65900915473609459064905231761800314477 805287212 0881773930441 m^{20}} - \frac{68959023210693123145537235058148 02026987520000000000}{41270658428739732431791780796711469 4267640 18413395218027142430247 m^{21}}$$

$$- \frac{782777318532469276486862637804979 48410245 15276800000000 0 0}{834877012333520324823723736934132653 479998475650 6270882908991689170 12984319 m^{22}} - \frac{716397801920915881840776686119097 51438505636381327360000000000 0 0}{}$$

$$+ \frac{691226377112842389324830972397948 29220987 158490869403746304000000 0 00}{1036124296109169767893285105039184 7414725499196030490784896250326 2685988469316323 m^{23}}$$

$$+ \frac{16803713227613198484486640938994 12298362 197822913035205072650240 000000000 0}{13001154796661517526746641181082603 23102059262348630812251712961147 4960711 914 035283132389 m^{24}}$$

$$+ \frac{81066481396382291906079349371487 65789528 569019936902544018485469511 68000000 00000000 0}{38378576214365891796971631483086625 326141834313069085956850526083920 723491012 50882473562857139 m^{25}}$$

$$+ \frac{12612647441574742503915449926135 35765980 12882397863045408572043288 505221120 00000000000 00}{48266000093934113745889798457010966 158090015715950117550413055296524784 5070609987 3160960534763749888133 m^{26}}$$

$$+ \frac{11561013789812686784132444574311 31144635 42069862611315175852319362 400174179 192896000 0000000000}{20363014574996859075185150195787552066 4346680781969055276784096345978369638 117207228 94991925 2640 1659632919767 m^{27}}$$

$$- \frac{76284523770537974141032764034898 07377708 15789521263077195224918255 757002350 2326446161 92000000000 000 00}{194847462058064240364505208843313707281 42909797 64614530058665885993706353306 69650031420 0776388768095981222339020882103 m^{28}}$$

$$- \frac{30591738580993796003110344490230 26182552 10676195489012602559631522572 172887936 4533467240 988672000000 000 00000}{4426651927625222930098133909462257664316 9589138145305361010620572156311213737 14143227639 18597941113478055905946141227203601513 m^{29}}$$

$$- \frac{14469138568371429822337676304990 67456982 00841449404846111740178640855 921597653 9836787900 3548238951219200 000 000000000}{410606111324971829385118363141538392575665 77862631099097745423350411114719242736523271 44975944 69825727628517254791228810 6951671340910783 m^{30}}{530508479811332428224367424109693885 70456229008378676749343916529042918663306528346 28440634 83371559 00014854144 00000000000 00000 00}$$

$$\bar{a}_5 = 0.0465655644734701 m^9 + 0.398893260844408 m^{10} + 1.68491755472327 m^{11} + 4.6532519666216 m^{12} + 9.28265251139782 m^{13} + 13.216756667873 m^{14} + 8.73014196639014 m^{15}$$
$$- 22.819469070537 m^{16} - 115.381539150257 m^{17} - 307.859437617562 m^{18} - 598.294335572118 m^{19} - 841.885858383822 m^{20} - 576.085776897675 m^{21}$$
$$+ 1207.81995591761 m^{22} + 6166.04367186253 m^{23} + 16037.6453655255 m^{24} + 30428.644257398 m^{25} + 41749.4528480927 m^{26} + 26693.5068686295 m^{27}$$
$$- 63692.8370521315 m^{28} - 305937.489416377 m^{29} - 773985.953006817 m^{30}$$

$$\bar{a}_5 = 1.36867236166344 \cdot 10^{-11}$$

$$\bar{a}_{-5} = \frac{447 m^9}{163840} + \frac{117747149 m^{10}}{5086642176} + \frac{199973223959 m^{11}}{2098239897600} + \frac{154681921246663037 m^{12}}{610713704595456000} + \frac{49052001826917693323 m^{13}}{100767761258250240000} + \frac{42495178607599701879125 75957 m^{14}}{6100528635099873489715200000}$$

$$+ \frac{78748427761833379030361428219703 m^{15}}{13085633922289228635439104000000} - \frac{128627713271033818570629269294798 63519 m^{16}}{24756579961239768763100886466560000000} - \frac{218198225473074769849884228849300117716357 m^{17}}{53102864016859303996851401470771200000000}$$

$$- \frac{7677791557629535654723987028225718017721 9463775131 m^{18}}{6429745753910780712870604667426388836352000000000} - \frac{8391817955462392716663859022617466051032379 917046749 m^{19}}{34479511605346561572768617529074010134937600 0000000}$$

$$- \frac{1919924307057214251328360263236290694064187 17406290970596367 m^{20}}{5218515456883128509912417585366386322734967685 120000000000} + \frac{1545737908719832620517300426514768920721980484559393469049021 3809 m^{21}}{44774862620057242615048542882443594649066022 73832960000000000000}$$

$$+ \frac{7356945682903532821187387554296302382422613256649921326026649922660210443 m^{22}}{460817584741894926216553981598632194806581056605796024975360000000000}$$

$$+ \frac{4185164952826120374021302536933469482652975218802671908030752962150387786 65401 m^{23}}{24005304610875997834980915627134461405174254613043360072466432000000000000}$$

$$+ \frac{27781523950378910565962403908091294925957346211410972471139926717112730016 5477936339403 m^{24}}{540443209309215279373862329143251052652379346624601696012323646341076244636573649217871151 m^{25}}$$

$$+ \frac{13317036225116561982197369863129793808328218663369186162351107535073410772 44636573649217871151 m^{25}}{126126474415747425039154494926135357659801288239786304540857204328850522112000 00000000000}$$

$$+ \frac{124273846840778115858504307462429492891544014384444209942243393862997519394 55743419446398795630 79762771 m^{26}}{770734252654179118942162971620754096423694713241740876783901546241600116119463 52640000000000000}$$

$$+ \frac{5086946685496912759065934919009302439144659569719413068914040384244329553274 184410712644918000850 7518097229 m^{27}}{33374479121103636867018342652679072774731907915552596272910901736893688528226 78201958400000000000000}$$

$$- \frac{22391588305991155479638251457997139940895191783257517701988477710532513665067 091321012425833438645 8432734579727093691 m^{28}}{3263118781972671573665103412291227928055580546085216134427302736240769844137988 83569839038791680000 00000000000000}$$

$$- \frac{21798383500354553928791633650948702050078991542850096426558583183146954170921 28726773941674188702 96379517836687228589005 2087 m^{29}}{289382771367428596446753526099813491396401682988809692223480357281711843195307 967357580070964779024 384000000000000000000}$$

$$- \frac{38881353196932130630041780704968390922213463656197418617881856893059395095293 61908794061356188626564559016155 80939861350962746651239 m^{30}}{17683615993711080940812247470323129523485409669459558916447972176347639554435509 44876146878277905196667161804800 00000000000000000000}$$

$$\bar{a}_{-5} = 0.002728271484375 m^9 + 0.0231483058815419 m^{10} + 0.0953052242442499 m^{11} + 0.253280579889928 m^{12} + 0.486782689368338 m^{13} + 0.69658190542866 m^{14} + 0.601792990920359 m^{15}$$
$$- 0.519569801129317 m^{16} - 4.10897283061419 m^{17} - 11.9410500064636 m^{18} - 24.3385638738604 m^{19} - 36.7906222166089 m^{20} - 34.5224489427558 m^{21}$$
$$+ 15.9649846848275 m^{22} + 174.343338718975 m^{23} + 514.050754488871 m^{24} + 1055.84781361762 m^{25} + 1612.40851062238 m^{26} + 1524.20257018492 m^{27}$$
$$- 686.202060117917 m^{28} - 7532.71640787392 m^{29} - 21987.2186835321 m^{30}$$

$$\bar{a}_{-5} = 7.94049498073308 \cdot 10^{-13}$$

$$\bar{a}_6 = \frac{31979701 m^{11}}{754974720} + \frac{8176206447469 m^{12}}{19397062164480} + \frac{8832390156215257 m^{13}}{4255230512332800} + \frac{105775817514605724187357 m^{14}}{15743059305555610828800} + \frac{281277749662514619497467285991 m^{15}}{1768260421200006208290816000}$$
$$+ \frac{192156353013064054611140188918121191 m^{16}}{681461042424164392582156124160000} + \frac{588186774436490141035751362666159368379 m^{17}}{1793946194181612763472525996851200000} - \frac{911412578381737223585914309693317311557 22029 m^{18}}{2765439816178723339403437725185998848000 0000}$$
$$- \frac{712842951210668199491837455605471257884530365057317 m^{19}}{46592130022979130822269118793937085911040000 0000} - \frac{3251119249994785723506908337190012266484824423097981 6020419 m^{20}}{6105008729827976192859662677776048377030089 1136000000000}$$
$$- \frac{135706528902185028995819162653626541800535282114089754 68727043053 m^{21}}{1092857612726506018283808215948690419972156 2522255360000000000}$$
$$- \frac{3876690105238384656752140000123464426453565743872814027163 849418507437 m^{22}}{17899976457092569217889298074539390744379120108371737313 2800000000}$$
$$- \frac{260983738812653646495096834386584335826505065335717438473367 46917321087343969 m^{23}}{102535557938692506102219623946345482788841801299187990286408 0896000000000000}$$
$$- \frac{42503370192520688549586658113678991088617207657208798410911833 41722089846002687337333 m^{24}}{163373603668937305976046445719505959003984463713418630988588065 304346624000000000000}$$
$$+ \frac{622632414682614747736942383935265956940167301736460697474462857712948163119 51718547082 3297983 m^{25}}{694580833828441094640836763395908090880952397671802891915094802290613367603 20000000000000 0}$$
$$+ \frac{543473464802264874867373993316535085042643448632774984291181650068018889071 570762040987359707892 6994663 m^{26}}{1729219483806530562327706339356204975842846724957790513454514315059761491072 4846387200000000000 0}$$
$$+ \frac{61570424981378066837392923864043604017164515513310305749176499555663578228 391964193299351228183214 288092627009 m^{27}}{8554769478004267175684904035425353893508264512667909114845613881073614815 739380404899020800000000 00000000}$$
$$+ \frac{576287931224125890632951609368045513605050003019003239142665684551814691 362195931462326261525220103714975507 35853982083 m^{28}}{4685525621900894607218485565293238894566848796307678885584252938178397231 03940045196646337698856960000000000 00000000000}$$
$$+ \frac{281427329087770173295575818066766203321748909946606367172685155656984531 25345330070500609628364919437133046966774 4434496369569 m^{29}}{19920453612441379661406801149774638994764775571892142936075381563503728597 798623029029769466474691297280000000000 000000000000}$$
$$+ \frac{231048835209219265877360904625283088258808426732333612731557858514519973 12823109858887519353412877297158912735469026369803660 494250373 m^{30}}{198374817358729833244962599789764365940934111927112065800375782723068540 7592216116334175831883709572076221235200000000 000000000000000}$$

$$\bar{a}_6 = 0.0423586381806268 m^{11} + 0.421517773059536 m^{12} + 2.07565492177606 m^{13} + 6.7188857935178 m^{14} + 15.9070319218947 m^{15} + 28.1977018568083 m^{16} + 32.7873141538014 m^{17}$$
$$- 3.29572378704348 m^{18} - 152.996428980408 m^{19} - 532.533104188762 m^{20} - 1241.75855410495 m^{21} - 2165.77714744653 m^{22} - 2545.2993036477 m^{23}$$
$$- 260.160572075341 m^{24} + 8964.14620672362 m^{25} + 31428.8307465699 m^{26} + 71972.0445298799 m^{27} + 122993.230157672 m^{28} + 141275.562576549 m^{29}$$
$$+ 11647.0849619689 m^{30}$$

$$\bar{a}_6 = 9.10170014968458 \cdot 10^{-14}$$

$$\bar{\bar{a}}_{-6} = \frac{24931 m^{11}}{11796480} + \frac{1794384055127 m^{12}}{87286779740160} + \frac{1667146256627447 m^{13}}{17020922049331200} + \frac{107756938499258356265887 m^{14}}{354218834375001243648000} + \frac{38136212223171656963007 1559 m^{15}}{552581381625001940090880000} + \frac{21918454479373554033216094676867 2021 m^{16}}{1839944814545243859971821535232 00000}$$

$$+ \frac{2585042621293296595447603637666620 2179 m^{17}}{1793946194181612763472525996851 2000000} + \frac{25539264940459435630860857620650617 1955377377 m^{18}}{74666875036825530163892818580021 9688960000000}$$

$$- \frac{33520370088526205489901376437279182 4707814574661 m^{19}}{72800203160904891909795498115521 4196736000000 00} - \frac{11513269811371083015325800220693730 53092428502174105 58366427 m^{20}}{659340942821421428828435691998132 2471924962426880000 00000}$$

$$- \frac{13076065093769632830240919492646120 2325564085410409152327984731 m^{21}}{3122450322075731480810880616996258 34277758929207296000000000 0}$$

$$- \frac{580550008721583312918852126186314267 5620177656451189984992764543956520 27 m^{22}}{7732698294639899021281767682010168 0157177988681659051933696000000 00}$$

$$- \frac{49086132189011042785526571284607318 637768801300640094375956279203005 31544803 m^{23}}{51267789693462530511098119731727 413944209006495939951432040448000 0000000000}$$

$$- \frac{47779111868493864615242177643644564 579365772571739344711990579914 826376853901300724673 m^{24}}{11027718247653268153383135086066652 23276895130065575759172969440804 33971200000000000000}$$

$$+ \frac{13129901130118400034594888763913829 01550431910997011339539835936220 70272957444935823448838 89 m^{25}}{63143712166221917694621523945082 55371645021797016389926500861839 00557606912000000000 00000}$$

$$+ \frac{39839424396730832277968679967493687 73715878574805903193484604020774 970438326931167897394605448846 102123 m^{26}}{46688926062776325182848071162617 53434775686157386034386327188650 66135602589570852454400000000000 00000}$$

$$+ \frac{12178234102130693675848164468815974 94343400970061965109656828240413 64721877094332904552829380810178 68250869739 m^{27}}{58814040161279336832833715243549 30801786931852459187516456359543 23811018582082402836807680000000 000000000}$$

$$+ \frac{23607212758013330763087199610927926 23043240410850816775493000135289 74403968403879286616592887167902 44483508390997 3212493 m^{28}}{6325459589566207719744955513145 87250766524587501536649553874146 6540836261903190610154725558934 5689600000000000000000}$$

$$+ \frac{19094329939531150470683226151697634 65691531099777099599443900589882 31005772379714218324711989703956 764946960795046896336246 11 m^{29}}{3984090722488275932281360229954927 79895295511437842858721507631270 07457195597246058059538932949382 594560000000000000000}$$

$$+ \frac{61698919332214916049504750408409946 49426245893072292383687406014827 08278457555536416870391875637515 175752538328383497 003098328 73113 m^{30}}{26780600343428527488069999597161818 94020261051101601288830507306676 14253002494917570511373730430079 22302898667520000000000000000000}$$

$$\bar{\bar{a}}_{-6} = 0.0021134270562066 m^{11} + 0.0205573405327659 m^{12} + 0.0979468827714274 m^{13} + 0.304210075924927 m^{14} + 0.690146528480976 m^{15} + 1.19125608040537 m^{16} + 1.44098113403706 m^{17} + 0.342042772352044 m^{18} - 4.60443359126878 m^{19} - 17.4617850396246 m^{20} - 41.8775760860687 m^{21} - 75.0772869444558 m^{22} - 95.7445844310903 m^{23} - 43.3263806668813 m^{24} + 207.936794966294 m^{25} + 853.294940713867 m^{26} + 2070.63382633392 m^{27} + 3732.09447056673 m^{28} + 4792.644362176 m^{29} + 2303.86617704613 m^{30}$$

$$\bar{\bar{a}}_{-6} = 4.44797524851968 \cdot 10^{-15}$$

$$\bar{a}_7 = \frac{75164925 m^{13}}{1879048192} + \frac{5485137547831499 m^{14}}{12103766790635520} + \frac{269685852617699521 m^{15}}{105907959418060800} + \frac{4635678939926234695659985 27 m^{16}}{4911834503333505785856000} + \frac{56699292473033854072566124578 31 m^{17}}{22005018574933410592063488000 0}$$
$$+ \frac{23387444829895429067114565114328122363731 m^{18}}{4337363242821321525906907299053568000 00} + \frac{164055294144938144291622177771632262816419 9 m^{19}}{198517779190668177531893064841297920000 00}$$
$$- \frac{30778791340595005986445215726750065486144761119849127 m^{20}}{5086825217842143771773515125213 2513401798656000000 0} - \frac{209408240063312671474553960073129813686025982666895972521 m^{21}}{13836164592530631059223961140580043645289234432000 000}$$
$$- \frac{15007172258892635705113848565012429235532949429 0615522307358742651429 m^{22}}{1813599636892109462910909253940664971255837683137 7263820800000000 00}$$
$$- \frac{96025797972383297633609179756834293082474287648542 7722642747670 1563398159 m^{23}}{41005487790130594956415658231598435000094490015 7439934988288000 00000000}$$
$$- \frac{186541522053273395281334564044804017224402339838679 3569116729734896 30820754699242159 m^{24}}{38391903679781077947969575598171528140255071215664 86442942069714452 4800000000000}$$
$$- \frac{827664266033208992089779942061559506153677688519977 312445931623774023349256178133034860 237 m^{25}}{111110924060158082206765978934715625616014939650383 13708478978511920263331840000 0000000}$$
$$- \frac{372823751982277748289408923830935237728572651821127 4446185658553873667591044722484114581304 142906469 m^{26}}{594442422639579433675661875716346350576929501232624 5643968763422959060534669593408000 0000000000}$$
$$+ \frac{350420634100611919703515863406374236503076475179955 9567256388046867463217857679575540735437 1014049417751 m^{27}}{42764728286893744076294093213466425866277560072 76704082140619 031598775049191668 10849280000000 0000000}$$
$$+ \frac{94119733191834286595296667001076818826068785797184760 65229342461668662174 7399524091533728662 63843960987 47 89364083849 m^{28}}{176171664773645702593401324700363965 26679120593331049558454515354314300588196421 534930046780 70272 000000000000 0}$$
$$+ \frac{38654499523845317864402924166488884092478088903129297 2446227226873332739359887287014765813821 005087116473561876004 492777521 m^{29}}{25492744579405627748075545289441467229965754663373 7619530660218983069655231437498179051 74893548063948 800000000 00000000}$$
$$+ \frac{21281483932659841079512261822419884051758741491236886 29637473901218557458319171110840762 5728004 19168320717 891360604 3646516129930949951 m^{30}}{68620220380634555887131055811125 750300028823184818 13307034106 320550222373793570246438 07120705927511546724876288000000000000 00000000}$$

$$\bar{a}_7 = 0.0400015951267311 m^{13} + 0.453176076729705 m^{14} + 2.54641722963561 m^{15} + 9.43777510577829 m^{16} + 25.766527885426 m^{17} + 53.9208812372436 m^{18} + 82.6401014628366 m^{19}$$
$$+ 60.5068781066779 m^{20} - 151.348474255908 m^{21} - 827.479888814369 m^{22} - 2341.77918974775 m^{23} - 4858.87659047016 m^{24} - 7449.10379687568 m^{25}$$
$$- 6271.8227667329 m^{26} + 8194.15083733869 m^{27} + 53425.0120828251 m^{28} + 151629.415198678 m^{29} + 310134.298820552 m^{30}$$

$$\bar{a}_7 = 6.27993800222651 \cdot 10^{-16}$$

$$\bar{a}_{-7} = \frac{228281 m^{13}}{132120576} + \frac{1727729817811273 m^{14}}{90778250929766400} + \frac{2952510841426631951 m^{15}}{28595149042876416000} + \frac{451016304701942402853151 87 m^{16}}{12279586258333376446464 0000} + \frac{8917565111375015998086 20144291 m^{17}}{928336721130003259352678400000}$$

$$+ \frac{4204625801639991758103388367503811553 m^{18}}{21686816214106607629534536495267 84000000} + \frac{10233778631940745199195435554926266 3200407131 m^{19}}{34839870247962265156847232879647784960000 00}$$

$$+ \frac{61473034322083526019248652651318540232 25702596935389 m^{20}}{25434126089210718858867575626066256700 89932800000000} - \frac{69499040774223690047090711481826169204 37386659696972391 m^{21}}{185726925283259181121685273833070188136 3398656000000000}$$

$$- \frac{858312879959562802540950215389349249922385 748013492223602397 56604613 m^{22}}{362719927378421892582181850788132994251167 53662754527641600000 00000}$$

$$- \frac{10809543432894429705973134456276670502016542 8643266264706663021967 24144067 m^{23}}{1581640243333608662604603960361653921432216043 4644111778119680 0000000000}$$

$$- \frac{31403263301877454800642257876151600514855349617 48514671211184568 6685100336088196107 m^{24}}{21938230674160615970268328913240873223002897 837522779673954684 08258560000000 00000}$$

$$- \frac{2108831288338403669333431877258492083179509392 9751541887260000 035310650079095806740 40923159 m^{25}}{9374842175758381861958794726166309113512605330107 7191529138119432722 186240000000000}$$

$$- \frac{22495263001381610141577970521811283842297615615 427546655711414 08499635952079606125 160466182434491057 m^{26}}{10402742396192640089324082825036061135092662715709 29876945335990178 355935671903846400000000000}$$

$$+ \frac{803916141015307016382604651883151848772269527698 4509998490049368 494367787079652668383050047 45230053487843 m^{27}}{6350562150603720995329672842199764241142217670805905 5619788192619 2418094804962714111180800000 00000000}$$

$$+ \frac{55400698784090784499256775372079981291153922199014 43345227660924 443049125360512931218504635017 192117664458150345829 m^{28}}{44042916193411425648350331175090991316697801483327623 8961362883857857 5147049105383732511695175680000000000000}$$

$$+ \frac{406304956642367555935218937428966100282486271221697 10615974804882 409240863157699478752972919550780 1187880763910116843 35531 m^{29}}{10754751619436749206219370668983118987641802748610805 823949727988348 251080076269454428745658215589478400000000000000}$$

$$+ \frac{87603239264897151382036013111873040607975374852751 11227392676622874 83845184622899327829230662700498362 09218213366482140414563 113423431 m^{30}}{1097923526090152894194096892978012004800461170957090129 125457011288035 579806971239430091393129484018474759802060 80000000000000000}$$

$$\bar{a}_{-7} = 0.00172782322717091 m^{13} + 0.0190324202120615 m^{14} + 0.103252157804792 m^{15} + 0.367289495927329 m^{16} + 0.960595967863929 m^{17} + 1.93879348638783 m^{18} + 2.93737564437092 m^{19}$$
$$+ 2.41695091494261 m^{20} - 3.74200136400406 m^{21} - 23.6632402901893 m^{22} - 68.3438821088116 m^{23} - 143.144010874427 m^{24} - 224.945790958642 m^{25}$$
$$- 216.243584091968 m^{26} + 126.5897603945 m^{27} + 1257.87989471002 m^{28} + 3777.91111333584 m^{29} + 7978.99281536152 m^{30}$$

$$\bar{a}_{-7} = 2.63862469941499 \cdot 10^{-17}$$

$$\bar{a}_8 = \frac{52553071771 m^{15}}{1352914698240} + \frac{146196445549002066149 m^{16}}{296300211034757529600} + \frac{384238055673929255372 3557 m^{17}}{123409037895976511 0784000} + \frac{4513594747561991206132016052775417 m^{18}}{347498537974225220053311160320000}$$

$$+ \frac{37137944559411368495649663029502 52333619 m^{19}}{926292102824094746574106228948 99200000} + \frac{23815116481340197434826455491205539332023804914083 m^{20}}{24778646613611684309304523029458908820 2752000000}$$

$$+ \frac{2173105620579778950035742076502819634 83583266788739971 m^{21}}{12255363748551003986355083 9371014421843279872000 0000} + \frac{2261696056396147312550722959252465377887387125894344880520 3196099 m^{22}}{1049073229563847675002499174454004402668872306759041 02400000000}$$

$$- \frac{3041405721462755298000807782322098065483284069376467540020 0823226961 m^{23}}{11352944960209926257958669698518651405839481708769 464811520 00000000}$$

$$- \frac{15863533662102433887575642003174612031412907609216048219037086837248438746 18645181 m^{24}}{14212942724746910609590095916139848108271956969159288764 179129222521600000 00}$$

$$- \frac{4133602988793503769432187712884824118683368366889979084 33235106661076463565485575 15797 m^{25}}{10224920204753152464453747639 30660800043728467958563923996320035567042560 0000000}$$

$$- \frac{59464337148883137087530019667379117012687889813 4172158053316110784883877518737696114 3556108290103 m^{26}}{60174463756488929423284687177246508745082110909776400929064 827863850972892836 65920000000 0000}$$

$$- \frac{5562948943189127684691641806439047500580299069541235202278869615 3085438382491092435391591831576237 79449 m^{27}}{30476199611966409 151434455806536975805069266219808995120137888979 237419135197625765068800000 00000000}$$

$$- \frac{4382439700701818074563181086913108726080437666707962379544990262616336196730 469343569736988421325986373709219354349 m^{28}}{18750733698575936113390005998320 83818250074372947625427999953890135129969161362617 095408781885440 0000 000000 00}$$

$$- \frac{152618517832862081568921398448031860797898542444 5426345265011358508002372158949250417607999994641947040967638232 2836097 m^{29}}{232165334411 201048977290590935875844762996708610465155078660 9574968980 0901499604 80400298640104488960000 000000 00}$$

$$+ \frac{30906505622862111065414885847557814289911520967044733328077996940816 272347991186545017919419527909776829596932697938 17913642772379647 m^{30}}{41995442411249054480069318433695491047018838 1132893451503051190378636698 7630952 4017216276761889329510446478131 2000000000000 000000}$$

$\bar{a}_8 = 0.0388443350045395 m^{15} + 0.49340648472185 m^{16} + 3.1135325436846 m^{17} + 12.9888165109252 m^{18} + 40.0931244541377 m^{19} + 96.1114497200094 m^{20} + 177.318736935631 m^{21}$
$\quad + 215.589912377847 m^{22} - 26.7895751465575 m^{23} - 1116.13294792792 m^{24} - 4042.67505860042 m^{25} - 9881.98871028091 m^{26} - 18253.4207480543 m^{27}$
$\quad - 23372.0971731077 m^{28} - 6573.69965330616 m^{29} + 73594.904228425 m^{30}$

$$\bar{a}_8 = 4.45392389724284 \cdot 10^{-18}$$

$$\bar{a}_{-8} = \frac{1386117 m^{15}}{939524096} + \frac{2443758832628437 m^{16}}{134376512940933120} + \frac{9115047895180925941003 m^{17}}{8227269193065100738560 0} + \frac{343056104165304476836260537887 3 m^{18}}{7722189732760560445629136896000} + \frac{16924320163843548436084154545564 69 m^{19}}{128651680947790937024181420687360 0}$$
$$+ \frac{8353960619642714835783845331528033227286459031 m^{20}}{27531829570679649232560581143843232022528000 00} + \frac{32363692058698510549276021489419486604721636266 9237 m^{21}}{594199454475200193277822251495827499846205440000 00}$$
$$+ \frac{1109579111319056697553261047155729372225400035605549932 72523 m^{22}}{16651956024822978968293637689746101629664639789826048000 00000} + \frac{38318024179909444107306802587854346541219869159389737598 9515531799 m^{23}}{11069121336204678101509702956055685120693494666050228191 23200000000}$$
$$- \frac{8947862914954125746263338998241946114386903105625419979039186550230473306020 9 m^{24}}{3158431716610245799089102035866329129493237709242863921595362050048000000000}$$
$$- \frac{8187192193295558341611764467670410040855173044542545582363412577218499379788189 m^{25}}{78107029341864359103466127800258811114451480191279188638607780494704640000000 00}$$
$$- \frac{8571903084320700859737709821107145596167992913398313345888045688996351062976687901 84493374067 m^{26}}{334302576424938496796026039873591715250456171720980005161471265910283182737981440 0000000000}$$
$$- \frac{1207495049514145155957774630101116690670238105909585399397008101082480020578623915295530385135772503 m^{27}}{253968330099720076261953798378081317089105518317416260011490748269784927933135480422400000000000}$$
$$- \frac{911176305434609349236408824269339435621774603631070416540039600713114869554294737277393572732186563403201 9113 m^{28}}{1449332073319879119875556019193881212173970529814589702801928803102232268147738173302060582174720000000000}$$
$$- \frac{8120553153398565700919396239793254877081720280129542768015132187363289860689578428702020057477699536923114241986439 m^{29}}{28263605928320127701583202374802276753756121048230540618053263830397019240182560584835146170561986560000000000000}$$
$$+ \frac{347878042768325628089979388550766823148200857552943006650201341951276647594180204426000278251435431227524435713648937188130069 29 m^{30}}{2318588953002018190756070030845851817640791614260281305744934108370665001387492196092291383088415897918269030400000000000000 0}$$

$$\bar{a}_{-8} = 0.00147533948932375 m^{15} + 0.0181859074859505 m^{16} + 0.110790684992587 m^{17} + 0.444247183813583 m^{18} + 1.31551488788644 m^{19} + 3.03429185415973 m^{20} + 5.44660413518593 m^{21}$$
$$+ 6.66335600253215 m^{22} + 0.346170423252834 m^{23} - 28.3300818817664 m^{24} - 104.820171273718 m^{25} - 256.4115172545 m^{26} - 475.451033221357 m^{27}$$
$$- 628.687049854244 m^{28} - 287.314830740043 m^{29} + 1500.38687244631 m^{30}$$

$$\bar{a}_{-8} = 1.63768169685695 \cdot 10^{-19}$$

$$\bar{a}_9 = \frac{7509100901081 m^{17}}{194819716546560} + \frac{276853885173236720976701 m^{18}}{51042649687587563765760 0} + \frac{1591773575683033417921182751 m^{19}}{4193003546395060752875520 00} + \frac{20277803868640570348377234757850 42683 m^{20}}{11525510264436219259413686393398886 4}$$
$$+ \frac{1275011512166342374048036862792233768182818 53 m^{21}}{21070309201246054074464133445154022359040 00} + \frac{13629402696061564961442878460175219587067495336 1557230157 m^{22}}{836510873762320781483258981748966792506520895488 000000}$$
$$+ \frac{8503973345673074324120840389915791625957912974389639526762509 m^{23}}{24531936139390700398168794528261074640442850415288320000000}$$
$$+ \frac{13500675664775403729076016131195343944719241556338164478898752699988859339 m^{24}}{247912173861315549708587926777423491405180995638982906656849920000 0}$$
$$+ \frac{435928377888757522194259585816737997738787653783034929398068970021984620870461 m^{25}}{1107869922551446928537737272718295009839147283331148681326813092249600000000 0}$$
$$- \frac{298762139093314686817960752137655010894488676111124356260201969751220359833227165318477659 m^{26}}{2575032033255861958339501646029934605580020957630303286765368973584325463244800000000 00}$$
$$- \frac{472014010422038573087409469292253003490080384123493601003765570260676823661678824539242110 73141 m^{27}}{74437867252945492353779706080769263610753473483374587377733424187233756754502221824000000 00000}$$
$$- \frac{3202652027326584754256877558372596969182592160950230940529935580736952306308694681746347135 7559047836159117 37 m^{28}}{173016604893595711816190832668803442388468378926484193417391610302585737170418848637250437 12000000000}$$
$$- \frac{4267686794707626218571022850013463170073813190529210352120748588059797480941202236249608678 17282915326527861132200 79 m^{29}}{106698371344893071341058715838188137621691073861040813769022268680586828832086914960903997 59694561280000000000}$$
$$- \frac{915778949658926081917340831689933026442815543441179753024242849290987082086672981370817160 0213464689331721146853740553464567 59 m^{30}}{141714277109134258573729048854011178603988421201760176474294536111105737611795418690043885 2569910041243339980800000000000000}$$

$$\bar{\bar{a}}_9 = 0.0385438447103294 m^{17} + 0.542397165640406 m^{18} + 3.79626098110879 m^{19} + 17.593844787255 m^{20} + 60.5122354868405 m^{21} + 162.931566385521 m^{22} + 346.649090285961 m^{23}$$
$$+ 544.574937749036 m^{24} + 393.483358483824 m^{25} - 1160.22688352953 m^{26} - 6341.0469407741 m^{27} - 18510.6627730685 m^{28} - 39997.6751370713 m^{29}$$
$$- 64621.5023877717 m^{30}$$

$$\bar{\bar{a}}_9 = 3.2269895420955 \cdot 10^{-20}$$

$$\bar{\bar{a}}_{-9} = \frac{1134559157 m^{17}}{869730877440} + \frac{545378014613241049877 m^{18}}{306255898125525382594560} + \frac{25753684335415054352685144539 m^{19}}{21384318086614809839665152000} + \frac{348452663289591140609710080239835619229 m^{20}}{648309952374537333342019859628687360000}$$
$$+ \frac{18927350449970492251815087877329598701671 m^{21}}{10651351158718133953083890987899551744 0000} + \frac{101115510116665811161944108122691739213246533316087587 17 m^{22}}{21958410436260920513935548270910378303296173506560000 00}$$
$$+ \frac{38769011458551140147410600579152509909947165134334068 9006161 m^{23}}{40886560232317833996479908804351244007374750692147200 00000} + \frac{27162841505931105144566790103765152300649289055930012901 378871394002633 m^{24}}{18593413039598666228144094508306761855388574672923717999 26374400000000}$$
$$+ \frac{2873547995520973198875235852754341284533053036151044205618 24144457099791759 m^{25}}{25965701309799537387603227980850392931050144530737972185 97181849600000 0000}$$
$$- \frac{6068470415676368376198528472387790470579855849460184864 56568618783542064283808226289 69217 m^{26}}{23175288299302757625055514814269411450220188618672729580 8883207622589291692032000000000}$$
$$- \frac{11068888458497664911423039323430448605309602485315312854 36408182643851687943199411795 4326105919 m^{27}}{74437866725294592353797060807692636107534734833745873777 33424187233756754502221824000000 0000}$$
$$- \frac{44894279655262988476863173040500547947442558899873355322 637566967839698401856267798960 219499033876810250033 m^{28}}{10380996293615742708971449960128206543308102735589051605 04349661855134423022513091823502622 7200000000000}$$
$$- \frac{20653486179633076817198889241345708236913924751179475783 21529509594080348579640767481069 3746906171523427907550 3407 m^{29}}{22228827363519389862720565799622528671185640387716836201 87963930845558934001810728352166616 603033600000000000}$$
$$- \frac{11212132774978992128217122517591916592476926574303386795 50918684849051668402057091685335 3185528845867610061137043 05444382002623 m^{30}}{74399995482295485751207750648355868767093921130924092649 00463145833051224619254812273039759 9202771652753489920000000 0000}$$

$$\bar{\bar{a}}_{-9} = 0.00130449451253186 m^{17} + 0.0178079187356485 m^{18} + 0.120432572276107 m^{19} + 0.53747850393678 m^{20} + 1.776990558562 m^{21} + 4.60486474693492 m^{22} + 9.48209172849593 m^{23}$$
$$+ 14.6088517735189 m^{24} + 11.0667066575109 m^{25} - 26.1850913667337 m^{26} - 148.699699050817 m^{27} - 432.466002158894 m^{28} - 929.130711300063 m^{29}$$
$$- 1507.0071849194 m^{30}$$

$$\bar{\bar{a}}_{-9} = 1.05360059412311 \cdot 10^{-21}$$

$$\bar{\bar{a}}_{10} = \frac{1069535470353 m^{19}}{27487790694400} + \frac{8878126743606798340880 87 m^{20}}{147780623857396375093248 0} + \frac{3224223426189956650221144 1397 m^{21}}{6982634477261978723155968 000} + \frac{7360399672834286446343486043 4421542674019 m^{22}}{3128352780612595132694291639 22554880000}$$
$$+ \frac{6747413531040308012335961125346946450567 00449 m^{23}}{7568111060039399014501403033361444765696 000} + \frac{274250143763636086351280771216232003396293568316 7687388259 m^{24}}{1031549211524893633885528536222376998469234463866880 00}$$
$$+ \frac{898711694508695858899059617054029214104512546009458883274 0552379 m^{25}}{14125054930870944773476236522938893588891509754523811840 000}$$
$$+ \frac{518281468616380207116973441115709284068492386873735013549312 23338905907688259 m^{26}}{21876797300488438377925087045290495615625130869653331750916611171928011862797783 7263}$$
$$+ \frac{436652197851747806268825880357465301170361190894099775090163554713600000 0}{15185016832492381710804688815311213313500480774533213778535527787201536000 0 m^{27}}$$
$$- \frac{37721457210739553437240621452844951314127572441291676816515610291757872243438235420824179 72301 m^{28}}{9779565175106228328135717944576139763744467322343008813830317685571009440884195328000000 0}$$
$$- \frac{11927485532037700285476653929060865261557444985842195049916810389602836120930056711326485864 628086471 m^{29}}{13603766341179676785369909089623193456959103824072018980390525113336696972647551069061120000000 0}$$
$$- \frac{936079682067481880467080974623841170516753344755068541913116648153704064425688883426603666870 7085846722082212761 m^{30}}{292039890435093852132297941997632335970538480290844127057844262461914647986178934584394949812019 200000000000}$$

$$\bar{\bar{a}}_{10} = 0.0389094737457708m^{19} + 0.600763923704498m^{20} + 4.61748848044275m^{21} + 23.5280359223851m^{22} + 89.1558471791927m^{23} + 265.862394832549m^{24} + 636.253592575078m^{25}$$
$$+ 1186.94345560662m^{26} + 1440.68311163655m^{27} - 385.717120703476m^{28} - 8767.78182813401m^{29} - 32053.1445438111m^{30}$$

$$\bar{\bar{a}}_{10} = 2.37819310532035 \cdot 10^{-22}$$

$$\bar{\bar{a}}_{-10} = \frac{144484346557m^{19}}{1217623228416000} + \frac{35472468885361335067067m^{20}}{19950384220748510637588848} + \frac{17398623509246548342028708990m^{21}}{131595803609937291321016320000} + \frac{160129760295115729230740358503006889824830m^{22}}{2463577819023241866699675466589011968000000}$$
$$+ \frac{687396009745950274348864606844541619916644230m^{23}}{289716751517133243523881834870867807436800000} + \frac{1196346645588791840759201050048064096155923476856257156739470m^{24}}{175466520880384407123928404011663274396167823037562880000000}$$
$$+ \frac{982401632880598812984655147524237766459890399063784574024699585390m^{25}}{622914922451408664510302030661605207270115580174500102144000000000}$$
$$+ \frac{7872681285432342605926598898440049517148333140818441039598809632595664150788890m^{26}}{275090884646601179493603046252031397373275502632828583068030394695680000000000}$$
$$+ \frac{10883187185053770119569722828375833755318748610810240381919820907408291366644235910m^{27}}{3170735048830085385636774511151631189040583343454599776341579912424980480000000000}$$
$$- \frac{2475983715337572720355754961160590021228491441237147423255837901710057837562820205634841869138019m^{28}}{3943120678602831261904321475253099552741769224368701153763840908222310065646507755624960000000000}$$
$$- \frac{48606082421539375158397880571538166656247169051169729461993405916955379814498624032796576039305879m^{29}}{2579661573985327495452498240679320740677229440323254373216469545485673961531462857523200000000000}$$
$$- \frac{96748236996937000020254535113720762641664982138362208059034735816374604989495995202341399698429900483486500932121m^{30}}{14130058058811580941569103625613442943598533830392202243566736794957278328163281570931362598770473697280000000000}$$

$$\bar{\bar{a}}_{-10} = 0.00118660964397796m^{19} + 0.0177803437231398m^{20} + 0.132212601252983m^{21} + 0.649988642772421m^{22} + 2.37264847871699m^{23} + 6.81809065106125m^{24} + 15.7710402732764m^{25}$$
$$+ 28.6184738383352m^{26} + 34.3238618725629m^{27} - 6.27924914591985m^{28} - 188.420383943804m^{29} - 684.698085416601m^{30}$$

$$\bar{\bar{a}}_{-10} = 6.97771810697177 \cdot 10^{-24}$$

$$\bar{\bar{a}}_{11} = \frac{109266428162197927m^{21}}{2743061608975564800} + \frac{240833265104892047633948656907 9m^{22}}{3597453282685371438169954713600} + \frac{53559858435737774605631154870121 3871m^{23}}{95566346454536892254984846966784000} + \frac{5016735508198707513426533692565670889237222782643m^{24}}{1611421127181238936665174987370902223123906560000}$$
$$+ \frac{588141435472635942832167563124275322362106866586902131 01m^{25}}{45661229059807586509443984421418853944390162841600000} + \frac{4860613969723212692680930263826724392210093607228179637204989228393687m^{26}}{115489613127105359559570968488529532294940895451171698937692160000000}$$
$$+ \frac{44523273932767997877396999705601771127249120801193556825467094494150546 1m^{27}}{39947676728145232769531286170544101893425844891411148206768128000000}$$
$$+ \frac{613226654200425545485000049889782174943153595607119999308337781455248914681132715180238 1m^{28}}{25865853972130873158550490834224814533950448941724385686081847307716335239168000000}$$
$$+ \frac{27241436738639405961645305531821870486429417095578310707224544849435284682370219731190711326 71m^{29}}{73293483815430042182068670827859434463401992121270219280081522531145007533706444800000000}$$
$$+ \frac{45043244605832925974389266192486324772903344754978160797808157996963938529340863811210275905903308545609809m^{30}}{185379068081910343731311293960658842165719986174782365526540901340443703114147084320138644684800000000}$$

$$\bar{\bar{a}}_{11} = 0.039833749196397m^{21} + 0.669454878716642m^{22} + 5.60446856270868m^{23} + 31.1323677192577m^{24} + 128.805432438597m^{25} + 420.870227037104m^{26} + 1114.5397574873m^{27}$$
$$+ 2370.79608839185m^{28} + 3716.76107077126m^{29} + 2429.79129585064m^{30}$$

$$\bar{\bar{a}}_{11} = 1.77719418660699 \cdot 10^{-24}$$

$$\bar{\bar{a}}_{-11} = \frac{83522332107 m^{21}}{75591424409600} + \frac{85821346759884353873042479 m^{22}}{4758536088208163278002585600} + \frac{232979321937082660167662935 6807853 m^{23}}{159277244090894820424974474494464000} + \frac{976595399826981154385370642502364388700059863 m^{24}}{12433804993682399202663387248232270240015360000}$$
$$+ \frac{4475992990616208085078276303831212643010843070981 5393 m^{25}}{142691340811898707841701245131693391857621925888 00000} + \frac{423171287840394567795419675262934605440995262886 55112446157599 8683 m^{26}}{427739307878167998368781364772331601092373686856191477547008000 00}$$
$$+ \frac{510921648522310613628428411147831439315361555704472159598033132 5105261 m^{27}}{201708190185506500769230951867047141710394798662746626053898240 000000}$$
$$+ \frac{569051339679376701094986400131073405001912442119307850151527430 81294650131719995487 m^{28}}{1088630217682275806336300119285556167253806773641598724161693910 2574215168 00000}$$
$$+ \frac{915778174909198310567245686638656758903768024782555535501541194 2697937123156064102959 72231 m^{29}}{11452106846160944090948229816853036634906561268948471762512737895 491407427141632000000}$$
$$+ \frac{19157102252869391166315945738609480409703606392916144373315609679157480696598975644207284928296 8875663 m^{30}}{366180875223526604901355642391424873413767873925496030669710422400876450595846092484224483 3280000000}$$

$\bar{\bar{a}}_{-11} = 0.00110491808772415 m^{21} + 0.0180352413366272 m^{22} + 0.146272823382183 m^{23} + 0.785435673410664 m^{24} + 3.13683574991186 m^{25} + 9.89320551201073 m^{26} + 25.3297423397844 m^{27}$
$+ 52.2722344499037 m^{28} + 79.9659125793252 m^{29} + 52.3159551715402 m^{30}$

$$\bar{\bar{a}}_{-11} = 4.73252341203497 \cdot 10^{-26}$$

$$\bar{\bar{a}}_{12} = \frac{217295418508894375 m^{23}}{5266678289233084416} + \frac{110279988206983628880575573392111 m^{24}}{14709586755869074324961592606 7200} + \frac{1427501501240259821315662743430061 5529 m^{25}}{2102459621999811629609666633269248 0000}$$
$$+ \frac{7061702240940323682723160578024079275557563613276 18377 m^{26}}{172959200984119645868728781977810171948632637440 00000} + \frac{5517138781591273133614408345289860347882993814568 057238327383 m^{27}}{301364111794730070961673029718136443603297507475 45600000000}$$
$$+ \frac{402829523424050605240950095309475117830453473563 47840912495651597290707 40151 m^{28}}{619794257115465429636364197555108489982849472254 09656214509621450 96200000000}$$
$$+ \frac{151252744929019672670399173359118938882482740795 58556480427724142484043836896041 m^{29}}{803518983331835539135715013258944220942051280101 5270953589932032 00000000 0000}$$
$$+ \frac{1547642780403533934790552603147202923357681449053187889719700540977162166977435353502562305 69767 m^{30}}{34703354079275588154388575202584959499716852330146884128826478471186083112550400000000000 0}$$

$\bar{\bar{a}}_{12} = 0.0412585327175046 m^{23} + 0.749715067032609 m^{24} + 6.78967380064333 m^{25} + 40.8287168347216 m^{26} + 183.072189609266 m^{27} + 649.940716293221 m^{28} + 1882.37923517179 m^{29}$
$+ 4459.63458421953 m^{30}$

$$\bar{\bar{a}}_{12} = 1.34339815485257 \cdot 10^{-26}$$

$$\bar{\bar{a}}_{-12} = \frac{863391067766779 m^{23}}{8229184826926694 40} + \frac{204470213450734807424689817180 69 m^{24}}{1103219006690180574372119445504 000} + \frac{152159912659879575537254324126 880761 m^{25}}{934426498665829464931851703418 880000}$$
$$+ \frac{615013488524814408228235801289 55343975041074017841 m^{26}}{648597003690448672007732932416 7881448073723904000} + \frac{138308549107596560439278716664 046338906383038734252070559 m^{27}}{336343877090409705543864923899 15665215373967360000 0}$$
$$+ \frac{656148043925871317730826716811 5021888534683398826324022547531 9614268066931 m^{28}}{464845692836599072227273148166 331367487137104190966088224210944 00000000 00}$$
$$+ \frac{476847147009744527565959141116 7380774615979745737163982827037099479774135935417 m^{29}}{120527847499775330870357251988 8416331413076920152290643038489804 8000000000000}$$
$$+ \frac{10752968901925804523190388528702651550459767999204378503692314563521823783388278770805473 8291 m^{30}}{118306888906621323253597415463357816476307451125500741348272085697225283338240000000000 00}$$

$\bar{\bar{a}}_{-12} = 0.00104918176699796 m^{23} + 0.0185339639918075 m^{24} + 0.162837754362733 m^{25} + 0.948221291534577 m^{26} + 4.11211737397637 m^{27} + 14.1153947221044 m^{28} + 39.5632342982507 m^{29}$
$+ 90.8904713943838 m^{30}$

$$\bar{\bar{a}}_{-12} = 3.27391754177529 \cdot 10^{-28}$$

$$\bar{\bar{a}}_{13} = \frac{4863942539190675027 m^{25}}{112702580674946662400} + \frac{330704111093741726593983556104449 m^{26}}{392255646823175315332091361792000} + \frac{14370201547115219900490022320793757689 m^{27}}{17499463840574662623401787103641600000}$$
$$+ \frac{51416742415589565718752180556519558171923274425979 2131 m^{28}}{9676039215894805363285526264492876752371056640000000} + \frac{48258436217015921566808973719674529067815389800011 39160463658287 m^{29}}{18805120575991156428008397054411714080845764466468454400000000}$$
$$+ \frac{12722446803726753791265538678711034870635587832772 105116823234247568398017804 7469 m^{30}}{1294130408857091817080728444495066527084189698067649589616203268096000000000000}$$

$$\bar{\bar{a}}_{13} = 0.0431573306490568 m^{25} + 0.843083111159951 m^{26} + 8.21179533157818 m^{27} + 53.138212101422 m^{28} + 256.623912736983 m^{29} + 983.088467487798 m^{30}$$

$$\bar{\bar{a}}_{13} = 1.02034804512926 \cdot 10^{-28}$$

$$\bar{\bar{a}}_{-13} = \frac{3152346664059167 m^{25}}{3112128080001368064} + \frac{30591024823358244603127117381268 63 m^{26}}{158863536963386002709585200152576000} + \frac{1247170123211394403795501351571132368399 m^{27}}{6844877238928082383703123369530490880000}$$
$$+ \frac{17801885762102555344578906480375989114113507415170 63343 m^{28}}{1556632808857076812818559037800291547537693736960000000} + \frac{26205327753762549236113954768184583695901109359223 0706122693 m^{29}}{4897166816664363653127186732919717208553584496476160 0000000}$$
$$+ \frac{19270110893444459972751004450229208034748493580715 7048979031579434679829362914 1 m^{30}}{97059780664281886281054633337129989531314227355073 71922121524510720 0000000000}$$

$$\bar{\bar{a}}_{-13} = 0.00101292317765334 m^{25} + 0.0192561650131261 m^{26} + 0.182204892750816 m^{27} + 1.14361496563683 m^{28} + 5.35112009347722 m^{29} + 19.8538578611644 m^{30}$$

$$\bar{\bar{a}}_{-13} = 2.29330375198912 \cdot 10^{-30}$$

$$\bar{\bar{a}}_{14} = \frac{13964132737914394304046 1 m^{27}}{306731343564934836387 8400} + \frac{4207848391002200221992344255532360867 83 m^{28}}{44227608690606663154348519722477158400 0} + \frac{8925004033914884717928143622548094961723552 91 m^{29}}{899976985946989107515609777452950434611 20000}$$
$$+ \frac{1430845088707157421438158239490986943096628095352 2466401418470 9 m^{30}}{2082635935554664944703917881028277607473196574210 785280000000}$$

$$\bar{\bar{a}}_{14} = 0.0455256139643851 m^{27} + 0.951407619715124 m^{28} + 9.91692473616274 m^{29} + 68.7035628397569 m^{30}$$

$$\bar{\bar{a}}_{14} = 7.18839466958989 \cdot 10^{-31}$$

$$\bar{\bar{a}}_{-14} = \frac{1956386521290599 07 m^{27}}{19722951618115665920 0} + \frac{826964698111865150435275678210169 11 m^{28}}{4095148952833950292069307381710848000} + \frac{348054863760834241631216003714372758335953 77 m^{29}}{169995652901097942530726291296668415426560 0}$$
$$+ \frac{2391402222822150795918413196616079711911436703458 21510158163 m^{30}}{1735529946295554120586598234190231339560997145175 65440000000}$$

$$\bar{\bar{a}}_{-14} = 0.000991933945370349 m^{27} + 0.0201937635880029 m^{28} + 0.204743390681484 m^{29} + 1.37790893664874 m^{30}$$

$$\bar{\bar{a}}_{-14} = 1.50826965458942 \cdot 10^{-32}$$

$$\bar{\bar{a}}_{15} = \frac{356117996563085075664453 53 m^{29}}{7361552245558430733081600 0} + \frac{33095518413573819997191852587857650833 9 m^{30}}{30732846541337499652760826458362871808000}$$

$$\bar{\bar{a}}_{15} = 0.0483753948466436 m^{29} + 1.0768777428104 m^{30}$$

$$\bar{\bar{a}}_{15} = 2.84643294327497 \cdot 10^{-33}$$

$$\bar{\bar{a}}_{-15} = \frac{21753628667269520848 9 m^{29}}{22120048814778954547200 0} + \frac{9700471846967451239966488048538316914 03 m^{30}}{45440708814691874486582079120579389030400}$$

$$\bar{\bar{a}}_{-15} = 0.000983434930429963 m^{29} + 0.0213475363831277 m^{30}$$

$$\bar{\bar{a}}_{-15} = 5.69405171111742 \cdot 10^{-35}$$

## C

$$C = m^{\frac{-2}{3}}\left(-\frac{1}{2} - \frac{4m}{3} - \frac{7m^2}{36} + \frac{70m^3}{81} + \frac{39533m^4}{15552} + \frac{1271m^5}{729} + \frac{424859m^6}{419904} + \frac{421861m^7}{314928} + \frac{1890215575m^8}{967458816} - \frac{10199959357m^9}{1632586752} - \frac{9665055967729m^{10}}{391820820480} - \frac{61517689176353m^{11}}{1469328076800} - \frac{355429192472810887m^{12}}{8463329722368000}\right.$$

$$-\frac{10556781029241299m^{13}}{396718580736000} + \frac{128198709231235333793m^{14}}{6664872156364800000} + \frac{18757719443128394 0375411m^{15}}{1049717364627456000000} + \frac{10255140782539863250420585693m^{16}}{18427038537917399040000000}$$

$$+\frac{639060826412948052999716 2919849m^{17}}{59254445798490636288000000 00} + \frac{24649181034360950572831229960608277m^{18}}{17918544409463568413491200000000} + \frac{20635202471360270602846624734 72768691m^{19}}{235180895374209335427072000000000 0}$$

$$-\frac{40132809601796205503760881990065295289221m^{20}}{506853021992757049896474776667756620891 50800439} - \frac{65439887395717893184337278982858 605898597m^{21}}{31524796941509014220374587997380961347703 02383531}$$

$$-\frac{316083123382937346813984768000000000000}{28749735591199295051149502054400000000 0000000}m^{22} - \frac{97232601431274672115630080000000000}{9487412745095767366879335677952000 00000000000}m^{23}$$

$$\left. -\frac{52004009513230191904964485064561276449582019433989298 5887m^{24}}{1178200044197364927916330428767875891200000000000000}\right)$$

$$C = m^{-0.666666666666667}(-0.5 - 1.33333333333333m - 0.194444444444444m^2 + 0.864197530864197m^3 + 2.54198816872428m^4 + 1.74348422496571m^5 + 1.01180031626276m^6$$
$$+ 1.33954745211604m^7 + 1.95379435665818m^8 - 6.24772885392126m^9 - 24.6670300876018m^{10} - 41.867905573771m^{11} - 41.9963778007415m^{12} - 26.6102510491345m^{13}$$
$$+ 19.2349839912246m^{14} + 178.693047054485m^{15} + 556.526799541761m^{16} + 1078.50274827687m^{17} + 1375.62407253028m^{18} + 877.418314039772m^{19}$$
$$- 1269.6916296026m^{20} - 6730.2413421461m^{21} - 17629.8324687171m^{22} - 33228.0230538139m^{23} - 44138.5227995449m^{24})$$

$$C = -3.25443973748474$$

## $A_i$

$$A_0 = m^{\frac{2}{3}} \cdot \left(1 - \frac{2m}{3} - \frac{115m^2}{144} - \frac{599m^3}{648} - \frac{14347m^4}{62208} - \frac{76249m^5}{93312} - \frac{31682233m^6}{26873856} + \frac{51841729m^7}{100776960} + \frac{299491849721m^8}{48372940800} + \frac{76605292015663m^9}{6530347008000} + \frac{5372696615725109m^{10}}{391820820480000} + \frac{1065650143979656267m^{11}}{82282372300800000} \right.$$
$$+ \frac{27402387238566958543m^{12}}{32398684093440000000} - \frac{20332917769649336560659 73m^{13}}{87087662843166720000000} - \frac{17199927460523885354702331667m^{14}}{14630727357652008960000000000} - \frac{250016484538188917007535181 74163m^{15}}{92173582353207656448000000000000}$$
$$+ \frac{6328802293735169819846881322077 0177m^{16}}{15485161835338886283264000000000000} + \frac{66881603729723995467060779693984406 13m^{17}}{162594199271058305974272000000000000} - \frac{110578480143322577709677706157 13562840959m^{18}}{1229212146489200793165496320000000000000}$$
$$+ \frac{14188283252512347309636269385147785 1300478819m^{19}}{141974002919502691610614824960000000000000} + \frac{188436209816073211508350336775625172 32410578832347m^{20}}{5247359147904819481928323930521600000000000000}$$
$$+ \frac{91001993552462034951790319955751149860188205438668 09m^{21}}{11363812154681374690551026512035840000000000000000} + \frac{10728097148331172957411332285157269341 91963188879296734559m^{22}}{8400129944740472171255318797696892928000000000000000}$$
$$+ \frac{6712162730763564659074072160425507430800261673349792 67932 75587m^{23}}{5045118044811127586055944469896753892556800000000000000000} + \left. \frac{26733980204116826197243341417026632325575670182403203 84047775548817m^{24}}{36361174772562758738222402983439884654435368960000000000000000} \right)$$

$$A_0 = m^{0.666666666666667}$$
$$\cdot (1.0 - 0.666666666666667 m - 0.798611111111111 m^2 - 0.924382716049383 m^3 - 0.230629501028807 m^4 - 0.817140346364883 m^5 - 1.17892396982406 m^6$$
$$+ 0.514420448880379 m^7 + 6.19130953727337 m^8 + 11.7306617736879 m^9 + 13.7121263978348 m^{10} + 12.9511353912349 m^{11} + 8.4578704368167 m^{12}$$
$$- 23.3476443227857 m^{13} - 117.560303326466 m^{14} - 271.245272403681 m^{15} - 408.701075328262 m^{16} - 411.340650709357 m^{17} - 89.9588248124215 m^{18}$$
$$+ 999.357837403296 m^{19} + 3591.06751614881 m^{20} + 8008.0515511666 m^{21} + 12771.3466564268 m^{22} + 13304.2729053029 m^{23} + 735.234226378451 m^{24})$$

$$A_0 = 0.175827099959204$$

$$A_1 = m^{\frac{2}{3}} \cdot \left(\frac{3m^2}{16} + \frac{3m^3}{8} + \frac{31m^4}{96} + \frac{2381m^5}{17280} - \frac{165871m^6}{1382400} - \frac{46898921m^7}{62208000} - \frac{209730782347m^8}{89579520000} - \frac{1821549712907m^9}{391910400000} - \frac{95132047577539547m^{10}}{15801827328000000} - \frac{30166270980514716431m^{11}}{7466363412480000000} \right.$$
$$+ \frac{17845547547717110460817m^{12}}{4181163510988800000000} + \frac{64364880175519604068305313m^{13}}{263413301192294400000000000} + \frac{84841560959941366460308840 4681m^{14}}{1327603038009163776000000000000} + \frac{34978209410820454746271776358 43m^{15}}{29041316456450457600000000000000}$$
$$+ \frac{37829364700189629842941077757315278 1m^{16}}{23418917590481649008640000000000000} + \frac{4414106577262384519455717539941627 55131m^{17}}{40573274725509456907468800000000000000} - \frac{420533096932195628602284984 3152065423077637m^{18}}{2499313723091382545500078080000000000000000}$$
$$- \frac{147495506584328692309330012707490121 5909090 99973m^{19}}{1732024410232810403155410944000000000000000} - \frac{6081145954099720472810376373 76369509976732694152578 33m^{20}}{2880702998821921026252807948206080000000000000000}$$
$$- \frac{5519274352534229171283765406236789730892330814917611 78787m^{21}}{144179185094053714736395303780771430400000000000000000} - \frac{51964117797562713473418148042 71005614563767126584875892 1744967m^{22}}{10391282222809863932848148233408775853178880000000000000000}$$
$$- \frac{139101598044503160611028437 8749570920206888587222375447382710262467m^{23}}{46807530796470320855144837173898308306442649600000000000000000000}$$
$$+ \left. \frac{26624081938952122022544521213446726277994 4733776920909103168481 0011674017m^{24}}{3748347066181343294079998560885776529179927379968000000000000000000000} \right)$$

$$A_1 = m^{0.666666666666667}$$
$$\cdot (0.1875 m^2 + 0.375 m^3 + 0.322916666666667 m^4 + 0.137789351851852 m^5 - 0.119987702546296 m^6 - 0.753904980066872 m^7 - 2.34128048852014 m^8$$
$$- 4.64787286304982 m^9 - 6.02031939742631 m^{10} - 4.04028967168835 m^{11} + 4.2680817195539 m^{12} + 24.4349392700305 m^{13} + 63.9058201366934 m^{14}$$
$$+ 120.442919532497 m^{15} + 161.533361027603 m^{16} + 108.793451037047 m^{17} - 168.25942779686 m^{18} - 851.578682864029 m^{19} - 2110.99372495513 m^{20}$$
$$- 3828.06599228161 m^{21} - 5000.74164640131 m^{22} - 2971.77816640976 m^{23} + 7102.88601051973 m^{24})$$

$$A_1 = 0.000268869268448282$$

$$A_2 = m^{\frac{2}{3}}\Bigg(\frac{25m^4}{256} + \frac{113m^5}{320} + \frac{23533m^6}{38400} + \frac{96142847m^7}{145152000} + \frac{30380129071m^8}{81285120000} - \frac{5269223469151m^9}{8534937600000} - \frac{54691607012857417m^{10}}{16131032064000000} - \frac{81591250948534450631m^{11}}{9033377955840000000} - \frac{27788404820492619078611m^{12}}{1686230551756800000000}$$
$$- \frac{4455648363224671183015 30181m^{13}}{21512086264037376000000000} - \frac{28829247887695513221597 1165841m^{14}}{24093536615721861120000000000} + \frac{37262600617375465101011 5673591053m^{15}}{13914017395579374796800000 0000000}$$
$$+ \frac{28376206730794903727408547 1985792969563m^{16}}{23141793732327616162037760 0000000000} + \frac{53536058234197296807610538 3645531305550153m^{17}}{17819181173892226444476907 52000000000000}$$
$$- \frac{88247475549783482702361764986 548682531 75283641m^{18}}{16464923404676452346966625 4848000000000000000} + \frac{3005361604396690777411557 1656984 7039690234 4777601371m^{19}}{4449974848 5190477581466986977 6896000000000000000}$$
$$+ \frac{12251357928385120271441 728059072268920 53952285564 81964379m^{20}}{35635398587443893444 72387631173 383168000 00000000000000} - \frac{22389974776988238530717 39764917441 18281036241921 004955701741m^{21}}{1981724110335074296 564922232971697586176000000 0000000000000}$$
$$- \frac{16223807762519638536875 96206548138779872 3652468886827937 4193856680683m^{22}}{34706917279456882352592 5792674608937887030 312960000000000000000000}$$
$$- \frac{19096795852788175504778566 6285136831552983883047260 5766740779809657799237m^{23}}{173708120983681696174725859 233641773412458671636480000000 00000000000000}$$
$$- \frac{795496319303642300108032091520847086384 28368879605296069878829 4960207858539m^{24}}{4173163898511969068901614042229009644609 072739479552000000 00000000000000000}\Bigg)$$

$$A_2 = m^{0.666666666666667}(0.09765625m^4 + 0.353125m^5 + 0.612838541666667m^6 + 0.662359781470459m^7 + 0.373747729855108m^8 - 0.617371059531941m^9 - 3.39045925864309m^{10}$$
$$- 9.03219718552642m^{11} - 16.4795998931114m^{12} - 20.7123024170527m^{13} - 11.9655525660286m^{14} + 26.7806195421418m^{15} + 122.61887327755m^{16}$$
$$+ 300.440619082068m^{17} + 535.972584753835m^{18} + 675.365975462631m^{19} + 343.797415323478m^{20} - 1129.8229990855m^{21} - 4674.51708023707m^{22}$$
$$- 10993.611435462m^{23} - 19062.1873151757m^{24})$$

$$A_2 = 1.043130540764 \cdot 10^{-6}$$

$$A_3 = m^{\frac{2}{3}}\Bigg(\frac{833m^6}{12288} + \frac{55583m^7}{161280} + \frac{232353563m^8}{270950400} + \frac{939047456477m^9}{682795008000} + \frac{848348752689791m^{10}}{573547806720000} + \frac{47086774900464097m^{11}}{120445039411200000} - \frac{145906781338568783459m^{12}}{37940187414528000000} - \frac{3402995638140810754060913m^{13}}{23371155447349248000000000}$$
$$- \frac{13111650211696478684309115 05117m^{14}}{3887090574003126927360000000 0} - \frac{7575033594411624030409562 340831281m^{15}}{13468768838920834803302400 0000000} - \frac{588893108974225356315623 33797271893471m^{16}}{9333856805372138518688563 20000000000}$$
$$- \frac{52122253065932552557580997 47259384226 31939m^{17}}{420443579797987957432629344 00000000000} + \frac{209505095114577025144235 696050467338079764819 42891m^{18}}{12120883873280235865523 38883392307200 0000000000}$$
$$+ \frac{327661663049561232729484041 53870291723083439135528369m^{19}}{545985214071908224563805105 0024064778240000000000000} + \frac{10486945381750750008549529 76989575076630863924755187 8140954207m^{20}}{78700492697181139121525123 0554668793394626560000000 0000000}$$
$$+ \frac{3876956541909721411694997 80966712832414111724579430 456259936521619m^{21}}{1772531846772262205864549 58401675278992304766976 00000000000000}$$
$$+ \frac{3059577651577031913838346 674398899165825304364747 88962410015344534070941m^{22}}{12774991526057048170106981761 92554070753338916549427200000 00000000}$$
$$+ \frac{1922403052724167774009381 045448930598772148681693 019818904539 94161107400863m^{23}}{115089986582479469644937 986931871962341683029919 37896448000000 00000000000000}$$
$$- \frac{3163802833248364068024577 25649391144931281131948069 4668554730549410297124914 1727159m^{24}}{4147379588048623016812498 529707693803494488966617 47403640012800000 000000000000}\Bigg)$$

$$A_3 = m^{0.666666666666667}(0.0677897135416667m^6 + 0.344636656746032m^7 + 0.857550175235025m^8 + 1.37529924131636m^9 + 1.47912474383142m^{10} + 0.390939926879924m^{11}$$
$$- 3.84570533994512m^{12} - 14.5606649436187m^{13} - 33.7312701159763m^{14} - 56.2414700631135m^{15} - 63.0921516425328m^{16} - 12.3969672913012m^{17}$$
$$+ 172.846384228149m^{18} + 600.129187759299m^{19} + 1332.51330739463m^{20} + 2187.24224840844m^{21} + 2394.974310031m^{22} + 167.034907071438m^{23} - 7628.4380681368m^{24})$$

$$A_3 = 5.32887300001772 \cdot 10^{-9}$$

$$A_4 = m^{\frac{2}{3}} \cdot \left(\frac{3537m^8}{65536} + \frac{16905377m^9}{48168960} + \frac{4880409049m^{10}}{4335206400} + \frac{15870928675919m^{11}}{6729627598848} + \frac{47710796855984309m^{12}}{13596594536448000} + \frac{50839452119613445889 0233m^{13}}{1615948462359 57657600000} - \frac{186410882174130929960126887m^{14}}{79989448886799040512000000} \right.$$

$$- \frac{8221731388379258679429390015210713m^{15}}{40354996921185663134466048 0000000} - \frac{25170844117757102222967448278111 4054559m^{16}}{41549504830052758763246243020800 0000000} - \frac{818240788366452496658282417104684484753024281m^{17}}{6550591057744042814716494559051776000000000}$$

$$- \frac{31747940697236635687034267032550238042956070881117m^{18}}{1686122138263316 620508025699499927142400000000} + \frac{3577036821014700349279776773902041620385622888 8040452653m^{19}}{2126638408105990720781952493751278107623424000000000}$$

$$+ \frac{2618731507421662657907673501704338008795429920808 79264486351m^{20}}{21895869049859280461170982875663159396090773504 00000000000} + \frac{137508063725173828054951164815456726049652900460 2932714109888513573m^{21}}{138081918989127580372282569 308794582099567244948275200000000000 00}$$

$$+ \frac{5047498111112763151483860614717705702582694084398170 59698709920 20901461m^{22}}{177711429739007195939127666700418627162143044248430 18240000000000000}$$

$$+ \frac{870224850884015394586146213981983172391323137629 2090849221781852871024 7401397389m^{23}}{152415416 1533817468461252294137475984322766844723670 3526060032 000000000000}$$

$$\left. + \frac{112639874655425296593415445382229588 7587256698437877851013539534213681447 89167782138797m^{24}}{1333878756007935695698545 577373534824399126318283673457866697605120 000000000000000} \right)$$

$A_4 = m^{0.666666666666667}$
$\cdot (0.0539703369140625m^8 + 0.350959975054475m^9 + 1.12576163593964m^{10} + 2.35836655785171m^{11} + 3.50902549370634m^{12} + 3.14610603641274m^{13}$
$- 2.33044338682642m^{14} - 20.3735150926576m^{15} - 60.5803708629303m^{16} - 124.910986070354m^{17} - 188.289685407586m^{18} - 168.20145857332m^{19}$
$+ 119.59934092858m^{20} + 995.844095532892m^{21} + 2840.27770106047m^{22} + 5709.55926143503m^{23} + 8444.53621800955m^{24})$

$A_4 = 3.10986256277136 \cdot 10^{-11}$

$$A_5 = m^{\frac{2}{3}} \cdot \left( \frac{732413m^{10}}{15728640} + \frac{28066785623m^{11}}{76299632640} + \frac{36237718586269m^{12}}{25178878771200} + \frac{352689501599570366629m^{13}}{95271337916891136000} + \frac{45224967070533 96773344319m^{14}}{65394246346154075750 4000} + \frac{70391598994086172122571969 60927m^{15}}{77324200449890896482140 1600000} \right.$$

$$+ \frac{546786879475783066714029534 30636773m^{16}}{1326883279720127783633525145 6000000} - \frac{4580824846750735836835599757 15439541938 3611m^{17}}{200825376645577004206274391016734720 000000} - \frac{671666887597144032912377450233212896 176767544489m^{18}}{689232692647620278435933709969433559 0400000000000}$$

$$- \frac{160153194111253621262695939952472287581598394784 8939223m^{19}}{651976219446553164285079313277035828006092 8000000000} - \frac{12740712257328858983613964107931 33965641327855855786029 88107m^{20}}{2796977981425713074782990253958 4837021461381120000000000}$$

$$- \frac{862770451049010874355998285287128618923710547626 063273792920172011m^{21}}{1439308311132021908061924069384732278900885967297 5769600000000}$$

$$- \frac{31596549523998852564399790773613439984235633384301 18202647273807682276 81m^{22}}{41686178460126839673693225884773088730026709575428 3332363715783148103073586 20249m^{23}}$$

$$+ \frac{104968755130858357 754956122380228525100241613 59501228769280000000000 0}{33760187484568335137014069522888192539822338078525 52548732336640000000000}$$

$$\left. + \frac{132223534953610513326969678724282980060594061310088 2248021703054634546478463460 84487m^{24}}{2512378033928131307696363357453301920335962363335373 199271421542400000000000} \right)$$

$A_5 = m^{0.666666666666667}$
$\cdot (0.0465655644734701m^{10} + 0.367849551195428m^{11} + 1.43921097184511m^{12} + 3.70194760891502m^{13} + 6.91574100130198m^{14} + 9.1034370332355m^{15}$
$+ 4.12083630740387m^{16} - 22.8099900683125m^{17} - 97.4513970045444m^{18} - 245.642692684718m^{19} - 455.51707385392m^{20} - 599.434078422321m^{21}$
$- 301.009090606146m^{22} + 1234.77331040242m^{23} + 5263.35399173004m^{24})$

$A_5 = 1.96333204070725 \cdot 10^{-13}$

$$A_6 = m^{\frac{2}{3}}\Bigg(\frac{31979701 m^{12}}{754974720} + \frac{68656059199267 m^{13}}{174573559480320} + \frac{41141695074528329 m^{14}}{22694562732441600} + \frac{41673728188288091099639 m^{15}}{755666846666693197824} + \frac{10891349120155520280396668 2663 m^{16}}{884130210600003104145408 0000}$$
$$+ \frac{10487862329618598467621955511896 2317 m^{17}}{51109578181812329443661709312 00000} + \frac{7538012159390218489904608223148479 72033 m^{18}}{358789238836322552694505199370240 00000} - \frac{28870253524369227305406337165692578 0424501182369 m^{19}}{22565988900018382449532051837517775 0599680000000}$$
$$- \frac{53860502306563875230543101325787367131316224000 0000 m^{20}}{6134342635464442047951542051114679659001144295193 77340263404533420 9} - \frac{46352469152745600579360525704486920088475570560866 517836499521 m^{21}}{105860851375217107184186550832636678857701745229824 000000000}$$
$$- \frac{63167170015592047856804114881834306274390631378635980 80000000000 m^{22}}{151730801978964166531605170236192340306109683271959310 29256199525807912 8833}$$
$$- \frac{943561029685948567241293029820396279429031857446644120 37316608000000 m^{23}}{145457355732741622690547415240436236667373782872039579 29881523543491882166490 76249763}$$
$$- \frac{855794756948680375310776917358622299101661030194387070 87253566390272000000000 0 m^{24}}{}\Bigg)$$

$A_6 = m^{0.666666666666667}(0.0423586381806268 m^{12} + 0.393278680939119 m^{13} + 1.81284370003379 m^{14} + 5.51482817753823 m^{15} + 12.3187161682489 m^{16} + 20.5203460930749 m^{17}$
$+ 21.0095826280593 m^{18} - 12.7937019078945 m^{19} - 137.61034445283 m^{20} - 437.862236611457 m^{21} - 971.128298758715 m^{22} - 1608.06558564066 m^{23}$
$- 1699.67570555547 m^{24})$

$$A_6 = 1.3055741726478 \cdot 10^{-15}$$

$$A_7 = m^{\frac{2}{3}}\Bigg(\frac{75164925 m^{14}}{1879048192} + \frac{5162357562053249 m^{15}}{12103766790635520} + \frac{5747831759060530169 m^{16}}{2541791026033459200} + \frac{10596802544052932539954870 3709 m^{17}}{133601898490667135737528320 00} + \frac{196071967808509459995641850829 8911 m^{18}}{9539175552233633491659522048 0000}$$
$$+ \frac{5112691069661020936671747044794792820 549 m^{19}}{125349797717536192098709620942648115 200000} + \frac{12944737881716614211095431620530134423884 241569 m^{20}}{22374938892580210289619667338262688563200 0000}$$
$$+ \frac{6430203089673988139528865925994490954583113942700020 66341 m^{21}}{22345405816936969160646697242036770487142113607680000 000} - \frac{104656931233332848210086172645991069293570621449266001 51323563 m^{22}}{666236868818828403589743065462573249293560248795136000 00000}$$
$$- \frac{53563981075382967357912264545249309385088742529354948512863004856175636929 m^{23}}{75684414606926755262412900621750254314460115857630342286409728000000000}$$
$$- \frac{12072496137350222630732211372788410048483594530336675493667329932166289377 5769 m^{24}}{6417092303484802261811833811466649687687287073278832646608964812800000000 0}\Bigg)$$

$A_7 = m^{0.666666666666667}(0.0400015951267311 m^{14} + 0.426508346645218 m^{15} + 2.26133136052109 m^{16} + 7.93162572071772 m^{17} + 20.5543934834702 m^{18} + 40.7873898702412 m^{19}$
$+ 57.8537351268889 m^{20} + 28.776398792454 m^{21} - 157.086670119111 m^{22} - 707.728022388386 m^{23} - 1881.30317695356 m^{24})$

$$A_7 = 9.00786265693113 \cdot 10^{-18}$$

$$A_8 = m^{\frac{2}{3}}\Bigg(\frac{52553071771 m^{16}}{1352914698240} + \frac{138523389109435488089 m^{17}}{296300211034757529600} + \frac{518504093065027273943 3713 m^{18}}{18511355684396476661760 00} + \frac{587357432342879682592884783010 674929 m^{19}}{5281977772082233448103296368640 00}$$
$$+ \frac{1642279205820234837737335266213920358310481 m^{20}}{50158717367924730526987852297587916800} + \frac{191373206374141833041974409825424717901313873 2138433 m^{21}}{255574040786109086733112366103847602403409920 00}$$
$$+ \frac{55368134813936388845572405152354296200966114516018602 0681 m^{22}}{42472188606978359415112178892418758034007072440320000 0} + \frac{83949325693679353753983393188876162566956906872572506650 783298050519 m^{23}}{60594469739607841708144352316463294298154064438402209546 2400000000}$$
$$- \frac{11016372143813783240798086799437999606243686798065876241718099653555157718 83 m^{24}}{1150834407082527972857760796935197470628261253439910124586008576000000000}\Bigg)$$

$A_8 = m^{0.666666666666667}(0.0388443350045395 m^{16} + 0.467510261385491 m^{17} + 2.80100551199544 m^{18} + 11.1200284650445 m^{19} + 32.7416507438532 m^{20} + 74.8797513963098 m^{21}$
$+ 130.363272131541 m^{22} + 138.542883623595 m^{23} - 95.7250849993381 m^{24})$

$$A_8 = 6.38850877515203 \cdot 10^{-20}$$

$$A_9 = m^{\frac{2}{3}} \cdot \left( \frac{7509100901081 m^{18}}{194819716546560} + \frac{23736421687694033734 25249 m^{19}}{459383847188288073 8918400} + \frac{110690750348273790371760059629 m^{20}}{320764771299222214759497728000} + \frac{56640144440716925445571206350495 7304031 m^{21}}{3704628299283070476240113483592499 2000} \right.$$

$$+ \frac{19455633785882528095990203275504 33940578843 59 m^{22}}{386289002284432469850911316115707658 24000} + \frac{380508418869157903159302948007308073124 3178291241592747 87 m^{23}}{29277880581681227351914064361213837737728 23134208000000}$$

$$\left. + \frac{323300992477921366696683429536370735659062216 7329315072054507 m^{24}}{1226596806969535019908439726413053732022124252076 44160000000} \right)$$

$$A_9 = m^{0.666666666666667}$$
$$\cdot (0.0385438447103294 m^{18} + 0.516701269166854 m^{19} + 3.45083875326873 m^{20} + 15.2890222351533 m^{21} + 50.3654871974118 m^{22} + 129.964468502968 m^{23} + 263.57560254594 m^{24})$$

$$A_9 = 4.62819175388996 \cdot 10^{-22}$$

$$A_{10} = m^{\frac{2}{3}} \cdot \left( \frac{1069535470353 m^{20}}{27487790694400} + \frac{424739449499963264 5185343 m^{21}}{738903119286981875 4662400} + \frac{7389749261283848589876318593 m^{22}}{17456586193154946807 88992000} + \frac{992877672378024296939933292221016946312123 m^{23}}{479680760529393125367979138468125 08160000} \right.$$

$$\left. + \frac{107785669924303908818198613870338726728 63153859 m^{24}}{14298324109860150280968722159457872 43233280000} \right)$$

$$A_{10} = m^{0.666666666666667} \cdot (0.0389094737457708 m^{20} + 0.574824274540651 m^{21} + 4.23321557807305 m^{22} + 20.6987178573151 m^{23} + 75.3834289222572 m^{24})$$

$$A_{10} = 3.38690017224207 \cdot 10^{-24}$$

$$A_{11} = m^{\frac{2}{3}} \cdot \left( \frac{109266428162197927 m^{22}}{2743061608975564800} + \frac{23127992831767576470432693 63403 m^{23}}{3597453282685371438169954713600} + \frac{123631922131857269591160588680188349 m^{24}}{238915866136342230637462117416 96000} \right)$$

$$A_{11} = m^{0.666666666666667} \cdot (0.039833749196397 m^{22} + 0.642899045919044 m^{23} + 5.17470539446318 m^{24})$$

$$A_{11} = 2.18641247838221 \cdot 10^{-26}$$

$$A_{12} = m^{\frac{2}{3}} \cdot \frac{217295418508894375 m^{24}}{52666782892330844 16}$$

$$A_{12} = 0.0412585327175046 m^{24.6666666666667}$$

$$A_{12} = 4.69331574678584 \cdot 10^{-29}$$

## $B_i$

$$B_0 = m^{\frac{2}{3}} \cdot \left(1 - \frac{2m}{3} + \frac{227m^2}{144} + \frac{535m^3}{648} + \frac{53477m^4}{62208} - \frac{18073m^5}{93312} - \frac{4862647m^6}{26873856} - \frac{236184929m^7}{100776960} - \frac{302769729871m^8}{48372940800} - \frac{72827805074513m^9}{6530347008000} - \frac{4062893273303209m^{10}}{391820820480000} - \frac{60188566076031467m^{11}}{82282372300800000}\right.$$
$$+ \frac{14427807250661313 7739m^{12}}{6479736818688000000} + \frac{5815522714826227513979723m^{13}}{870876628431 66720000000} + \frac{211907883409273211769 73114067m^{14}}{14630727357652008960 0000000} + \frac{2158648792755150194 1804000961363m^{15}}{9217358235320765644 8000000000}$$
$$+ \frac{37024991117836509562 9575405365 32127m^{16}}{154851618353888628326 40000000000} - \frac{27165703021349783229 6215111735371237m^{17}}{162594199271058305974 272000000000 00} - \frac{92473525718715669807 91638503483536 5600041m^{18}}{122921214648920079316 54963200000 0000000}$$
$$- \frac{311614657713763729521 45334224040208640781 9619m^{19}}{757201435678760319429 816870491219413435114 4500586459m^{21}} - \frac{23128656604539887016 361092303433776732018 58087397m^{20}}{524735914790481948192 832393052160000000000 0}$$
$$- \frac{14197400291950269161 06148249600000000000 0 m^{21}}{113638121546813746905 510265120358400000000 00000} - \frac{527406529435222356900 92301976178476793337 6029488132315259m^{22}}{840012994474047217125 531879769689292800000 0000000000}$$
$$+ \frac{120388928385448228540 209538088987958928490 99684323096584780413m^{23}}{504511804481112758605 59444698967538925568 00000000000000}$$
$$\left. + \frac{100348796245110274552 201551236536143701455 24819423921265272417 4746033m^{24}}{363611747725627587382 2240298343988465443536 896000000000000000000}\right)$$

$$B_0 = m^{0.666666666666667}$$
$$\cdot (1.0 - 0.666666666666667m + 1.57638888888889m^2 + 0.825617283950617m^3 + 0.859648276748971m^4 - 0.193683556241427m^5 - 0.180943404623438m^6$$
$$- 2.34364014354075m^7 - 6.25907221813977m^8 - 11.1522105923767m^9 - 10.3692633493186m^{10} - 0.731487977230407m^{11} + 22.2660389678953m^{12}$$
$$+ 66.7778021015354m^{13} + 144.837558809708m^{14} + 234.19386961475m^{15} + 239.099800903219m^{16} - 16.7076704723409m^{17} - 752.299153427932m^{18}$$
$$- 2194.87125322828m^{19} - 4407.67554737982m^{20} - 6663.26955577867m^{21} - 6278.55203317949m^{22} + 2386.246016766m^{23} + 27597.7871652351m^{24})$$

$$B_0 = 0.178911819843216$$

$$B_1 = m^{\frac{2}{3}} \cdot \left(\frac{3m^2}{16} + \frac{3m^3}{8} + \frac{31m^4}{96} + \frac{1139m^5}{17280} - \frac{444079m^6}{1382400} - \frac{62642729m^7}{62208000} - \frac{233316418453m^8}{89579520000} - \frac{1986826581893m^9}{391910400000} - \frac{102616704110645003m^{10}}{15801827328000000} - \frac{26008140328342680719m^{11}}{7466363412480000000}\right.$$
$$+ \frac{31124423014119484532983m^{12}}{418116351098880000 00} + \frac{807366073202537480846 57887m^{13}}{263413301192294400 0000000} + \frac{958363284161044834125 118604969m^{14}}{132760303800916377600 00000000} + \frac{377448208697678021344 5019975607m^{15}}{290413164564504576000 00000000}$$
$$+ \frac{394515124226624129465 194702350915619m^{16}}{234189175904816490086 4000000000000} + \frac{384722016017301089441 527223789153603269m^{17}}{405732747255094569074 6880000000000000} - \frac{603130595691267954731 289344200402026 9620413m^{18}}{249931372309138254550 00780800000000000 00}$$
$$- \frac{177033524744306765107 46866193390367388308 7793477m^{19}}{173202441010232810403 15541094400000 00000000} - \frac{686343726887469857499 2523108475509799657 5638641233367m^{20}}{288070299888219210262 52807948206080000 00000000000}$$
$$- \frac{596780063840647706285 641066478341004535575 076433430098013m^{21}}{144179185094053714736 3953037807714304000 00000000000000} - \frac{535452753268661265411 77112738147936817804 80120161676 38248402583m^{22}}{113955914002885337064 8046617298407604378 022384969829171212 45510083m^{23}}$$
$$- \frac{103912822280986393284 814823340877585317888 00000000000000000 00}{113955914002885337064 8046617298407604378 022384969829171212 45510083m^{23}}$$
$$\left. + \frac{352022773114408747590 19168765826019959905 84447313176633650847 90760133424783m^{24}}{374834706618134329407 9998560885776529179 927379968000000 00000000000000}\right)$$

$$B_1 = m^{0.666666666666667}$$
$$\cdot (0.1875m^2 + 0.375m^3 + 0.322916666666667m^4 + 0.0659143518518518m^5 - 0.321237702546296m^6 - 1.00698831340021m^7 - 2.60457321554078m^8$$
$$- 5.06959392221538m^9 - 6.4939770559835m^{10} - 3.48337455485627m^{11} + 7.44396217280651m^{12} + 30.6501634332107m^{13} + 72.1874880309237m^{14}$$
$$+ 129.969386637031m^{15} + 168.460016438578m^{16} + 94.821534278429m^{17} - 241.318482797454m^{18} - 1022.11910935971m^{19} - 2382.55636611547m^{20}$$
$$- 4139.15547831225m^{21} - 5152.90357354327m^{22} - 2434.56367092704m^{23} + 9391.41351905374m^{24})$$

$$B_1 = 0.00026881116558826$$

$$B_2 = m^{\frac{2}{3}}\left(\frac{25m^4}{256} + \frac{113m^5}{320} + \frac{7711m^6}{12800} + \frac{87056897m^7}{145152000} + \frac{16909308271m^8}{81285120000} - \frac{7521567147551m^9}{8534937600000} - \frac{596011713556313317m^{10}}{16131032064000000} - \frac{84151805073784068281m^{11}}{9033377955840000000} - \frac{25010437442592996562429m^{12}}{15176074965811200000000}\right.$$
$$- \frac{41926082211540183018577258 1m^{13}}{2151208626403737600000000 0} - \frac{190514951579970346995111729791m^{14}}{240935366157218611200000000000} + \frac{161403625033118845877075841380 01m^{15}}{4638005798526458265600000000000 0}$$
$$+ \frac{3094902527751244888251794334488369701 63m^{16}}{2314179373232761620377600000000000 00} + \frac{5535160214052620039645410418177479868973 53m^{17}}{17819181173892264444769075200000000000 00}$$
$$+ \frac{879876357833252974845093788262318255512 2325591m^{18}}{1646492340467645234696662548480000000000000 00} + \frac{2828302934461127849283291191295740348054 251073714021m^{19}}{44499748485819047758146698697768960000000000000 00000}$$
$$+ \frac{77916228291146223373515379033396036995171096170315 781179m^{20}}{356353985874438934447238763171733831680000000000000000 00} - \frac{74317903620205707622553047383294741165708643075201770 69007807m^{21}}{53506550979047006007252900290235834826752000000000000000000 0}$$
$$- \frac{17568499180179983161730652815730607309155723908138763072209 4336422533m^{22}}{34706917279456882352592579267460893788703031296000000000000000000 0000}$$
$$- \frac{1973146876354001421487622500200977481860703919137339696122879599412 337787m^{23}}{17370812098368169617472585923336417734124586716364800000000000000000000000 0}$$
$$\left.- \frac{79155399605836285190698938752110341375858967636318879214829759333590167451 29339m^{24}}{4173163898511969068901614042229009964460907127394795520000000000000000000000000 0}\right)$$

$$B_2 = m^{0.666666666666667}(0.09765625m^4 + 0.353125m^5 + 0.602421875m^6 + 0.599763675319665m^7 + 0.20802464548247m^8 - 0.881267971724949m^9 - 3.69481451150572m^{10}$$
$$- 9.3156519615545m^{11} - 16.4801752092927m^{12} - 19.48954726982m^{13} - 7.90730537482209m^{14} + 34.8002206216298m^{15} + 133.736501307929m^{16} + 310.629324660689m^{17}$$
$$+ 534.394443391912m^{18} + 635.577285422734m^{19} + 218.648398445584m^{20} - 1388.94961944582m^{21} - 5061.95898607763m^{22} - 11358.9788731833m^{23}$$
$$- 18967.71886531m^{24})$$

$$B_2 = 1.04225774375648 \cdot 10^{-6}$$

$$B_3 = m^{\frac{2}{3}}\left(\frac{833m^6}{12288} + \frac{55583m^7}{161280} + \frac{230324963m^8}{270950400} + \frac{904852584227m^9}{682795008000} + \frac{757145763542591m^{10}}{573547806720000} + \frac{8529232716945697m^{11}}{120445039411200000} - \frac{163543745569252647959m^{12}}{37940187414528000000} - \frac{1055476701986206779299 3389m^{13}}{70113466342047744000000 0}\right.$$
$$- \frac{13158376247864680790414893 35517m^{14}}{38870905740031269273600000 000} - \frac{9107693717719147693766033 5645601m^{15}}{16628109677680042967040000 00000} - \frac{53748376172674277661183622 2113205918 71m^{16}}{93338568053721385186885632 00000000000}$$
$$+ \frac{25596543039516661491587937 26582403174011m^{17}}{42044357979879795743263293 4400000000000} + \frac{2341178482229141877493649 1517659818585990063132491m^{18}}{1212088387328023586555233 888339230720000000000}$$
$$+ \frac{34020307482676412620124666 249826342545431529807817 44369m^{19}}{54598521407190822456380510 500240647782400000000000} + \frac{1054034378636681239500512 4701119952763133137789674 968483013057m^{20}}{78700492697181139121525123 0554668793394626560000000 0000}$$
$$+ \frac{37774788547394977891958122 3704838694913434191141193 530929457337069m^{21}}{17725318467722620586454958 40167527899230476697600000 0000000}$$
$$+ \frac{27945188762350771300193679 856013343511970824032299002 3464527702355509941m^{22}}{12774991526057048170106981761 92554070753338916549427200000 00000000}$$
$$+ \frac{3532915923105265984011450599 823713654877323484979409135 2850490388080429223 37m^{23}}{11508989865824794696449379869 31871962341683029919378964480 00000000000000000}$$
$$\left.- \frac{34941715908769730102489937606 07845302690083892656501400972 4935330256381106573083209m^{24}}{41473795880486230168124985297 07693803494488966617474036400 128000000000000000000}\right)$$

$$B_3 = m^{0.666666666666667}(0.0677897135416667m^6 + 0.344636656746032m^7 + 0.850063196068358m^8 + 1.32521851159609m^9 + 1.32010924751426m^{10} + 0.0708143129732961m^{11}$$
$$- 4.31056767807712m^{12} - 15.0538371164974m^{13} - 33.8514783675686m^{14} - 54.7728749344517m^{15} - 57.5843162086429m^{16} + 0.0608798523022164m^{17}$$
$$+ 193.152455440162m^{18} + 623.09942844342m^{19} + 1339.2983226831m^{20} + 2131.12044312106m^{21} + 2187.4917650905m^{22} - 306.970113302127m^{23} - 8425.01033892827m^{24})$$

$$B_3 = 5.32451552019803 \cdot 10^{-9}$$

$$B_4 = m^{\frac{2}{3}} \cdot \left( \frac{3537 m^8}{65536} + \frac{16905377 m^9}{48168960} + \frac{4856753809 m^{10}}{4335206400} + \frac{77919250272181 m^{11}}{33648137994240} + \frac{45509942889388469 m^{12}}{13596594536448000} + \frac{444205443663851424011833 m^{13}}{161594846235957657600000} - \frac{24315652554262727282494467187 m^{14}}{799894488867990405120000000} \right.$$

$$- \frac{86024833214317748585181244473050063 m^{15}}{4035499692118566313446604800000000} - \frac{254455877446105844708326409769335929961759 m^{16}}{41549504830052758762462430200000000000} - \frac{809585430660338896403864839812767301593823481 m^{17}}{65505910577440428147164945590517760000000000}$$

$$- \frac{30525859568790525699132450903264788496784973946317 m^{18}}{1686122138263316620508025699499927142400000000} + \frac{3163625611986995998345763957834243772442298244001923 4003 m^{19}}{212663840810599072078195249375127810762342400000000}$$

$$+ \frac{3436215441772298517840574069538922718511340820140957130582351 m^{20}}{218958690498592804611709828756631593960907735040000000000000} + \frac{144776090535042897079065321991524935674234251719225192267854 6743973 m^{21}}{1380819189891275803722825693087945820995672449482752000000000000}$$

$$+ \frac{512052782205094528306023096190569361841810869866382991372951 09087867881 m^{22}}{17771142973900719539127666700418627162143044248430182400000000000}$$

$$+ \frac{86244066462416607821258328136250173456074091028365169606986003070363076088813939 m^{23}}{1524154161533817468461252294137475984322768447236703526060032000000000000}$$

$$+ \left. \frac{10841715945529925317080052023389643680838595917017247475061512290106754663046320 93597 m^{24}}{1333878756007935695698549557737353482439912631828367345786669760512000000000000000} \right)$$

$$B_4 = m^{0.666666666666667}$$

$$\cdot \left( 0.0539703369140625 m^8 + 0.350959975054475 m^9 + 1.12030509297089 m^{10} + 2.31570764140112 m^{11} + 3.3471574641276 m^{12} + 2.74888372996272 m^{13} \right.$$
$$\left. - 3.03985749278936 m^{14} - 21.3170213796141 m^{15} - 61.2416141905524 m^{16} - 123.589676645019 m^{17} - 181.041805193494 m^{18} - 148.761801720893 m^{19} \right.$$
$$\left. + 156.934416896067 m^{20} + 1048.47971113758 m^{21} + 2881.37225026613 m^{22} + 5658.48709002151 m^{23} + 8127.96207803569 m^{24} \right)$$

$$B_4 = 3.10758520789195 \cdot 10^{-11}$$

$$B_5 = m^{\frac{2}{3}} \cdot \left( \frac{732413 m^{10}}{15728640} + \frac{28066785623 m^{11}}{76299632640} + \frac{36131291138989 m^{12}}{25178878771200} + \frac{349040916290743627579 m^{13}}{95271337916891136000} + \frac{44112429439991228546318 39 m^{14}}{6539424634615407575040 00} + \frac{66574840757326003275944167981 27 m^{15}}{77324200449890896482140 1600000} \right.$$

$$+ \frac{407442963103908004596431948172646 73 m^{16}}{13268832797201277836335251456000000} - \frac{49222371267562606523868181929912109632 5061 m^{17}}{20082537664557700420627439101673472000000 0} - \frac{68433713915208127334400504113962231687452060 2089 m^{18}}{6892326926476202784593337099694335590400000000}$$

$$- \frac{15994565239445245015290753296162282521031095008817 18423 m^{19}}{6519762194465531642850793132770358280060928000000} - \frac{27969779814257130747829902539584837021461381120000000000 00}{1249720328199684060235219856963277333180977243859790 9998707 m^{20}}$$

$$- \frac{82073987403352248450160820645218461629815543176347316549995 1839381 m^{21}}{14393083111320219080619240693847322789008859672975769600000 00000}$$

$$- \frac{246824484758741736704362450485621411512208118654099634263651 863570527081 m^{22}}{1049687551308583577549561223802285251002416135950122876928068 5233218761849 m^{23}}$$

$$+ \frac{454483345150692756050870116845097324757326999455285121768646 2668068523218761849 m^{23}}{33760187484568335137014069522888192539822338078525525483723366 400000000000}$$

$$+ \left. \frac{276474118140907764917835111243782372690563060047737958059907283 7508938651190 21533 m^{24}}{5127302110057410832033394607047554939461147680272525142708453 37600000000000} \right)$$

$$B_5 = m^{0.666666666666667}$$

$$\cdot \left( 0.0465655644734701 m^{10} + 0.367849551195428 m^{11} + 1.4349841177327 m^{12} + 3.6636508305911 m^{13} + 6.74561324653687 m^{14} + 8.6098324159807 m^{15} \right.$$
$$\left. + 3.07067674550732 m^{16} - 24.5100355790353 m^{17} - 99.2897096223435 m^{18} - 245.324365557729 m^{19} - 446.810928258599 m^{20} - 570.232150878089 m^{21} \right.$$
$$\left. - 235.140908788563 m^{22} + 1346.21096330829 m^{23} + 5392.1948074523 m^{24} \right)$$

$$B_5 = 1.96205634968797 \cdot 10^{-13}$$

$$B_6 = m^{\frac{2}{3}}\left(\frac{31979701 m^{12}}{754974720} + \frac{68656059199267 m^{13}}{174573559480320} + \frac{41063270689289129 m^{14}}{22694562732441600} + \frac{1035087339800718436008641 m^{15}}{18891671166667329945600} + \frac{10730020450248895862618413703 m^{16}}{884130210600003104145408000}\right.$$

$$+ \frac{101753058991146103328644546542141997 m^{17}}{51109578181812329443661709312000000} + \frac{6995886308985133392795638935409743246 13 m^{18}}{358789238836322552694505199370240000 00} - \frac{353748759896162108669484475643287076706641708639 m^{19}}{22565988900018382449532051837517775059968000000 0}$$

$$- \frac{76295691644994332927680828035786521874000583870466 6137 m^{20}}{53860502306563875230543101325787367131316224000000 00} - \frac{46609509662463364372609253350205498429225472486028 7158 53082561 m^{21}}{10586085137521710718418655083263667885770174522982 4000000000}$$

$$- \frac{608029783362969235385562814836974391075028278186624362477 9250320249 m^{22}}{631671700155920478568041148818343062743906313786359808000 0000000}$$

$$- \frac{73922255566672716452822020078984805107375205691785006956014 2209608679104991 m^{23}}{471780514842974283620646514910198139714515928723322060186583 040000000000}$$

$$\left. - \frac{136012257430919039815840390389795151682195173891219415607223 27055024571467 9993319203 m^{24}}{855794756948680375310776917358622299101661030194387070872535 66390272000000000000}\right)$$

$$B_6 = m^{0.666666666666667}(0.0423586381806268 m^{12} + 0.393278680939119 m^{13} + 1.80938805357945 m^{14} + 5.479067101417 m^{15} + 12.136244550412 m^{16} + 19.9088043006693 m^{17}$$
$$+ 19.4985956983415 m^{18} - 15.6761913454577 m^{19} - 141.65425196136 m^{20} - 440.290334500135 m^{21} - 962.572461632972 m^{22} - 1566.87809777979 m^{23}$$
$$- 1589.30930958105 m^{24})$$

$$B_6 = 1.30481740811 \cdot 10^{-15}$$

$$B_7 = m^{\frac{2}{3}}\left(\frac{75164925 m^{14}}{1879048192} + \frac{5162357562053249 m^{15}}{12103766790635520} + \frac{5740331749711898369 m^{16}}{2541791026033459200} + \frac{1055083721749085221339164 33059 m^{17}}{13360189849066713573752832000} + \frac{194178622900261707210103756 8683111 m^{18}}{9539175552233633491659522048 0000}\right.$$

$$+ \frac{50180805099613562762231976432771409753 93349 m^{19}}{12534979771753619209870962094264811520 0000} + \frac{124694935962440838313363190422299837 16068864569 m^{20}}{22374938892580210289619667338262688563 2000000}$$

$$+ \frac{53891093767786522896381457640560787574337926019 5426812891 m^{21}}{22345405819369691606466972420367704871421113074 807680000000} - \frac{10990427156374512396116809666841664126235645070616 000284683163 m^{22}}{66623686881882840358974306546257324929356024879513 600000000}$$

$$- \frac{542023542383086530820261225612742258628881687728 604820633743946646456699 0529 m^{23}}{756844146069267552624129006217502543144601158576 3034228640972800 0000000}$$

$$\left. - \frac{120464352471276367142874110228698019935367291800 4171526359507287611204681835969 m^{24}}{641709230348480226181183811466649687687287073278 83264660896481280000000000}\right)$$

$$B_7 = m^{0.666666666666667}(0.0400015951267311 m^{14} + 0.426508346645218 m^{15} + 2.25838068154244 m^{16} + 7.89722102506492 m^{17} + 20.3559125038635 m^{18} + 40.0326175337684 m^{19}$$
$$+ 55.7297325195338 m^{20} + 24.1173036682731 m^{21} - 164.962758303356 m^{22} - 716.162693730447 m^{23} - 1877.24200890578 m^{24})$$

$$B_7 = 9.00316573625593 \cdot 10^{-18}$$

$$B_8 = m^{\frac{2}{3}}\left(\frac{52553071771 m^{16}}{1352914698240} + \frac{138523389109435488089 m^{17}}{296300211034757529600} + \frac{5180211338268308606550613 m^{18}}{185113556843964766176000} + \frac{5855680825361549511056995802 05873679 m^{19}}{52819777720822334481032963686 40000}\right.$$

$$+ \frac{16313377908833375019088258355053316391 30281 m^{20}}{50158717367924730526987852297587916800 000} + \frac{18900121856626574631192844956145193402 67897019632833 m^{21}}{25557404078610908673311236610384760240340 992000000}$$

$$+ \frac{541236679886065938906671870958066391246950765494 123763681 m^{22}}{424721886069783594151121788924187580340070724403 20000000} + \frac{7955232683112156186899790813592650295497320115188 52299519994758 08169 m^{23}}{60594469739607841708144352316463294298154064438402 2095462400000000}$$

$$\left. - \frac{12651634528727730464128186273230635748984012967098 26162290794632847711 83483 m^{24}}{11508344070825279728577607969351974706282612534399 101245860085760000000}\right)$$

$$B_8 = m^{0.666666666666667}(0.0388443350045395 m^{16} + 0.467510261385491 m^{17} + 2.79839652297038 m^{18} + 11.0861519535899 m^{19} + 32.5235148841054 m^{20} + 73.9516493869742 m^{21}$$
$$+ 127.433197496476 m^{22} + 131.286447712112 m^{23} - 109.934448004564 m^{24})$$

$$B_8 = 6.3854872088242 \cdot 10^{-20}$$

$$B_9 = m^{\frac{2}{3}} \cdot \left( \frac{7509100901081 m^{18}}{194819716546560} + \frac{237364216876940337342 5249 m^{19}}{459383847188288073891 8400} + \frac{110614625834059382233085744509 m^{20}}{3207647712992221475949 7728000} + \frac{565142665752178549833142185112912938401 m^{21}}{3704628299283070476240 1134835924992000} \right.$$
$$+ \frac{19362290526630484955651295024791542 9286387383 m^{22}}{386289002028443246985091131611570765824000} + \frac{37717833581672911509340435978182622506913865546223 3853187 m^{23}}{292778805816812273519140643612138377377282313420 8000000}$$
$$\left. + \frac{31841856341688632328656887290672061739549889041210987431969807 m^{24}}{1226596806969535019908439726413053732022124252076441600 00000} \right)$$

$$B_9 = m^{0.666666666666667}$$
$$\cdot (0.0385438447103294 m^{18} + 0.516701269166854 m^{19} + 3.44846553398078 m^{20} + 15.255043693899 m^{21} + 50.123846201258 m^{22} + 128.827062725546 m^{23}$$
$$+ 259.595134772591 m^{24})$$

$$B_9 = 4.62620269376059 \cdot 10^{-22}$$

$$B_{10} = m^{\frac{2}{3}} \cdot \left( \frac{1069535470353 m^{20}}{27487790694400} + \frac{42473944949996326451 85343 m^{21}}{7389031192869818754662400} + \frac{7385891641716902071064099813 m^{22}}{1745658619315494680788992000} + \frac{99121810844837302347401011764810482264 9173 m^{23}}{47968076052939312536797913846812508160000} \right.$$
$$\left. + \frac{7518037314893698141456149516475265752169 19852773 m^{24}}{100088268769021051966781055116205107026 32960000} \right)$$

$$B_{10} = m^{0.666666666666667} \cdot (0.0389094737457708 m^{20} + 0.574824274540651 m^{21} + 4.2310057418976 m^{22} + 20.6641205987588 m^{23} + 75.1140708832068 m^{24})$$

$$B_{10} = 3.38572241460503 \cdot 10^{-24}$$

$$B_{11} = m^{\frac{2}{3}} \cdot \left( \frac{109266428162197927 m^{22}}{2743061608975564800} + \frac{2312799283176757647043269363403 m^{23}}{3597453282685371438169954713600} + \frac{12358178889773791424074978503795 2399 m^{24}}{23891586613634223063746211741696000} \right)$$

$$B_{11} = m^{0.666666666666667} \cdot (0.039833749196397 m^{22} + 0.642899045919044 m^{23} + 5.17260703092918 m^{24})$$

$$B_{11} = 2.1861737815123 \cdot 10^{-26}$$

$$B_{12} = m^{\frac{2}{3}} \cdot \frac{2172954185088 94375 m^{24}}{52666782892 33084416}$$

$$B_{12} = 0.0412585327175046 m^{24.6666666666667}$$

$$B_{12} = 4.69331574678584 \cdot 10^{-29}$$

$R_i$

$$R_0 = -\frac{9m^4}{32} + 4m^5 + \frac{34m^6}{3} + 15m^7 + \frac{2704801m^8}{221184} + \frac{122957m^9}{12960} + \frac{1260881m^{10}}{207360} - \frac{291394307m^{11}}{11664000} - \frac{39665655102569m^{12}}{358318080000} - \frac{332504601769m^{13}}{1530900000} - \frac{1278479163300757m^{14}}{5038848000000} - \frac{416971254603120521m^{15}}{2592487296000000}$$

$$+ \frac{11103376680448272785217 61m^{16}}{9910906100121600000000} + \frac{15284355354140738398495691m^{17}}{18292590360576000000000} + \frac{15302889517265221792461076 0877m^{18}}{6146310361153536000000000000} + \frac{3355236291866552442084685548 89m^{19}}{67225269575116800000000000}$$

$$+ \frac{9328314401393881173026714361 5892557m^{20}}{13877877090655792005120000000000} + \frac{10227133735968054283807111560 97576661m^{21}}{22361813671466852352000000000000} - \frac{54132797503072695543223378498538 767700483m^{22}}{96424140551365067341824000000000000}$$

$$- \frac{504565001182638539410413023000565702423068287m^{23}}{1670548235052399791697100800000000000000} - \frac{12251043443545137459551439853290462377984225130695423m^{24}}{158063264684893942530879604654080000000000000000}$$

$$- \frac{150417071231193877987531582647338838975298657 06356823m^{25}}{10300867705050705499570995068928000000000000000} - \frac{260150283772228136943801374283113890248625123004154207 5677m^{26}}{131787716669541026052972915374161920000000000000000}$$

$$- \frac{836039859047609468863326567034934919168000000000000000000}{10837656008604102598166779877830905239535624932225000266041960003m^{27}}$$

$$+ \frac{2828005779402385639043858779755390954369995503653358698326223835098853m^{28}}{1224229886400595497445946358640595900836085760000000000000000000}$$

$$+ \frac{6304207327491693672078050984045725080525718612807237966638269293085078263m^{29}}{1776162016926786068583405147499535589557911361363979140893410484816691751 7648687m^{30}}$$

$$+ \frac{5654561229937555214021024063595109811603472384000000000000000000000}{6520568591424951670158771940982828069470167386554368000000000000000000}$$

$$R_0 = -0.28125m^4 + 4.0m^5 + 11.3333333333333m^6 + 15.0m^7 + 12.2287371600116m^8 + 9.48742283950617m^9 + 6.08063753858025m^{10} - 24.9823651406036m^{11} - 110.699563646269m^{12}$$
$$- 217.195507067085m^{13} - 253.724494825158m^{14} - 160.838301983764m^{15} + 112.031902716867m^{16} + 835.548987478636m^{17} + 2489.76843310483m^{18}$$
$$+ 4991.03434329474m^{19} + 6721.71567773488m^{20} + 4573.48133126502m^{21} - 5614.02955665819m^{22} - 30203.5577659817m^{23} - 77507.2150253769m^{24}$$
$$- 146023.68998239m^{25} - 197401.010008055m^{26} - 129630.853018779m^{27} + 231002.83785075m^{28} + 1114888.8607156m^{29} + 2723937.32543904m^{30}$$

$$R_0 = 5.32841568758022 \cdot 10^{-6}$$

$$R_1 = \frac{3m^2}{2} + \frac{19m^3}{4} + \frac{20m^4}{3} + \frac{43m^5}{9} + \frac{18709m^6}{13824} + \frac{759413m^7}{414720} + \frac{6675059m^8}{1555200} - \frac{41991161m^9}{11664000} - \frac{4528082998913m^{10}}{179159040000} - \frac{1391582016661969m^{11}}{37623398400000} - \frac{6944620829982767m^{12}}{493807104000000} + \frac{2445280490863011629m^{13}}{103699491840000000}$$

$$+ \frac{2594335112595348085 90987 m^{14}}{5574884681318400000000} + \frac{14231974272408875947 0285481 m^{15}}{11707257830768640000000000} + \frac{22424955312454904160585209707 m^{16}}{61463103611535360000000} + \frac{4208733959395124572 8660 94307983 m^{17}}{6453625879211212800 000 00000}$$

$$+ \frac{44132941499925409566697523857815131 m^{18}}{8673673181659870003 2000000000000} - \frac{88108296106525800859627026875623134817 m^{19}}{20036185049634299707392000000000000} - \frac{49694969145779250028599997789319 6197633553 m^{20}}{23141793732327616162037760 000000000000}$$

$$- \frac{130898978418785758848285139101632593067288 8777 m^{21}}{26728771760838396667153612800000000000000} - \frac{47868601326812649292358381111546768662898797001213 m^{22}}{493947702140293570408998764544000000000 0000}$$

$$- \frac{45117326504191255751075889848958707911027818 87794432659 m^{23}}{29666498990546031838764465798512640000000 000000000} - \frac{11463463816766773817762875948866109599 29956495832745 59904 481 m^{24}}{89088496468609733611809690792933457920000000 0000000000}$$

$$+ \frac{41619260338147489888162741073738453067106709742836730224 2614887 m^{25}}{33441594361904378754533062681397396766720000000 0000000000} + \frac{573908145684498396093745636196154818318438567522493998 06899308346179019 m^{26}}{7712648284323751633909462059437541752673402880000 0000000000 0}$$

$$+ \frac{627123798204960301509753671181438372833171847426 486024965446555753185 0765269 m^{27}}{34741624196736339234945171846728354682491734327296 000000000000 00} + \frac{20450633823029379068294483796830690331 73098175296106013616920292768435944481 885451 m^{28}}{625974584776795360335242106334351494669136069109 219328000000000000 00}$$

$$+ \frac{12704795562438368197645558843972326235 2858896675095502050023127490181072273 8382463 49201 m^{29}}{281970251712707470063009806798308630773712 3423302478 4629 76000 000000000000 00} + \frac{52324747435493006616989 94439872655 261770289308515004889076299486596596070363 8181197 4324502103 m^{30}}{16257727985150602225904994236454159706989 7491404982226885017 60000000000 00000000}$$

$R_1 = 1.5m^2 + 4.75m^3 + 6.66666666666667m^4 + 4.77777777777778m^5 + 1.35337094907407m^6 + 1.83114631558642m^7 + 4.2920904063786m^8 - 3.60006524348423m^9 - 25.2740972429468m^{10} - 36.9871429972144m^{11} - 14.0634283584198m^{12} + 23.5804481533611m^{13} + 46.5361215683804m^{14} + 121.565395399466m^{15} + 364.852309674876m^{16} + 652.1502854624m^{17} + 508.814899704115m^{18} - 439.745869227405m^{19} - 2147.41215484773m^{20} - 4897.30615346015m^{21} - 9691.02621986017m^{22} - 15208.1735423411m^{23} - 12867.5017215112m^{24} + 12445.3576847278m^{25} + 74411.2948662508m^{26} + 180510.788630277m^{27} + 326700.705114434m^{28} + 450572.196367118m^{29} + 321845.386288202m^{30}$

$R_1 = 0.0126168462489296$

$$R_2 = \frac{33m^4}{16} + \frac{2937m^5}{320} + \frac{23051m^6}{1200} + \frac{97051m^7}{4000} + \frac{167206573m^8}{8640000} + \frac{1059187213m^9}{172800000} - \frac{12134100557m^{10}}{486000000} - \frac{7476551840719m^{11}}{68040000000} - \frac{11893216016466239209m^{12}}{43893964800000000} - \frac{89543441888089926385lm^{13}}{2048385024000000000}$$

$$- \frac{52221193797291368137531m^{14}}{120982740480000000000} - \frac{14344322156303508126916 79m^{15}}{33875167334400000000} + \frac{17302029580914565641742708 1827m^{16}}{170730843365376000000000} + \frac{85348379941702431789789157 2501649m^{17}}{2629254987826790400000000000}$$

$$+ \frac{10065460848885718081631500857094485 7m^{18}}{14234950832530982400000000000000} + \frac{249706951106163335695671476731301 50759m^{19}}{219218242820977128960000000000000} + \frac{244124489292947939344096049609021 6646975550727m^{20}}{2074190401193808559708569600000000000000}$$

$$- \frac{23334972204747009648173033436260642054208130231 27m^{21}}{86244836881638559912682323968000000000000000000} - \frac{35669596328055661935097783558621881379007339761 38032813m^{22}}{97122466933335223281669382078464000000000000000000}$$

$$- \frac{9739480535524585513095513991709313076116239836703546199 62501m^{23}}{874976304602417026544559463144882176000000000000000000000} - \frac{14456699628642098687426742423593513943178873978366795172 05334293141m^{24}}{63061292225305399937119489627777948188672000000000000000000000000}$$

$$- \frac{1136238363315552696067018964113303070463492096000000000000000000 0m^{25}}{402291688712359781122360586935526810924463055491595749615804274073288189}$$

$$- \frac{799716516805454237411544831851308387641062522880000000000000000000 m^{26}}{2783236502342611504764879456509699828598618674074908090169674875564203 05001}$$

$$+ \frac{5763716879920269779872485912118749811406665814900736000000000000000000000 m^{27}}{49125527893920559522432451579786496483096708068467718744150773130839100645279359}$$

$$+ \frac{5317153318011055149759613499625248466018575598227530815897600000000000000000000 m^{28}}{74535747394301875371714698779493562737227445543002178425915764007958455379027278690 10357}$$

$$+ \frac{16286759642266942587022681726162113566322858171906820540943302656000000000000000000000000 m^{29}}{66444591368835013163490669620343198444272270190760679346972877989863526618754322233241194813651}$$

$$+ \frac{5087775047887069092583450354521120182492244297525762737788072460344016238735951559454477501678 75616971m^{30}}{6235915248730272645075711936017265447557611744005113215767724079185920000000000000000000000000000}$$

$$R_2 = 2.0625m^4 + 9.178125m^5 + 19.2091666666667m^6 + 24.26275m^7 + 19.3526126157407m^8 + 6.12955563078704m^9 - 24.9672850967078m^{10} - 109.884653743665m^{11} - 270.953331982128m^{12}$$
$$- 437.141654713103m^{13} - 431.641683682345m^{14} - 42.3446532815705m^{15} + 1013.40971788484m^{16} + 3246.10508820398m^{17} + 7070.94879870131m^{18}$$
$$+ 11390.7924766137m^{19} + 11769.6277618699m^{20} - 270.566599097076m^{21} - 36726.4109472623m^{22} - 111311.363339721m^{23} - 229248.388646893m^{24}$$
$$- 354055.717269103m^{25} - 348027.88786463m^{26} + 85232.3750756476m^{27} + 1401797.97226879m^{28} + 4079669.18087253m^{29} + 8158826.4832865m^{30}$$

$$R_2 = 0.000125776675006612$$

$$R_3 = \frac{1393 m^6}{512} + \frac{562041 m^7}{35840} + \frac{40986497 m^8}{940800} + \frac{11343987331 m^9}{148176000} + \frac{4562756952133 m^{10}}{49172480000} + \frac{4728398067354011 m^{11}}{69701990400000} - \frac{8887204170411416783 m^{12}}{175649015808000000} - \frac{2418055378954711805497 m^{13}}{6147715553280000000}$$
$$- \frac{9240786345947890301360368 3 m^{14}}{8262529703608320000000} - \frac{13762402831903321529743411 9979 m^{15}}{63621478717784064000000000} - \frac{584482368061771568734250541461 9599 m^{16}}{198403581381409603584000000000}$$
$$- \frac{15084958681950626735632589216204328149 m^{17}}{687468409486584276418560000000000} + \frac{14136901273256133407190630130567712792266859 m^{18}}{6098119779509797165543194624000000000000}$$
$$+ \frac{78044817571835685401249308766865937491507170374 79 m^{19}}{54937961093603762664378640367616000000000000} + \frac{4637088778945426433461318442472364273112970566228732 69 m^{20}}{123734022873069074460846792767963136000000000000}$$
$$+ \frac{395769989726762387683694399339406008192017287366392561 87467 m^{21}}{5573599060317396459088843780232899461120000000000} + \frac{62588186726536147991084945262104303271841663202705115155 4272920799 m^{22}}{642720690360312636159121838286312847938945024000000000000}$$
$$+ \frac{400541968616693535658121031719446626321998092371638918260577121186 74009 m^{23}}{5790270699456056539157528641121392447081955721216000000000000}$$
$$- \frac{6676323242990476649450011059123608617934252249678240738194668861353 49181583 m^{24}}{16055751857933986211318523633687243693585360082072172891661466457294 8048676251009 m^{25}}$$
$$- \frac{29371901220073694798230188208157149038928368992734150656000000000000000}{46870569356042415455944130728921493560131191022573819347783402915687 25806163145083002843 m^{26}}$$
$$- \frac{33870266635730421303968737992612832728538326560709371297267712000000000000000000}{165141058980703855352863078468792855376558980424825797938186165868248 4644838565756478022062557 m^{27}}$$
$$- \frac{6484166182577526517458405162228291463592287909690402927863052369920 00000000000000000}{269611398660649025582317889005400791092301713644823809272249532356138 027346685811601184964839 79958153 m^{28}}$$
$$- \frac{79445560268823673498264570064859961801883973618625142368880804768828 620800000000000000000000000}{129978626333140620063598062224043950268633117729170082023342551286396 4935862729256027733075141250 3986000153 m^{29}}$$
$$+ \frac{60836629459255760336398568495717488649219681058066482146106009463820 488069120000000000000000000000000}{2781942766994988582285332318129371658214354364758209824755545652259703 89793 67394948028110414235372434869 6278181109 m^{30}}$$
$$R_3 = 2.720703125 m^6 + 15.6819475446429 m^7 + 43.5655792942177 m^8 + 76.557521670176 m^9 + 92.7908649743312 m^{10} + 67.8373463974138 m^{11} - 50.5963789750232 m^{12}$$
$$- 393.325839167136 m^{13} - 1118.39674741652 m^{14} - 2163.16928013437 m^{15} - 2945.92649987586 m^{16} - 2194.27663493896 m^{17} + 2318.23935645825 m^{18}$$
$$+ 14205.9909065177 m^{19} + 37476.2629652987 m^{20} + 71007.9762544356 m^{21} + 97380.0714139899 m^{22} + 69174.9987879359 m^{23} - 102388.666714899 m^{24}$$
$$- 546636.451540324 m^{25} - 1383826.40621428 m^{26} - 2546835.69684604 m^{27} - 3393662.24806462 m^{28} - 2136519.19063319 m^{29} + 4665278.38508789 m^{30}$$

$$R_3 = 1.20593989996351 \cdot 10^{-6}$$

$$R_4 = \frac{7177 m^8}{2048} + \frac{99759441 m^9}{4014080} + \frac{466011937 m^{10}}{5419008} + \frac{114711969628733 m^{11}}{597445632000} + \frac{32898091945274873 m^{12}}{107540213760000} + \frac{141364189729430825 0959 m^{13}}{4173420615598080000} + \frac{493449557112643031622011 m^{14}}{5164608011802624000000}$$
$$- \frac{21778534864722558980169656 3401 m^{15}}{25053513465254529024000 0000} - \frac{26020188051488715297690854466226877 m^{16}}{79369530657926347948032 00000000} + \frac{15230481851451358117366 3979934539277639443 m^{17}}{200211228475232371225869680640 00000000}$$
$$- \frac{207103498415327875931026940769040118801995211 m^{18}}{1610449069047650386048089243648000000000 0} - \frac{15138438115590085972581589223239083383944465571 24953 m^{19}}{101559749641351976295350651972173824000000000 0}$$
$$- \frac{39966948006350239093975229475856470544130652712 70767 m^{20}}{9336242699172856678008306363441979392000000 0000} + \frac{301501850581618403492362922706844341073287531 3901628593237369 m^{21}}{8242800626731130049593329538768381649567744 0000000000}$$
$$+ \frac{703846794923483479383395891294720900994670722 6988083810848 7973643 m^{22}}{53042422033014821869133075581974535914968432 64000000000 00} $$
$$+ \frac{80560094446010445320685353742025106106833229804 16520810083398 026756941 m^{23}}{2676011408534410969226511316340848126724523 41406105600000000 00}$$
$$+ \frac{3569147634561809345412806388546285597970956844768322162166544 012400689293607631 m^{24}}{70533666463411956068239770913396153120654574 26220193939456 00000000000000}$$
$$+ \frac{175526681732328936828488239400378095469272431575 56777662641073928557836936572 63219083751 m^{25}}{30246839335638607661635517771582109088853080335 412381474661 33504000000000 0000}$$
$$+ \frac{5026674756209375416109281797100723484667154390342 4187969055100788313542325394 47094527316221 m^{26}}{4136066236402731856427768120827778102971885340 41155468392757209333760000000 0000 00000}$$
$$- \frac{14108704771862043078656483704634780658074623943132 54908938805624558007373005506943455359103763 495311 m^{27}}{886831333225302628810474759412791379726234417756 071168310583228335306833920000000 00000000 0}$$
$$- \frac{2137015457180710859255555689931470210628422658575266 5753159368287071309563150461 8673846231315280 9688166733 m^{28}}{38805965479272792431488754522384925194060565652170162182934500 9054963564386713 6000000000 00000000}$$
$$- \frac{385058355651809445730905094774650350313574714348822111497665296259760605 7625152045654043464330947 1363947361175 83923 m^{29}}{316180901651703910096621212176714741446684955625866563563178772339485210 1077611257856 0000000000 00000000}$$
$$- \frac{658210064456520360040259711200917494643391424591160658568745550136361019 86892065836718698 701602453688270300031 6134515757 m^{30}}{328591792493787417827688836307681236835320956780807089084944244710074085 3320649351937549926400000 000000000 000000000}$$

$$R_4 = 3.50439453125 m^8 + 24.8523798728476 m^9 + 85.9958016301138 m^{10} + 192.004030968851 m^{11} + 305.914325395468 m^{12} + 338.724999826485 m^{13} + 95.5444355089424 m^{14}$$
$$- 869.280665760758 m^{15} - 3278.35982344822 m^{16} - 7607.20663243694 m^{17} - 12859.9843606231 m^{18} - 14905.94272736 m^{19} - 4280.83858722858 m^{20}$$
$$+ 36577.5983473212 m^{21} + 132695.070840731 m^{22} + 301045.407314356 m^{23} + 506020.431592513 m^{24} + 580314.127319455 m^{25} + 121532.743164704 m^{26}$$
$$- 1590911.85023316 m^{27} - 5506925.11006366 m^{28} - 12178419.1783974 m^{29} - 20031238.7434012 m^{30}$$

$$R_4 = 1.13029247232401 \cdot 10^{-8}$$

$$R_5 = \frac{582527 m^{10}}{131072} + \frac{34020726071 m^{11}}{908328960} + \frac{81354695374793 m^{12}}{524559974400} + \frac{2282895448460627797 m^{13}}{5452800933888000} + \frac{33059130768273548816 4143 m^{14}}{40307104503300096 0000} + \frac{32261413404165268102114167233 m^{15}}{272345028352672923648000 00}$$

$$+ \frac{32186866291276856681538813826 59587 m^{16}}{3271408480572307158859776000 0} + \frac{48239127098405859287871689231 3782758461 m^{17}}{4420817850221387279125158297 60000000} - \frac{6437227004187126955026506845 6486735761575 03219 m^{18}}{8496458242697488639496224235 32339200000000}$$

$$- \frac{32714084805723071588597760000 00 m^{19}}{4960441473940301953860421436 903302255169498207151191} - \frac{4420817850221387279125158297 60000000 m^{20}}{1506945251384234540009605964 95982997805417255637934004573}$$

$$- \frac{2296337769253850254596644524 0808531558400 0000 m^{21}}{3036916674981946649864782463 84351252801588377174416722857 12091} - \frac{3447951160534656157276861752 9074010134937 60000000 m^{22}}{1278980005115986439750837245 60563589337390903019279204739 58156228381}$$

$$+ \frac{4659388800788507598136087129 79141635958479257600000000 00 m^{23}}{1853369347779470191627015521 76599234541121839286314069843 175806855074771593} + \frac{2238743131002862130752427144 12217973245330113691648000 00000000 m^{24}}{7416024210740301213774016518 16584714413893999790785921956 8065460736528933 5307901}$$

$$+ \frac{4114442720909776126933517692 844930310773045148266035937 280000000000 m^{25}}{4819531382147490672315088181 05319903926266459771472280801 87580655668955551658115722917 79}$$

$$+ \frac{2100464153451649810560830117 374265372952747278641294006 3408128000000000 00 m^{26}}{4825385797403707851552342224 493312969957481559486251514 296717541376000000000000 3158316152593271468914241617 2232586093942821990230593336 602744530863320868937801219971 1390485207}$$

$$+ \frac{1576580930196842812989431186 576691970747516102997328806760 7150541106315264000000000 00 m^{27}}{4083259628104257423579824292 694883408525012177536855237 71918699135918514271426013103 2999775665856928 0359}$$

$$+ \frac{1376311165453891283852910207 513466007565977022173944228 28395618288571635927613440000 000000000 m^{28}}{1059516015159742692226785604 597835519939894703228851467 29603529426388669066936436 23177499220547241304170603961}$$

$$- \frac{4171809890137954608377292831 584884096841488494440745341 1386271711711706602834775244 800000000000000 m^{29}}{3261930362360794337189649593 935566218529588398093278502 651391521705234390273731694 5598587939592701004254707721735857}$$

$$- \frac{1456749456237799809671921166 20144103931052702950232863 1440760150107486537561602158 793924280320000000000000 m^{30}}{2940848191332390991524795950 80338255308028245644914551246 9854578849116801843542770240543 48151044387546706211541275872 62117293023}$$

$$-\frac{1808642321046428727792209538 123834321227510518117560576 39675223301699019970674795984 8754435298689024 0000000000000000 00}{}$$

$R_5 = 4.44432830810547 m^{10} + 37.4541906832961 m^{11} + 155.091313377174 m^{12} + 418.664733251661 m^{13} + 820.18123543374 m^{14} + 1184.57875288946 m^{15} + 983.884051240401 m^{16}$
$- 1091.18106044541 m^{17} - 7576.36513981549 m^{18} - 21601.5324067596 m^{19} - 43705.5277531994 m^{20} - 65178.4344433332 m^{21} - 57129.37707789 m^{22}$
$+ 45045.4526529332 m^{23} + 353065.973468469 m^{24} + 998786.746697152 m^{25} + 2003269.28488152 m^{26} + 2966814.28633012 m^{27} + 2539703.49335527 m^{28}$
$- 2239184.19766222 m^{29} - 16259976.6527132 m^{30}$

$$R_5 = 1.04374473026266 \cdot 10^{-10}$$

$$R_6 = \frac{29229663 m^{12}}{5242880} + \frac{792236855589589 m^{13}}{14547796623360} + \frac{2327178857896901 m^{14}}{8865063567360} + \frac{244987075491155476336 1359 m^{15}}{295182361979167703 0400} + \frac{126469211749803229236 34645427 m^{16}}{657834978125002309 6320000}$$
$$+ \frac{647967463863340760809113596100808 1009 m^{17}}{19166091818179623541373140992 00000} + \frac{4820425873572407092401323816230 936273 m^{18}}{116793372017032080955242577920 00000} + \frac{21678737584043951421589613350 2885287907845837 m^{19}}{2430562338438331060022552688151 75680000000}$$
$$- \frac{26673738914361024678107738673428 6929471085410908577 m^{20}}{194133875095746378426121328 30805711912960000000} - \frac{35670658314669496749134911731 1551935930829739390 2305933740459 m^{21}}{6868134821056473216967120512 49805442415885025 28000000000}$$
$$- \frac{56909216316178963184126950090878 070645692827143015772872933342507 m^{22}}{45535733863604417428492008997 862100832173177177606400000000 00}$$
$$- \frac{18098853039050884517608903211 291516650126799073602911908518 30723567514191 m^{23}}{805489405691656148050184133 542725834970604048767281 7909760000000000}$$
$$- \frac{60771005140171748843871799825472 692403112461951871308699005630554455 551653773 m^{24}}{2136157903894272104629088322 15530891434204193733083130966 8352000000000 00}$$
$$- \frac{325477820449273540734297351032 4709219942649074216661047318672 68335577979256681558250649 m^{25}}{28718016269930385816101914286 631906856169144012124368728462 74585427968000000000}$$
$$+ \frac{154746580037003287785066875597 047314417391311239963053448154 1419914862067263378522725481 52319 m^{26}}{2261005318451956688283973839 179388316669766919504566705 4527825595397570560000000 00000000}$$
$$+ \frac{826600513176632413098539419055 0512555095978478408249287996 5036928371913884815502071405 70966260883905957 m^{27}}{3039643623878667004091671299 64957905909875400871486613693 17634444409869960258519040 000000000000000}$$
$$+ \frac{12818265647969768320316485398 29602612521014179533664356780 0519278756839878735053443232 016856659985773627938247 71 m^{28}}{1960468005375977894427790508 11831026726231061748639583881 87865144127033952736080094560 2560000000000000000}$$
$$+ \frac{30806206194262390664734326689 7599984016040529761639776234651 125118832431235236897109483953 86377976521438559286210386 083489 m^{29}}{26356081623192532165603981 30477446878193852447923069373 14114227772534844245966275423 11356495560704000000000000 000}$$
$$+ \frac{69750447647831337722425085141 2133091262744181270701034009 66194781209733831537588844303 94116262785758576185848234 34946952 52801 m^{30}}{4742965145819376109858762178 51777118922970846949812927049 4138467500887761380624530721 37368249397411840000000 0000000000}$$

$R_6 = 5.57511577606201 m^{12} + 54.4575151894454 m^{13} + 262.511243175432 m^{14} + 829.951606351213 m^{15} + 1922.50664612382 m^{16} + 3380.80120877186 m^{17} + 4127.31115672409 m^{18}$
$+ 891.922714394268 m^{19} - 13739.8683775336 m^{20} - 51936.4561762964 m^{21} - 124977.048764915 m^{22} - 224693.868239146 m^{23} - 284487.420285666 m^{24}$
$- 113335.760168807 m^{25} + 684414.931597568 m^{26} + 2719399.42789039 m^{27} + 6538370.23242391 m^{28} + 11688462.1298008 m^{29} + 14706084.8021015 m^{30}$

$R_6 = 9.53733002810468 \cdot 10^{-13}$

$$R_7 = \frac{3491520547 m^{14}}{503316480} + \frac{388748860895310229 m^{15}}{5043236162764800} + \frac{1010188457276086430137 m^{16}}{2382929086906368000} + \frac{23561696970421974493692020891 m^{17}}{15349482822916720558080000} + \frac{12676620473596066599046343681 22103 m^{18}}{3094455737100010864508928 00000}$$

$$+ \frac{11434992393159180051410168878753942 44741517 m^{19}}{1355426013381662976845908530954240 00000} + \frac{1889344995639628865799912761390356 6429844367641 m^{20}}{145166126033176104820196803665 1991040000000}$$

$$+ \frac{17379874888625315493616680357232 2840001456167227 1832321 m^{21}}{1589632880575669928679223476629141 04380620800000000} - \frac{90061687171064569812830104713869321 3158759026251451309567859 m^{22}}{5447989808308935979569434699103392 18533263605760000000 00}$$

$$- \frac{3088300648539777656642600405648 9644065332 08997321272676638820280 6833 m^{23}}{28915810537182867712227507237574377 73048210591737446400000000}$$

$$- \frac{130776133140110961727281054860248592 566349394216732502731934208981 2898161739 m^{24}}{41937430694451744841788741373225 67215918754660701090244198400000000}$$

$$- \frac{632819101994975642807029168477048454 4189571621161478068320307865544094 80759257272 48181 m^{25}}{959797591994526948699239899542882 035063767803916216107355174 28613120000000000}$$

$$- \frac{16303043806694202914891208617023997 41577884953984516012277905585824 461933378903350115430 25871 m^{26}}{15624736959597303103264657876943 84852252100888335128652548 56353238787031 04000000000000}$$

$$- \frac{2603638120261792998024435281324278972 5744275779602765724548539342396548068 73737807206991303261968907647 m^{27}}{2600685599048160022331020706259015 2837740665678927324692363339975445889 83917975961600000000000}$$

$$+ \frac{5298382359975726882989356468496706 58057688133455820263185600733340801065 989179766413567126435992268 23813 m^{28}}{830790443564066064276513977263182 135152759902463442726044208031451 870068676440883 2000000000000000}$$

$$+ \frac{6684053030151605968304039205230 0881097902497867410914464107606569877 3625373562897357808487008 686816802066465376643 75159 m^{29}}{11010729048352856412087582793772 747829174450370831905974034072096 44643786762276345933127923 7939200000000000000}$$

$$+ \frac{261304048596404308628292740731569776 28394103040798083521002773696909306 3460234529032788870525535529395575 1664336124235724 30147 m^{30}}{14339668825916656082924942253108253 168557369981477410985996373177976 6810676835927257166087762078597120 0000000000000000}$$

$R_7 = 6.93702806433042 m^{14} + 77.0832156870858 m^{15} + 423.927200698851 m^{16} + 1535.01569025143 m^{17} + 4096.55899149362 m^{18} + 8436.45634676136 m^{19} + 13015.0541814958 m^{20} + 10933.2633345703 m^{21} - 16531.1776159544 m^{22} - 106803.184526629 m^{23} - 311836.302259243 m^{24} - 659325.577885576 m^{25} - 1043412.36904345 m^{26} - 1001135.28571647 m^{27} + 637751.962726705 m^{28} + 6070490.88284622 m^{29} + 18222460.4883592 m^{30}$

$R_7 = 8.64668742089923 \cdot 10^{-15}$

$$R_8 = \frac{60441179201 m^{16}}{7046430720} + \frac{1319281872968288692247 m^{17}}{12345842126448230400} + \frac{25461942126523350106434571 m^{18}}{38565324342492659712000} + \frac{974723240621191448915936 1067921369 m^{19}}{361977643723151270888865 7920000}$$
$$+ \frac{16706720181941002964083628361842791049 m^{20}}{20558684810548977862528991232 00000} + \frac{12333293240390659673359141593088724708615371826989 m^{21}}{64527725556280427888138620558825750528000000}$$
$$+ \frac{535829384120635452219543652454818477108366528144 2645587 m^{22}}{15319204685688754982943854921 376802730409984 0000000} + \frac{11961837132979224195756558624676319219874013281123865471146045437 m^{23}}{273196153532251998698567493347396979861685496551833600000000}$$
$$+ \frac{4315281030454429172589602532275838307427408 67600 480389 4896310316357501 m^{24}}{103773012526918857201653465213022048006501512494220 889292 8000 000000}$$
$$- \frac{27019219538448738479002133858937373138613856999747152134404675838 92003576123436147 m^{25}}{14805148671611365218323016579312341779449955176207592463247825960960000000}$$
$$- \frac{12112426431538440865714880293781568937300881784954221474903572469470371690563298 81557283 m^{26}}{17574081601919480798279878755058232500751583043037817443686750611308544 00000000000}$$
$$- \frac{76510599274637952286876405400362772381595489925588977827920526140014892282398885083 83671728 42809 m^{27}}{4477266648548283439232491605449889043532895156977410783412561597012721197838368000000000}$$
$$- \frac{23854359494714944534817152307112268100125318679033523006811569086337734195144119558488196865 22201974669 m^{28}}{74404784208902366092369276871428163586594888231955544925241430157163553105410785280000000000}$$
$$- \frac{65006769517682509 1132898056423672131035122496341177148761067176385822823611741446134898949005459907104 8725997051 m^{29}}{1516460735197976199646577866065026379925331 88805935027497411635369234682074 6098 814141385932800 0000000}$$
$$- \frac{12729266732148880405829966356354 62996311285450348902510940557 7943510944465740384527083001 8969312500306793707763014888 3 m^{30}}{63695025379327210865947721698021957828476835294990701513040708 752397128696 324881125680167151206400 0000000000 0000}$$

$R_8 = 8.57755956209841 m^{16} + 106.860419844671 m^{17} + 660.228911869113 m^{18} + 2692.77193639804 m^{19} + 8126.35649405381 m^{20} + 19113.1690045912 m^{21} + 34977.6241726967 m^{22}$
$\quad + 43784.7933739926 m^{23} + 4158.38465644913 m^{24} - 182498.805907013 m^{25} - 689221.019106654 m^{26} - 1708868.49679694 m^{27} - 3206024.95502713 m^{28}$
$\quad - 4286742.67713208 m^{29} - 1998471.09822807 m^{30}$

$$R_8 = 7.79163539918136 \cdot 10^{-17}$$

$$R_9 = \frac{3172750609873 m^{18}}{300647710720} + \frac{309866619771715974715 5683 m^{19}}{2126777070316148490 2400} + \frac{3560696435924612672603777911 9 m^{20}}{3564053014435801639944 1920} + \frac{67992909887937841157 16619812534929538557 m^{21}}{150071748234846 604940282374914 0480000}$$
$$+ \frac{92013232076042293439077979534933464477856317033 m^{22}}{6035765656069425734142048931430793216 00000} + \frac{98164357303545 349798781476430559408302614 0800437488758577 m^{23}}{243982338180676894599 2838696767819811 4773526118400000}$$
$$+ \frac{169169075476245577154478162 440175841304714432565991524997 64177 m^{24}}{2004243148643031078281764258844859039251837013196 80000000} + \frac{685954322456029390994534073 1327498404880306162442931934655361453600183509 m^{25}}{516483695544407395226224847452965607094127074247881055535 10400000}$$
$$+ \frac{146464010074901554092222747675353775836574803 7080171962550208 6165029960554150196 3 m^{26}}{138483740318930866067217215897 68676229893 41041639358516585163653120000 00000}$$
$$- \frac{4047027751665516281703065230445622 388350 0024726152707676 3146598245320774871066703462262379 m^{27}}{1788216689761015248846876143076343476097236776 132155 06025372845378003793920000000000}$$
$$- \frac{83510618387683178018458012577315655936611811458207851123830017330 34116120068387192348 26159 45485749 m^{28}}{620315560441212436281642173397438 63422945612361454 89481444 5201560281 306287518515 200000000}$$
$$- \frac{287332963442855053311191 48893 58139464635896629 15117883144135586549578089912830409 871133141056 1557778163894251 m^{29}}{7209025203899821325674618 027866810099 519515788 603508059057983762607322382100 7853598854348800 00000}$$
$$- \frac{643456202974246334973372785389167685566420517956923925001146265 9347918 251145 99462858435506 20756021460783567 1340129 m^{30}}{7409609121173129954240188599874176223728546 79590561206729321310 2818529780006 035761173888722010112000 000000000}$$

$R_9 = 10.553050952142 m^{18} + 145.697743358524 m^{19} + 999.058213079998 m^{20} + 4530.69353077276 m^{21} + 15244.6660972203 m^{22} + 40234.2063099877 m^{23} + 84405.4652704095 m^{24}$
$\quad + 132812.386600702 m^{25} + 105759.781192879 m^{26} - 226316.406442129 m^{27} - 1346260.25386635 m^{28} - 3985739.47678 m^{29} - 8684077.55998296 m^{30}$

$$R_9 = 6.98707157982718 \cdot 10^{-19}$$

$$R_{10} = \frac{239916034627727 m^{20}}{1855425871872} + \frac{22806537483190236860377761929 m^{21}}{11637724128769964538593280} + \frac{732785984940995312253225647 m^{22}}{4962134374432024559616000} + \frac{224150235899994109760250588296432238612 9 m^{23}}{30414540975595578601230561315913728000}$$
$$+ \frac{3960864928623957965591930576476832740161069612 37 m^{24}}{144858375758566621761940917435433903718 40000} + \frac{340839900948578933997689995338329 8 76771448138049924493 6070757 m^{25}}{42648112713982321175954820419501490304624 12365 4963200000}$$
$$+ \frac{76289131123782134675409345 816853106362879071497 623247135255906 33 m^{26}}{40554356930430251595722788454531589 0149814830842773504000}$$
$$+ \frac{8654630315986322920676944552062829498715465 9432268331469088194265 09183040841 m^{27}}{2487439278126818557846501597088425 3810160549611480293177336791040000000}$$
$$+ \frac{39985682615954560031006273847793754849222467324100184014816072656470744 4 59240262901267 m^{28}}{9300822809901583797867189937811815451904447440159343935301076446609408000 00000}$$
$$- \frac{169412702175584359351648344142507232597784 62273574520024687242771421456819212030 232122246116844 9 m^{29}}{39118260700424913312542871778 3045590549778692893720352553212 70742284037635367813120000 0000}$$
$$- \frac{3765432928870710279358682135784436903197444809904265350761067045188894698 1117309952222508 8231882243927 m^{30}}{166646137679452156206 2313863478 84119 84774 9021844882232509783932 63837453791493250059599 87200000 0000}$$

$$R_{10} = 12.9305103623036 m^{20} + 195.970769119793 m^{21} + 1476.75562499226 m^{22} + 7369.8378706373 m^{23} + 27343.016293552 m^{24} + 79919.1052683636 m^{25} + 188115.746119841 m^{26}$$
$$+ 347933.330155651 m^{27} + 429915.540089487 m^{28} - 43307.8309572553 m^{29} - 2259538.07349175 m^{30}$$

$$R_{10} = 6.24071528740691 \cdot 10^{-21}$$

$$R_{11} = \frac{1640613010307 30749 m^{22}}{10390384882483200} + \frac{3005131754279905698481692526949 m^{23}}{11530298982965934096698572800} + \frac{56830761542516043717470 2735513423663 m^{24}}{2654620734848247007082912415744 00}$$
$$+ \frac{3010564063948238367483541286127058963426647137773 m^{25}}{258240565253403675476241966940548 43449344000} + \frac{224814050689808369299433541627903918577087203284290 14259 m^{26}}{47563780270632902613900415043897797285873975296000 00}$$
$$+ \frac{700684324477577177175001059782755809337247588142730254361642310296 1 m^{27}}{4626987705412874982354606109316087031047311516473225117696 00}$$
$$+ \frac{359808173817774945768671123317851209537376204387061841632267085204150761069 m^{28}}{9130897537861767490178579696124366147068764546608262447261286 40000}$$
$$+ \frac{857448195945000340348746182388705449248313272517439621157905748311544896333617709180233 m^{29}}{103629222644755100795474722893528904382814298644729109319238170303350702080 00000 00}$$
$$+ \frac{237234569049311356052425697677607197232489482779255642372463631928487323539306737543909143546 47 m^{30}}{183233709538575105455 171677069648586158504980303175548200203 80632786251883426611200 00000 0}$$

$$R_{11} = 15.7897231802564 m^{22} + 260.62912667915 m^{23} + 2140.82414095831 m^{24} + 11657.9827843626 m^{25} + 47265.8080183367 m^{26} + 151434.230883709 m^{27} + 394055.647131961 m^{28}$$
$$+ 827419.307085189 m^{29} + 1294710.28909868 m^{30}$$

$$R_{11} = 5.55562482796697 \cdot 10^{-23}$$

$$R_{12} = \frac{3255392585602 18599 m^{24}}{16932479067750400} + \frac{5010121819676300911850407 44723683 m^{25}}{14592844003838367385874595840 00} + \frac{13372694784873542313 00930190643290774373 m^{26}}{43801242124996075616868054859776000 0}$$
$$+ \frac{10817925147884079742469011265300086591116317608167 88811 m^{27}}{60055278119485988148864160408961865259941888000000} + \frac{155427063329932 4071192873522363186150665868833198401429937397 m^{28}}{19620059361636072328233921205607841380423014809600000000}$$
$$+ \frac{148451903878439949327963217958390838212638679865486933278651854994656916223 7 m^{29}}{538015848190508185448232810377698341999001277998803342852096000000000}$$
$$+ \frac{13690085869107772593689333353577179931518378509342547364464804386182948127030 73 m^{30}}{1743747779368887920819382598363486159058604878494255012013670400 0000000000}$$

$$R_{12} = 19.2257292778965 m^{24} + 343.327306065801 m^{25} + 3053.04008199396 m^{26} + 18013.2795761277 m^{27} + 79218.4470317381 m^{28} + 275924.77875461 m^{29} + 785095.523484308 m^{30}$$

$$R_{12} = 4.92896891413673 \cdot 10^{-25}$$

$$R_{13} = \frac{186342648094524945833 m^{26}}{7979815589747097600} + \frac{395909332095663002907881489508677747 m^{27}}{882575205352144459497695556403200} + \frac{281631207032191691293273711740865072269 25499 m^{28}}{655967402063941228438215989 5800053760000}$$
$$+ \frac{70778790880938741050477366523016655532348 4565719749650287 m^{29}}{259438801476179468803093172966715257922948956160000000} + \frac{7600156629992418430163557847274993859148054927519869927 6805047547 m^{30}}{587660017999723638375262407950366065026 43013957713920000 0000}$$

$$R_{13} = 23.3517486712285 m^{26} + 448.584244940007 m^{27} + 4293.37198992 m^{28} + 27281.4977860732 m^{29} + 129329.142653976 m^{30}$$

$$R_{13} = 4.28006930409465 \cdot 10^{-27}$$

$$R_{14} = \frac{36910325662253363212769 m^{28}}{1304129862095811379200} + \frac{1532112285737528705444400204750059206997 m^{29}}{2632595755393253759187411888242688000} + \frac{20131045936730775291084899256016554211 8342509377 m^{30}}{33749136973012091531835366654485641297920000}$$

$$R_{14} = 28.302645875263 m^{28} + 581.977799895322 m^{29} + 5964.9068812719 m^{30}$$

$$R_{14} = 2.97232914714305 \cdot 10^{-29}$$

$$R_{15} = \frac{34575097372036180448315 m^{30}}{1009815122847522095104}$$

$$R_{\{15\}} = \ 34.2390370175283 \ m^{\{30\}}$$

$$5.8176883515273 \ \cdot 10^{\{-32\}}$$

$$R_{16} = 0$$
$$R_{16} = 0$$
$$R_{16} = 0$$

$U_i$

$$U_0 = 1$$
$$U_0 = 1.0$$
$$U_0 = 1.0$$

$$U_1 = \frac{9m^2}{8} + 3m^3 + \frac{7m^4}{2} + \frac{11m^5}{6} - \frac{38273m^6}{18432} - \frac{611123m^7}{69120} - \frac{20324711m^8}{1036800} - \frac{1014978499m^9}{31104000} - \frac{9546264328121m^{10}}{238878720000} - \frac{20964609276307m^{11}}{783820800000} + \frac{19903067874778561m^{12}}{658409472000000} + \frac{5747537759755617467m^{13}}{34566497280000000}$$
$$+ \frac{3102924009599298706448929m^{14}}{74331795750912000000} + \frac{2981146963577387845508211063m^{15}}{390241927692288000000000} + \frac{20914818553852579204215677011m^{16}}{204877012038451200000000000} + \frac{6112253156451466499535323613811m^{17}}{86048345056149504000000000000}$$
$$- \frac{118984186633146875869206734608446073m^{18}}{11564897575546493337600000000000000} - \frac{71813045531654162770210060848466410207m^{19}}{1335745669975619980492800000000000000} - \frac{1023745360520713644561940506810152531222969m^{20}}{77139312441092053873459200000000000000000}$$
$$- \frac{846156328042925538211104918122714431554926 1859m^{21}}{3563836234778452888953815040000000000000000} - \frac{10081302986378043012092231515230504815064491 7728923m^{22}}{32929846809352904693932509696000000000000000000}$$
$$- \frac{7106329867434570273362998731717596425934221263140370253m^{23}}{39555331987394709118352621064683520000000000000000000} + \frac{31888192186448958167839895497886088572627177274974588929m^{24}}{7332386540626315523605735867731148800000000000000000000}$$
$$+ \frac{103586543738537204862312689518384227720987886200722641686545293373m^{25}}{535065509790470060072529002902358348267520000000000000000000000}$$
$$+ \frac{1424501621551335692380133134644028539878949964143341472662977959157352369m^{26}}{3085059313729500653563784823774301670106936115200000000000000000000000}$$
$$+ \frac{2245573164471250607434509886852377564690989118051859000324006530974475647041 13m^{27}}{27793299357389071387956137477382683745993387461836800000000000000000000000}$$
$$+ \frac{6300154823398346100398500049420302917905279239919978846158509278966087253880063619m^{28}}{62597458477679536033524210633435149446691360691092193280000000000000000000}$$
$$+ \frac{53938756098455288827337370905034328504170413433379965795216943078979470420261506312 5101m^{29}}{1127881006850829880252039227193234523094849369320991385192099138019017954124239653878524029754749261978599542320905116557333232632745893328 2697m^{30}}$$
$$- \frac{12005472931221831390179541242396538785240297547492619785995423209051165573332326327458933282697m^{30}}{650309119406024089036199769458166388387935869965619928907540070400000000000000000000000}$$

$U_1 = 1.125m^2 + 3.0m^3 + 3.5m^4 + 1.83333333333333m^5 - 2.07644314236111m^6 - 8.84147858796296m^7 - 19.603309220679m^8 - 32.6317675861626m^9 - 39.9628076042981m^{10}$
$- 26.7466865848763m^{11} + 30.2290120680077m^{12} + 166.274809773136m^{13} + 417.442358045175m^{14} + 763.922775086348m^{15} + 1020.84749995901m^{16}$
$+ 710.327799153256m^{17} - 1028.83908703813m^{18} - 5376.25141865254m^{19} - 13271.3830099305m^{20} - 23742.8510262489m^{21} - 30614.4849222763m^{22}$
$- 17965.5422174163m^{23} + 43489.5132843408m^{24} + 193596.002439218m^{25} + 461742.053130664m^{26} + 807954.872717998m^{27} + 1006455.37001231m^{28}$
$+ 478230.910626453m^{29} - 1846117.8803993m^{30}$

$$U_1 = 0.00909409327603924$$

$$U_{-1} = -\frac{19m^2}{8} - \frac{10m^3}{3} - \frac{43m^4}{18} - \frac{28m^5}{27} - \frac{292693m^6}{165888} - \frac{738053m^7}{124416} - \frac{22461197m^8}{1866240} - \frac{950625733m^9}{55987200} - \frac{7675497823457m^{10}}{429981696000} - \frac{53834460450451m^{11}}{5643509760000} + \frac{20709486101413687m^{12}}{1185137049600000}$$

$$+ \frac{84847797213022081 03m^{13}}{124439390208000000} + \frac{17342923087757157 30265393m^{14}}{133797232351641600 00} + \frac{12590716974442795 2353534771m^{15}}{70243546984611840 0000} + \frac{7656637160770847 66 0903200837m^{16}}{368778621669212160 00000}$$

$$+ \frac{289457025809993435 2230715012687m^{17}}{15488702110106910720 000000000} - \frac{72865801916416720235 17366727659441m^{18}}{20816815635983688007 6800000000000} - \frac{182942949576769445 482389251893454608319m^{19}}{24043422059561159648 8704000000000000}$$

$$- \frac{306950395748731146710 23902874874526081 1073m^{20}}{138850762393965969722 26560000000000000} - \frac{263766662548906392213 32271360290150488779 53203m^{21}}{641490522260121520011 686707200000000000000}$$

$$- \frac{314627775613937603400 89904259684674805924 671854641m^{22}}{592737242568352284490 79851745280000000000 0000} - \frac{225457057868312037319 04804089871644550426 06137582015501m^{23}}{711995975773104764130 34717916430336000000 000000000}$$

$$+ \frac{789866907155599009064 74856000448387873780 1567842198303 11833m^{24}}{1069061957623316803341 71628951520149504000 00000000000000} + \frac{107882707598108648472 82445931619576376814 61745528022054331066 847m^{25}}{321039305874282036043 51740174141500896051 20000000000000000000}$$

$$+ \frac{147904502075951083399 65736691996096769145 267070297448442639759 0498882291m^{26}}{185103558823770039213 82708942645810020641 61669120000000000000 00}$$

$$+ \frac{227194450992530370976 82339547733882121809 21597206122984294861 5532071207412707m^{27}}{166759796144334428327 73682486429610247596 03247710208000000000 00000000}$$

$$+ \frac{606261862492620755651 75479853254551672380 16616385734468984422 17522666179514697541m^{28}}{375584750866077216201 14526380061089680148 16414655315968000000 0000000000000}$$

$$+ \frac{424286108363642651566 17389457993599831328 66146890494587581634 84537190429028245720 39m^{29}}{667286041104979281512 23536315940713856909 62159259483111424000 00000000000000000000}$$

$$- \frac{126599612351726374551 86801941020156410327 91535804255488409974 184470077826194910313 5502 73172883m^{30}}{390185471643614453421 71986167489983303276 15219793719573445240 42240000000000000000 0000000}$$

$$U_{-1} = -2.375m^2 - 3.33333333333333m^3 - 2.38888888888889m^4 - 1.03703703703704m^5 - 1.76440128279321m^6 - 5.9321389531893m^7 - 12.0355350865912m^8 - 16.9793405099737m^9$$
$$- 17.8507547992392m^{10} - 9.53918088917259m^{11} + 17.4743386078458m^{12} + 68.1840348712729m^{13} + 129.620940455458m^{14} + 179.243752841824m^{15}$$
$$+ 207.621502735555m^{16} + 186.882686329872m^{17} - 35.0033372974019m^{18} - 760.885655642434m^{19} - 2210.64969652677m^{20} - 4111.77801379816m^{21}$$
$$- 5308.04803576445m^{22} - 3166.54960898486m^{23} + 7388.41094777697m^{24} + 33604.205349346m^{25} + 79903.6512403121m^{26} + 136240.542532139m^{27}$$
$$+ 161418.124962373m^{28} + 62696.6417240971m^{29} - 324460.08770762m^{30}$$

$$U_{-1} = -0.0173921860782608$$

$$U_2 = \frac{169 m^4}{128} + \frac{479 m^5}{96} + \frac{6143 m^6}{720} + \frac{354899 m^7}{43200} + \frac{12122443 m^8}{4608000} - \frac{122035633 m^9}{11340000} - \frac{2453358364771 m^{10}}{57153600000} - \frac{2560301152364071 m^{11}}{24004512000000} - \frac{196857851760234829 6529 m^{12}}{10323860520960000000} - \frac{3988340209132984310309 m^{13}}{16937583667200000000}$$
$$- \frac{58013218936341079153 2881 m^{14}}{474252342 6816000000000} + \frac{21741296433103343708766967 m^{15}}{62245619976960000000} + \frac{59370171404045249204339980329787 m^{16}}{4015589435953643520000000000} + \frac{10172482413678377454590521881149963 m^{17}}{289875362407903641600 0000000000}$$
$$+ \frac{40806691298346750241614815386058093809 m^{18}}{6696120871622574120960000000000000} + \frac{57121744829779071086463536679529035810431 m^{19}}{7734019606724073109708800000000000000}$$
$$+ \frac{4439951023515275474141444426115420563507 35635811 m^{20}}{146354874708235131973036670976000000000000000} - \frac{239007358768997641055657861239418267439621040 17225391 m^{21}}{1648138832808112879931359211028480000000000000000}$$
$$- \frac{20460564118520741324346123197460059255088599779 0614 3772633 m^{22}}{3712020686192072233825403 7830388940800000000000000000} - \frac{83732177470795739231225074332373071601595 45528194995 39913753907 m^{23}}{66883188723808757509066125362794793533440000000000000 00000}$$
$$- \frac{40621419846672678957818162506229270255658123193963073193 08061844618423 m^{24}}{192816207108093790 4773655148589385438168350720000000000000000000}$$
$$- \frac{133304522271236008358441575154622353744277670555531933280 85960465989138 7503 m^{25}}{5428378780740053005460183101051305419139333488640000000000000000000000}$$
$$- \frac{32975202515625876015370120164100489830152270595959431468993 99888258900 95381131 m^{26}}{489042644356871375261907895573712105210262553991577600000000000000000000}$$
$$+ \frac{53434542359628244308080228722532380801519326456050081353235743 64658625160 09418228101 m^{27}}{881157036602210843946905646244714471167851069782024519680000000000000000000}$$
$$+ \frac{34761307971863116518784484606425830864715790216852755554209995313875636901 3153585031342295321 m^{28}}{1625772798515060222590499423645415970 96983967491404982226885017600000000000000 000000}$$
$$+ \frac{29387049052600211722773243137585655025667170474503156403368198892661 6184784805270627721622107555041 m^{29}}{622479953527442545676006895573915980504859633027 8681074853027512320000000000000000000000000}$$
$$+ \frac{460908577638592763917389081 550037617550147524024398716254100786330646232422 347755525 39144275517788 4509 m^{30}}{59584170201617755 12369842754864497135141298021396885677666060357662146560000000000000000000000000}$$

$$U_2 = 1.3203125 m^4 + 4.98958333333333 m^5 + 8.53194444444444 m^6 + 8.21525462962963 m^7 + 2.63073849826389 m^8 - 10.761519664903 m^9 - 42.9257013516384 m^{10} - 106.659162759238 m^{11}$$
$$- 190.682401569224 m^{12} - 235.472797507504 m^{13} - 122.325634931633 m^{14} + 349.282350166177 m^{15} + 1478.49207073994 m^{16} + 3509.26078338592 m^{17}$$
$$+ 6094.07925583916 m^{18} + 7385.77709062394 m^{19} + 3033.68851387186 m^{20} - 14501.6520460097 m^{21} - 55119.7470278915 m^{22} - 125191.664854025 m^{23}$$
$$- 210674.301999417 m^{24} - 245092.002762235 m^{25} - 67428.0717563819 m^{26} + 606413.387625828 m^{27} + 2138140.58173523 m^{28} + 4720963.12275937 m^{29}$$
$$+ 7735419.92913545 m^{30}$$

$$U_2 = 7.62192021039069 \cdot 10^{-5}$$

$$U_{-2} = \frac{361m^4}{128} + \frac{4007m^5}{480} + \frac{15269m^6}{1200} + \frac{2759051m^7}{216000} + \frac{2701212463m^8}{207360000} + \frac{5775890957m^9}{226800000} + \frac{16223813466421m^{10}}{285768000000} + \frac{11546915103133721m^{11}}{120022560000000} + \frac{232902190843842108977m^{12}}{19118260224000000000}$$

$$+ \frac{3858605520366478614501 1m^{13}}{3387516733440000000000} + \frac{338638935829965139256809 3m^{14}}{711378514022400000000000} - \frac{2873721194594984325038420 11m^{15}}{2489824799078400000000000} - \frac{7568554149116134787142446603 7337m^{16}}{20077947179768217600000000000}$$

$$- \frac{9194251507788082518922892331806413m^{17}}{1449376812039518208000000000000} - \frac{83977766923467928581258157756560230 53m^{18}}{1116020145270429020160000000000000} - \frac{17693832445789683117577333440077717558 41m^{19}}{2416881127101272846784000000000000000}$$

$$- \frac{40999173810309090039039630513897173006961457096 1m^{20}}{73177437354117565986518335488000000000000000} + \frac{297127733700897348916903022469960218445984397817304 1m^{21}}{8240694164040564399656796055142400000000000000000}$$

$$+ \frac{59535608994542856455378179283974041294875972704416697558 3m^{22}}{185601034309603611691270189151944704000000000000000000} + \frac{269355454186436701039942877611204533323810840442782679950972957m^{23}}{33441594361904378754533062681397396766720000000000000000000}$$

$$+ \frac{110077788459487474828041025988978409578850777049058764480902554234737 3m^{24}}{96408103554046895423868275742946927190841753600000000000000000000} + \frac{14012451975934614392927968519057873896608814624290152018072698616334462 03m^{25}}{27141893903700265027300915505256527095696674432000000000000000000000}$$

$$- \frac{57593262908413817058790415716339068675157575919536740156060501003185698044821 9m^{26}}{2445213221784356876309539477868560526051312769957880000000000000000000000} - \frac{40536679589271751102800820506250443615215989275977755191705089270667133154992691725 1m^{27}}{4405785183011054219734528231223572355839255348910122598400000000000000000000000}$$

$$- \frac{1727829819879103069435880332529435866293813325617532567367679632639919990545621035839823319 71m^{28}}{8128863992575301112952497118227079854849198374570249111344250880000000000000000000000000} - \frac{11126933204356404103706301495858835168995803686580586508361237544664082388441811128329830329947269 1m^{29}}{31123997676372127283800344778695799025242981966513934053742651375616000000000000000000000000000}$$

$$- \frac{1191051326675339214796474519747409590075032712491827197559271637812107330134417450029090101947406712985 9m^{30}}{2979208510080887756184921377432248567570649010698442838833030178831073280000000000000000000000000000}$$

$U_{-2} = 2.8203125m^4 + 8.34791666666667m^5 + 12.7241666666667m^6 + 12.7733842592593m^7 + 13.0266804735725m^8 + 25.4668913447972m^9 + 56.772673869786m^{10} + 96.2062057594316m^{11}$
$+ 121.821854141032m^{12} + 113.906611361535m^{13} + 47.6031998654519m^{14} - 115.418610805816m^{15} - 376.958564605782m^{16} - 634.35894871605m^{17}$
$- 752.475367755291m^{18} - 732.093616329864m^{19} - 560.270696716358m^{20} + 360.561534970508m^{21} + 3207.71967764095m^{22} + 8054.50395909586m^{23}$
$+ 11417.8979153736m^{24} + 5162.66551834995m^{25} - 23553.4727177641m^{26} - 92007.8440174146m^{27} - 212554.893458331m^{28} - 357503.342599317m^{29}$
$- 399787.837153769m^{30}$

$U_{-2} = 0.000153209408076593$

$$U_3 = \frac{3185 m^6}{2048} + \frac{19887 m^7}{2560} + \frac{4924057 m^8}{268800} + \frac{4535613997 m^9}{169344000} + \frac{18807789518657 m^{10}}{758661120000} - \frac{80290182259901 m^{11}}{159318835200000} - \frac{1073693539509141017 m^{12}}{12546358272000000} - \frac{1031889595119933417017 m^{13}}{3512980316160000000}$$
$$- \frac{30805905500112452818050181 m^{14}}{47214455449190400000000} - \frac{188517607019194530982220227 m^{15}}{181775653479383040000000000} - \frac{1498624913467962986582892091742410 m^{16}}{141716843843864002560000000000}$$
$$+ \frac{3058820479222785753380448748185604 1 m^{17}}{39283909113519101509632000000000000} + \frac{13061532464038202872181521168240788111260383 m^{18}}{348463987400559838031039692800 0000000000}$$
$$+ \frac{9223739753228508843439913587 68723346818986185681 m^{19}}{7848280156229108952054091481088 0000000000000000} + \frac{4379257299009910701420728655253770363675938308 7442997 m^{20}}{17676288981867010637263827538280448000000000000000}$$
$$+ \frac{12291546235314269572499763311096521317626263656 7379694 47357 m^{21}}{31849137487527979766221964458473711206400000000000 0000} + \frac{140011273151329202524819547188079583287733773177162296175354 411233 m^{22}}{367268965920178649233783907592178770250825728000000000000000 00000}$$
$$- \frac{39335704517040548018553728312757696515244398627368649736679396 6421047 m^{23}}{4135907642468611836839490293724231764871112294400000000000000 0000000}$$
$$- \frac{18502858438058122540380611631408830364985723708076414457266959 7204922483 m^{24}}{1164387248468741369671209275175505012405387033312800000000000000 0000000}$$
$$- \frac{7956604748289482777601308715156716119818432741603962237188762235 9931082492364159 m^{25}}{16783943554327825598988678976089799450816210852990943232000000000000 000000000000}$$
$$- \frac{1877940930384852334020915877221750088272880446266242622239827349414 91479088817708450771 m^{26}}{1935443807756024074512499313863590441630761577548212169867264000 0000000000000000000000}$$
$$- \frac{43626921293296116392511628915698804429780255681890785391936572444254 5249611773083499870658421 m^{27}}{29641902548925835508381280741615046690707601872870413384516810833920000000000000000000000}$$
$$- \frac{775724198155713038645508252000937726204785095817298888318371053510203333 654224545928416523894498167 m^{28}}{567468287634454810701889786177571155727742668704465302634862891205918720000 0000000000000000000}$$
$$+ \frac{237543427551728514519878451399559982233201007765438338683333118591309874432963 7817553844702655677564147407 m^{29}}{34763788262431863049370610568981422085268389176037989797774862550744027889664000 0000000000000000000}$$
$$+ \frac{233810983964135749215294809455768878148982584096701665882757688891831290721080 71880445264756840098407091165 78181843 m^{30}}{340747421680398535782416455710157551144070148785757600959112655509824066166854 7788800000000000000000000000}$$

$$U_3 = 1.55517578125 m^6 + 7.768359375 m^7 + 18.3186644345238 m^8 + 26.7834348840231 m^9 + 24.7907649711336 m^{10} - 0.503959134267516 m^{11} - 85.5781029229277 m^{12} - 293.736230280926 m^{13}$$
$$- 652.467664977394 m^{14} - 1037.08941990174 m^{15} - 1057.47832990062 m^{16} + 77.8644627850981 m^{17} + 3748.31630708052 m^{18} + 11752.5617964946 m^{19}$$
$$+ 24774.7550603088 m^{20} + 38593.0270172233 m^{21} + 38122.2717254468 m^{22} - 9510.77923334916 m^{23} - 158906.398729382 m^{24} - 474060.50446576 m^{25}$$
$$- 970289.565038914 m^{26} - 1471798.9582918 m^{27} - 1366991.27521186 m^{28} + 683307.083101856 m^{29} + 6861709.55633633 m^{30}$$

$$U_3 = 6.47424628215118 \cdot 10^{-7}$$

Hill's Lunar Equations, Series, Convergence, Motion of the Perigee    209$$U_{-3} = -\frac{6539m^6}{2048} - \frac{247773m^7}{17920} - \frac{57423407m^8}{1881600} - \frac{53420639383m^9}{1185408000} - \frac{33104764593493m^{10}}{590069760000} - \frac{98668689946741951m^{11}}{1115231846400000} - \frac{16687818677941891873m^{12}}{878824507904000000} - \frac{96545449908635817857347m^{13}}{24590862213120000000}$$
$$-\frac{22136695639944219208726199 m^{14}}{3305011881443328000000} - \frac{10674678344117445456740752178 33m^{15}}{11451866169201131520000000} - \frac{34495961305904778132223632141 2633m^{16}}{33067263563568267264000000000}$$
$$-\frac{7441901901208669964697612838986023663 m^{17}}{9166245459821123685580800000000000} - \frac{29733836415034968814866614266015966761 m^{18}}{813082637267972955405759283200000000000} + \frac{29061270260194221871160329282240658922835866 3797m^{19}}{183126536978679208881262134558720000000000000}$$
$$+ \frac{16687516943536823874213942466255676601209252941983491 m^{20}}{41244674291023024820282264255987712000000000000000} + \frac{5555270239766531743715833186022802152988993625733631363 69m^{21}}{74314654137565286121184583736438659481600000000000000}$$
$$+ \frac{99526242473362508322579162619061589040015172739257211044753567009 m^{22}}{8569609204804168482121624510484171305852600320000000000000000}$$
$$+ \frac{133534940726845436095734162233704947194846190936430435481519968804 2091m^{23}}{96504511657600942319292144018689874118032595353600000000000000000}$$
$$+ \frac{131520097188693569250255037865179691143671360401559452878899936038 23092373m^{24}}{21735228638083172233862573136609426898233891288514560000000000000000}$$
$$- \frac{98987313260935087781721139184352410462854503442918075031944112484 32921028870493 m^{25}}{39162534960098259730973584277542865385237825323645534208000000000 00000000}$$
$$- \frac{40879174593862154309660550226553808775001928317015196625389197407 68852264403329464 96123 m^{26}}{45160355514307228405291650656817110304717768747612495063023616000 00000000000000000}$$
$$- \frac{12304846882707348157740512591877456447792309280090391725311538742 4293630860477979641 7099232397 m^{27}}{69164439280826949519556321730435108944984404370030964563872558612 4800000000000000000000}$$
$$- \frac{29770155479048376651184248796609462304788868744502526746614706919 7307577905646753483 724599305412861m^{28}}{13240926711470612249710761677476660300313995603104190394813467461 471436800000000000000000000000}$$
$$- \frac{64756086286017827831911432524538854638771387524513121208159217886 892559074361026518933 9727629726585660441m^{29}}{81115505945674347115198091327623318198959574440886428614746792850 6939840921600000000000000000000000000}$$
$$+ \frac{42088866240442169399484955267364839002672353486650532847660487266 9001240766061469441412 05400623130118295891582579m^{30}}{79507731725426325015897172999036761933601636805001010689045961961 89589487722661150720000000000000000000000000000}$$

$$U_{-3} = -3.19287109375 m^6 - 13.8266183035714 m^7 - 30.5183923256803 m^8 - 45.0651922232683 m^9 - 56.1031370146693 m^{10} - 88.4737019170025 m^{11} - 190.01323293702 m^{12}$$
$$- 392.607010887354 m^{13} - 669.791711316842 m^{14} - 932.134395075811 m^{15} - 1043.20580502798 m^{16} - 811.88114957527 m^{17} - 0.036569267442413 m^{18}$$
$$+ 1586.95024433175 m^{19} + 4045.98102188648 m^{20} + 7475.33619611963 m^{21} + 11613.8601066625 m^{22} + 13837.1707636456 m^{23} + 6051.01052207254 m^{24}$$
$$- 25276.0229545383 m^{25} - 90520.0460189275 m^{26} - 177907.12988717 m^{27} - 224834.380007998 m^{28} - 79831.9452379266 m^{29} + 529368.217745071 m^{30}$$

$$U_{-3} = -1.26705630577254 \cdot 10^{-6}$$

$$U_4 = \frac{60049 m^8}{32768} + \frac{13808173 m^9}{1204224} + \frac{781936613 m^{10}}{22579200} + \frac{8899653977189 m^{11}}{132765696000} + \frac{17285051375740277 m^{12}}{191182602240000} + \frac{153061664705700624257 m^{13}}{2318567008665600000} - \frac{299530726458451736191 m^{14}}{28692266732236800000}$$

$$- \frac{1085195718248415198172694893 m^{15}}{17398273239760089600000} - \frac{347235953729042596502408238328 1627 m^{16}}{20157341119473358209024000000} - \frac{18652181018018507612864343961417975027 49 m^{17}}{5561423013200899200718602240000000}$$

$$- \frac{83911405242076533152098495006077972705878 93 m^{18}}{178937854497389317831210270720000000} - \frac{3078067732990571090732207960226958973669002236751 m^{19}}{123846075088398998746150878620014990852395994880778687 7}$$

$$+ \frac{90275333014535090040311690641932288000000000 0 m^{20}}{54096321225010889941998660519758486688411587395443 1126694597 m^{21}} + \frac{m^{21}}{}$$

$$+ \frac{2323687071794133217637622917123337093120000000 000 m^{22}}{45747428034136842407900120733013657927164606405 3549665477626 4207} + \frac{18317334726069177887985176752818625887928320000000 0000 m^{22}}{}$$

$$+ \frac{58936024481127579854592306202193928794409369600000 0000000 0 m^{23}}{43838015155919797410031394263071050591962308216107 820395717978484384007}$$

$$+ \frac{29733460094826788546961236848231645852494704600 678400000000000000 0 m^{24}}{50872134223917457768392532524229652604869435 1968192230592839070297739788336 8789}$$

$$+ \frac{25078636964768695490929696324763076665121626 40433846734028800000000000 00 m^{25}}{1154435672418168455086576770927582068790528517 32634799272666852702157838549087426 7207}$$

$$+ \frac{8401899815455168794892653271432836363579252 23153923281874059264000000 0000000 00 m^{26}}{306178004642903030882684909111767498634386143 07083551471667896420543350193499780526298376 7}$$

$$- \frac{11489072878896477378966022557854939174921903 72336542967757658914816000000 00000000 m^{27}}{132457525672887791476164398894603908248061 967676767941070388222831026563286069382147520598190497 7}$$

$$- \frac{98536814802811403201163862156976819969581601 972896796478953692037256314880000000 00000000 0 m^{28}}{2356728112149564889220856207838548973195949 8541094184178117668894593625758834670091120986457874976680072 19}$$

$$- \frac{6898838307426274210042445248423986701166322782 60802883252168904986601892243046400000000 0000000000 m^{29}}{4443937810245413350815952671832483241701372060007 44355018807363314356662634588810937254383275561033731314664 1789}$$

$$- \frac{70262422589267535577026936039269942543707767907 9027970347458484171630996713357950247239680000000 000000000000 m^{30}}{1099226491489817256465928665727574617518931746519414 44820521581547453628951494895541624705239649308847302364036 06415467}$$

$$\phantom{U_4 =} \frac{}{1303935684499156419951146175824131892203654590 40002813128946128853204002115898783807045632000000 00000000000000000}$$

$$U_4 = 1.83255004882813 m^8 + 11.4664489330889 m^9 + 34.6308378064768 m^{10} + 67.0327821517164 m^{11} + 90.4112151064959 m^{12} + 66.0156312643264 m^{13} - 104.394235998761 m^{14}$$
$$- 623.73759929717 m^{15} - 1722.62775963834 m^{16} - 3353.85043967069 m^{17} - 4689.39168393126 m^{18} - 3409.64428510607 m^{19} + 5329.72260299992 m^{20}$$
$$+ 29532.8561900554 m^{21} + 77622.1817418072 m^{22} + 147436.642140237 m^{23} + 202850.475069217 m^{24} + 137401.742198187 m^{25} - 266494.96253549 m^{26}$$
$$- 1344244.03648481 m^{27} - 3416123.13135772 m^{28} - 6324771.68660025 m^{29} - 8430066.79360901 m^{30}$$

$$U_4 = 5.52309198157658 \cdot 10^{-9}$$

$$U_{-4} = \frac{387251 m^8}{98304} + \frac{19858463 m^9}{860160} + \frac{13935272173 m^{10}}{203212800} + \frac{69778908700217 m^{11}}{512096256000} + \frac{1096891228338029527 m^{12}}{5161930260480000} + \frac{2888540112211495990933 m^{13}}{8943044176281600000} + \frac{22618999090044066116 73223 m^{14}}{38734560088519680 00000}$$
$$+ \frac{15647683221745409921502789929 m^{15}}{13421525070672069120000000} + \frac{820864075314083168682740779944812369 m^{16}}{380973747158046470155360 0000000} + \frac{737501859084793874319344321634299 0547167 m^{17}}{2145120305091775405991460864000000000}$$
$$+ \frac{22889724487739892780222966606826514 6428136489 m^{18}}{4831347207142951158144267730944000000 0000} + \frac{99726221237141587709687383331399462851535727543 593 m^{19}}{1741024279566033879348868319522979840000000 00000}$$
$$+ \frac{38619471815088603791061260841740439498476578386 297692591 m^{20}}{6273955093844159687621581876233010151424000000000 00000} + \frac{426040897035978933509561244875501537232432 93723970111 7788261 m^{21}}{70652576800552543282228538903728985567723520 00000000 0000000}$$
$$+ \frac{8967014084634297957051053127442568977923255991927 87554064 1580217 m^{22}}{15912726609904446560739922674592360774490529792 00000000 000000} + \frac{4999996409347317017688787612823336 7293865903350 6801653361161706683 4363 m^{23}}{11468620322290332725256477070032206257390814631690240 000000000 00}$$
$$− \frac{617124 7161412463163775392083506501370937967391936209100015979 9818411 48983048493 m^{24}}{338561 599024377389127550900384301534979141956458 569309093 88800000 000000000 0} − \frac{686481 6391791 404589421976249834735892468787980997157 56045631852 0614343 76497419 6705993 m^{25}}{324073 278596127939231573769040980831166628003593 704087228 51430400000000000000000}$$
$$− \frac{1911083935639606237505579106384806578210560359948 9417817912 653387616543789 1407019 048118103 m^{26}}{310204967730204889232082609062083357722891400 5308666012 9456790700 032000000000 00000} − \frac{47114774 18913012902587957339827261239945745940 568827 5524545678 7032222 52128 237577997547254889 4831329 m^{27}}{380057241 3822725552044891826054820198826718932 40305 0070453566928656007 168000000000 00000000 0}$$
$$− \frac{38055724 66696626805096862738584217121727296513 103550382511665109658990270814706521 15137914 7203220 56386659 3 m^{28}}{18626863430050940367114602170744764093149 0715130416778478085604346382510905622 528000000 000000 00000} − \frac{207421875103350290 827291961157307556989 4026213389501364892 26191346 8423267630856286407896686 4887 8807951866 30903149 m^{29}}{67753050353936552163561 6883235817303100039190491 25712076 35383083 58461116 452309 16698112 0000000000000000000}$$
$$− \frac{108750527 9836808571036188 15323822092673678839 09711602 185209063078604492216 8002643028717157 1385149679361020 63718895143807 m^{30}}{24644384437034 0563370666272307609 27626490717585 60531 6813708183532555 63999 0487013953 162444800 000000 000000 0000000}$$

$U_{-4} = 3.93932088216146 m^8 + 23.0869408017113 m^9 + 68.5747756686587 m^{10} + 136.261313928093 m^{11} + 212.496328502515 m^{12} + 322.992937893829 m^{13} + 583.948779548628 m^{14}$
$+ 1165.86476867206 m^{15} + 2154.64735152354 m^{16} + 3438.04427814244 m^{17} + 4737.75191604908 m^{18} + 5728.02013203396 m^{19} + 6155.52251130726 m^{20}$
$+ 6030.08292590182 m^{21} + 5635.12106030466 m^{22} + 4359.71918926408 m^{23} - 1822.7841489395 m^{24} - 21182.9140049112 m^{25} - 61607.1351024184 m^{26}$
$- 123963.226139238 m^{27} - 204305.597718457 m^{28} - 306143.965503833 m^{29} - 441279.141142829 m^{30}$

$U_{-4} = 1.156764614328 \cdot 10^{-8}$

$$U_5 = \frac{3395915 m^{10}}{1572864} + \frac{5649213295 m^{11}}{346816512} + \frac{114518728333 m^{12}}{1907490816} + \frac{1198624098229403963 m^{13}}{8327914153574400} + \frac{7279878452179894438783 m^{14}}{29314257820581888000} + \frac{485793068006605074565 82720641 m^{15}}{16637805368454200426496 0000}$$
$$+ \frac{4563888685281556437816 8917151 m^{16}}{106214561057542440222720 0000} - \frac{703811224920353227885811 1 31494586542067 m^{17}}{67517945348835732990275 1449088000000} - \frac{2375202393133143206640388 67145773740362152227 m^{18}}{617924235832544628326998126 20533760000}$$
$$- \frac{15861731295932035079500012983566826793962 01675863923 m^{19}}{17535670237938492853283467274799242280960 0000000} - \frac{24603456838369167312419083105358510305455 1343949582947 m^{20}}{15672505275157527987622098876851822788608 000000000}$$
$$- \frac{26447614729289899353694145770665707943270 4418624402055 1420403 m^{21}}{14232314882408532299761138869181053607459 00277760000000} + \frac{25387718709755392609780479111513040055259 99446220128068165329637 m^{22}}{65127072901904438037069714653725013077323 9671029760000000}$$
$$+ \frac{35078106470007499096655164666266987435150 40090971240837459515597704 74069907 m^{23}}{62838761555712944484075542945268026564533 780446244912496640000000000} + \frac{76697309434750238150108649063429030301532 216352201164038764486789225 81116580959 m^{24}}{38190257335484542010196911224986643144595 405066205345569832960000000000}$$
$$+ \frac{33965995296232078201256634143145292800269 285938771509839796627999830 2405053046346024607 m^{25}}{73696801269438447187344863064988779904805 172908517295854713504268288000000000} + \frac{90113528437793076455130413547248394402366 384295610311437986697933782293498053041976 06470342599 m^{26}}{11466043128704311367195863175103214332709 208021798754958259745848077320192000000000000}$$
$$+ \frac{23867914103531829548231818499554652591423 83508441456029144746543609630 8868725036279697064 03705506552391 m^{27}}{26275031340483379054846464941616616923535 04704233207534490573453096364032225443840000 0000000000} + \frac{16604114133159480707677105171902704420838 21671459274864144437183853982134108910338463 1927020815 4932863875397 m^{28}}{12136174225855867951643033691883299190811 602878382762281058509722506795828446102552 5760000000000000}$$
$$- \frac{31564835498541813840011985000373280131418 877146999455267103799234621740590686992352 9474113295103049599 1152591606453741 m^{29}}{11124268574906834910221943450992824820076 60952563236821911259328093533559561325576 2445126860800000000000000000} - \frac{18446837260295199159009878752353877716256 30068848067137146341788022884491299194012 19267911605392756243165072550591657484520731 m^{30}}{18791089049833025743295683512974902038727 38200641621378074547774556570410359142645 1790913699011624960000000000000000000000}$$

$U_5 = 2.15906461079915 m^{10} + 16.2887668249198 m^{11} + 60.0363196364663 m^{12} + 143.928488709859 m^{13} + 248.33916985845 m^{14} + 291.981458640984 m^{15} + 42.9685783177039 m^{16} - 1042.40616518182 m^{17} - 3843.84080668559 m^{18} - 9045.40920347322 m^{19} - 15698.4836861839 m^{20} - 18582.7920108624 m^{21} - 3898.18205832711 m^{22} + 55822.4026087898 m^{23} + 200829.517227387 m^{24} + 460888.325017677 m^{25} + 785916.531328939 m^{26} + 908387.654965695 m^{27} + 136815.060695032 m^{28} - 2837475.136994 m^{29} - 9816800.51186768 m^{30}$

$U_5 = 4.72072488602269 \cdot 10^{-11}$

$$U_{-5} = -\frac{6953929 m^{10}}{1572864} - \frac{40794344441 m^{11}}{1271660544} - \frac{37069948123261 m^{12}}{314735984640} - \frac{1771135402332851497 m^{13}}{6107137045954560} - \frac{533447379472214598633157 m^{14}}{96737050807920230400} - \frac{57428712450892176055978504 2667 m^{15}}{61005286350998734897152 0000}$$
$$- \frac{33629692958565841851693522850147 m^{16}}{1962845088343384295315865 6000} - \frac{86229273937225534219145839029766 89813023 m^{17}}{24756579961239768763100886466 56000} - \frac{14463576519125610496471017806 2143157 66223288591 m^{18}}{20391499782473972734790938164 77614080000000}$$
$$- \frac{8452318556804115251548958330616427383505806871862911 m^{19}}{6429745753910780712870604667 4263888363520000 0000} - \frac{15910954706404075693782210270391866 87871264988449576747 m^{20}}{73884667725742631941647037562301450289152000 0000}$$
$$- \frac{8167233098303847180937284190 6868556187455413569963596390109931 m^{21}}{26092577284415642549562087926831931613 6748384 2560000 0000} - \frac{21854031528843411738820481 47358531738458637363744519348 0066480796497 m^{22}}{537298351440686911380582514589 323135788792272859955 2000 00000000}$$
$$- \frac{2160461581612474766361208970268960560726 9297493607570957461224699869 3658861 m^{23}}{46081758474189492621655398 159863214806581056605 79602 49753600000 0000}$$
$$- \frac{572970998543723312933095818 71100712705531361452314129819385 09519326685407 1685201 m^{24}}{12602784920709898863364980704245592 23771648367184776403804487 680000000000}$$
$$- \frac{7762389616832288088307855313116292 78240346547890837463264602 776556937735729975171876469 m^{25}}{270221604654607639686931164571625526 317618967331230084800616182317050000 00000000}$$
$$+ \frac{121143846416033040930801996541 6662067793140452773384883123 71353155478543734858489089 3238914023 m^{26}}{756758846494484550234926969556 8121459588077294387178272451432 25973103132672 00000000 00000}$$
$$+ \frac{983040760129946861814371452837 95886103940195914614214441475207813 0042153990314048360467442349518574149 m^{27}}{963417815817723898677703714 52594262052961839155217609597 98769328020001451493294 0 80000 0000000000 0}$$
$$+ \frac{87507038395707824613615996157602693 2240548633724112389600788657621781 8194615021349450219601290022707260620821 m^{28}}{4004937494532436424042201118321488 7329678289498663115527493 0820842724262338721384235008 0000 000000000 0}$$
$$+ \frac{1139514933818547232529285976003232414353 9218468425645827416556631200422 40609905924823175838973163897557411663 66778 17033 m^{29}}{40788984774658394670813792 65364034910069475682606520168034128 42030096230517248604462298798489 60000000 00000 00000}$$
$$+ \frac{35911385141804637491461021920884382942 121676849962562811812625685551073 48991026464478923910 93970199790 78800 2112445566766205351 m^{30}}{4340741570511428946701302891 49720237094602524348214 58 33522053592256 776479296 195103 6370 1064471685365760 000 000 000 00 00 00}$$

$U_{-5} = -4.42118899027507 m^{10} - 32.0795865166042 m^{11} - 117.78109251048 m^{12} - 290.010751192504 m^{13} - 551.440606279615 m^{14} - 941.372721709256 m^{15} - 1713.31365670578 m^{16}$
$\qquad - 3483.08506555553 m^{17} - 7092.94395871594 m^{18} - 13145.6497353152 m^{19} - 21534.853165294 m^{20} - 31300.9826866812 m^{21} - 40673.9225427456 m^{22}$
$\qquad - 46883.2278356425 m^{23} - 45463.8401074489 m^{24} - 28726.0140681721 m^{25} + 16008.2497848826 m^{26} + 102036.805214731 m^{27} + 218497.887957485 m^{28}$
$\qquad + 279368.300072649 m^{29} + 82730.9909112455 m^{30}$

$$U_{-5} = -9.50486985713594 \cdot 10^{-11}$$

$$U_6 = \frac{106666653 m^{12}}{41943040} + \frac{43713663183 m^{13}}{1942814720} + \frac{763308798778727 m^{14}}{7793462476800} + \frac{8814061288843242767969 m^{15}}{31536577134526464000} + \frac{16509739570596328765958 15979 m^{16}}{283375067500000994918400 00} + \frac{182724297267807271755861 4097091469 m^{17}}{204765938228414781424926720 0000}$$

$$+ \frac{53310807100058081236030197370558847 89 m^{18}}{718728443181735882801492787200 0000} - \frac{953651487939209624074737620833194346 9057861 m^{19}}{8309614832267798495803598933852160000 000}$$

$$- \frac{43854181057026465234002690272151451749 6673332190059 m^{20}}{597335000294604241311142548640175751 1680000000000} - \frac{49856098557294904515767484302775544745 8824462200808120 62809 m^{21}}{23480802807030677664844856452984801 4501 1572736000000000}$$

$$- \frac{38590569501569203559215020994736347534892844056624210 0934244601 m^{22}}{87568718968470033516330786534350193908025 34072320000000000}$$

$$- \frac{185929836384311031664785169822246586649005 35345414948480068944939564 7749 m^{23}}{275380993398856802752199703775290883750633862 8264211251200000000000}$$

$$- \frac{3895606952656581094529933686476415098221700 192095874667816976 2034453598424789 m^{24}}{65727935504439141680895025297086428133601290379410194143641 6000000000000000}$$

$$+ \frac{113083094415477746954933038301244905483893243 0396334793787357756165476984517 69131547 m^{25}}{191760258212676187340424107148984420781043963756172333924030087168 0000000000000}$$

$$+ \frac{14931421078196329455463907120467361393522147634492420801905709450008349 3501 7783362251901111 m^{26}}{347846972069531798197534436796828971795348 75684685641622350500915996264000000000 00000}$$

$$+ \frac{25533681892425982835439290864923318266197329 80882958218408913200027188918173 25434749725142825267964253 m^{27}}{207838880265208000279772396557236174981111385 211272898251744509216059484462120960 00000000000000}$$

$$+ \frac{39292030564703231432455047262817061278920749552826584258935119980878 67744976729907331751403603405 0044134929569 m^{28}}{15668075967040782373049274790156325812286198741 15001669385643567962200515 70318322434048000000000000000}$$

$$+ \frac{170224 4760827548746103025618463581903995334194961565018578060801147225961440 087345060592713908545963733327392077957072619 m^{29}}{450531309798162943001777458201272970631427768875 78354383101244055615118369173120381390709325824 00000000000000000}$$

$$+ \frac{2067575952 92009295402776770015161267655347388704736430421261529438068439289366079624977528318530410 6672584495248978 32616597 m^{30}}{63847607732183909171175644710816150624246075509363555643441716778965660185853302212492611049580544000000000000000 00}$$

$U_6 = 2.54313118457794 m^{12} + 22.5001708773341 m^{13} + 97.9421920681579 m^{14} + 279.486935162458 m^{15} + 582.610873858745 m^{16} + 892.356897092815 m^{17} + 741.737823315531 m^{18}$
$- 1147.64824506185 m^{19} - 7341.63928706633 m^{20} - 21232.7061246675 m^{21} - 44068.8980678867 m^{22} - 67517.3090522677 m^{23} - 59268.6644234167 m^{24}$
$+ 58971.0795497888 m^{25} + 429252.581655696 m^{26} + 1228532.49882045 m^{27} + 2507776.36305553 m^{28} + 3778311.45417638 m^{29} + 3238298.23286845 m^{30}$

$$U_6 = 4.03903869309804 \cdot 10^{-13}$$

$$U_{-6} = \frac{229075037 m^{12}}{41943040} + \frac{7914003284093 m^{13}}{165315870720} + \frac{2383140315152479 m^{14}}{11257223577600} + \frac{420609986241536549204947 m^{15}}{670869004498108416000} + \frac{27583766751688186977796017627 m^{16}}{1949140675925932769280000}$$

$$+ \frac{59084029640409558556778105315689519 m^{17}}{2177964979338593584246947840000} + \frac{1478940769172056117511119647965904151 m^{18}}{29661808766230369766410813440000000} + \frac{4246692977615319533578133623111126383111127397 m^{19}}{44192042517060564727682776148213760000000}$$

$$+ \frac{23644928665721118355963607451674157779412091 93648367 m^{20}}{12325960323539452598483893860829023436800 0000000} + \frac{92348736576966364754488947571658145547476464 9883130346879033 m^{21}}{24975035712932629879880438227265652451486728 192000000000}$$

$$+ \frac{23286853146001074771827718927960770117702730 345490323526801 m^{22}}{35410024809250407812158432324171356432236544 0000000000}$$

$$+ \frac{31448947510485674340133393925943749811520611 7068712863471825378366658 7749 m^{23}}{29290523843332950838188513947008212180749238 136992065126400000000000}$$

$$+ \frac{31801553117122983753178297492743437288835769 4549932686230148412722947333 8663 m^{24}}{19375581894732626799356810178279445935075210 315925907570688000000000000}$$

$$+ \frac{98684331251374363879612767987866621714858729 688612579504524038928074509607920 91291721 m^{25}}{40792636747060207125144764611693049511603897 744494841943839127633920000000000000}$$

$$+ \frac{25263409709313133596799879001783563211962394 7869517316718822014436375947046701753499894 0219 m^{26}}{71777946617522434548697582196171057672056092 682684657315961351096500224000000000000000}$$

$$+ \frac{11166369100784584677482216595170763011822470 1178947323529201167472757271534824780603805712 132432106883 m^{27}}{22106499082753941847939427633815120429809120 0633808446322310068686617236074607411200000 00000000000}$$

$$+ \frac{11410516644690105909176872871321401524016950 021013608849850781059924368134555753300012948547751 23562506913049 m^{28}}{16596554542865421328489231814758182156717973 4813663139794182985347107165737448297096806 4000000000000000}$$

$$+ \frac{38881998212936821551760271433278156510545520 8749084833495334476388208347993014925356443 900150766120013781569049428836069 m^{29}}{47920148405804603974617841904990341489791354 1678307951165712322279062590175686440566102 71920128000000000000000000}$$

$$+ \frac{86596100175108268141067473534815395086832706 3868254961865454061723304709538158573820845124167885267181855174304795843580979 m^{30}}{13174903182831600305163228273660475525638079 081939247973594829076391354892723957029781593562483261440000000000000000000}$$

$$U_{-6} = 5.46157448291779 m^{12} + 47.8720116200892 m^{13} + 211.698763795944 m^{14} + 626.962914401156 m^{15} + 1415.17577937696 m^{16} + 2712.8089845757 m^{17} + 4986.01006037033 m^{18}$$
$$+ 9609.63272058734 m^{19} + 19183.0316219381 m^{20} + 36976.4182275608 m^{21} + 65763.4477000357 m^{22} + 107369.016951344 m^{23} + 164132.118921127 m^{24}$$
$$+ 241917.020131056 m^{25} + 351966.180419347 m^{26} + 505117.02730424 m^{27} + 687523.221474622 m^{28} + 811391.439852616 m^{29} + 657280.732718802 m^{30}$$

$$U_{-6} = 8.67242834781383 \cdot 10^{-13}$$

$$U_7 = \frac{6029347343 m^{14}}{2013265920} + \frac{39463562703482629 m^{15}}{1296832156139520} + \frac{10403396529426494453 m^{16}}{68083688197324800} + \frac{26594004326627738383338733841 m^{17}}{526267982500018477056000} + \frac{43524578087347632548900802675473 m^{18}}{353652084240001241658163200000}$$

$$+ \frac{10583851080124858818483932845352151 2804833 m^{19}}{4647174903022844492043114963271680 0000} + \frac{919025394864139300477584801466660507 2453961 m^{20}}{31904643084214528531911385420922880000 00}$$

$$+ \frac{4195294084694006121229239399595469601 829267827289489 m^{21}}{10900339752518879510943246696885538586 099712000000} - \frac{10310474021432433713925457113461568717 16601745944388645 1043 m^{22}}{88946772380554056809296893046585994862 5736499200000000}$$

$$- \frac{489620091148378526924716560989662122745 4535540324720989238889 82115689 m^{23}}{1110367124627822120149536277922856104850 51286722717941760000 00000} $$

$$- \frac{1432115123035403109530630833697655182218 594344772436086064306693628240 79607 m^{24}}{1318033536111340552170503300301378267860 1800362203426481766400 00000000}$$

$$- \frac{6603759850265833994250021847121789574824 4654507547498141358386963522680759188 47113211 m^{25}}{3290734601124092395540249336986130983450 4346756284169510932026123878400 000000000}$$

$$- \frac{4635996305901891092456823188922623497471 19781992036014704577573962260299348870645 4085626081 m^{26}}{1785684223953977497515960900022215411685 954387238300417434121546558613749760 00000000}$$

$$- \frac{8493763838199676306835514071727858066386 8494573127880366012717110830518191450570 9040275058556846 63317 m^{27}}{8916636339593691505134928135745195258653 9425184893684659531451344385908200448901 120000 00000000}$$

$$+ \frac{1071571051363505681246611319147982860673 3459487624197385770856629285014137320311 372346336606533091920006 659 m^{28}}{1512038607286600236983255438618991485986 2423025728346576140045861724240352499112 2407424 000000000000}$$

$$+ \frac{1066905250166804562405891193448189934922 3108176724650251871271973502622950001845 5616122120335002278396391174 393967311 m^{29}}{3775107102292407912715742672150656398574 0972699995106196688247187816358403278046 146278671672934 4000000000000000}$$

$$+ \frac{1133526623223094871973403965827012224023 3818848410855743089925479004137314340863 11368230563171983122386130549 31353473203269 m^{30}}{1638819294390361783804856482892665750497 7985140740275541256728363197334979163839 168653326717280898252800 00000000000000000}$$

$U_7 = 2.99480922172467 m^{14} + 30.4307404135936 m^{15} + 152.803069353039 m^{16} + 505.331983152284 m^{17} + 1230.71742050897 m^{18} + 2277.48068471458 m^{19} + 2880.53808481201 m^{20}$

$\quad + 384.877368957655 m^{21} - 11591.7348606194 m^{22} - 44095.3339025137 m^{23} - 108655.438863919 m^{24} - 200677.376048802 m^{25} - 259620.16372842 m^{26}$

$\quad - 95257.4885271896 m^{27} + 708692.917098508 m^{28} + 2826158.89101248 m^{29} + 6916727.34817892 m^{30}$

$$U_7 = 3.45788508476525 \cdot 10^{-15}$$

$$U_{-7} = -\frac{37073479519 m^{14}}{6039797760} - \frac{31162415271001211 m^{15}}{498781598515200} - \frac{327506777260098218743 m^{16}}{1021255322959872000} - \frac{74442609684177968972 9741189 m^{17}}{67470254166669035520000} - \frac{15301567991947999829 51113165500133 m^{18}}{530478126360001862487244800000}$$

$$- \frac{37480283340757379152783818760762 04021353 m^{19}}{59579165423369801180039935426560 00} - \frac{78507842549318902203823956401615 29251026623589 m^{20}}{62214054014218330637227201570799 6160000000}$$

$$- \frac{253915566284084280245932911795445 53655359482920045067 m^{21}}{998199611036527427742055558322851 5188736000000000} - \frac{494347863977071117145649352872690 0654203704783168021567083911 m^{22}}{933941109995817596497617376989152 94605702332416000000000}$$

$$- \frac{1088364249369521556284888008227967 5444496876378272604241951991299543057 m^{23}}{99648331697368651808291717249487086 332738334238336614400000000} - \frac{6011979047308074214072704558753498 937933203286239165880480798060402 47016007 m^{24}}{28243575773814440403653642149315248 597003857919007342460928000000000}$$

$$- \frac{16169600723442723771387616271787067 12706273269010526910640268689167123 1564429128151781 m^{25}}{42188905142616569173592940217770910 04423634199523611475760516169728000 0000000000} - \frac{17098499984052284584693013177112260 02427746707897676275593119696431993 2912988805568833805 06699 m^{26}}{26785263359306624627394135033323117 52893158085745062615118231983792062 464000000000000}$$

$$- \frac{11376861947098582859649658828864899 58195183713921637409247172509385374 80685085342047573674079397542787 m^{27}}{11431585050761142955301189917622045 20340249040831970316147839119799819 33359031910400000000000000} - \frac{33323285626939657107679450380484758 23000797911744766063974392429594088 32155526481371880559177039264732481 579 m^{28}}{22680579109299003554748831579284872 28979363453859251986421006879258636 05287486683611136000000000000000}$$

$$- \frac{99738159462384101342775101928553924 26923255881686055363652176054795488 51201184005149689595589809261104694 84557173431 m^{29}}{48398809003748819397915727198802102 38129452179480905380369547676687638 97856159762343419445248000000000000 00000} - \frac{66814318874498964036264776413777241 81072002829167136148222578479357074 56709987908030786719957111597009342 8028129764 3873511177 m^{30}}{24582289415855426757072847243389986 25746697771110413311885092544796002 46874575875297999007592134737920000 00000000000000}$$

$$U_{-7} = -6.13819882588254 m^{14} - 62.4770748635619 m^{15} - 320.690399253828 m^{16} - 1103.33969544991 m^{17} - 2884.48613271639 m^{18} - 6290.83725400018 m^{19} - 12618.9883931011 m^{20}$$
$$- 25437.3537593768 m^{21} - 52931.3742254354 m^{22} - 109220.518881829 m^{23} - 212861.823710083 m^{24} - 383266.659060778 m^{25} - 638354.740483818 m^{26}$$
$$- 995212.990725296 m^{27} - 1469243.15584504 m^{28} - 2060756.48379403 m^{29} - 2717986.0160382 m^{30}$$

$$U_{-7} = -7.14698254913953 \cdot 10^{-15}$$

$$U_8 = \frac{477035249183 m^{16}}{135291469824} + \frac{44104188882644296267 m^{17}}{1089339011157196800} + \frac{21778143279350859529577 m^{18}}{9452285378061926400{0}} + \frac{1106506922101207928136703538743871 m^{19}}{1277568154317004485490114560000} + \frac{6837738103188768035855485003039355841 m^{20}}{28379047267895059637687078092800000}$$
$$+ \frac{11804192900837389566659505705916575461781961962788 1 m^{21}}{2277449137280485690193430425501737943040000} + \frac{19147359752612310697046015310897028624790883008387 m^{22}}{585059155747288289255402027682601643962134603122507 m^{22}}$$
$$+ \frac{7040075682761376370838168621956251254784000000 0}{336769448944477620249441021125193372248324245905901725874693499 m^{23}}$$
$$- \frac{48211085917456235064453087061305349387356264097382400000 0}{250627259529631083758157768768131342654217950325504424402 06029252676939 m^{24}}$$
$$- \frac{1953374353447884370854653462833356197769440235185334386680000000 0}{21080977784273480528946170030826110847170616978122515134741393454381996990260 4609 m^{25}}$$
$$- \frac{261267329499024092088053233752570737284410973697781043469079281664000000 00}{23538485754136283210435983925914205853261562820480126782164122614352147151296 239579 m^{26}}$$
$$- \frac{97894839582884808368314832637356464649709394108612825517310083072000000000 }{1442039990089142655484532553942192055816926605718651073601386414713557180227960 8684892358090023 m^{27}}$$
$$- \frac{27653705770445280065847742268955197033585528910742831309312880452137395630899200 0000}{1184608077022194767115357250333856851608097296462325042704228816456248025578499628296895235328940724609 m^{28}}$$
$$- \frac{1400560643932279832326951094050412491041786131425045731623983868531131394080773242880000000 00}{70749751249271564330460698046043113842432764135996185897970253030393955959105891793390054375243210409950041236099 m^{29}}$$
$$+ \frac{8617065118830852993285848344816561664752180022737249209558802339216511934572438497681106534400000000000 0}{32707481640820366833461996581603817236824812945034617452324725698428280573887299770820772758533806451697923032176853 m^{30}}$$
$$\frac{}{538597641075193327175293234946509202226090886685583137794094228421005132358629509518619060469760000000000000 0}$$

$U_8 = 3.52598171786863 m^{16} + 40.4871104687537 m^{17} + 230.400822745961 m^{18} + 866.104026123564 m^{19} + 2409.43187367824 m^{20} + 5183.07641106437 m^{21} + 8310.41003124283 m^{22}$
$+ 6985.31141823006 m^{23} - 12830.477634113 m^{24} - 80687.3856930215 m^{25} - 240446.645139113 m^{26} - 521463.561542017 m^{27} - 845809.913447399 m^{28}$
$- 821042.318627282 m^{29} + 607271.16397181 m^{30}$

$$U_8 = 2.96150342821939 \cdot 10^{-17}$$

$$U_{-8} = \frac{5117470980763 m^{16}}{676457349120} + \frac{545070155618532566489 m^{17}}{6172921063224115200} + \frac{75031933056785587446566 77 m^{18}}{1446199662843474739200 0} + \frac{1477606604587972527740967 22206729 17 m^{19}}{72395528744630254177773158400 00}$$
$$+ \frac{263748300062630348627925462750288372703 61 m^{20}}{434199423198794412456612294819840000 0} + \frac{1914735975261231069704601531089702862479088300838 7 m^{21}}{12905545111256085577762772411176515010560000 00}$$
$$+ \frac{52333577096485724326936160147614379095065071865608 27 m^{22}}{16320175446401372496033936350898582454272000000 0} + \frac{18037666918926812781130228233750273700729088220700046854678946393 m^{23}}{27319615353225199869856749334739697986168549655183360000000 0}$$
$$+ \frac{10418951291847946701635400342270539188525670225083181804661867059695527 79 m^{24}}{766323784814170022412210204650008969894165015341938874777600000 00}$$
$$+ \frac{414952873095800629521872828167001025080848446042692497479427062404246587976939323 7 m^{25}}{148051486716113652183230165793123417794499551762057924632478259609600000000 0}$$
$$+ \frac{9257347861037298200878066655190053720816813095589929174678424382166496894076149842777 1 m^{26}}{16475701501799513248387386332867092969454609102847953853456328698101760000 00}$$
$$+ \frac{167590603560892891330765386402653302257928737756638615871293337207663897479920855263431057056945 9 m^{27}}{1567043326991899203713720619074611652365133049420937741943965589544524190842880000000 0}$$
$$+ \frac{13743147719093889279286660660203056132453330301831666215293885269547214087096080646538788930002314425607 7 m^{28}}{71428592840546271448674505796571037043131092702677332312823177729508770109811943538688000000 0}$$
$$+ \frac{16153682939146400196717620082099818842401978598661301104396857337626293784735662312616342947762358951303557432761 27 m^{29}}{4883003567337483362861980728729384943359568679551107885416654658893567628024381815352627036160000000000 0}$$
$$+ \frac{1973227988921123154484734327244204214365433101350377317854445227964247455966483438092894521285419647484805063170426600 3 m^{30}}{35709023603285317591721941476953560107589825787254162035748447344312640275377136481084443709145088000000000000 0}$$

$$U_{-8} = 7.56510515765745m^{16} + 88.3001985665831m^{17} + 518.821397795516m^{18} + 2041.01914884843m^{19} + 6074.35860046906m^{20} + 14836.53699828m^{21} + 32066.7980980715m^{22}$$
$$+ 66024.6005872019m^{23} + 135960.171122373m^{24} + 280276.059565323m^{25} + 561878.828651162m^{26} + 1069470.13317494m^{27} + 1924040.10390817m^{28}$$
$$+ 3308144.81627635m^{29} + 5525852.54316386m^{30}$$

$$U_{-8} = 6.49716657867964 \cdot 10^{-17}$$

$$U_9 = \frac{4991522108059m^{18}}{1202590842880} + \frac{11335921726155109949m^{19}}{21321073386628055040} + \frac{31097947079481477962 6871523m^{20}}{9195183215778642001 92000} + \frac{44988020517743711136 0977669172150055787m^{21}}{315940522599677063021 73420871680000}$$
$$+ \frac{571285212124679773581 3977623753309568763 0667m^{22}}{12835227339940335084347 059847194214400000} + \frac{31012717971496780767715 2972678035909471822914 24508106341m^{23}}{285359459860440812396823 2393880491007575851008 000000}$$
$$+ \frac{7395482888009330040780237244095067664796 394233819077997 3473m^{24}}{35865403712559503506094 72884248695122871083 394048000000} + \frac{9589663016220821891 39572190083021872627084276200018106691378 09038066331m^{25}}{362444698627654312439 456033300326741820440052 103776179322 8800000000}$$
$$+ \frac{8003952526948469333779 070464308554222610404528014202159274923552 768448253313m^{26}}{9718157215363569548576 4467296750008630830950290481299 4094851686400 0000000}$$
$$- \frac{28367562055316293184701 06004705285687051609266561790888958897 970111245381 45842305863581347m^{27}}{22588000291718087353855 277596753812329649306645879 53392678675206880 04792320000000000}$$
$$- \frac{174458697335044492731 6864404786979914038210158344176708869588 645710560234546044 29854769695569043m^{28}}{362757637684919553381077294384467037561085 4524061689755230713541536918 4963461120 0000000000}$$
$$- \frac{117605872019969980075565745671302988547869 1011024960143939140928344825584144363436 7578036282896823637 06451m^{29}}{963612391498722984217158633632990489493 0012750013043353795132848 932641254081 97632212336640000000000}$$
$$- \frac{18297642345058221310334212119289281206 279014548048197143952601 1541331531621 5921201827395591157716258 618211116451029m^{30}}{7799588548603294688673882736 7096591828721542200590743925601 3792440557663 1642481696567249705369600000 00000000}$$

$$U_9 = 4.1506403758282m^{18} + 53.1676877640978m^{19} + 338.198232158315m^{20} + 1423.93954873422m^{21} + 4450.91619333433m^{22} + 10867.9480913876m^{23} + 20620.1021666446m^{24}$$
$$+ 26458.2791596366m^{25} + 823.608051359255m^{26} - 125586.867757024m^{27} - 480923.567725331m^{28} - 1220468.65583635m^{29} - 2345975.33331869m^{30}$$

$$U_9 = 2.53705034790077 \cdot 10^{-19}$$

$$U_{-9} = -\frac{10251752972473m^{18}}{1202590842880} - \frac{9313313770936307770 9231m^{19}}{83403022365339156 4800} - \frac{72888279725446367799657679073m^{20}}{99001472623216712220 672000} - \frac{11480730415568546914772547069106948 56403m^{21}}{353109995846697893977 134999779776 0000}$$
$$- \frac{8734887340640161277962281200215780951 7279842283m^{22}}{80476875421425900978856 065241907724 28800000} - \frac{3144393405441519057140244222882427473712 08065919274 2719m^{23}}{10631038700683089092934148007 31241008615915520 00000}$$
$$- \frac{239546173935113110146114917656057078936760 97957700844242049 2409m^{24}}{340721335269315283307 8999240036260366728 12292243456 0000000}$$
$$- \frac{630034444625314807251585746175691109 1085816846539603738323300771 05305203m^{25}}{4050852514073783491970390960415416 5262284476411598514159616000 00000}$$
$$- \frac{76838796017079498172203993829677818675 18924021487740482250597218322 5710541521m^{26}}{1838031076223914617828122019 88927160298430300161378160444820 6232225017 16820234702304203513 3}$$
$$- \frac{22795677418754052027525467375092612 837751611792486998614841753600000 00}{183800876521781571018527 2400185205164978 6235938363372 2833792743658324947338011 4985794560818799291m^{28}/ 319008765217815710185 272400 18520516497 8962359383633723283379274 365832494733801 14985794 560818799291m^{28}}$$
$$- \frac{2067718534804041454272140577991462114098187 078715163160481506718676043542917283840000 0000}{590215743485718366028 00843160595782632 9730144290117720072213 75073565690586832907490 9637284101124 0315991027m^{29}}$$
$$- \frac{188471247160779642501297203342923139856719 36702231393618452245131000607158719159571 2118784000000000}{80949818243276308142279928 28494239990273630 97598634359880599498428 7183945796239771 510901129405070 136441 6157 10179169m^{30}}{1347201658395114537134579745431668 40431428123561920219405331147 323973268727382 46838497979449456384000 0000000000}$$

$$U_{-9} = -8.52472229700486m^{18} - 111.66638218625m^{19} - 736.234298280058m^{20} - 3251.31844201683m^{21} - 10853.9096416194m^{22} - 29577.4805639592m^{23} - 70305.5984873297m^{24}$$
$$- 155531.321477739m^{25} - 337076.168457553m^{26} - 728065.388521755m^{27} - 1542805.56008097m^{28} - 3131595.68038631m^{29} - 6008738.01919971m^{30}$$

$$U_{-9} = -5.37483044657163 \cdot 10^{-19}$$

$$U_{10} = \frac{24171220943377 m^{20}}{4947802324992} + \frac{160790515917264061960473179 m^{21}}{232754482575399290 7718656} + \frac{89040904926697732191920655337 m^{22}}{1832941550281269414828 44160} + \frac{22313333652230432016856679 39304220594395943 m^{23}}{98543112760929674667987018 6635604787200}$$
$$+ \frac{2228628307781013776172695192702560720159833691 m^{24}}{28380416475147746304380261375105417871 3600} + \frac{10271776551172802260715538104545021963191 23309169529312748873 m^{25}}{4797912680323011132294917297193917659270 2139111833600000}$$
$$+ \frac{28675419351634262076137313456014899986784336070650728246569863 m^{26}}{617971153225603833839585347878576594514003551760416768000}$$
$$+ \frac{1291328834894714245628316210886297696701623296185257038893781648824408642120637 m^{27}}{3462115816541093352141965753507124611487829734279064087628076172648297696389767 67181 m^{28}}$$
$$+ \frac{1719318029041256987183501903907519623358297189145517864417518996684 8000000}{57979155178607275623072448203915541 74245638113912681624531746970056785920000000}$$
$$- \frac{28746211156406204604335038214116921956702346049742240337041512447140 97134731673933835539485660 22283 m^{29}}{19715603393014156309521607376265497763708846121843505786681920454111 1550328225377812480 0000}$$
$$- \frac{15416072141290911957889229446040293897866676101746641024229200652278731 10224834 77132624776512170889 3587 m^{30}}{1785494332279844530781050568013044141225882376909452491176256421125441 47765999107781427200000000}$$

$$U_{10} = 4.88524386297427 m^{20} + 69.0815979731677 m^{21} + 485.78147466314 m^{22} + 2264.32198324845 m^{23} + 7852.69768585879 m^{24} + 21408.8442945178 m^{25} + 46402.5208975502 m^{26}$$
$$+ 75107.0373882369 m^{27} + 59713.1125121761 m^{28} - 145804.369175898 m^{29} - 863406.389064623 m^{30}$$

$$U_{10} = 2.17384216569511 \cdot 10^{-21}$$

$$U_{-10} = \frac{3266218223100271 m^{20}}{311711546474496} + \frac{802090735962983985 2688853 m^{21}}{5250101110723292 2730496} + \frac{108031176561780120226504 8523 m^{22}}{9647060790954049 551728640} + \frac{28491388661896767088462378948150 1801901557 m^{23}}{5186479618996298666736158877029 4988800}$$
$$+ \frac{52229145865421904036470113896453315231015 1 m^{24}}{2569153990834079834915744335370649600 0} + \frac{220125092234691493873856733464839457 40176297781662516832499 m^{25}}{360745314310000837014655435879241 929268437136179200000}$$
$$+ \frac{1151326642557374836640096750232474542 48913724413561813750369 1501 m^{26}}{73180794460926769796793001722463017771 39515744531251200000}$$
$$+ \frac{332665088504464712455613158058147435376424413864211 7436670653163342582135 34471 m^{27}}{9049042258111878879913167915302734859780511521818515078816789 2992000000}$$
$$+ \frac{391602419351305431700675199970659378094420463575273 57065907429660263850880 7441597699 m^{28}}{479526847341864685604358594167722552689488 665642067509011219298543 20640000000}$$
$$+ \frac{1843946240187973790265872245050846747482615612318 3604946478703076898673442 77955002680 27163457 11991 m^{29}}{10376633364744292794485056513823946191425708485180 79250983258971269008159 6222388305920000 0000}$$
$$+ \frac{1191894559851438076089649663920771854405920287821707385080668 33831154549633 478010359599403280367 4556121 m^{30}}{31324461969821833873351764351106037565366357489639517389057130195183 183818596334698496000000000}$$

$$U_{-10} = 10.4783356922183 m^{20} + 152.776245456439 m^{21} + 1119.83513841934 m^{22} + 5493.39643744913 m^{23} + 20329.3169859646 m^{24} + 61019.5291533379 m^{25} + 157326.338288402 m^{26}$$
$$+ 367624.638072888 m^{27} + 816643.367356914 m^{28} + 1777017.82010818 m^{29} + 3804996.11134492 m^{30}$$

$$U_{-10} = 4.86789184446773 \cdot 10^{-21}$$

$$U_{11} = \frac{716830175607798991 m^{22}}{124684618589798400} + \frac{105425519251868223171571270327 m^{23}}{1184931911292941843929497600} + \frac{32362797389719515384782077925975491 m^{24}}{472165743352455001259806556160000} + \frac{2789484732011286646937645215910536212723088176721 m^{25}}{79615668338994018609939475660617698770944000000}$$
$$+ \frac{30108967109251547731068862466149898894861163247062316 21 m^{26}}{22559895780537345113312449823192631123734691840000} + \frac{22900029596883689214601326756520665435663793424320035954142 1579076 93 m^{27}}{491146748239782131064914690775002067849675865794608924 7967844753875624471 m^{28}}$$
$$+ \frac{5706008553710739108674454964848297050145301158654728 2083840000 00}{5052670574310859480731229871373167037903664175988761828 5240320000000}$$
$$+ \frac{23633309255556681281703741552074545788655566506464893 698728657147026058571839700176 9697 m^{29}}{12779572120618020335252218791613050659066427342749202 41407205894650016563200000}$$
$$+ \frac{11668259662713666403291524473212436447251925351348885 04129625154861586534275388613711355111 m^{30}}{501206858975546331099940306822349347370661386631714873 4276243918066255134720 00000000}$$

$U_{11} = 5.74914679705689 m^{22} + 88.9717951277326 m^{23} + 685.411803913141 m^{24} + 3503.68814356239 m^{25} + 13346.2350190584 m^{26} + 40133.1848372209 m^{27} + 97268.8551865175 m^{28}$
$\quad + 184930.364119372 m^{29} + 232803.271818013 m^{30}$

$$U_{11} = 1.86289273184418 \cdot 10^{-23}$$

$$U_{-11} = -\frac{1476046950617733941 m^{22}}{124684618589798400} - \frac{22149172473785998884433189 15103 m^{23}}{1168004312560185531873361920 0} - \frac{16583598888681631782141976628738 71367 m^{24}}{1085981209710646502897555079168 000}$$
$$- \frac{5582775451625412093246370136923555500992133408607 m^{25}}{6782075451099490474180029408126692858265600 00} - \frac{5788349370298757625343490113769890598107166240 0794473931 m^{26}}{17295920098411964586872878197781017194863263744 00000}$$
$$- \frac{32119571495501549593042050227745553905832961446946388110 0072671579157 m^{27}}{291640437189659998878054759811351825629820592201305528 729600000}$$
$$- \frac{61670266225309072010795880000977059314976909040337264710132 42698395792463 m^{28}}{19921958264425674524025992064271344320877304465327118066751897 600000}$$
$$- \frac{21055280934151375819475900228110190315566353097526940081220 65826009727574206203443461 3 m^{29}}{26990831843362210074453721965757590923648101825824761756 07505562621706240 0000000}$$
$$- \frac{18414194787649193748545994404374810196604434259395470675574 66191280152664477209176953271183 46193 m^{30}}{99945659748313693884639096583444683359184534710823026291020 25799701591936414515 2000000 000}$$

$U_{-11} = -11.8382441018952 m^{22} - 189.632625801154 m^{23} - 1527.06131012158 m^{24} - 8231.66225719348 m^{25} - 33466.5593814244 m^{26} - 110134.149451345 m^{27} - 309559.258215256 m^{28}$
$\quad - 780090.108239085 m^{29} - 1842420.65478585 m^{30}$

$$U_{-11} = -4.04128867941581 \cdot 10^{-23}$$

$$U_{12} = \frac{1832800525780261109 m^{24}}{270919665084006400} + \frac{17428261945234168240703 1070964941 m^{25}}{15322486204030285755168 32563200} + \frac{208649815719672459620809 21960 9826489 m^{26}}{2190062106249803780843 402742988 800}$$
$$+ \frac{12729981217086442596915890926681940238744641234800319 m^{27}}{240221112477943952595456641635847461039767552000 0} + \frac{276128113858058226725807574741716385828575494404508282811327 m^{28}}{1255683799144708629006970957158901848347072947814400 0}$$
$$+ \frac{777383790254430728593302921137490818278316429427673901 7686584602505427581 m^{29}}{1076031696381016370896465620755396683998002555997606685 70419 2000000}$$
$$+ \frac{157038936072191184099109949398295753317229681195720378 81405098460567520542131 m^{30}}{81374897038814363049045212569626874227348943297319005 60637 9520000000}$$

$U_{12} = 6.765107011379 m^{24} + 113.743042174513 m^{25} + 952.711866591578 m^{26} + 5299.27660636209 m^{27} + 21990.2585385062 m^{28} + 72245.4359726559 m^{29} + 192982.039654424 m^{30}$

$$U_{12} = 1.59593936165853 \cdot 10^{-25}$$

$$U_{-12} = \frac{3932424662846267381 m^{24}}{270919665084006400} + \frac{8455969014716890950341832936451 m^{25}}{33309752617457142946018099200} + \frac{20349823713725029431596678479757373839 m^{26}}{912525877604084908684751142912000}$$

$$+ \frac{380222090650394917057376532811117077856117702029753577 m^{27}}{29012211651925598139547903579208630560358400000} + \frac{607135595000127665669949942987850751554282421826160124976718 1 m^{28}}{10464031659539238575058091309657515402892274565120000000}$$

$$+ \frac{8081491958402652820641678844053942466045175424035138361963585860893664774 1 m^{29}}{389866556659788540179878848099781407245653099999132857139200000000}$$

$$+ \frac{11499693027792975285866206674715808073876601984174420861481998068707486388348211 m^{30}}{183093518337332316860351728281660467011535122418967762614353920000000000}$$

$U_{-12} = 14.515094951217 m^{24} + 253.858655489985 m^{25} + 2230.05442510356 m^{26} + 13105.5471077371 m^{27} + 58021.1924766731 m^{28} + 207288.668913832 m^{29} + 628077.560157312 m^{30}$

$U_{-12} = 3.64503146179747 \cdot 10^{-25}$

$$U_{13} = \frac{111792382780331305141 m^{26}}{14044475437954891776} + \frac{294296094366039142606565153231273819 m^{27}}{2036712012351102598840835899392000} + \frac{32988045173347564378979574456878801459197 9 m^{28}}{252295154639977395553159995992309760000}$$

$$+ \frac{570535438964381074293821911786177011178318192074866972 5 9 m^{29}}{7257029411921104022464144698369657564278294248000000} + \frac{638234214283393775693542428773900078103777630174681559806683788 1 m^{30}}{18081846707683804257700381783088186616197850448527360000000}$$

$U_{13} = 7.95988310664952 m^{26} + 144.495683523914 m^{27} + 1307.5179830711 m^{28} + 7861.83170247545 m^{29} + 35296.9596856597 m^{30}$

$U_{13} = 1.34620985509513 \cdot 10^{-27}$

$$U_{-13} = -\frac{173140896903399972155 17 m^{26}}{1053335657846616883200} - \frac{57635302164492644072939082932072131 m^{27}}{185155637486463872621894172672000} - \frac{22410340146037277056791698982180373888658 7 1 m^{28}}{75688546391993218665947998797692928000}$$

$$- \frac{90253480231084250635254414728759919562049222101780883818 9 m^{29}}{4789639411867928654826335500923973992423673036800000 0} - \frac{4900359111733403605846330042865384357084072319829439111595573248 9 m^{30}}{54245540123051412773101145349264559848593551345582080000000}$$

$U_{-13} = -16.4373906469055 m^{26} - 311.280298817292 m^{27} - 2960.86280082239 m^{28} - 18843.4812039193 m^{29} - 90336.6267644742 m^{30}$

$U_{-13} = -2.97367565148244 \cdot 10^{-27}$

$$U_{14} = \frac{29311427084360848282935 7 m^{28}}{31299116690299473100800} + \frac{14205617078606222425990223905309842 9 1 m^{29}}{778107119820666135720417306869760 0} + \frac{627059083915489221822485289462989768855525745 7 m^{30}}{3532853008258654901731534686738028953600 00}$$

$U_{14} = 9.36493747551838 m^{28} + 182.566342303618 m^{29} + 1774.93680730455 m^{30}$

$U_{14} = 9.28708107854666 \cdot 10^{-30}$

$$U_{-14} = \frac{629416946475841385108461 m^{28}}{31299116690299473100800} + \frac{15427732688481786612882314005050700712 9 m^{29}}{37608510791332196559820169832038400 0} + \frac{1844375157755213817887807694195203511350303 7 m^{30}}{43908315959786139492949073963744074137600 00}$$

$U_{-14} = 20.1097351309891 m^{28} + 410.219186132664 m^{29} + 4200.51445253424 m^{30}$

$U_{-14} = 2.09859260439762 \cdot 10^{-29}$

$$U_{15} = \frac{311199967652831601233 m^{30}}{28246576862867750912}$$

$U_{15} = 11.0172630532773 m^{30}$

$U_{15} = 1.87198614546171 \cdot 10^{-32}$

$$U_{-15} = -\frac{92176469683447808831365 m^{30}}{4039260491390088380416}$$

$$U_{-15} = -22.8201349925134 m^{30}$$
$$U_{-15} = -3.87745816152074 \cdot 10^{-32}$$

$\theta_i$

$\theta_0 = 1 + 2m - \frac{m^2}{2} + \frac{255m^4}{32} + 19m^5 + \frac{80m^6}{3} + \frac{533m^7}{18} + \frac{11230225m^8}{221184} + \frac{1576037m^9}{10368} + \frac{49539583m^{10}}{124416} + \frac{720508007m^{11}}{933120} + \frac{3367704784027m^{12}}{2866544640} + \frac{19381937097637m^{13}}{11757312000} + \frac{26266300871718563m^{14}}{9876142080000}$
$+ \frac{51635998792150519 13m^{15}}{1036994918400000} + \frac{32412659763218896266846193m^{16}}{35679261960437760000 00} + \frac{88151686411545316416063 79m^{17}}{585362891538432000000} + \frac{599751374867525145849887 22277m^{18}}{2458524144461414400000000}$
$+ \frac{281379732947214285166743924 3979m^{19}}{64536258792112128000000000} + \frac{12132985666428439668135761023 22746757m^{20}}{13877877090655792005120000000000} + \frac{45017058353619822964315302310 870006621m^{21}}{25045231312042874634240000000000 0}$
$+ \frac{326976625544075492074540772313150 1455499m^{22}}{925671749293104646481510400000000 0} + \frac{87069079467524161766663531030014868 4363246571m^{23}}{1336438588041 9198333576806400000000000 0}$
$+ \frac{71846218250118518364403567194018864 4680300776933113m^{24}}{6322530587395757701235184186163200 000000000} + \frac{14099962183469371888363273718721684 62192390850352189673m^{25}}{74166247476365079596911164496281600 0000000000}$
$+ \frac{27547812376526614423191198621439419966 70705329537111 09184979m^{26}}{8908849646860973361180969079293345792 00000000000000} + \frac{66269986023078463691310769942844518007 35637125093150066 50497131m^{27}}{13376637744761751 501813225072558958706688000000000000 000}$
$+ \frac{8198414151785249582375868466904589084318 894231141788214682046 6994315833m^{28}}{102835310457650021785459494125810055 670231203840000000000000000}$
$+ \frac{301246838565051697872123889140247967522 9111862642990633691720529178 274687m^{29}}{2261824491975022085608409625438043924641388953600 00000000000000000}$
$+ \frac{31237635844642555465400582764658454190629237862 8120287323617417496419139889 54059m^{30}}{13041137182849903340317543881965656138940334773 108736000000000000000000000}$

$\theta_0 = 1.0 + 2.0m - 0.5m^2 + 7.96875m^4 + 19.0m^5 + 26.6666666666667m^6 + 29.6111111111111m^7 + 50.7732250072338m^8 + 152.009741512346m^9 + 398.176946694959m^{10}$
$+ 772.149355924211m^{11} + 1174.83074815364m^{12} + 1648.50070302098m^{13} + 2659.57098014112m^{14} + 4979.38783266371m^{15} + 9084.45354031118m^{16}$
$+ 15059.3226331563m^{17} + 24394.7726207469m^{18} + 43600.2548355973m^{19} + 87426.813100959m^{20} + 179743.032886159m^{21} + 353231.721497143m^{22}$
$+ 651500.789086712m^{23} + 1136352.24467474m^{24} + 1901129.24183776m^{25} + 3092185.12698023m^{26} + 4954158.68229141m^{27} + 7972372.63669423m^{28}$
$+ 13318753.9366506m^{29} + 23953153.3229498m^{30}$

$\theta_0 = 1.15884393959659$

$$\theta_1 = -\frac{15m^2}{2} - \frac{57m^3}{4} - 11m^4 - \frac{23m^5}{6} - \frac{68803m^6}{4608} - \frac{1792417m^7}{27648} - \frac{7172183m^8}{51840} - \frac{596404499m^9}{3110400} - \frac{2813929549973m^{10}}{11943936000} - \frac{1038170530002499m^{11}}{2508226560000} - \frac{61742169209753239m^{12}}{65840947200000}$$
$$- \frac{26225428721736277907m^{13}}{13826598912000000} - \frac{1198213185342582797674073m^{14}}{371658978754560000000} - \frac{39439844465868357901549349m^{15}}{78048385538457600000000} - \frac{3417965815630477600021503395m^{16}}{40975402407690240000000000}$$
$$- \frac{206916258233784461755446571951m^{17}}{1594828334824252076893220251396420828496663m^{20}} - \frac{17350634366455236377088798428909464m^{18}}{31827880077677243396939667146939719910540472m^{21}} - \frac{7646604911652445797840544511584146006007m^{19}}{13357456997561998049280000000000000}$$
$$- \frac{15427862488218410774691840000000000000m^{22}}{100763657220235102547642872819680682253178024277m^{23}} - \frac{178191811738922644476907520000000000000m^{23}}{10557417454154808034488374842819893767394008039232017239m^{23}}$$
$$- \frac{3292984680935290469393325096960000000000000000m^{24}}{5596630018365554863943128004598969774074419745083908934129m^{24}} - \frac{19777665993697354559176310532341760000000000000000m^{25}}{244984791701296697485829637080316888251955640729703903487943665m^{25}}$$
$$- \frac{59392330979073155741206460528622305280000000000000m^{26}}{14661872808498199603992414037956568297581520679406939712412410939393320m^{26}} - \frac{148629308275130572242369167472877318963200000000000000m^{26}}{\cdots}$$
$$- \frac{514176552288250108927297470629050278351156019200000000000000m^{27}}{5779363406351404654875989188161844022274153458849651236506424013855662199413m^{27}}$$
$$- \frac{115805413989121113078315057282242784894163911442432000000000000000m^{28}}{37672654054721580828633091563501549703801296245674064372774680550057035684488967m^{28}}$$
$$- \frac{417316389851196906890161404222900996446090712739479552000000000000000000m^{29}}{31610467745282561920801903765675694209411461690057489254589022413956393751616872576467m^{29}}$$
$$- \frac{18798016780847164670867320453220575384914156155349856419840000000000000000000m^{30}}{3422045642633981269940116688932269263896727014635540364760512300952097329183864060559885037037m^{30}}$$
$$- \frac{108384853243373481726999615763610647313226449942699885125667840000000000000000000}{\cdots}$$

$$\theta_1 = -7.5m^2 - 14.25m^3 - 11.0m^4 - 3.83333333333333m^5 - 14.9312065972222m^6 - 64.8298972800926m^7 - 138.352295524691m^8 - 191.745273598251m^9 - 235.594828201775m^{10}$$
$$- 413.906202317903m^{11} - 937.747280916293m^{12} - 1896.73750490986m^{13} - 3223.95866597339m^{14} - 5053.25564312086m^{15} - 8341.50640333674m^{16}$$
$$- 15389.7678314684m^{17} - 30005.6861777011m^{18} - 57245.9644341726m^{19} - 103373.253167258m^{20} - 178615.839679041m^{21} - 305994.905792893m^{22}$$
$$- 533805.023177113m^{23} - 942315.2663157m^{24} - 1648293.9639859m^{25} - 2851524.97585667m^{26} - 4990581.36167479m^{27} - 9027360.28847431m^{28} - 16815852.4985943m^{29}$$
$$- 31573098.4590183m^{30}$$

$$\theta_1 = -0.0570440187469028$$

$$\theta_2 = \frac{111 m^4}{16} + \frac{1397 m^5}{64} + \frac{8807 m^6}{240} + \frac{319003 m^7}{7200} + \frac{126191191 m^8}{1728000} + \frac{149693929741 m^9}{725760000} + \frac{2508021553069 m^{10}}{4762800000} + \frac{2011436476413019 m^{11}}{2000376000000} + \frac{646554057549386707253 m^{12}}{4301608550400000000}$$

$$+ \frac{36234933712232765428 4153 m^{13}}{180667559116800000000} + \frac{34307610923478020461 26827 m^{14}}{118563085670400000000 0} + \frac{47703357587514432952 79460979 m^{15}}{99592991963136000000 0000} + \frac{13447181432478280306 404930216841 m^{16}}{16731622649806848000 00000000}$$

$$+ \frac{903517288567783816285 8246325411141 m^{17}}{702728151291887616000 0000000000} + \frac{541951645217551887817 4473170678737117 m^{18}}{253640942106915686400 00000000000000} + \frac{150457894094941945783 2000843421071 19891 m^{19}}{366194110166859522240 0000000000000000}$$

$$+ \frac{16258464227626129987760593882024517402813933231 m^{20}}{184791508469993853501308928000000000000000000} + \frac{18726846922930786997491191407977350789379103714080147 m^{21}}{99887201988370477571597527941120000000000000000000}$$

$$+ \frac{42033181277875138898796344806019091542769845559356 7817193 m^{22}}{11248547533915370405531526615269376000000000000000 0000} + \frac{69732960111008388015390114047946460133396569615055 5639515 02261 m^{23}}{10133816473304357198343352327696180838400000000000 000000000}$$

$$+ \frac{862627196940779667468842261220299696047760988024331504253 59778144551 m^{24}}{7303644208639916319990020889617191453851648000000000000 00000000000}$$

$$+ \frac{25282918164269107787900127293519740362634848060473 5177241271588232210 65329 m^{25}}{13159706135127401225358019638912255561549899366400 0000000000000000000000}$$

$$+ \frac{55951971651468099584802687780553623882867925645629 93139746351 71702603579 36897 m^{26}}{18524342589275430881132874832337579742812975529984 00000000000000000000000 00}$$

$$+ \frac{31313339809231572465008086147684584962910813323176 27101331168789846002981 6265991801 m^{27}}{66754320954712942723250427745811702361200838619850 34240000000000000000000000 0000}$$

$$+ \frac{45673904898893138922226718360218763995813844951336 99234602786609607026046 42028072254 8816073 m^{28}}{61582302974055311461761341804750604960978775564926 12963139584000000000000000 0000000}$$

$$+ \frac{23532758277391416870372357612214715040753451351611 54470853289865161692971 22422382314 66541952695089 m^{29}}{18886302889477098623260626956284593880317756482819 02630628743341670400000000 00000000000 000}$$

$$+ \frac{16727747080807569686855180525810741226694359855156 80891248426175117985370 47151416933 2131496407162126456 9 m^{30}}{72223236608021521362058700058963601638076339653295 58397170982251711692800000 00000000000 00000000000000}$$

$\theta_2 = 6.9375 m^4 + 21.828125 m^5 + 36.6958333333333 m^6 + 44.3059722222222 m^7 + 73.0273096064815 m^8 + 206.258170388283 m^9 + 526.585528065214 m^{10} + 1005.52919871715 m^{11}$
$+ 1503.05182346093 m^{12} + 2005.6137299562 m^{13} + 2893.61656956632 m^{14} + 4789.83075487597 m^{15} + 8036.98584048184 m^{16} + 12857.2803993517 m^{17}$
$+ 21366.8834658841 m^{18} + 41086.9235516608 m^{19} + 87982.745323312 m^{20} + 187479.942877077 m^{21} + 373676.522690075 m^{22} + 688121.403172308 m^{23}$
$+ 1181091.48296179 m^{24} + 1921237.29091344 m^{25} + 3020456.53613971 m^{26} + 4690833.39645908 m^{27} + 7416725.69766278 m^{28} + 12475598.9129163 m^{29}$
$+ 23161170.6514821 m^{30}$

$$\theta_2 = 0.000383237997558365$$

$$\theta_3 = -\frac{11669 m^6}{512} - \frac{528813 m^7}{5120} - \frac{5165893 m^8}{22400} - \frac{28363226117 m^9}{84672000} - \frac{76670095052989 m^{10}}{189665280000} - \frac{12489618650011331 m^{11}}{19914854400000} - \frac{32897028677070454973 m^{12}}{25092716544000000} - \frac{22503210096331249900189 m^{13}}{878245079040000000}$$
$$- \frac{18360846063213973082343 89 m^{14}}{43717088378880000 0} - \frac{17346316206575052627277177 2839 m^{15}}{27266348021907456000000 0} - \frac{29793542414112517609386969 43605649 m^{16}}{28343368768772800512000000 0}$$
$$- \frac{19682123670403960959050517006783 144143 m^{17}}{98209772783797753774080000000000 0} - \frac{34980613230893345729768292622108880 419261201 m^{18}}{87115996850139959507759923200000000000 0}$$
$$- \frac{46542641124669606332617221436907598710 6023520839 m^{19}}{60371385817146991938877626777600000000000 0} - \frac{18510030764773006514630262488526451219162913776 282327 m^{20}}{13597145370666931259433713490984960000000000000 0}$$
$$- \frac{66957321508542596651721407609849294308562606098 2593361691 m^{21}}{30624170661084595929059581210070876160000000000000 0} - \frac{22777849558011164420764004242331743898727608597605 7522068924 35937 m^{22}}{70628647292342047929573828383111301971312640000000000 0000}$$
$$- \frac{28674711368954147236474172699969165919124779984816118266668 752798687009 m^{23}}{636293483456709509797530619903449719459555573760000000000 00000000}$$
$$- \frac{22472425309092559711274264921621614063123810845882247506255767950964959603 m^{24}}{3582729995288434983603720846693861576631960102502400000000000000000}$$
$$- \frac{60179918192648760533022265986288045100080661933489414787917119407195136004672717 m^{25}}{645536290551070215345718422157299978877546571268882432000000000000000000000}$$
$$- \frac{6025579106164935432514986725754347552668076568855600120604543048107998250005766668411 m^{26}}{37220073226077386048317294497376739262130029187592715711283200000000000000000000}$$
$$- \frac{976509451900602303269161560925732526652517606663750664994539350785110406 70526002563568213217 m^{27}}{2850182937396714952728969302078369874106500180083693594665077964800000000000000000000000}$$
$$- \frac{7093564574383964182914418927522316088350673887597109344542066714354741654541019240989112 60651486791 m^{28}}{87302813482223817031059967104241716265806564416071585020748137108602880000000000000000000000}$$
$$- \frac{127907166617153080741307320386967851039596879533713026133720591821428192519646707141836194033 9718074343377 m^{29}}{6685343896621512124879635709579657856285363800073057303413197212969284403200000000000000000000000}$$
$$- \frac{27415415974565616832599410663634509492357712065242141032374456196748515961124948999755718271 64350676174815197 11147 m^{30}}{6552835032315356457354162609810722137384750286126456924921370491365046281090105344000000000000000000000000}$$

$\theta_3 = -22.791015625 m^6 - 103.2837890625 m^7 - 230.620223214286 m^8 - 334.977632712113 m^9 - 404.238957456995 m^{10} - 627.150889439158 m^{11} - 1311.0190209731 m^{12} - 2562.2927623949 m^{13}$
$- 4199.92427310969 m^{14} - 6361.80400566954 m^{15} - 10511.6447720701 m^{16} - 20040.9013405752 m^{17} - 40154.0641164541 m^{18} - 77093.8756742111 m^{19}$
$- 136131.74133377 m^{20} - 218642.072791307 m^{21} - 322501.568856761 m^{22} - 450652.287261795 m^{23} - 627243.061538143 m^{24} - 932246.863166677 m^{25}$
$- 1618905.76344789 m^{26} - 3426129.03574716 m^{27} - 8125241.66340678 m^{28} - 19132473.7507957 m^{29} - 41837488.4143524 m^{30}$

$$\theta_3 = -9.17328891116338 \cdot 10^{-6}$$

$$\theta_4 = \frac{37303 m^8}{2048} + \frac{12647675 m^9}{114688} + \frac{49303273 m^{10}}{143360} + \frac{1376892000103 m^{11}}{1896652800} + \frac{60964655772059 m^{12}}{48771072000} + \frac{153574057736105965289 m^{13}}{66244771676160000} + \frac{879605988694110119823 53 m^{14}}{16395580989849600000}$$
$$+ \frac{1015101523751081383945 8236561 m^{15}}{79534963381760409600 0000} + \frac{66333130390949441667 99906717 78151 m^{16}}{25196676399341697761 280000000} + \frac{28954220857440693780 5381324727 7313121559 m^{17}}{63559120150867419436 784025600000000}$$
$$+ \frac{34693821661418156720 22905966512388882974 591 m^{18}}{51125367271353980509 463150592000000000} + \frac{29128710666331305041 30684066945858677127 583534601 m^{19}}{32241190362333960728 68274665783296000000 0}$$
$$+ \frac{22688705202765808362 07055915867263771271 2797491760887 m^{20}}{20747205998161903728 90734747431550976000 00000000} + \frac{61883685718553795644 42692169039557637293 89192609116 99744673 m^{21}}{52335242074483365394 24336215091035967979 52000000000000}$$
$$+ \frac{19518358064160854566 10318412256908609798 1091881641189 2380141793 m^{22}}{16838864137465022815 59780177205540822697 4105600000000000000}$$
$$+ \frac{30916563840376164403 03858161993625731758 32576707044540423846 442595051 m^{23}}{21238185782019134676 40088346302260418035 3360429056000000000 00}$$
$$+ \frac{84652163258900207805 72111387469264145195 82695834084654845340 35283407032799043 m^{24}}{22391640147114906683 30086004252747022430 02357530220298240000 000000000000}$$
$$+ \frac{11983051958677028264 15534602120176203034 15590468996461181427 113530283857134312 4468869 m^{25}}{96021712176630500513 05889453066098701233 43112175912322141782 016000000000000000}$$
$$+ \frac{68076666533222024119 25854037261211191655 85049633384665086989 4714043557466419563 55712401 m^{26}}{18757670006361595720 76085315568153334681 12713850864158001250 4350720000000000000 0}$$
$$+ \frac{25133467749742451211 74564317808986200088 28432746169828285884 21535157577234279350 47395195 0755051923 m^{27}}{28153375657946115200 33253204485051999130 90291351133704255819 77249303756800000000 000000000}$$
$$+ \frac{23130218596943594362 13717185620331910232 41015501436097961292 16491462653753777430 50079699133 15019558689863 m^{28}}{12319354120404061089 36150937218569053779 70049689429086295030 16160475033790054400 00000000000000000}$$
$$+ \frac{34008709045205818899 68314972882555946957 75354383303926837417 33016748450337422672 32931228342 52229800069773 21999631 m^{29}}{10037488941323933653 86099086275284893481 53954414685290677978 34530901423876225643 2103424000000000000 0000000}$$
$$+ \frac{27015870240455219046 86899892176042973191 40473985723224926270 45089168523144634247 86110865534 17793046202162474 8889150989 m^{30}}{52157427379966256798 04584703296527568814 61836160011252515784 51541281600846359513 5228182528000000000 00000000000000}$$

$\theta_4 = 18.21435546875 m^8 + 110.278974260603 m^9 + 343.912339564732 m^{10} + 725.95891040416 m^{11} + 1250.01672655584 m^{12} + 2318.28193909186 m^{13} + 5364.89673186128 m^{14}$
$+ 12762.9595914779 m^{15} + 26326.142916485 m^{16} + 45554.7855110539 m^{17} + 67860.2883716738 m^{18} + 90346.2630845081 m^{19} + 109357.882718164 m^{20}$
$+ 118244.768277714 m^{21} + 115912.557431556 m^{22} + 145570.644111 m^{23} + 378052.535244085 m^{24} + 1247952.33151377 m^{25} + 3628791.13611857 m^{26} + 8927337.1886574 m^{27}$
$+ 18775512.3936521 m^{28} + 33881054.4184121 m^{29} + 51796784.4610987 m^{30}$

$$\theta_4 = 5.44474786780049 \cdot 10^{-8}$$

$$\theta_5 = -\frac{6372443 m^{10}}{131072} - \frac{68644425419 m^{11}}{192675840} - \frac{62291190081847 m^{12}}{47687270400} - \frac{921058150219046957 m^{13}}{289163685888000} - \frac{31082504180640669207151 m^{14}}{5234688896532480000} - \frac{112090777652654589326371910759 m^{15}}{11554031505870972518400000}$$

$$- \frac{4819794254089495573717498139910403 m^{16}}{297400770961118832623616000000} - \frac{27751049026753208274175726517502642 93161 m^{17}}{937749240956051847087154790400 00000} - \frac{42686128252887158809901706593014 76093905267859 m^{18}}{7724052947906807854087476577566 7200000000}$$

$$- \frac{4805518970746258807681431407021603 9338709623 5748841 m^{19}}{4871019510538470237023185354110900 6336000000} - \frac{531810781844293564237032734860953 2806950904139 8123889143 m^{20}}{3134501055031505597524419775370364 5577216000000 0}$$

$$- \frac{2946748756854881184180105625000409940 2496958004 7031884229 17341 m^{21}}{9883552001672591874834124214709065005 179863040000 0000000} - \frac{11395036231228060718382420659489274 9518758095431 40374020495 736983301 m^{22}}{2035221028184420118865842858292890 665866637397196 80000000000000}$$

$$- \frac{9606502710714490014629559821498916538 5273101954 5511222238974019428629335907 m^{23}}{8727605771626797845010492075731670 356185247284200 68229120000000000}$$

$$- \frac{40613591687062388938854203651726112635147984046 7441790335071921397944370018036349 m^{24}}{190951286677422710050984556124933215722977025331026 72784916480000 00000000}$$

$$- \frac{38999870129394635727625457722899024459269476073 769283041108460742581995926329 00468336121 m^{25}}{10235666842977562109353453203470663875667385126182 95775599097815040000000000}$$

$$- \frac{8686669937943420205510882386347742020315567109 80056440948898578011294083182026 78923723994416633 m^{26}}{14332553910880389208994828968879017915886510027 248443697824682310096650240000000000 00}$$

$$- \frac{2388464322222307993237655748046000689467728752 001269488330702088051117409609062 04975883950716145 87909091 m^{27}}{291944792672037545053849610462406854705944 96713702305938784149478848492469493760000 0000000000}$$

$$- \frac{18467433064731107858686696432354333134280522 066710596359976101484199149734962657041 1914220435 6968989628 3808113 m^{28}}{216717396890283356279339887355058914121635 765685406469304616245044764211222251831 2960000000000 00000000}$$

$$- \frac{43218049431941314400824966566706571860550647348 4273333298954377352399426706319183808 7800631495690 2238527579 776449751 m^{29}}{96564831379399608595676592456534917378538723546555 8751902019038897008121489698400 71566950400000 0000000000000}$$

$$+ \frac{10276767349193946988598256897210269217438477005207 0843745846078677019368290352215977 3476665307 66159373 6812949816 34868509569457 m^{30}}{16442202918603897525383723073853039283886459 2556141870581522930273699910906424981453 170494866351 718400000 0000 00000 00000}$$

$\theta_5 = -48.6178817749023 m^{10} - 356.268982239808 m^{11} - 1306.24356477839 m^{12} - 3185.24833915623 m^{13} - 5937.79397305351 m^{14} - 9701.442963497 m^{15} - 16206.3946186596 m^{16}$
$- 29593.2513882502 m^{17} - 55263.8990705714 m^{18} - 98655.3012228651 m^{19} - 169663.61551885 m^{20} - 298146.73473223 m^{21} - 559891.828623319 m^{22}$
$- 1100703.09797275 m^{23} - 2126908.51126191 m^{24} - 3810193.38824529 m^{25} - 6060796.97446597 m^{26} - 8181219.12832143 m^{27} - 8521435.44068155 m^{28}$
$- 4475547.54816888 m^{29} + 6250237.51383464 m^{30}$

$\theta_5 = -1.04932440527466 \cdot 10^{-9}$

$$\theta_6 = \frac{39092253 m^{12}}{1048576} + \frac{3107423715553 m^{13}}{9248440320} + \frac{388587923799599 m^{14}}{252565913600} + \frac{8889273555389142711247 m^{15}}{1876556655938764800} + \frac{14727696213531248018157557541 m^{16}}{131192160879630090240000} + \frac{28437433563857311655640147511511501 m^{17}}{121844194648312927790039040000000}$$

$$+ \frac{61417628437651540463215473732076578 1 m^{18}}{12477924360794025743081472000000} + \frac{86500425921940863737239699166718583606791 61 m^{19}}{77258815589266721551892965294080000000}$$

$$+ \frac{42464502889113578559748188568732140771963837969869 m^{20}}{16592638897072340036420626351115993088000 0000} + \frac{235276101544390260647627702547494885635877025805388722 18997 m^{21}}{4366264984778431797181894794976512666343833 6000000000}$$

$$+ \frac{13077441691673931272402266912657118961997221069565227 58592304923 m^{22}}{129731435508844494098267831902741028011889393 6640000000000}$$

$$+ \frac{4284470353526474223407509262601558083806666124415023164505955 3441045419 m^{23}}{25603604758158173809605344359272912745410 1731966713856000000000}$$

$$+ \frac{3239149635198537285514139104087152015846701309108 480193030351733599981907639 m^{24}}{1304125704453157573033631454307270399476216 078956551471104000000000}$$

$$+ \frac{6013559427543960424503117051497468156218484137566 19218891723676352362958038 3641070283 m^{25}}{18256844418264708083980873672366120061137408780 7529362545853433816000000000}$$

$$+ \frac{742794005290457519969809986632115550643801975424480261016047 18759791777789962209 11720345243 m^{26}}{19324831781640655455418579822046053988630486491492 023123528056044423680000 00000000 00}$$

$$+ \frac{68827080417807157845422660501834960777500450765569822688610858 3336856608647689695 2602553558 16587373 m^{27}}{17567148031431930902685495576776160544984996 87172667246680785669791936078151680 00000000 00000}$$

$$+ \frac{756577575122010335922067949024494859828754499475203910251693772 6586978106689468816 99993242546 932773791774609 m^{28}}{1861792977565031238772830492040180690657464973 87122111948602707921434320538804179 43552 0000000000 00000000}$$

$$+ \frac{124871014649385411503168186573677337810220281088803925458128612 3807165223739584418137 59577779451 6256062167920048850869 m^{29}}{16755296645386225152958665800873788163978718677 1968809481318644483493225839775127414566792 724480000000000000000}$$

$$+ \frac{668057583562781068789391727830948488175356458415585361190443743 7743197560857034660183 53544146712320 24272127128512663538 52187 m^{30}}{2837671454763729296496695320480717805522047802263 8380250819631856842918230482368987221893826887 0246400000000000000000}$$

$$\theta_6 = 37.2812776565552 m^{12} + 335.994352348591 m^{13} + 1538.56044254263 m^{14} + 4737.01314972673 m^{15} + 11226.0489611449 m^{16} + 23339.1780756877 m^{17} + 49221.0296053945 m^{18}$$
$$+ 111961.884559305 m^{19} + 255923.745177183 m^{20} + 538849.800377678 m^{21} + 1008039.54264288 m^{22} + 1673385.6009714 m^{23} + 2483771.0230984 m^{24}$$
$$+ 3293865.73592521 m^{25} + 3843728.18187292 m^{26} + 3917942.75853193 m^{27} + 4063704.09728105 m^{28} + 7452629.29640641 m^{29} + 23542457.0536974 m^{30}$$

$$\theta_6 = 6.05146535096331 \cdot 10^{-12}$$

$$\theta_7 = -\frac{45769552751 m^{14}}{503316480} - \frac{215134672356593317 m^{15}}{232764745973760} - \frac{106949006928845196377 m^{16}}{22694562732441600} - \frac{189444565802976546463143 6001 m^{17}}{118072944791667081216000} - \frac{121749528259299231993280214201507 m^{18}}{29471007020000103 47151360000}$$

$$- \frac{3658662429688744200743228 8660829572 23137 m^{19}}{417054157963588608260279547985920000} - \frac{2100542373335084081292907298455654176 70210817 m^{20}}{125684957604481476034802427 4157568000000}$$

$$- \frac{75749729701561920942457329129411662066 7603204645212093 m^{21}}{244558 90470394921979680361178909862 21240320000000} - \frac{6064121670365327608085734866889087617570357852 1261978771349 m^{22}}{1037712344439797329441 79708554350327339669258240000000}$$

$$- \frac{194356789317273778124241204332617342546 3899390879856192606835 96640309 m^{23}}{174384580470395140664510505186 6024010822920849170890752000000000}$$

$$- \frac{921897412495206930158517572726 1602909018479373523535560675417609813 17295853 m^{24}}{4393445120371135173901677667671260 89286726678740114216058880 0000000000}$$

$$- \frac{57516088083422726857528644897168187866370 7525139179769988430215329143811936670002931 m^{25}}{14766116799915799210757529076 219818515482719698332640165161806594048000000000 00}$$

$$- \frac{2216375457450710264000519665380 85162880222851762612045846236741576568 4092415983171777726901 m^{26}}{30368779318945195535985 72959561486593298544000651367643437441 2680036352000000000000}$$

$$- \frac{56353674759379894276071822977440226937815 77907223582125607862949381 61501565202594079706473925 894658221 m^{27}}{400105476776640003435541647116771582119 0871642911896106517436919299367667566116864 0000000000000000}$$

$$- \frac{13963544597356246379057601008229 120752162344592150840975 7385249399261999956424778779 5053648841985 40871377779 m^{28}}{5040128690955334123277518128729971 6199541410085761155253800152872 4141345083303741358 08000000000000000}$$

$$- \frac{181155172619216276727265485779943126090716 171423732033859235754534424516471245 6022 736903050917942449118626502457 09103 m^{29}}{3387916630262417357565410090 39161471 6669061652563663376625868337368134728499311 833640393611673600000 0000000}$$

$$- \frac{132348922709371317844670 043287083599703911220121922470885922951459057 4273546670145447069148514746712657724461 41724989 79774999 m^{30}}{13656827453253014865040470 690772214587481654283950 2296177139403026644458159698 659738777722644007485440000000000000000}$$

$$\theta_7 = -90.9359311083953 m^{14} - 924.257973244994 m^{15} - 4712.53877810842 m^{16} - 16044.705765342 m^{17} - 41311.6281288507 m^{18} - 87726.3146722582 m^{19} - 167127.587371696 m^{20}$$
$$- 309740.223089651 m^{21} - 584374.051523788 m^{22} - 1114530.18614064 m^{23} - 2098347.39535185 m^{24} - 3895139.72175479 m^{25} - 7298203.96853439 m^{26}$$
$$- 14084704.6667245 m^{27} - 27704738.2191138 m^{28} - 53470965.3127399 m^{29} - 96910445.0959774 m^{30}$$

$$\theta_7 = -1.05572669732099 \cdot 10^{-13}$$

$$\theta_8 = \frac{482330627647m^{16}}{7046430720} + \frac{198257785762767188493m^{17}}{2420753358127 10400} + \frac{160846261564938267377 1853m^{18}}{324078355819266048000} + \frac{476571533679088961403804307799777m^{19}}{235866952438897195 3520640000} + \frac{73721786362960996467145705693700 2957m^{20}}{118773914346659010764090 8800000}$$
$$+ \frac{66995215264557399824848991647431 0150182842508057m^{21}}{4217498402371269796654500787966181376 000000} + \frac{475486232387947819365522756106509198145498035381374741m^{22}}{128732812484779453638183654801485737230 3360000000}$$
$$+ \frac{15096512535959585394169550035764836586577817042276105 8941838441m^{23}}{17855957747206012986834476689372351624946764480512 00000000}$$
$$+ \frac{171035456787319310355868548143541572583 3657385358768685641481632 6133013m^{24}}{87204212207494837984582743876489115971850010499345285 12000000000}$$
$$+ \frac{432991531980737792165494839178915737260545809435089149661377324669488361 082496709m^{25}}{9676567759223114521779493982433606401633693962141127210770 104320000000}$$
$$+ \frac{14187766031350180202757078735556312462677073686306408 94431810305666129794189347 1524331m^{26}}{14768135799932336805277209037864060925001330288267073482089706 39605760000000 0}$$
$$+ \frac{38760513565025724541480375939327063543172075552599943092625598550 858304527020968210505850 9787601m^{27}}{20484226496626133382109438717744590395248539933885378 476391707052869597265920 00000000}$$
$$+ \frac{2123345694057479136999879748765134674914445132690 70012942499216238412437758 299945285360929912596468963m^{28}}{62525028746976778 2288817452701077004929368808671895415903 5642702282651214630912000 00000000 00}$$
$$+ \frac{8878145379712784102743332051371302367994321152980 08116739880166090425444713793254 4855737142615439167670457 71637m^{29}}{159575279978349129502934878669733641 6207740782846757795479263591141688767328229351422427 13600 000000000 00}$$
$$+ \frac{84985700786943872245415 47489836363595696210271381767149087876190014362538 21100518596585261148779751612042142 46684298413m^{30}}{10184391758512746550223726624443083096638809493691026605561054 50105173341187226708907934223433728000 00000000 00000 }$$

$\theta_8 = 68.4503469647396m^{16} + 820.099185636835m^{17} + 4963.19049627122m^{18} + 20143.632049461m^{19} + 62069.0046029747m^{20} + 158850.600220476m^{21} + 369359.002736125m^{22}$
$+ 845460.811997954m^{23} + 1961321.04697369m^{24} + 4474639.59075817m^{25} + 9607012.16697587m^{26} + 18922127.0187624m^{27} + 33959931.5123889m^{28}$
$+ 55636094.6439659m^{29} + 83447006.7551236m^{30}$

$$\theta_8 = 6.03772704734079 \cdot 10^{-16}$$

$$\theta_9 = -\frac{476967599208 53m^{18}}{300647710720} - \frac{2472027914563825558841m^{19}}{1197172569837404160} - \frac{4685517417382687687892905781m^{20}}{34737358815163758673 9200} - \frac{1492606472755264518003244457375473 3241m^{21}}{25342822668423293826110 16744960 00}$$
$$- \frac{13689487434892541318465323859 4552301906093011m^{22}}{70593750369671842963908 82915956817920000} - \frac{50845066345443896388771533237633212692901019 9219704831m^{23}}{9809912676638530601877040315097180698 31270400000}$$
$$- \frac{285385398770712491824146 3781345595676506494366523330255105 23m^{24}}{2391026914170635 67072981922832463415247805559603200000} - \frac{72865820438567943011015977655896234013913414017266414635470 87020907813m^{25}}{29073104167993661425624815505373802819 821394553778837913600 00000}$$
$$- \frac{336528664563223998228358 10423281794373047984700291586968252651185776 25546539m^{26}}{19836933163830216750112716061 007983206378217488679418 4857697439308134522729296572018721}$$
$$- \frac{66654027540216526396273297185699594397003395260961107060 9408020000m^{27}}{1969425740384627472145019553309155922719946 52225178912398 3841711829811 2000000}$$
$$- \frac{63207542288834487805372245837071026 35125553253342999 3862997149314293532630802 0323325099 47344689m^{28}}{31544142407384308986588951638666989183552567309712152628 75699264079869247488 00000000}$$
$$- \frac{698035829793154218666909597776431385 930787443888 8912101274252873977763440663418571558971224 16664769616367m^{29}}{17643449390961272961427863845 29132784179815168713153136738820534418613110344002872279 0400000000000}$$
$$- \frac{4387014162973617324560748020143280717884796643 39576716975887491131345134911812652 3540265798327781881191 26207856109m^{30}}{56518757598574599193289005338475791180233003782651 5031830145926988446207 475670157221501809459 2000000000000}$$

$\theta_9 = -158.646675893947m^{18} - 2064.88853557643m^{19} - 13488.4101071534m^{20} - 58896.6151199497m^{21} - 193919.254370338m^{22} - 518302.945412828m^{23} - 1193568.32446908m^{24}$
$- 2506296.54190093m^{25} - 5048887.17129922m^{26} - 10072445.3616393m^{27} - 20037806.5355291m^{28} - 39563455.7804075m^{29} - 77620498.9170579m^{30}$

$$\theta_9 = -9.91956649282977 \cdot 10^{-18}$$

$$\theta_{10} = \frac{219377139570217 m^{20}}{1854425871872} + \frac{277666044519762625862956693 m^{21}}{1570543067310386577408 0} + \frac{18978880674483576230481157319863 m^{22}}{14256212057743206559776768 00} + \frac{111158769499930403205724283351900050892521 7 m^{23}}{166233321121676239318466630674022400 0}$$
$$+ \frac{4572595507421996144045126169303204167825805869 9 m^{24}}{1788375009365020021752357005375727206406 40} + \frac{822491251751763343260468219651556362583882180128903458845 81 m^{25}}{10359865436594895832215745850891050276426912628736000 0}$$
$$+ \frac{96592423969849188533896644268172187655616788600882726078638243223 m^{26}}{4506039658936694621746976494947954334997942564919705600000 0}$$
$$+ \frac{38790347357410276702685685685927010955647368970658635292983670030697789255267 m^{27}}{725083514271785166659708967572334524020874320658534861849493504000 0}$$
$$+ \frac{53472985277638192368178103013571933683465971756297601860971746221928883193685417935 7 m^{28}}{41336990266229261323857208441680251311957532195626375082356033976270848000 0}$$
$$+ \frac{102421182932028016616664964910559591822343904834858095693446863106451010369615452677190634993286433 m^{29}}{332584402716163230592469760058459813827747066832717708648480439509297487058409881600000000 0}$$
$$+ \frac{244110342636463817127297333843121694454623521668356572727005484053162215562062213135503459590814600693 59 m^{30}}{34289328740628015680294523939893851820524490091539554014358833876208752657393518641152000000000 0}$$

$$\theta_{10} = 118.235464372867 m^{20} + 1767.96198906688 m^{21} + 13312.7092930518 m^{22} + 67127.1525145228 m^{23} + 255684.377352463 m^{24} + 793920.786699042 m^{25} + 2143621.25682317 m^{26}$$
$$+ 5349776.48697035 m^{27} + 12935868.0768115 m^{28} + 30795546.0615623 m^{29} + 71191344.8300397 m^{30}$$

$$\theta_{10} = 5.66838640516016 \cdot 10^{-20}$$

$$\theta_{11} = -\frac{78686026266850619 m^{22}}{296868139499520} - \frac{7835194121072514829053860291 m^{23}}{186203586060319432617492480} - \frac{5276333328116014096424280666400420 3 m^{24}}{15738858111748500041993551872000} - \frac{8688632096949003728436150519725941008649377014719 m^{25}}{486540195404963447060741240148219270266880 00}$$
$$- \frac{40408608611647355143552106244174453623750605641345478649 m^{26}}{5639973945134336278328112455798157780933672960 00} - \frac{8717513068169184749363750640740453826610876770166948739200000 0 m^{27}}{2022380170793153572148544337229435813474277275629718356011484503883243 }$$
$$- \frac{161048782776534394169217086054898943780210941624590976027767312395913644333 m^{28}}{252633528715542974036561493568658351895183208799438091426201600000 0}$$
$$- \frac{947304787611667847899291038254583809943323475670672975514921507252221842884847383477 71 m^{29}}{610135821730895068092986402245814116741783887597275458107319809526988800000 0}$$
$$- \frac{280884903595095127296475102574543322868527915729032456890341121168223064607865561838142509319 m^{30}}{8047154569107382760437930481758831188340139670758697769003241384622859852185600000 0}$$

$$\theta_{11} = -265.053792567652 m^{22} - 4207.86424517858 m^{23} - 33524.244838178 m^{24} - 178579.944247303 m^{25} - 716468.001532317 m^{26} - 2319904.94878362 m^{27} - 6374798.45194654 m^{28}$$
$$- 15526129.6562503 m^{29} - 34904872.4220357 m^{30}$$

$$\theta_{11} = -8.93135341059547 \cdot 10^{-22}$$

$$\theta_{12} = \frac{475056800868533919 m^{24}}{2418925581107200} + \frac{3127150294091409605700545165759 1 m^{25}}{888260069798857145227149312 00} + \frac{252626751481205989088116386651944775621 1 m^{26}}{79638622045447410212487372472320 00}$$
$$+ \frac{3994202965352248318476252276150813392040842388895201 m^{27}}{20888792389864306604744905770302140034580480 00} + \frac{6197015549916319763221033183246094585764451281777941645352 9 m^{28}}{71345670405949353920850622565846695928810962944000 0}$$
$$+ \frac{110587860364023076831448718721409447213838620036767425443507925821902156 9 m^{29}}{3465480503642564801598923094220279175516916444436736507904000000 0}$$
$$+ \frac{19028357314006568398014324756811851273852684727731752864223412008673847059403 3 m^{30}}{1902270320387868227120537436692576280693325947210054676512768000000 0}$$

$$\theta_{12} = 196.391656105058 m^{24} + 3520.53458262459 m^{25} + 31721.7335412826 m^{26} + 191212.727423233 m^{27} + 868590.275297149 m^{28} + 3191126.31705139 m^{29} + 10002972.2958232 m^{30}$$

$$\theta_{12} = 5.11263907210428 \cdot 10^{-24}$$

$$\theta_{13} = -\frac{3773957372114969392919 9m^{26}}{87777971487218073600} - \frac{118517771257918339839357429624 12505571m^{27}}{14709586755869074324961592606 72000} - \frac{4247524051266901615636230045413456842318801m^{28}}{5606558991999497678959111022 0513280000}$$
$$- \frac{8237912954845147404874000742979558303564194337385101415927m^{29}}{1729592009841196458687287819778101719486326374400 0000} - \frac{12578431877340592254681197831194868185776209170 44838020666436907m^{30}}{55808168850875939066976486984 84008214875879768064000000 00}$$

$$\theta_{13} = -429.943561940768 m^{26} - 8057.17884702846 m^{27} - 75759.9100861701 m^{28} - 476292.26476373 m^{29} - 2253869.30557625 m^{30}$$

$$\theta_{13} = -7.65360976053436 \cdot 10^{-26}$$

$$\theta_{14} = \frac{41399088458282094127659 1m^{28}}{1304129862095811379200} + \frac{249389615732484096769486582274374999 m^{29}}{37589715933365513802918710476800} + \frac{30267787012346884552873405 1179613526990 3999379m^{30}}{4348126779395267571361888845 2160356352000}$$

$$\theta_{14} = 317.446058567752 m^{28} + 6634.51716886001 m^{29} + 69611.0958764558 m^{30}$$

$$\theta_{14} = 3.40229700650567 \cdot 10^{-28}$$

$$\theta_{15} = -\frac{68930918648965378437341 5m^{30}}{10098151228475220951 04}$$

$$\theta_{15} = -682.609292427617 m^{30}$$

$$\theta_{15} = -1.15984807842797 \cdot 10^{-30}$$

# q

$q_{1,24}$ is too long for this document. As an example, we display $q_{1,10}$.

$$q_1 = m^{\frac{2}{3}}\left(1 - \frac{2m}{3} - \frac{115m^2}{144} - \frac{599m^3}{648} - \frac{14347m^4}{62208} - \frac{76249m^5}{93312} - \frac{31682233m^6}{26873856} + \frac{51841729m^7}{100776960} + \frac{299491849721m^8}{48372940800} + \frac{76605292015663m^9}{6530347008000} + \frac{5372696615725109m^{10}}{391820820480000}\right)\cos\left(\frac{t}{m}\right)$$
$$+ m^{\frac{2}{3}}\left(\frac{3m^2}{16} + \frac{3m^3}{8} + \frac{31m^4}{96} + \frac{2381m^5}{17280} - \frac{165871m^6}{1382400} - \frac{46898921m^7}{62208000} - \frac{209730782347m^8}{89579520000} - \frac{1821549712907m^9}{391910400000} - \frac{95132047577539547m^{10}}{15801827328000000}\right)\cos\left(\frac{3t}{m}\right)$$
$$+ m^{\frac{2}{3}}\left(\frac{25m^4}{256} + \frac{113m^5}{320} + \frac{23533m^6}{38400} + \frac{96142847m^7}{145152000} + \frac{30380129071m^8}{81285120000} - \frac{5269223469151m^9}{8534937600000} - \frac{54691607012857417m^{10}}{16131032064000000}\right)\cos\left(\frac{5t}{m}\right)$$
$$+ m^{\frac{2}{3}}\left(\frac{833m^6}{12288} + \frac{55583m^7}{161280} + \frac{232353563m^8}{270950400} + \frac{939047456477m^9}{682795008000} + \frac{848348752689791m^{10}}{573547806720000}\right)\cos\left(\frac{7t}{m}\right) + m^{\frac{2}{3}}\left(\frac{3537m^8}{65536} + \frac{16905377m^9}{48168960} + \frac{4880409049m^{10}}{4335206400}\right)\cos\left(\frac{9t}{m}\right)$$
$$+ m^{\frac{2}{3}}\left(\frac{732413m^{10}}{15728640}\right)\cos\left(\frac{11t}{m}\right)$$

$$q_1 = +m^{0.666666666666667}(1.0 - 0.666666666666667m - 0.798611111111111m^2 - 0.924382716049383m^3 - 0.230629501028807m^4 - 0.817140346364883m^5 - 1.17892396982406m^6$$
$$+ 0.514420448880379m^7 + 6.19130953727337m^8 + 11.7306617736879m^9 + 13.7121263978348m^{10})\cos\left(\frac{t}{m}\right)$$
$$+ m^{0.666666666666667}(0.1875m^2 + 0.375m^3 + 0.322916666666667m^4 + 0.137789351851852m^5 - 0.119987702546296m^6 - 0.753904980066872m^7 - 2.34128048852014m^8$$
$$- 4.64787286304982m^9 - 6.02031939742631m^{10})\cos\left(\frac{3t}{m}\right)$$
$$+ m^{0.666666666666667}(0.09765625m^4 + 0.353125m^5 + 0.612838541666667m^6 + 0.662359781470459m^7 + 0.373747729855108m^8 - 0.617371059531941m^9$$
$$- 3.39045925864309m^{10})\cos\left(\frac{5t}{m}\right)$$
$$+ m^{0.666666666666667}(0.0677897135416667m^6 + 0.344636656746032m^7 + 0.857550175235025m^8 + 1.37529924131636m^9 + 1.47912474383142m^{10})\cos\left(\frac{7t}{m}\right)$$
$$+ m^{0.666666666666667}(0.0539703369140625m^8 + 0.350959975054475m^9 + 1.12576163593964m^{10})\cos\left(\frac{9t}{m}\right) + m^{10.6666666666667}0.0465655644734701\cos\left(\frac{11t}{m}\right)$$

$$q_1 = 0.175827099956786\cos(12.3687469072992t) + 0.000268869269078907\cos(37.1062407218977t) + 1.04313243563513 \cdot 10^{-6}\cos(61.8437345364962t) + 5.328879422264$$
$$\cdot 10^{-9}\cos(86.5812283510947t) + 3.0618751075018 \cdot 10^{-11}\cos(111.318722165693t) + 1.03894057349971 \cdot 10^{-13}\cos(136.056215980292t)$$

$$q_2 = m^{\frac{2}{3}}\left(1 - \frac{2m}{3} + \frac{227m^2}{144} + \frac{535m^3}{648} + \frac{53477m^4}{62208} - \frac{18073m^5}{93312} - \frac{4862647m^6}{26873856} - \frac{236184929m^7}{100776960} - \frac{302769729871m^8}{48372940800} - \frac{72827805074513m^9}{6530347008000} - \frac{4062893273303209m^{10}}{391820820480000}\right)\sin\left(\frac{t}{m}\right)$$
$$+ m^{\frac{2}{3}}\left(\frac{3m^2}{16} + \frac{3m^3}{8} + \frac{31m^4}{96} + \frac{1139m^5}{17280} - \frac{444079m^6}{1382400} - \frac{62642729m^7}{62208000} - \frac{233316418453m^8}{89579520000} - \frac{1986826581893m^9}{391910400000} - \frac{102616704110645003m^{10}}{15801827328000000}\right)\sin\left(\frac{3t}{m}\right)$$
$$+ m^{\frac{2}{3}}\left(\frac{25m^4}{256} + \frac{113m^5}{320} + \frac{7711m^6}{12800} + \frac{87056897m^7}{145152000} + \frac{16909308271m^8}{81285120000} - \frac{7521567147551m^9}{8534937600000} - \frac{59601171355631317m^{10}}{16131032064000000}\right)\sin\left(\frac{5t}{m}\right)$$
$$+ m^{\frac{2}{3}}\left(\frac{833m^6}{12288} + \frac{55583m^7}{161280} + \frac{230324963m^8}{270950400} + \frac{904852584227m^9}{682795008000} + \frac{757145763542591m^{10}}{573547806720000}\right)\sin\left(\frac{7t}{m}\right) + m^{\frac{2}{3}}\left(\frac{3537m^8}{65536} + \frac{16905377m^9}{48168960} + \frac{4856753809m^{10}}{4335206400}\right)\sin\left(\frac{9t}{m}\right)$$
$$+ m^{\frac{2}{3}}\left(\frac{732413m^{10}}{15728640}\right)\sin\left(\frac{11t}{m}\right)$$

$$\begin{aligned}
q_2 = {} & 0.0465655644734701\, m^{10.6666666666667} \sin\left(\frac{11t}{m}\right) \\
& + m^{0.666666666666667}\big(1.0 - 0.666666666666667\, m + 1.57638888888889\, m^2 + 0.825617283950617\, m^3 + 0.859648276748971\, m^4 - 0.193683556241427\, m^5 \\
& \quad - 0.180943404623438\, m^6 - 2.34364014354075\, m^7 - 6.25907221813977\, m^8 - 11.1522105923767\, m^9 - 10.3692633493186\, m^{10}\big) \sin\left(\frac{t}{m}\right) \\
& + m^{0.666666666666667}\big(0.1875\, m^2 + 0.375\, m^3 + 0.322916666666667\, m^4 + 0.0659143518518518\, m^5 - 0.321237702546296\, m^6 - 1.00698831340021\, m^7 - 2.60457321554078\, m^8 \\
& \quad - 5.06959392221538\, m^9 - 6.4939770559835\, m^{10}\big) \sin\left(\frac{3t}{m}\right) \\
& + m^{0.666666666666667}\big(0.09765625\, m^4 + 0.353125\, m^5 + 0.602421875\, m^6 + 0.599763675319665\, m^7 + 0.20802464548247\, m^8 - 0.881267971724949\, m^9 \\
& \quad - 3.69481451150572\, m^{10}\big) \sin\left(\frac{5t}{m}\right) \\
& + m^{0.666666666666667}\big(0.0677897135416667\, m^6 + 0.344636656746032\, m^7 + 0.850063196068358\, m^8 + 1.32521851159609\, m^9 + 1.32010924751426\, m^{10}\big) \sin\left(\frac{7t}{m}\right) \\
& + m^{0.666666666666667}\big(0.0539703369140625\, m^8 + 0.350959975054475\, m^9 + 1.12030509297089\, m^{10}\big) \sin\left(\frac{9t}{m}\right)
\end{aligned}$$

$$\begin{aligned}
q_2 = {} & 0.178911819842929 \sin(12.3687469072992\, t) + 0.000268811166063915 \sin(37.1062407218977\, t) + 1.04225968786912 \cdot 10^{-6} \sin(61.8437345364962\, t) + 5.32458704526969 \\
& \cdot 10^{-9} \sin(86.5812283510947\, t) + 3.0606576791413 \cdot 10^{-11} \sin(111.318722165693\, t) + 1.03894057349971 \cdot 10^{-13} \sin(136.056215980292\, t)
\end{aligned}$$

$\dot{q}$

$$\dot{q}_1 = -\frac{\left(1 - \frac{2m}{3} - \frac{115m^2}{144} - \frac{599m^3}{648} - \frac{14347m^4}{62208} - \frac{76249m^5}{93312} - \frac{31682233m^6}{26873856} + \frac{51841729m^7}{100776960} + \frac{299491849721m^8}{48372940800} + \frac{76605292015663m^9}{6530347008000} + \frac{5372696615725109m^{10}}{391820820480000}\right)\sin\left(\frac{t}{m}\right)}{\sqrt[3]{m}}$$

$$-\frac{3\left(\frac{3m^2}{16} + \frac{3m^3}{8} + \frac{31m^4}{96} + \frac{2381m^5}{17280} - \frac{165871m^6}{1382400} - \frac{46898921m^7}{62208000} - \frac{209730782347m^8}{89579520000} - \frac{1821549712907m^9}{391910400000} - \frac{95132047577539547m^{10}}{158018273280000000}\right)\sin\left(\frac{3t}{m}\right)}{\sqrt[3]{m}}$$

$$-\frac{5\left(\frac{25m^4}{256} + \frac{113m^5}{320} + \frac{23533m^6}{38400} + \frac{96142847m^7}{145152000} + \frac{30380129071m^8}{81285120000} - \frac{5269223469151m^9}{8534937600000} - \frac{546916070712857417m^{10}}{16131032064000000}\right)\sin\left(\frac{5t}{m}\right)}{\sqrt[3]{m}}$$

$$-\frac{7\left(\frac{833m^6}{12288} + \frac{55583m^7}{161280} + \frac{232353563m^8}{270950400} + \frac{939047456477m^9}{682795008000} + \frac{848348752689791m^{10}}{573547806720000}\right)\sin\left(\frac{7t}{m}\right)}{\sqrt[3]{m}} - \frac{9\left(\frac{3537m^8}{65536} + \frac{16905377m^9}{48168960} + \frac{4880409049m^{10}}{4335206400}\right)\sin\left(\frac{9t}{m}\right)}{\sqrt[3]{m}}$$

$$-\frac{\left(\frac{8056543m^{10}}{15728640}\right)\sin\left(\frac{11t}{m}\right)}{\sqrt[3]{m}}$$

$\dot{q}_1$
$= -0.512221209208171 m^{9.66666666666667} \sin\left(\frac{11t}{m}\right)$

$$-\frac{\left(1.0 - 0.666666666666667m - 0.798611111111111m^2 - 0.924382716049383m^3 - 0.230629501028807m^4 - 0.817140346364883m^5 - 1.17892396982406m^6 + 0.514420448880379m^7 + 6.19130953727337m^8 + 11.7306617736879m^9 + 13.7121263978348m^{10}\right)\sin\left(\frac{t}{m}\right)}{m^{0.333333333333333}}$$

$$-\frac{3.0\left(0.1875m^2 + 0.375m^3 + 0.322916666666667m^4 + 0.137789351851852m^5 - 0.119987702546296m^6 - 0.753904980066872m^7 - 2.34128048852014m^8 - 4.64787286304982m^9 - 6.02031939742631m^{10}\right)\sin\left(\frac{3t}{m}\right)}{m^{0.333333333333333}}$$

$$-\frac{5.0(0.09765625m^4 + 0.353125m^5 + 0.612838541666667m^6 + 0.662359781470459m^7 + 0.373747729855108m^8 - 0.617371059531941m^9 - 3.39045925864309m^{10})\sin\left(\frac{5t}{m}\right)}{m^{0.333333333333333}}$$

$$-\frac{7.0(0.0677897135416667m^6 + 0.344636656746032m^7 + 0.857550175235025m^8 + 1.37529924131636m^9 + 1.47912474383142m^{10})\sin\left(\frac{7t}{m}\right)}{m^{0.333333333333333}}$$

$$-\frac{9.0(0.0539703369140625m^8 + 0.350959975054475m^9 + 1.12576163593964m^{10})\sin\left(\frac{9t}{m}\right)}{m^{0.333333333333333}}$$

$\dot{q}_1 = -2.17476089880989 \sin(12.3687469072992t) - 0.00997672782116261 \sin(37.1062407218977t) - 6.45112054358276 \cdot 10^{-5} \sin(61.8437345364962t) - 4.61380926114488 \cdot 10^{-7} \sin(86.5812283510947t) - 3.40844024398045 \cdot 10^{-9} \sin(111.318722165693t) - 1.41354323058764 \cdot 10^{-11} \sin(136.056215980292t)$

$$\dot{q}_2 = \frac{\left(1 - \frac{2m}{3} + \frac{227m^2}{144} + \frac{535m^3}{648} + \frac{53477m^4}{62208} - \frac{18073m^5}{93312} - \frac{4862647m^6}{26873856} - \frac{236184929m^7}{100776960} - \frac{302769729871m^8}{48372940800} - \frac{72827805074513m^9}{6530347008000} - \frac{4062893273303209m^{10}}{391820820480000}\right)\cos\left(\frac{t}{m}\right)}{\sqrt[3]{m}}$$

$$+ \frac{3\left(\frac{3m^2}{16} + \frac{3m^3}{8} + \frac{31m^4}{96} + \frac{1139m^5}{17280} - \frac{444079m^6}{1382400} - \frac{62642729m^7}{62208000} - \frac{233316418453m^8}{89579520000} - \frac{1986826581893m^9}{391910400000} - \frac{102616704110645003m^{10}}{15801827328000000}\right)\cos\left(\frac{3t}{m}\right)}{\sqrt[3]{m}}$$

$$+ \frac{5\left(\frac{25m^4}{256} + \frac{113m^5}{320} + \frac{7711m^6}{12800} + \frac{87056897m^7}{145152000} + \frac{16909308271m^8}{81285120000} - \frac{7521567147551m^9}{8534937600000} - \frac{596011713556313 17m^{10}}{16131032064000000}\right)\cos\left(\frac{5t}{m}\right)}{\sqrt[3]{m}}$$

$$+ \frac{7\left(\frac{833m^6}{12288} + \frac{55583m^7}{161280} + \frac{230324963m^8}{270950400} + \frac{904852584227m^9}{682795008000} + \frac{757145763542591m^{10}}{573547806720000}\right)\cos\left(\frac{7t}{m}\right)}{\sqrt[3]{m}} + \frac{9\left(\frac{3537m^8}{65536} + \frac{16905377m^9}{48168960} + \frac{4856753809m^{10}}{4335206400}\right)\cos\left(\frac{9t}{m}\right)}{\sqrt[3]{m}}$$

$$+ \frac{\left(\frac{8056543m^{10}}{15728640}\right)\cos\left(\frac{11t}{m}\right)}{\sqrt[3]{m}}$$

$\dot{q}_2$
$= 0.512221209208171 m^{9.66666666666667} \cos\left(\frac{11t}{m}\right)$

$$+ \frac{\left(1.0 - 0.666666666666667m + 1.57638888888889m^2 + 0.825617283950617m^3 + 0.859648276748971m^4 - 0.193683556241427m^5 - 0.180943404623438m^6 - 2.34364014354075m^7\right.}{\left.-6.25907221813977m^8 - 11.1522105923767m^9 - 10.3692633493186m^{10}\right)\cos\left(\frac{t}{m}\right)}{m^{0.333333333333333}}$$

$$+ \frac{3.0\left(0.1875m^2 + 0.375m^3 + 0.322916666666667m^4 + 0.0659143518518518m^5 - 0.321237702546296m^6 - 1.00698831340021m^7 - 2.60457321554078m^8 - 5.06959392221538m^9\right.}{\left.-6.4939770559835m^{10}\right)\cos\left(\frac{3t}{m}\right)}{m^{0.333333333333333}}$$

$$+ \frac{5.0(0.09765625m^4 + 0.353125m^5 + 0.602421875m^6 + 0.599763675319665m^7 + 0.20802464548247m^8 - 0.881267971724949m^9 - 3.69481451150572m^{10})\cos\left(\frac{5t}{m}\right)}{m^{0.333333333333333}}$$

$$+ \frac{7.0(0.0677897135416667m^6 + 0.344636656746032m^7 + 0.850063196068358m^8 + 1.32521851159609m^9 + 1.32010924751426m^{10})\cos\left(\frac{7t}{m}\right)}{m^{0.333333333333333}}$$

$$+ \frac{9.0(0.0539703369140625m^8 + 0.350959975054475m^9 + 1.12030509297089m^{10})\cos\left(\frac{9t}{m}\right)}{m^{0.333333333333333}}$$

$\dot{q}_2 = 2.21291501836151 \cos(12.3687469072992 t) + 0.00997457183670166 \cos(37.1062407218977 t) + 6.44572314546692 \cdot 10^{-5} \cos(61.8437345364962 t) + 4.61009286841776$
$\cdot 10^{-7} \cos(86.5812283510947 t) + 3.40708501828626 \cdot 10^{-9} \cos(111.318722165693 t) + 1.41354323058764 \cdot 10^{-11} \cos(136.056215980292 t)$

$\ddot q$

$$\ddot q_1 = -\frac{\left(1 - \frac{2m}{3} - \frac{115m^2}{144} - \frac{599m^3}{648} - \frac{14347m^4}{62208} - \frac{76249m^5}{93312} - \frac{31682233m^6}{26873856} + \frac{51841729m^7}{100776960} + \frac{299491849721m^8}{48372940800} + \frac{76605292015663m^9}{6530347008000} + \frac{5372696615725109m^{10}}{391820820480000}\right)\cos\left(\frac{t}{m}\right)}{m^{\frac{4}{3}}}$$

$$-\frac{9\left(\frac{3m^2}{16} + \frac{3m^3}{8} + \frac{31m^4}{96} + \frac{2381m^5}{17280} - \frac{165871m^6}{1382400} - \frac{46898921m^7}{62208000} - \frac{209730782347m^8}{89579520000} - \frac{1821549712907m^9}{391910400000} - \frac{95132047577539547m^{10}}{15801827328000000}\right)\cos\left(\frac{3t}{m}\right)}{m^{\frac{4}{3}}}$$

$$-\frac{25\left(\frac{25m^4}{256} + \frac{113m^5}{320} + \frac{23533m^6}{38400} + \frac{96142847m^7}{145152000} + \frac{30380129071m^8}{81285120000} - \frac{5269223469151m^9}{8534937600000} - \frac{546916070128574177m^{10}}{16131032064000000}\right)\cos\left(\frac{5t}{m}\right)}{m^{\frac{4}{3}}}$$

$$-\frac{49\left(\frac{833m^6}{12288} + \frac{55583m^7}{161280} + \frac{232353563m^8}{270950400} + \frac{939047456477m^9}{682795008000} + \frac{848348752689791m^{10}}{573547806720000}\right)\cos\left(\frac{7t}{m}\right)}{m^{\frac{4}{3}}} - \frac{81\left(\frac{3537m^8}{65536} + \frac{16905377m^9}{48168960} + \frac{4880409049m^{10}}{4335206400}\right)\cos\left(\frac{9t}{m}\right)}{m^{\frac{4}{3}}}$$

$$-\frac{\left(\frac{88621973m^{10}}{15728640}\right)\cos\left(\frac{11t}{m}\right)}{m^{\frac{4}{3}}}$$

$\ddot q_1$
$= -5.63443330128988 m^{8.66666666666667}\cos\left(\frac{11t}{m}\right)$

$$-\frac{\left(\begin{array}{l}1.0 - 0.666666666666667m - 0.798611111111111m^2 - 0.924382716049383m^3 - 0.230629501028807m^4 - 0.817140346364883m^5 - 1.17892396982406m^6 + 0.514420448880379m^7 \\ + 6.19130953727337m^8 + 11.7306617736879m^9 + 13.7121263978348m^{10}\end{array}\right)\cos\left(\frac{t}{m}\right)}{m^{1.33333333333333}}$$

$$-\frac{9.0\left(\begin{array}{l}0.1875m^2 + 0.375m^3 + 0.322916666666667m^4 + 0.137789351851852m^5 - 0.119987702546296m^6 - 0.753904980066872m^7 - 2.34128048852014m^8 - 4.64787286304982m^9 \\ - 6.02031939742631m^{10}\end{array}\right)\cos\left(\frac{3t}{m}\right)}{m^{1.33333333333333}}$$

$$-\frac{25.0(0.09765625m^4 + 0.353125m^5 + 0.612838541666667m^6 + 0.662359781470459m^7 + 0.373747729855108m^8 - 0.617371059531941m^9 - 3.39045925864309m^{10})\cos\left(\frac{5t}{m}\right)}{m^{1.33333333333333}}$$

$$-\frac{49.0(0.0677897135416667m^6 + 0.344636656746032m^7 + 0.857550175235025m^8 + 1.37529924131636m^9 + 1.47912474383142m^{10})\cos\left(\frac{7t}{m}\right)}{m^{1.33333333333333}}$$

$$-\frac{81.0(0.0539703369140625m^8 + 0.350959975054475m^9 + 1.12576163593964m^{10})\cos\left(\frac{9t}{m}\right)}{m^{1.33333333333333}}$$

$\ddot q_1 = -26.8990671412702\cos(12.3687469072992t) - 0.370198864148914\cos(37.1062407218977t) - 0.00398961386360269\cos(61.8437345364962t) - 3.99469273207581 \cdot 10^{-5}\cos(86.5812283510947t) - 3.79423212538027 \cdot 10^{-7}\cos(111.318722165693t) - 1.92321343078311 \cdot 10^{-9}\cos(136.056215980292t)$

$$\ddot{q}_2 = -\frac{\left(1 - \frac{2m}{3} + \frac{227m^2}{144} + \frac{535m^3}{648} + \frac{53477m^4}{62208} - \frac{18073m^5}{93312} - \frac{4862647m^6}{26873856} - \frac{236184929m^7}{100776960} - \frac{302769729871m^8}{48372940800} - \frac{72827805074513m^9}{6530347008000} - \frac{4062893273303209m^{10}}{391820820480000}\right) sin\left(\frac{t}{m}\right)}{m^{\frac{4}{3}}}$$

$$-\frac{9\left(\frac{3m^2}{16} + \frac{3m^3}{8} + \frac{31m^4}{96} + \frac{1139m^5}{17280} - \frac{444079m^6}{1382400} - \frac{62642729m^7}{62208000} - \frac{233316418453m^8}{89579520000} - \frac{1986826581893m^9}{391910400000} - \frac{102616704110645003m^{10}}{15801827328000000}\right) sin\left(\frac{3t}{m}\right)}{m^{\frac{4}{3}}}$$

$$-\frac{25\left(\frac{25m^4}{256} + \frac{113m^5}{320} + \frac{7711m^6}{12800} + \frac{87056897m^7}{145152000} + \frac{16909308271m^8}{81285120000} - \frac{7521567147551m^9}{8534937600000} - \frac{596011171355631317m^{10}}{16131032064000000}\right) sin\left(\frac{5t}{m}\right)}{m^{\frac{4}{3}}}$$

$$-\frac{49\left(\frac{833m^6}{12288} + \frac{55583m^7}{161280} + \frac{230324963m^8}{270950400} + \frac{904852584227m^9}{682795008000} + \frac{757145763542591m^{10}}{573547806720000}\right) sin\left(\frac{7t}{m}\right)}{m^{\frac{4}{3}}} - \frac{81\left(\frac{3537m^8}{65536} + \frac{16905377m^9}{48168960} + \frac{4856753809m^{10}}{4335206400}\right) sin\left(\frac{9t}{m}\right)}{m^{\frac{4}{3}}}$$

$$-\frac{\left(\frac{88621973m^{10}}{15728640}\right) sin\left(\frac{11t}{m}\right)}{m^{\frac{4}{3}}}$$

$$\ddot{q}_2 = -5.63443330128988 m^{8.66666666666667} sin\left(\frac{11t}{m}\right)$$

$$-\frac{\left(1.0 - 0.666666666666667 m + 1.57638888888889 m^2 + 0.825617283950617 m^3 + 0.859648276748971 m^4 - 0.193683556241427 m^5 - 0.180943404623438 m^6 - 2.34364014354075 m^7 - 6.25907221813977 m^8 - 11.1522105923767 m^9 - 10.3692633493186 m^{10}\right) sin\left(\frac{t}{m}\right)}{m^{1.33333333333333}}$$

$$-\frac{9.0\left(0.1875 m^2 + 0.375 m^3 + 0.322916666666667 m^4 + 0.0659143518518518 m^5 - 0.321237702546296 m^6 - 1.00698831340021 m^7 - 2.60457321554078 m^8 - 5.06959392221538 m^9 - 6.4939770559835 m^{10}\right) sin\left(\frac{3t}{m}\right)}{m^{1.33333333333333}}$$

$$-\frac{25.0(0.09765625 m^4 + 0.353125 m^5 + 0.602421875 m^6 + 0.599763675319665 m^7 + 0.20802464548247 m^8 - 0.881267971724949 m^9 - 3.69481451150572 m^{10}) sin\left(\frac{5t}{m}\right)}{m^{1.33333333333333}}$$

$$-\frac{49.0(0.0677897135416667 m^6 + 0.344636656746032 m^7 + 0.850063196068358 m^8 + 1.32521851159609 m^9 + 1.32010924751426 m^{10}) sin\left(\frac{7t}{m}\right)}{m^{1.33333333333333}}$$

$$-\frac{81.0(0.0539703369140625 m^8 + 0.350959975054475 m^9 + 1.12030509297089 m^{10}) sin\left(\frac{9t}{m}\right)}{m^{1.33333333333333}}$$

$$\ddot{q}_2 = -5.63443330128988 m^{8.66666666666667} \sin\left(\frac{11t}{m}\right)$$

$$- \frac{\left(1.0 - 0.666666666666667m + 1.57638888888889m^2 + 0.825617283950617m^3 + 0.859648276748971m^4 - 0.193683556241427m^5 - 0.180943404623438m^6 - 2.34364014354075m^7 -6.25907221813977m^8 - 11.1522105923767m^9 - 10.3692633493186m^{10}\right) \sin\left(\frac{t}{m}\right)}{m^{1.33333333333333}}$$

$$- \frac{9.0\left(0.1875m^2 + 0.375m^3 + 0.322916666666667m^4 + 0.0659143518518518m^5 - 0.321237702546296m^6 - 1.00698831340021m^7 - 2.60457321554078m^8 - 5.06959392221538m^9 - 6.4939770559835m^{10}\right)\sin\left(\frac{3t}{m}\right)}{m^{1.33333333333333}}$$

$$- \frac{25.0\left(0.09765625m^4 + 0.353125m^5 + 0.602421875m^6 + 0.599763675319665m^7 + 0.20802464548247m^8 - 0.881267971724949m^9 - 3.69481451150572m^{10}\right)\sin\left(\frac{5t}{m}\right)}{m^{1.33333333333333}}$$

$$- \frac{49.0\left(0.0677897135416667m^6 + 0.344636656746032m^7 + 0.850063196068358m^8 + 1.32521851159609m^9 + 1.32010924751426m^{10}\right)\sin\left(\frac{7t}{m}\right)}{m^{1.33333333333333}}$$

$$- \frac{81.0\left(0.0539703369140625m^8 + 0.350959975054475m^9 + 1.12030509297089m^{10}\right)\sin\left(\frac{9t}{m}\right)}{m^{1.33333333333333}}$$

## A2. Appendix 2 – Eliminate all $w_i$ from Equation 0

Now testing solution of Hill literal perigee. This document looks only at elimination of $w_i$ (function cElim). The other document looks at solving for c (cSolve).

solve for c, maxLines=0, maxTheta=1, maxK=3
eqs

$$0: s_0 w_0$$

checking out equation 0

$$s_0$$

solve for c, maxLines=1, maxTheta=1, maxK=4
eqs

$$-1: -w_0 \theta_1 + s_{-1} w_{-1}$$
$$0: -w_1 \theta_1 - w_{-1} \theta_1 + s_0 w_0$$
$$1: -w_0 \theta_1 + s_1 w_1$$

eqSol

$$-1: \frac{w_0 \theta_1}{s_{-1}}$$

eqs

$$-1: 0$$
$$0: -\frac{w_0 \theta_1^2}{s_{-1}} - w_1 \theta_1 + s_0 w_0$$
$$1: -w_0 \theta_1 + s_1 w_1$$

eqSol

$$-1: \frac{w_0 \theta_1}{s_{-1}}$$
$$1: \frac{w_0 \theta_1}{s_1}$$

eqs

$$-1: 0$$
$$0: -\frac{w_0 \theta_1^2}{s_{-1}} - \frac{w_0 \theta_1^2}{s_1} + s_0 w_0$$
$$1: 0$$

checking out equation 0

$$s_0 - \left(\frac{1}{s_1} + \frac{1}{s_{-1}}\right) \theta_1^2$$

solve for c, maxLines=2, maxTheta=2, maxK=8
eqs

$$-2: -w_0 \theta_2 - w_{-1} \theta_1 + s_{-2} w_{-2}$$
$$-1: -w_1 \theta_2 - w_0 \theta_1 - w_{-2} \theta_1 + s_{-1} w_{-1}$$
$$0: -w_2 \theta_2 - w_1 \theta_1 - w_{-2} \theta_2 - w_{-1} \theta_1 + s_0 w_0$$

## A2. Appendix 2 – Eliminate all $w_i$ from Equation 0

$$1: -w_2\theta_1 - w_0\theta_1 - w_{-1}\theta_2 + s_1 w_1$$

$$2: -w_1\theta_1 - w_0\theta_2 + s_2 w_2$$

eqSol

$$-2: \frac{w_0\theta_2 + w_{-1}\theta_1}{s_{-2}}$$

eqs

$$-2: 0$$

$$-1: \frac{-\theta_1(w_0\theta_2 + w_{-1}\theta_1) + s_{-2}(-w_1\theta_2 - w_0\theta_1 + s_{-1}w_{-1})}{s_{-2}}$$

$$0: \frac{-\theta_2(w_0\theta_2 + w_{-1}\theta_1) + s_{-2}(-w_2\theta_2 - w_1\theta_1 - w_{-1}\theta_1 + s_0 w_0)}{s_{-2}}$$

$$1: -w_2\theta_1 - w_0\theta_1 - w_{-1}\theta_2 + s_1 w_1$$

$$2: -w_1\theta_1 - w_0\theta_2 + s_2 w_2$$

eqSol

$$-2: \frac{w_0\theta_2 + w_{-1}\theta_1}{s_{-2}}$$

$$-1: \frac{w_0\theta_1\theta_2 + s_{-2}w_1\theta_2 + s_{-2}w_0\theta_1}{-\theta_1^2 + s_{-1}s_{-2}}$$

eqs

$$-2: 0$$

$$-1: 0$$

$$0: \frac{-\theta_2\left(\theta_1(w_0\theta_1\theta_2 + s_{-2}w_1\theta_2 + s_{-2}w_0\theta_1) + w_0\theta_2(-\theta_1^2 + s_{-1}s_{-2})\right) - s_{-2}\left((-\theta_1^2 + s_{-1}s_{-2})(w_2\theta_2 + w_1\theta_1 - s_0 w_0) + \theta_1(w_0\theta_1\theta_2 + s_{-2}w_1\theta_2 + s_{-2}w_0\theta_1)\right)}{s_{-2}(-\theta_1^2 + s_{-1}s_{-2})}$$

$$1: \frac{(-\theta_1^2 + s_{-1}s_{-2})(-w_2\theta_1 - w_0\theta_1 + s_1 w_1) - \theta_2(w_0\theta_1\theta_2 + s_{-2}w_1\theta_2 + s_{-2}w_0\theta_1)}{-\theta_1^2 + s_{-1}s_{-2}}$$

$$2: -w_1\theta_1 - w_0\theta_2 + s_2 w_2$$

eqSol

$$-2: \frac{w_0\theta_2 + w_{-1}\theta_1}{s_{-2}}$$

$$-1: \frac{w_0\theta_1\theta_2 + s_{-2}w_1\theta_2 + s_{-2}w_0\theta_1}{-\theta_1^2 + s_{-1}s_{-2}}$$

$$1: \frac{\theta_1(-w_2\theta_1^2 + w_0\theta_2^2 - w_0\theta_1^2 + s_{-2}w_0\theta_2 + s_{-1}s_{-2}w_2 + s_{-1}s_{-2}w_0)}{-s_1\theta_1^2 - s_{-2}\theta_2^2 + s_{-1}s_{-2}s_1}$$

eqs

$$-2: 0$$

$$-1: 0$$

$$0: \frac{\begin{matrix}-w_2\theta_1^2\theta_2^2 + w_2\theta_1^4 + w_0\theta_2^4 - 2w_0\theta_1^2\theta_2^2 + w_0\theta_1^4 + s_1w_2\theta_1^2\theta_2 - 2s_1w_0\theta_1^2\theta_2 - s_0s_1w_0\theta_1^2 \\ +s_{-2}w_2\theta_2^3 - s_{-2}w_2\theta_1^2\theta_2 - 2s_{-2}w_0\theta_1^2\theta_2 - s_{-2}s_1w_0\theta_1^2 - s_{-2}s_0w_0\theta_2^2 - s_{-1}s_1w_0\theta_2^2 - s_{-1}s_{-2}w_2\theta_1^2 \\ -s_{-1}s_{-2}w_0\theta_1^2 - s_{-1}s_{-2}s_1w_2\theta_2 + s_{-1}s_{-2}s_0s_1w_0\end{matrix}}{-s_1\theta_1^2 - s_{-2}\theta_2^2 + s_{-1}s_{-2}s_1}$$

$$1: 0$$

$$2: \frac{(-w_0\theta_2 + s_2w_2)(s_1\theta_1^2 + s_{-2}\theta_2^2 - s_{-1}s_{-2}s_1)}{+\theta_1^2(-w_2\theta_1^2 + w_0\theta_2^2 - w_0\theta_1^2 + s_{-2}w_0\theta_2 + s_{-1}s_{-2}w_2 + s_{-1}s_{-2}w_0)}{s_1\theta_1^2 + s_{-2}\theta_2^2 - s_{-1}s_{-2}s_1}$$

eqSol

$$-2: \frac{w_0\theta_2 + w_{-1}\theta_1}{s_{-2}}$$

$$-1: \frac{w_0\theta_1\theta_2 + s_{-2}w_1\theta_2 + s_{-2}w_0\theta_1}{-\theta_1^2 + s_{-1}s_{-2}}$$

$$1: \frac{\theta_1(-w_2\theta_1^2 + w_0\theta_2^2 - w_0\theta_1^2 + s_{-2}w_0\theta_2 + s_{-1}s_{-2}w_2 + s_{-1}s_{-2}w_0)}{-s_1\theta_1^2 - s_{-2}\theta_2^2 + s_{-1}s_{-2}s_1}$$

$$2: \frac{w_0(\theta_1^2\theta_2^2 - \theta_1^4 - s_1\theta_1^2\theta_2 - s_{-2}\theta_2^3 + s_{-2}\theta_1^2\theta_2 + s_{-1}s_{-2}\theta_1^2 + s_{-1}s_{-2}s_1\theta_2)}{\theta_1^4 - s_1s_2\theta_1^2 - s_{-2}s_2\theta_2^2 - s_{-1}s_{-2}\theta_1^2 + s_{-1}s_{-2}s_1s_2}$$

eqs

$$-2: 0$$

$$-1: 0$$

$$0: \frac{w_0 \begin{pmatrix}-2\theta_1^2\theta_2^3 + 4\theta_1^4\theta_2 + s_2\theta_2^4 - 2s_2\theta_1^2\theta_2^2 + s_2\theta_1^4 + s_1\theta_1^2\theta_2^2 - 2s_1s_2\theta_1^2\theta_2 + s_0\theta_1^4 - s_0s_1s_2\theta_1^2 \\ +s_{-2}\theta_2^4 - 2s_{-2}\theta_1^2\theta_2^2 + s_{-2}\theta_1^4 - 2s_{-2}s_2\theta_1^2\theta_2 - s_{-2}s_1s_2\theta_1^2 - s_{-2}s_0s_2\theta_2^2 + s_{-1}\theta_1^2\theta_2^2 \\ -s_{-1}s_1s_2\theta_2^2 - 2s_{-1}s_{-2}\theta_1^2\theta_2 - s_{-1}s_{-2}s_2\theta_1^2 - s_{-1}s_{-2}s_1\theta_2^2 - s_{-1}s_{-2}s_0\theta_1^2 + s_{-1}s_{-2}s_0s_1s_2\end{pmatrix}}{\theta_1^4 - s_1s_2\theta_1^2 - s_{-2}s_2\theta_2^2 - s_{-1}s_{-2}\theta_1^2 + s_{-1}s_{-2}s_1s_2}$$

$$1: 0$$
$$2: 0$$

checking out equation 0

$$s_0 - \left(\frac{1}{s_1} + \frac{1}{s_{-1}}\right)\theta_1^2 + \left(\frac{1}{s_{-1}s_1s_2} + \frac{1}{s_{-1}s_{-2}s_1}\right)\theta_1^4 - 2\left(\frac{1}{s_1s_2} + \frac{1}{s_{-1}s_1} + \frac{1}{s_{-1}s_{-2}}\right)\theta_1^2\theta_2 - \left(\frac{1}{s_2} + \frac{1}{s_{-2}}\right)\theta_2^2$$

solve for c, maxLines=3, maxTheta=3, maxK=12

eqs

$$-3: -w_0\theta_3 - w_{-2}\theta_1 - w_{-1}\theta_2 + s_{-3}w_{-3}$$

$$-2: -w_1\theta_3 - w_0\theta_2 - w_{-3}\theta_1 - w_{-1}\theta_1 + s_{-2}w_{-2}$$

$$-1: -w_2\theta_3 - w_1\theta_2 - w_0\theta_1 - w_{-3}\theta_2 - w_{-2}\theta_1 + s_{-1}w_{-1}$$

$$0: -w_3\theta_3 - w_2\theta_2 - w_1\theta_1 - w_{-3}\theta_3 - w_{-2}\theta_2 - w_{-1}\theta_1 + s_0w_0$$

$$1: -w_3\theta_2 - w_2\theta_1 - w_0\theta_1 - w_{-2}\theta_3 - w_{-1}\theta_2 + s_1w_1$$

$$2: -w_3\theta_1 - w_1\theta_1 - w_0\theta_2 - w_{-1}\theta_3 + s_2w_2$$

$$3: -w_2\theta_1 - w_1\theta_2 - w_0\theta_3 + s_3w_3$$

eqSol

$$-3: \frac{w_0\theta_3 + w_{-2}\theta_1 + w_{-1}\theta_2}{s_{-3}}$$

eqs

$$-3: 0$$

$$-2: \frac{-\theta_1(w_0\theta_3 + w_{-2}\theta_1 + w_{-1}\theta_2) + s_{-3}(-w_1\theta_3 - w_0\theta_2 - w_{-1}\theta_1 + s_{-2}w_{-2})}{s_{-3}}$$

$$-1: \frac{-\theta_2(w_0\theta_3 + w_{-2}\theta_1 + w_{-1}\theta_2) + s_{-3}(-w_2\theta_3 - w_1\theta_2 - w_0\theta_1 - w_{-2}\theta_1 + s_{-1}w_{-1})}{s_{-3}}$$

$$0: \frac{-\theta_3(w_0\theta_3 + w_{-2}\theta_1 + w_{-1}\theta_2) + s_{-3}(-w_3\theta_3 - w_2\theta_2 - w_1\theta_1 - w_{-2}\theta_2 - w_{-1}\theta_1 + s_0 w_0)}{s_{-3}}$$

$$1: -w_3\theta_2 - w_2\theta_1 - w_0\theta_1 - w_{-2}\theta_3 - w_{-1}\theta_2 + s_1 w_1$$

$$2: -w_3\theta_1 - w_1\theta_1 - w_0\theta_2 - w_{-1}\theta_3 + s_2 w_2$$

$$3: -w_2\theta_1 - w_1\theta_2 - w_0\theta_3 + s_3 w_3$$

eqSol

$$-3: \frac{w_0\theta_3 + w_{-2}\theta_1 + w_{-1}\theta_2}{s_{-3}}$$

$$-2: \frac{w_0\theta_1\theta_3 + w_{-1}\theta_1\theta_2 + s_{-3}w_1\theta_3 + s_{-3}w_0\theta_2 + s_{-3}w_{-1}\theta_1}{-\theta_1^2 + s_{-2}s_{-3}}$$

eqs

$$-3: 0$$

$$-2: 0$$

$$-1: \frac{-\theta_2\left((-\theta_1^2 + s_{-2}s_{-3})(w_0\theta_3 + w_{-1}\theta_2) + \theta_1(w_0\theta_1\theta_3 + w_{-1}\theta_1\theta_2 + s_{-3}w_1\theta_3 + s_{-3}w_0\theta_2 + s_{-3}w_{-1}\theta_1)\right) - s_{-3}\binom{(-\theta_1^2 + s_{-2}s_{-3})(w_2\theta_3 + w_1\theta_2 + w_0\theta_1 - s_{-1}w_{-1})}{+\theta_1(w_0\theta_1\theta_3 + w_{-1}\theta_1\theta_2 + s_{-3}w_1\theta_3 + s_{-3}w_0\theta_2 + s_{-3}w_{-1}\theta_1)}}{s_{-3}(-\theta_1^2 + s_{-2}s_{-3})}$$

$$0: \frac{-\theta_3\left((-\theta_1^2 + s_{-2}s_{-3})(w_0\theta_3 + w_{-1}\theta_2) + \theta_1(w_0\theta_1\theta_3 + w_{-1}\theta_1\theta_2 + s_{-3}w_1\theta_3 + s_{-3}w_0\theta_2 + s_{-3}w_{-1}\theta_1)\right) - s_{-3}\binom{(-\theta_1^2 + s_{-2}s_{-3})(w_3\theta_3 + w_2\theta_2 + w_1\theta_1 + w_{-1}\theta_1 - s_0 w_0)}{+\theta_2(w_0\theta_1\theta_3 + w_{-1}\theta_1\theta_2 + s_{-3}w_1\theta_3 + s_{-3}w_0\theta_2 + s_{-3}w_{-1}\theta_1)}}{s_{-3}(-\theta_1^2 + s_{-2}s_{-3})}$$

$$1: \frac{(-\theta_1^2 + s_{-2}s_{-3})(-w_3\theta_2 - w_2\theta_1 - w_0\theta_1 - w_{-1}\theta_2 + s_1 w_1) - \theta_3(w_0\theta_1\theta_3 + w_{-1}\theta_1\theta_2 + s_{-3}w_1\theta_3 + s_{-3}w_0\theta_2 + s_{-3}w_{-1}\theta_1)}{-\theta_1^2 + s_{-2}s_{-3}}$$

$$2: -w_3\theta_1 - w_1\theta_1 - w_0\theta_2 - w_{-1}\theta_3 + s_2 w_2$$

$$3: -w_2\theta_1 - w_1\theta_2 - w_0\theta_3 + s_3 w_3$$

eqSol

$$-3: \frac{w_0\theta_3 + w_{-2}\theta_1 + w_{-1}\theta_2}{s_{-3}}$$

$$-2: \frac{w_0\theta_1\theta_3 + w_{-1}\theta_1\theta_2 + s_{-3}w_1\theta_3 + s_{-3}w_0\theta_2 + s_{-3}w_{-1}\theta_1}{-\theta_1^2 + s_{-2}s_{-3}}$$

$$-1: \frac{\begin{array}{c}-w_2\theta_1^2\theta_3 + w_1\theta_1\theta_2\theta_3 - w_1\theta_1^2\theta_2 + w_0\theta_1\theta_2^2 + w_0\theta_1^2\theta_3 - w_0\theta_1^3 + s_{-3}w_1\theta_1\theta_3 + s_{-3}w_0\theta_1\theta_2 \\ + s_{-2}w_0\theta_2\theta_3 + s_{-2}s_{-3}w_2\theta_3 + s_{-2}s_{-3}w_1\theta_2 + s_{-2}s_{-3}w_0\theta_1\end{array}}{-2\theta_1^2\theta_2 - s_{-3}\theta_1^2 - s_{-2}\theta_2^2 - s_{-1}\theta_1^2 + s_{-1}s_{-2}s_{-3}}$$

eqs

$$-3: 0$$
$$-2: 0$$
$$-1: 0$$

$$0: \frac{\begin{array}{c}2w_3\theta_1^2\theta_2\theta_3 - w_2\theta_1\theta_2^2\theta_3 - w_2\theta_1^2\theta_3^2 + 2w_2\theta_1^2\theta_2^2 + w_2\theta_1^3\theta_3 + w_1\theta_2^3\theta_3 - w_1\theta_1\theta_2\theta_3^2 - w_1\theta_1\theta_2^3 \\ -2w_1\theta_1^2\theta_2\theta_3 + 3w_1\theta_1^3\theta_2 + w_0\theta_2^4 - 2w_0\theta_1\theta_2^2\theta_3 + w_0\theta_1^2\theta_3^2 - 2w_0\theta_1^2\theta_2^2 - 2w_0\theta_1^3\theta_3 + w_0\theta_1^4 \\ -2s_0w_0\theta_1^2\theta_2 + s_{-3}w_3\theta_1^2\theta_3 - s_{-3}w_2\theta_1\theta_2\theta_3 + s_{-3}w_2\theta_1^2\theta_2 - s_{-3}w_1\theta_1\theta_2^2 - s_{-3}w_1\theta_1^2\theta_3 + s_{-3}w_1\theta_1^3 \\ -2s_{-3}w_0\theta_1^2\theta_2 - s_{-3}s_0w_0\theta_1^2 + s_{-2}w_3\theta_2^2\theta_3 - s_{-2}w_2\theta_2\theta_3^2 + s_{-2}w_2\theta_2^3 - s_{-2}w_1\theta_2^2\theta_3 + s_{-2}w_1\theta_1\theta_2^2 \\ -2s_{-2}w_0\theta_1\theta_2\theta_3 - s_{-2}s_0w_0\theta_2^2 - s_{-2}s_{-3}w_2\theta_1\theta_3 - s_{-2}s_{-3}w_1\theta_1\theta_2 - s_{-2}s_{-3}w_0\theta_1^2 + s_{-1}w_3\theta_1^2\theta_3 \\ +s_{-1}w_2\theta_1^2\theta_2 - s_{-1}w_1\theta_1\theta_3^2 + s_{-1}w_1\theta_1^3 - 2s_{-1}w_0\theta_1\theta_2\theta_3 - s_{-1}s_0w_0\theta_1^2 - s_{-1}s_{-3}w_1\theta_2\theta_3 \\ -s_{-1}s_{-3}w_0\theta_2^2 - s_{-1}s_{-2}w_0\theta_3^2 - s_{-1}s_{-2}s_{-3}w_3\theta_3 - s_{-1}s_{-2}s_{-3}w_2\theta_2 - s_{-1}s_{-2}s_{-3}w_1\theta_1 + s_{-1}s_{-2}s_{-3}s_0w_0\end{array}}{-2\theta_1^2\theta_2 - s_{-3}\theta_1^2 - s_{-2}\theta_2^2 - s_{-1}\theta_1^2 + s_{-1}s_{-2}s_{-3}}$$

$$1: \frac{\begin{array}{c}2w_3\theta_1^2\theta_2^2 - w_2\theta_1\theta_2\theta_3^2 + w_2\theta_1^2\theta_2\theta_3 + 2w_2\theta_1^3\theta_2 + w_1\theta_2^2\theta_3^2 - 2w_1\theta_1\theta_2^2\theta_3 + w_1\theta_1^2\theta_2^2 + w_0\theta_2^3\theta_3 \\ -w_0\theta_1\theta_2\theta_3^2 - w_0\theta_1\theta_2^3 - 2w_0\theta_1^2\theta_2\theta_3 + 3w_0\theta_1^3\theta_2 - 2s_1w_1\theta_1^2\theta_2 + s_{-3}w_3\theta_1^2\theta_2 - s_{-3}w_2\theta_1\theta_3^2 \\ +s_{-3}w_2\theta_1^3 - 2s_{-3}w_1\theta_1\theta_2\theta_3 - s_{-3}w_0\theta_1\theta_2^2 - s_{-3}w_0\theta_1^2\theta_3 + s_{-3}w_0\theta_1^3 - s_{-3}s_1w_1\theta_1^2 + s_{-2}w_3\theta_2^3 \\ +s_{-2}w_2\theta_1\theta_2^2 - s_{-2}w_0\theta_2^2\theta_3 + s_{-2}w_0\theta_1\theta_2^2 - s_{-2}s_1w_1\theta_2^2 - s_{-2}s_{-3}w_2\theta_2\theta_3 - s_{-2}s_{-3}w_1\theta_2^2 \\ -s_{-2}s_{-3}w_0\theta_1\theta_2 + s_{-1}w_3\theta_1^2\theta_2 + s_{-1}w_2\theta_2^3 - s_{-1}w_0\theta_1\theta_3^2 + s_{-1}w_0\theta_1^3 - s_{-1}s_1w_1\theta_1^2 - s_{-1}s_{-3}w_1\theta_3^2 \\ -s_{-1}s_{-3}w_0\theta_2\theta_3 - s_{-1}s_{-2}s_{-3}w_3\theta_2 - s_{-1}s_{-2}s_{-3}w_2\theta_1 - s_{-1}s_{-2}s_{-3}w_0\theta_1 + s_{-1}s_{-2}s_{-3}s_1w_1\end{array}}{-2\theta_1^2\theta_2 - s_{-3}\theta_1^2 - s_{-2}\theta_2^2 - s_{-1}\theta_1^2 + s_{-1}s_{-2}s_{-3}}$$

$$2: \frac{\begin{array}{c}(-w_3\theta_1 - w_1\theta_1 - w_0\theta_2 + s_2w_2)(2\theta_1^2\theta_2 + s_{-3}\theta_1^2 + s_{-2}\theta_2^2 + s_{-1}\theta_1^2 - s_{-1}s_{-2}s_{-3}) \\ +\theta_3\begin{pmatrix}-w_2\theta_1^2\theta_3 + w_1\theta_1\theta_2\theta_3 - w_1\theta_1^2\theta_2 + w_0\theta_1\theta_2^2 + w_0\theta_1^2\theta_3 - w_0\theta_1^3 + s_{-3}w_1\theta_1\theta_3 + s_{-3}w_0\theta_1\theta_2 \\ +s_{-2}w_0\theta_2\theta_3 + s_{-2}s_{-3}w_2\theta_3 + s_{-2}s_{-3}w_1\theta_2 + s_{-2}s_{-3}w_0\theta_1\end{pmatrix}\end{array}}{2\theta_1^2\theta_2 + s_{-3}\theta_1^2 + s_{-2}\theta_2^2 + s_{-1}\theta_1^2 - s_{-1}s_{-2}s_{-3}}$$

$$3: -w_2\theta_1 - w_1\theta_2 - w_0\theta_3 + s_3w_3$$

eqSol

$$-3: \frac{w_0\theta_3 + w_{-2}\theta_1 + w_{-1}\theta_2}{s_{-3}}$$

$$-2: \frac{w_0\theta_1\theta_3 + w_{-1}\theta_1\theta_2 + s_{-3}w_1\theta_3 + s_{-3}w_0\theta_2 + s_{-3}w_{-1}\theta_1}{-\theta_1^2 + s_{-2}s_{-3}}$$

$$-1: \frac{\begin{array}{c}-w_2\theta_1^2\theta_3 + w_1\theta_1\theta_2\theta_3 - w_1\theta_1^2\theta_2 + w_0\theta_1\theta_2^2 + w_0\theta_1^2\theta_3 - w_0\theta_1^3 + s_{-3}w_1\theta_1\theta_3 + s_{-3}w_0\theta_1\theta_2 \\ +s_{-2}w_0\theta_2\theta_3 + s_{-2}s_{-3}w_2\theta_3 + s_{-2}s_{-3}w_1\theta_2 + s_{-2}s_{-3}w_0\theta_1\end{array}}{-2\theta_1^2\theta_2 - s_{-3}\theta_1^2 - s_{-2}\theta_2^2 - s_{-1}\theta_1^2 + s_{-1}s_{-2}s_{-3}}$$

$$1: \frac{\begin{aligned}&-2w_3\theta_1^2\theta_2^2 + w_2\theta_1\theta_2\theta_3^2 - w_2\theta_1^2\theta_2\theta_3 - 2w_2\theta_1^3\theta_2 - w_0\theta_2^3\theta_3 + w_0\theta_1\theta_2\theta_3^2 + w_0\theta_1\theta_2^3 + 2w_0\theta_1^2\theta_2\theta_3\\&-3w_0\theta_1^3\theta_2 - s_{-3}w_3\theta_1^2\theta_2 + s_{-3}w_2\theta_1\theta_3^2 - s_{-3}w_2\theta_1^3 + s_{-3}w_0\theta_1\theta_2^2 + s_{-3}w_0\theta_1^2\theta_3 - s_{-3}w_0\theta_1^3\\&-s_{-2}w_3\theta_2^3 - s_{-2}w_2\theta_1\theta_2^2 + s_{-2}w_0\theta_2^2\theta_3 - s_{-2}w_0\theta_1\theta_2^2 + s_{-2}s_{-3}w_2\theta_2\theta_3 + s_{-2}s_{-3}w_0\theta_1\theta_2\\&-s_{-1}w_3\theta_1^2\theta_2 - s_{-1}w_2\theta_1^3 + s_{-1}w_0\theta_1\theta_3^2 - s_{-1}w_0\theta_1^3 + s_{-1}s_{-3}w_0\theta_2\theta_3 + s_{-1}s_{-2}s_{-3}w_3\theta_2\\&+s_{-1}s_{-2}s_{-3}w_2\theta_1 + s_{-1}s_{-2}s_{-3}w_0\theta_1\end{aligned}}{\begin{aligned}&\theta_2^2\theta_3^2 - 2\theta_1\theta_2^2\theta_3 + \theta_1^2\theta_2^2 - 2s_1\theta_1^2\theta_2 - 2s_{-3}\theta_1\theta_2\theta_3 - s_{-3}s_1\theta_1^2 - s_{-2}s_1\theta_2^2 - s_{-2}s_{-3}\theta_2^2 - s_{-1}s_1\theta_1^2\\&-s_{-1}s_{-3}\theta_3^2 + s_{-1}s_{-2}s_{-3}s_1\end{aligned}}$$

eqs

$$-3: 0$$

$$-2: 0$$

$$-1: 0$$

$$0: \frac{\begin{aligned}&-w_3\theta_2^2\theta_3^3 + w_3\theta_2^4\theta_3 + w_3\theta_1\theta_2^2\theta_3^2 - w_3\theta_1\theta_2^4 - 3w_3\theta_1^2\theta_2^2\theta_3 + 3w_3\theta_1^3\theta_2^2 + w_2\theta_2\theta_3^4 - w_2\theta_2^3\theta_3^2\\&-w_2\theta_1\theta_2\theta_3^3 + 3w_2\theta_1\theta_2^3\theta_3 - 2w_2\theta_1^2\theta_2\theta_3^2 - 2w_2\theta_1^2\theta_2^3 - w_2\theta_1^3\theta_2\theta_3 + 3w_2\theta_1^4\theta_2 - 2w_0\theta_2^3\theta_3^2\\&+4w_0\theta_1\theta_2\theta_3^3 + 4w_0\theta_1\theta_2^3\theta_3 - 4w_0\theta_1^2\theta_2\theta_3^2 - 2w_0\theta_1^2\theta_2^3 - 4w_0\theta_1^3\theta_2\theta_3 + 4w_0\theta_1^4\theta_2 + 2s_1w_3\theta_1^2\theta_2\theta_3\\&-s_1w_2\theta_1\theta_2^2\theta_3 - s_1w_2\theta_1^2\theta_3^2 + 2s_1w_2\theta_1^2\theta_2^2 + s_1w_2\theta_1^3\theta_3 + s_1w_0\theta_2^4 - 2s_1w_0\theta_1\theta_2^2\theta_3 + s_1w_0\theta_1^2\theta_3^2\\&-2s_1w_0\theta_1^2\theta_2^2 - 2s_1w_0\theta_1^3\theta_3 + s_1w_0\theta_1^4 + s_0w_0\theta_2^2\theta_3^2 - 2s_0w_0\theta_1\theta_2^2\theta_3 + s_0w_0\theta_1^2\theta_2^2 - 2s_0s_1w_0\theta_1^2\theta_2\\&+2s_{-3}w_3\theta_1\theta_2\theta_3^2 - s_{-3}w_3\theta_1\theta_2^3 - s_{-3}w_3\theta_1^2\theta_2\theta_3 + s_{-3}w_3\theta_1^3\theta_2 - s_{-3}w_2\theta_2^2\theta_3^2 + s_{-3}w_2\theta_1\theta_3^3\\&+2s_{-3}w_2\theta_1\theta_2^2\theta_3 - s_{-3}w_2\theta_1^2\theta_3^2 - s_{-3}w_2\theta_1^2\theta_2^2 - s_{-3}w_2\theta_1^3\theta_3 + s_{-3}w_2\theta_1^4 + s_{-3}w_0\theta_2^4 - 2s_{-3}w_0\theta_1\theta_2^2\theta_3\\&+s_{-3}w_0\theta_1^2\theta_3^2 - 2s_{-3}w_0\theta_1^2\theta_2^2 - 2s_{-3}w_0\theta_1^3\theta_3 + s_{-3}w_0\theta_1^4 + s_{-3}s_1w_3\theta_1^2\theta_3 - s_{-3}s_1w_2\theta_1\theta_2\theta_3\\&+s_{-3}s_1w_2\theta_1^2\theta_2 - 2s_{-3}s_1w_0\theta_1^2\theta_2 - 2s_{-3}s_0w_0\theta_1\theta_2\theta_3 - s_{-3}s_0s_1w_0\theta_1^2 - s_{-2}w_3\theta_2^2\theta_3 + s_{-2}w_3\theta_1\theta_2^3\\&-s_{-2}w_2\theta_1\theta_2\theta_3^2 + s_{-2}w_2\theta_1^2\theta_2^2 + s_{-2}w_0\theta_2^2\theta_3^2 - 2s_{-2}w_0\theta_1\theta_2^2\theta_3 + s_{-2}w_0\theta_1^2\theta_2^2 + s_{-2}s_1w_3\theta_2^2\theta_3\\&-s_{-2}s_1w_2\theta_2\theta_3^2 + s_{-2}s_1w_2\theta_2^3 - 2s_{-2}s_1w_0\theta_1\theta_2\theta_3 - s_{-2}s_0s_1w_0\theta_2^2 + s_{-2}s_{-3}w_3\theta_2^2\theta_3 - s_{-2}s_{-3}w_3\theta_1\theta_2^2\\&+s_{-2}s_{-3}w_2\theta_2^3 - s_{-2}s_{-3}w_2\theta_1\theta_2\theta_3 - s_{-2}s_{-3}w_2\theta_1^2\theta_2 - 2s_{-2}s_{-3}w_0\theta_1^2\theta_2 - s_{-2}s_{-3}s_1w_2\theta_1\theta_3\\&-s_{-2}s_{-3}s_1w_0\theta_1^2 - s_{-2}s_{-3}s_0w_0\theta_2^2 - s_{-1}w_3\theta_1\theta_2\theta_3^2 + s_{-1}w_3\theta_1^3\theta_2 - s_{-1}w_2\theta_1^2\theta_3^2 + s_{-1}w_2\theta_1^4\\&+s_{-1}w_0\theta_3^4 - 2s_{-1}w_0\theta_1^2\theta_3^2 + s_{-1}w_0\theta_1^4 + s_{-1}s_1w_3\theta_1^2\theta_3 + s_{-1}s_1w_2\theta_1^2\theta_2 - 2s_{-1}s_1w_0\theta_1\theta_2\theta_3\\&-s_{-1}s_0s_1w_0\theta_1^2 + s_{-1}s_{-3}w_3\theta_3^3 - s_{-1}s_{-3}w_3\theta_2^2\theta_3 + s_{-1}s_{-3}w_2\theta_2\theta_3^2 - s_{-1}s_{-3}w_2\theta_1\theta_2\theta_3\\&-2s_{-1}s_{-3}w_0\theta_1\theta_2\theta_3 - s_{-1}s_{-3}s_1w_0\theta_2^2 - s_{-1}s_{-3}s_0w_0\theta_3^2 - s_{-1}s_{-2}s_1w_0\theta_3^2 - s_{-1}s_{-2}s_{-3}w_3\theta_1\theta_2\\&-s_{-1}s_{-2}s_{-3}w_2\theta_1^2 - s_{-1}s_{-2}s_{-3}w_0\theta_1^2 - s_{-1}s_{-2}s_{-3}s_1w_3\theta_3 - s_{-1}s_{-2}s_{-3}s_1w_2\theta_2 + s_{-1}s_{-2}s_{-3}s_0s_1w_0\end{aligned}}{\begin{aligned}&\theta_2^2\theta_3^2 - 2\theta_1\theta_2^2\theta_3 + \theta_1^2\theta_2^2 - 2s_1\theta_1^2\theta_2 - 2s_{-3}\theta_1\theta_2\theta_3 - s_{-3}s_1\theta_1^2 - s_{-2}s_1\theta_2^2 - s_{-2}s_{-3}\theta_2^2 - s_{-1}s_1\theta_1^2\\&-s_{-1}s_{-3}\theta_3^2 + s_{-1}s_{-2}s_{-3}s_1\end{aligned}}$$

$$1: 0$$

$$2:\frac{\begin{array}{l}-2w_3\theta_1\theta_2^2\theta_3^2+3w_3\theta_1^2\theta_2^2\theta_3+w_3\theta_1^3\theta_2^2-2w_2\theta_1^2\theta_2\theta_3^2+2w_2\theta_1^3\theta_2\theta_3+2w_2\theta_1^4\theta_2+w_0\theta_2\theta_3^4\\-w_0\theta_2^3\theta_3^2-w_0\theta_1\theta_2\theta_3^3+3w_0\theta_1\theta_2^3\theta_3-2w_0\theta_1^2\theta_2\theta_3^2-2w_0\theta_1^2\theta_2^3-w_0\theta_1^3\theta_2\theta_3+3w_0\theta_1^4\theta_2\\+s_2w_2\theta_2^2\theta_3^2-2s_2w_2\theta_1\theta_2^2\theta_3+s_2w_2\theta_1^2\theta_2^2+2s_1w_3\theta_1^3\theta_2+s_1w_2\theta_1^2\theta_3^2-s_1w_0\theta_1\theta_2^2\theta_3-s_1w_0\theta_1^3\theta_3\\+2s_1w_0\theta_1^2\theta_2^2+s_1w_0\theta_1^3\theta_3-2s_1s_2w_2\theta_1^2\theta_2-s_{-3}w_3\theta_1\theta_2^2\theta_3^2+2s_{-3}w_3\theta_1^2\theta_2\theta_3+s_{-3}w_3\theta_1^3\theta_2\\+s_{-3}w_2\theta_3^4-2s_{-3}w_2\theta_1^2\theta_3^2+s_{-3}w_2\theta_1^4-s_{-3}w_0\theta_2^2\theta_3^2+s_{-3}w_0\theta_1\theta_2^3+2s_{-3}w_0\theta_1\theta_2^2\theta_3-s_{-3}w_0\theta_1^2\theta_3^2\\-s_{-3}w_0\theta_1^2\theta_2^2-s_{-3}w_0\theta_1^3\theta_3+s_{-3}w_0\theta_1^4-2s_{-3}s_2w_2\theta_1\theta_2\theta_3+s_{-3}s_1w_3\theta_1^3-s_{-3}s_1w_0\theta_1\theta_2\theta_3\\+s_{-3}s_1w_0\theta_1^2\theta_2-s_{-3}s_1s_2w_2\theta_1^2+s_{-2}w_3\theta_1\theta_2^3+s_{-2}w_2\theta_1^2\theta_2^2-s_{-2}w_0\theta_1\theta_2^2\theta_3+s_{-2}w_0\theta_1^2\theta_2^2\\+s_{-2}s_1w_3\theta_1\theta_2^2-s_{-2}s_1w_0\theta_2\theta_3^2+s_{-2}s_1w_0\theta_2^3-s_{-2}s_1s_2w_2\theta_2^2-s_{-2}s_{-3}w_3\theta_2^2\theta_3+s_{-2}s_{-3}w_3\theta_1\theta_2^2\\-2s_{-2}s_{-3}w_2\theta_1\theta_2\theta_3+s_{-2}s_{-3}w_0\theta_2^3-s_{-2}s_{-3}w_0\theta_1\theta_2\theta_3-s_{-2}s_{-3}w_0\theta_1^2\theta_2-s_{-2}s_{-3}s_2w_2\theta_2^2\\-s_{-2}s_{-3}s_1w_2\theta_3^2-s_{-2}s_{-3}s_1w_0\theta_1\theta_3+s_{-1}w_3\theta_1^3\theta_2+s_{-1}w_2\theta_1^4-s_{-1}w_0\theta_1^2\theta_3^2+s_{-1}w_0\theta_1^4\\+s_{-1}s_1w_3\theta_1^3+s_{-1}s_1w_0\theta_1^2\theta_2-s_{-1}s_1s_2w_2\theta_1^2+s_{-1}s_{-3}w_3\theta_1\theta_3^2+s_{-1}s_{-3}w_0\theta_2\theta_3^2-s_{-1}s_{-3}w_0\theta_1\theta_2\theta_3\\-s_{-1}s_{-3}s_2w_2\theta_3^2-s_{-1}s_{-2}s_{-3}w_3\theta_1\theta_2-s_{-1}s_{-2}s_{-3}w_2\theta_1^2-s_{-1}s_{-2}s_{-3}w_0\theta_1^2-s_{-1}s_{-2}s_{-3}s_1w_3\theta_1\\-s_{-1}s_{-2}s_{-3}s_1w_0\theta_2+s_{-1}s_{-2}s_{-3}s_1s_2w_2\end{array}}{\theta_2^2\theta_3^2-2\theta_1\theta_2^2\theta_3+\theta_1^2\theta_2^2-2s_1\theta_1^2\theta_2-2s_{-3}\theta_1\theta_2\theta_3-s_{-3}s_1\theta_1^2-s_{-2}s_1\theta_2^2-s_{-2}s_{-3}\theta_2^2-s_{-1}s_1\theta_1^2\\-s_{-1}s_{-3}\theta_3^2+s_{-1}s_{-2}s_{-3}s_1}$$

$$3:\frac{(-w_2\theta_1-w_0\theta_3+s_3w_3)\begin{pmatrix}-\theta_2^2\theta_3^2+2\theta_1\theta_2^2\theta_3-\theta_1^2\theta_2^2+2s_1\theta_1^2\theta_2+2s_{-3}\theta_1\theta_2\theta_3+s_{-3}s_1\theta_1^2\\+s_{-2}s_1\theta_2^2+s_{-2}s_{-3}\theta_2^2+s_{-1}s_1\theta_1^2+s_{-1}s_{-3}\theta_3^2-s_{-1}s_{-2}s_{-3}s_1\end{pmatrix}+\theta_2\begin{pmatrix}-2w_3\theta_1^2\theta_2^2+w_2\theta_1\theta_2\theta_3^2-w_2\theta_1^2\theta_2\theta_3-2w_2\theta_1^3\theta_2-w_0\theta_2^3\theta_3+w_0\theta_1\theta_2\theta_3^2+w_0\theta_1\theta_2^3\\+2w_0\theta_1^2\theta_2\theta_3-3w_0\theta_1^3\theta_2-s_{-3}w_3\theta_1^2\theta_2+s_{-3}w_2\theta_1\theta_3^2-s_{-3}w_2\theta_1^3+s_{-3}w_0\theta_1\theta_2^2\\+s_{-3}w_0\theta_1^2\theta_3-s_{-3}w_0\theta_1^3-s_{-2}w_3\theta_2^3-s_{-2}w_2\theta_1\theta_2^2+s_{-2}w_0\theta_2^2\theta_3-s_{-2}w_0\theta_1\theta_2^2\\+s_{-2}s_{-3}w_2\theta_2\theta_3+s_{-2}s_{-3}w_0\theta_1\theta_2-s_{-1}w_3\theta_1^2\theta_2-s_{-1}w_2\theta_1^3+s_{-1}w_0\theta_1\theta_3^2-s_{-1}w_0\theta_1^3\\+s_{-1}s_{-3}w_0\theta_2\theta_3+s_{-1}s_{-2}s_{-3}w_3\theta_2+s_{-1}s_{-2}s_{-3}w_2\theta_1+s_{-1}s_{-2}s_{-3}w_0\theta_1\end{pmatrix}}{-\theta_2^2\theta_3^2+2\theta_1\theta_2^2\theta_3-\theta_1^2\theta_2^2+2s_1\theta_1^2\theta_2+2s_{-3}\theta_1\theta_2\theta_3+s_{-3}s_1\theta_1^2+s_{-2}s_1\theta_2^2+s_{-2}s_{-3}\theta_2^2\\+s_{-1}s_1\theta_1^2+s_{-1}s_{-3}\theta_3^2-s_{-1}s_{-2}s_{-3}s_1}$$

eqSol

$$-3:\frac{w_0\theta_3+w_{-2}\theta_1+w_{-1}\theta_2}{s_{-3}}$$

$$-2:\frac{w_0\theta_1\theta_3+w_{-1}\theta_1\theta_2+s_{-3}w_1\theta_3+s_{-3}w_0\theta_2+s_{-3}w_{-1}\theta_1}{-\theta_1^2+s_{-2}s_{-3}}$$

$$-1:\frac{\begin{array}{l}-w_2\theta_1^2\theta_3+w_1\theta_1\theta_2\theta_3-w_1\theta_1^2\theta_2+w_0\theta_1\theta_2^2+w_0\theta_1^2\theta_3-w_0\theta_1^3+s_{-3}w_1\theta_1\theta_3+s_{-3}w_0\theta_1\theta_2+s_{-2}w_0\theta_2\theta_3+s_{-2}s_{-3}w_2\theta_3\\+s_{-2}s_{-3}w_1\theta_2+s_{-2}s_{-3}w_0\theta_1\end{array}}{-2\theta_1^2\theta_2-s_{-3}\theta_1^2-s_{-2}\theta_2^2-s_{-1}\theta_1^2+s_{-1}s_{-2}s_{-3}}$$

$$1:\frac{\begin{array}{l}-2w_3\theta_1^2\theta_2^2+w_2\theta_1\theta_2\theta_3^2-w_2\theta_1^2\theta_2\theta_3-2w_2\theta_1^3\theta_2-w_0\theta_2^3\theta_3+w_0\theta_1\theta_2\theta_3^2+w_0\theta_1\theta_2^3+2w_0\theta_1^2\theta_2\theta_3-3w_0\theta_1^3\theta_2\\-s_{-3}w_3\theta_1^2\theta_2+s_{-3}w_2\theta_1\theta_3^2-s_{-3}w_2\theta_1^3+s_{-3}w_0\theta_1\theta_2^2+s_{-3}w_0\theta_1^2\theta_3-s_{-3}w_0\theta_1^3-s_{-2}w_3\theta_2^3-s_{-2}w_2\theta_1\theta_2^2+s_{-2}w_0\theta_2^2\theta_3\\-s_{-2}w_0\theta_1\theta_2^2+s_{-2}s_{-3}w_2\theta_2\theta_3+s_{-2}s_{-3}w_0\theta_1\theta_2-s_{-1}w_3\theta_1^2\theta_2-s_{-1}w_2\theta_1^3+s_{-1}w_0\theta_1\theta_3^2-s_{-1}w_0\theta_1^3+s_{-1}s_{-3}w_0\theta_2\theta_3\\+s_{-1}s_{-2}s_{-3}w_3\theta_2+s_{-1}s_{-2}s_{-3}w_2\theta_1+s_{-1}s_{-2}s_{-3}w_0\theta_1\end{array}}{\theta_2^2\theta_3^2-2\theta_1\theta_2^2\theta_3+\theta_1^2\theta_2^2-2s_1\theta_1^2\theta_2-2s_{-3}\theta_1\theta_2\theta_3-s_{-3}s_1\theta_1^2-s_{-2}s_1\theta_2^2-s_{-2}s_{-3}\theta_2^2-s_{-1}s_1\theta_1^2-s_{-1}s_{-3}\theta_3^2+s_{-1}s_{-2}s_{-3}s_1}$$

$$2:\frac{\begin{array}{l}2w_3\theta_1\theta_2^2\theta_3^2-3w_3\theta_1^2\theta_2^2\theta_3-w_3\theta_1^3\theta_2^2-w_0\theta_2\theta_3^4+w_0\theta_2^3\theta_3^2+w_0\theta_1\theta_2\theta_3^3-3w_0\theta_1\theta_2^3\theta_3+2w_0\theta_1^2\theta_2\theta_3^2+2w_0\theta_1^2\theta_2^3\\+w_0\theta_1^3\theta_2\theta_3-3w_0\theta_1^4\theta_2-2s_1w_3\theta_1^3\theta_2+s_1w_0\theta_1\theta_2^2\theta_3+s_1w_0\theta_1^2\theta_3^2-2s_1w_0\theta_1^2\theta_2^2-s_1w_0\theta_1^3\theta_3+s_{-3}w_3\theta_1\theta_2^2\theta_3^2\\-2s_{-3}w_3\theta_1^2\theta_2\theta_3-s_{-3}w_3\theta_1^3\theta_2+s_{-3}w_0\theta_2^2\theta_3^2-s_{-3}w_0\theta_1\theta_2^3-2s_{-3}w_0\theta_1\theta_2^2\theta_3+s_{-3}w_0\theta_1^2\theta_3^2+s_{-3}w_0\theta_1^2\theta_2^2+s_{-3}w_0\theta_1^3\theta_3\\-s_{-3}w_0\theta_1^4-s_{-3}s_1w_3\theta_1^3+s_{-3}s_1w_0\theta_1\theta_2\theta_3-s_{-3}s_1w_0\theta_1^2\theta_2-s_{-2}w_3\theta_1\theta_2^3+s_{-2}w_0\theta_1\theta_2^2\theta_3-s_{-2}w_0\theta_1^2\theta_2^2-s_{-2}s_1w_3\theta_1\theta_2^2\\+s_{-2}s_1w_0\theta_2\theta_3^2-s_{-2}s_1w_0\theta_2^3+s_{-2}s_{-3}w_3\theta_2^2\theta_3-s_{-2}s_{-3}w_3\theta_1\theta_2^2-s_{-2}s_{-3}w_0\theta_2^3+s_{-2}s_{-3}w_0\theta_1\theta_2\theta_3+s_{-2}s_{-3}w_0\theta_1^2\theta_2\\+s_{-2}s_{-3}s_1w_0\theta_1\theta_3-s_{-1}w_3\theta_1^3\theta_2+s_{-1}w_0\theta_1^2\theta_3^2-s_{-1}w_0\theta_1^4-s_{-1}s_1w_3\theta_1^3-s_{-1}s_1w_0\theta_1^2\theta_2-s_{-1}s_{-3}w_3\theta_1\theta_3^2\\-s_{-1}s_{-3}w_0\theta_2\theta_3^2+s_{-1}s_{-3}w_0\theta_1\theta_2\theta_3+s_{-1}s_{-2}s_{-3}w_3\theta_1\theta_2+s_{-1}s_{-2}s_{-3}w_0\theta_1^2+s_{-1}s_{-2}s_{-3}s_1w_3\theta_1+s_{-1}s_{-2}s_{-3}s_1w_0\theta_2\end{array}}{\begin{array}{l}-2\theta_1^2\theta_2\theta_3^2+2\theta_1^3\theta_2\theta_3+2\theta_1^4\theta_2+s_2\theta_2^2\theta_3^2-2s_2\theta_1\theta_2^2\theta_3+s_2\theta_1^2\theta_2^2+s_1\theta_1^2\theta_3^2-2s_1\theta_1^2\theta_2^2+s_{-3}\theta_3^4-2s_{-3}\theta_1^2\theta_3^2\\+s_{-3}\theta_1^4-2s_{-3}s_2\theta_1\theta_2\theta_3-s_{-3}s_1\theta_1^2+s_{-2}\theta_1^2\theta_2^2-s_{-2}s_1\theta_2^2-2s_{-2}s_{-3}\theta_1\theta_2\theta_3-s_{-2}s_{-3}s_2\theta_2^2-s_{-2}s_{-3}s_1\theta_3^2+s_{-1}\theta_1^4\\-s_{-1}s_1s_2\theta_1^2-s_{-1}s_{-3}s_2\theta_3^2-s_{-1}s_{-2}s_{-3}\theta_1^2+s_{-1}s_{-2}s_{-3}s_1s_2\end{array}}$$

eqs

$$-3 : 0$$

$$-2 : 0$$

$$-1 : 0$$

$$0 : \frac{\begin{aligned}&2w_3\theta_1\theta_2\theta_3^4 - 2w_3\theta_1\theta_2^3\theta_3^2 + w_3\theta_1^2\theta_2\theta_3^3 + 3w_3\theta_1^2\theta_2^3\theta_3 - 6w_3\theta_1^3\theta_2^2\theta_3 - 2w_3\theta_1^4\theta_2\theta_3 + 3w_3\theta_1^5\theta_2 - w_0\theta_3^6 + 2w_0\theta_2^2\theta_3^4 - w_0\theta_2^4\theta_3^2\\ &-4w_0\theta_1\theta_2^2\theta_3^3 + 4w_0\theta_1\theta_2^4\theta_3 + 4w_0\theta_1^2\theta_3^4 - 4w_0\theta_1^2\theta_2^4 - 2w_0\theta_1^3\theta_3^3 - 2w_0\theta_1^3\theta_2^2\theta_3 - 4w_0\theta_1^4\theta_3^2 + 8w_0\theta_1^4\theta_2^2 + 4w_0\theta_1^5\theta_3 - w_0\theta_1^6\\ &-s_2w_3\theta_2^2\theta_3^3 + s_2w_3\theta_2^4\theta_3 + s_2w_3\theta_1\theta_2^2\theta_3^2 - s_2w_3\theta_1\theta_2^4 - 3s_2w_3\theta_1^2\theta_2^2\theta_3 + 3s_2w_3\theta_1^3\theta_2^2 - 2s_2w_0\theta_2^3\theta_3^2 + 4s_2w_0\theta_1\theta_2^2\theta_3^3 + 4s_2w_0\theta_1\theta_2^3\theta_3\\ &-4s_2w_0\theta_1^2\theta_2\theta_3^2 - 2s_2w_0\theta_1^2\theta_3^3 - 4s_2w_0\theta_1^3\theta_2\theta_3 + 4s_2w_0\theta_1^4\theta_2 - s_1w_3\theta_2^2\theta_3^3 - s_1w_3\theta_1^2\theta_2\theta_3^3 - s_1w_3\theta_1^3\theta_3^3 + 2s_1w_3\theta_1^3\theta_2^2 + s_1w_3\theta_1^4\theta_3\\ &+2s_1w_0\theta_1\theta_2\theta_3^3 - 2s_1w_0\theta_1\theta_2^3\theta_3 - 2s_1w_0\theta_1^2\theta_2\theta_3^2 + 2s_1w_0\theta_1^2\theta_3^3 + 2s_1w_0\theta_1^3\theta_2\theta_3 + 2s_1s_2w_3\theta_1^2\theta_2\theta_3 + s_1s_2w_0\theta_2^4 - 2s_1s_2w_0\theta_1\theta_2^2\theta_3\\ &+s_1s_2w_0\theta_1^2\theta_3^2 - 2s_1s_2w_0\theta_1^2\theta_2^2 - 2s_1s_2w_0\theta_1^3\theta_3 + s_1s_2w_0\theta_1^4 - 2s_0w_0\theta_1^2\theta_2\theta_3^2 + 2s_0w_0\theta_1^3\theta_2\theta_3 + 2s_0w_0\theta_1^4\theta_2 + s_0s_2w_0\theta_2^2\theta_3^2\\ &-2s_0s_2w_0\theta_1\theta_2^2\theta_3 + s_0s_2w_0\theta_1^2\theta_2^2 + s_0s_1w_0\theta_1^2\theta_3^2 - 2s_0s_1s_2w_0\theta_1^2\theta_2 - s_{-3}w_3\theta_3^5 + s_{-3}w_3\theta_2^2\theta_3^3 - 2s_{-3}w_3\theta_1\theta_2^2\theta_3^2 + 3s_{-3}w_3\theta_1^2\theta_3^3\\ &+s_{-3}w_3\theta_1^2\theta_2^2\theta_3 - s_{-3}w_3\theta_1^3\theta_3^2 - 2s_{-3}w_3\theta_1^4\theta_3 + s_{-3}w_3\theta_1^5 - 2s_{-3}w_0\theta_2^2\theta_3^2 + 4s_{-3}w_0\theta_1\theta_2^2\theta_3^3 + 4s_{-3}w_0\theta_1\theta_2^3\theta_3 - 4s_{-3}w_0\theta_1^2\theta_2\theta_3^2\\ &-2s_{-3}w_0\theta_1^2\theta_3^3 - 4s_{-3}w_0\theta_1^3\theta_2\theta_3 + 4s_{-3}w_0\theta_1^4\theta_2 + 2s_{-3}s_2w_3\theta_1\theta_2^2\theta_3 - s_{-3}s_2w_3\theta_1\theta_2^3 - s_{-3}s_2w_3\theta_1^2\theta_2\theta_3 + s_{-3}s_2w_3\theta_1^3\theta_2 + s_{-3}s_2w_0\theta_2^4\\ &-2s_{-3}s_2w_0\theta_1\theta_2^2\theta_3 + s_{-3}s_2w_0\theta_1^2\theta_3^2 - 2s_{-3}s_2w_0\theta_1^2\theta_2^2 - 2s_{-3}s_2w_0\theta_1^3\theta_3 + s_{-3}s_2w_0\theta_1^4 - s_{-3}s_1w_3\theta_1^2\theta_2\theta_3 + s_{-3}s_1w_3\theta_1^3\theta_2 + s_{-3}s_1w_0\theta_2^2\theta_3^2\\ &-2s_{-3}s_1w_0\theta_1\theta_2^2\theta_3 + s_{-3}s_1w_0\theta_1^2\theta_2^2 + s_{-3}s_1s_2w_3\theta_1^2\theta_3 - 2s_{-3}s_1s_2w_0\theta_1^2\theta_2 + s_{-3}s_0w_0\theta_3^4 - 2s_{-3}s_0w_0\theta_1^2\theta_3^2 + s_{-3}s_0w_0\theta_1^4 - 2s_{-3}s_0s_2w_0\theta_1\theta_2\theta_3\\ &-s_{-3}s_0s_1s_2w_0\theta_1^2 - s_{-2}w_3\theta_1\theta_2^2\theta_3^2 + s_{-2}w_3\theta_1\theta_2^4 - 2s_{-2}w_3\theta_1^2\theta_2^2\theta_3 + s_{-2}w_3\theta_1^3\theta_2^2 + 2s_{-2}w_0\theta_1\theta_2^3 - 2s_{-2}w_0\theta_1\theta_2^3\theta_3 - 2s_{-2}w_0\theta_1^2\theta_2\theta_3^2\\ &+2s_{-2}w_0\theta_1^2\theta_3^3 + 2s_{-2}w_0\theta_1^3\theta_2\theta_3 - s_{-2}s_2w_3\theta_2^3\theta_3 + s_{-2}s_2w_3\theta_1\theta_2^3 + s_{-2}s_2w_0\theta_2^2\theta_3^2 - 2s_{-2}s_2w_0\theta_1\theta_2^2\theta_3 + s_{-2}s_2w_0\theta_1^2\theta_2^2 - s_{-2}s_1w_3\theta_1\theta_2\theta_3^2\\ &+s_{-2}s_1w_3\theta_1\theta_2^3 + s_{-2}s_1w_0\theta_3^4 - 2s_{-2}s_1w_0\theta_2^2\theta_3^2 + s_{-2}s_1w_0\theta_2^4 + s_{-2}s_1s_2w_3\theta_2^2\theta_3 - 2s_{-2}s_1s_2w_0\theta_1\theta_2\theta_3 + s_{-2}s_0w_0\theta_1^2\theta_3^2 - s_{-2}s_0s_1s_2w_0\theta_2^2\\ &-s_{-2}s_{-3}w_3\theta_2^3\theta_3 + 3s_{-2}s_{-3}w_3\theta_1\theta_2\theta_3^2 + s_{-2}s_{-3}w_3\theta_1\theta_2^3 - 2s_{-2}s_{-3}w_3\theta_1^2\theta_2\theta_3 - s_{-2}s_{-3}w_3\theta_1^3\theta_2 + s_{-2}s_{-3}w_0\theta_2^4 - 2s_{-2}s_{-3}w_0\theta_1\theta_2^2\theta_3\\ &+s_{-2}s_{-3}w_0\theta_1^2\theta_3^2 - 2s_{-2}s_{-3}w_0\theta_1^2\theta_2^2 - 2s_{-2}s_{-3}w_0\theta_1^3\theta_3 + s_{-2}s_{-3}w_0\theta_1^4 + s_{-2}s_{-3}s_2w_3\theta_2^2\theta_3 - s_{-2}s_{-3}s_2w_3\theta_1\theta_2^2 - 2s_{-2}s_{-3}s_2w_0\theta_1^2\theta_2\\ &+s_{-2}s_{-3}s_1w_3\theta_3^3 - s_{-2}s_{-3}s_1w_3\theta_1^2\theta_3 - 2s_{-2}s_{-3}s_1w_0\theta_1\theta_2\theta_3 - s_{-2}s_{-3}s_1s_2w_0\theta_1^2 - 2s_{-2}s_{-3}s_0w_0\theta_1\theta_2\theta_3 - s_{-2}s_{-3}s_0s_2w_0\theta_2^2 - s_{-2}s_{-3}s_0s_1w_0\theta_3^2\\ &-s_{-1}w_3\theta_1^2\theta_3^3 + s_{-1}w_3\theta_1^2\theta_2^2\theta_3 - s_{-1}w_3\theta_1^4\theta_3 + s_{-1}w_3\theta_1^5 - 2s_{-1}w_0\theta_1\theta_2^2\theta_3^3 + 2s_{-1}w_0\theta_1\theta_2^3\theta_3 + 2s_{-1}w_0\theta_1^4\theta_3 - s_{-1}s_2w_3\theta_1\theta_2^2\theta_3 + s_{-1}s_2w_3\theta_1^3\theta_2\\ &+s_{-1}s_2w_0\theta_3^4 - 2s_{-1}s_2w_0\theta_1^2\theta_3^2 + s_{-1}s_2w_0\theta_1^4 + s_{-1}s_1w_3\theta_1^3\theta_2 + s_{-1}s_1w_0\theta_1^2\theta_2^2 + s_{-1}s_1s_2w_3\theta_1^2\theta_3 - 2s_{-1}s_1s_2w_0\theta_1^2\theta_2 + s_{-1}s_0w_0\theta_1^4\\ &-s_{-1}s_0s_1s_2w_0\theta_1^2 + s_{-1}s_{-3}w_3\theta_1\theta_2\theta_3^2 - s_{-1}s_{-3}w_3\theta_1^2\theta_2\theta_3 + s_{-1}s_{-3}w_0\theta_2^2\theta_3^2 - 2s_{-1}s_{-3}w_0\theta_1\theta_2^2\theta_3 + s_{-1}s_{-3}w_0\theta_1^2\theta_2^2 + s_{-1}s_{-3}s_2w_3\theta_3^3\\ &-s_{-1}s_{-3}s_2w_3\theta_2^2\theta_3 - 2s_{-1}s_{-3}s_2w_0\theta_1\theta_2\theta_3 - s_{-1}s_{-3}s_1s_2w_0\theta_2^2 - s_{-1}s_{-3}s_0w_0\theta_3^2 + s_{-1}s_{-2}w_0\theta_1^2\theta_3^2 - s_{-1}s_{-2}s_1s_2w_0\theta_3^2 - s_{-1}s_{-2}s_{-3}w_3\theta_1\theta_2^2\\ &+s_{-1}s_{-2}s_{-3}w_3\theta_1^2\theta_3 - s_{-1}s_{-2}s_{-3}w_3\theta_1^3 - 2s_{-1}s_{-2}s_{-3}w_0\theta_1^2\theta_2 - s_{-1}s_{-2}s_{-3}w_3\theta_1\theta_2 - s_{-1}s_{-2}s_{-3}w_0\theta_1^2 - s_{-1}s_{-2}s_{-3}s_1w_3\theta_1\theta_2\\ &-s_{-1}s_{-2}s_{-3}s_1w_0\theta_2^2 - s_{-1}s_{-2}s_{-3}s_1s_2w_3\theta_3 - s_{-1}s_{-2}s_{-3}s_0w_0\theta_1^2 + s_{-1}s_{-2}s_{-3}s_0s_1s_2w_0\end{aligned}}{\begin{aligned}&-2\theta_1^2\theta_2\theta_3^2 + 2\theta_1^3\theta_2\theta_3 + 2\theta_1^4\theta_2 + s_2\theta_2^2\theta_3^2 - 2s_2\theta_1\theta_2^2\theta_3 + s_2\theta_1^2\theta_2^2 + s_1\theta_1^2\theta_3^2 - 2s_1s_2\theta_1^2\theta_2 + s_{-3}\theta_3^4 - 2s_{-3}\theta_1^2\theta_3^2 + s_{-3}\theta_1^4 - 2s_{-3}s_2\theta_1\theta_2\theta_3\\ &-s_{-3}s_1s_2\theta_1^2 + s_{-2}\theta_1^2\theta_2^2 - s_{-2}s_1s_2\theta_2^2 - 2s_{-2}s_{-3}\theta_1\theta_2\theta_3 - s_{-2}s_{-3}s_2\theta_2^2 - s_{-2}s_{-3}s_1\theta_3^2 + s_{-1}\theta_1^4 - s_{-1}s_1s_2\theta_1^2 - s_{-1}s_{-3}s_2\theta_3^2 - s_{-1}s_{-2}s_{-3}\theta_1^2\\ &+s_{-1}s_{-2}s_{-3}s_1s_2\end{aligned}}$$

$$1 : 0$$

$$2 : 0$$

$$3 : \frac{\begin{aligned}&-4w_3\theta_1^2\theta_2^2\theta_3^2 + 4w_3\theta_1^3\theta_2^2\theta_3 + 3w_3\theta_1^4\theta_2^2 + 2w_0\theta_1\theta_2\theta_3^4 - 2w_0\theta_1\theta_2^3\theta_3^2 + w_0\theta_1^2\theta_2\theta_3^3 + 3w_0\theta_1^2\theta_2^3\theta_3 - 6w_0\theta_1^3\theta_2^2\theta_3 - 2w_0\theta_1^4\theta_2\theta_3 + 3w_0\theta_1^5\theta_2\\ &-2s_3w_3\theta_1^2\theta_2\theta_3^2 + 2s_3w_3\theta_1^3\theta_2\theta_3 + 2s_3w_3\theta_1^4\theta_2 + 2s_2w_3\theta_1^2\theta_2^2 - s_2w_0\theta_2^2\theta_3^3 + s_2w_0\theta_2^4\theta_3 + s_2w_0\theta_1\theta_2^2\theta_3^2 - s_2w_0\theta_1\theta_2^4 - 3s_2w_0\theta_1^2\theta_2^2\theta_3\\ &+3s_2w_0\theta_1^3\theta_2^2 + s_2s_3w_3\theta_2^2\theta_3^2 - 2s_2s_3w_3\theta_1\theta_2^2\theta_3 + s_2s_3w_3\theta_1^2\theta_2^2 + 2s_1w_3\theta_1^4\theta_2 - s_1w_0\theta_1^2\theta_3^3 - s_1w_0\theta_1^2\theta_2^2\theta_3 - s_1w_0\theta_1^3\theta_3^2 + 2s_1w_0\theta_1^3\theta_2^2\\ &+s_1w_0\theta_1^4\theta_3 + s_1s_3w_3\theta_1^2\theta_3^2 + 2s_1s_2w_0\theta_1^2\theta_2\theta_3 - 2s_1s_2s_3w_3\theta_1^2\theta_2 - 2s_{-3}w_3\theta_1^2\theta_2^2\theta_3 + 2s_{-3}w_3\theta_1^3\theta_2^2 + 2s_{-3}w_3\theta_1^4\theta_2 - s_{-3}w_0\theta_3^5 + s_{-3}w_0\theta_2^2\theta_3^3\\ &-2s_{-3}w_0\theta_1\theta_2^2\theta_3^2 + 3s_{-3}w_0\theta_1^2\theta_3^3 + s_{-3}w_0\theta_1^2\theta_2^2\theta_3 - s_{-3}w_0\theta_1^3\theta_3^2 - 2s_{-3}w_0\theta_1^4\theta_3 + s_{-3}w_0\theta_1^5 + s_{-3}s_3w_3\theta_3^4 - 2s_{-3}s_3w_3\theta_1^2\theta_3^2 + s_{-3}s_3w_3\theta_1^4\\ &+s_{-3}s_2w_3\theta_1\theta_2^2 + 2s_{-3}s_2w_0\theta_1\theta_2^2\theta_3 - s_{-3}s_2w_0\theta_1\theta_2^3 - s_{-3}s_2w_0\theta_1^2\theta_2\theta_3 + s_{-3}s_2w_0\theta_1^3\theta_2 - 2s_{-3}s_2s_3w_3\theta_1\theta_2\theta_3 + s_{-3}s_1w_3\theta_1^4 - s_{-3}s_1w_0\theta_1^2\theta_2\theta_3\\ &+s_{-3}s_1w_0\theta_1^3\theta_2 + s_{-3}s_1s_2w_0\theta_1^2\theta_3 - s_{-3}s_1s_2s_3w_3\theta_1^2 + 2s_{-2}w_3\theta_1^2\theta_2^2 - s_{-2}w_0\theta_1\theta_2^2\theta_3^2 + s_{-2}w_0\theta_1\theta_2^4 - 2s_{-2}w_0\theta_1^2\theta_2^2\theta_3 + s_{-2}w_0\theta_1^3\theta_2^2\\ &+s_{-2}s_3w_3\theta_2^2\theta_3^2 + s_{-2}s_2w_3\theta_2^4 - s_{-2}s_2w_0\theta_2^3\theta_3 + s_{-2}s_2w_0\theta_1\theta_2^3 - s_{-2}s_1w_3\theta_1^3\theta_2 - s_{-2}s_1w_3\theta_1\theta_2\theta_3^2 + s_{-2}s_1w_0\theta_1\theta_2^3 - s_{-2}s_1s_2w_0\theta_2^2\theta_3\\ &-s_{-2}s_1s_2s_3w_3\theta_2^2 + s_{-2}s_{-3}w_3\theta_2^2\theta_3^2 - 2s_{-2}s_{-3}w_3\theta_1\theta_2^2\theta_3 + s_{-2}s_{-3}w_3\theta_1^2\theta_2^2 - s_{-2}s_{-3}w_0\theta_2^3\theta_3 + 3s_{-2}s_{-3}w_0\theta_1\theta_2\theta_3^2 + s_{-2}s_{-3}w_0\theta_1\theta_2^3\\ &-2s_{-2}s_{-3}w_0\theta_1^2\theta_2\theta_3 - s_{-2}s_{-3}w_0\theta_1^3\theta_2 - 2s_{-2}s_{-3}s_3w_3\theta_1\theta_2\theta_3 + s_{-2}s_{-3}s_2w_0\theta_2^2\theta_3 - s_{-2}s_{-3}s_2w_0\theta_1\theta_2^2 - s_{-2}s_{-3}s_2s_3w_3\theta_2^2 + s_{-2}s_{-3}s_1w_0\theta_3^3\\ &-s_{-2}s_{-3}s_1w_0\theta_1^2\theta_3 - s_{-2}s_{-3}s_1s_2s_3w_3 + 2s_{-1}w_3\theta_1^4\theta_2 - s_{-1}w_0\theta_1^2\theta_3^3 + s_{-1}w_0\theta_1^2\theta_2^2\theta_3 - s_{-1}w_0\theta_1^4\theta_3 + s_{-1}w_0\theta_1^5 + s_{-1}s_3w_3\theta_1^4 + s_{-1}s_2w_3\theta_1^3\theta_2\\ &-s_{-1}s_2w_0\theta_1\theta_2^2\theta_3 + s_{-1}s_2w_0\theta_1^3\theta_2 + s_{-1}s_1w_3\theta_1^4 + s_{-1}s_1w_0\theta_1^3\theta_2 - s_{-1}s_1s_2s_3w_3\theta_1^2 + s_{-1}s_{-3}w_0\theta_1\theta_2\theta_3^2 - s_{-1}s_{-3}w_0\theta_1^2\theta_2\theta_3\\ &-s_{-1}s_{-3}s_3w_3\theta_3^2 + s_{-1}s_{-3}s_2w_0\theta_3^3 - s_{-1}s_{-3}s_2w_0\theta_2^2\theta_3 - s_{-1}s_{-3}s_2s_3w_3\theta_2^2 - 2s_{-1}s_{-2}s_{-3}w_3\theta_1^2\theta_2 - s_{-1}s_{-2}s_{-3}w_0\theta_1\theta_2^2 + s_{-1}s_{-2}s_{-3}w_0\theta_1^2\theta_3\\ &-s_{-1}s_{-2}s_{-3}w_0\theta_1^3 - s_{-1}s_{-2}s_{-3}s_3w_3\theta_1^2 - s_{-1}s_{-2}s_{-3}s_2w_3\theta_2^2 - s_{-1}s_{-2}s_{-3}s_1w_0\theta_2^2 - s_{-1}s_{-2}s_{-3}s_1s_2w_3\theta_3 - s_{-1}s_{-2}s_{-3}s_1s_2w_0\theta_1\\ &-s_{-1}s_{-2}s_{-3}s_1s_2w_0\theta_3 + s_{-1}s_{-2}s_{-3}s_1s_2s_3w_3\end{aligned}}{\begin{aligned}&-2\theta_1^2\theta_2\theta_3^2 + 2\theta_1^3\theta_2\theta_3 + 2\theta_1^4\theta_2 + s_2\theta_2^2\theta_3^2 - 2s_2\theta_1\theta_2^2\theta_3 + s_2\theta_1^2\theta_2^2 + s_1\theta_1^2\theta_3^2 - 2s_1s_2\theta_1^2\theta_2 + s_{-3}\theta_3^4 - 2s_{-3}\theta_1^2\theta_3^2 + s_{-3}\theta_1^4 - 2s_{-3}s_2\theta_1\theta_2\theta_3\\ &-s_{-3}s_1s_2\theta_1^2 + s_{-2}\theta_1^2\theta_2^2 - s_{-2}s_1s_2\theta_2^2 - 2s_{-2}s_{-3}\theta_1\theta_2\theta_3 - s_{-2}s_{-3}s_2\theta_2^2 - s_{-2}s_{-3}s_1\theta_3^2 + s_{-1}\theta_1^4 - s_{-1}s_1s_2\theta_1^2 - s_{-1}s_{-3}s_2\theta_3^2 - s_{-1}s_{-2}s_{-3}\theta_1^2\\ &+s_{-1}s_{-2}s_{-3}s_1s_2\end{aligned}}$$

eqSol

$$-3 : \frac{w_0\theta_3 + w_{-2}\theta_1 + w_{-1}\theta_2}{s_{-3}}$$

$$-2 : \frac{w_0\theta_1\theta_3 + w_{-1}\theta_1\theta_2 + s_{-3}w_1\theta_3 + s_{-3}w_0\theta_2 + s_{-3}w_{-1}\theta_1}{-\theta_1^2 + s_{-2}s_{-3}}$$

$$-1 : \frac{-w_2\theta_1^2\theta_3 + w_1\theta_1\theta_2\theta_3 - w_1\theta_1^2\theta_2 + w_0\theta_1\theta_2^2 + w_0\theta_1^2\theta_3 - w_0\theta_1^3 + s_{-3}w_1\theta_1\theta_3 + s_{-3}w_0\theta_1\theta_2 + s_{-2}w_0\theta_2\theta_3 + s_{-2}s_{-3}w_2\theta_3 + s_{-2}s_{-3}w_1\theta_2 + s_{-2}s_{-3}w_0\theta_1}{-2\theta_1^2\theta_2 - s_{-3}\theta_1^2 - s_{-2}\theta_2^2 - s_{-1}\theta_1^2 + s_{-1}s_{-2}s_{-3}}$$

$$1: \frac{\begin{array}{c}-2w_3\theta_1^2\theta_2^2 + w_2\theta_1\theta_2\theta_3^2 - w_2\theta_1^2\theta_2\theta_3 - 2w_2\theta_1^3\theta_2 - w_0\theta_2^3\theta_3 + w_0\theta_1\theta_2^2\theta_3 + w_0\theta_1\theta_2^3 + 2w_0\theta_1^2\theta_2\theta_3 - 3w_0\theta_1^3\theta_2 - s_{-3}w_3\theta_1^2\theta_2 + s_{-3}w_2\theta_1\theta_3^2 \\ -s_{-3}w_2\theta_1^3 + s_{-3}w_0\theta_1\theta_2^2 + s_{-3}w_0\theta_1^2\theta_3 - s_{-3}w_0\theta_1^3 - s_{-2}w_3\theta_2^3 - s_{-2}w_2\theta_1^2\theta_2 + s_{-2}w_0\theta_2^2\theta_3 - s_{-2}w_0\theta_1\theta_2^2 + s_{-2}s_{-3}w_2\theta_2\theta_3 + s_{-2}s_{-3}w_0\theta_1\theta_2 \\ -s_{-1}w_3\theta_1^2\theta_2 - s_{-1}w_2\theta_1^3 + s_{-1}w_0\theta_1\theta_3^2 - s_{-1}w_0\theta_1^3 + s_{-1}s_{-3}w_0\theta_2\theta_3 + s_{-1}s_{-2}s_{-3}w_3\theta_2 + s_{-1}s_{-2}s_{-3}w_2\theta_1 + s_{-1}s_{-2}s_{-3}w_0\theta_1\end{array}}{\theta_2^2\theta_3^2 - 2\theta_1\theta_2^2\theta_3 + \theta_1^2\theta_2^2 - 2s_1\theta_1^2\theta_2 - 2s_{-3}\theta_1\theta_2\theta_3 - s_{-3}s_1\theta_1^2 - s_{-2}s_1\theta_2^2 - s_{-2}s_{-3}\theta_2^2 - s_{-1}s_1\theta_1^2 - s_{-1}s_{-3}\theta_3^2 + s_{-1}s_{-2}s_{-3}s_1}$$

$$2: \frac{\begin{array}{c}2w_3\theta_1\theta_2^2\theta_3^2 - 3w_3\theta_1^2\theta_2^2\theta_3 - w_3\theta_1^3\theta_2^2 - w_0\theta_2\theta_3^4 + w_0\theta_2^3\theta_3^2 + w_0\theta_1\theta_2^3 - 3w_0\theta_1\theta_2^3\theta_3 + 2w_0\theta_1^2\theta_2\theta_3^2 + 2w_0\theta_1^2\theta_3^3 + w_0\theta_1^3\theta_2\theta_3 \\ -3w_0\theta_1^4\theta_2 - 2s_1w_3\theta_1^3\theta_2 + s_1w_0\theta_1\theta_2^2\theta_3 + s_1w_0\theta_1^2\theta_3^2 - 2s_1w_0\theta_1^2\theta_2 - s_1w_0\theta_1^3\theta_3 + s_{-3}w_3\theta_1\theta_2\theta_3^2 - 2s_{-3}w_3\theta_1^2\theta_2\theta_3 - s_{-3}w_3\theta_1^3\theta_2 \\ +s_{-3}w_0\theta_2^2\theta_3^2 - s_{-3}w_0\theta_1\theta_3^3 - 2s_{-3}w_0\theta_1\theta_2^2\theta_3 + s_{-3}w_0\theta_1^2\theta_3^2 + s_{-3}w_0\theta_1^2\theta_2^2 + s_{-3}w_0\theta_1^3\theta_3 - s_{-3}w_0\theta_1^4 - s_{-3}s_1w_3\theta_1^3 + s_{-3}s_1w_0\theta_1\theta_2\theta_3 \\ -s_{-3}s_1w_0\theta_1^2\theta_2 - s_{-2}w_3\theta_1\theta_2^3 + s_{-2}w_0\theta_1\theta_2^2\theta_3 - s_{-2}w_0\theta_1^2\theta_2^2 - s_{-2}s_1w_3\theta_1\theta_2^2 + s_{-2}s_1w_0\theta_2^2\theta_3 - s_{-2}s_1w_0\theta_2^3 + s_{-2}s_{-3}w_3\theta_2^2\theta_3 - s_{-2}s_{-3}w_3\theta_1\theta_2^2 \\ -s_{-2}s_{-3}w_0\theta_2^3 + s_{-2}s_{-3}w_0\theta_1\theta_2\theta_3 + s_{-2}s_{-3}w_0\theta_1^2\theta_2 + s_{-2}s_{-3}s_1w_0\theta_1\theta_3 - s_{-1}w_3\theta_1^3\theta_2 + s_{-1}w_0\theta_1^2\theta_3^2 - s_{-1}w_0\theta_1^4 - s_{-1}s_1w_3\theta_1^3 - s_{-1}s_1w_0\theta_1^2 \\ -s_{-1}s_{-3}w_3\theta_1\theta_3^2 - s_{-1}s_{-3}w_0\theta_2\theta_3^2 + s_{-1}s_{-3}w_0\theta_1\theta_2\theta_3 + s_{-1}s_{-2}s_{-3}w_3\theta_1\theta_2 + s_{-1}s_{-2}s_{-3}w_0\theta_1^2 + s_{-1}s_{-2}s_{-3}s_1w_3\theta_1 + s_{-1}s_{-2}s_{-3}s_1w_0\theta_2\end{array}}{\begin{array}{c}-2\theta_1^2\theta_2\theta_3^2 + 2\theta_1^3\theta_2\theta_3 + 2\theta_1^4\theta_2 + s_2\theta_2^2\theta_3^2 - 2s_2\theta_1\theta_2^2\theta_3 + s_2\theta_1^2\theta_2^2 + s_1\theta_1^2\theta_3^2 - 2s_1s_2\theta_1^2\theta_2 + s_{-3}\theta_3^4 - 2s_{-3}\theta_1^2\theta_3^2 + s_{-3}\theta_1^4 - 2s_{-3}s_2\theta_1\theta_2\theta_3 \\ -s_{-3}s_1s_2\theta_1^2 + s_{-2}\theta_1^2\theta_2^2 - s_{-2}s_1s_2\theta_2^2 - 2s_{-2}s_{-3}\theta_1\theta_2\theta_3 - s_{-2}s_{-3}s_2\theta_2^2 - s_{-2}s_{-3}s_1\theta_3^2 + s_{-1}\theta_1^4 - s_{-1}s_1s_2\theta_1^2 - s_{-1}s_{-3}s_2\theta_3^2 - s_{-1}s_{-2}s_{-3}\theta_1^2 \\ +s_{-1}s_{-2}s_{-3}s_1s_2\end{array}}$$

$$3: \frac{w_0 \begin{pmatrix}-2\theta_1\theta_2\theta_3^4 + 2\theta_1\theta_2^3\theta_3^2 - \theta_1^2\theta_2\theta_3^3 - 3\theta_1^2\theta_2^3\theta_3 + 6\theta_1^3\theta_2\theta_3^2 + 2\theta_1^4\theta_2\theta_3 - 3\theta_1^5\theta_2 + s_2\theta_2^2\theta_3^3 - s_2\theta_2^4\theta_3 - s_2\theta_1\theta_2^2\theta_3^2 + s_2\theta_1\theta_2^4 \\ +3s_2\theta_1^2\theta_2^2\theta_3 - 3s_2\theta_1^3\theta_2^2 + s_1\theta_1^2\theta_3^3 + s_1\theta_1^2\theta_2^2\theta_3 + s_1\theta_1^3\theta_3^2 - 2s_1\theta_1^3\theta_2^2 - s_1\theta_1^4\theta_3 - 2s_1s_2\theta_1^2\theta_2\theta_3 + s_{-3}\theta_3^5 - s_{-3}\theta_2^2\theta_3^3 + 2s_{-3}\theta_1\theta_2\theta_3^2 \\ -3s_{-3}\theta_1^2\theta_3^3 - s_{-3}\theta_1^2\theta_2^2\theta_3 + s_{-3}\theta_1^3\theta_3^2 + 2s_{-3}\theta_1^4\theta_3 - s_{-3}\theta_1^5 - 2s_{-3}s_2\theta_1\theta_2^2\theta_3 + s_{-3}s_2\theta_1^2\theta_3 + s_{-3}s_2\theta_1^2\theta_2\theta_3 - s_{-3}s_2\theta_1^3\theta_2 + s_{-3}s_1\theta_1^2\theta_2\theta_3 \\ -s_{-3}s_1\theta_1^3\theta_2 - s_{-3}s_1s_2\theta_1^2\theta_3 + s_{-2}\theta_1\theta_2^2\theta_3^2 - s_{-2}\theta_1\theta_2^4 + 2s_{-2}\theta_1^2\theta_2^2\theta_3 - s_{-2}\theta_1^3\theta_2^2 + s_{-2}\theta_2^3\theta_3 - s_{-2}s_2\theta_1\theta_2^3 + s_{-2}s_1\theta_1\theta_2\theta_3^2 \\ -s_{-2}s_1\theta_1\theta_2^3 - s_{-2}s_1s_2\theta_2^2\theta_3 + s_{-2}s_{-3}\theta_2\theta_3^3 - 3s_{-2}s_{-3}\theta_1\theta_2\theta_3^2 - s_{-2}s_{-3}\theta_1^3\theta_3 + 2s_{-2}s_{-3}\theta_1^2\theta_2\theta_3 + s_{-2}s_{-3}\theta_1^3\theta_2 - s_{-2}s_{-3}s_2\theta_2^2\theta_3 \\ +s_{-2}s_{-3}s_2\theta_1\theta_2^2 - s_{-2}s_{-3}s_1\theta_3^3 + s_{-2}s_{-3}s_1\theta_1^2\theta_3 + s_{-1}\theta_1^3\theta_3^2 - s_{-1}\theta_1^3\theta_2^2 + s_{-1}\theta_1^4\theta_3 - s_{-1}\theta_1^5 + s_{-1}s_2\theta_1\theta_2\theta_3^2 - s_{-1}s_2\theta_1^2\theta_2 - s_{-1}s_1\theta_1^3\theta_2 \\ -s_{-1}s_1s_2\theta_1^2\theta_3 - s_{-1}s_{-3}\theta_1\theta_2\theta_3^2 + s_{-1}s_{-3}\theta_1^2\theta_2\theta_3 - s_{-1}s_{-3}\theta_3^3 + s_{-1}s_{-3}s_2\theta_2^2\theta_3 + s_{-1}s_{-2}s_{-3}\theta_1\theta_2^2 - s_{-1}s_{-2}s_{-3}\theta_1^2\theta_3 + s_{-1}s_{-2}s_{-3}\theta_1^3 \\ +s_{-1}s_{-2}s_{-3}s_2\theta_1\theta_2 + s_{-1}s_{-2}s_{-3}s_1\theta_1\theta_2 + s_{-1}s_{-2}s_{-3}\theta_2\theta_3\end{pmatrix}}{\begin{array}{c}-4\theta_1^2\theta_2^2\theta_3^2 + 4\theta_1^3\theta_2^2\theta_3 + 3\theta_1^4\theta_2^2 - 2s_3\theta_1^2\theta_2\theta_3^2 + 2s_3\theta_1^3\theta_2\theta_3 + 2s_3\theta_1^4\theta_2 + 2s_2\theta_1^2\theta_3^3 + s_2s_3\theta_2^2\theta_3^2 - 2s_2s_3\theta_1\theta_2^2\theta_3 + s_2s_3\theta_1^2\theta_2^2 + 2s_1\theta_1^4\theta_2 \\ +s_1s_3\theta_1^2\theta_3^2 - 2s_1s_2s_3\theta_1^2\theta_2 - 2s_{-3}\theta_1^2\theta_2\theta_3^2 + 2s_{-3}\theta_1^4\theta_2 + s_{-3}s_3\theta_3^4 - 2s_{-3}s_3\theta_1^2\theta_3^2 + s_{-3}s_3\theta_1^4 + s_{-3}\theta_1^2\theta_2^2 \\ -2s_{-3}s_2s_3\theta_1\theta_2\theta_3 + s_{-3}s_1\theta_1^4 - s_{-3}s_1s_2s_3\theta_1^2 + 2s_{-2}\theta_1^2\theta_2^3 + s_{-2}s_3\theta_1^2\theta_2^2 + s_{-2}s_2\theta_2^4 + s_{-2}s_1\theta_1^2\theta_2^2 - s_{-2}s_1s_2s_3\theta_2^2 + s_{-2}s_{-3}\theta_2^2\theta_3^2 \\ -2s_{-2}s_{-3}\theta_1\theta_2^2\theta_3 + s_{-2}s_{-3}\theta_1^2\theta_2^2 - 2s_{-2}s_{-3}s_3\theta_1\theta_2\theta_3 - s_{-2}s_{-3}s_2s_3\theta_2^2 - s_{-2}s_{-3}s_1s_3\theta_3^2 + 2s_{-1}\theta_1^4\theta_2 + s_{-1}s_3\theta_1^4 + s_{-1}s_2\theta_1^2\theta_2^2 + s_{-1}s_1\theta_1^4 \\ -s_{-1}s_1s_2s_3\theta_1^2 + s_{-1}s_{-3}\theta_1^2\theta_3^2 - s_{-1}s_{-3}s_2s_3\theta_3^2 - 2s_{-1}s_{-2}s_{-3}\theta_1^2\theta_2 - s_{-1}s_{-2}s_{-3}s_3\theta_1^2 - s_{-1}s_{-2}s_{-3}s_2\theta_2^2 - s_{-1}s_{-2}s_{-3}s_1\theta_1^2 + s_{-1}s_{-2}s_{-3}s_1s_2s_3\end{array}}$$

eqs

$$-3 : 0$$

$$-2 : 0$$

$$-1 : 0$$

$$0: \frac{w_0 \begin{pmatrix} 4\theta_1\theta_2\theta_3^5 - 8\theta_1\theta_2^3\theta_3^3 + 4\theta_1\theta_2^5\theta_3 + 2\theta_1^2\theta_2\theta_3^4 + 12\theta_1^2\theta_2^3\theta_3^2 - 6\theta_1^2\theta_2^5 - 16\theta_1^3\theta_2\theta_3^3 - 8\theta_1^3\theta_2^3\theta_3 - 2\theta_1^4\theta_2\theta_3^2 + 12\theta_1^4\theta_2^3 + 16\theta_1^5\theta_2\theta_3 \\ -6\theta_1^6\theta_2 - s_3\theta_3^6 + 2s_3\theta_2^2\theta_3^4 - s_3\theta_2^4\theta_3^2 - 4s_3\theta_1\theta_2^2\theta_3^3 + 4s_3\theta_1\theta_2^4\theta_3 + 4s_3\theta_1^2\theta_3^4 - 4s_3\theta_1^2\theta_2^4 - 2s_3\theta_1^3\theta_3^3 - 2s_3\theta_1^3\theta_2^2\theta_3 - 4s_3\theta_1^4\theta_3^2 \\ +8s_3\theta_1^4\theta_2^2 + 4s_3\theta_1^5\theta_3 - s_3\theta_1^6 - s_2\theta_2^2\theta_3^4 + 2s_2\theta_2^4\theta_3^2 - s_2\theta_2^6 - 6s_2\theta_1^2\theta_2^2\theta_3^2 + 2s_2\theta_1^2\theta_2^4 + 8s_2\theta_1^3\theta_2^2\theta_3 - s_2\theta_1^4\theta_2^2 - 2s_2s_3\theta_2^3\theta_3^2 \\ +4s_2s_3\theta_1\theta_2^2\theta_3^3 + 4s_2s_3\theta_1\theta_2^3\theta_3 - 4s_2s_3\theta_1^2\theta_2\theta_3^2 - 2s_2s_3\theta_1^2\theta_2^3 - 4s_2s_3\theta_1^3\theta_2\theta_3 + 4s_2s_3\theta_1^4\theta_2 - s_1\theta_1^2\theta_3^4 - 2s_1\theta_1^2\theta_2^2\theta_3^2 - s_1\theta_1^2\theta_2^4 \\ -2s_1\theta_1^3\theta_3^3 + 6s_1\theta_1^3\theta_2^2\theta_3 + s_1\theta_1^4\theta_3^2 + 2s_1\theta_1^4\theta_2^2 + 2s_1\theta_1^5\theta_3 - s_1\theta_1^6 + 2s_1s_3\theta_1\theta_2^2\theta_3^3 - 2s_1s_3\theta_1\theta_2^3\theta_3^2 - 2s_1s_3\theta_1^2\theta_2^2\theta_3^2 + 2s_1s_3\theta_1^2\theta_2^3 \\ +2s_1s_3\theta_1^3\theta_2\theta_3 + 2s_1s_2\theta_1^2\theta_2\theta_3^2 + s_1s_2s_3\theta_2^4 - 2s_1s_2s_3\theta_1\theta_2^2\theta_3 + s_1s_2s_3\theta_1^2\theta_3^2 - 2s_1s_2s_3\theta_1^2\theta_2^2 - 2s_1s_2s_3\theta_1^3\theta_3 + s_1s_2s_3\theta_1^4 - 4s_0\theta_1^2\theta_2^2\theta_3^3 \\ +4s_0\theta_1^3\theta_2^2\theta_3 + 3s_0\theta_1^4\theta_2^2 - 2s_0s_3\theta_1^2\theta_2\theta_3^2 + 2s_0s_3\theta_1^3\theta_2\theta_3 + 2s_0s_3\theta_1^4\theta_2 + 2s_0s_2\theta_1^2\theta_3^2 + s_0s_2s_3\theta_2^2\theta_3 - 2s_0s_2s_3\theta_1\theta_2^2\theta_3 + s_0s_2s_3\theta_1^2\theta_2^2 \\ +2s_0s_1\theta_1^4\theta_2 + s_0s_1s_3\theta_1^2\theta_3^2 - 2s_0s_1s_2s_3\theta_1^2\theta_2 - s_{-3}\theta_3^6 + 2s_{-3}\theta_2^2\theta_3^4 - s_{-3}\theta_2^4\theta_3^2 - 4s_{-3}\theta_1\theta_2^2\theta_3^3 + 4s_{-3}\theta_1\theta_2^4\theta_3 + 4s_{-3}\theta_1^2\theta_3^4 - 4s_{-3}\theta_1^2\theta_2^4 \\ -2s_{-3}\theta_1^3\theta_3^3 - 2s_{-3}\theta_1^3\theta_2^2\theta_3 - 4s_{-3}\theta_1^4\theta_3^2 + 8s_{-3}\theta_1^4\theta_2^2 + 4s_{-3}\theta_1^5\theta_3 - s_{-3}\theta_1^6 - 2s_{-3}s_3\theta_2^2\theta_3^2 + 4s_{-3}s_3\theta_1\theta_2^2\theta_3 + 4s_{-3}s_3\theta_1\theta_2^3\theta_3 \\ -4s_{-3}s_3\theta_1^2\theta_2\theta_3^2 - 2s_{-3}s_3\theta_1^2\theta_2^3 - 4s_{-3}s_3\theta_1^3\theta_2\theta_3 + 4s_{-3}s_3\theta_1^4\theta_2 + 2s_{-3}s_2\theta_1\theta_2\theta_3^3 - 2s_{-3}s_2\theta_1\theta_2^3\theta_3 - 2s_{-3}s_2\theta_1^2\theta_2\theta_3^2 + 2s_{-3}s_2\theta_1^3\theta_2\theta_3 \\ +2s_{-3}s_2\theta_1^3\theta_2\theta_3 + s_{-3}s_2s_3\theta_2^4 - 2s_{-3}s_2s_3\theta_1\theta_2^2\theta_3 + s_{-3}s_2s_3\theta_1^2\theta_3^2 - 2s_{-3}s_2s_3\theta_1^2\theta_2^2 - 2s_{-3}s_2s_3\theta_1^3\theta_3 + s_{-3}s_2s_3\theta_1^4 - 2s_{-3}s_1\theta_1^2\theta_2\theta_3^2 \\ +2s_{-3}s_1\theta_1^3\theta_2\theta_3 + 2s_{-3}s_1\theta_1^4\theta_2 + s_{-3}s_1s_3\theta_2^2\theta_3^2 - 2s_{-3}s_1s_3\theta_1\theta_2^2\theta_3 + s_{-3}s_1s_3\theta_1^2\theta_2^2 + s_{-3}s_1s_2\theta_1^2\theta_3^2 - 2s_{-3}s_1s_2s_3\theta_1^2\theta_2 - 2s_{-3}s_0\theta_1^2\theta_2\theta_3^2 \\ +2s_{-3}s_0\theta_1^3\theta_2\theta_3 + 2s_{-3}s_0\theta_1^4\theta_2 + s_{-3}s_0s_3\theta_3^4 - 2s_{-3}s_0s_3\theta_1^2\theta_3^2 + s_{-3}s_0s_3\theta_1^4 + s_{-3}s_0s_2\theta_1^2\theta_2^2 - 2s_{-3}s_0s_2s_3\theta_1\theta_2\theta_3 + s_{-3}s_0s_1\theta_1^4 \\ -s_{-3}s_0s_1s_2s_3\theta_1^2 - s_{-2}\theta_2^2\theta_3^4 + 2s_{-2}\theta_2^4\theta_3^2 - s_{-2}\theta_2^6 - 6s_{-2}\theta_1^2\theta_2^2\theta_3^2 + 2s_{-2}\theta_1^2\theta_2^4 + 8s_{-2}\theta_1^3\theta_2^2\theta_3 - s_{-2}\theta_1^4\theta_2^2 + 2s_{-2}s_3\theta_1\theta_2\theta_3^3 \\ -2s_{-2}s_3\theta_1\theta_2^3\theta_3 - 2s_{-2}s_3\theta_1^2\theta_2\theta_3^2 + 2s_{-2}s_3\theta_1^2\theta_2^3 + 2s_{-2}s_3\theta_1^3\theta_2\theta_3 - 2s_{-2}s_2\theta_1^2\theta_2\theta_3^2 + 4s_{-2}s_2\theta_1^2\theta_2^3 + s_{-2}s_3s_2\theta_2^2\theta_3^2 - 2s_{-2}s_2s_3\theta_1\theta_2^2\theta_3 \\ +s_{-2}s_2s_3\theta_1^2\theta_2^2 - 2s_{-2}s_1\theta_1\theta_2\theta_3^3 + 2s_{-2}s_1\theta_1\theta_2^3\theta_3 + 2s_{-2}s_1\theta_1^3\theta_2\theta_3 + s_{-2}s_1s_3\theta_2^4 - 2s_{-2}s_1s_3\theta_1\theta_2^2\theta_3 + s_{-2}s_1s_3\theta_1^2\theta_2^2 + 4s_{-2}s_1\theta_1^3\theta_2\theta_3 \\ +s_{-2}s_1s_2\theta_1^2\theta_2^2 - 2s_{-2}s_1s_2s_3\theta_1\theta_2\theta_3 + 2s_{-2}s_0\theta_1^4\theta_2 + s_{-2}s_0s_3\theta_1^4 + s_{-2}s_0s_2\theta_1^2\theta_2^2 - s_{-2}s_0s_1s_2s_3\theta_1^2 + 2s_{-2}s_{-3}\theta_1\theta_2\theta_3^3 \\ -2s_{-2}s_{-3}\theta_1\theta_2^3\theta_3 - 2s_{-2}s_{-3}\theta_1^2\theta_2\theta_3^2 + 2s_{-2}s_{-3}\theta_1^2\theta_2^3 + 2s_{-2}s_{-3}\theta_1^3\theta_2\theta_3 - 2s_{-2}s_{-3}s_3\theta_1^2\theta_2^2 + s_{-2}s_{-3}s_3\theta_1^2\theta_2^2 \\ +s_{-2}s_{-3}\theta_2^4 - 2s_{-2}s_{-3}\theta_2^2\theta_3^2 + s_{-2}s_{-3}\theta_2^4 - 2s_{-2}s_{-3}s_2\theta_1\theta_2\theta_3 + s_{-2}s_{-3}\theta_1^2\theta_2^2 - s_{-2}s_{-3}s_1s_2s_3\theta_1^2 + s_{-2}s_{-3}s_0\theta_1^4\theta_2 \\ -s_{-2}s_{-3}s_0s_2s_3\theta_1^2 + 2s_{-1}\theta_1^2\theta_2\theta_3^2 + s_{-1}s_2s_3\theta_1^2\theta_2^2 + s_{-1}s_2\theta_1^2\theta_2^2 + s_{-1}s_2\theta_1^2\theta_2^2 - s_{-1}s_2s_3\theta_1^2 + s_{-1}s_2s_3\theta_1^4 \\ -2s_{-1}s_2s_3\theta_1\theta_2\theta_3 + s_{-1}s_2s_3\theta_1^2\theta_2^2 - 2s_{-1}s_2s_3\theta_1^2\theta_2^2 - 2s_{-1}s_2s_3\theta_1^3\theta_3 + s_{-1}s_2s_3\theta_1^4 - 2s_{-1}s_2s_3\theta_1^2 \\ -2s_{-1}s_2s_3\theta_1\theta_2\theta_3 + s_{-1}s_2s_3\theta_1^2\theta_2^2 - 2s_{-1}s_2s_3\theta_1^2 - 2s_{-1}s_2s_3s_3\theta_1^2\theta_2 \\ -s_{-1}s_2s_3s_0s_3\theta_1^2 - s_{-1}s_2s_3s_0s_2\theta_2^2 - s_{-1}s_2s_3s_0s_1\theta_1^2 + s_{-1}s_{-2}s_{-3}s_0s_1s_2s_3 \end{pmatrix}}{\begin{matrix} -4\theta_1^2\theta_2^2\theta_3^2 + 4\theta_1^3\theta_2^2\theta_3 + 3\theta_1^4\theta_2^2 - 2s_3\theta_1^2\theta_2\theta_3^2 + 2s_3\theta_1^3\theta_2\theta_3 + 2s_3\theta_1^4\theta_2 + 2s_2\theta_1^2\theta_3^2 + s_2s_3\theta_2^2\theta_3 - 2s_2s_3\theta_1\theta_2^2\theta_3 + s_2s_3\theta_1^2\theta_2^2 + 2s_1\theta_1^4\theta_2 \\ +s_1s_3\theta_1^2\theta_3^2 - 2s_1s_2s_3\theta_1^2\theta_2 - 2s_{-3}\theta_1^2\theta_2\theta_3^2 + 2s_{-3}\theta_1^3\theta_2\theta_3 + 2s_{-3}\theta_1^4\theta_2 + s_{-3}s_3\theta_3^4 - 2s_{-3}s_3\theta_1^2\theta_3^2 + s_{-3}s_3\theta_1^4 + s_{-3}\theta_1^2\theta_2^2 - 2s_{-3}s_3\theta_1\theta_2\theta_3 \\ +s_{-3}s_1\theta_1^4 - s_{-3}s_1s_2s_3\theta_1^2 + 2s_{-2}\theta_1^2\theta_2^2 + s_{-2}s_3\theta_2^2\theta_3^2 + s_{-2}s_2\theta_2^4 + s_{-2}s_1\theta_1^2\theta_2^2 - s_{-2}s_1s_2\theta_2^2 + s_{-2}s_{-3}\theta_2^2\theta_3^2 - 2s_{-2}s_{-3}\theta_1\theta_2^2\theta_3 \\ +s_{-2}s_{-3}\theta_1^2\theta_2^2 - 2s_{-2}s_{-3}s_3\theta_1\theta_2\theta_3 - s_{-2}s_{-3}s_2s_3\theta_2^2 - s_{-2}s_{-3}s_1s_3\theta_3^2 + 2s_{-1}\theta_1^4\theta_2 + s_{-1}s_3\theta_1^4 + s_{-1}s_2\theta_1^2\theta_2^2 + s_{-1}s_1\theta_1^4 - s_{-1}s_1s_2s_3\theta_1^2 \\ +s_{-1}s_{-3}\theta_1^2\theta_3^2 - s_{-1}s_{-3}s_2s_3\theta_3^2 - 2s_{-1}s_{-2}s_{-3}\theta_1^2\theta_2 - s_{-1}s_{-2}s_{-3}s_3\theta_1^2 - s_{-1}s_{-2}s_{-3}s_2\theta_2^2 - s_{-1}s_{-2}s_{-3}s_1\theta_1^2 + s_{-1}s_{-2}s_{-3}s_1s_2s_3 \end{matrix}}$$

$$1: 0$$

$$2: 0$$

$$3: 0$$

checking out equation 0

divide by w_0

(leave out due to size)

eliminate higher-order terms

$$s_0 + \theta_1^6 \left( -\frac{1}{s_{-2}s_{-3}s_1s_2s_3} - \frac{1}{s_{-1}s_{-2}s_1s_2s_3} - \frac{1}{s_{-1}s_{-2}s_{-3}s_2s_3} - \frac{1}{s_{-1}s_{-2}s_{-3}s_1s_2} \right)$$
$$+ \theta_1^4 \left( \theta_2 \left( \frac{2}{s_{-2}s_{-3}s_1s_2} + \frac{4}{s_{-1}s_1s_2s_3} + \frac{2}{s_{-1}s_{-2}s_2s_3} + \frac{4}{s_{-1}s_{-2}s_1s_2} + \frac{4}{s_{-1}s_{-2}s_{-3}s_1} \right) \right.$$
$$+ \frac{1}{s_1s_2s_3} + \frac{1}{s_{-2}s_{-3}s_1} + \frac{1}{s_{-1}s_2s_3} + \frac{1}{s_{-1}s_1s_2} + \frac{1}{s_{-1}s_{-2}s_1} + \frac{1}{s_{-1}s_{-2}s_{-3}} \right)$$
$$+ \theta_1^2 \left( \theta_2^2 \left( -\frac{2}{s_1s_2s_3} + \frac{1}{s_{-2}s_2s_3} + \frac{1}{s_{-2}s_1s_2} + \frac{1}{s_{-2}s_{-3}s_2} + \frac{1}{s_{-1}s_1s_3} - \frac{2}{s_{-1}s_1s_2} \right. \right.$$
$$+ \frac{1}{s_{-1}s_{-3}s_1} + \frac{1}{s_{-1}s_{-2}s_2} - \frac{2}{s_{-1}s_{-2}s_1} - \frac{2}{s_{-1}s_{-2}s_{-3}} \right) + \theta_2 \left( -\frac{2}{s_1s_2} - \frac{2}{s_{-1}s_1} - \frac{2}{s_{-1}s_{-2}} \right)$$
$$\left. -\frac{1}{s_1} - \frac{1}{s_{-1}} \right) + \theta_2^2 \left( -\frac{1}{s_2} - \frac{1}{s_{-2}} \right) + \theta_3^2 \left( -\frac{1}{s_3} - \frac{1}{s_{-3}} \right)$$
$$+ \theta_3 \left( \theta_1^3 \left( -\frac{2}{s_1s_2s_3} - \frac{2}{s_{-1}s_1s_2} - \frac{2}{s_{-1}s_{-2}s_1} - \frac{2}{s_{-1}s_{-2}s_{-3}} \right) \right.$$
$$\left. \left. + \theta_1 \theta_2 \left( -\frac{2}{s_2s_3} - \frac{2}{s_1s_3} - \frac{2}{s_{-2}s_1} - \frac{2}{s_{-2}s_{-3}} - \frac{2}{s_{-1}s_2} - \frac{2}{s_{-1}s_{-3}} \right) \right) \right)$$

reformat manually

$$s_0 - \theta_1^2 \left( \frac{1}{s_1} + \frac{1}{s_{-1}} \right) + \theta_1^4 \left( \frac{1}{s_{-1}s_1s_2} + \frac{1}{s_{-1}s_{-2}s_1} + \frac{1}{s_1s_2s_3} + \frac{1}{s_{-2}s_{-3}s_1} + \frac{1}{s_{-1}s_2s_3} + \frac{1}{s_{-1}s_{-2}s_{-3}} \right)$$
$$- 2\theta_1^2 \theta_2 \left( \frac{1}{s_1s_2} + \frac{1}{s_{-1}s_1} + \frac{1}{s_{-1}s_{-2}} \right) - \theta_2^2 \left( \frac{1}{s_2} + \frac{1}{s_{-2}} \right)$$
$$- \theta_1^6 \left( \frac{1}{s_{-2}s_{-3}s_1s_2s_3} + \frac{1}{s_{-1}s_{-2}s_1s_2s_3} + \frac{1}{s_{-1}s_{-2}s_{-3}s_2s_3} + \frac{1}{s_{-1}s_{-2}s_{-3}s_1s_2} \right)$$
$$+ 2\theta_1^4 \theta_2 \left( \frac{1}{s_{-2}s_{-3}s_1s_2} + \frac{2}{s_{-1}s_1s_2s_3} + \frac{1}{s_{-1}s_{-2}s_2s_3} + \frac{2}{s_{-1}s_{-2}s_1s_2} + \frac{2}{s_{-1}s_{-2}s_{-3}s_1} \right)$$
$$+ \theta_1^2 \theta_2^2 \left( -\frac{2}{s_1s_2s_3} + \frac{1}{s_{-2}s_2s_3} + \frac{1}{s_{-2}s_1s_2} + \frac{1}{s_{-2}s_{-3}s_2} + \frac{1}{s_{-1}s_1s_3} - \frac{2}{s_{-1}s_1s_2} \right.$$
$$\left. + \frac{1}{s_{-1}s_{-3}s_1} + \frac{1}{s_{-1}s_{-2}s_2} - \frac{2}{s_{-1}s_{-2}s_1} - \frac{2}{s_{-1}s_{-2}s_{-3}} \right) - \theta_3^2 \left( \frac{1}{s_3} + \frac{1}{s_{-3}} \right)$$
$$- 2\theta_1^3 \theta_3 \left( \frac{1}{s_{-1}s_{-2}s_{-3}} + \frac{1}{s_{-1}s_{-2}s_1} + \frac{1}{s_{-1}s_1s_2} + \frac{1}{s_1s_2s_3} \right)$$
$$- 2\theta_1 \theta_2 \theta_3 \left( \frac{1}{s_{-1}s_2} + \frac{1}{s_{-1}s_{-3}} + \frac{1}{s_2s_3} + \frac{1}{s_1s_3} + \frac{1}{s_{-2}s_1} + \frac{1}{s_{-2}s_{-3}} \right)$$

## A3. Appendix 3 – Solve for $c$ up to $m^7$

Now testing solution of Hill literal perigee, function cSolve.

Documentation, page 34 and page 39

Document some expressions for use in eliminating higher-order terms.

$$(-\theta_0 + c^2)^3 = -\theta_0^3 + 3c^2\theta_0^2 - 3c^4\theta_0 + c^6$$

$$(-\theta_0 + c^2)^4 = \theta_0^4 - 4c^2\theta_0^3 + 6c^4\theta_0^2 - 4c^6\theta_0 + c^8$$

$$(-\theta_0 + c^2)^5 = -\theta_0^5 + 5c^2\theta_0^4 - 10c^4\theta_0^3 + 10c^6\theta_0^2 - 5c^8\theta_0 + c^{10}$$

$$(-\theta_0 + c^2)^6 = \theta_0^6 - 6c^2\theta_0^5 + 15c^4\theta_0^4 - 20c^6\theta_0^3 + 15c^8\theta_0^2 - 6c^{10}\theta_0 + c^{12}$$

$$(-\theta_0 + c^2)^7 = -\theta_0^7 + 7c^2\theta_0^6 - 21c^4\theta_0^5 + 35c^6\theta_0^4 - 35c^8\theta_0^3 + 21c^{10}\theta_0^2 - 7c^{12}\theta_0 + c^{14}$$

New version of function c (2024-08-11)

### 1x1 matrix, order 3

eliminate w_i, maxLines=0, maxTheta=1, maxK=3

For eqs (equations) and eqSol (solutions), look at results 4.
Compared with results 5, we will now leave out some unused elements, such as the solution for c before replacing thetas. This should make it easier to do the calculations for one additional approximation.

checking out equation 0

$$s_0$$

solve for c, maxLines=0, maxTheta=1, maxK=3
new equation

$$s_0$$

replace square backets

$$-\theta_0 + c^2$$

replace thetas

$$-1 - 2m + \frac{m^2}{2} + c^2$$

solve for c**2
convert to series

$$c^2 = 1 + 2m - \frac{m^2}{2}$$

$$c = 1 + m - \frac{3m^2}{4} + \frac{3m^3}{4} + O(m^4)$$

calculate numeric result

$$c = 1.07634287606451$$

$$\frac{1}{n}\frac{d\omega}{dt} = 0.00416899864805931$$

### 3x3 matrix, order 6

eliminate w_i, maxLines=1, maxTheta=1, maxK=6
For eqs (equations) and eqSol (solutions), look at results 4.

checking out equation 0

$$s_0 - \left(\frac{1}{s_1} + \frac{1}{s_{-1}}\right)\theta_1^2$$

solve for c, maxLines=1, maxTheta=1, maxK=6
new equation

$$-s_1\theta_1^2 - s_{-1}\theta_1^2 + s_{-1}s_0 s_1$$

replace square backets

$$(-\theta_0 + c^2)((-2+c)^2 - \theta_0)((2+c)^2 - \theta_0) - \theta_1^2((2+c)^2 - \theta_0) - \theta_1^2((-2+c)^2 - \theta_0)$$

eliminate higher order terms. We can neglect $(c^2 - \theta_0)^3$ and $\theta_1^2(c^2 - \theta_0)$ because $c^2 - \theta_0 = s_0 = [0]$ is of order 3, so $(c^2 - \theta_0)^3$ is of order 9, and $\theta_1^2$ is of order 4, so $\theta_1^2(c^2 - \theta_0)$ is of order 7.

$$-8\theta_1^2 - 16\theta_0 + 8\theta_0^2 + 16c^2 - 8c^4$$

replace thetas

$$-8 + 32m^2 - 16m^3 - 448m^4 - 1455m^5 - \frac{9601m^6}{4} + 16c^2 - 8c^4$$

solve for c

$$c^2 = 1 - \frac{\sqrt{1024m^2 - 512m^3 - 14336m^4 - 46560m^5 - 76808m^6}}{16}$$

$$c^2 = 1 + \frac{\sqrt{1024m^2 - 512m^3 - 14336m^4 - 46560m^5 - 76808m^6}}{16}$$

convert to series

$$c^2 = 1 - 2m + \frac{m^2}{2} + \frac{225m^3}{16} + \frac{3135m^4}{64} + \frac{139973m^5}{1024} + \frac{1550723m^6}{4096} + O(m^7)$$

$$c^2 = 1 + 2m - \frac{m^2}{2} - \frac{225m^3}{16} - \frac{3135m^4}{64} - \frac{139973m^5}{1024} - \frac{1550723m^6}{4096} + O(m^7)$$

$$c = 1 - m - \frac{m^2}{4} + \frac{217m^3}{32} + \frac{3999m^4}{128} + \frac{207429m^5}{2048} + \frac{2256067m^6}{8192} + O(m^7)$$

$$c = 1 + m - \frac{3m^2}{4} - \frac{201m^3}{32} - \frac{2367m^4}{128} - \frac{111749m^5}{2048} - \frac{1378947m^6}{8192} + O(m^7)$$

First, we should look at Hill numeric perigee page 22 and neglect $(c^2 - \theta_0)^3$ and $\theta_1^2(c^2 - \theta_0)$. This is in our Hill Perigee Notes, yielding

$$c = 1 + m - \frac{3m^2}{4} - \frac{201m^3}{32} - \frac{2367m^4}{128} - \frac{111749m^5}{2048} - \frac{5007241m^6}{24576} + O(m^7) \quad (o31)$$

This agrees with Hill literal perigee up to $m^5$. The difference in $m^6$ comes from the difference in the main equation in the coefficient of $\theta_1^4$. This means that we calculate the numerical values based on one additional term.

calculate numeric result

$$c = 1.07160144326069$$

$$c = 0.922862293738814$$

$$\frac{1}{n}\frac{d\omega}{dt} = 0.0085576598945951$$

$$\frac{1}{n}\frac{d\omega}{dt} = 0.146169029850306$$

solve for c, maxLines=1, maxTheta=1, maxK=7

eliminate higher order terms

$$-8\theta_1^2 - 16\theta_0 + 2\theta_0\theta_1^2 + 8\theta_0^2 + 16c^2 - 2c^2\theta_1^2 - 8c^4$$

convert to series

$$c^2 = 1 - 2m + \frac{m^2}{2} + \frac{225m^3}{16} + \frac{2235m^4}{64} + \frac{85253m^5}{1024} + \frac{3784381m^6}{12288} + \frac{122194727m^7}{98304} + O(m^8)$$

$$c^2 = 1 + 2m - \frac{m^2}{2} - \frac{225m^3}{16} - \frac{3135m^4}{64} - \frac{139973m^5}{1024} - \frac{4915069m^6}{12288} - \frac{131312423m^7}{98304} + O(m^8)$$

$$c = 1 - m - \frac{m^2}{4} + \frac{217m^3}{32} + \frac{3099m^4}{128} + \frac{138309m^5}{2048} + \frac{5027773m^6}{24576} + \frac{133457143m^7}{196608} + O(m^8)$$

$$c = 1 + m - \frac{3m^2}{4} - \frac{201m^3}{32} - \frac{2367m^4}{128} - \frac{111749m^5}{2048} - \frac{4399741m^6}{24576} - \frac{42332413m^7}{65536} + O(m^8)$$

calculate numeric result

$$c = 0.922440836180662$$
$$c = 1.07158387041564$$
$$c = 0$$
$$\frac{1}{n}\frac{d\omega}{dt} = 0.146558961824117$$
$$\frac{1}{n}\frac{d\omega}{dt} = 0.00857202436239157$$
$$\frac{1}{n}\frac{d\omega}{dt} = 1$$

## 5x5 matrix, order 6

eliminate w_i, maxLines=2, maxTheta=2, maxK=6

For eqs (equations) and eqSol (solutions), look at results 4.

checking out equation 0

$$s_0 - \left(\frac{1}{s_1} + \frac{1}{s_{-1}}\right)\theta_1^2 + \left(\boxed{\frac{1}{s_{-1}s_1s_2} + \frac{1}{s_{-1}s_{-2}s_1}}\right)\theta_1^4 - 2\left(\frac{1}{s_1s_2} + \frac{1}{s_{-1}s_1} + \frac{1}{s_{-1}s_{-2}}\right)\theta_1^2\theta_2 - \left(\frac{1}{s_2} + \frac{1}{s_{-2}}\right)\theta_2^2$$

solve for c, maxLines=2, maxTheta=2, maxK=6

Note that equation 0 differs from Hill's version in the coefficients of $\theta_1^4$ (marked in yellow). Where we have $s_{-1}s_1s_2$, Hill has $s_{-1}{}^2s_2$. In order to improve our confidence in this result, we have calculated it manually. We also argue that each $s_i$ comes from eliminating $w_i$, and this is done only once, so the power of 2 is not plausible.

new equation

$$-s_1\theta_1^2 - s_{-1}\theta_1^2 + s_{-1}s_0s_1$$

replace square backets

$$(-\theta_0 + c^2)((-2+c)^2 - \theta_0)((2+c)^2 - \theta_0) - \theta_1^2((2+c)^2 - \theta_0) - \theta_1^2((-2+c)^2 - \theta_0)$$

expand

$$-8\theta_1^2 - 16\theta_0 + 2\theta_0\theta_1^2 + 8\theta_0^2 - \theta_0^3 + 16c^2 - 2c^2\theta_1^2 + 3c^2\theta_0^2 - 8c^4 - 3c^4\theta_0 + c^6$$

eliminate higher order terms
$$-8\theta_1^2 - 16\theta_0 + 8\theta_0^2 + 16c^2 - 8c^4$$

replace thetas
$$-8 + 32m^2 - 16m^3 - 448m^4 - 1455m^5 - \frac{9601m^6}{4} + 16c^2 - 8c^4$$

solve for c
$$c^2 = 1 - \frac{\sqrt{2}m\sqrt{128 - 64m - 1792m^2 - 5820m^3 - 9601m^4}}{8}$$

$$c^2 = 1 + \frac{\sqrt{2}m\sqrt{128 - 64m - 1792m^2 - 5820m^3 - 9601m^4}}{8}$$

convert to series
$$c^2 = 1 - 2m + \frac{m^2}{2} + \frac{225m^3}{16} + \frac{3135m^4}{64} + \frac{139973m^5}{1024} + \frac{1550723m^6}{4096} + O(m^7)$$

$$c^2 = 1 + 2m - \frac{m^2}{2} - \frac{225m^3}{16} - \frac{3135m^4}{64} - \frac{139973m^5}{1024} - \frac{1550723m^6}{4096} + O(m^7)$$

$$c = 1 - m - \frac{m^2}{4} + \frac{217m^3}{32} + \frac{3999m^4}{128} + \frac{207429m^5}{2048} + \frac{2256067m^6}{8192} + O(m^7)$$

$$c = 1 + m - \frac{3m^2}{4} - \frac{201m^3}{32} - \frac{2367m^4}{128} - \frac{111749m^5}{2048} - \frac{1378947m^6}{8192} + O(m^7)$$

calculate numeric result
$$c = 0.922862293738814$$
$$c = 1.07160144326069$$
$$c = 0$$
$$\frac{1}{n}\frac{d\omega}{dt} = 0.146169029850306$$
$$\frac{1}{n}\frac{d\omega}{dt} = 0.00855576598945951$$
$$\frac{1}{n}\frac{d\omega}{dt} = 1$$

This is the same as ours in Hill Perigee Notes.

5x5 matrix, order 7

solve for c, maxLines=2, maxTheta=2, maxK=7

new equation
$$-s_1\theta_1^2 - s_{-1}\theta_1^2 + s_{-1}s_0s_1$$

replace square backets
$$(-\theta_0 + c^2)((-2+c)^2 - \theta_0)((2+c)^2 - \theta_0) - \theta_1^2((2+c)^2 - \theta_0) - \theta_1^2((-2+c)^2 - \theta_0)$$

eliminate higher order terms
$$-8\theta_1^2 - 16\theta_0 + 2\theta_0\theta_1^2 + 8\theta_0^2 + 16c^2 - 2c^2\theta_1^2 - 8c^4$$

replace thetas

$$-8 + 32m^2 - 16m^3 - \frac{671m^4}{2} - \frac{1605m^5}{2} - \frac{6923m^6}{8} - \frac{1597m^7}{6} + 16c^2 - \frac{225c^2m^4}{2} - \frac{855c^2m^5}{2}$$
$$- \frac{5889c^2m^6}{8} - 742c^2m^7 - 8c^4$$

solve for c

$$c^2 = 1 - \frac{\sqrt{-256 + \left(16 - \frac{225m^4}{2} - \frac{855m^5}{2} - \frac{5889m^6}{8} - 742m^7\right)^2 + 1024m^2 - 512m^3 - 10736m^4 - 25680m^5 - 276\ldots}}{16}$$
$$- \frac{225m^4}{32} - \frac{855m^5}{32} - \frac{5889m^6}{128} - \frac{371m^7}{8}$$

$$c^2 = 1 + \frac{\sqrt{-256 + \left(16 - \frac{225m^4}{2} - \frac{855m^5}{2} - \frac{5889m^6}{8} - 742m^7\right)^2 + 1024m^2 - 512m^3 - 10736m^4 - 25680m^5 - 276\ldots}}{16}$$
$$- \frac{225m^4}{32} - \frac{855m^5}{32} - \frac{5889m^6}{128} - \frac{371m^7}{8}$$

convert to series

$$c^2 = 1 - 2m + \frac{m^2}{2} + \frac{225m^3}{16} + \frac{2235m^4}{64} + \frac{85253m^5}{1024} + \frac{3784381m^6}{12288} + \frac{122194727m^7}{98304} + O(m^8)$$

$$c^2 = 1 + 2m - \frac{m^2}{2} - \frac{225m^3}{16} - \frac{3135m^4}{64} - \frac{139973m^5}{1024} - \frac{4915069m^6}{12288} - \frac{131312423m^7}{98304} + O(m^8)$$

$$c^2 = 0$$

$$c = 1 - m - \frac{m^2}{4} + \frac{217m^3}{32} + \frac{3099m^4}{128} + \frac{138309m^5}{2048} + \frac{5027773m^6}{24576} + \frac{133457143m^7}{196608} + O(m^8)$$

$$c = 1 + m - \frac{3m^2}{4} - \frac{201m^3}{32} - \frac{2367m^4}{128} - \frac{111749m^5}{2048} - \frac{4399741m^6}{24576} - \frac{42332413m^7}{65536} + O(m^8)$$

$$c = 0$$

calculate numeric result

$$c = 0.922440836180662$$
$$c = 1.07158387041564$$
$$c = 0$$
$$\frac{1}{n}\frac{d\omega}{dt} = 0.146558961824117$$
$$\frac{1}{n}\frac{d\omega}{dt} = 0.00857202436239157$$
$$\frac{1}{n}\frac{d\omega}{dt} = 1$$

### 7x7 matrix, order 10

eliminate w_i, maxLines=3, maxTheta=3, maxK=10

For eqs (equations) and eqSol (solutions), look at results 4.

solve for c, maxLines=3, maxTheta=3, maxK=10

Here we note that $\theta_1^2$ has order 4, $\theta_1^4, 2\theta_1^2\theta_2, \theta_2^2$ have order 8, and
$\theta_1^6, 2\theta_1^4\theta_2, \theta_1^2\theta_2^2, \theta_3^2, 2\theta_1^3\theta_3, 2\theta_1\theta_2\theta_3$ have order 12.
In order to arrive at our new equation, we eliminate all terms of order 12 and keep all coefficients of $\theta_1^4$. The yellow coefficients differ from Hill's version and the green coefficients appear for the first time in this approximation.

new equation

$$s_1s_2s_3\theta_1^4 + s_{-3}s_2s_3\theta_1^4 - 2s_{-3}s_1s_2s_3\theta_1^2\theta_2 + s_{-2}s_{-3}s_3\theta_1^4 - 2s_{-2}s_{-3}s_2s_3\theta_1^2\theta_2 + s_{-2}s_{-3}s_1\theta_1^4 \\ - s_{-2}s_{-3}s_1s_2s_3\theta_1^2 + s_{-1}s_2s_3\theta_1^4 - s_{-1}s_{-3}s_1s_2s_3\theta_2^2 + s_{-1}s_{-2}s_{-3}\theta_1^4 \\ - 2s_{-1}s_{-2}s_{-3}s_3\theta_1^2\theta_2 - s_{-1}s_{-2}s_{-3}s_2s_3\theta_1^2 - s_{-1}s_{-2}s_{-3}s_1s_3\theta_2^2 \\ + s_{-1}s_{-2}s_{-3}s_0s_1s_2s_3$$

reformat manually

$$s_{-1}s_{-2}s_{-3}s_0s_1s_2s_3 - (s_{-2}s_{-3}s_1s_2s_3 + s_{-1}s_{-2}s_{-3}s_2s_3)\theta_1^2 \\ + (s_1s_2s_3 + s_{-3}s_2s_3 + s_{-2}s_{-3}s_3 + s_{-2}s_{-3}s_1 + s_{-1}s_2s_3 + s_{-1}s_{-2}s_{-3})\theta_1^4 \\ - 2(s_{-3}s_1s_2s_3 + s_{-2}s_{-3}s_2s_3 + s_{-1}s_{-2}s_{-3}s_3)\theta_1^2\theta_2 \\ - (s_{-1}s_{-3}s_1s_2s_3 + s_{-1}s_{-2}s_{-3}s_1s_3)\theta_2^2$$

Separate document.

### 7x7 matrix, order 11

solve for c, maxLines=3, maxTheta=3, maxK=11

Separate document.

### Alternative 7x7 matrix, order 10

Alternative: eliminate w_i, maxLines=3, maxTheta=3, maxK=10

For eqs (equations) and eqSol (solutions), look at results 4.

solve for c, maxLines=3, maxTheta=3, maxK=10

Separate document.

In order to arrive at our new equation, we eliminate all terms of order 12 and keep all the first two coefficients of $\theta_1^4$ (yellow coefficients differ from Hill's version) and discard the other four (the green coefficients that appear for the first time in this approximation).

new equation

$$s_2\theta_1^4 - 2s_1s_2\theta_1^2\theta_2 + s_{-2}\theta_1^4 - 2s_{-2}s_2\theta_1^2\theta_2 - s_{-2}s_1s_2\theta_1^2 - s_{-1}s_1s_2\theta_2^2 - 2s_{-1}s_{-2}\theta_1^2\theta_2 - s_{-1}s_{-2}s_2\theta_1^2 \\ - s_{-1}s_{-2}s_1\theta_2^2 + s_{-1}s_{-2}s_0s_1s_2$$

reformat manually

$$+s_{-1}s_{-2}s_0s_1s_2 - (s_{-2}s_1s_2 + s_{-1}s_{-2}s_2)\theta_1^2 + \text{\colorbox{yellow}{$(s_2 + s_{-2})\theta_1^4$}} - 2(s_1s_2 + s_{-2}s_2 + s_{-1}s_{-2})\theta_1^2\theta_2 \\ - (s_{-1}s_1s_2 + s_{-1}s_{-2}s_1)\theta_2^2$$

### Alternative 7x7 matrix, order 11

solve for c, maxLines=3, maxTheta=3, maxK=11

Separate document.

## Summary of numeric values

| method | perigee | |
|---|---|---|
| maxLines=0, maxTheta=1, maxK=3 | 1.0763428761 | 0.0041689986 |
| Observation (according to Hill) | | 0.008452 |
| maxLines=1, maxTheta=1, maxK=6 | 1.0716014433 | 0.0085557660 |
| maxLines=2, maxTheta=2, maxK=6 | 1.0716014433 | 0.0085557660 |
| maxLines=1, maxTheta=1, maxK=7 | 1.0715838704 | 0.0085720244 |
| maxLines=2, maxTheta=2, maxK=7 | 1.0715838704 | 0.0085720244 |
| Hill's calculation (literal perigee) | | 0.008572573 |
| Hill's calculation (numeric perigee) | | 0.008572573 |
| maxLines=3, maxTheta=3, maxK=10 | | 0.008579088<br>0.008579347 |
| maxLines=3, maxTheta=3, maxK=11 | 1.0715759535 | 0.0085793491 |
| Alt: maxLines=3, maxTheta=3, maxK=10 | | 0.008579451<br>0.008579446 |
| Alt: maxLines=3, maxTheta=3, maxK=11 | 1.0715759652<br>1.0715758431 | 0.008579338<br>0.008579451 |
| Hill's estimate | | 0.008591 |

# A4. Appendix 4 – Solve for $c$ and $m^{11}$

Function cSolve – cubic equation

## 7x7 matrix, order 11

This document is maxLines=3, maxTheta=3, maxK=11, normal
Previous file name: Hill-Perigee-Results-3-3-11.docx

Here, the equation 0 for order 11 leads to a cubic equation in $c^2$ and is solved using the trigonometric method.

coefficients of non-reduced polynomial (estimates using m=config.mEarth)

$$a = 626688 - 466944m + 166912m^2 - 25088m^3 - 2220944m^4 - 5371008m^5 - 7585552m^6$$
$$- \frac{23850152m^7}{3} - \frac{444079561m^8}{36} - \frac{7983096391m^9}{216} - \frac{625435419923m^{10}}{6480}$$
$$- \frac{35253284387077m^{11}}{194400}$$

$$a = 589898.071738499$$

$$b = -3649536 + 884736m + 1545216m^2 - 1184256m^3 + 9012912m^4 + 42155424m^5$$
$$+ 72222600m^6 + 78948664m^7 + \frac{1503552725m^8}{12} + \frac{28718558867m^9}{72}$$
$$+ \frac{2384926666549m^{10}}{2160} + \frac{144363435183911m^{11}}{64800}$$

$$b = -3567978.53578651$$

$$c = 5419008 - 368640m - 2558976m^2 - 1316352m^3 - 10336944m^4 - 70790592m^5$$
$$- 163300512m^6 - 200897256m^7 - \frac{1167121947m^8}{4} - \frac{7863838087m^9}{8}$$
$$- \frac{239275843909m^{10}}{80} - \frac{46942328358593m^{11}}{7200}$$

$$c = 5371044.25476846$$

$$d = -2396160 - 49152m + 7924736m^2 - 6026752m^3 - 89229424m^4 - 196328288m^5$$
$$- 181379032m^6 - \frac{11426248m^7}{3} + \frac{3842914237m^8}{36} + \frac{14024108299m^9}{216}$$
$$+ \frac{813714695317m^{10}}{1296} + \frac{113187154453487m^{11}}{38880}$$

$$d = -2356059.61123989$$

check plausibility: this should be zero

$$plaus = 0$$

$$aNum = 589898.071738499$$

$$bNum = -3567978.53578651$$

$$cNum = 5371044.25476846$$

$$dNum = -2356059.61123989$$

coefficients of reduced polynomial

$$A = -\frac{99}{17} - \frac{846m}{289} + \frac{54109m^2}{29478} + \frac{18424m^3}{751689} - \frac{4198100735m^4}{613378224} + \frac{81320586391m^5}{41709719232}$$
$$+ \frac{751722575307223m^6}{25526348169984} + \frac{8728204637390743m^7}{162730469583648}$$
$$+ \frac{20858549457458219495m^8}{265576126360513536} + \frac{18251103641990874842015m^9}{72236706370059681792}$$
$$+ \frac{2070022421973935930915 49163m^{10}}{22104432149238262 6283520}$$
$$+ \frac{14320041199195637741486 9861977m^{11}}{563663019805575697 02297600}$$

$$A = -6.04846617888402$$

$$B = \frac{147}{17} + \frac{1692m}{289} - \frac{29833m^2}{14739} - \frac{3624494m^3}{751689} + \frac{3475047011m^4}{306689112} - \frac{176351776663m^5}{20854859616}$$
$$- \frac{1562453891448175m^6}{12763174084992} - \frac{21395193831458323m^7}{81365234791824}$$
$$- \frac{109003568496397303219m^8}{265576126360513536} - \frac{433750899552101 44163735m^9}{36118353185029840896}$$
$$- \frac{48207341071096116724 2023419m^{10}}{1105221607461913131417 60}$$
$$- \frac{34102036641537370595 9947022221m^{11}}{2818315099027878485 1148800}$$

$$B = 9.10503782386652$$

$$C = -\frac{65}{17} - \frac{846m}{289} + \frac{338485m^2}{29478} - \frac{326642m^3}{751689} - \frac{97792200359m^4}{613378224} - \frac{19797102710633m^5}{41709719232}$$
$$- \frac{16114988246226041m^6}{25526348169984} - \frac{55897431046891547m^7}{162730469583648}$$
$$- \frac{31777002681814676281m^8}{66394031590128384} - \frac{244015984271395629545425m^9}{72236706370059681792}$$
$$- \frac{44992887472681209627 0820945m^{10}}{44208864298476525 256704}$$
$$- \frac{195494612882718892744 620569059m^{11}}{112732603961115139 40459520}$$

$$C = -3.99401137653237$$

check plausibility: this should be zero

$$plaus = 0$$
$$ANum = -6.04846617869297$$
$$BNum = 9.10503782278752$$
$$CNum = -3.9940113794512$$

$$p = -\frac{768}{289} - \frac{27072m}{4913} + \frac{562720m^2}{250563} - \frac{14624200m^3}{12778713} - \frac{42533852783m^4}{2606857452} - \frac{1897684547407m^5}{132949730052}$$
$$+ \frac{36995509838267m^6}{9040581643536} + \frac{41355009184511m^7}{86450561966313} - \frac{237156346840083664453m^8}{4514794148128730112}$$
$$- \frac{2854135261682226193 2025m^9}{230254501554565235712} - \frac{11559137921652674596766237m^{10}}{5871489789641413 5106560}$$
$$- \frac{51238646532905056939259 15587m^{11}}{9981532642390402968115200}$$

$$p = -3.08960988211272$$

$$q = -\frac{8192}{4913} - \frac{433152m}{83521} + \frac{45660160m^2}{4259571} - \frac{692647936m^3}{217238121} - \frac{213560837312m^4}{1231016019}$$
$$- \frac{287444641694213m^5}{565036352721} - \frac{249255662231617313m^6}{345802247865252}$$
$$- \frac{3241902018017128327m^7}{5878638213709284} - \frac{47703919850664 0889493m^8}{532996531376308416}$$
$$- \frac{32974536838878017986 94917m^9}{733936223705176688832}$$
$$- \frac{68921746823119294234539 27931m^{10}}{49907663211952014840 5760}$$
$$- \frac{7755196857383157107089 21370387m^{11}}{282810091534394750 76326400}$$

$$q = -2.02771008986013$$

p must be negative

check plausibility: this should be zero

$$plaus = 0$$

$$pNum = -3.0896098821434$$

$$qNum = -2.02771009369001$$

$$Disc = -\frac{786432m^2}{83521} - \frac{35389440m^3}{1419857} + \frac{4132093952m^4}{24137569} + \frac{2907032121344m^5}{3693048057}$$
$$+ \frac{173153781830144m^6}{188345450907} - \frac{333918266116480m^7}{3201872665419} + \frac{3201920988725891056m^8}{489886517809107}$$
$$+ \frac{14741331980984822252m^9}{308447066768697} + \frac{74196972843959175855 3995m^{10}}{5096779331285949228}$$
$$+ \frac{69164737784801672205 2123941m^{11}}{2599357458955834106280}$$

$$Disc = -0.0644111574326681$$

$$DiscNum = -0.0644111202032174$$

α1 = (9/4)*q/(p**2)
α2 = sqrt(-(4/3)*p)

$$\alpha_1 = -\frac{17}{32} + \frac{141m}{256} + \frac{725m^2}{288} - \frac{1581131m^3}{110592} + \frac{8966461m^4}{1327104} - \frac{8627315177m^5}{63700992}$$
$$+ \frac{171699270187m^6}{764411904} - \frac{5728567796147m^7}{9172942848} + \frac{395295869907401m^8}{220150628352}$$
$$- \frac{24599313171391673m^9}{5283615080448} + \frac{766487099465477351m^{10}}{79254226206720}$$
$$- \frac{68130110502310514281m^{11}}{2377626786201600}$$

$$\alpha_2 = \frac{32}{17} + \frac{564m}{289} - \frac{319609m^2}{176868} + \frac{164443007m^3}{72162144} + \frac{300154163803m^4}{117768619008} + \frac{24561482314217m^5}{5338844061696}$$
$$- \frac{2019872624155174 21m^6}{39208470789095424} + \frac{104022754069193110943m^7}{15997056081950932992}$$
$$- \frac{197463215600450706 85357m^8}{52214391051487845285888} + \frac{35891452669140313191 4718543m^9}{7101157183002346958880768}$$
$$+ \frac{90753125307360152 8125035621021m^{10}}{86918163919948726776700600320}$$
$$+ \frac{396730973202192618802317852 19953457m^{11}}{17731305439669540262446922 4652800}$$

$$\alpha_1 = -0.477948489885453$$

$$\alpha_2 = 2.02965017746998$$

$$\alpha 1 Num = -0.477948485836843$$

$$\alpha 2 Num = 2.02965017745864$$

the following must be between -1 and 1 ($arg = \alpha_1 * \alpha_2$):

$$arg = -1 + \frac{867m^2}{128} - \frac{12393m^3}{512} - \frac{649831m^4}{32768} - \frac{124581731m^5}{589824} + \frac{14200641281m^6}{113246208}$$
$$- \frac{25578912841m^7}{50331648} + \frac{241589534597669m^8}{173946175488} - \frac{2019731412177637m^9}{521838526464}$$
$$+ \frac{2581283623490644391m^{10}}{601157982486528} - \frac{80638403236591 4837207m^{11}}{36069478949191680}$$

$$arg = -0.970068231127763$$

$$argNum = -0.970068229094834$$

calculate arcsin
α = arcsin(arg)
Starting here, we display floating-point coefficients.

$$\alpha = -1.5707963267949 + 3.68060796608386m - 6.57638040998808m^2$$
$$- 9.18573299378651m^3 - 88.6478237539325m^4 - 117.668197577857m^5$$
$$- 695.887238731462m^6 - 1550.52951880806m^7 - 9003.57951335859m^8$$
$$- 30727.7775730884m^9 - 156887.129144476m^{10} - 587800.245832013m^{11}$$

$$\alpha = -1.3255112035197$$

$$\alpha Num = -1.32551162359817$$

calculate the three solutions for Y

$$sol_1 = -0.5 + 1.0625m - 1.52213541666667m^2 - 4.26296657986111m^3$$
$$- 24.8857789216218m^4 - 48.9381783803304m^5 - 177.17569465647m^6$$
$$- 540.070531883339m^7 - 2367.24519696651m^8 - 9719.94454489321m^9$$
$$- 43442.2006948374m^{10} - 177445.977568016m^{11}$$

$$sol_1 = -0.427600841645389$$

$$sol_2 = 1.0 + 1.55078425594432 \cdot 10^{-17}m - 0.752604166666666m^2 + 2.689453125m^3$$
$$+ 1.44825688114871m^4 + 28.8663034321349m^5 - 20.481088097694m^6$$
$$+ 101.982710240718m^7 - 399.08146881301m^8 + 647.496170668718m^9$$
$$- 2881.9425421264m^{10} + 7438.82687621927m^{11}$$

$$sol_2 = 0.996659385755349$$

$$sol_3 = -0.5 - 1.0625m + 2.27473958333333m^2 + 1.5735134548611m^3$$
$$+ 23.4375220404731m^4 + 20.0718749481955m^5 + 197.656782754164m^6$$
$$+ 438.087821642619m^7 + 2766.32666577953m^8 + 9072.44837422441m^9$$
$$+ 46324.1432369641m^{10} + 170007.150691795m^{11}$$

$$sol_3 = -0.56905854410996$$

$$sol1Num = -0.427600962159898$$
$$sol2Num = 0.996659384631507$$
$$sol3Num = -0.569058422471608$$

sol1 = α2*sol1

$$sol_1 = -0.941176470588235 + 1.0242214532872m + 0.111857430400073m^2$$
$$- 14.054329954498m^3 - 51.2657845070755m^4 - 136.04248963102m^5$$
$$- 390.173043101294m^6 - 1357.24032857753m^7 - 5369.42278304507m^8$$
$$- 22596.819913533m^9 - 98222.7329996352m^{10} - 408903.338592038m^{11}$$

$$sol_1 = -0.867880101761458$$

$$sol_2 = 1.88235294117647 + 1.95155709342561m - 3.22371486080015m^2$$
$$+ 5.87254879788494m^3 + 11.8834208660028m^4 + 55.1884466617307m^5$$
$$+ 14.2234874210821m^6 + 113.029718285326m^7 - 429.833295356065m^8$$
$$+ 321.053879215539m^9 - 3106.29584699083m^{10} + 6509.83548346616m^{11}$$

$$sol_2 = 2.02286989765472$$

$$sol_3 = -0.941176470588234 - 2.9757785467128m + 3.11185743040007m^2$$
$$+ 8.18178115661309m^3 + 39.3823636410727m^4 + 80.8540429692887m^5$$
$$+ 375.949555680212m^6 + 1244.2106102922m^7 + 5799.25607840115m^8$$
$$+ 22275.7660343173m^9 + 101329.028846626m^{10} + 402393.503108569m^{11}$$

$$sol_3 = -1.15498979589326$$

$$sol1Num = -0.867880368729321$$
$$sol2Num = 2.02286989688315$$
$$sol3Num = -1.15498952815383$$

check plausibility: this should be zero

$$plaus_1 = -1.77635683940025 \cdot 10^{-15}m - 7.105427357601 \cdot 10^{-15}m^2 + 5.6843418860808$$
$$\cdot 10^{-14}m^4 + 1.13686837721616 \cdot 10^{-13}m^5 + 1.81898940354586 \cdot 10^{-12}m^6$$
$$- 1.81898940354586 \cdot 10^{-12}m^7 + 4.72937244921923 \cdot 10^{-11}m^8$$
$$- 1.01863406598568 \cdot 10^{-10}m^9 + 1.28056854009628 \cdot 10^{-9}m^{10}$$
$$- 2.3283064365387 \cdot 10^{-9}m^{11}$$

$$plaus_2 = 2.66453525910038 \cdot 10^{-15} + 5.32907051820075 \cdot 10^{-15}m + 6.21724893790088 \\ \cdot 10^{-15}m^2 + 4.08562073062058 \cdot 10^{-14}m^3 - 1.4210854715202 \cdot 10^{-13}m^4 \\ - 7.67386154620908 \cdot 10^{-13}m^5 + 5.6843418860808 \cdot 10^{-14}m^6 \\ + 1.59161572810262 \cdot 10^{-12}m^7 + 5.6843418860808 \cdot 10^{-12}m^8 \\ + 7.76481101638637 \cdot 10^{-11}m^9 + 5.59339241590351 \cdot 10^{-11}m^{10} \\ - 3.25826476910152 \cdot 10^{-10}m^{11}$$

$$plaus_3 = -5.32907051820075 \cdot 10^{-15}m - 2.04281036531029 \cdot 10^{-14}m^2 + 3.5527136788005 \\ \cdot 10^{-14}m^3 + 1.70530256582424 \cdot 10^{-13}m^4 + 9.09494701772928 \cdot 10^{-13}m^5 \\ + 5.22959453519434 \cdot 10^{-12}m^6 + 5.45696821063757 \cdot 10^{-12}m^7 \\ + 3.63797880709171 \cdot 10^{-12}m^8 + 1.01863406598568 \cdot 10^{-10}m^9 \\ - 2.3283064365387 \cdot 10^{-10}m^{10} + 6.98491930961609 \cdot 10^{-10}m^{11}$$

calculate solution of cubic equation

convert to original variable X ($X = Y - \frac{A}{3}$).

check plausibility: this should be zero

$$plaus_1 = -4.65661287307739 \cdot 10^{-10} - 1.86264514923096 \cdot 10^{-9}m - 1.86264514923096 \\ \cdot 10^{-9}m^2 - 1.49011611938477 \cdot 10^{-8}m^3 - 4.76837158203125 \cdot 10^{-7}m^5 \\ + 9.5367431640625 \cdot 10^{-7}m^6 + 1.52587890625 \cdot 10^{-5}m^8 - 6.103515625 \\ \cdot 10^{-5}m^9 + 0.000732421875m^{10} - 0.00244140625m^{11}$$

$$plaus_2 = 7.45058059692383 \cdot 10^{-9} + 1.49011611938477 \cdot 10^{-8}m^2 + 1.49011611938477 \\ \cdot 10^{-8}m^3 - 5.96046447753906 \cdot 10^{-8}m^4 - 3.57627868652344 \cdot 10^{-7}m^5 \\ + 4.76837158203125 \cdot 10^{-7}m^6 + 2.02655792236328 \cdot 10^{-6}m^7 \\ + 1.9073486328125 \cdot 10^{-6}m^8 + 5.340576171875 \cdot 10^{-5}m^9 \\ - 0.00022125244140625m^{11}$$

$$plaus_3 = -2.3283064365387 \cdot 10^{-10} - 4.65661287307739 \cdot 10^{-9}m - 1.30385160446167 \\ \cdot 10^{-8}m^2 + 3.72529029846191 \cdot 10^{-8}m^3 + 9.68575477600098 \cdot 10^{-8}m^4 \\ + 5.96046447753906 \cdot 10^{-7}m^5 + 2.74181365966797 \cdot 10^{-6}m^6 \\ + 9.5367431640625 \cdot 10^{-7}m^7 - 1.9073486328125 \cdot 10^{-6}m^8 + 4.57763671875 \\ \cdot 10^{-5}m^9 + 0.0001220703125m^{10} + 0.00030517578125m^{11}$$

convert solution for c**2 to series

$$sol_1 = 1.0 + 2.0m - 0.499999999999999m^2 - 14.0625m^3 - 48.984375m^4 \\ - 136.6923828125m^5 - 399.989339192708m^6 - 1375.11898125543m^7 \\ - 5395.60303469056m^8 - 22681.0388864029m^9 - 98534.8909729017m^{10} \\ - 409750.182699009m^{11}$$

$$sol_1 = 1.14827529119988$$

$$sol_2 = 3.82352941176471 + 2.92733564013841m - 3.83557229120022m^2 \\ + 5.86437875238296m^3 + 14.1648303730782m^4 + 54.5385534802505m^5 \\ + 4.4071913296686m^6 + 95.1510656074298m^7 - 456.013547001555m^8 \\ + 236.834906345556m^9 - 3418.45382025733m^{10} + 5662.99137649511m^{11}$$

$$sol_2 = 4.03902529061606$$

$$sol_3 = 1.0 - 2.0m + 2.5m^2 + 8.17361111111112m^3 + 41.6637731481482m^4 \\ + 80.2041497878085m^5 + 366.133259588798m^6 + 1226.33195761431m^7 \\ + 5773.07582675566m^8 + 22191.5470614474m^9 + 101016.87087336m^{10} \\ + 401546.659001598m^{11}$$

$$sol_3 = 0.861165597068084$$

$$sol1Num = 1.14827502416834$$

$$sol2Num = 4.03902528978081$$

$$sol3Num = 0.861165864743827$$

insert solutions into equation

check plausibility: this should be zero

first, the original equation

$$Plaus_1 = -4.65661287307739 \cdot 10^{-10} - 1.86264514923096 \cdot 10^{-9}m - 1.86264514923096 \\ \cdot 10^{-9}m^2 - 1.49011611938477 \cdot 10^{-8}m^3 - 4.76837158203125 \cdot 10^{-7}m^5 \\ + 9.5367431640625 \cdot 10^{-7}m^6 + 1.52587890625 \cdot 10^{-5}m^8 - 6.103515625 \\ \cdot 10^{-5}m^9 + 0.000732421875m^{10} - 0.00244140625m^{11}$$

$$Plaus_1 = -6.37659902310186 \cdot 10^{-10}$$

$$Plaus_2 = 7.45058059692383 \cdot 10^{-9} + 1.49011611938477 \cdot 10^{-8}m^2 + 1.49011611938477 \\ \cdot 10^{-8}m^3 - 5.96046447753906 \cdot 10^{-8}m^4 - 3.57627868652344 \cdot 10^{-7}m^5 \\ + 4.76837158203125 \cdot 10^{-7}m^6 + 2.02655792236328 \cdot 10^{-6}m^7 \\ + 1.9073486328125 \cdot 10^{-6}m^8 + 5.340576171875 \cdot 10^{-5}m^9 \\ - 0.00022125244140625m^{11}$$

$$Plaus_2 = 7.55226564602548 \cdot 10^{-9}$$

$$Plaus_3 = -2.3283064365387 \cdot 10^{-10} - 4.65661287307739 \cdot 10^{-9}m - 1.30385160446167 \\ \cdot 10^{-8}m^2 + 3.72529029846191 \cdot 10^{-8}m^3 + 9.68575477600098 \cdot 10^{-8}m^4 \\ + 5.96046447753906 \cdot 10^{-7}m^5 + 2.74181365966797 \cdot 10^{-6}m^6 \\ + 9.5367431640625 \cdot 10^{-7}m^7 - 1.9073486328125 \cdot 10^{-6}m^8 + 4.57763671875 \\ \cdot 10^{-5}m^9 + 0.0001220703125m^{10} + 0.00030517578125m^{11}$$

$$Plaus_3 = -6.67862921962778 \cdot 10^{-10}$$

$$Plaus1Num = -1.86264514923096 \cdot 10^{-9}$$

$$Plaus2Num = -5.12227416038513 \cdot 10^{-9}$$

$$Plaus3Num = -4.65661287307739 \cdot 10^{-10}$$

then the reduced form

$$Plaus_1 = 8.88178419700125 \cdot 10^{-16}m - 7.105427357601 \cdot 10^{-15}m^2 - 2.8421709430404 \\ \cdot 10^{-14}m^3 + 1.36424205265939 \cdot 10^{-12}m^6 - 2.72848410531878 \cdot 10^{-12}m^7 \\ + 4.00177668780088 \cdot 10^{-11}m^8 - 1.30967237055302 \cdot 10^{-10}m^9 \\ + 1.04773789644241 \cdot 10^{-9}m^{10} - 3.02679836750031 \cdot 10^{-9}m^{11}$$

$$Plaus_1 = 1.0725896330793 \cdot 10^{-17}$$

$$Plaus_2 = 5.32907051820075 \cdot 10^{-15} + 1.77635683940025 \cdot 10^{-14}m + 6.21724893790088 \\ \cdot 10^{-15}m^2 + 2.75335310107039 \cdot 10^{-14}m^3 - 1.13686837721616 \cdot 10^{-13}m^4 \\ - 4.2632564145606 \cdot 10^{-13}m^5 + 5.11590769747272 \cdot 10^{-13}m^6 \\ + 2.27373675443232 \cdot 10^{-12}m^7 + 4.77484718430787 \cdot 10^{-12}m^8 \\ + 7.76481101638637 \cdot 10^{-11}m^9 + 4.13820089306682 \cdot 10^{-11}m^{10} \\ - 3.25826476910152 \cdot 10^{-10}m^{11}$$

$$Plaus_2 = 6.81431061479013 \cdot 10^{-15}$$

$$Plaus_3 = -5.32907051820075 \cdot 10^{-15}m - 2.04281036531029 \cdot 10^{-14}m^2 \\ + 3.5527136788005 \cdot 10^{-14}m^3 + 1.70530256582424 \cdot 10^{-13}m^4 \\ + 9.09494701772928 \cdot 10^{-13}m^5 + 5.22959453519434 \cdot 10^{-12}m^6 \\ + 5.45696821063757 \cdot 10^{-12}m^7 + 3.63797880709171 \cdot 10^{-12}m^8 \\ + 1.01863406598568 \cdot 10^{-10}m^9 - 2.3283064365387 \cdot 10^{-10}m^{10} \\ + 6.98491930961609 \cdot 10^{-10}m^{11}$$

$$Plaus_3 = -5.33572617408437 \cdot 10^{-16}$$

$$Plaus1Num = 0$$
$$Plaus2Num = -4.44089209850063 \cdot 10^{-15}$$
$$Plaus3Num = 8.88178419700125 \cdot 10^{-16}$$

check if c**2 is negative for m-earth

$$c^2 = 1.14827529119988$$
$$c^2 = 4.03902529061606$$
$$c^2 = 0.861165597068084$$

convert solution for c to series

$$c = 1.0 + 1.0m - 0.75m^2 - 6.28125m^3 - 18.4921875m^4 - 54.56494140625m^5 \\ - 179.025919596354m^6 - 665.611329820423m^7 - 2680.17616469771m^8 \\ - 11293.0834607446m^9 - 48964.6126025335m^{10} - 203292.296203175m^{11}$$

$$c = 1.95538472218761 + 0.748531889127021m - 1.1240428112076m^2 \\ + 1.92983571074577m^3 + 2.56017930320913m^4 + 14.0750403411534m^5 \\ - 3.74166913667155m^6 + 31.3270903897775m^7 - 146.314772337405m^8 \\ + 119.842348133255m^9 - 1080.77297913867m^{10} + 2060.98639335335m^{11}$$

$$c = 1.0 - 0.999999999999997m + 0.750000000000001m^2 + 4.83680555555556m^3 \\ + 25.3874421296296m^4 + 61.8619128568671m^5 + 214.190617062917m^6 \\ + 658.166040093482m^7 + 2762.58583794809m^8 + 10758.2207393619m^9 \\ + 48660.0956118288m^{10} + 198043.275487354m^{11}$$

calculate numeric result

$$c = 1.07157608626447$$
$$c = 2.00973264167593$$
$$c = 0.927990095914759$$
$$\frac{1}{n}\frac{d\omega}{dt} = 0.00857922624873186$$
$$\frac{1}{n}\frac{d\omega}{dt} = -0.8594019745153$$
$$\frac{1}{n}\frac{d\omega}{dt} = 0.141424794078265$$

purely numeric solution

$$c = 1.07157595352282$$
$$c = 2.0097326413682$$
$$c = 0.927990228797603$$
$$\frac{1}{n}\frac{d\omega}{dt} = 0.008579349061132$$
$$\frac{1}{n}\frac{d\omega}{dt} = -0.859401974230588$$
$$\frac{1}{n}\frac{d\omega}{dt} = 0.141424671135235$$